\documentclass[
  paper=A4, 	
  pagesize, 		
  DIV=calc, 	
  headings=small,
  english, 
  12pt, 		
  listof=totoc,  
  index=totoc,
  BCOR5mm, 			
]{scrbook}

\setcapindent{0pt}
\usepackage[T1]{fontenc}	
\usepackage[utf8]{inputenc}	
\usepackage{fixltx2e}  		
\usepackage[english]{babel} 
\usepackage[english]{translator}
\usepackage{xcolor}
\usepackage{tensor}
\usepackage{bm}
\usepackage{gensymb}
\usepackage{graphicx} 	
\usepackage[pdftex,a4paper,breaklinks,bookmarksnumbered]{hyperref}  %
\usepackage{acronym}
\usepackage{comment}
\usepackage{dirtytalk}
\usepackage{amssymb}
\usepackage{romannum}
\usepackage[section]{placeins}
\usepackage{longtable}
\usepackage{amsmath}
\usepackage{cleveref}
\usepackage{tabularx}
\usepackage{cite} 
\usepackage{yfonts}
\usepackage[title]{appendix}
\usepackage{lscape}

\usepackage{
  ellipsis, 
  ragged2e, 
  marginnote,
}
\usepackage[tracking=true]{microtype}%
\DeclareMicrotypeSet*[tracking]{my}%
  { font = */*/*/sc/* }%
\SetTracking{ encoding = *, shape = sc }{ 45 }
\usepackage{%
  lmodern, 
}

\graphicspath{{./gfx/}}

\newcommand{\roberto}[1]{\color{black}{#1}\color{black}}
\newcommand{\robert}[1]{\color{black}{#1}\color{black}}

\DeclareMathOperator{\rank}{rank}
\DeclareMathOperator{\range}{range}
\newcommand{\lm}{{\ell,m}}
\newcommand{\IMRP}{\texttt{IMRPhenomPv2}}
\newcommand{\IMRPD}{\texttt{IMRPhenomD}}
\newcommand{\IMRPHM}{\texttt{IMRPhenomHM}}
\newcommand{\SEOBP}{\texttt{SEOBNRv3}}
\newcommand{\SEOBA}{\texttt{SEOBNRv4}}
\newcommand{\SEOBHM}{\texttt{SEOBNRv4HM}}
\newcommand{\NRSur}{\texttt{NRSur7dq2}}
\begin{document}
\pdfinfo{                               
    /Author (Dein Name)
    /CreationDate (D:20130310111111)    
                                        %
                                        %
    /ModDate (D:20130310111111)         
    /Creator (TeX \& TXC)               
    /Producer (pdfTeX)                  
    /Title () 													
    /Subject ()                         
    /Keywords ()                        
}
\frontmatter
\pagenumbering{Roman}
\thispagestyle{empty}
\begin{center}
		
    \includegraphics[width=3cm]{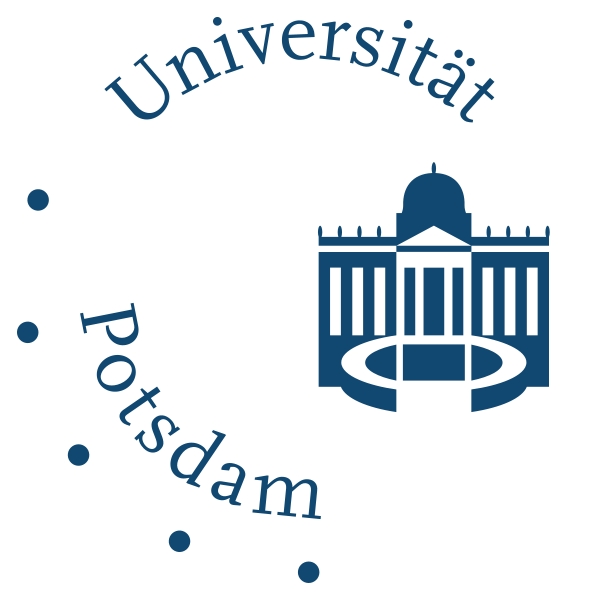}\\
    \vspace{.5cm}
    
    {\Huge Universität Potsdam}\\
    \vspace{1cm}
    
    {\Large Institut für Physik und Astronomie\\[1mm]}
    
    \vspace{2cm}
    {\Large \textbf{Multipolar gravitational waveforms for spinning binary black holes and their impact on source characterization}}\\
    \vspace*{3mm}
    
    \vspace{1.0cm}
    {\normalsize \textbf{Dissertation}}\\ \vspace{0.5cm}
    {\normalsize zur Erlangung des wissenschaftlichen Grades}\\
    {\normalsize doctor rerum naturalium}\\
    {\normalsize (Dr. rer. nat.)}\\
    \vspace{1cm}
    
    \vspace{0.5cm}

    \parbox{1cm}{
      \begin{large}
        \begin{tabbing}
	        Kandidat: \hspace{1.5cm} \=Roberto Cotesta\\[2mm]
	    	  Gutachterin: \>Prof. Dr. Alessandra Buonanno\\
	    		
        \end{tabbing}
      \end{large}
    }\\
    
    \vspace{5cm} 
\end{center}

\thispagestyle{empty}
\cleardoublepage
    
\thispagestyle{empty}
\cleardoublepage

\begin{center}
    \textbf{Acknowledgements}
\end{center}
\vspace*{1cm}
First and foremost, I would like to thank my supervisor, Alessandra Buonanno, for her guidance and patience during the course of my PhD.  Thanks to her encouragement and her attention to details I have been able to produce my highest-quality work. This thesis would not have been possible without her support.

I would also like to thank my collaborators: Stanislav Babak, Alejandro Boh\'{e}, Juan Calder\'{o}n Bustillo, Katerina Chatziioannou, Sudarshan Ghonge, Ian Harry, Ian Hinder, Jacob Lange, Lionel London, Sylvain Marsat, Ken K.-Y. Ng, Frank Ohme, Serguei Ossokine, Harald Pfeiffer, Michael P{\"u}rrer, Andrea Taracchini and Salvatore Vitale. My gratitude also goes to the members of the LIGO Scientific and Virgo collaboration for the interesting discussions during meetings and teleconferences. 

A special thanks goes to the wondeful members of the Astrophysical and Cosmological Relativity division of the Albert Einstein Institute (AEI) during my PhD studies. They are too many for me to list them here. Although I have learned a lot from each of them, my gratitude goes particularly to Alejandro Boh\'{e}, Sylvain Marsat, Serguei Ossokine, Michael P{\"u}rrer and Andrea Taracchini for helping me to understand the exciting field of gravitational-waveform modeling; and Ian Harry, Jonathan Gair, Serguei Ossokine, Harald Pfeiffer, Michael P{\"u}rrer and Vivien Raymond for helping me to understand the details of the data-analysis techniques used to analyze the data from LIGO and Virgo detectors.

I would like to thank my friend at the AEI: Andrea, Cristi\'{a}n, Hugo, Lorenzo C., Lorenzo S., Luca, Matteo, Mohammed, Nils, Niko, Noah, Ollie, Riccardo, Serena,  Stefano and Valentino. My PhD would have been much less fun without them. A special thanks also to Darya for her invaluable help.

I am also grateful to my friends: Assunta, Ciro, Danio, Emanuele, Fabrizio, Gianmarco, Giuseppe, Graziano, Leonardo, Lorenzo, Marco, Matteo, Mirko, Oliviero, Raffaele, Roberto, Sommo and Stefano. My life would have been much less fun without them.
Among my friends, a special thanks goes to Lorenzo C., Ollie and Marco who spend their time reading this thesis.

Last but not least, I would like to thank my parents, my family and my partner for all their unconditional support during my PhD, and in general during my entire life. In particular, I would like to thank my dad for feeding me with science for my entire childhood, my mum for constantly reminding me about the important things in life, and my partner for patiently listening to my rants during the PhD and conforting me when needed.

This thesis is dedicated to my grandmother Erasmina and my uncle Pietro who passed away during my PhD.


\begin{center}
    \textbf{Abstract}
\end{center}
\vspace*{1cm}
\noindent 
In the last five years, gravitational-wave astronomy has gone from a purerly theoretical field into a thriving experimental science. Many gravitational-wave signals, emitted by stellar-mass binary black holes and binary neutron stars, have been detected, and many more are expected in the future as consequence of the planned upgrades in the gravitational-wave detectors.
The observation of the gravitational-wave signals from these systems, and the characterization of their sources, heavily relies on the precise models for the emitted gravitational waveforms. To take full advantage of the increased detector sensitivity, it is then necessary to also improve the accuracy of the gravitational-waveform models.

In this work, I present an updated version of the waveform models for spinning binary black holes within the effective-one-body formalism. This formalism is based on the notion that the solution to the relativistic two-body problem varies smoothly with the mass ratio of the binary system, from the equal-mass regime to the test-particle limit. For this reason, it provides an elegant method to combine, under a unique framework, the solution to the relativistic two-body problem in different regimes. The main two regimes that are combined under the effective-one-body formalism are the slow-motion, weak field limit (accessible through the post-Newtonian theory), and the extreme mass-ratio regime (\robert{described } using the black-hole-perturbation theory). This formalism is nevertheless flexible enough to integrate information about the solution to the relativistic two-body problem obtained using other techniques, such as numerical relativity.

The novelty of the waveform models presented in this work is the inclusion of beyond-quadupolar terms in the waveforms emitted by spinning binary black holes. In fact, while the time variation of the source quadupole moment is the leading contribution to the waveforms emitted by binary black holes observable by LIGO and Virgo detectors, beyond-quadupolar terms can be important for binary systems with asymmetric masses, large total mass, or observed with large inclination angle with respect to the orbital angular momentum of the binary. For this purpose, I combine the approximate analytic expressions of these beyond-quadupolar terms, with their calculations from numerical relativity, to develop \robert{an accurate } waveform model including inspiral, merger and ringdown for spinning binary black holes. I first construct this model in the simplified case of black holes with spins aligned with the orbital angular momentum of the binary, then I extend it to the case of generic spin orientations. Finally, I test the accuracy of both these models against a large number of waveforms obtained from numerical relativity. \roberto{The waveform models I present in this work are the state of the art for spinning binary black holes, without restrictions in the allowed values for the masses and the spins of the system. }

The measurement of the source properties of a binary system emitting gravitational waves requires to compute $\mathcal{O}(10^7-10^9)$ different waveforms. Since the waveform models mentioned before can require $\mathcal{O}(1-10)$s to generate a single waveform, they can be difficult \robert{to use in data-analysis studies } \roberto{given the increasing number of sources observed by the LIGO and Virgo detectors}. To overcome this obstacle, I use the reduced-order-modeling technique to develop a faster version of the waveform model for black holes with spins aligned to the orbital angular momentum of the binary. This version of the model is as accurate as the original and reduces the time for evaluating a waveform by two orders of magnitude.

The waveform models developed in this thesis have been used \robert{by the LIGO and Virgo collaborations } in the \roberto{inference of the source parameters } of the gravitational-wave signals detected during the second observing run (O2), and first half of the third observing run (O3a) of LIGO and Virgo detectors. Here, I present a study on the source properties of the signals GW170729 and GW190412, for which I have been directly involved in the analysis. \roberto{In addition, these models have been used by the LIGO and Virgo collaborations to perform tests on General Relativity employing the gravitational-wave signals detected during O3a, and to analyze the population of the observed binary black holes. } 

\newpage
I declare that this thesis is an original report of my research, has been written by me and has not been submitted for any previous degree. 
\newpage

\tableofcontents 
\newpage

\section*{List of Acronyms}
\begin{acronym}[ICANN]

\acro{ADM}[ADM]{Arnowitt-Deser-Misner}
\acrodef{aLIGO}{Advanced Laser interferometer Gravitational-Wave Observatory}
\acrodef{AZDHP}{aLIGO zero detuned high power density}
\acro{BBH}[BBH]{binary black hole}
\acro{BH}[BH]{black hole}
\acro{BNS}{binary neutron star}
\acro{EOB}[EOB]{effective-one-body}
\acro{FTA}[FTA]{Flexible Theory Agnostic}
\acro{GR}[GR]{General relativity}
\acro{GW}[GW]{gravitational wave}
\acro{HM}[HM]{higher-order mode}
\acro{LAL}[LAL]{LIGO Algorithm Library}
\acrodef{LSC}[LSC] {LIGO Scientific Collaboration}
\acro{NR}[NR]{numerical relativity}
\acro{NS}{neutron star}
\acro{NSBH}{neutron star black hole}
\acrodef{ODE}{ordinary differential equation}
\acrodef{PDE}{partial differential equation}
\acrodef{PE}{parameter estimation}
\acro{PM}[PM]{post-Minkowskian}
\acro{PN}[PN]{post-Newtonian}
\acro{PSD}{power spectral density}
\acro{PTA}[PTA]{pulsar-timing array}
\acro{QNM}[QNM]{quasi-normal mode}
\acro{ROM}[ROM]{reduced-order model}
\acro{SNR}[SNR]{signal-to-noise ratio}
\acro{SPA}[SPA]{stationary phase approximation}
\acro{SVD}[SVD]{single value decomposition}
\acrodef{TPI}{tensor-product interpolant}

\end{acronym}

\mainmatter 
\pagenumbering{arabic}

\chapter{Introduction}
\label{chap:introduction}


\section{Introduction to general relativity}
\label{sec:GR_intro}
\ac{GR} was proposed by Albert Einstein in two seminal papers~\cite{Einstein:1915by,Einstein:1915ca} as an attempt to overcome two fundamental issues in Newtonian gravity. The first problem dates back to the 19th century, when Urbain Le Verrier found a discrepancy between the observed motion of Mercury and the Newtonian prediction~\cite{VerrierThorieDM}. 
The second problem is theoretical, and arose when Einstein tried to incorporate Newton's theory of gravitation into the framework of special relativity. In fact, Newton's theory is inconsistent with special relativity as it implies the instantaneous influence of one body on another. Both these problems are solved by \ac{GR}. From the experimental side, the equations of motion for Mercury's orbit include correction terms, with respect to the Newtonian equations, which resolve the discrepancy with the observations. In addition to this, other \ac{GR} predictions have been confirmed by multiple experiments~\cite{Will:2014kxa} over the years. From the theoretical perspective, \ac{GR} is a local theory, hence action at a distance is not possible. 

In \ac{GR}, spacetime is not a static and absolute entity, as in the case of the Newtonian theory, but rather deformed by the presence of matter and energy.
In this framework, gravity is not considered as a force between two massive objects, but as spacetime curvature. The relation between mass-energy content in a system, and resulting spacetime curvature, is given by Einstein's field equations
\begin{equation}
\label{eq:einstein_eq}
R\indices{_\mu_\nu} - \frac{1}{2}g\indices{_\mu_\nu}R = \frac{8\pi G}{c^4} T\indices{_\mu_\nu}.
\end{equation}
The quantities $R\indices{_\mu_\nu}$ and $R$ are the Ricci tensor and scalar respectively; they are functions of the metric tensor $g\indices{_\mu_\nu}$, which describes spacetime geometry. $T\indices{_\mu_\nu}$ is the energy-momentum tensor, which accounts for the mass-energy content in a system. A solution to these equations is a metric tensor $g\indices{_\mu_\nu}$, describing a spacetime geometry, whose curvature depends on the mass-energy content of the system.

The equations of motion of a point particle in a generic spacetime $g\indices{_\mu_\nu}$ can be found using the equivalence principle, one of the foundational ideas in \ac{GR}. From the equivalence principle, there always exists a local inertial frame, where the equations of motion of freely falling bodies take the same form they would take in the absence of gravity. Mathematically, this translates into the equation
\begin{equation}
\label{eq:free_falling}
\frac{d^2\tilde{x}^\mu}{d\tau^2} = 0,
\end{equation}
where $\tilde{x}^\mu(\tau)$ are the coordinates of a point particle in the locally inertial frame and $\tau$ is an affine parameter of the worldline. The metric tensor in this frame is locally flat, namely $g\indices{_\mu_\nu} \approx \eta_{\mu \nu} = \textrm{diag}(-1, 1, 1, 1)$ in the spacetime region close to the origin of the frame. Starting from this local inertial frame, one can obtain the equations of motion in any generic frame $x^\nu = x^\nu(\tilde{x}^\nu)$, by applying the coordinate transformation to Eq.~\eqref{eq:free_falling}. The result is the well-known geodesic equation
\begin{equation}
\label{eq:geodesic}
\frac{d^2x^\alpha}{d\tau^2} + \Gamma\indices{^\alpha_\mu_\nu}\frac{dx^\mu}{d\tau}\frac{dx^\nu}{d\tau} = 0,
\end{equation}
which is the general relavitistic generalization of the Newtonian law of gravitation.
The quantities $\Gamma\indices{^\alpha_\mu_\nu}$ are called \roberto{affine connections}: they depend on the derivatives of $g\indices{_\mu_\nu}$ in the new coordinates, and carry the gravitational effects (as well as other inertial forces). 

General relativity predicts a plethora of phenomena that are absent in Newtonian theory. I will now introduce two of them that are relevant for this thesis: \textit{\acp{BH}} and \textit{\acp{GW}}.

The first exact and non-trivial solution to Einstein's field equations was found by Karl Schwarzschild~\cite{Schwarzschild:1916ae,Schwarzschild:1916uq}, just one year after the publication of Einstein's first article on \ac{GR}. This solution describes the gravitational field outside a spherically-symmetric non-rotating body.
An interesting property of this solution appears when the radius of the body is smaller than a characteristic length called the Schwarzschild radius $R_s = 2GM/c^2$, where $M$ is the mass of the body. In this case the Schwarzschild solution features an event horizon, which is the defining property of a \ac{BH}, the so-called Schwarzschild \ac{BH}. 
The event horizon is a hypersurface that divides two regions of the spacetime: the interior and the exterior of the \ac{BH}. The gravity is so strong in the interior that no particles nor radiation can escape from it.  Generalizations to the Schwarzschild \ac{BH} were found by: Hans Reissner and Gunnar Nordstr{\"o}m~\cite{reissner_h_1916_1447315,1918KNAB...20.1238N} in the case of a non-rotating \ac{BH} with an electric charge; Roy Kerr in the case of a rotating \ac{BH}~\cite{Kerr:1963ud} (Kerr \ac{BH}); and finally by Ezra Newman in the case of an electrically-charged rotating \ac{BH}~\cite{Newman:1965my}. Theoretical arguments, usually referred to as the \say{no-hair conjecture}~\cite{Israel:1967wq,Israel:1967za,Carter:1971zc}, suggest that isolated \acp{BH} are very simple objects that can be completely characterized by their mass, angular momentum and electric charge. This is in stark contrast to regular stars, for which many more properties need to be given to characterize them, even approximately.
\acp{BH} are also very compact objects, the ratio between their mass $M$ and their radius $R \approx R_\mathrm{s}$ is $M/R_\mathrm{s} = c^2/(2 G) \approx 10^6 M_\odot/R_\odot$. There is observational evidence for two different classes of astrophysical \acp{BH} depending on their mass: supermassive~\cite{Schodel:2003gy,Ghez:2003qj,Barth:2003qu,Willott:2005pj} and stellar-mass \acp{BH}~\cite{1972Natur.235..271B,Webster:1972bsw,doi:10.1146/annurev.astro.44.051905.092532,Abbott:2016blz,Abbott:2016nmj, Abbott:2017vtc,
Abbott:2017gyy, Abbott:2017oio, LIGOScientific:2018mvr,Abbott:2020tfl,LIGOScientific:2020stg,Abbott:2020khf,Abbott:2020niy}. Supermassive \acp{BH} are approximately in the range $10^5 \lesssim M \lesssim 10^9 M_\odot$. Their formation remains an open question, but the most accredited theory suggests that they grow by accretion and mergers with other \robert{\acp{BH}}, starting from a \say{seed} \ac{BH} with lower mass (see Ref.~\cite{2010A&ARv..18..279V} for a review on the subject). Stellar-mass \acp{BH} have masses approximately in the range $1 \lesssim M \lesssim 100 M_\odot$, and are thought to form as a result of the gravitational collapse of a star (see Refs.~\cite{Oppenheimer:1939ue,Penrose:1969pc,May:1966zz}  for more details). Some theories also predict the existence of a class of \acp{BH} that is not a product of stellar evolution: the so-called primordial \acp{BH}.  They are formed as a result of fluctuations or phase transitions in the early universe (see Ref.~\cite{1967SvA....10..602Z,Hawking:1971ei,1975ApJ...201....1C,Sasaki:2018dmp}). The expected mass spectrum for this class of \acp{BH} is broader than the other two, but no observations have so far confirmed their existence. For the work of this thesis, I focus on stellar-mass \acp{BH}, as they are sources for ground-based \ac{GW} detectors. Observations indicate that stellar-mass \acp{BH} have non-negligible angular momentum, while there are strong theoretical arguments suggesting that they have negligible electric charge~\cite{Gibbons:1975kk,Eardley:1975kp}. For this reason, in the rest of the thesis I will consider \acp{BH} as neutral in charge.

The second \ac{GR} prediction that is widely used in this thesis are \acp{GW}. \roberto{They are perturbations of the gravitational field that propagate as waves \robert{far from the source}, and carry both energy and momentum. They were predicted by Einstein~\cite{Einstein:1916cc,1918SPAW.......154E} one year after the publication of \ac{GR}.
The first evidence for the existence of \acp{GW} was obtained by Russell Hulse and Joseph Taylor, by measuring the variation in the time of arrival of \robert{radio } pulses from binary pulsars~\cite{Hulse:1974eb}. This variation is consistent with that predicted by \ac{GR} when the orbit of the system shrinks as a consequence of the energy and angular momentum lost by \robert{the emission of } \acp{GW}. The existence of \acp{GW} has been recently confirmed by the LIGO Scientific Collaboration~\cite{Abbott:2016blz}, which, using laser-interferometry techniques, was able to measure the \ac{GW} signal passing through Earth emitted by a binary black-hole merger (see Sec.~\ref{sec:strategy_detection} for more details on this topic).}

The equations for the generation and propagation of \acp{GW} can be perturbatively derived by linearizing the Einstein's field equations~\eqref{eq:einstein_eq} around the flat metric $\eta\indices{_\mu_\nu},$\footnote{Linearization about a generic metric tensor is also possible but it requires a more careful treatment (see \robert{sections } 1.4.1 and 1.4.2 in Ref.~\cite{maggiore2008gravitational}). For simplicity here I use the flat metric as background.} using the ansatz $g\indices{_\mu_\nu} = \eta\indices{_\mu_\nu}+h\indices{_\mu_\nu} + \mathcal{O}(h^2)$ with $|h\indices{_\mu_\nu}|\ll 1$. When written as a function of ${\bar{h}}\indices{_\mu_\nu} \equiv h\indices{_\mu_\nu} - \frac{1}{2}\eta\indices{_\mu_\nu} \eta\indices{^\alpha^\beta} h\indices{_\alpha_\beta}$, and using the Lorenz gauge ($\partial\indices{_\mu} {\bar{h}}\indices{^\mu^\nu} = 0$), the linearized Einstein equations are
\begin{equation}
\label{eq:wave_eq_GW}
\Box {\bar{h}}\indices{_\mu_\nu} = -\frac{16\pi G}{c^2} T\indices{_\mu_\nu} + \robert{\mathcal{O}(\bar{h}^2)}.
\end{equation}
The $\Box$ symbol in the equation above is the d'Alambertian operator in flat spacetime, \roberto{and $T\indices{_\mu_\nu}$ is the stress-energy tensor associated with the source of \acp{GW}}. The $10$ degrees of freedom of the symmetric tensor $h\indices{_\mu_\nu}$ are reduced to 6 in ${\bar{h}}\indices{_\mu_\nu} $, by imposing the 4 conditions defining the Lorenz gauge. By computing the coordinate divergence of Eqs.~\eqref{eq:wave_eq_GW}, and using the Lorenz gauge conditions, it is straightforward to derive the equations
\robert{
\begin{equation}
\label{eq:cons_T_linear}
 \partial^\mu T_{\mu\nu} = 0 + \mathcal{O}(\bar{h}^2),
\end{equation}
namely the conservation of the energy-momentum tensor in the linearized theory}. 

In vacuum $(T_{\mu\nu} = 0)$, \roberto{the Lorenz gauge conditions } do not fix the gauge completely. In fact, Eq.~\eqref{eq:wave_eq_GW} and the Lorenz gauge conditions are invariant under the coordinate transformation $x'^\mu = x^\mu + \zeta^\mu(x)$ with $\Box \zeta^\mu = 0$. One can use this additional freedom to impose other $4$ conditions on $h\indices{_\mu_\nu}$, and reduce the number of degrees of freedom to $2$. A very common choice for these 4 conditions is $h\indices{^\mu^0} = 0$ and $h\indices{^i_i} = 0$  ($i = 0,1,2$) that, together with the 4 conditions of the Lorenz gauge, define the transverse-traceless (or TT) gauge. The metric tensor pertubation in the TT-gauge $h_{i j}^\mathrm{TT}$ can be computed directly from $h\indices{_\mu_\nu}$ by using the projection operator $\Lambda_{ij,kl}(\bm{\hat{N}})$ (see Eq. (1.36) in Ref.~\cite{maggiore2008gravitational} for its definition) where $\bm{\hat{N}} = \bm{x}/r$ is the direction of propagation of the \ac{GW}. For example, the metric tensor in the TT-gauge associated with a \ac{GW} propagating in the $z$ direction is
\begin{equation}
\label{eq:projection_operator}
h_{ij}^\mathrm{TT} = \Lambda_{ij}^{kl}(\bm{\hat{N}} = \bm{\hat{z}})h_{kl} =  \Lambda_{ij}^{kl}(\bm{\hat{N} = \bm{\hat{z}}})\bar{h}_{kl} = \begin{pmatrix} 0 & 0 & 0 & 0\\ 0 & h_+ & h_\times & 0\\
0 & h_\times & -h_+ & 0\\ 0 & 0 & 0 & 0 \end{pmatrix},
\end{equation}
where $h_+$ and $h_\times$ are the two physical degrees of freedom or \ac{GW} polarizations, usually referred to as plus and cross polarizations respectively.
The power radiated as consequence of the \ac{GW} emission can be easily computed in the TT-gauge as (see Sec.1.4.3 in Ref.~\cite{maggiore2008gravitational} for the details on the calculation)
\begin{equation}
\label{eq:GW_power}
\frac{dE_\mathrm{GW}}{dt} = \frac{D_\mathrm{L}^2c^3}{32\pi G}\int d\Omega \left\langle \sum_{i,j} \dot{h}_{ij}^\mathrm{TT} \dot{h}_{ij}^\mathrm{TT} \right\rangle,
\end{equation}
where $D_\mathrm{L}$ is the luminosity distance between source and the observer, the integral is performed over the solid angle, and the average $\langle\cdot\rangle$ is computed over several characteristic periods of the \ac{GW}.

Using the method of Green's functions, one can formally write a solution to Eq.~\eqref{eq:wave_eq_GW} as:
\begin{equation}
\label{eq:Green_gw_sol}
{\bar{h}}\indices{_\mu_\nu}(t, \bm{x}) = \frac{4G}{c^4}\int d^3x' \frac{1}{|\bm{x -x'}|} T_{\mu \nu}\left(t -  \frac{|\bm{x -x'}|}{c}, \bm{x'}\right) + \roberto{\mathcal{O}(G^2)}.
\end{equation}
When the observer of the \ac{GW} signal is at a distance $D_\mathrm{L} \equiv |\bm{x}| \gg d$, where $d$ is the typical size of the \ac{GW} source, the term $|\bm{x -x'}|$ can be expanded as $|\bm{x -x'}| = D_\mathrm{L} - \bm{x'}\cdot\bm{\hat{n}} + \mathcal{O}(d^2/D_\mathrm{L})$. At leading order in $d/D_\mathrm{L}$ the energy-momentum tensor is then
\begin{align}
\label{eq:PN_expansion_stress_energy}
T_{\mu \nu}\left(t -  \frac{D_\mathrm{L}}{c} + \frac{\bm{x'}\cdot\bm{\hat{n}}}{c}, \bm{x'}\right) &\approx T_{\mu \nu}\left(t -  \frac{D_\mathrm{L}}{c}, \bm{x'}\right) + \frac{\bm{x'}\cdot\bm{\hat{n}}}{c} \partial_t T_{\mu \nu}\left(t -  \frac{D_\mathrm{L}}{c}, \bm{x'}\right)+ \nonumber \\
&+\left(\frac{\bm{x'}\cdot\bm{\hat{n}}}{c}\right)^2 \partial_t^2 T_{\mu \nu}\left(t -  \frac{D_\mathrm{L}}{c}, \bm{x'}\right) + ...
\end{align}
where each time derivative of $T_{\mu \nu}$ carries a factor $v/d$, with $v$ being the typical source velocity. The energy-momentum tensor is now a series in $v/c$. At leading order in $d/D_\mathrm{L}$ and $v/c$, and when using the TT-gauge, Eq.~\eqref{eq:Green_gw_sol} is then
\begin{align}
h_{ij}&^\mathrm{TT}(t, r) = \frac{4G}{c^4D_\mathrm{L}}\Lambda_{ij}^{kl}(\bm{\hat{n}})\left[\int d^3x' T_{k l}\left(t -  \frac{D_\mathrm{L}}{c}, \bm{x'}\right) + \mathcal{O}\left(\frac{v}{c}\right)^3\right] + \roberto{\mathcal{O}(G^2)} \label{eq:Green_gw_sol_exp} \\
& = \frac{2G}{c^4D_\mathrm{L}}\Lambda_{ij}^{kl}(\bm{\hat{n}})\left[\frac{1}{c^2}\frac{d^2}{dt^2}\int d^3x' T_{0 0}\left(t -  \frac{D_\mathrm{L}}{c}, \bm{x'} \right)x'_k x'_l + \mathcal{O}\left(\frac{v}{c}\right)^3\right] + \roberto{\mathcal{O}(G^2)} \label{eq:Green_gw_sol_exp_2} \\
& = \frac{2G}{c^4D_\mathrm{L}}\Lambda_{ij}^{kl}(\bm{\hat{n}})\left[\frac{d^2}{dt^2}\roberto{Q_{k l}\left(t -  \frac{D_\mathrm{L}}{c}, \bm{x'} \right)} + \mathcal{O}\left(\frac{v}{c}\right)^3\right] + \roberto{\mathcal{O}(G^2)}, \label{eq:Green_gw_sol_exp_3}
\end{align}
where Eq.~\eqref{eq:Green_gw_sol_exp_2} is obtained integrating by parts Eq.~\eqref{eq:Green_gw_sol_exp} two times, and employing the identity $\partial_\mu T^{\mu \nu} = 0$. The quantity $Q_{k l}$, used in Eq.~\eqref{eq:Green_gw_sol_exp_3}, is the quadrupole moment of the source, defined as
\begin{equation}
\label{eq:quad_moment} 
Q_{k l} = \int d^3x' \frac{T_{0 0}}{c^2}x'_k x'_l.
\end{equation}
For this reason, Eq~\eqref{eq:Green_gw_sol_exp_3} is called \textit{quadrupole formula}. 

Since Eq.~\eqref{eq:Green_gw_sol_exp_3} contains a second time-derivative of the quadupole moment, one should not expect a \ac{GW} emission from objects for which the quadupole moment is constant over time, as for example spherical or axisymmetric stationary distributions of matter. This consideration already allows one to exclude isolated stars as source of \acp{GW}, as they are approximately spherically symmetric\footnote{Real stars are not exactly spherically symmetric, therefore they are expected to emit \acp{GW} with a magnitude proportional to their deformation~\cite{1971ApJ...166..175I}. Since the latter is expected to be small (see Ref.~\cite{Lasky:2015uia} for a review on the mechanism originating the deformation), the \ac{GW} emission from these systems is difficult to detect. However, there are good chances to observe such \ac{GW} signals in the future, because the sensitivity of the experiments increases with the observation time\robert{, and it will increase even more with future ground-based \ac{GW} detectors, like Einstein Telescope~\cite{Punturo:2010zz} and Cosmic Explorer~\cite{Reitze:2019iox}.}}. On the other hand, the quadupole moment of binary systems is not constant, therefore they are the natural candidate as \ac{GW} source.

It is instructive to use the dimensional analysis in Eq.~\eqref{eq:Green_gw_sol_exp_3} to estimate the expected \ac{GW} amplitude as a function of the parameters of the binary system. A binary system with total mass $M$, and typical binary separation and velocity respectively $d$ and $v$, has a quadupole moment $Q_{\mu\nu} \propto M d^2$. Considering that every time derivative gives a factor $v/d$, with $v^2\propto GM/d$, one can derive that dimensionally Eq.~\eqref{eq:Green_gw_sol_exp_3} reads
\begin{equation}
\left|h_{ij}^\mathrm{TT}\right| \propto \roberto{\frac{G^2}{c^4} \frac{1}{D_\mathrm{L}} \frac{M^2}{d} = \left(\frac{GM}{c^2D_\mathrm{L}}\right) \left(\frac{GM}{c^2d} \right)}.
\end{equation}
Since the \ac{GW} amplitude is suppressed by the factor $G^2/c^4 \approx 10^{-12} (\mathrm{R_\odot}/\mathrm{M_\odot})^2$, the only hope to detect \acp{GW} is by using \roberto{compact objects, i.e. systems with large total mass and small radius, to be able to reach separations $d$ of the order of $GM/c^2$}. As discussed before, \acp{BH} are extremely compact objects, therefore binary black-hole (BBH)\acused{BBH} systems are perfect candidates for detecting \acp{GW}. \roberto{For a \ac{BBH} system with masses $\sim 10 \, M_\odot$ located in the Virgo cluster, $(GM/c^2D_\mathrm{L}) \sim 10^{-20}$. Since, for \acp{BBH}, $(GM/c^2d)$ can reach values close to $1$, also $|h_{ij}^\mathrm{TT}| \sim 10^{-20}$.}

Models for the \ac{GW} signal emitted by \ac{BBH} systems are crucial for their detection and the source characterization.
The main goals of the work summarized in this thesis are (i) improving these models by including the effect on the waveforms of higher-order corrections to the quadupole formula and, (ii) testing the consequences of these improvements on the measurement of \acp{BBH} properties. 

In the next section of this chapter, I will outline the anatomy of the gravitational waveforms emitted by \ac{BBH} systems and summarize the methods used for their detection and characterization. Finally, in the remaining two sections, I will give a comprehensive introduction to the work discussed in this thesis.

\section{Binary black-holes as sources of gravitational waves}
Binary systems composed of two stellar-mass \acp{BH} are the main sources of \acp{GW} for ground-based \ac{GW} detectors, such as LIGO~\cite{TheLIGOScientific:2014jea} and Virgo~\cite{TheVirgo:2014hva}. There are two canonical channels typically considered for the formation of these systems: isolated binary evolution~\cite{Postnov:2014tza} and dynamical formation~\cite{Benacquista:2006ika}. In the first case, the progenitor of the \ac{BBH} system is a binary system composed of two massive stars that collapse into \acp{BH} during the final stage of their lifetime. In the second formation scenario, the \ac{BBH} mergers originate from the dynamical interactions in globular clusters or nuclear star clusters.

In Sec.~\ref{sec:waveform_anatomy}, I describe the \ac{GW} signal emitted by \ac{BBH} systems and discuss the main techniques used for its computation. In Sec.~\ref{sec:strategy_detection}, I summarize the experimental methods and data analysis approaches used for the detection of these \ac{GW} signals. Finally, in Sec.~\ref{sec:source_characterization}, I introduce the methods adopted to measure the parameters of the \ac{BBH} system from the \ac{GW} signal. 

\subsection{Anatomy of the gravitational waveforms}
\label{sec:waveform_anatomy}

The coalescence of a \ac{BBH} system is conventionally divided into three different regimes: \textit{inspiral}, \textit{merger} and \textit{ringdown}.

\subsubsection{Inspiral}
The \textit{inspiral} begins with the two \acp{BH} well separated. Here, I outline their motion and the emitted \acp{GW} in this regime. While in principle the two \acp{BH} could follow a generic elliptic orbit, it is common to approximate their motion with a \robert{quasi-}circular trajectory. This approximation is motivated by the fact that ground-based detectors are currently only able to observe the latest stage of the inspiral. This is when the orbital eccentricity has been reduced as a consequence of the \ac{GW} emission during the earlier inspiral phase (see Sec.4.1.3 in Ref.~\cite{maggiore2008gravitational} for the explicit calculation of this effect)\footnote{Within the dynamical formation scenario it is possible that $5-10 \%$ of the \ac{BBH} systems have non-negligible eccentricity even at the frequencies for which the \ac{GW} signal is observable by ground-base detectors~\cite{Samsing:2013kua,Samsing:2017xmd}.}. For this reason, throughout this thesis, I will restrict my focus on \ac{BBH} systems in \robert{quasi-}circular orbits.

\begin{figure}[h!]
\centering
\includegraphics[angle=0,width=\linewidth]{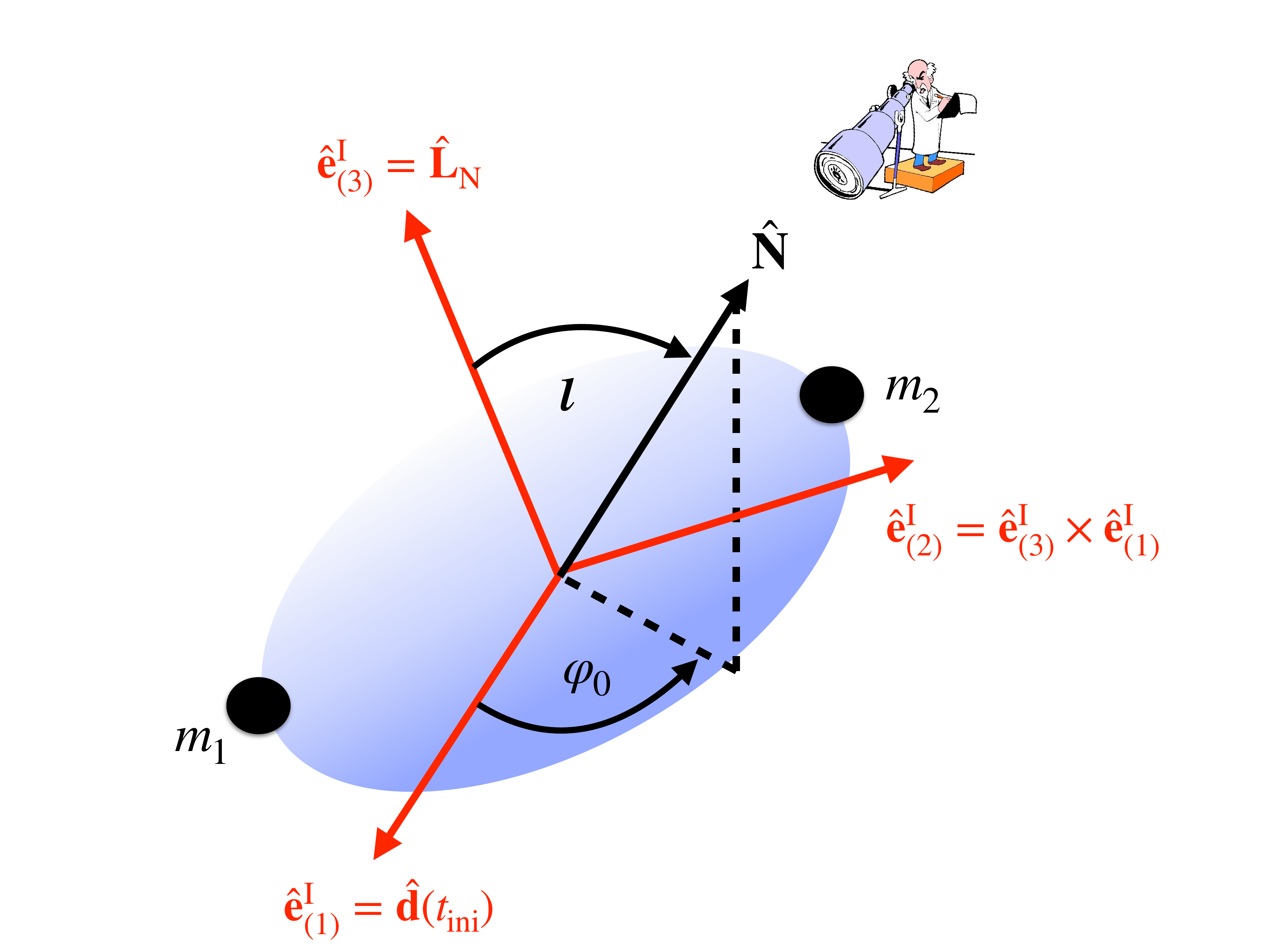}
\caption{The inertial frame, defined by the directions of the Newtonian orbital angular momentum $\bm{\hat{L}}_N$ and separation
  $\bm{\hat{d}}(t_\mathrm{ini})$. In this frame, the observer is in the direction $\bm{\hat{N}}$, defined by the angles $\iota$ and $\varphi_0$.}
\label{fig:frame_easy}
\end{figure}

\robert{I begin by discussing systems in circular orbits, which serve as a baseline for the generalization to the case of interest of quasi-circular orbits. To describe the motion of the two \acp{BH} with masses $m_1$ and $m_2$ and relative distance and velocity respectively $\bm{d}$ and $\bm{v}$, I use the center of mass frame that I show in Fig.~\ref{fig:frame_easy}. This is an inertial frame, whose basis $\bm{\hat{e}}_{(3)}^\mathrm{I}$ is aligned with $\bm{\hat{L}_\mathrm{N}} \equiv \bm{L_\mathrm{N}}/|\bm{L_\mathrm{N}}|$, the direction of the Newtonian orbital angular momentum $\bm{L_\mathrm{N}} \equiv \mu\, \bm{d} \times \bm{v}$, where $\mu \equiv m_1m_2/(m_1+m_2)$ is the reduced mass. The basis $\bm{\hat{e}}_{(1)}^\mathrm{I}$ of this frame is conventionally defined as $\bm{\hat{d}}(t = t_\mathrm{ini}) \equiv \bm{d}(t = t_\mathrm{ini})/|\bm{d}(t = t_\mathrm{ini})|$, the direction of the separation between the two \acp{BH} at a conventionally chosen initial time $t_\mathrm{ini}$. The frame is completed by the basis $\bm{\hat{e}}_{(2)}^\mathrm{I} = \bm{\hat{e}}_{(3)}^\mathrm{I} \times \bm{\hat{e}}_{(1)}^\mathrm{I}$. In this frame, $\bm{\hat{N}}$ is the direction of an observer, defined by the angles $\iota$ and $\varphi_0$.
Under the assumption of circular orbits, and in the frame defined above, the equations of motion of the \ac{BBH} system at leading (Newtonian) order are}
\begin{equation}
\label{eq:circular_orbit}
\bm{x_1} = \frac{m_2 \, d}{M}\left\{\cos(\omega_\mathrm{orb}t), \sin(\omega_\mathrm{orb}t), 0 \right\},\qquad \bm{x_2} = -\frac{m_1}{m_2}\bm{x}_1,
\end{equation}
where $M = m_1+m_2$, $\omega_\mathrm{orb} = \sqrt{GM/d^3}$ and $d = |\bm{d}|$. 

At leading order, the \ac{GW} signal emitted by the binary during the inspiral can be calculated \robert{by inserting in Eq.~\eqref{eq:Green_gw_sol_exp_3} the quadrupole moment of the binary, computed using Eq.~\eqref{eq:quad_moment}}\footnote{\robert{The conservation of the energy-momentum tensor in Eq.~\eqref{eq:cons_T_linear}, used to derive Eq.~\eqref{eq:Green_gw_sol_exp_3}, implies that in the linearized theory self-gravitating sources of \acp{GW} move following geodesics in a flat spacetime, i.e. straight lines. To generalize Eq.~\eqref{eq:Green_gw_sol_exp_3} to self-gravitating systems moving in Newtonian orbits, one should include, in the derivation of this equation, the terms $\mathcal{O}(\hat{h}^2) \propto \mathcal{O}(G^2)$ in Eq.~\eqref{eq:cons_T_linear}. These terms give corrections at higher orders in $G$ in the waveform, compared to its leading order expression which I am considering here. See Sec. 4.2 in Ref.~\cite{Flanagan:2005yc} for more details.} }. 
The result is
\begin{align}
h_+(t) =& \frac{4}{D_\mathrm{L}}\left(\frac{G\mathcal{M}}{c^2}\right)^{5/3} \left(\frac{\pi f_{\mathrm{GW}}}{c}\right)^{2/3} \frac{1+(\bm{\hat{L}_\mathrm{N}}\cdot \bm{\hat{N}})^2}{2}\cos\left(\robert{2\pi t_\mathrm{ret} f_{\mathrm{GW}} + 2{\varphi}_0 + \Phi_\mathrm{GW}^\mathrm{c}} \right) \label{eq:quad_plus} \\
h_\times (t) =& \frac{4}{D_\mathrm{L}}\left(\frac{G\mathcal{M}}{c^2}\right)^{5/3} \left(\frac{\pi f_{\mathrm{GW}}}{c}\right)^{2/3}\bm{\hat{L}_\mathrm{N}}\cdot \bm{\hat{N}} \sin\left(\robert{2\pi t_\mathrm{ret} f_{\mathrm{GW}} + 2{\varphi}_0 + \Phi_\mathrm{GW}^\mathrm{c}} \right)\label{eq:quad_cross}
\end{align}
\robert{where $\mathcal{M} \equiv (m_1 m_2)^{3/5}/(m_1 + m_2)^{1/5}$ is the chirp mass, $D_\mathrm{L}$ the luminosity distance of the source from the observer, $t_\mathrm{ret} \equiv (t - D_\mathrm{L}/c)$ the retarded time and $\Phi_\mathrm{GW}^\mathrm{c}$ an integration constant}. The scalar product $\bm{\hat{L}_\mathrm{N}}\cdot \bm{\hat{N}}$ coincides with $\cos\iota$.
Under the approximations discussed before, the frequency of the \ac{GW} signal $f_\mathrm{GW} \equiv \omega_\mathrm{GW}/2\pi$ is $f_\mathrm{GW} = 2 f_\mathrm{orb}$, with $f_\mathrm{orb} \equiv \omega_\mathrm{orb}/2\pi$ being the orbital frequency.  The \ac{GW} signal predicted by Eqs.~\eqref{eq:quad_plus} and~\eqref{eq:quad_cross} implicitly assumes that the binary remains \roberto{at a fixed separation $d$. } This is not the case, since some of the energy of the binary is lost through the emission of \acp{GW}. Since the energy of the binary, at leading Newtonian order, is $E = -G\,m_1\,m_2/d$, and $d^3 = G\,M/\omega_\mathrm{orb}^2$, a lost of energy induces the binary to shrink and the orbital frequency to increase. As a consequence, the \ac{GW} frequency increases too. Its variation can be computed by inserting the expressions for $h_+$ and $h_\times$ in Eq.~\eqref{eq:GW_power}, to obtain the radiated power in \acp{GW}, then equating the latter with $-dE/dt$. The result is
\begin{equation}
\label{eq:GW_freq_evolution_0}
\dot{f}_\mathrm{GW} = \frac{96}{5}\pi^{8/3} \left( \frac{G\mathcal{M}}{c^3} \right)^{5/3}f_\mathrm{GW}^{11/3},
\end{equation}
which can be integrated to obtain the \ac{GW} frequency time evolution
\begin{equation}
\label{eq:GW_freq_evolution}
f_\mathrm{GW}(t) = \frac{1}{\pi} \left( \frac{5}{256} \frac{1}{t_\mathrm{c} - t} \right)^{3/8} \left( \frac{G\mathcal{M}}{c^3} \right)^{-5/8},
\end{equation}
with $t_\mathrm{c}$ being the coalescence time.
By computing the time derivative of the orbital energy and using Eq.~\eqref{eq:GW_freq_evolution_0}, one can also find the expression for the relative radial velocity of the two \acp{BH}
\begin{equation}
\label{eq:radial_velocity}
\dot{d} = -\frac{2}{3} (\omega_\mathrm{orb}d)\frac{\dot{\omega}_\mathrm{orb}}{\omega_\mathrm{orb}^2},
\end{equation}
with $\dot{\omega}_\mathrm{orb} = \dot{\omega}_\mathrm{GW}/2 = \pi\dot{f}_\mathrm{GW}$.

Since $|h_{+,\times}| \propto f_\mathrm{GW}^{2/3}$, the \ac{GW} emission will increase as a consequence of the growing \ac{GW} frequency, inducing an even larger loss of energy. This process over a long period of time leads to the coalescence of the system, if one can assume that the binary is on a circular orbit during every step of this process, and therefore Eq.~\eqref{eq:circular_orbit} holds. This assumption, also called \textit{adiabatic approximation}, is only true if the tangential velocity $\omega_\mathrm{orb}d$ of the \acp{BH} is much larger than their radial velocity $\dot{d}$. From Eq.~\eqref{eq:radial_velocity}, one can observe that the condition above is fulfilled as long as $\dot{\omega}_\mathrm{orb}/\omega_\mathrm{orb}^2 \sim \mathcal{O}(v^5/c^5) \ll 1$.

Under this approximation, the motion of the binary system can be described as a sequence of \textit{quasi-circular} orbits, whose equations of motion are
\begin{equation}
\label{eq:circular_orbit_RR}
\bm{x_1} = \frac{m_2 \, d(t)}{M}\left\{\cos(\Phi_\mathrm{orb}(t)), \sin(\Phi_\mathrm{orb}(t)), 0 \right\},\qquad \bm{x_2} = -\frac{m_1}{m_2}\bm{x}_1,
\end{equation}
where 
\begin{equation}
\Phi_\mathrm{orb}(t) = \int_{t_\mathrm{c}}^{t} dt' \omega_\mathrm{orb}(t') = \pi\int_{t_\mathrm{c}}^{t} dt' f_\mathrm{gw}(t') = - \left(\frac{5G \mathcal{M}}{c^3} \right)^{-5/8}(t_\mathrm{c} - t)^{5/8} + \Phi_\mathrm{orb}(t_\mathrm{c}),
\end{equation}
with $\Phi_\mathrm{orb}(t_\mathrm{c})$ being the coalescence phase. Similarly, the explicit expression of $d(t)$ can be computed by integrating Eq.~\eqref{eq:radial_velocity}.

The \ac{GW} signal from this system is obtained, as before, by inserting the quadrupole moment of the binary in Eq.~\eqref{eq:Green_gw_sol_exp_3}. \roberto{ The result of this calculation at leading order in $v/c$ and $G$ is }
\begin{align}
h_+(t) =& \frac{4}{D_\mathrm{L}}\left(\frac{G\mathcal{M}}{c^2}\right)^{5/3} \left[\frac{5}{c(t_\mathrm{c} - t)}\right]^{1/4} \frac{1+(\bm{L_\mathrm{N}}\cdot \bm{\hat{N}})^2}{2}\cos\left(\Phi_\mathrm{GW}(t) +2{\varphi}_0 \right) \label{eq:quad_plus_evol} \\
h_\times (t) =& \frac{4}{D_\mathrm{L}}\left(\frac{G\mathcal{M}}{c^2}\right)^{5/3}  \left[\frac{5}{c(t_\mathrm{c} - t)}\right]^{1/4}\bm{L_\mathrm{N}}\cdot \bm{\hat{N}} \sin\left(\Phi_\mathrm{GW}(t) +2{\varphi}_0 \right),\label{eq:quad_cross_evol}
\end{align}
\robert{where I redefined the coalescence time as $t_\mathrm{c} \rightarrow t_\mathrm{c} + D_\mathrm{L}/c$ to incorporate the delay due to the propagation of the wave from the source to the observer. The phase of the \ac{GW} signal, at this order in $G$ and $v/c$, is simply $\Phi_\mathrm{GW}(t) = 2\Phi_\mathrm{orb}(t)$, and its instantaneous frequency is $f_\mathrm{GW}(t) = 2 f_\mathrm{orb}(t)$, as before. It is important to highlight that, at leading order in $G$ and $v/c$, the phase $\Phi_\mathrm{GW}(t)$\footnote{The phase of the \ac{GW} signal is more relevant than the amplitude in data-analysis applications because \ac{GW} detectors are more sensitive to it. The reason will be clear when I will introduce \ac{GW} detectors in Sec.~\ref{sec:strategy_detection}.} depends on the parameters of the \ac{BBH} system only through the chirp mass $\mathcal{M}$}. 

A more accurate description of the \ac{GW} signal can be obtained perturbatively solving Einstein's equations~\eqref{eq:einstein_eq} for the binary system at higher orders in the expansion parameters $G$ and $v/c$. In the case of a binary system, and in general for any self-gravitating system, these two expansion parameters coincide, by virtue of the virial theorem. For this reason, one can define a unique expansion parameter $\epsilon \sim (GM/c^3d)^{1/2} \sim v/c$. The \ac{GW} signal, and the two-body dynamics obtained with this procedure are a perturbative series in $\epsilon$, commonly referred to as \ac{PN} expansions. In the following, I discuss some properties of the \ac{PN} expansion of the \ac{GW} signal that are relevant for this thesis. The interested reader can find more informations about the \ac{PN} calculations in Ref.~\cite{Blanchet:2013haa} and in Sec. 5 of Ref.~\cite{maggiore2008gravitational}.

\robert{The first \ac{PN} correction to the \ac{GW} phase is proportional to $\epsilon^2$ relatively to the leading-order term (also referred to as $1$\ac{PN}\footnote{The $n$\ac{PN} term corresponds to $\epsilon^{2n}$ corrections.} correction). This new term depends on the parameters of the \ac{BBH} system through the symmetric mass ratio $\nu \equiv m_1 m_2/(m_1+m_2)^2$~\cite{1976ApJ...210..764W}. }
Starting from the term proportional to $\epsilon^3$ relatively to the leading-order term, the phase of the \ac{GW} signal also depends on the spins of the two \acp{BH}, $\bm{S_\mathrm{i}} \equiv (Gm_\mathrm{i}^2/c^2)\bm{\chi_\mathrm{i}}$, with $|\bm{\chi_\mathrm{i}}| < 1$ for Kerr \acp{BH}. 
In particular, the $1.5$\ac{PN} correction to the \ac{GW} phase depends on $\bm{S_\mathrm{i}}\cdot\bm{L}$~\cite{Kidder:1992fr}, namely the projection of the spins \robert{along } the orbital angular momentum of the binary $\bm{L}$.
Because of \roberto{this term, and its sign}, the coalescence of \ac{BBH} systems with spins aligned with $\bm{L}$ has a longer duration compared to systems with spins of the same magnitude but anti-aligned with $\bm{L}$, or in another generic \robert{direction}. In fact, \roberto{when the spins are aligned with $\bm{L}$, the total angular momentum of the binary $\bm{J} \equiv \bm{L} + \bm{S_1} + \bm{S_2}$ has the largest possible magnitude $|\bm{J}| = |\bm{L} + \bm{S_1} + \bm{S_2}| = |\bm{L}| + |\bm{S_1}| + |\bm{S_2}|$ } and, before the \ac{BBH} system can merge into a single \ac{BH}, it has to lose enough angular momentum by emitting \acp{GW} to allow the spin of the final \ac{BH} to respect the Kerr bound $c^2|\bm{S_\mathrm{final}}|/(Gm_\mathrm{final}^2) \leq 1$. Since, in this case, the binary coalescence has a longer duration, also the waveform will last longer. It is also interesting to discuss the effect of \ac{BH} spins on the two-body dynamics and the emitted waveform, in the case where \robert{the spins } are not aligned nor anti-aligned with $\bm{L}$. 
In this case, the interaction between \roberto{$\bm{L}$ } and the \ac{BH} spins induces a precession of the orbital plane of the system~\cite{Apostolatos:1994mx,Racine:2008qv,Kidder:1995zr}. This in turn causes a modulation in the amplitude and the phase of the waveform. I leave the detailed discussion \robert{of precessional effects } to Sec.~\ref{sec:precessing_modes}.

\robert{Today, relativistic corrections of } the two-body dynamics and waveforms are known to \robert{much higher } \ac{PN} orders than 1.5\ac{PN} for non-spinning~\cite{Damour:2014jta,Bernard:2016wrg,Bernard:2016wrg,Bernard:2017ktp,Foffa:2019rdf,Foffa:2019yfl,Blanchet:2013haa} and spinning~\cite{Kidder:1992fr,Hartung:2011te,Hartung:2013dza,Marsat:2012fn,Bohe:2012mr,Levi:2015uxa,Levi:2020kvb,Hartung:2011ea,Levi:2011eq,Levi:2014sba,Levi:2020uwu,Levi:2015ixa,Levi:2015msa,Levi:2016ofk,Kidder:1995zr,Blanchet:2006gy,Buonanno:2012rv,Marsatetal2017} binaries. 
While these \ac{PN} corrections improve the accuracy of Eqs.~\eqref{eq:circular_orbit_RR},~\eqref{eq:quad_plus_evol} and~\eqref{eq:quad_cross_evol}, the \ac{PN} series converges slowly, and it is inaccurate when approaching the plunge and the merger, where $\epsilon \rightarrow 1$~\cite{1993PhRvD..47.1511C,Poisson:1995vs}.
The validity of the \ac{GW} signal and \ac{BBH} dynamics obtained using the \ac{PN} expansion can be extended to larger values of $\epsilon$ by using the \textit{\ac{EOB}} formalism~\cite{Buonanno:1998gg,Buonanno:2000ef}. The \ac{EOB} formalism includes some non-perturbative strong-field effects in the two-body dynamics, which are especially relevant during the late stage of the inspiral. Since the waveform models I describe in this thesis are based on the \ac{EOB} formalism, I will provide an extensive introduction \robert{of this approach } in Sec.~\ref{sec:EOB}.

\begin{figure}[h!]
  \centering
  \includegraphics[trim=6cm 0cm 6cm 0cm,clip,width=\textwidth]{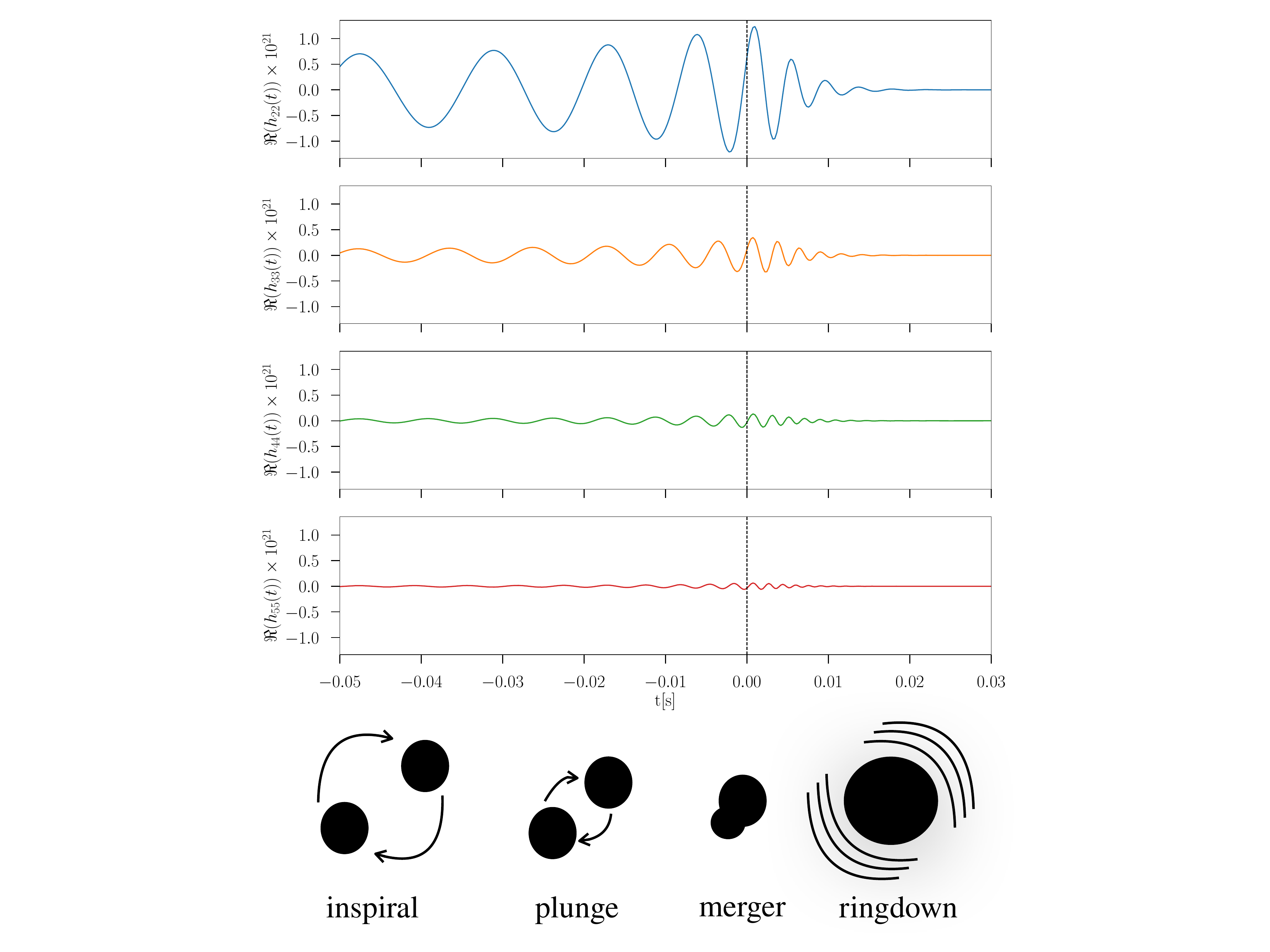}
\caption{Real part of the \ac{GW} modes $(\ell, m) = (2,2),(3,3),(4,4),(5,5)$ for a non-spinning \ac{BBH} system. The vertical line shows the peak of the amplitude of the mode $(\ell, m) = (2, 2)$. These \ac{GW} modes have been generated with the waveform model described in Sec.~\ref{sec:intro_SEOBNRv4HM}.}
\label{fig:IMR_modes}
\end{figure}

When perturbatively solving the Einstein's equations~\eqref{eq:einstein_eq} at higher orders in $\epsilon$, one finds that the \ac{PN} corrections to the quadrupole formula Eq.~\eqref{eq:Green_gw_sol_exp_3} depend on higher-order multiple moments of the binary. Correction terms corresponding to different multiple moments of the binary are proportional to distinct functions of the angles $\iota$ and $\varphi_0$.
For this reason, it is useful to decompose $h_+$ and $h_\times$ into a set of orthonormal bases on a sphere. This separates their dependence on $\iota$ and $\varphi_0$ from that on the other parameters of the binary, such as masses and spins. For this purpose, the most commonly used set of orthonormal bases are the $-2$-spin-weighted spherical harmonics~\cite{Thorne:1980ru}, which are defined as
\begin{align}
{}_{-2}Y_{l m} (\iota, \varphi_0) &= \left(-1\right)^m \sqrt{ \frac{(l+m)! (l-m)! (2l+1)} {4\pi (l-2)! (l+2)!} } \sin^{2l} \left( \frac{\iota}{2} \right)\nonumber \\
& \times\sum_{r=0}^{l+2} {l+2 \choose r} {l-2 \choose r-2-m} \left(-1\right)^{l+2-r} e^{i m \varphi_0} \cot^{2r-2-m} \left( \frac{\iota} {2} \right)\,. \label{eq:sph_harm_def}
\end{align}
When decomposed in $-2$-spin-weighted spherical harmonics, the combination $h_+ - ih_\times$  reads
\begin{equation}
\label{eq:sph_harm_expansion}
h_+(t; \bm{\lambda})-ih_\times(t; \bm{\lambda}) = \sum_{\ell = 2}^\infty \sum_{m = -\ell}^\ell {}_{-2}Y_{\ell m}(\iota, \varphi_0)\, h_{\ell m}(t;\bm{\lambda}),
\end{equation}
where the functions $h_{\ell m}(t;\bm{\lambda})$ are the \ac{GW} modes, not dependent on $\iota$ and $\varphi_0$\robert{, and $\bm{\lambda}$ is a vector including all the parameters of the binary, defined as
\begin{equation}
\bm{\lambda} \equiv \left\{m_1, m_2, \bm{\chi_1}, \bm{\chi_1}, D_\mathrm{L}, \iota, \varphi_0, \psi, \theta, \phi, t_\mathrm{c} \right\}.
\end{equation} 
The angle $\psi$ included in the definition of $\bm{\lambda}$ is the polarization angle, which describes the orientation of the projection of the binary’s orbital momentum vector $\bm{L}$ onto the plane on the sky. The angles $\theta$ and $\phi$ define the position in the sky of the \ac{GW} source}. 
\robert{Before my work}, most of the waveform models used to analyze the \ac{GW} signals detected by LIGO and Virgo only included the modes $(\ell, |m|) = (2,2)$\roberto{, for which leading-order expressions are in Eqs.~\eqref{eq:quad_plus_evol} and~\eqref{eq:quad_cross_evol}. } The other modes are typically referred to as \textit{\acp{HM}} \roberto{or \textit{higher harmonics}, } and they are the main topic of this thesis.
In Sec.~\ref{sec:HM_importance}, I will examine the reasons they were neglected in the past, and I will provide the motivations to include them \robert{for current and future data-analysis studies with LIGO and Virgo detectors}.
\robert{In Fig.~\ref{fig:IMR_modes}, I show the most important \ac{GW} modes emitted during the inspiral, as well as during the other phases of the \ac{BBH} coalescence, which I describe below. }

\subsubsection{Merger}
The inspiral phase ends when the relative radial velocity of the two \acp{BH} becomes comparable to their relative tangential velocity, and the condition $\dot{\omega}_\mathrm{orb}/\omega_\mathrm{orb}^2 \sim \mathcal{O}(v^5/c^5) \ll 1$ is violated. In this regime, the \acp{BH} plunge into each other with a non-negligible radial velocity. 
After the plunge, the two \acp{BH} go through the non-linear phase of the binary coalescence called \textit{merger}. No analytical techniques are available to compute the \ac{BBH} dynamics and the emitted waveform in this phase. The waveform in this regime can be computed through solving numerically the fully non-linear Einstein's equations~\eqref{eq:einstein_eq} \roberto{on a supercomputer}. An entire field, called \textit{\ac{NR}}, is devoted to this goal (see Ref.~\cite{Baumgarte:2002jm} for an extensive review of this topic). \ac{NR} is based on the $3+1$ decomposition of spacetime, originally developed in the \ac{ADM} formalism~\cite{PhysRev.116.1322,Arnowitt:1962hi}. In this formulation, the Einstein's equations are divided in two different sets: 4 constraint equations and $12$ evolution equations. The constraint equations are solved on a spacelike hypersurface and the evolution equations are used to \roberto{connect } nearby hypersurfaces by conserving the constraint equations. In practice, \ac{NR} simulations require a reformulation of the $3+1$ decomposition such that the equations are a well-defined initial value problem, that can be solved numerically~\cite{Pretorius:2005gq,Campanelli:2005dd,Baker:2005vv}. This was a formidable task and, for this reason, the first \ac{BBH} \ac{NR} simulation including the merger was successfully performed as late as 2005~\cite{Pretorius:2005gq,Campanelli:2005dd,Baker:2005vv}. Since then, many simulations contributed to shed light on the merger. \robert{During the merger phase, the amplitude and the instantaneous frequency of each \ac{GW} mode increase, as it is clear from Fig.~\ref{fig:IMR_modes}}. \ac{NR} codes currently provide the \say{gold standard} solution to the relativistic two-body problem in \ac{GR}. They are now capable to generate more accurate and longer waveforms than in the past in reasonable timescales ($\mathcal{O}(\textrm{months}))$. There are currently thousands \ac{NR} simulations of \ac{BBH} systems performed in different regions of the binary parameter space~\cite{Mroue:2013xna,Jani:2016wkt,Healy:2017psd,Boyle:2019kee,Healy:2019jyf,Healy:2020vre}. \roberto{However, \ac{NR} waveforms alone cannot be used in data analysis applications for the LIGO and Virgo detectors, as their time-duration is still too short, and they cover a limited region of the binary parameter space. } 

\subsubsection{Ringdown}
The final product of the \ac{BBH} merger is a perturbed Kerr \ac{BH}, which reaches the equilibrium state by emitting \acp{GW}. At linear order in the perturbation, this \ac{BH} can be described by the metric tensor
\begin{equation}
g^{\mu\nu} = g^{\mu\nu}_{\mathrm{Kerr}} + h^{\mu \nu}, \quad |h^{\mu \nu}| \ll 1,
\end{equation}
where $g^{\mu\nu}_{\mathrm{Kerr}}$  is the Kerr metric, and the tensor perturbation $h^{\mu \nu}$ can be computed within the \ac{BH} perturbation theory framework, by providing appropriate boundary conditions. The \ac{GW} signal emitted as a result of this process is called \textit{ringdown}. The \ac{GW} signal in this phase is a superposition of the \acp{QNM} of the \ac{BH} remnant~\cite{Vishveshwara:1970zz,Press:1971wr,Chandrasekhar:1975zza}, whose decomposition in $-2$-spin-weighted spherical harmonics\footnote{The natural bases for the decomposition of the ringdown \ac{GW} signal are the $-2$-spin-weighted spheroidal harmonics~\cite{Teukolsky:1973ha}.} reads
\begin{equation}
h_{\ell m}(t; \bm{\lambda}) = \sum_{n = 0}^\infty  A_{\ell m n}(\bm{\lambda}) e^{-i\sigma_{\ell m n}(\bm{\lambda})t}.
\end{equation}
In the equation above, the quantities $\sigma_{\ell m n}$ are the complex \ac{QNM} frequencies of the remnant \ac{BH} that depend on its mass and spin. The factors $A_{\ell m n}$ are instead \roberto{complex } constants, \roberto{called excitation coefficients. } The latter cannot be computed within the \ac{BH} perturbation theory framework, because they depend on the details of the merger \roberto{of the two \acp{BH}}~\cite{Buonanno:2000ef,Buonanno:2006ui,Berti:2006hb,Berti:2006wq,Dorband:2006gg,Zhang:2013ksa,Taracchini:2014zpa,Hughes:2019zmt,Lim:2019xrb}. Each mode $h_{\ell m}(t; \bm{\lambda})$ is a damped sinusoid, \robert{as can also be seen in Fig.~\ref{fig:IMR_modes}}, oscillating at the frequency $\Re{(\sigma_{\ell m n})}$, which is the \roberto{asymptotic } frequency the modes \roberto{approach at the end } of the merger.
This prediction of the \ac{GW} signal is also confirmed by \ac{NR} simulations~\cite{Buonanno:2006ui}.

While different techniques are able to predict the \ac{GW} signal in different regimes, for data analysis purposes one needs \roberto{the complete } signal for the entire binary coalescence, as shown in Fig.~\ref{fig:IMR_modes}. Over the years, three main approaches have been developed to produce a smooth signal incorporating inspiral, merger and ringdown\robert{: the \ac{EOB} formalism, the phenomenological approach, and the \ac{NR} surrogate method}. Since the work on this thesis revolves around the approach based on the \ac{EOB} formalism, I will provide, in Secs.~\ref{sec:EOB_dynamics} and~\ref{sec:EOB_waveform}, an extensive introduction of this method, and summarize here the other two approaches. \roberto{In Sec.~\ref{sec:waveform_modeling_intro}, I will also provide a more extensive description of some waveform models developed within the other two approches, which I use as comparison with the waveform models I describe in this thesis.}

In the phenomenological approach~\cite{Ajith:2007qp,Ajith:2007kx,Ajith:2009bn,Hannam:2013oca,Schmidt:2014iyl,Khan:2015jqa,Husa:2015iqa,London:2017bcn,Khan:2018fmp,Garcia-Quiros:2020qpx,Pratten:2020ceb}, the starting point are hybrid \ac{GW} modes constructed by smoothly blending \ac{EOB} inspiral modes with \ac{NR} \ac{GW} modes including the late inspiral, merger and ringdown. These hybrid modes are first converted to frequency domain, and then used to construct phenomenological fits for the amplitude and phase of each mode. 

The \ac{NR} surrogate method is based on \ac{NR} simulations. The waveform models developed using this method~\cite{Field:2013cfa, Blackman:2015pia, Blackman:2017dfb,Blackman:2017pcm,Varma:2018mmi,Varma:2019csw,Rifat:2019ltp}, are capable of generating new waveforms through interpolating available \ac{NR} waveforms. Although in the last few years these models have been proven capable to produce very accurate waveforms, they are still limited to regions of the parameter space where the \ac{NR} waveforms can be generated with reasonable timescales.
\FloatBarrier

\subsection{Strategy for detection}
\label{sec:strategy_detection}
In this section, I describe the main experimental and data analysis tools used for the detection of \acp{GW}. 

\roberto{In the proper detector frame, where the coordinates are marked using rigid rods starting from a conventionally chosen origin, the effect of a \ac{GW} traveling through a region of spacetime can be described as a variation in the proper distance of two nearby geodesics parametrized by $x^\mu$ and $x^\mu + \zeta^\mu$, respectively. In the idealized case, where other gravitational effects and external forces are absent, the proper distance\footnote{In the proper detector frame, at leading order, proper distances coincide with coordinate distances, see Sec.1.3 of Ref.~\cite{maggiore2008gravitational} for more details.} between these two nearby geodesic in this frame changes according to the equation for the \textit{geodesics deviation}}
\begin{equation}
\label{eq:geodesic_deviation}
\ddot{\zeta}^i = \frac{1}{2} \ddot{h}_{ij}^{TT}\zeta^j,
\end{equation}
where $h_{ij}^{TT}$ is the metric tensor perturbation associated with the \ac{GW} in the TT-gauge. The goal of a \ac{GW} detector is to track the geodesic motion of freely falling bodies, and observe if their proper distance changes according to Eq.~\eqref{eq:geodesic_deviation}. Different experimental methods have been proposed over the years to achieve this purpose (e.g. resonant mass detectors~\cite{PhysRev.117.306,Weber1961-WEBGRA,Astone:2003mp}, \ac{PTA}~\cite{1978SvA....22...36S,1979ApJ...234.1100D,1990ApJ...361..300F}, and atom interferometry~\cite{Stodolsky:1978ks,Dimopoulos:2008sv,Graham:2012sy}). I will focus here on the \textit{laser-interferometry} technique, which is used in the ground-based LIGO and Virgo detectors. 

A schematic example of a laser-interferometric \ac{GW} detector is provided by the Michelson interferometer. \roberto{Real \ac{GW} detectors are more sophisticated, see Chapter 9 of Ref.~\cite{maggiore2008gravitational} for an extensive review on the topic. } In this \roberto{schematic example of a \ac{GW} } detector, a laser beam is divided by a beam-splitter into two orthogonal beams with equal probability amplitudes. The two beams travel through the orthogonal interferometer arms until they are reflected by a mirror placed at the end of each arm. \roberto{The reflected beams are finally recombined at the beam-splitter, and the intensity of the recombined beam is measured by a photodetector. } 
The observed light intensity depends on the relative phase between the two beams, which is a function of the difference in their travel path. When a \ac{GW} travels through the experimental apparatus, it changes the length of the intereferometer arms according to Eq.~\eqref{eq:geodesic_deviation}. This causes a variation in the light intensity which is observed by the photodetector. The fractional variation of the arm length caused by an incoming \ac{GW} is $\Delta L / L ~\sim h$, where $h$ is the amplitude of the \ac{GW} signal, $\mathcal{O}(10^{-20})$ for typical sources detectable by ground-based interferometers. The obstacle for the \ac{GW} detection is that, in a real interferometer, there are other \roberto{non-astrophysical forces } that can cause a change in the arm length of the interferometer which is much larger than the variation caused by a \ac{GW}. These forces are usually referred to as \textit{noise sources}. In the case of LIGO and Virgo interferometers, these noise sources limit the detectors to be most sensitive only in the frequency range $20\,\mathrm{Hz} \lesssim f \lesssim 1\,\mathrm{kHz}$. Below $20$ Hz the sensitivity of these detectors is limited by the \textit{seismic noise} caused by mirrors movements as consequence of ground vibrations. \roberto{At high frequencies, starting from $\sim 100\,\mathrm{Hz}$, } the limiting noise source consists of \textit{quantum shot noise}, specifically the statistical uncertainty in the light intensity measured by the photodetector due to the discrete nature of light. In the intermediate frequency regime, the most important noise source is
the \textit{thermal noise} due to the atomic motion in the suspensions that sustain the mirrors of the interferometer, \robert{and on the surface of the mirrors itself}. See Refs.~\cite{TheLIGOScientific:2014jea,Martynov:2016fzi} for a detailed discussion of LIGO and Virgo noise sources.

Although LIGO and Virgo interferometers use various techniques to reduce the effect of these noise sources (see Refs.~\cite{TheLIGOScientific:2014jea,TheVirgo:2014hva, Meers:1988wp,Mizuno:1993cj} for details), the detector noise is still \roberto{usually } much larger than the typical \ac{GW} signals.
\roberto{Nevertheless, it is possible to detect a \ac{GW} signal buried in the noise by exploiting knowledge of the signal and the noise. Here I provide a basic introduction to the data analysis techniques used to detect \ac{GW} signals. However, more sophisticated methods are employed in practice, see Ref.~\cite{LIGOScientific:2019hgc} for a complete review on the topic. Since the discussion of these techniques is beyond the scope of this thesis, I will only outline them and provide the relevant references for the interested reader. } 

The detector output in presence of a \ac{GW} signal is
\begin{equation}
\label{eq:LIGOsignal}
s(t) =  h(t) + n(t),
\end{equation}
where $n(t)$ is the detector noise and $h(t)$ is the detector response to the \ac{GW} signal, typically $|h(t)|\ll|n(t)|$. \roberto{The detector response to the \ac{GW} signal is given by $h(t) = D^{ij}h_{ij}$,  where $h_{ij}$ is the \ac{GW} and $D^{ij}$ is the detector tensor, which depends on the geometry of the interferometer. In the case of LIGO and Virgo detectors, $h(t)$ is simply $h(t) \equiv F_+(\theta,\phi,\psi)\,h_+ + F_\times(\theta,\phi,\psi)\,h_\times$, where $F_+(\theta,\phi,\psi)$ and $F_\times(\theta,\phi,\psi)$ are the so-called antenna patterns~\cite{Hawking:1987en,1987MNRAS.224..131S,1988MNRAS.234..663D}, which depend } on the wave polarization $\psi$ and the sky-position of the source, as defined by the angles $(\theta, \phi)$.
The expressions of $F_\times(\theta,\phi,\psi)$ and $F_+(\theta,\phi,\psi)$ are known and also $h_{ij}$ can be \roberto{predicted } for expected sources, as discussed in the previous section in the case of \ac{BBH} systems. \roberto{For this reason, one can use the best available model of $h(t)$ } as a filter to extract the true \roberto{signal } from the detector output. \roberto{This is a commonly used technique in signal processing, its name is \textit{matched filtering}.}

A simplified version of this \roberto{matched filtering } procedure can be illustrated by computing the average value of the detector output $s(t)$ over a time period $T$, when filtered by the estimated $h(t)$. Under the assumption that the model of  $h(t)$ exactly matches its true expression in \ac{GR}, this is
\begin{equation}
\label{eq:simple_match_filter}
\frac{1}{T}\int_{0}^{T} dt\, h(t) s(t) =  \frac{1}{T}\int_{0}^{T} dt\, h^2(t) + \frac{1}{T}\int_{0}^{T} dt\, h(t) n(t). 
\end{equation}
Since, at first approximation, $h(t)$ is an oscillating function with some typical amplitude $h_0$ and characteristic frequency $\omega$, for large $T$ the first integral grows linearly in $T$. Therefore the first term in the RHS of the equation is proportional to $h_0^2$. The noise $n(t)$ is an oscillating function with a typical amplitude $n_0$ and timescale $\tau_0$, it is not correlated with the signal $h(t)$, and it arises from an underlying random process. For this reason, the second integral grows proportionally to $T^{1/2}$, and the second term in the RHS of the equation above scales as $T^{-1/2}$. In summary, the equation above reads
\begin{equation}
\frac{1}{T}\int_{0}^{T} dt\, h(t) s(t) \sim h_0^2 + \left(\frac{\tau_0}{T}\right)^{1/2} n_0 h_0,
\end{equation}
and one can conclude whether a \ac{GW} signal is present in the detector data if $h_0 > (\tau_0/T)^{1/2} n_0$, which is possible to fulfill even in the case $n_0 \gg h_0$.

In the more realistic case of the LIGO and Virgo detectors, one starts from the assumption that the noise $n(t)$ is a realization from a stationary Gaussian random process, therefore it can be completely described by providing its mean value $\langle n(t)\rangle$ and its \ac{PSD} $S_n(f)$. Without loss of generality, one can assume that $\langle n(t) \rangle = 0$, while the \ac{PSD} can be estimated from the data as
\begin{equation}
S_n(f) = 2\int_{-\infty}^{\infty} \langle n(t)\,n(t+\tau) \rangle \, e^{-2i\pi f \tau} d\tau,
\end{equation}
where $\langle n(t)\,n(t+\tau) \rangle$ is the autocorrelation function. Henceforth, for convenience, I will work in the Fourier domain. Therefore I will consider $\tilde{h}(f)$ and $\tilde{s}(f)$, the Fourier transform of $h(t)$ and $s(t)$, respectively.

A particular realization of the noise occurs with probability
\begin{equation}
\label{eq:noise_probability}
P(n) = N e^{-\frac{1}{2} (n|n)},
\end{equation}
where $N$ is a normalization factor and $(\cdot|\cdot)$ is the inner product
\begin{equation}
(a|b) \equiv \int_{-\infty}^{\infty} df \frac{\tilde{a}^*(f) \tilde{b}(f) + \tilde{a}(f) \tilde{b}^*(f)}{S_n(f)} = 4 \Re{\left(\int_{0}^{\infty} df \frac{\tilde{a}^*(f) \tilde{b}(f)}{S_n(f)}\right)}.
\end{equation}
Given a detector output $s(t)$, the detection of a \ac{GW} signal consists in a comparison between the null hypothesis $\mathcal{H}_0$, that $s(t)$ contains only noise, with the signal hypothesis $\mathcal{H}_1$, for which also a \ac{GW} signal is present. Using Bayes' theorem, one can write the probability of the signal hypothesis given a particular detector output $s(t)$ as
\begin{equation}
p(\mathcal{H}_1|s) = \frac{p(\mathcal{H}_1)p(s|\mathcal{H}_1)}{p(\mathcal{H}_1)p(s|\mathcal{H}_1) + p(\mathcal{H}_0)p(s|\mathcal{H}_0)} = \frac{p(s|\mathcal{H}_1)}{p(s|\mathcal{H}_0)} \left[ \frac{p(s|\mathcal{H}_1)}{p(s|\mathcal{H}_0)} + \frac{p(\mathcal{H}_0)}{p(\mathcal{H}_1)}  \right]^{-1},
\end{equation}
where $p(s|\mathcal{H}_0)$ $\left(p(s|\mathcal{H}_1)\right)$ is the probability of the detector output $s(t)$ under the null (signal) hypothesis, and $p(\mathcal{H}_0)$ $\left(p(\mathcal{H}_1)\right)$ is the prior probability of the null (signal) hypothesis. The probability $p(\mathcal{H}_1|s)$ depends on the detector output only through the ratio $p(s|\mathcal{H}_1)/p(s|\mathcal{H}_0)$ which, using Eqs.\eqref{eq:LIGOsignal} and \eqref{eq:noise_probability}, can be expressed as
\begin{equation}
\label{eq:mon_incr_sh}
\frac{p(s|\mathcal{H}_1)}{p(s|\mathcal{H}_0)} = \frac{e^{-\frac{1}{2} (s-h|s-h)}}{e^{-\frac{1}{2} (n|n)}} = e^{(s|h)} e^{-(h|h)/2}.
\end{equation}
\roberto{The function $p(\mathcal{H}_1|s)$ is monotonically increasing in the ratio $p(s|\mathcal{H}_1)/p(s|\mathcal{H}_0)$, which in turn is a monotonic function of $(s|h)$, because of Eq.~\eqref{eq:mon_incr_sh}. Thus, any choice } for the threshold on the probability $p(\mathcal{H}_1|s)$ to accept the signal hypothesis can be directly translated into a threshold in $(s|h)$, or in its normalized form, the \textit{\ac{SNR}}
\begin{equation}
\rho \equiv \frac{(s|h)}{(h|h)^{1/2}} = \frac{4 \Re{\left( \displaystyle\int_{0}^{\infty}df \frac{\tilde{s}^*(f) \tilde{h}(f)}{S_n(f)}\right)}}{\sqrt{4 \Re{\left( \displaystyle\int_{0}^{\infty}df \frac{|h(f)|^2}{S_n(f)}\right)}}}.
\end{equation}
For this reason, one can directly use the \ac{SNR} to establish whether there is a statistically significant evidence for a \ac{GW} signal in a given data stream. The filtering procedure, introduced for a simplified case in Eq.~\eqref{eq:simple_match_filter}, appears clearly in the numerator of the \ac{SNR}, where the detector output $\tilde{s}(f)$ is filtered using the noise-weighted \ac{GW} signal $\tilde{h}(f)/S_n(f)$. \roberto{In signal processing, the function $\tilde{W}_\mathrm{h}(f) \equiv \tilde{h}(f)/S_n(f)$ is typically called \textit{Wiener filter}. With this choice of the Wiener filter, the \ac{SNR} is maximized for $\tilde{s}(f) = \tilde{h}(f)$. Its maximum value $\rho_\mathrm{opt} \equiv (h|h)^{1/2} $ is typically called \textit{optimal \ac{SNR}}. }
The function used for the \roberto{Wiener filter, } $\tilde{h}(f)$, usually referred to as \textit{template}, depends on $\bm{\lambda}$, the parameters of the binary system, which are unknown at this stage. Since for the detection purpose, one is interested only in finding the signal regardless of the binary parameters, the \ac{SNR} for a given detector output $\tilde{s}(f)$ is computed against a discrete set of pre-computed templates spanning the binary parameter space $\bm{\lambda}$, typically referred to as a \textit{template bank}.
The largest \ac{SNR} obtained in this process, $\rho_{\mathrm{max}}^\mathrm{template} \equiv \rho(\bm{\lambda}_\mathrm{max}^\mathrm{template})$, is compared with the \ac{SNR} distribution expected in the case of only noise being present in the detector output. 
An \ac{SNR} threshold $\hat{\rho}$ is set such that \ac{SNR} values larger than the threshold are unlikely to be due to noise. Therefore, a candidate \ac{GW} signal \roberto{(or trigger) } is recorded if, for a detector output $s(f)$, $\rho_{\mathrm{max}}^\mathrm{template}$ is larger than the threshold.
Because of the discreteness of the template bank, in practice the $\rho_{\mathrm{max}}^\mathrm{template}$ value is smaller than the maximum one would obtain when maximizing over the templates in the continuous binary parameter space $\bm{\lambda}$. The quantity $\rho_{\mathrm{max}}^\mathrm{template}$ is related to the \say{true} maximum of the \ac{SNR} $\rho_{\mathrm{max}} \equiv \rho(\bm{\lambda}_{\mathrm{max}})$ through the equation
\begin{equation}
\label{eq:SNR_reduction_discreteness}
\rho_{\mathrm{max}}^\mathrm{template} \approx \rho_{\mathrm{max}}\left[1-F\left(h(f,\bm{\lambda}_\mathrm{max}),h(f,\bm{\lambda}_\mathrm{max}^\mathrm{template})\right) \right],
\end{equation}
where the function $F\left(h_1(f),h_2(f)\right)$ is the so-called \textit{faithfulness} between two waveforms $h_1(f)$ and $h_2(f)$, defined as
\begin{equation}
\label{eq:faithfulness_intro}
F\left(h_1(f),h_2(f)\right) \equiv \max_{t_\mathrm{c},\varphi_0} \frac{(h_1(f)|h_2(f))}{\sqrt{(h_1(f)|h_1(f))(h_2(f)|h_2(f))}}.
\end{equation}
The maximization in the equation above is performed over the time of coalescence $t_\mathrm{c}$ and the phase $\varphi_0$.
The faithfulness function attains its maximum value, $F\left(h_1(f),h_2(f)\right) = 1$, when the two waveforms are exactly the same, while its value decreases proportionally to the difference between the two waveforms.
The template banks used for searching \ac{GW} signals in the data of LIGO and Virgo detectors are built to ensure that the minimum faithfulness between a given waveform and the best matching template in the template bank is always larger than $0.97$~\cite{Sathyaprakash:1991mt,Owen:1995tm,TheLIGOScientific:2016pea,DalCanton:2017ala}. Median faithfulness between randomly chosen waveforms and best matching templates in the template bank are as large as $0.99$. A reduction in the maximum \ac{SNR}, similar to that decribed above, can also be caused by the inaccuracy of the waveform models in representing the \say{true} \ac{GR} waveforms. The reduction in the maximum \ac{SNR} in this case can be quantified using Eq.~\eqref{eq:SNR_reduction_discreteness}, and substituting $F\left(h(f,\bm{\lambda}_\mathrm{max}),h(f,\bm{\lambda}_\mathrm{max}^\mathrm{template})\right)$ with $F\left(h^\mathrm{model}(f,\bm{\lambda}_{\mathrm{max}}^\mathrm{model}),h^\mathrm{GR}(f,\bm{\lambda}_\mathrm{max}^\mathrm{GR})\right)$, where $h^\mathrm{model}(f)$ and $h^\mathrm{GR}(f)$ are respectively the approximate and true waveform. In practice, it is desirable that the decrease in \ac{SNR} due to the inaccuracy of the waveforms is negligible with respect to that caused by the discreteness of the template bank. For this reason, a typical requirement for the accuracy of the waveform models is that the median faithfulness between them and the \ac{NR} waveforms (i.e. the best representation of the true \ac{GR} waveforms) is larger than $0.99$, which is the median faithfulness between randomly chosen waveforms and best matching templates in the template bank. 

The method for the detection of \ac{GW} signals described so far assumes stationary and Gaussian noise. In reality, the detector noise is non-Gaussian, therefore \ac{SNR} values larger than the threshold could be obtained also due to non-Gaussian noise. Such non-Gaussian artifacts are typically referred to as \textit{glitches}, and their effect can be tamed by using more sophisticated strategies like \textit{signal-based vetoes} \roberto{(also known as $\chi^2$-vetoes), see Refs.~\cite{Allen:2004gu,Babak:2012zx} for more details. }
\roberto{Another powerful tool to detect \ac{GW} signals in presence of non-Gaussian noise consists in the analysis of \textit{coincident signals} in multiple \ac{GW} detectors. In fact, } while it is possible to have a glitch in one of the detectors, it is unlikely to have glitches in all detectors appearing with time delays between detectors compatible with the passage of a \ac{GW}. 
\roberto{In the coincidence analysis, \ac{GW} candidates from each detector,  identified using the methods described before, are compared with those from other detectors. \ac{GW} candidates are flagged as possible \ac{GW} signals, if they are measured by different detectors with consistent parameters and a separation in time compatible with that due to the travel time from one detector to another. The statistical significance of these potential \ac{GW} signals is determined by comparing them with a background distribution of coincidences due to noise. The latter is obtained using the \textit{time-slides method}. The first step of this method consists of time shifting the \ac{GW} candidates obtained from single detectors by an unphysical delay, much larger than the light travel time between detectors and the duration of the typical signal. The coincidence analysis is performed on this new list of single-detector \ac{GW} candidates and a new list of triggers is recorded. These triggers cannot originate from \ac{GW} signals, because they are computed using single-detector \ac{GW} candidates that are time shifted by unphysical delays. This procedure is repeated multiple times using different time delays to time shift the list of single-detector \ac{GW} candidates. All the triggers obtained with this procedure are used as noise background to estimate the significance of the potential \ac{GW} signals.
More details about this method can be found in Refs.~\cite{Bose:1999pj,Finn:2000hj,PhysRevD.72.063006,Harry:2010fr}}.

\roberto{The experimental and data analysis tools, outlined in this section, allowed the detection of \ac{GW} signals emitted by $10$ \acp{BBH}~\cite{Abbott:2016blz,Abbott:2016nmj, Abbott:2017vtc,
Abbott:2017gyy, Abbott:2017oio, LIGOScientific:2018mvr} during the first two LIGO and Virgo observation runs (henceforth O1 and O2, respectively). During O1 and O2 also a system composed of two \acp{NS} (a \ac{BNS}) was detected~\cite{TheLIGOScientific:2017qsa}. In addition, another $39$ \ac{GW} signals have been recently detected during the first half of the third LIGO and Virgo observing run (henceforth O3a)~\cite{Abbott:2020tfl,LIGOScientific:2020stg,Abbott:2020khf,Abbott:2020niy,Abbott:2020uma}. Among these \ac{GW} signals, $36$ of them were likely coming from \acp{BBH}~\cite{Abbott:2020niy,LIGOScientific:2020stg,Abbott:2020tfl}, detected up to a distance of $\sim 5\,\mathrm{ Gpc}$. Among the remaining $3$ signals, one of them likely originated from a \ac{BNS} merger at a distance of $\sim 100\,\mathrm{ Mpc}$~\cite{Abbott:2020uma}. The sources of the other two signals are still unclear~\cite{Abbott:2020niy,Abbott:2020khf}. They could either originate from \acp{BBH}, or from mixed systems composed of a \ac{BH} and a \ac{NS}, typically referred to as \ac{NSBH} systems.}

\subsection{Methods for source characterization}
\label{sec:source_characterization}
After a \ac{GW} signal is identified, the next goal is to measure the parameters of the emitting source. In this section, I will provide a summary of the most important techniques used for this purpose.

The parameters of the source are measured within a Bayesian framework. For this purpose, the starting point is the determination of $p(\bm{\lambda}|s,h)$, the multi-dimensional posterior probability-density function of the source parameters $\bm{\lambda}$ given the detector output $s(t)$, a model \roberto{$M_\mathrm{h}$ } for the waveform $h(\bm{\lambda})$, and the prior probability $p(\bm{\lambda})$ for the source parameters $\bm{\lambda}$. Using Bayes' theorem, one can write this posterior probability as
\begin{equation}
\label{eq:posterior_Bayes}
p(\bm{\lambda}|s,M_\mathrm{h}) = \frac{p(s|\bm{\lambda},M_\mathrm{h})\,p(\bm{\lambda})}{p(s|M_\mathrm{h})},
\end{equation}
where $p(s|\bm{\lambda},M_\mathrm{h})$ is the \textit{likelihood function} and $p(s|M_\mathrm{h} )$ is the evidence\roberto{, which I define below. }
Under the assumption of stationary and Gaussian noise, the likelihood can be computed using the expression for the noise probability, in Eq.~\eqref{eq:noise_probability}, and the detector output definition in Eq.~\eqref{eq:LIGOsignal}. The result is
\begin{equation}
\label{eq:LIGO_likelihood}
p(s|\bm{\lambda},M_\mathrm{h}) = N e^{-\frac{1}{2}\left(s-h(\bm{\lambda})|s-h(\bm{\lambda})\right)},
\end{equation}
where $N$ is a normalization. This likelihood is associated to the data of only one detector. The generalization to multiple detectors is trivial, since the noise realizations of different detectors are uncorrelated, and the joint likelihood is simply the product of the single-detector likelihoods. The evidence $p(s|M_\mathrm{h})$ is a normalization factor for the posterior $p(\bm{\lambda}|s,M_\mathrm{h})$, whose expression is
\begin{equation}
p(s|M_\mathrm{h}) = \int_{\bm{\lambda}} p(s|\bm{\lambda},M_\mathrm{h}) p(\bm{\lambda}) d\bm{\lambda}.
\end{equation}
Within the Bayesian framework, the evidence is often used in the context of hypothesis testing. In fact, the odds ratio between two hypotheses $\mathcal{H}_1$ and $\mathcal{H}_2$ is defined as
\begin{equation}
\mathcal{O}_{\mathcal{H}_1, \mathcal{H}_2} \equiv \frac{p(\mathcal{H}_1|s,M_\mathrm{h})}{p(\mathcal{H}_2|s,M_\mathrm{h})} = \frac{p(\mathcal{H}_1)}{p(\mathcal{H}_2)}\frac{p(s|\mathcal{H}_1,M_\mathrm{h})}{p(s|\mathcal{H}_2,M_\mathrm{h})},
\end{equation}
where \roberto{$p(\mathcal{H}_1|s,M_\mathrm{h})$ and $p(\mathcal{H}_2|s,M_\mathrm{h})$ are the probabilities of the hypotheses $\mathcal{H}_1$ and $\mathcal{H}_2$ given the detector output $s(t)$, and the model for the waveform $M_\mathrm{h}$. The functions $p(\mathcal{H}_1)$ and $p(\mathcal{H}_2)$ are the prior probabilities of the two hypotheses $\mathcal{H}_1$ and $\mathcal{H}_2$, while $p(s|\mathcal{H}_1,M_\mathrm{h})$ and  $p(s|\mathcal{H}_2,M_\mathrm{h})$ are their evidence}. The ratio of the two evidence $p(s|\mathcal{H}_1,M_\mathrm{h})/p(s|\mathcal{H}_2,M_\mathrm{h})$ is usually referred to as Bayes factor.

It is instructive to study the behaviour of the posterior $p(\bm{\lambda}|s,M_\mathrm{h})$ around its maximum $\bm{\lambda}_{\mathrm{max}}$ by computing the Taylor expansion around this point.
Under the assumption that the prior $p(\bm{\lambda})$ is approximately constant over the relevant binary parameter space region (as it is expected in absence of prior knowledge), the Taylor expansion of the posterior $p(s|\bm{\lambda},M_\mathrm{h})$ will be equal to the Taylor-expanded likelihood (up to a normalisation factor).
The likelihood peak $\bm{\lambda}_{\mathrm{max}}$ can be found by imposing the condition $\left .\frac{\partial}{\partial \lambda^i} p(s|\bm{\lambda},M_\mathrm{h})\right|_{\bm{\lambda} = \bm{\lambda}_{\mathrm{max}}} = 0$ which maximizes the likelihood. In the case of the likelihood in Eq.~\eqref{eq:LIGO_likelihood}, the condition above reduces to
\begin{equation}
\label{eq:max_likelihood}
\left(s-h(\bm{\lambda}_{\mathrm{max}})\big|\partial_i h(\bm{\lambda}_\mathrm{max}) \right) = 0,
\end{equation}
with $\partial_i h(\bm{\lambda}_\mathrm{max}) = \frac{\partial h(\bm{\lambda})}{\partial \lambda^i}\bigg|_{\bm{\lambda} = \bm{\lambda}_\mathrm{max}}$.
The Taylor expanded likelihood reads
\begin{equation}
\label{eq:likelihood_linear_signal_approx}
p(s|\bm{\lambda},M_\mathrm{h}) = N e^{-\frac{1}{2} \left. \Gamma_{ij}\right|_{\bm{\lambda} = \bm{\lambda}_{\mathrm{max}}}(\lambda^i - \lambda_{\mathrm{max}}^i)(\lambda^j - \lambda_{\mathrm{max}}^j)(1+\mathcal{O}(\rho^{-1}))},
\end{equation}
where $\Gamma_{ij} \equiv \left(\partial_i h|\partial_j h\right)$ is the Fisher information matrix~\cite{fisher1922mathematical}. The accuracy of this approximation increases with the \ac{SNR}, \roberto{since the deviations from Eq.~\eqref{eq:likelihood_linear_signal_approx} are proportional to $\rho^{-1}$, as indicated in the exponent of the formula}. 
\roberto{The posterior distribution in Eq.~\eqref{eq:likelihood_linear_signal_approx} has the form of a multivariate Gaussian distribution, with a width proportional to $\sqrt{\Gamma_{ii}^{-1}}$. }The latter quantifies the statistical uncertainty of the measurement and is proportional to $\rho^{-1}$. For this reason, signals with larger \ac{SNR} allow for more precise measurements of the binary parameters.

The expression in Eq.~\eqref{eq:likelihood_linear_signal_approx} implicitly assumes that the waveforms generated using the waveform model $M_\mathrm{h}$ are an exact representation of the true waveforms predicted by \ac{GR}. This is typically not the case since, as discussed in Sec.~\ref{sec:waveform_anatomy}, waveform models employ various approximations. In a more realistic case, the maximum likelihood is reached by the approximated waveform model $h^{\mathrm{model}}(\bm{\lambda})$ in a point $\bm{\lambda}_{\mathrm{max}}^\mathrm{model}$, which is different from the point $\bm{\lambda}_{\mathrm{max}}^\mathrm{GR}$ that would be reached when using the true \ac{GR} waveforms $h^{\mathrm{GR}}(\bm{\lambda})$. In this case, Eq.~\eqref{eq:max_likelihood} becomes
\begin{equation}
\label{eq:likelihood_linear_signal_approx_systematics}
\left(s-h^\mathrm{model}(\bm{\lambda}_{\mathrm{max}}^\mathrm{model})\big|\partial_i h^\mathrm{model}(\bm{\lambda}_\mathrm{max}^\mathrm{model}) \right) = 0.
\end{equation}
The error introduced in the measured binary parameters, due to this incorrect representation of the waveform, can be estimated by the bias $\Delta\lambda^i \equiv (\lambda_{\mathrm{max}}^\mathrm{model})^i - (\lambda_{\mathrm{max}}^\mathrm{GR})^i$. Assuming that $|\Delta\lambda^i/\lambda^i| \ll 1$, its value can be obtained from Eq.~\ref{eq:likelihood_linear_signal_approx_systematics} by using the fact that $s(t) = n(t) + h^\mathrm{GR}(t;\bm{\lambda}_{\mathrm{max}}^\mathrm{GR})$, and computing $h^{\mathrm{model}}(\bm{\lambda}_{\mathrm{max}}^\mathrm{model})$ as a Taylor expansion around $\bm{\lambda}_{\mathrm{max}}^\mathrm{GR}$ (see Ref.~\cite{Cutler:2007mi} for the detailed calculation). The result is
\begin{align}
\label{eq:systematic_error_fisher}
\Delta\lambda^i &=  \left(\Gamma^{-1}(\bm{\lambda}_{\mathrm{max}}^\mathrm{model})\right)^{ij}\left( \partial_j h^{\mathrm{model}}(\bm{\lambda}_{\mathrm{max}}^\mathrm{model}) \big| \partial_j h^{\mathrm{GR}}(\bm{\lambda}_{\mathrm{max}}^\mathrm{model}) - \partial_j h^{\mathrm{model}}(\bm{\lambda}_{\mathrm{max}}^\mathrm{model}) \right) + \nonumber \\
&+ \mathcal{O}(\rho^{-1}).
\end{align}  
To assess the relevance of this bias, one has to compare it with the statistical uncertainty, as quantified by the width of the posterior distribution $\sqrt{\Gamma_{ii}^{-1}}$. Since $\sqrt{\Gamma_{ii}^{-1}}\propto \rho^{-1}$, while the bias in Eq.~\eqref{eq:systematic_error_fisher} is independent of the \ac{SNR}, \roberto{there is always a value } of $\rho$ for which this bias is dominant with respect to the statistical uncertainty. \roberto{In practice, computing Eq.~\eqref{eq:systematic_error_fisher}, to estimate at which \ac{SNR} the systematic bias is larger than the statistical uncertainty, is difficult. In fact, it usually involves the computation and the numerical invertion of the Fisher matrix $\Gamma_{ij}$, which is not trivial, as it is a $15$D matrix~\cite{Vallisneri:2007ev}. In addition, to compute the Fisher matrix, and also part of Eq.~\eqref{eq:systematic_error_fisher}, it is necessary to evaluate the derivatives of the waveform with respect to the parameters of the \ac{BBH} system. For waveform models that are not entirely analytical, these derivatives have to be computed numerically, and this adds additional obstacles to the use of Eq.~\eqref{eq:systematic_error_fisher} for the estimation of the systematic bias. For this reason, other methods are typically used to estimate for which values of the \ac{SNR} the systematic bias is larger than the statistical uncertainty. A very popular method for this purpose is the so-called Lindblom's criterium~\cite{Flanagan:1997kp,Lindblom:2008cm,McWilliams:2010eq,Chatziioannou:2017tdw}. According to Lindblom's criterium, a sufficient but not necessary condition such that all the parameters of the binary have biases smaller than the statistical uncertainty is that
\begin{equation}
\label{eq:lindblom_crit}
1 - F\left(h^{\mathrm{model}}(\bm{\lambda}_{\mathrm{max}}^\mathrm{GR}),h^{\mathrm{GR}}(\bm{\lambda}_{\mathrm{max}}^\mathrm{GR})\right) \leq \frac{N_\mathrm{intr} -1}{2\rho^2},
\end{equation}
where $F(\cdot,\cdot)$ is the faithfulness function, defined in Eq.~\eqref{eq:faithfulness_intro}, and $N_\mathrm{intr}$ is the number of the intrinsic parameters (masses and spins) of the system. For \ac{BBH} systems with precessing spins this number is $8$, while it is just $4$ for \acp{BBH} with non-precessing spins. Given a \ac{GR} waveform $h^{\mathrm{GR}}$ (typically a \ac{NR} waveform), and a waveform model $h^{\mathrm{model}}$, using Eq.~\eqref{eq:lindblom_crit} it is possible to compute the \ac{SNR} threshold below which the parameter estimation of the binary system is unbiased. Lindblom's criterium is easy to use, as it only requires the computation of a faithfulness, but it has some limitations. First of all, when the condition in Eq.~\eqref{eq:lindblom_crit} is violated, it is unknown what is (or are) the biased binary parameter (or parameters). In addition, being a sufficient and not necessary condition, it is often too conservative, as discussed in Ref.~\cite{Purrer:2019jcp}. The latter limitation makes Lindblom's criterium sometimes too strict for practical applications.
}

While the approximation of the likelihood function described in Eq.~\eqref{eq:likelihood_linear_signal_approx} is useful to qualitatively understand certain properties of the distribution, it is inappropriate to estimate the parameters on current LIGO-Virgo signals, for which the \ac{SNR} is not large enough to successfully use the likelihood approximation in Eq.~\eqref{eq:likelihood_linear_signal_approx}. For this reason, one has to compute the posterior distribution in Eq.~\eqref{eq:posterior_Bayes} numerically. Since this function has a very large dimensionality ($\mathcal{O}(15)$ parameters) and, considering that each evaluation of the likelihood in Eq.~\eqref{eq:LIGO_likelihood} takes $\mathcal{O}(1\,\mathrm{ms})$, it is impractical to compute directly the posterior distribution. For this reason, \textit{stochastic sampling methods} are used as alternative.
The first stochastic sampling method I use for the analyses in this thesis is the \textit{Markov Chain Monte Carlo (MCMC) method} described in Ref.~\cite{Veitch:2014wba}, and included in the \ac{LAL}~\cite{lalsuite} under the name \texttt{LALInference MCMC}. As alternative to the MCMC method, I also use the \textit{dynamic nested sampling method} \texttt{dynesty} described in~\cite{2020MNRAS.493.3132S} and implemented in the \texttt{bilby} software~\cite{Ashton:2018jfp}. The advantages of this method with respect to the MCMC method is that it provides a more accurate estimate of the evidence, and it converges faster to the true posterior distribution, especially when used in its highly parallelized implementation included in the software \texttt{parallel bilby}~\cite{Ashton:2018jfp}.

\section{Research overview}
\label{sec:research_overview}

In Secs.~\ref{sec:waveform_modeling_intro} and~\ref{sec:PE}, I summarize the reaserch work presented in this thesis that has been published in international peer-reviewed journals. I report in Chapters~\ref{chap:two}, \ref{chap:three}, \ref{chap:four} and \ref{chap:five} the published version of these articles. In the following, I briefly summarize my contribution to each of these publications. \robert{In addition, I summarize my contribution to some articles published by the LIGO Scientific and Virgo collaborations. }
\begin{enumerate}
\item Chapter~\ref{chap:two} consists of the paper:\newline \textbf{Roberto Cotesta} et al. \textit{Enriching the Symphony of Gravitational Waves from Binary Black Holes by Tuning Higher Harmonics}. Phys. Rev. D98(8):084028, 2018. \newline\newline I was the main developer of the waveform model \texttt{SEOBNRv4HM} for \robert{spinning, } non-precessing \ac{BBH} systems, which is described in the paper. I also: produced all the comparisons between the waveforms computed with this model and the \ac{NR} waveforms; made all the plots of the paper and wrote the article. In addition, I impletemented the computer code of the model in the \ac{LAL} software package, such that it could be used by the members of the LIGO Scientific and Virgo collaborations, and the rest of the community, as the \ac{LAL} software package is publicly available.
Finally, I led the review of \robert{\texttt{SEOBNRv4HM} within } the LIGO Scientific and Virgo collaborations, as it is a requirement to use the model in the analyses produced by the collaborations. During the review, a set of tests was performed on the waveform model to ensure that it always returned sensical waveforms. The computer code of the model was inspected to ensure the compatibility with the code standards of the collaborations, and with other softwares developed by the collaborations. Additional tests were also performed to confirm the accuracy of the waveform models against the \ac{NR} waveforms. All the tests were performed under the supervision of members of the collaboration with expertise in waveform modeling, but who were not involved in the development \robert{\texttt{SEOBNRv4HM}}.

\item Chapter~\ref{chap:three} consists of the paper:\newline Serguei Ossokine, Alessandra Buonanno, Sylvain Marsat, \textbf{Roberto Cotesta} et al. \textit{Multipolar Effective-One-Body Waveforms for Precessing Binary Black Holes: Construction and Validation}. Phys. Rev. D, 102(4):044055, 2020. \newline\newline In this work, I implemented the \acp{HM} in the waveform model \texttt{SEOBNRv4PHM} for precessing \acp{BBH} systems, which is presented in the paper. In addition, I performed the parameter-estimation studies in Sec.~\ref{sec:peEOBNR}, whose results are shown in Figs.~\ref{fig:PE_q_3} and~\ref{fig:PE_q_6} of the paper. I also wrote Secs.~\ref{sec:unfaith_sec} and~\ref{sec:peEOBNR} of the paper. In addition, I implemented part of the computer code of the model in the \ac{LAL} software package.
Finally, also for this waveform model, I was one of the three people responsible for its review, which followed the same procedure described before for \texttt{SEOBNRv4HM}.

\item Chapter~\ref{chap:four} consists of the paper:\newline \textbf{Roberto Cotesta} et al. \textit{Frequency domain reduced order model of aligned-spin effective-one-body waveforms with higher-order modes}. Phys. Rev. D, 101(12):124040, 2020. \newline\newline I was the main developer of \texttt{SEOBNRv4HM\_ROM}, the reduced-order model of \texttt{SEOBNRv4HM}, which is described in the paper. In Sec.~\ref{sec:results}, I compared its accuracy and speed against the original waveform model \texttt{SEOBNRv4HM}. The results of these analyses are summarized in Figs.~\ref{fig:average_unfaith_M} and~\ref{fig:speedup}, which I produced. In addition, I performed the parameter-estimation studies described in Sec.~\ref{sec:PEsec}, whose results are summarized in Figs.~\ref{fig:PE_all_EOBinj} and~\ref{fig:PE_all_NRinj}. I wrote the entire paper, with the exception of the Secs.~\ref{sec:intro_v4HMROM},\ref{sec:preparation_and_decomposition_of_wf_data},\ref{sec:SVD},\ref{sec:TPI} and \ref{sec:conclusion}, which were written by my collaborators. I also produced all the figures, with the exception of Figs.~\ref{fig:PE_all_EOBinj} and~\ref{fig:PE_all_NRinj}, which were produced by one of my collaborators. In addition, also in this case, I implemented the computer code of the model in the \ac{LAL} software package.
Finally, as for the waveform model \texttt{SEOBNRv4HM}, also for \texttt{SEOBNRv4HM\_ROM} I led its review, which consisted in a set of tests similar to those performed to \texttt{SEOBNRv4HM}. 

\item Chapter~\ref{chap:five} consists of the paper: \newline Katerina Chatziioannou, \textbf{Roberto Cotesta} et al. \textit{On the properties of the massive binary black hole merger GW170729}. Phys. Rev. D, 100(10):104015, 2019.
\newline\newline In this work, I performed the parameter-estimation study to estimate the \ac{BBH} parameters of the source of the \ac{GW} signal GW170729, using the waveform models \texttt{SEOBNRv4HM} and \texttt{SEOBNRv4\_ROM}. These studies are discussed in Sec.~\ref{CBCresults} of the paper, and the main results are summarized in Figs.~\ref{fig:intrinsic_IMREOB}, \ref{fig:systematics}, \ref{fig:extrinsic} and~\ref{fig:prior}. In addition, I also contributed to the writing of Secs.~\ref{analysis} and~\ref{CBCresults}.
\end{enumerate}

In addition to the publications listed above, I significatively contributed also to the article published by the LIGO Scientific and Virgo collaborations \robert{that presented the discovery } of the \ac{GW} signal GW190412
\begin{enumerate}
\setcounter{enumi}{4}
\item R. Abbott,..., \textbf{Roberto Cotesta} et al. \textit{GW190412: Observation of a Binary-Black-Hole Coalescence with Asymmetric Masses}. Phys. Rev. D, 102(4):043015, 2020.
\newline\newline
I was one of the five members in the editorial team of the paper. As member of the editorial team, I was responsible of the parameter-estimation analyses to measure the \ac{BBH} parameters of the source of the signal, using the waveform models \texttt{SEOBNRv4\_ROM}, \texttt{SEOBNRv4HM\_ROM}, \texttt{SEOBNRv4PHM} and \texttt{NRSurHyb3dq8}. The results of these analyses are described in Secs. 3 and 4 of the paper, and summarized in Figs.~2, 3, 4, 5 and 6. I also contributed to the writing of Secs. 1, 3, 4 and 7 of the paper. 
I summarize my contribution to this article in Sec.~\ref{sec:pe_GW190412} of this thesis. 
\end{enumerate}

I also contributed to other articles published by the LIGO Scientific and Virgo collaborations, describing the sources of the \ac{GW} signals detected during O2 and O3a
\begin{enumerate}
\setcounter{enumi}{5}
\item R. Abbott,..., \textbf{Roberto Cotesta} et al. \textit{GWTC-1: A Gravitational-Wave Transient Catalog of Compact Binary Mergers Observed by LIGO and Virgo during the First and Second Observing Runs}. Phys.Rev.X 9 (2019) 3, 031040, 2019.
\newline\newline
In this publication, I contributed to the early studies of the \ac{GW} signal GW170729, by performing the parameter-estimation analysis using the model \texttt{SEOBNRv4\_ROM}. In addition, I contributed to the writing of Appendix C3, about the impact of the \acp{HM} on the parameter estimation of the detected \ac{GW} signals. The discussions with other members of the LIGO Scientific and Virgo collaborations, during the analysis of the \ac{GW} signal GW170729, led to the publication I report in chapter~\ref{chap:five}.

\item I.M. Romero-Shaw,..., \textbf{Roberto Cotesta} et al., \textit{Bayesian inference for compact binary coalescences with bilby: validation and application to the first LIGO–Virgo gravitational-wave transient catalogue}. Mon.Not.Roy.Astron.Soc. 499 (2020) 3, 3295-3319, (2020)
\newline\newline

In this publication, I performed some tests for the LIGO and Virgo review of the parameter-estimation software \texttt{bilby}. These tests were necessary to ensure the correctness of this parameter-estimation software in its parallelized version called \texttt{parallel bilby}.

\item R. Abbott,..., \textbf{Roberto Cotesta} et al. \textit{GW190814: Gravitational Waves from the Coalescence of a 23 Solar Mass Black Hole with a 2.6 Solar Mass Compact Object}. Astrophys.J.Lett. 896 (2020) 2, L44, 2020.
\newline\newline
In this publication, I contributed to the parameter-estimation study on the \ac{GW} signal by performing the analysis using the model \texttt{SEOBNRv4HM\_ROM}. The results of this study are discussed in Sec. 4.1 and summarized in Fig. 4. My study contributed to the measurement of the mass of the lighter object in the binary system, whose nature is still under debate. In addition, \texttt{SEOBNRv4PHM} was used for measuring the parameters of this binary system. The results obtained by this analysis were combined together with those measured by the waveform model \texttt{IMRPhenomPv3HM}, which I will introduce later, and reported as the official measurements by the LIGO Scientific and Virgo collaborations.

\item R. Abbott,..., \textbf{Roberto Cotesta} et al. \textit{Properties and Astrophysical Implications of the 150 $M_\odot$ Binary Black Hole Merger GW190521}. Astrophys.J.Lett. 900 (2020) 1, L13, 2020.
\item R. Abbott et al. \textit{GW190521: A Binary Black Hole Merger with a Total Mass of 150 $M_{\odot}$}. Phys.Rev.Lett. 125 (2020) 10, 101102, 2020.
\newline\newline
In \robert{the above } two publications, I performed some of the early analyses on the \ac{GW} signal to measure the \ac{BBH} parameters, using the waveform models \texttt{SEOBNRv4\_ROM} and \texttt{SEOBNRv4HM\_ROM}. In addition, the waveform model \texttt{SEOBNRv4PHM}, which I developed, was used in the parameter-estimation study of the second publication to measure the parameters of the \ac{BBH} system.

\item R. Abbott,..., \textbf{Roberto Cotesta} et al. \textit{GWTC-2: Compact Binary Coalescences Observed by LIGO and Virgo During the First Half of the Third Observing Run}. arXiv:2010.14527
\newline\newline
In this publication, I was responsible for the early analyses of the \ac{GW} signal GW190602\_175927, which I performed using the waveform model \texttt{SEOBNRv4\_ROM}. In addition, \texttt{SEOBNRv4PHM}, the waveform model I developed, was used for the measurement of the \ac{BBH} parameters for all the \ac{GW} signals discussed in the paper. For many signals, the \ac{BBH} parameters measured with \texttt{SEOBNRv4PHM} were directly reported as the official measurements by the LIGO Scientific and Virgo collaborations. In other cases, the results obtained with this waveform model were first combined with those obtained with the waveform model \texttt{NRSur7dq4}, which I will introduce later, and then reported as the official measurements by the collaborations. See Table VIII in the paper for more details. The \ac{BBH} parameters measured from the \ac{GW} signals were later used, by the LIGO Scientific and Virgo collaborations, as input to analyze the properties of the observed population of \acp{BBH}~\cite{Abbott:2020gyp}. I will discuss in Sec.~\ref{sec:pe_GW190412} how the improved waveform models used to analyze the \ac{GW} signals also had an impact on this study.

\item R. Abbott,..., \textbf{Roberto Cotesta} et al. \textit{Tests of General Relativity with Binary Black Holes from the second LIGO-Virgo Gravitational-Wave Transient Catalog}. arXiv:2010.14529
\newline\newline
In this publication, the waveform model \texttt{SEOBNRv4HM} was used as a baseline to measure \robert{possible deviations from \ac{GR} } of the complex part of the dominant \ac{QNM} frequency of the \ac{BH} remnant from a \ac{BBH} merger. The \robert{parametrized } \texttt{SEOBNRv4HM} model to perform this measurement is called \texttt{pSEOBNRv4HM}, and it is described in Ref.~\cite{Ghosh:2021aa}. The results of this analysis are decribed in Sec. 7A of the paper. In addition, the waveform model \texttt{SEOBNRv4HM\_ROM} was used as baseline for the \acp{FTA} infrastructure~\cite{Abbott:2018lct}, which tested the departure from the \ac{GR} predictions of the \ac{PN} coefficients of the inspiral waveform. The results of this analysis can be found in Appendix C2 of the paper. 
\end{enumerate}

Finally, I served as waveform expert during the LIGO and Virgo reviews of the waveform models \texttt{NRHybSur3dq8}, \texttt{NRSur7dq4} and \texttt{IMRPhenomXHM}. I will describe these models later. For this task, I performed some of the sanity tests on these waveform models, and I supervised that the other tests were performed according to the guidelines of the LIGO Scientific and Virgo collaborations. 


\section{Gravitational waveform models with higher-order modes for spinning binary black-holes}
\label{sec:waveform_modeling_intro}

Accurate waveform models are crucial to detect \ac{GW} signals and correctly measure the parameters of the source, as already discussed in Secs.~\ref{sec:strategy_detection} and~\ref{sec:source_characterization}. I begin this section by introducing, in Sec.~\ref{sec:HM_importance}, the \acp{HM} of a \ac{GW} signal, and highlighting the importance of including them in waveform models to improve their accuracy. I introduce then, in Secs.~\ref{sec:intro_SEOBNRv4HM} and~\ref{sec:intro_SEOBNRv4PHM} respectively, the waveform models I developed within the \ac{EOB} formalism. They include these \acp{HM} for spinning \ac{BBH} systems, first in the case of \ac{BH} spins aligned with the \roberto{orbital } angular momentum of the binary, and then extended to generic spin directions. Finally, in Sec.~\ref{sec:ROM}, I introduce a method to reduce the time \roberto{to generate such waveforms}, which is \roberto{crucial } in data analysis applications \roberto{(detection and parameter inference)}.

\subsection{Motivations for including higher-order modes in gravitational waveforms}
\label{sec:HM_importance}

In this section, I describe what are the \acp{HM} and why it is important to include them in waveform models.

As already mentioned in Sec.~\ref{sec:waveform_anatomy}, the combination of the \ac{GW} polarizations $h_+ - ih_\times$ can be decomposed \roberto{in } $-2$-spin-weighted spherical harmonics ${}_{-2}Y_{\ell m}(\iota, \varphi_0)$ as
\begin{equation}
\label{eq:sph_harm_expansion_2}
h_+(t; \bm{\lambda})-ih_\times(t; \bm{\lambda}) = \sum_{\ell = 2}^\infty \sum_{m = -\ell}^\ell {}_{-2}Y_{\ell m}(\iota, \varphi_0)\, h_{\ell m}(t; \bm{\lambda}), 
\end{equation}
where $h_{\ell m}(t; \bm{\lambda})$ are the \ac{GW} modes, which are functions of time and the parameters of the binary system $\bm{\lambda}$, while the angles $(\iota,\varphi_0)$, used in the $-2$-spin-weighted spherical harmonics, define the direction of the observer in the source frame.
For simplicity, I restrict the discussion here to \ac{BBH} systems with spins aligned with the \roberto{orbital } angular momentum. \roberto{As already discussed in Sec.~\ref{sec:waveform_anatomy}, when the \ac{BH} spins are aligned with the orbital angular momentum of the binary, there is no precession of the orbital plane, and the direction of the orbital angular momentum stays constant over time. } For this reason, it is convenient to use the frame defined in Sec.~\ref{sec:waveform_anatomy} for the harmonic decomposition in Eq.~\eqref{eq:sph_harm_expansion_2}. In this frame, the $z$-axis is aligned with the constant direction of the orbital angular momentum of the binary, and the binary system is invariant under reflection across the orbital plane. This invariance implies \roberto{that the \ac{GW} modes with negative $m$ are linked to those with positive $m$ by the simple relation } $h_{\ell m} = (-1)^l h_{\ell -m}^*$, therefore I will only focus on modes with $m > 0$.
All the features I discuss here \roberto{in the simplified case of \acp{BH} with non-precessing spins, } are straightforward to generalize for systems with precessing spins by using the appropriate frame for the harmonic decomposition, described in Sec.~\ref{sec:precessing_modes}. 

\begin{figure}[h]
  \centering
  \includegraphics[width=0.7\textwidth]{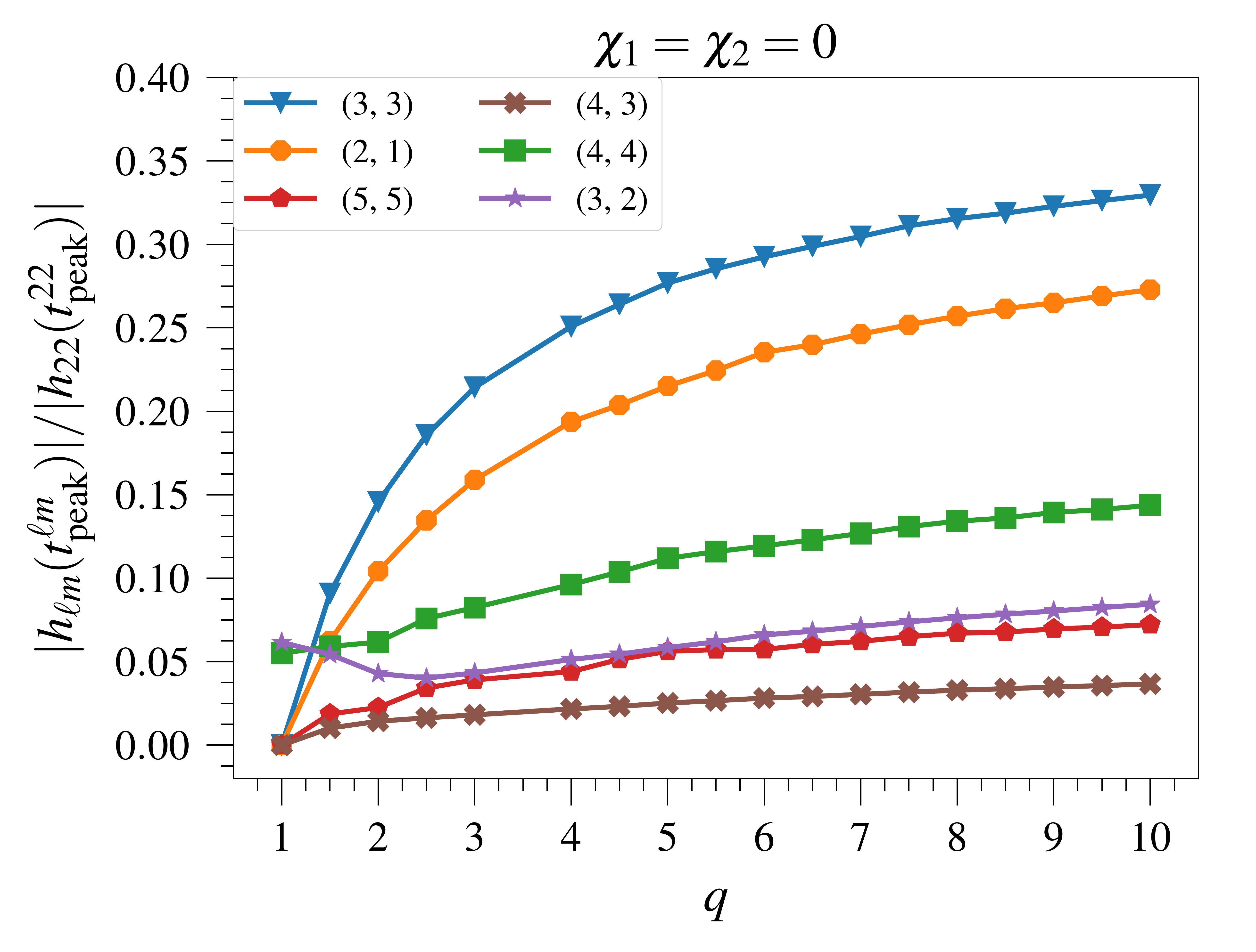}
\caption{Amplitude ratio between the $(\ell,m)$ mode and the dominant $(2,2)$ mode, both 
evaluated at their peak, as function of the mass ratio. 
Only nonspinning NR waveforms are used. 
The markers represent the \ac{NR} data, and are connected by a line. Figure taken from Ref.~\cite{Cotesta:2018fcv}}
\label{fig:rel_amp_hm_0spin_intro}
\end{figure}

For \ac{BBH} systems with comparable masses, as expected in the case of LIGO-Virgo sources, the leading term in Eq.~\eqref{eq:sph_harm_expansion_2} is the mode $(\ell, m) = (2, 2)$. The other modes, usually referred to as \acp{HM}, are one or two order of magnitude smaller. For this reason, they are typically neglected in waveform models. 
This approximation degrades when increasing the mass ratio $q \equiv m_1/m_2 \geq 1$ of the binary, \robert{because in this case the \acp{HM} become more and more relevant as the binary evolves toward merger}~\cite{Berti:2007fi,Kidder:2007rt,Blanchet:2008je,
Healy:2013jza,
Blanchet:2013haa}. A useful quantity \roberto{that illustrates this behaviour } is the ratio $|h_{\ell m}(t^{\ell m}_\mathrm{peak})|/|h_{2 2}(t^{2 2}_\mathrm{peak})|$, where $t^{\ell m}_\mathrm{peak}$  is the time for which the $(\ell, m)$ mode reaches its maximum value.
In Fig.~\ref{fig:rel_amp_hm_0spin_intro} I show this ratio for the largest \acp{HM} as a function of $q$. \roberto{The ratio presented in the figure is computed from } \ac{NR} simulations of nonspinning \acp{BBH}. The figure shows that this ratio is an increasing function of $q$ and, for $q \gtrsim 2$, the hierarchy of the largest \acp{HM} is $(\ell,m) = (3,3),(2,1),(4,4),(3,2),(5,5),(4,3)$. When $q$ approaches $1$ all the modes with odd $m$ vanish and the only \acp{HM} contributing to the waveform are $(\ell,m) = (4,4),(3,2)$. The hierarchy of the modes changes, in this case, because when the two \acp{BH} have the same masses and spins, the system is invariant under the rotation $\varphi_0 \rightarrow \varphi_0 + \pi$, and the modes with odd $m$ have to vanish as a consequence of this symmetry. These considerations on the mode hierarchy are roughly the same even when considering spinning \acp{BBH} (see Sec.~\ref{sec:intro_SEOBNRv4HM} for more details).

\begin{figure}[h]
  \centering
  \includegraphics[width=0.7\textwidth]{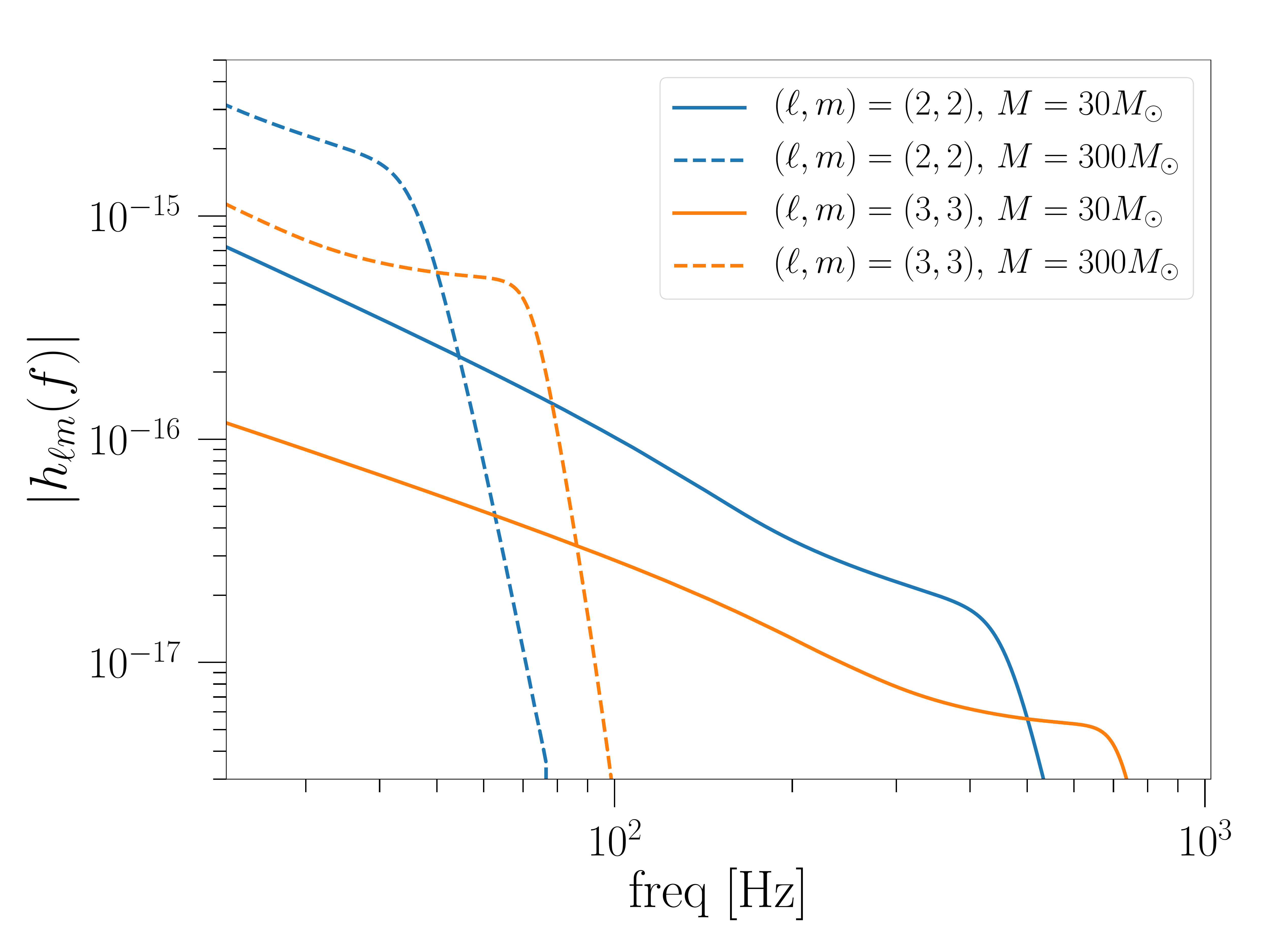}
\caption{Amplitude of the modes $(\ell, m) = (2,2), (3,3)$ for two different values of the total mass $M = 30 M_\odot$ and $M = 300M_\odot$. All the other parameters of the binary are fixed to be the same.}
\label{fig:modes_M}
\end{figure}

The contribution of the \acp{HM} in the waveform is also more relevant when increasing the total mass $M$ of the binary. This is easy to explain by looking at the mode amplitudes in Fourier domain in the frequency range $20 \lesssim f \lesssim 1$ kHz, where the LIGO-Virgo detectors are most sensitive. In the following, I explain the reason for this effect using the mode $(\ell, m) = (3,3)$ but the same is true for all the \acp{HM} with $\ell > 2$. In Fig.~\ref{fig:modes_M}, I show the frequency domain amplitude of the modes $(\ell, m) = (2,2),(3,3)$ for two values of the total mass $M = 30 M_\odot$ and $M = 300 M_\odot$, while all the other parameters of the binary are fixed to be the same. In both cases $|h_{22}(f)| > |h_{33}(f)|$ up to a cutoff frequency above which the opposite is true. This happens because the frequency for which the mode amplitude decays is approximately the ringdown frequency of each mode $f_{\ell m}^\mathrm{RD}$, which scales roughly as $f_{\ell m}^\mathrm{RD} \sim (\ell/2) f_{22}^\mathrm{RD}$. Since $f_{\ell m}^\mathrm{RD} \propto 1/M$, when increasing the total mass of the system a large part of the signal for which $|h_{22}(f)| > |h_{33}(f)|$ shifts to frequencies \roberto{lower than $20$ Hz, } where the detector is not sensitive. Also the region where $|h_{33}(f)| > |h_{22}(f)|$ shifts to lower frequencies but still within the \roberto{bandwidth } where the detector is sensitive. This means that for a larger fraction of the detectable signal $|h_{33}(f)| > |h_{22}(f)|$ with respect to the case with lower total mass, hence the mode $(\ell, m) = (3,3)$ is more relevant for larger total masses. 

\begin{figure}[h]
  \centering
  \includegraphics[width=\textwidth]{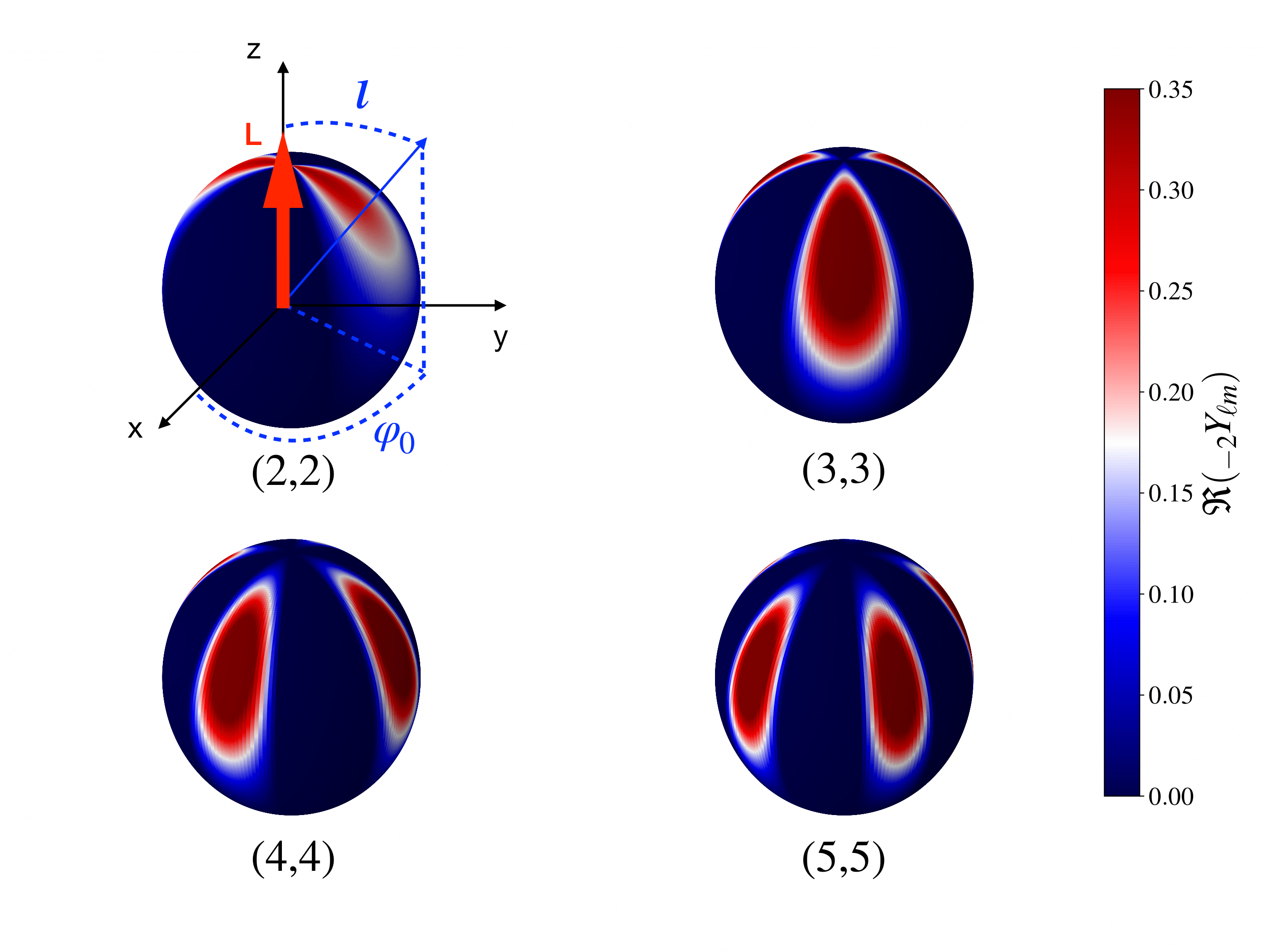}
\caption{Real part of the $-2$-spin-weighted spherical harmonics $(\ell, m) = (2,2),(3,3),(4,4),(5,5)$ as a function of the angles $(\iota, \varphi_0)$.}
\label{fig:spherical_harm}
\end{figure}

The impact of \acp{HM} on the waveform also depends on the direction of the observer with respect to the binary system. In fact, in Eq.~\eqref{eq:sph_harm_expansion_2} the \acp{HM} are multiplied by the $-2$-spin-weighted spherical harmonics ${}_{-2}Y_{\ell m}(\iota, \varphi_0)$, which can enhance or decrease the impact of \acp{HM} on the signal observed from a particular direction.
To better understand \roberto{in which direction the emission of \acp{HM} is more prominent, } it is instructive to look at the functions ${}_{-2}Y_{\ell m}(\iota, \varphi_0)$ when varying the angles $(\iota, \varphi_0)$. \roberto{Since they are complex functions, to show their dependece on $(\iota, \varphi_0)$ it is useful to plot their real and imaginary part. Since the only complex term in their definition in Eq.~\eqref{eq:sph_harm_def} is $e^{im\varphi_0}$, in practice one only needs to show their real part, because their imaginary part coincides with the real part rotated by $\pi/2$ around the $z$-axis. In Fig.~\ref{fig:spherical_harm}, I show the real part of the most important ${}_{-2}Y_{\ell m}(\iota, \varphi_0)$ when varying the angles $(\iota, \varphi_0)$. The real part of each ${}_{-2}Y_{\ell m}(\iota, \varphi_0)$ with a certain $m$ has exactly $m$ different maxima when varying $\varphi_0$, because the functions ${}_{-2}Y_{\ell m}(\iota, \varphi_0)$ depend on $\varphi_0$ only through the term $e^{im\varphi_0}$, which real part is $\cos(m\varphi_0)$. The dependence of the functions ${}_{-2}Y_{\ell m}(\iota, \varphi_0)$ on the angle $\iota$ is more relevant compared to that on $\varphi_0$ for the importance of the \acp{HM} when varying the direction of the observer with respect to the binary system. In fact, when varying $\iota$ from a face-on ($\iota = 0$) to an edge-on ($\iota = \pi/2$) orientation, one finds that the function ${}_{-2}Y_{2 2}(\iota, \varphi_0)$ associated with the mode $(\ell, m) = (2,2)$ has its maximum amplitude for $\iota = 0$, while the others have their maximum close to $\iota = \pi/2$\footnote{The situation is analogous when considering the modes with $m<0$. The only difference is that the function ${}_{-2}Y_{\ell m}(\iota, \varphi_0)$ associated with them are a reflection about the orbital plane of those described in Fig.~\ref{fig:spherical_harm}.}.} 
This means that neglecting the \acp{HM} in the waveform has a larger impact if the signal is observed close to an edge-on inclination. 

\roberto{The improvements in the sensitivity of the LIGO and Virgo detectors are opening the possibility to probe a wider portion of the \ac{BBH} population, and not just the \say{tip of the iceberg}, as happened with the first few detections of \ac{GW} signals from \ac{BBH} systems in the $2015-2016$ biennium. }
Since the waveform models are key \roberto{ingredients } for the detection and the source characterization of \acp{BBH}, their inaccuracy could cause a loss of interesting signals or biases in the measured source parameters. \roberto{For this reason, it is important that the accuracy of the waveform models improve at the same pace of the sensitivity of \ac{GW} detectors.}

From the detection point of view, the authors in Refs.~\cite{Bustillo:2015qty,Brown:2012nn,Pekowsky:2012sr,Varma:2014jxa,Capano:2013raa} showed that using waveform models without \acp{HM} \roberto{in } the template banks employed in the detection pipeline can cause a loss of $10 \%$ of the signals with mass ratio $q \geq 6$ ($q \geq 4$ in the case on nonspinning \acp{BBH}) and total mass $M \geq 100 M_\odot$~\cite{Harry:2017weg}. 
When waveform models without \acp{HM} are used for the parameter estimation of signals with \ac{SNR} larger than $25$, \ac{BBH} systems with mass ratio $q \geq 4$, total mass $M\geq 100$ and inclination $\pi/4 \leq \iota \leq 3\pi/4$ exhibit biases larger than the statistical uncertainty in the measured parameters~\cite{Shaik:2019dym,Kalaghatgi:2019log}. 
Signals with a larger \ac{SNR} yield smaller statistical errors and, at an \ac{SNR} of $48$, the systematic error from neglecting 
\acp{HM} can be larger than the statistical error~\cite{Littenberg:2012uj}, even for equal-mass systems. \roberto{Missing the detection of certain \ac{GW} signals from the \ac{BBH} population, or inferring biased measurements of their parameters, could have the effect of mischaracterizing the \ac{BBH} population, and obtaining a distorted understanding of the formation scenario of these \ac{BBH} systems.

Finally, including \acp{HM} in the waveform models allows to perform a wider set of tests of \ac{GR}, using \ac{BBH} systems as laboratories. For example, having multiple \ac{GW} modes in the ringdown part of the signal allows to test the Kerr nature of the \ac{BH} remnant of the \ac{BBH} coalescence, see Refs.~\cite{Dreyer:2003bv,Berti:2005ys,Meidam:2014jpa,Bhagwat:2016ntk,Berti:2016lat,Maselli:2017kvl,Gossan:2011ha,Brito:2018rfr}. In addition, using waveform models with \acp{HM} is important to avoid false deviation from \ac{GR} due to the missing of these effects in the waveform used to perform other tests of \ac{GR}~\cite{Pang:2018hjb}.}

For these reasons, waveform models including the effect of \acp{HM} have been developed over the years. The first of such models for a complete \ac{BBH} coalescence covering inspiral, merger and ringdown was developed within the \ac{EOB} framework in Ref.~\cite{Pan:2011gk} for nonspinning \acp{BBH}. I have \roberto{extended and improved } this model for spinning non-precessing~\cite{Cotesta:2018fcv} and precessing~\cite{Ossokine:2020kjp} \acp{BBH}. In Secs.~\ref{sec:intro_SEOBNRv4HM} and~\ref{sec:intro_SEOBNRv4PHM}, I provide an introduction to these models, while the complete discussion can be found in Chapters 2 and 3. Waveform models including \acp{HM} have been also developed within the phenomenological approach and the \ac{NR} surrogate method, outlined in Sec.~\ref{sec:waveform_anatomy}.
The models in the phenomenological approach are described in Ref.~\cite{Mehta:2017jpq} for nonspinning \acp{BBH}, and in Refs.~\cite{London:2017bcn} and \cite{Khan:2019kot} for spinning non-precessing and precessing \acp{BBH}, respectively. The names of the latter two models are respectively \texttt{IMRPhenomHM} and \texttt{IMRPhenomPv3HM}. For these two models, the mode $(\ell, m) = (2,2)$ is built as described in Sec.~\ref{sec:waveform_anatomy}, and the \acp{HM} are constructed by rescaling the mode $(\ell, m) = (2,2)$. In particular, during the inspiral regime the phases of the \acp{HM} are obtained using the approximate leading-order \ac{PN} rescaling $\phi_{\ell m}(t) \propto (m/2)\phi_{22}(t)$~\cite{Blanchet:2013haa}. A similar rescaling is also performed for the ringdown signal. The amplitudes are built in a similar fashion. New versions of these models that are not using this approximation have been recently developed for spinning non-precessing and precessing \acp{BBH} under the names \texttt{IMRPhenomXHM}~\cite{Garcia-Quiros:2020qpx} and \texttt{IMRPhenomXPHM}~\cite{Pratten:2020ceb}, respectively. In this case, each \ac{GW} mode is built separately by constructing phenomenological fits for the amplitude and the phase of hybrid \ac{GW} modes. The latter are assembled by smoothly blending \ac{EOB} inspiral modes with \ac{NR} modes including the late inspiral, merger and ringdown. 
The \ac{NR} surrogate models are described in Ref.~\cite{Varma:2018mmi}, in the case of \acp{BBH} with non-precessing spins and, in Refs.~\cite{Blackman:2017pcm,Varma:2019csw}, for precessing \ac{BBH} systems.
Because of the computational cost of producing \ac{NR} waveforms to construct the surrogate models, these waveform models can be used only in limited regions of the binary parameter space. In particular, the \ac{NR} surrogate for \ac{BBH} binaries with non-precessing spins called \texttt{NRHybSur3dq8}~\cite{Varma:2018mmi}, can only be used for \ac{BBH} with $q\leq8$ and dimensionless spins $|\bm{\chi_i}| \leq 0.8$. While in general also the time-duration of the waveforms generated by \ac{NR} surrogates is limited by the duration of the underlying \ac{NR} waveforms, this is not the case for this waveform model. In fact, in this case, the waveforms used to construct the surrogate are hybrids between \ac{NR} waveforms and \ac{EOB} inspiral waveforms.
Similar limitations are also present in the \ac{NR} surrogate for \acp{BBH} with precessing spins. In the case of the \ac{NR} surrogate \texttt{NRSur7dq4}~\cite{Varma:2019csw}, the model is limited to binary systems with $q\leq4$ and dimensionless spins $|\bm{\chi_i}| \leq 0.8$. In this case, also the duration of the waveform is limited and, for this reason, waveforms starting from a frequency of $20$ Hz can only be generated for \ac{BBH} systems with total masses $M \gtrsim 50 M_\odot$. \roberto{In Secs.~\ref{sec:intro_SEOBNRv4HM},\ref{sec:intro_SEOBNRv4PHM} and \ref{sec:PE}, I will compare these waveform models with \acp{HM} built within these other two approaches against the waveform models I developed. For these comparisons, I will also use two waveform models of the \say{previous generation}, which only include the mode $(\ell, m) = (2,2)$, and are limited to spins aligned with the orbital angular momentum. These models are \texttt{SEOBNRv4} (and its fast version \texttt{SEOBNRv4\_ROM})~\cite{Bohe:2016gbl}, belonging to the \ac{EOB} family, and \texttt{IMRPhenomD}~\cite{Husa:2015iqa, Khan:2015jqa}, developed within the phenomenological approach. I summarize all the waveform models used in this thesis in Table~\ref{tbl:wf_models}.}

\begin{table}
\centering
	\begin{tabular}{llll}
		\hline
		\hline
		Model name & \acp{HM} & Precession & Reference\\
		\hline
		\texttt{SEOBNRv4} & $\times$ & $\times$ & \cite{Bohe:2016gbl} \\
		\texttt{SEOBNRv4\_ROM} & $\times$ & $\times$ & \cite{Bohe:2016gbl} \\
		\textbf{\texttt{SEOBNRv4HM}} & \checkmark & $\times$ & Chapter~\ref{chap:two} \\
		\textbf{\texttt{SEOBNRv4HM\_ROM}} & \checkmark & $\times$ & Chapter~\ref{chap:four} \\
		\textbf{\texttt{SEOBNRv4PHM}} & \checkmark & \checkmark & Chapter~\ref{chap:three} \\
		\hline
		\texttt{IMRPhenomD} & $\times$ & $\times$ &\cite{Husa:2015iqa, Khan:2015jqa} \\
		\texttt{IMRPhenomHM} & \checkmark & $\times$ &\cite{London:2017bcn} \\
		\texttt{IMRPhenomXHM} & \checkmark & $\times$ &\cite{Garcia-Quiros:2020qpx} \\
		\texttt{IMRPhenomPv3HM} & \checkmark & \checkmark &\cite{Khan:2019kot} \\
		\texttt{IMRPhenomXPHM} & \checkmark & \checkmark &\cite{Pratten:2020ceb} \\
		\hline
		\texttt{NRHybSur3dq8} & \checkmark & $\times$ &\cite{Varma:2018mmi} \\
		\texttt{NRSur7dq4} & \checkmark & \checkmark &\cite{Varma:2019csw} \\
		\hline
		\hline
	\end{tabular}
	\caption{The waveform models used in this thesis. I also specify whether they include the effects of \acp{HM} and spin precession. \robert{I highlight in boldface the waveform models I developed.}}
	\label{tbl:wf_models}
\end{table}


\subsection{Effective-one-body waveform models with higher-order modes}
\label{sec:EOB}
In this section, after a general introduction \roberto{on } the \ac{BBH} dynamics within the \ac{EOB} formalism in Sec.~\ref{sec:EOB_dynamics}, I describe the expression of the \ac{GW} modes, and delineate the construction of the \ac{EOB} waveform model for spinning non-precessing \acp{BBH} with \acp{HM}, in Secs.~\ref{sec:EOB_waveform} and~\ref{sec:intro_SEOBNRv4HM}. Finally, in Secs.~\ref{sec:precessing_modes} and~\ref{sec:intro_SEOBNRv4PHM}, I outline the main features of the waveforms emitted by \acp{BBH} with precessing spins, and describe the \ac{EOB} waveform model for precessing \acp{BBH} with \acp{HM}.

\subsubsection{Two-body dynamics}
\label{sec:EOB_dynamics}

The \ac{EOB} formalism, proposed by Buonanno and Damour in Refs.~\cite{Buonanno:1998gg,Buonanno:2000ef}, reduces the relativistic two-body problem for \acp{BH} with generic masses and spins, to the problem of an \textit{effective body} moving in a central potential, similarly to what is done in the Newtonian case.

The natural starting point for this approach is the relativistic two-body problem in the \roberto{test-mass } limit, in which a \roberto{non-spinning } test-particle of mass $\mu$ is orbiting in the potential generated by a massive central object of mass $M$ $(M \gg \mu)$ and spin $S = |\bm{S}|$.
As discussed in Sec.~\ref{sec:GR_intro}, the metric tensor associated with a massive spinning object is the Kerr metric $g^{\mu\nu}_{\mathrm{Kerr}}$, which, in the Boyer-Lindquist coordinates, is defined by the line element
\begin{equation}
ds^2 = g^{\mu\nu}_{\mathrm{Kerr}} dx_\mu dx_\nu = -\frac{\Lambda}{\Delta \Sigma}dt^2 + \frac{\Delta}{\Sigma}dr^2 + \frac{1}{\Sigma}d\theta^2 + \frac{\Sigma - 2M r}{\Sigma \Delta \sin^2(\theta)}d\phi^2 - \frac{4M ra}{\Sigma \Delta}dtd\phi,
\end{equation}
with the Kerr spin $a \equiv S/M$, $\Delta \equiv r^2 -2 M r + a^2$, $\Sigma \equiv r^2 +a^2\cos^2\theta$ and $\Lambda \equiv (r^2 + a^2)^2 - a^2\Delta\sin^2\theta$. 

The dynamics of a non-spinning test-particle with mass $\mu$, in this gravitational background, is determined by its Hamiltonian. The latter can be obtained using the mass-shell constraint
\begin{equation}
\label{eq:mass_shell_constraint}
 g^{\mu\nu}_{\mathrm{Kerr}} p_\mu p_\nu = -\mu^2,
\end{equation}
where $p_\mu = (p_t \equiv -H^\mathrm{Kerr},p_r, p_\theta,p_\phi)$. By \roberto{solving Eq.~\eqref{eq:mass_shell_constraint} with respect to } $H^\mathrm{Kerr}$, one obtains
\begin{equation}
H^\mathrm{Kerr} = H^\mathrm{Kerr}_\mathrm{even} + H^\mathrm{Kerr}_\mathrm{odd},
\end{equation}
\roberto{where $H^\mathrm{Kerr}_\mathrm{even} $ is even in $a$ and reads}
\begin{equation}
H^\mathrm{Kerr}_\mathrm{even} = \alpha^\mathrm{Kerr}\sqrt{\mu^2 + \gamma_\mathrm{Kerr}^{\phi\phi} p_\phi^2 + \gamma_\mathrm{Kerr}^{rr} p_r^2 + \gamma_\mathrm{Kerr}^{\theta\theta} p_\theta^2},
\end{equation}
\roberto{while $H^\mathrm{Kerr}_\mathrm{odd}$, the odd in $a$ part, is}
\begin{equation}
H^\mathrm{Kerr}_\mathrm{odd} = \beta^\mathrm{Kerr} p_\phi,
\end{equation}
\roberto{and}
\begin{subequations}
\begin{align}
\alpha^\mathrm{Kerr} \equiv& \frac{1}{\sqrt{-g_\mathrm{Kerr}^{tt}}} = \sqrt{\frac{\Delta \Sigma}{\Lambda}}, \\
\beta^\mathrm{Kerr} \equiv & \frac{g_\mathrm{Kerr}^{t\phi}}{g_\mathrm{Kerr}^{tt}} = \frac{2aMr}{\Lambda}, \\
\gamma_\mathrm{Kerr}^{\phi\phi} \equiv & g_\mathrm{Kerr}^{\phi\phi} - \frac{g_\mathrm{Kerr}^{t\phi}g_\mathrm{Kerr}^{t\phi}}{g_\mathrm{Kerr}^{tt}} = \frac{\Sigma}{\Lambda \sin^2\theta}, \\
\gamma_\mathrm{Kerr}^{rr} \equiv & g_\mathrm{Kerr}^{rr} = \frac{\Delta}{\Sigma}, \\
\gamma_\mathrm{Kerr}^{\theta\theta} \equiv & g_\mathrm{Kerr}^{\theta\theta} = \frac{1}{\Sigma}.
\end{align}
\end{subequations}

This Hamiltonian can be generalized for a spinning test-particle, at leading order in the spin of the test-particle, by substituting $p_\mu$ with $ P_\mu \equiv p_\mu - 1/2\, \omega_{\mu a b} S^{ab}_{*} + \mathcal{O}(S_{*}^2)$ in Eq.~\eqref{eq:mass_shell_constraint} (see Refs.~\cite{Barausse:2009aa,Vines:2016unv}), with the quantities $\omega_{\mu a b}$ and $S^{ab}_{*}$ being the Ricci rotation coefficients and the spin tensor in a local Lorentz frame.
The Hamiltonian for a spinning test-particle in a Kerr background is then
\begin{equation}
H^\mathrm{Kerr}_\mathrm{S} = H^\mathrm{Kerr} + H^\mathrm{Kerr}_\mathrm{S_*},
\end{equation}
with
\begin{align}
H^\mathrm{Kerr}_\mathrm{S_*} =& \left[\bm{F_t} + \left(\beta^\mathrm{Kerr} + \frac{\alpha^\mathrm{Kerr}\gamma_\mathrm{Kerr}^{\phi\phi}p_\phi}{\sqrt{q^\mathrm{Kerr}}}\right) \bm{F_\phi} \right] \cdot \bm{S_*} \nonumber \\ 
&+ \frac{\alpha^\mathrm{Kerr}}{\sqrt{q^\mathrm{Kerr}}}(\gamma_\mathrm{Kerr}^{rr} p_r \bm{F_r} + \gamma_\mathrm{Kerr}^{\theta\theta} p_\theta \bm{F_\theta}) \cdot \bm{S_*} + \mathcal{O}(S_*^2),
\end{align}
where
\begin{equation}
\sqrt{q^\mathrm{Kerr}} \equiv \frac{H_\mathrm{even}^\mathrm{Kerr}}{\alpha^\mathrm{Kerr}},
\end{equation}
and the explicit expression of the fictitious gravitomagnetic force $F_\mu$ are given in Eq.(6) of Ref.~\cite{Hinderer:2013uwa}.

\begin{figure}[h]
  \centering
  \includegraphics[trim=5cm 5cm 5cm 5cm, clip,width=\textwidth]{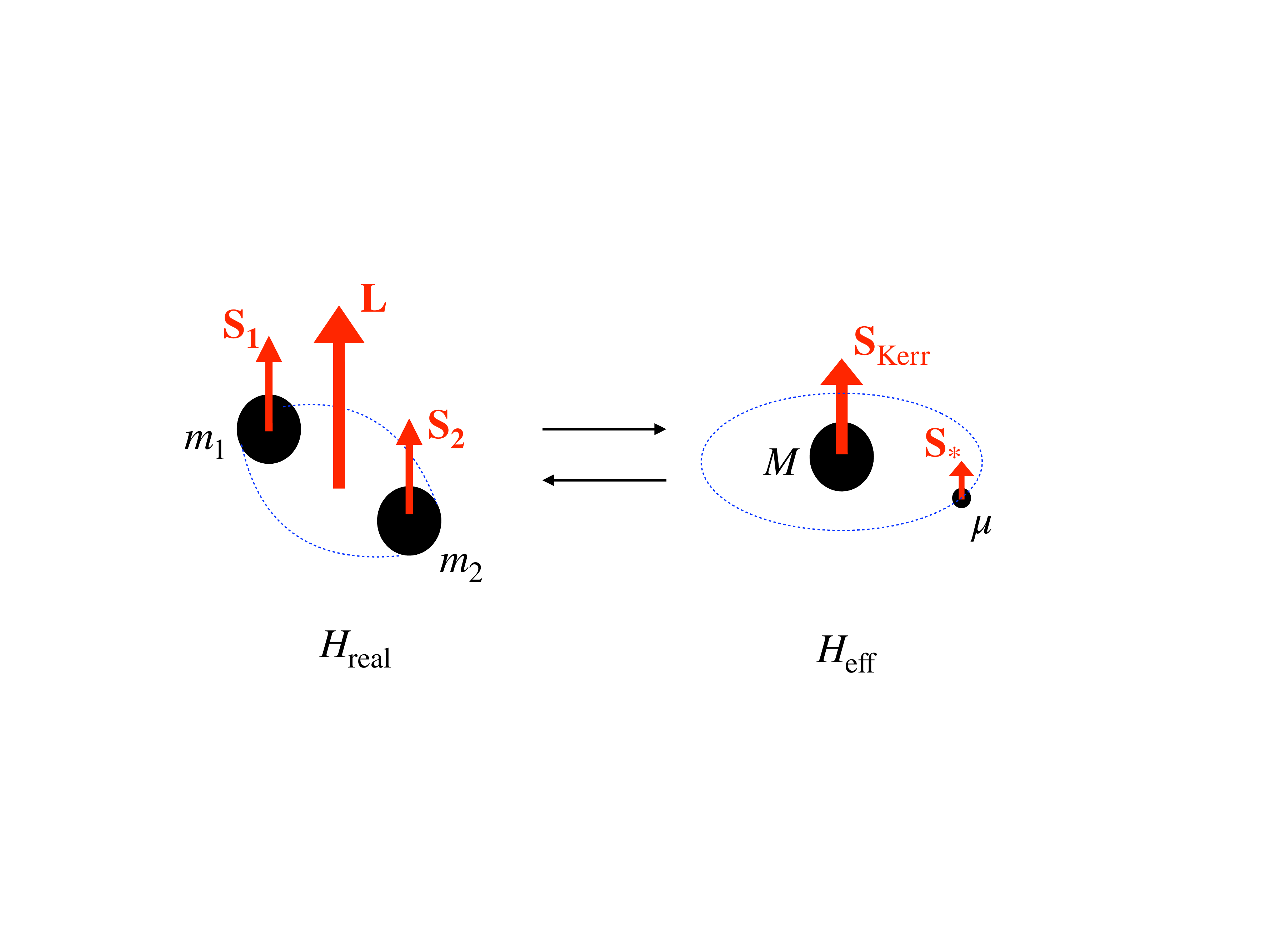}
\caption{The \ac{EOB} mapping between the real two-body problem of two objects with masses $m_1$ and $m_2$ and spins $\mathbf{S_1}$ and $\mathbf{S_2}$, and the effective problem of an effective particle with mass $\mu$ and spin $\mathbf{S_*}$ orbiting around a central object with mass $M$ and spin $\mathbf{S_\mathrm{eff}}$. The motion of the real two-body problem is determined by the Hamiltonian $H_\mathrm{real}$, while the Hamiltonian for the effective problem is $H_\mathrm{eff}$.}
\label{fig:EOB_map}
\end{figure}

In the \ac{EOB} formalism, one assumes that the solution for the relativistic two-body problem, with generic masses $m_1$ and $m_2$ and spins $\bm{S_1}$ and $\bm{S_2}$, can be obtained as a deformation of the solution in the \roberto{test-mass } limit, with the symmetric mass ratio $\nu = m_1 m_2/(m_1+m_2)^2$ as the deformation parameter. In practice, when $\nu \neq 0$, the solution of the relativistic two-body problem has to be described by the motion of an effective particle with mass $\mu = \mu(m_1,m_2)$ and spin $\bm{S_*} =  \bm{S_*}(m_1,m_2,\bm{S_1},\bm{S_2})$, whose orbit is determined by the Hamiltonian $H^\mathrm{eff}$, associated with an effective metric $g_{\mu\nu}^\mathrm{eff}$, of a central object with mass $M = M(m_1,m_2)$ and spin $\bm{S}_\mathrm{Kerr} = \bm{S}_\mathrm{Kerr}(m_1,m_2,\bm{S_1},\bm{S_2})$. 
The first trivial constraint that can be set to determine the functional form of $\mu(m_1,m_2)$, $\bm{S_*}(m_1,m_2,\bm{S_1},\bm{S_2})$, $M(m_1,m_2)$, $\bm{S}_\mathrm{Kerr}(m_1,m_2,\bm{S_1},\bm{S_2})$ and $H^\mathrm{eff}$ is obtained by requiring the correct limit in the \roberto{test-mass limit } case ($\nu \rightarrow 0$ or $m_2/m_1 \rightarrow 0$)

\begin{subequations}
\begin{align}
H^\mathrm{eff} &\xrightarrow{\nu \rightarrow 0} H^\mathrm{Kerr}_\mathrm{S} \\
M(m_1,m_2) &\xrightarrow{\nu \rightarrow 0} m_1\\
\mu(m_1,m_2) &\xrightarrow{\nu \rightarrow 0} m_2 \\
\bm{S}_\mathrm{Kerr}(m_1,m_2,\bm{S_1},\bm{S_2}) &\xrightarrow{\nu \rightarrow 0}  \bm{S_1}\\
\bm{S_*}(m_1,m_2,\bm{S_1},\bm{S_2}) &\xrightarrow{\nu \rightarrow 0} \bm{S_2}. 
\end{align}
\end{subequations}
The correction to these expressions will have to be $\mathcal{O}(\nu)$, to appropriately recover the test-particle limit. In addition, without loss of generality, one can assume the functions $M(m_1,m_2)$ and $\mu(m_1,m_2)$ to be the total mass $M = m_1 + m_2$ and the reduced mass $\mu = m_1 m_2/(m_1+m_2)$ respectively,  as in the Newtonian case. Finally, to find the expressions for $H^\mathrm{eff}$, $\bm{S}_\mathrm{Kerr}(m_1,m_2,\bm{S_1},\bm{S_2})$ and $\bm{S_*}(m_1,m_2,\bm{S_1},\bm{S_2})$, one can use the knowledge of the \ac{PN} Hamiltonian for the real relativistic two-body problem $H^\mathrm{real}$, at the highest known \ac{PN} order. For this purpose, it is necessary to find a map $H^\mathrm{eff} = f(H^\mathrm{real})$ between the Hamiltonian $H^\mathrm{real}$ and the one of the effective problem $H^\mathrm{eff}$. This map was found in Ref.~\cite{Buonanno:1998gg}, in the case of non-spinning black holes, using a $H^\mathrm{real}$ at 2\ac{PN} order and by demanding an identification, between the real and the effective problem, of the radial action integral and the orbital angular momentum, in the context of the Hamilton-Jacobi framework. The result is rather simple
\begin{equation}
\label{eq:EOB_hamiltonian_mapping}
H^\mathrm{EOB} = M\sqrt{1+2\nu\left(\frac{H^\mathrm{eff}}{\mu} -1 \right)}.
\end{equation}
Using this map, and the \ac{PN} Hamiltonian $H^\mathrm{real}$ up to a certain order, one can find the expression of $H^\mathrm{eff}$, $\bm{S}_\mathrm{Kerr}(m_1,m_2,\bm{S_1},\bm{S_2})$ and $\bm{S_*}(m_1,m_2,\bm{S_1},\bm{S_2})$ by demanding that, when expanded as \ac{PN} series, the $H^\mathrm{EOB}$ Hamiltonian agrees with the $H^\mathrm{real}$ Hamiltonian, up to a canonical transformation.
The explicit expressions of the Hamiltonian $H^\mathrm{real}$ used for the work in this thesis can be found in Sec.~2C of Ref.~\cite{Barausse:2011ys}. The expression for the spins $\bm{S}_\mathrm{Kerr}(m_1,m_2,\bm{S_1},\bm{S_2})$ and $\bm{S_*}(m_1,m_2,\bm{S_1},\bm{S_2})$ is given by
\begin{subequations}
\begin{align}
\bm{S}_\mathrm{Kerr} &= \bm{S_1} + \bm{S_2}, \\
\bm{S_*} &= \frac{m_2}{m_1}\bm{S_1} + \frac{m_1}{m_2}\bm{S_2} + \bm{\Delta}_{\sigma^*}^{(1)} + \bm{\Delta}_{\sigma^*}^{(2)},
\end{align}
\end{subequations}
where $\bm{\Delta}_{\sigma^*}^{(1)}$ and $\bm{\Delta}_{\sigma^*}^{(2)}$\footnote{The spin map is not unique and in Sec.2E of Ref.~\cite{Barausse:2011ys} the authors also explore the alternative map $\bm{S}_\mathrm{Kerr} = \bm{S_1} + \bm{S_2}$ and $\bm{S_*} = \bm{S_1}m_2/m_1 + \bm{S_2}m_1/m_2$.} are spin-orbit terms, explicitly given in Eqs~(51) and (52) of Ref.~\cite{Barausse:2011ys}. In Fig.~\ref{fig:EOB_map}, I show a schematic picture of the \ac{EOB} mapping between the real and effective two-body problem.

The expression of $H^\mathrm{EOB}$ can be finally used to compute the dynamics of the real relativistic two-body problem, by simply employing Hamilton's equations\footnote{Note that also the EOB spins are canonical variable~\cite{Damour:2001tu}, therefore also their evolution is determined by Hamilton's equations.}
\begin{subequations}
\begin{align}
\frac{d\bm{R}}{dt} =& \frac{\partial H^\mathrm{EOB}}{\partial \bm{p}} \label{eq:EOB_Hamiltonian_eqs_1} \\
\frac{d\bm{P}}{dt} =& -\frac{\partial H^\mathrm{EOB}}{\partial \bm{R}} + \bm{\mathcal{F}} \label{eq:EOB_Hamiltonian_eqs_2} \\
\frac{d\bm{S_i}}{dt} =& \frac{\partial H^\mathrm{EOB}}{\partial \bm{S_i}} \times \bm{S_i} \label{eq:EOB_Hamiltonian_eqs_3},
\end{align}
\end{subequations}
where $i = 1,2$, $\bm{R} = (R, \Phi, \Theta)$ are spherical coordinates, $\bm{P} = (P_R, P_\Phi, P_\Theta)$ the conjugate momenta and $\bm{\mathcal{F}}$ is a dissipative force, to account for radiation-reaction effects due to the emission of \acp{GW}. The expression for $\bm{\mathcal{F}}$ can be derived by using its relation with the gravitational-wave energy flux $F$ (see Ref.~\cite{Buonanno:2005xu})
\begin{equation}
\bm{\mathcal{F}} = \frac{F}{\Omega|\bm{R}\times\bm{P}|}\bm{P},
\end{equation}
where $\Omega \equiv  \bm{L}_N \cdot (\bm{R}\times\dot{\bm{R}})/R^2$ is the orbital frequency and $\bm{L}_N$ the Newtonian angular momentum. The gravitational-wave flux $F$ is available as \ac{PN} series (e.g. Refs.~\cite{Peters:1963ux,Arun:2007sg,Arun:2007rg,Arun:2009mc,Kidder:1995zr,Maia:2017yok,Bohe:2013cla,Bohe:2015ana,Blanchet:2013haa}). In the \ac{EOB} framework, the flux is used in a factorized form (see e.g. Ref.~\cite{Damour:2007xr,Damour:2008gu,Pan:2010hz,Nagar:2016ayt,Messina:2018ghh,Nagar:2019wrt}), to improve its accuracy with the numerical results, that can be computed in the test-particle limit~\cite{Poisson:1993zr,Poisson:1995vs} within the Regge-Wheeler-Zerilli and Teukolsky equations~\cite{1957PhRv..108.1063R,Zerilli:1970se,Zerilli:1971wd,Teukolsky:1973ha}. The expression of the factorized gravitational-wave flux is:
\begin{equation}
\label{eq:EOB_energy_flux}
F = \frac{1}{8\pi}\frac{c^5}{G}\frac{GM\Omega}{c^3} \sum_{\ell = 2}^{\infty} \sum_{m = 1}^{\ell} m^2 \left|\frac{D_\mathrm{L}c^2}{G M} h^F_{\ell m}\right|^2,
\end{equation}
where $D_\mathrm{L}$ is the source-observer distance, and $h^F_{\ell m}$ are the factorized gravitational-wave modes that I will describe in detail in the next section. Although the sum over $\ell$ has to be carried up to infinity, in practice it is limited to the maximum value of $\ell$ for which the modes are known in the \ac{PN} series.
Equations.~\eqref{eq:EOB_Hamiltonian_eqs_1}, \eqref{eq:EOB_Hamiltonian_eqs_2} and \eqref{eq:EOB_Hamiltonian_eqs_3} can be solved numerically by providing appropriate initial conditions~\cite{Buonanno:2005xu}.

The dynamics of the relativistic two-body problem, obtained within the \ac{EOB} formalism, has been proven to be more accurate than the \ac{PN} dynamics, when compared to \ac{NR} simulations~\cite{Ossokine:2017dge,Nagar:2015xqa,Antonelli:2020aeb,Antonelli:2019ytb,Dietrich:2016lyp}. This is not unexpected, since the \ac{EOB} Hamiltonian, in Eq.~\eqref{eq:EOB_hamiltonian_mapping}, includes non-perturbative information about the relativistic two-body problem in the \roberto{test-mass limit}, in addition to the \ac{PN} informations. Moreover the \ac{EOB} framework provides flexibility to include additional information about the relativistic two-body problem computed using other methods, like \ac{NR}~\cite{Taracchini:2012ig,Taracchini:2013rva,Bohe:2016gbl}, gravitational self-force~\cite{Antonelli:2019fmq} and \ac{PM} expansion~\cite{Antonelli:2019fmq}, to further improve the accuracy of the two-body dynamics. In particular, the \ac{EOB} dynamics used for the work in this thesis, includes \roberto{calibration parameters at \ac{PN} orders beyond the ones available, in the effective metric $g_{\mu\nu}^\mathrm{eff}$, and in the gyrogravito-functions entering in the definition of $\bm{S_*}$}~\cite{Bohe:2016gbl}.

\subsubsection{Gravitational-waveform modes for non-precessing binary black-hole systems}
\label{sec:EOB_waveform}

In this section, I describe how to compute the \ac{GW} modes in the \ac{EOB} formalism for non-precessing \acp{BBH}. The generalization to precessing systems will be discussed in Sec.~\ref{sec:precessing_modes}.

In the \ac{EOB} framework, the \ac{GW} modes are \roberto{obtained by combining their expressions in } the inspiral-plunge and merger-ringdown regimes (as defined in Sec.~\ref{sec:waveform_anatomy})
\begin{align}
\label{eq:GWmodes}
h_{\ell m}(t) = \begin{cases}
h_{\ell m}^{\mathrm{insp-plunge}}(t), &t \leq t_{\textrm{match}}^{\ell m}\\
h_{\ell m}^{\mathrm{merger-RD}}(t), &t > t_{\textrm{match}}^{\ell m}.\\
\end{cases}
\end{align}
The time $t_{\textrm{match}}^{\ell m}$, at which the transition between the two regimes occurs, depends on the mode, but typically corresponds to the time of the amplitude peak of the $(\ell, m) = (2,2)$ mode (see Sec.~\ref{subsec:waveform_inspiral} for more details). The continuity of $h_{\ell m}(t)$ for $t = t_{\textrm{match}}^{\ell m}$ is guaranteed by the definition of $h_{\ell m}^{\mathrm{merger-RD}}(t)$ (see Eqs.~\eqref{eq:continuity_amp} and~ \eqref{eq:continuity_phase}).

During the inspiral phase, the emission of \acp{GW} by the \ac{BBH} system can be computed as a \ac{PN} expansion (see Refs.~\cite{Peters:1963ux,Arun:2007sg,Arun:2007rg,Arun:2009mc,Kidder:1995zr,Maia:2017yok,Bohe:2013cla,Bohe:2015ana,Blanchet:2013haa}), which provides the natural starting point to derive $h_{\ell m}^{\mathrm{insp-plunge}}$.
Similarly to what is done for the dynamics, one can think to improve the accuracy of the \ac{PN} expression of the \ac{GW} modes by using results computed in the \roberto{test-mass } limit.
\roberto{For this purpose, Refs.~\cite{Damour:2007xr,Damour:2008gu,Pan:2010hz} proposed to introduce the factorized form of the \ac{PN} \ac{GW} modes $h^F_{\ell m}$, also used in the \ac{GW} energy flux in Eq.~\eqref{eq:EOB_energy_flux}, to improve the accuracy of the \ac{PN} \ac{GW} modes} 
\begin{equation}
\label{eq: factorized_modes_introduction}
h_{\ell m}^{\textrm{F}} = h_{\ell m}^{(N, \epsilon)}\, \hat{S}^{(\epsilon)}_{\textrm{eff}}\, T_{\ell m}\,f_{\ell m}\, e^{i \delta_{\ell m}}.
\end{equation}
\roberto{The quantity }$h_{\ell m}^{(N, \epsilon)}$ is simply the leading-order Newtonian \roberto{\ac{GW} mode}, the function $\hat{S}^{(\epsilon)}_{\textrm{eff}}$ is an effective source, inspired by the source term present in the right-hand side of the Regge-Wheeler-Zerilli equation~\cite{1957PhRv..108.1063R,Zerilli:1970se,Zerilli:1971wd}, \roberto{that describes the gravitational radiation in the test-mass limit at linear order in the perturbation theory. }
This term naturally reproduces the pole at the light ring that is observed in the test-particle limit for circular orbits~\cite{Damour:1997ub}.
The function $T_{\ell m}$ is a resummation of the \ac{PN} leading-order logarithmic terms in the orbital frequency, due to the back-scattering of the gravitational waves off the effective potential well. Finally, the functions $f_{\ell m}$ and $e^{i \delta_{\ell m}}$ are amplitude and phase corrections necessary to recover the known \ac{PN} series when computing the \ac{PN} expansion of Eq.~\eqref{eq: factorized_modes_introduction}. The explicit definition of each term is given in Sec.~\ref{subsec:waveform_inspiral}. The expression of $f_{\ell m}$ can be further resummed to improve the accuracy of $h_{\ell m}^{\textrm{F}}$ against the numerical results in the test-particle limit. In particular, for the work in this thesis, I use the resummation proposed in Refs.~\cite{Damour:2008gu,Pan:2010hz}. An alternative has been recently proposed in Refs.~\cite{Nagar:2016ayt,Messina:2018ghh,Nagar:2019wrt}.

The underlying assumption for the \ac{PN} computation of the \ac{GW} modes is the quasi-circularity of the orbit. Therefore, it is natural to expect that the accuracy of $h_{\ell m}^{\textrm{F}}$ degrades close to the plunge, when the orbit is no longer quasi-circular.
In the \ac{EOB} framework, the degradation of $h_{\ell m}^{\textrm{F}}$ in this regime is corrected through a phenomenological \roberto{function } $N_{\ell m}$, called the non-quasicircular (NQC) term, that multiplies the factorized expression of the modes
\begin{equation}
h_{\ell m}^{\mathrm{insp-plunge}} = N_{\ell m}\,h_{\ell m}^{\textrm{F}}.
\end{equation}
The function $N_{\ell m}$ is a polynomial in the canonical momentum $P_\mathrm{r}$ (see Eq.~\eqref{eq:NQC_corrections} for its explicit definition), \robert{to account for the \ac{GW} emission from the radial motion. }
The coefficients of \roberto{the polynomial } are tuned to reproduce the shape of numerical waveforms around the plunge and, by construction, $N_{\ell m} \rightarrow 1$ far away from this regime.

After the plunge, the \ac{BBH} system undergoes the fully non-linear merger phase, \roberto{which results in } the formation of a perturbed \ac{BH}, that relaxes to an equilibrium state by emitting \acp{GW}.
\roberto{As discussed in Sec.~\ref{sec:waveform_anatomy}}, this \ac{GW} signal can be accurately computed, within the black-hole perturbation theory framework, as a linear \roberto{superposition } of \acp{QNM}~\cite{Kokkotas:1999bd},
\begin{equation}
\label{eq:simple_RD}
h_{\ell m}^\mathrm{RD}(t) = \sum_{n = 0}^\infty A_{\ell m n} e^{-i\sigma_{\ell m n}t},
\end{equation}
where $A_{\ell m n}$ are complex functions, depending on the details of the merger, and, $\sigma_{\ell m n}$ are the complex \ac{QNM} frequencies, depending on the mass and spin of the perturbed \ac{BH}. The latters are typically computed from the masses and spins of the \ac{BBH} system using fitting formulae derived from \ac{NR} waveforms~\cite{Barausse:2012qz, Hemberger:2013hsa,Kumar:2015tha}. The expression in Eq.~\eqref{eq:simple_RD} was previously used by \ac{EOB} waveform models as starting point for building the merger-RD part of the signal~\cite{Taracchini:2012ig,Taracchini:2013rva}. In that case, the model included $N$ overtones and the values of $A_{\ell m n}$ were computed by requiring continuity conditions between the function $h_{\ell m}^{\mathrm{insp-plunge}}(t)$ and $h_{\ell m}^{\mathrm{merger-RD}}(t)$ \roberto{on a comb } for $t < t_{\textrm{match}}^{\ell m}$. This procedure was numerically unstable in certain circumstances. For this reason, a simpler ansatz for modeling the merger-RD signal was introduced in Ref.~\cite{Bohe:2016gbl}, based on the studies in Ref.~\cite{Nagar:2016iwa}.
The expression of this ansatz is
\begin{equation}
h_{\ell m}^{\textrm{merger-RD}}(t) = \nu \ \tilde{A}_{\ell m}(t)\ e^{i \tilde{\phi}_{\ell m}(t)} \ e^{-i \sigma_{\ell m 0}(t-t_{\textrm{match}}^{\ell m})},
\end{equation}
where $\sigma_{\ell m 0}$ is the least damped \ac{QNM}, that dominates the signal for $t \gg t_{\textrm{match}}^{\ell m}$, while $\tilde{A}_{\ell m}(t)$ and $e^{i \tilde{\phi}_{\ell m}(t)}$ are amplitude and phase corrections (see Eqs.~\eqref{eq:ansatz_amp} and \eqref{eq:ansatz_phase}), tuned to reproduce the shape of numerical waveforms for $t \sim t_{\textrm{match}}^{\ell m}$. The contribution of the overtones to the signal, which is relevant for $t \sim t_{\textrm{match}}^{\ell m}$ (see Refs.~\cite{Giesler:2019uxc}), is replaced by the phenomenological functions $\tilde{A}_{\ell m}(t)$ and $e^{i \tilde{\phi}_{\ell m}(t)}$ in this simplified version of the model. 

\subsubsection{\texttt{SEOBNRv4HM}: the inspiral-merger-ringdown waveform model including higher-order modes for binary black holes with non-precessing spins}
\label{sec:intro_SEOBNRv4HM}

\roberto{Chapter~\ref{chap:two} is the publication that describes the \ac{EOB} waveform model that I developed, henceforth \texttt{SEOBNRv4HM}, which includes the effect of \acp{HM} for \acp{BBH} with spins aligned (or anti-aligned) with the orbital angular momentum of the binary. } 

\texttt{SEOBNRv4HM} is based on the model described in Ref.~\cite{Bohe:2016gbl},  henceforth \texttt{SEOBNRv4}, with which it shares the two-body dynamics (also described in Sec.~\ref{sec:EOB_dynamics}), and the mode $(\ell, |m|) = (2,2)$. The novelty in \texttt{SEOBNRv4HM}, compared to \texttt{SEOBNRv4}, is the inclusion of the \acp{HM} $(\ell, |m|) = (2,1), (3,3), (4,4), (5,5)$. 
For these new \ac{GW} modes, I included new \ac{PN} terms in the factorized form, as defined in Eq.~\eqref{eq: factorized_modes_introduction}, and I tuned the coefficients of the phenomenological functions $N_{\ell m}$, $\tilde{A}_{\ell m}$ and $\tilde{\phi}_{\ell m}$ \roberto{using $157$ \ac{NR} waveforms produced by the \texttt{SXS} Collaboration, in the mass ratio and dimensionless spin ranges $1 \leq q \leq 10$, $-0.99 \leq \chi_\mathrm{i} \leq 0.99$ (see Appendix~\ref{sec:NRcatalog} for more details about these \ac{NR} waveforms). }

\begin{figure}[h]
  \centering
\includegraphics[width=\textwidth]{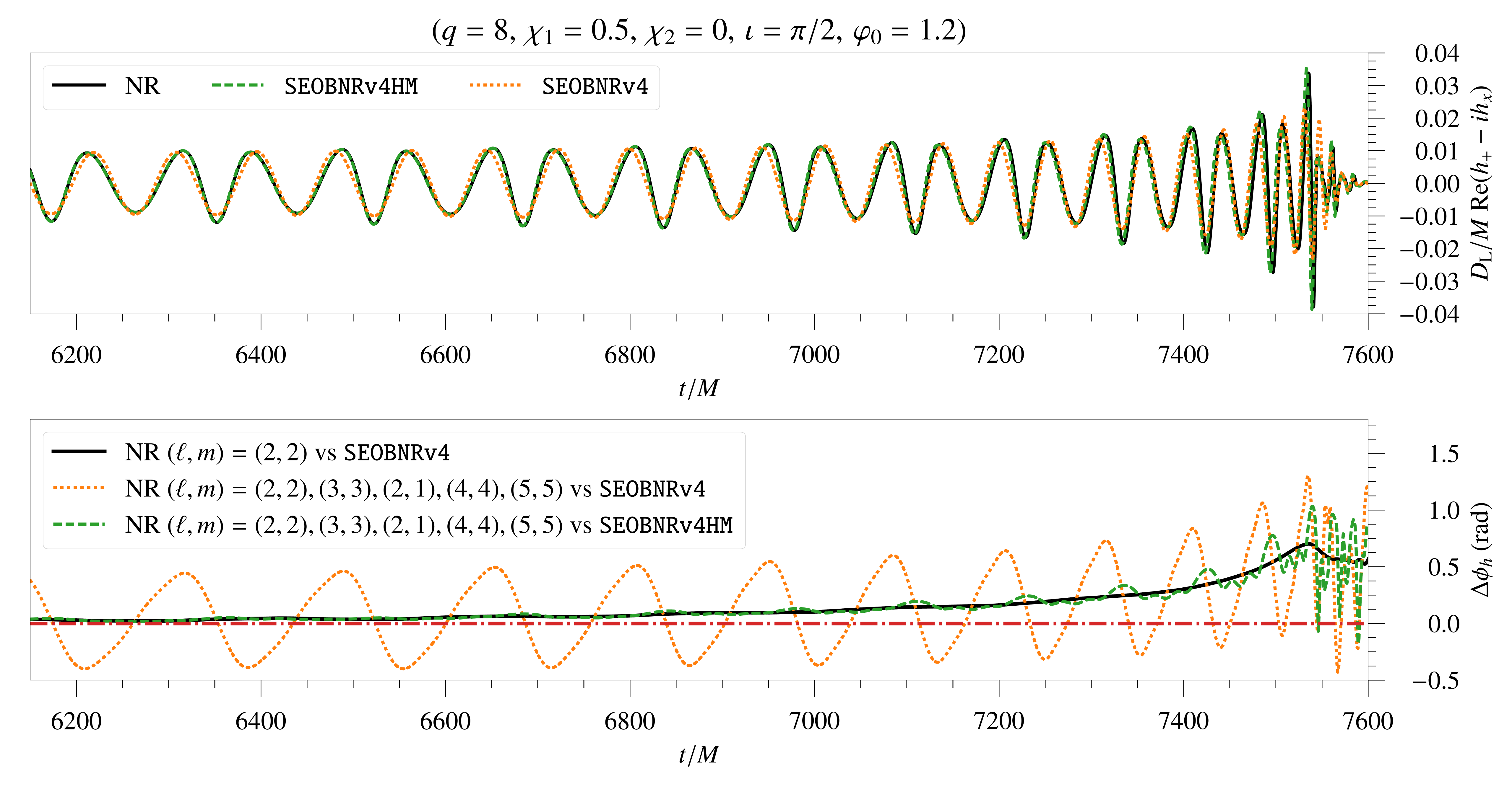}
\caption{Comparison between NR (solid black), \texttt{SEOBNRv4HM} (dashed green) and \texttt{SEOBNRv4} (dotted yellow) waveforms in an edge-on orientation $(\iota = \pi/2, \varphi_0 = 1.2)$ for the NR simulation \texttt{SXS:BBH:0065} $(q = 8,\, \chi_1 = 0.5,\, \chi_2 = 0)$. In the top panel is plotted the real part of the observer-frame's gravitational strain $h_+(\iota,\varphi_0;t) - i \ h_x(\iota,\varphi_0;t)$, while in the bottom panel the dephasing with the NR waveform $\Delta\phi_h$. The dotted-dashed red horizontal line in the bottom panel indicates zero dephasing with the NR waveform. Both \texttt{SEOBNRv4} and \texttt{SEOBNRv4HM} waveforms are phase aligned and time shifted at low frequency using as alignment window $t_\mathrm{ini} = 1000 M$ and $t_\mathrm{fin} = 3000 M$. Figure taken from Ref.~\cite{Cotesta:2018fcv}.}
\label{fig:wave_edge_on}
\end{figure}

In Fig.~\ref{fig:wave_edge_on}, I compare an \ac{NR} waveform with waveforms from the models \texttt{SEOBNRv4HM} and \texttt{SEOBNRv4}, emitted by the same \ac{BBH} system. \roberto{In particular, in the upper panel I show the real part of the function $h_+ - ih_\times$, while in the lower panel I plot the phase difference (or dephasing) between two waveforms. Both these quantities are shown as a function of time. } As it is clear from the plot \roberto{in the lower panel}, including \acp{HM} in the \ac{EOB} waveform allows to better track the phase modulations in the inspiral \ac{NR} waveform. \roberto{In fact, while the dephasing between the \ac{NR} and the \texttt{SEOBNRv4} waveform (dotted orange line) exhibits a periodic modulation, this effect disappears in the dephasing between the \ac{NR} and the \texttt{SEOBNRv4HM} waveform (dashed green line). The remaining phase difference between the \ac{NR} and \texttt{SEOBNRv4HM} waveform is simply due to the residual inaccuracy in the waveform model, and it is consistent with the inaccuracy between the \ac{NR} and the \texttt{SEOBNRv4} waveform when they only include the mode $(\ell, m) = (2,2)$ (solid black line). Also the post-merger amplitude of the \ac{NR} waveform (for $t \gtrsim 7500 M$) is better tracked by \texttt{SEOBNRv4HM} compared to \texttt{SEOBNRv4}, as it is evident from the upper panel of the figure. These modulations, in the amplitude and the phase of the waveform, are caused by the \acp{HM}. }  

I use the faithfulness function\roberto{, already defined in Eq.~\eqref{eq:faithfulness_intro}}, as a quantitative way of measuring the improved accuracy of the \ac{EOB} waveforms, when including \acp{HM}. In this case, I use \roberto{a slightly modified version of the faithfulness defined in Eq.~\eqref{eq:faithfulness_intro}. Its expression is}
\begin{equation}
\label{eq:faith}
\mathcal{F}(\iota_{\textrm{NR}},{\varphi_0}_{\textrm{NR}},\kappa_{\textrm{NR}}) \equiv  \max_{t_c, {\varphi_0}_{\mathrm{EOB}}, \kappa_{\textrm{EOB}}} \left . \frac{ \left( h_{\mathrm{NR}},\,h_{\mathrm{EOB}} \right)}{\sqrt{ \left( h_{\mathrm{NR}},\,h_{\mathrm{NR}} \right) \left( h_{\mathrm{EOB}},\,h_{\mathrm{EOB}} \right)}}\right \vert_{\substack{\iota_{\mathrm{NR}} = \iota_{\mathrm{EOB}} \\\bm{\lambda}_{\mathrm{NR}} = \bm{\lambda}_{\mathrm{EOB}}}} .
\end{equation}
\roberto{The functions } $h_\mathrm{NR,EOB}$ are defined as
\begin{align}
\label{eq:det_strain_intro}
h \equiv &  F_+(\theta,\phi,\psi) \ h_+(\iota,\varphi_0, D_{\mathrm{L}}, \bm{\lambda},t_{\mathrm{c}};t) 
+ F_\times(\theta,\phi,\psi)\ h_\times(\iota,\varphi_0, D_{\mathrm{L}}, \bm{\lambda},t_\mathrm{c};t)\, \nonumber\\
&= \mathcal{A}(\theta,\phi)\big[\cos\kappa(\theta,\phi,\psi) \ h_+(\iota, \varphi_0, D_{\mathrm{L}}, \bm{\lambda}, t_{\mathrm{c}};t) \nonumber \\
& + \sin\kappa(\theta,\phi,\psi) \ h_\times (\iota, \phi, D_{\mathrm{L}}, \bm{\lambda},t_{\mathrm{c}};t) \big],
\end{align}
where $F_+(\theta,\phi,\psi)$ and $F_\times(\theta,\phi,\psi)$ are the antenna patterns, already introduced in Sec.~\ref{sec:strategy_detection}, and the explicit definition of $\mathcal{A}(\theta,\phi)$ and the effective polarization $\kappa(\theta,\phi,\psi)$ can be found in Chapter 2, Eqs.~\eqref{eq:amp_decomp} and \eqref{eq:effective_pol} respectively. \roberto{As in the definition of the faithfulness in Eq.~\eqref{eq:faithfulness_intro}, also in Eq.~\eqref{eq:faith} there is a maximization over the coalescence time $t_c$ and the phase ${\varphi_0}_{\mathrm{EOB}}$. In addition, in Eq.~\eqref{eq:faith}, there is also the maximization over the effective polarization $\kappa_{\textrm{EOB}}$. The reason for this additional maximization is that the definition of the faithfulness, in Eq.~\eqref{eq:faithfulness_intro}, is the standard one used in literature, that does not consider waveforms with \acp{HM}. In this case, the angle $\kappa_{\textrm{EOB}}$ is degenerate with ${\varphi_0}_{\mathrm{EOB}}$, hence there is no need for the additional maximization over $\kappa_{\textrm{EOB}}$. When \acp{HM} are included in the waveforms, these two angles are not degenerate anymore, and the natural extension of the definition of the faithfulness in Eq.~\eqref{eq:faithfulness_intro} is to also include the maximization over $\kappa_{\textrm{EOB}}$, as done in Eq.~\eqref{eq:faith}. }
\roberto{As in the case of the faithfulness defined in Eq.~\eqref{eq:faithfulness_intro}, also this } faithfulness is $1$ when there is perfect agreement between two waveforms, and its value decreases proportionally to their difference.

\begin{figure}
\centering
\includegraphics[width=0.7\textwidth]{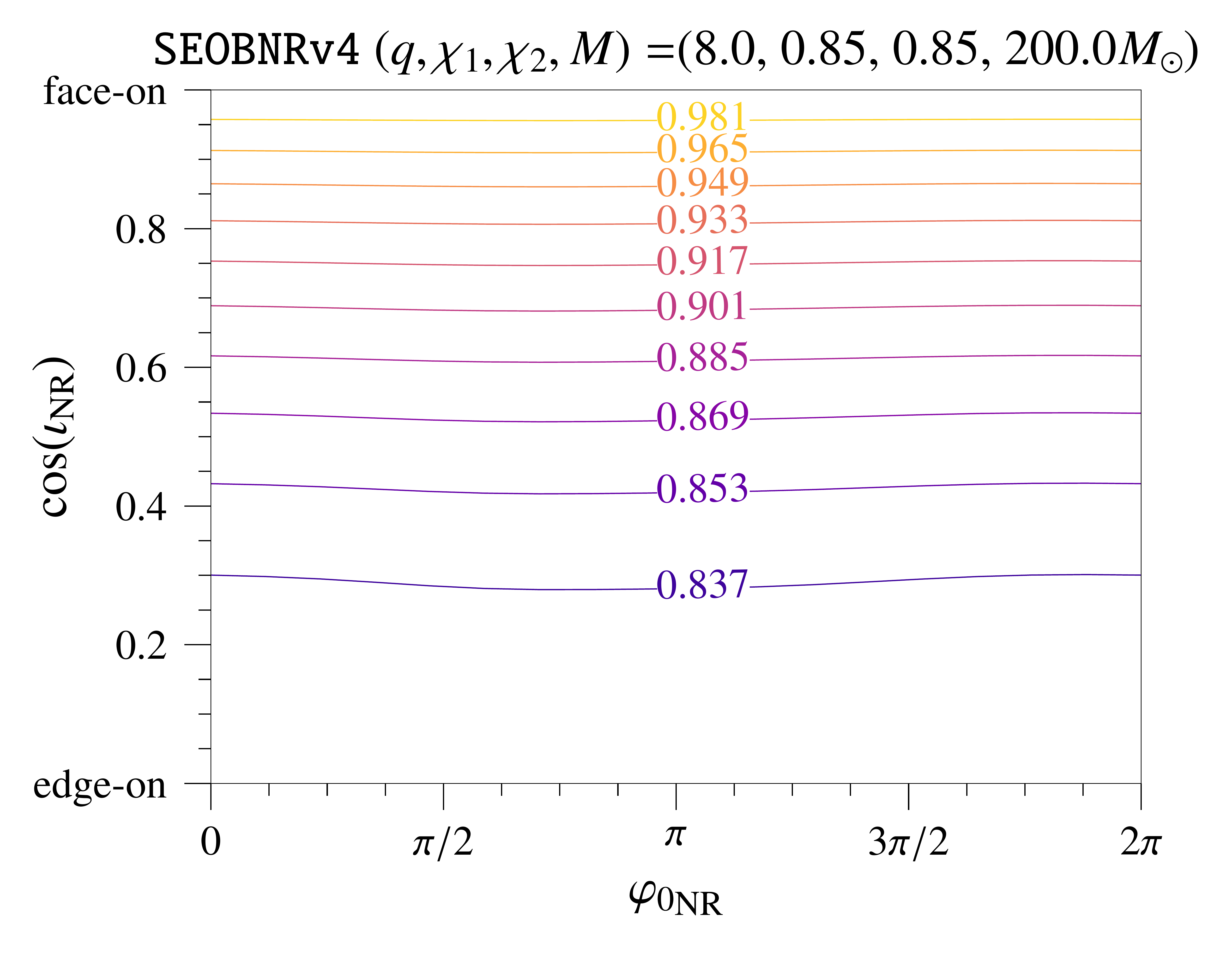} 
\includegraphics[width=0.7\textwidth]{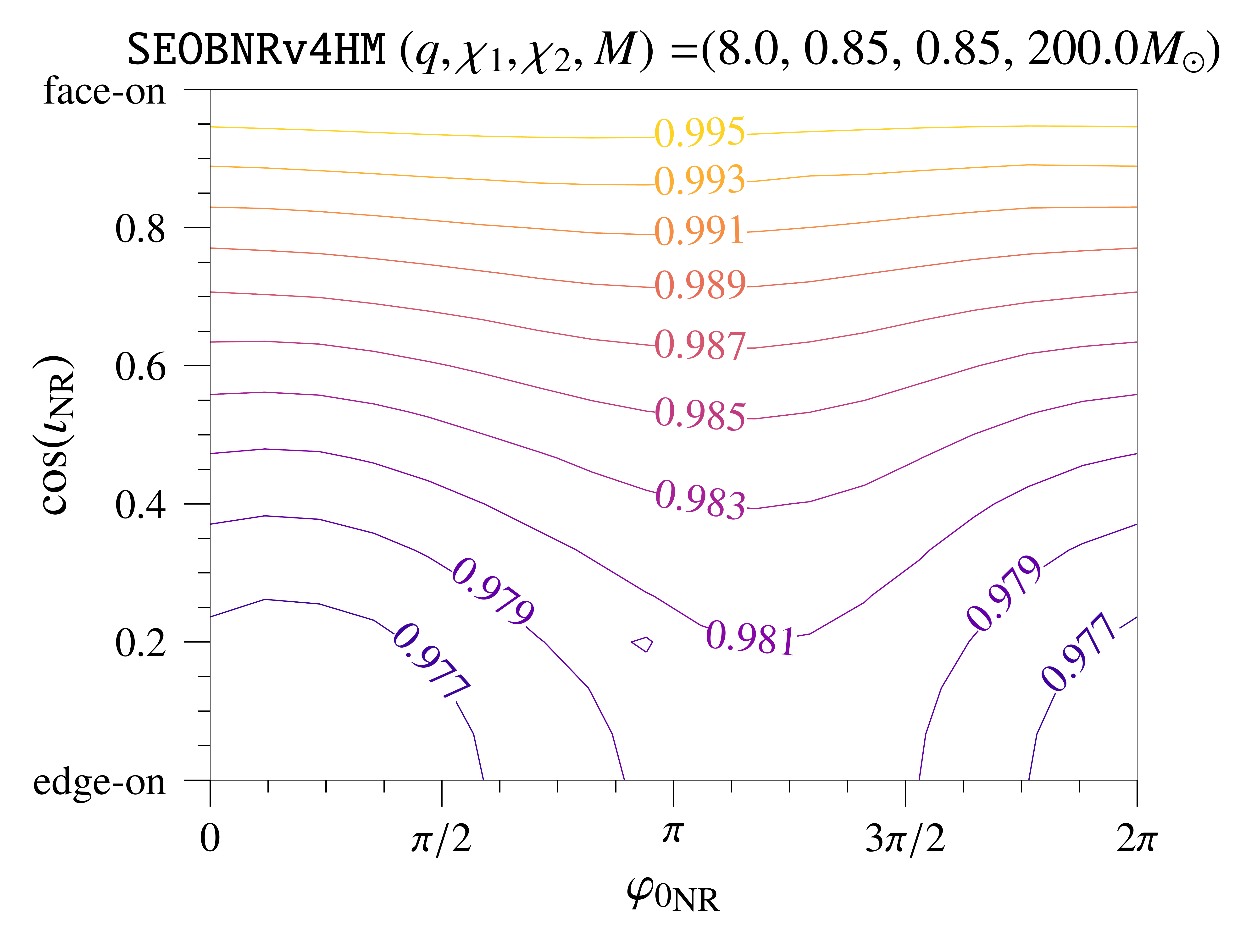} 
\caption{Faithfulness $\mathcal{F}(\cos(\iota_{\textrm{NR}}),{\varphi_0}_{\textrm{NR}},\kappa_{\textrm{NR}} = 0)$ for the configuration $(q = 8,\, M = 200 M_\odot,\, \chi_1 = 0.85, \, \chi_2 = 0.85)$: NR $(\ell \leq 5,\, m \neq 0)$ vs \texttt{SEOBNRv4} (left panel), NR $(\ell \leq 5,\, m \neq 0)$ vs \texttt{SEOBNRv4HM} (right panel). Figure taken from Ref.~\cite{Cotesta:2018fcv}.}
\label{fig:q8sv4vsHMthetaphi_intro}
\end{figure}

In Fig.~\ref{fig:q8sv4vsHMthetaphi_intro}, I show the faithfulness between an \ac{EOB} and an \ac{NR} waveform with $q = 8,\, M = 200 M_\odot,\, \chi_1 = 0.85, \, \chi_2 = 0.85$, as function of binary's orientation. As expected from the discussion in Sec.~\ref{sec:HM_importance}, the inclusion of \acp{HM} in the waveform enhances the accuracy of the \ac{EOB} model especially when the binary is observed from an edge-on inclination $(\iota = \pi/2)$. In fact, while for \texttt{SEOBNRv4} the faithfulness decreases from $0.995$ to $0.837$, when moving from face-on $(\iota = 0)$ to edge-on, for \texttt{SEOBNRv4HM} it only varies from $0.995$ to $0.977$. 

\begin{figure}
\centering
\includegraphics[width=0.7\textwidth]{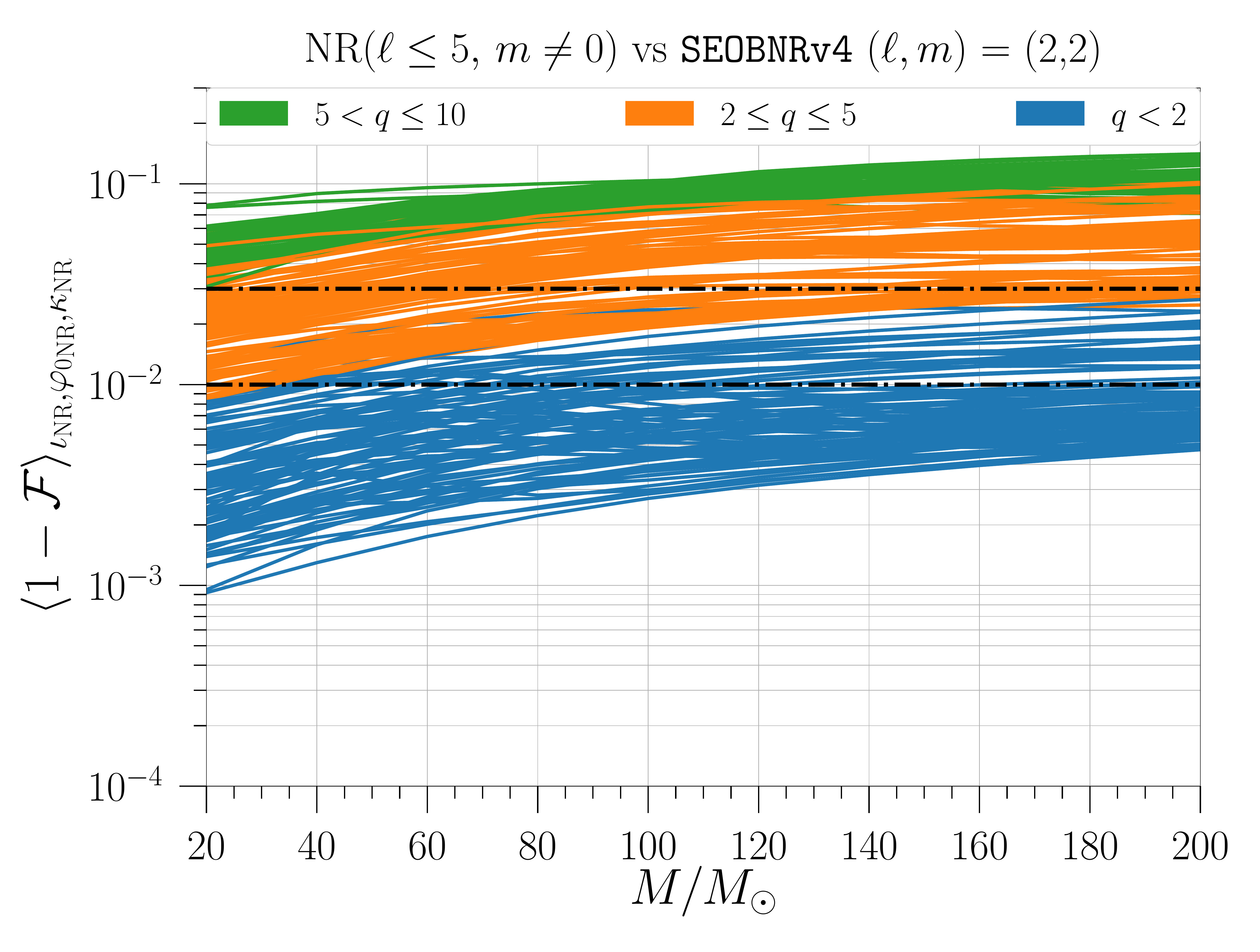} 
\includegraphics[trim=-2cm 0 2cm 0,clip,width=0.8\textwidth]{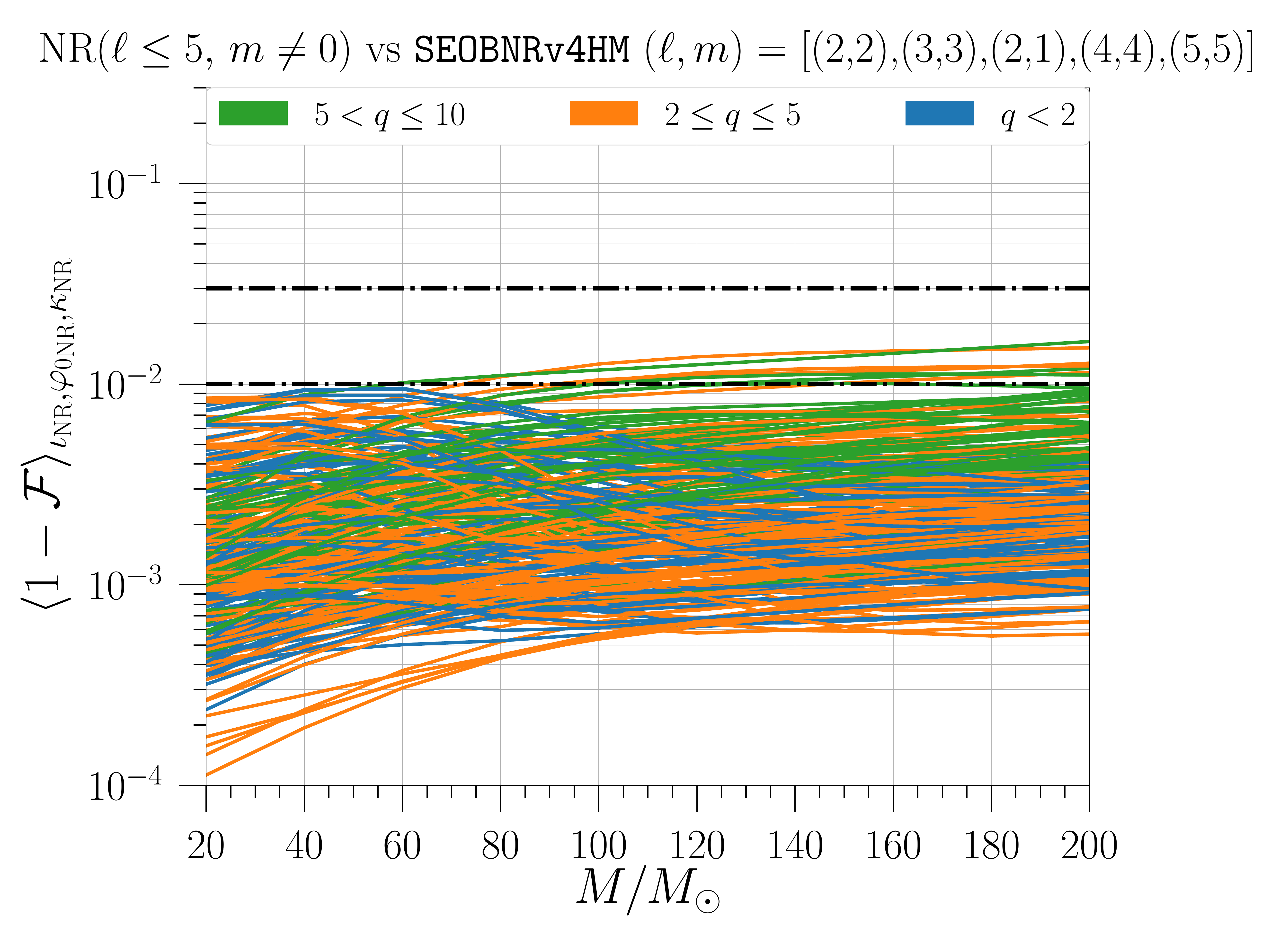} 
\caption{Unfaithfulness $(1-\mathcal{F})$ averaged over the three angles $(\iota_{\textrm{NR}},{\varphi_0}_{\textrm{NR}},\kappa_{\textrm{NR}})$ as a function of the total mass, in the range $20 M_\odot \leq M \leq 200 M_\odot$. Upper panel NR $(\ell \leq 5, \, m \neq 0)$ vs \texttt{SEOBNRv4}, lower panel NR $(\ell \leq 5,\, m \neq 0)$ vs \texttt{SEOBNRv4HM}. The horizontal dotted-dashed black lines represent the values of $1\%$ and $3\%$ unfaithfulness. This figure is an update of Fig.~\ref{fig:skyaverageall} in Ref.~\cite{Cotesta:2018fcv}.}
\label{fig:skyaverageallnew}
\end{figure}

Finally, I use the unfaithfulness ($1- \mathcal{F}$) averaged over binary's orientation, sky location and waveform polarization 
\begin{align}
\langle
1-\mathcal{F}&\rangle_{\iota_{\mathrm{NR}},{\varphi_0}_{\mathrm{NR}},\kappa_{\mathrm{NR}}} \equiv \nonumber \\
&1 - \frac{1}{8\pi^2}\int_{0}^{2\pi} d\kappa_{\mathrm{NR}} \int_{-1}^{1} d(\cos\iota_\mathrm{NR}) \int_{0}^{2\pi} d{\varphi_0}_{\mathrm{NR}} \ \mathcal{F}(\iota_{\textrm{NR}},{\varphi_0}_{\textrm{NR}},\kappa_{\textrm{NR}})\,, \label{eq:avg_unfaith_intro}
\end{align}
to \roberto{assess the accuracy of \ac{EOB} waveforms, when compared to several \ac{NR} waveforms. }
In Fig.~\ref{fig:skyaverageallnew}, I show an updated version of the plot in Fig.~\ref{fig:skyaverageall} of Chapter~\ref{chap:two}, including $91$ new \ac{NR} waveforms (described in Ref.~\cite{Varma:2019csw}), in addition to the $151$ already used in Chapter~\ref{chap:two}.
When using \texttt{SEOBNRv4}, that only includes the waveform mode $(\ell,m) = (2,2)$, the unfaithfulness against \ac{NR} waveforms can be as large as $0.2$ (upper panel). In this case, systems with larger total mass and mass ratio, for which \acp{HM} are more important, yield the largest unfaithfulness. When using \texttt{SEOBNRv4HM} (lower panel), the unfaithfulness is smaller than $0.01$ for most of the configurations and it is between $0.01$ and $0.015$ just for few cases. \roberto{Since the unfaithfulness of the waveform model against \ac{NR} waveforms is smaller than $1\%$, \texttt{SEOBNRv4HM} is accurate enough to be used in the construction of template banks for the detection of \ac{GW} signals. I will discuss, in Sec.~\ref{sec:synthetic_AS}, its suitability for parameter-estimation studies.}

\FloatBarrier

\subsubsection{Gravitational-waveform modes for precessing binary systems}
\label{sec:precessing_modes}

The morphology of the \ac{GW} modes, for a precessing \ac{BBH} system, is more complicated with respect to the non-precessing case, because of the modulations in the modes induced by the precession of the \roberto{orbital } angular momentum of the binary $\bm{L}$, as already mentioned in Sec.~\ref{sec:waveform_anatomy}.

There are two types of precession~\cite{PhysRevD.31.1815,Apostolatos:1994mx,Kidder:1995zr,Buonanno:2002fy}: \textit{simple} and \textit{transitional precession}. In the first case, the vectors $\bm{\hat{L}}  \equiv \bm{L}/|\bm{L}|$, $\bm{\hat{S}} \equiv (\bm{S_1}+\bm{S_2})/|\bm{S_1+S_2}|$ and $\bm{\hat{J}} \equiv (\bm{L} + \bm{S})/|\bm{L} + \bm{S}|$ precess on tight cones, with the angle formed by the $\bm{\hat{J}}$-cone being much smaller with respect to the opening angle of the $\bm{\hat{L}}$-cone.
The vectors $\bm{\hat{L}}$ and $\bm{\hat{S}}$ precess around $\bm{\hat{J}}$, and $\bm{\hat{L}}$ precess with an opening angle growing with time.
\roberto{A condition for the simple precession to occur is that $|\bm{L}| \gg |\bm{S}|$. Since at leading (Newtonian) order, $|\bm{L}| \propto d^{1/2}$, where $d$ is the binary separation, for every precessing \ac{BBH} system it exists a critical binary separation $d_\mathrm{crit}$ such that for $d < d_\mathrm{crit}$ the system undergoes simple precession. }  
 
The transitional precession regime occurs if, during the coalescence, the \acp{BH} reach a configuration in which $\bm{J} = \bm{L} + \bm{S} \approx 0$. When this condition is satisfied, the precessing dynamics is difficult to treat analytically. Analytical and numerical analyses~\cite{PhysRevD.31.1815,Apostolatos:1994mx,Kidder:1995zr,Buonanno:2002fy,Kesden:2014sla,Ossokine:2015phd,Zhao:2017tro,Gerosa:2015hba,Gerosa:2018mwg,Ossokine:2015phd} demonstrated that \robert{during transitional precession $\bm{J}$ may change its direction multiples times}. See Refs.~\cite{PhysRevD.31.1815,Apostolatos:1994mx,Kidder:1995zr,Buonanno:2002fy,Kesden:2014sla,Ossokine:2015phd,Zhao:2017tro,Gerosa:2015hba,Gerosa:2018mwg,Ossokine:2015phd} for more details on the simple and transitional precession. 

It is possible to understand the origin of the modulations on the waveform induced by precession, starting from the leading-order expressions for $h_+$ and $h_\times$ in Eqs.~\ref{eq:quad_cross_evol}, that I write here again for convenience 
\begin{align}
h_+(t) =& \frac{4}{D_\mathrm{L}}\left(\frac{G\mathcal{M}}{c^2}\right)^{5/3} \left[\frac{5}{c(t_\mathrm{c} - t)}\right]^{1/4} \frac{1+(\bm{L_\mathrm{N}}\cdot \bm{\hat{N}})^2}{2}\cos\left(\Phi_\mathrm{GW}(t) +2{\varphi}_0 \right)  \\
h_\times (t) =& \frac{4}{D_\mathrm{L}}\left(\frac{G\mathcal{M}}{c^2}\right)^{5/3}  \left[\frac{5}{c(t_\mathrm{c} - t)}\right]^{1/4}\bm{L_\mathrm{N}}\cdot \bm{\hat{N}} \sin\left(\Phi_\mathrm{GW}(t) +2{\varphi}_0 \right),
\end{align}
where I remind that $\bm{\hat{L}_\mathrm{N}}$ is the direction of the Newtonian orbital angular momentum, orthogonal to the orbital plane, and $\bm{\hat{N}}$ is the direction of the observer. In an \textit{inertial frame}, where the direction of the coordinate bases is constant over time, the direction $\bm{\hat{N}}$ is fixed over time, while, in the case of precession, the direction of $\bm{\hat{L}_\mathrm{N}}$ is a function of time. For this reason, in this frame the amplitude of $h_+(t)$ and $h_\times(t)$ is modulated over time by the variation of the factor $\bm{\hat{L}_\mathrm{N}}\cdot \bm{\hat{N}}$. Also $\Phi_\mathrm{GW}(t)$, the phase of $h_+(t)$ and $h_\times(t)$, is modulated as a consequence of the precession of the orbital plane, see e.g. Sec. IIIB in Ref.~\cite{Apostolatos:1994mx} for an explicit calculation of these modulations at leading (Newtonian) order. 

To simplify the modeling of \ac{GW} signals emitted by such systems, it is useful to find a frame, henceforth \textit{co-precessing frame}, where the modulations on the waveform due to precession can be factored out. \roberto{This co-precessing frame should ideally \say{follow} the precession of the orbital plane, such that the precession effects can be reabsorbed by the frame, instead of appearing in the waveform, as in the case of the waveform in the inertial frame}.  In the co-precessing frame, the waveform \roberto{should } be easier to model and more similar to that emitted by a non-precessing \ac{BBH} system. In this case, one could create a model of the waveform in the co-precessing frame, and then perform an instantenous rotation to compute the waveform in the conventional inertial frame.

\begin{figure}
\centering
\includegraphics[angle=0,width=\linewidth]{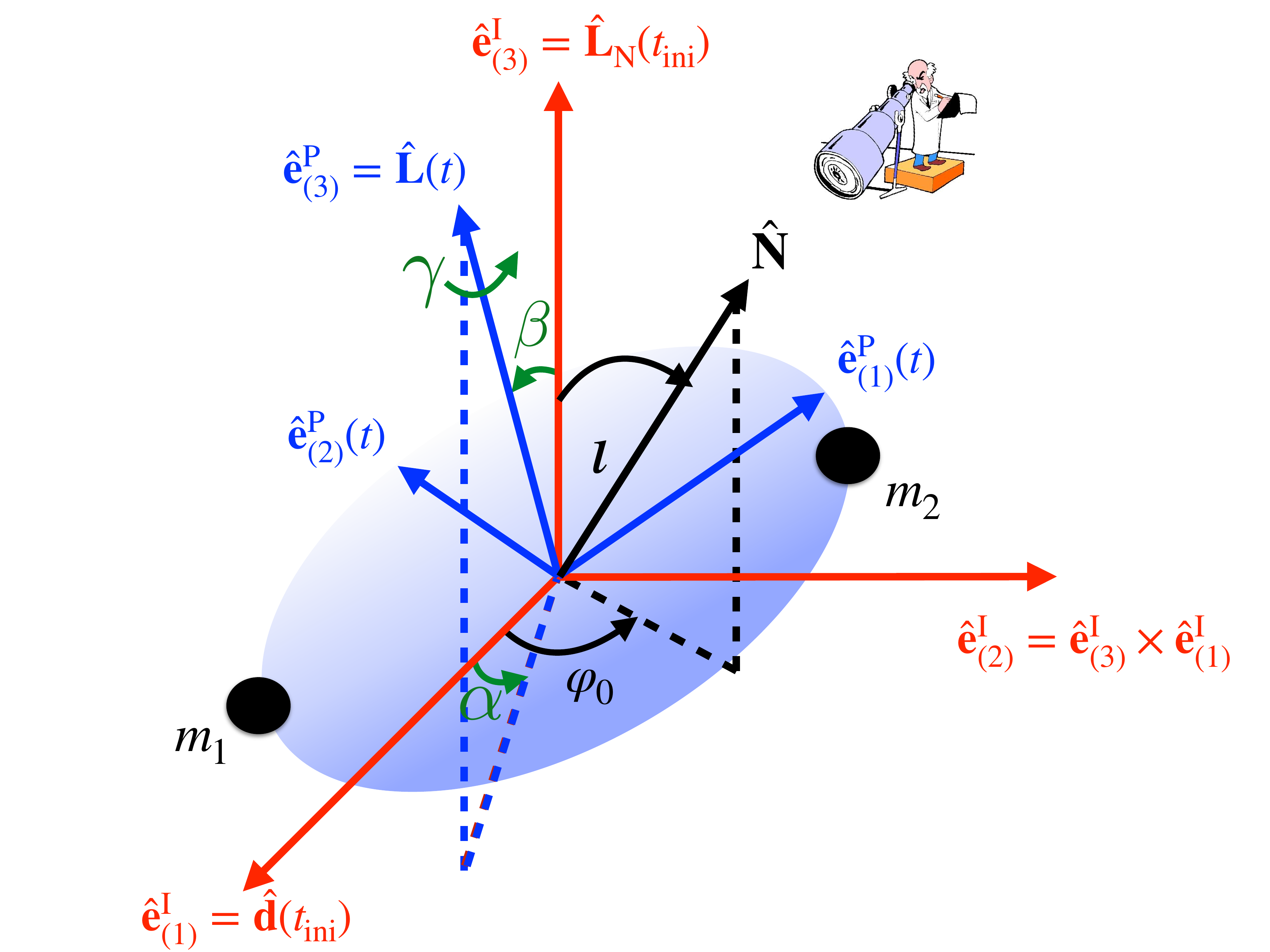}
\caption{The inertial frame (red), defined by the directions of the
  initial orbital angular momentum $\bm{\hat{L}}_N(t = t_\mathrm{ini})$ and separation
  $\bm{\hat{d}}(t_\mathrm{ini})$, and co-precessing frame (blue), instantaneously aligned
  with $\bm{\hat{L}}(t)$ and described by the Euler angles $(\alpha, \beta,
  \gamma)$. Figure adapted from Ref.~\cite{Ossokine:2020kjp}. \label{fig:Pframe_intro}}
\end{figure}

The co-processing frame used for the \ac{EOB} waveform models for \acp{BBH} with precessing spins~\cite{Pan:2013rra,Babak:2016tgq} was defined in Ref.~\cite{Buonanno:2002fy}. \roberto{In Fig.~\ref{fig:Pframe_intro}, I show the inertial frame, defined according to the convention used by the LIGO Scientific and Virgo collaborations, together with the co-precessing frame used in the precessing \ac{EOB} waveform model that I summarize in the next section. In the inertial frame (red), the basis $\bm{\hat{e}}_{(3)}^I$ is constant in time and aligned with $\bm{\hat{L}_\mathrm{N}}$, the direction of the Newtonian angular momentum of the binary, at the time $t = t_\mathrm{ini}$, i.e. the initial time of the waveform. The basis $\bm{\hat{e}}_{(1)}^I$ is also constant in time and it is aligned with $\bm{\hat{d}}(t = t_\mathrm{ini})$, the direction of the initial separation of the \acp{BH}. The frame is completed by the third basis $\bm{\hat{e}}_{(2)}^I = \bm{\hat{e}}_{(3)}^I \times \bm{\hat{e}}_{(1)}^I$. } The co-precessing frame (blue) is determined by the basis versors $\{\bm{\hat{e}}_{(i)}^P\}$. The basis versor $\bm{\hat{e}}_{(3)}^P(t)$ coincides, at each moment in time, with the orbital angular momentum $\bm{\hat{L}}(t)$~\footnote{In the original definition of the frame in Ref.~\cite{Buonanno:2002fy} the basis $\bm{\hat{e}}_{(3)}^P(t)$ coincided with $\bm{\hat{L}_\mathrm{N}}(t)$.}. The other two basis versors, $\bm{\hat{e}}_{(1)}^P(t)$ and $\bm{\hat{e}}_{(2)}^P(t)$, are initially aligned with the corresponding bases in the inertial frame, and their time evolution is determined by the equation
\begin{equation}
\frac{d\bm{\hat{e}}_{(1),(2)}^P}{dt} = \bm{\Omega}_e \times \hat{e}_{(1),(2)}^P,
\end{equation}
with $\bm{\Omega}_e \equiv \bm{\hat{L}}(t) \times \frac{d\bm{\hat{L}}(t)}{dt}$.
In Ref.~\cite{Buonanno:2002fy}, it is shown that, when this frame is used, the precessional modulations in both amplitude and phase are removed in leading-order \ac{PN} waveforms. 

\subsubsection{\texttt{SEOBNRv4PHM}: the inspiral-merger-ringdown waveform model including higher-order modes for binary black holes with precessing spins}
\label{sec:intro_SEOBNRv4PHM}

\roberto{Chapter~3 is the publication that describes } the generalization of the waveform model \texttt{SEOBNRv4HM}, introduced in Sec.~\ref{sec:intro_SEOBNRv4HM}, to \ac{BBH} systems with precessing spins, henceforth \texttt{SEOBNRv4PHM}. The \ac{EOB} dynamics, described in Sec.~\ref{sec:EOB_dynamics}, is generic and it can be used without any modification also for \texttt{SEOBNRv4PHM}. For the \ac{GW} modes in the co-precessing frame, they can be modeled using their expressions in the aligned-spin limit in Eq.~\eqref{eq:GWmodes}. The only difference is that, in the precessing case, the modes are functions of $\bm{\chi}_{1,2}(t)\cdot\bm{\hat{L}}(t)$, the time dependent projection of the dimensionless spins on the orbital angular momentum, instead of the constant projection $\bm{\chi}_{1,2}\cdot\bm{\hat{L}}$. Finally, the modes $h_{\ell m}^P$ in the co-precessing frame are rotated to the conventional inertial frame, defined in Fig.~\ref{fig:Pframe_intro}, using the transformation
\begin{equation}
\label{eq:rotation_inertial}
h_{\ell m}^I(t) = \sum_{m' = -\ell}^\ell D_{m' m}^{(\ell) *}[\alpha(t), \beta(t), \gamma(t)] \, h_{\ell m'}^P(t),
\end{equation}
where $D_{m' m}^{(\ell) *}$ is the complex conjugate of the Wigner D-matrix, and $(\alpha(t), \beta(t), \gamma(t))$ are the Euler angles, as defined in Fig.~\ref{fig:Pframe_intro}.

\begin{figure}
\centering
  \includegraphics[width=0.8\linewidth]{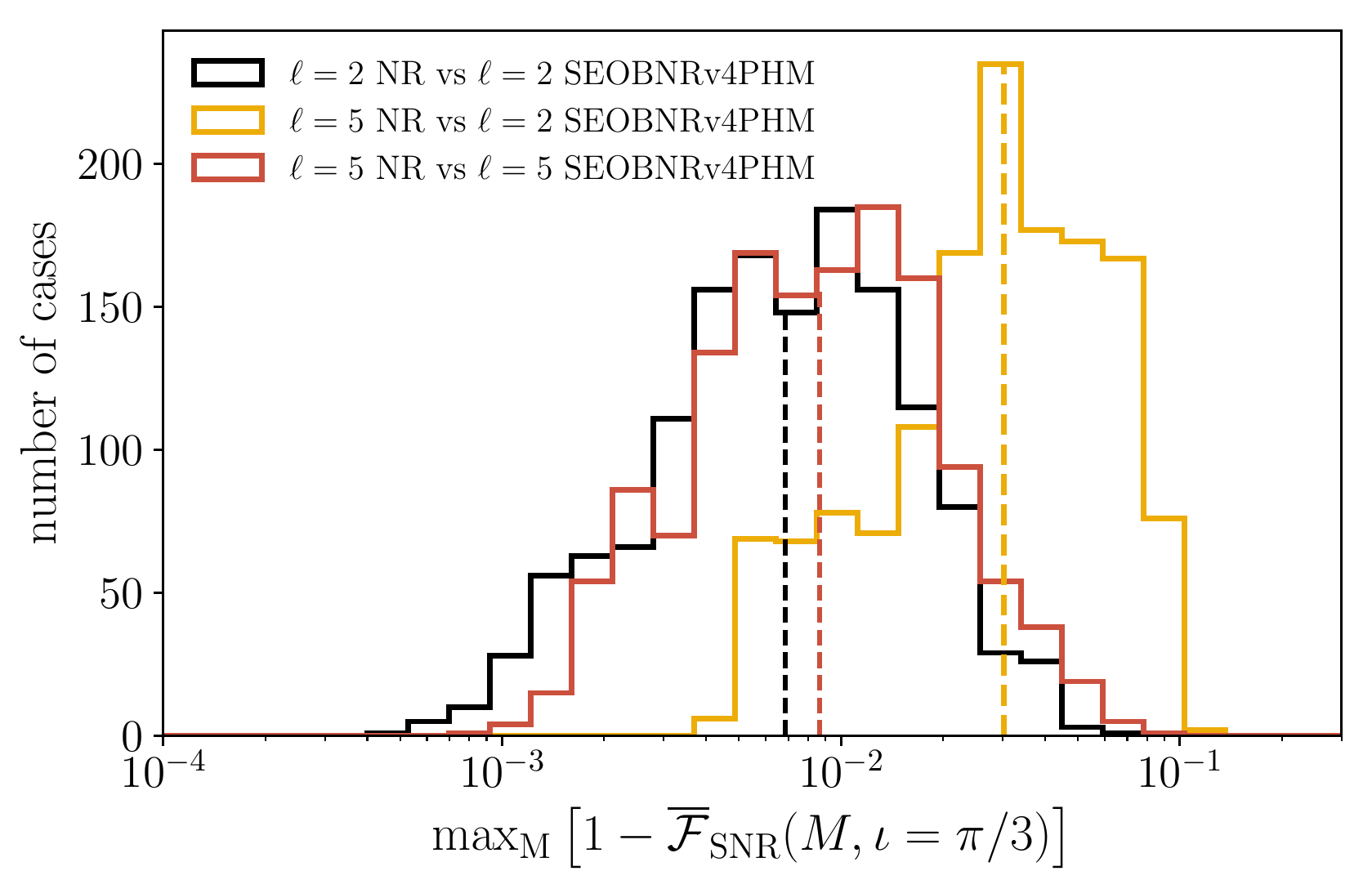}
  \caption{Sky-and-polarization averaged, SNR weighted unfaithfulness for
  an inclination $\iota=\pi/3$ between NR waveforms and \texttt{SEOBNRv4PHM},
  including (red histogram) and omitting (yellow histogram) \acp{HM}. The vertical dashed lines show the medians. Not including \acp{HM}
  in the model results in high unfaithfulness. However, when they are 
included, the unfaithfulness between \texttt{SEOBNRv4PHM} and NR is
  essentially at the same level as when only $\ell=2$ modes are compared (black histogram). Figure adapted from Ref.~\cite{Ossokine:2020kjp}.}
  \label{fig:higher_mode_effects_intro}
\end{figure}

\begin{figure}
\centering
	\includegraphics[width=0.8\linewidth]{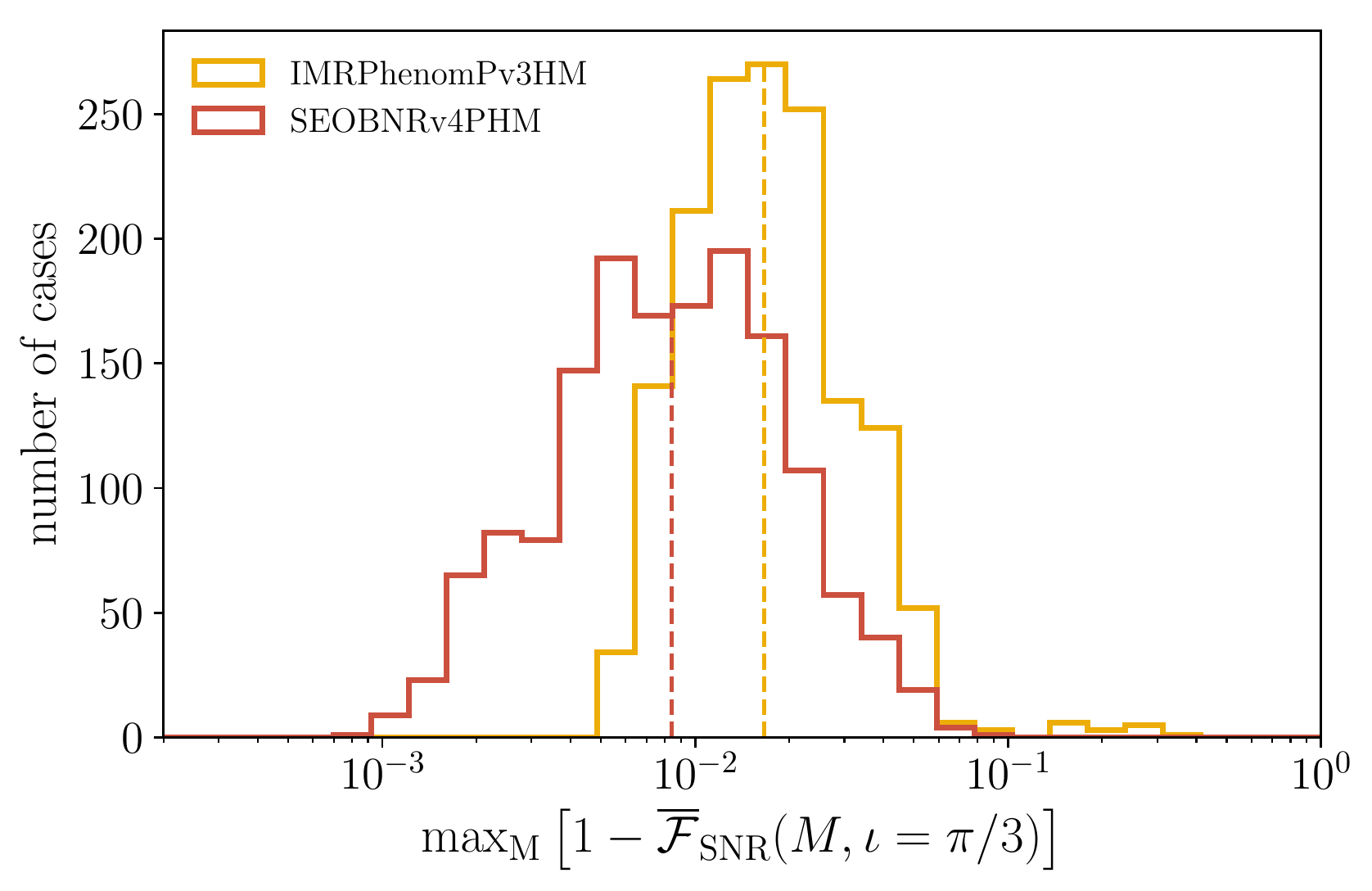}
	\caption{The median of unfaithfulness is 
		around 1\% for \texttt{SEOBNRv4PHM} (red histogram) and 2\% for \texttt{IMRPhenomPv3HM} (yellow histogram). Medians are shown as dashed vertical lines. 
		Note that for \texttt{SEOBNRv4PHM}, the worst unfaithfulness is below 10\% and the distribution is shifted to lower values. Figure adapted from Ref.~\cite{Ossokine:2020kjp}.}
	\label{fig:all_runs_hist_intro}
\end{figure}

Also in this case, I use the faithfulness to quantify the agreement of the \texttt{SEOBNRv4PHM} waveforms against \ac{NR} simulations. 
In particular, I use the sky-and-polarization-averaged, \ac{SNR}-weighted faithfulness defined as
\begin{equation}
\overline{\mathcal{F}}_{\mathrm{SNR}}(M,\iota) \equiv \sqrt[3]{\frac{\int_{0}^{2\pi} d\kappa \int_{0}^{2\pi} d{\varphi_0} \ \mathcal{F}^{3}(M,\iota,{\varphi_0},\kappa) \ \rho^3(\iota,{\varphi_0},\kappa)}{\int_{0}^{2\pi} d\kappa \int_{0}^{2\pi} d{\varphi_0} \ \rho^3(\iota,{\varphi_0},\kappa)}},
\end{equation}
\robert{where $\mathcal{F}(M,\iota,{\varphi_0},\kappa)$ is the faithfulness function defined in Eq.~\eqref{eq:faith}, and $\rho$ is the \ac{SNR} } defined in Sec.~\ref{sec:strategy_detection}.
In Fig.~\ref{fig:higher_mode_effects_intro} I show histograms of the maximum over total masses of the unfaithfulness $1- \overline{\mathcal{F}}_{\mathrm{SNR}}(M,\iota = \pi/3)$ between 1404 NR waveforms and the corresponding \texttt{SEOBNRv4PHM} waveforms, see Chapter 3 for details on the \ac{NR} waveforms. 
Including \acp{HM} in \texttt{SEOBNRv4PHM} improves the accuracy of the model by a factor of $5$ with respect to the same model without \acp{HM} (i.e. only the modes with $\ell = 2$ are included), respectively red and yellow histograms in the figure. \roberto{When computing the unfaithfulness, if I restrict both \ac{NR} and \texttt{SEOBNRv4PHM} waveforms to the $\ell = 2$ modes (black histogram), the accuracy is the same as in the case in which both waveforms include all modes up to $\ell = 5$ (red histogram)}. This means that the non-perfect modeling of \acp{HM} is not the limiting factor for the accuracy of the model.
Finally, in Fig.~\ref{fig:all_runs_hist_intro} I compare the unfaithfulness distribution computed before using \texttt{SEOBNRv4PHM}, with that obtained when using the model \texttt{IMRPhenomPv3HM}. As it is clear from the plot, the unfaithfulness computed with \texttt{SEOBNRv4PHM} (red histogram) are typically smaller than those obtained with \texttt{IMRPhenomPv3HM} (yellow histogram), therefore \texttt{SEOBNRv4PHM} is more accurate. \roberto{In particular, the median of the unfaithfulness distribution obtained with \texttt{SEOBNRv4PHM} is approximately $1\%$, while it is around $2\%$ for \texttt{IMRPhenomPv3HM}. In addition, all the unfaithfulness are lower than $10\%$ for \texttt{SEOBNRv4PHM}, while, in the case of \texttt{IMRPhenomPv3HM} some, configurations have unfaithfulness larger than $10\%$. In Sec.~\ref{sec:synthetic_PS}, I will discuss the impact of the different accuracy of the two waveform models on the measurement of the \ac{BBH} parameters. }

\FloatBarrier
\subsection{\texttt{SEOBNRv4HM\_ROM}: reduced-order-modeling techniques applied to waveform models with higher-order modes for binary black holes with non-precessing spins}
\label{sec:ROM}
Data analysis applications of waveform models, such as the construction of template banks and the Bayesian
inference of binary parameters, require to compute $\mathcal{O}(10^7-10^8)$ different waveforms~\cite{Harry:2009ea,Manca:2009xw, Veitch:2014wba,Ashton:2018jfp}. Since generating a single waveform, using the models introduced in Secs.~\ref{sec:intro_SEOBNRv4HM} and~\ref{sec:intro_SEOBNRv4PHM}, takes $\mathcal{O}(1-10)$s, they are \roberto{difficult } to use for these purposes. While it is possible to directly optimize these waveform models, \roberto{using analytical approximations to accelerate the integration of the \ac{EOB} equations of motion}~\cite{Devine:2016ovp,Nagar:2018gnk}, a better established method to speed up the waveform computation is the construction of a \ac{ROM}~\cite{Field:2013cfa,Purrer:2014fza,Purrer:2015tud,Blackman:2015pia,Blackman:2017pcm,Blackman:2017dfb,Lackey:2018zvw,Doctor:2017csx,Setyawati:2019xzw} of the original waveform model. The basic idea of a \ac{ROM} is that, given a large enough set of waveforms in a chosen region of the binary parameter space, one can compute new waveforms in the same region as an interpolation of the waveforms in the set. The method to produce these interpolated waveforms is called  \ac{ROM}. It is a useful replacement of the original model if the interpolated waveforms are an accurate representation of those generated with the original model, and they are computed faster.

\roberto{Chapter 4 is the publication that } describes the \ac{ROM} of \texttt{SEOBNRv4HM}, henceforth referred to as \texttt{SEOBNRv4HM\_ROM}. In particular, I costruct a \ac{ROM} for each \ac{GW} mode, $\tilde{h}_{\ell m}(f)$, expressed in frequency domain. Then, from the \ac{GW} modes, I obtain the two polatizations $h_+$ and $h_\times$ using Eq.~\eqref{eq:sph_harm_expansion_2}. Since $\tilde{h}_{\ell m}(f)$ are complex functions, \roberto{to build } the \ac{ROM} for each mode one has to interpolate two real functions over the binary parameter space. Typically, one chooses the amplitude $\tilde{A}_{\ell m}(f)$ and the phase $\Psi_{\ell m}(f)$ of $\tilde{h}_{\ell m}(f) = \tilde{A}_{\ell m}(f)\exp{(-i\Psi_{\ell m}(f))}$. \roberto{When the \acp{HM} are included in the waveform model, } this choice is not possible because the \roberto{amplitude of the } modes with odd $m$ vanishes in the limit of equal masses and \ac{BH}'s spins, and the phase is not a well-defined function in this limit. While using real and imaginary part of $\tilde{h}_{\ell m}(f)$ seems a valid alternative, it is also unfeasible as these are oscillatory functions,  difficult to interpolate in practice. The solution to this problem is based on the fact that, during the inspiral regime, the time-domain phase $\phi_{\ell m}(t)$ of each mode $h_{\ell m}(t) = A_{\ell m}(t)\exp{(-i\phi_{\ell m}(t))}$ can be approximated as $\phi_{\ell m}(t) \propto m \phi_\mathrm{orb}(t)$, where $\phi_\mathrm{orb}(t)$ is the orbital phase of the binary (see e.g. Ref.~\cite{Blanchet:2013haa} Eq.(327)). This means that the phase of all modes, including those with odd $m$, can be approximated as a function of $\phi_{22}(t)/2$, which is always well defined because the mode $(\ell, m) = (2,2)$ never vanishes.
In particular
\begin{equation}
\label{eq:phase_hm_22_TD}
\phi_{\ell m}(t) \propto -m \arg{k(t)},
\end{equation}
where $k(t) = A(t)\exp{(-i\phi_{22}/2)}$ and $A(t)$ is a non-vanishing amplitude whose expression is irrelevant for this discussion.
The expression equivalent to Eq.~\eqref{eq:phase_hm_22_TD}, for the phases $\Psi_{\ell m}(f)$ of the modes in frequency domain $\tilde{h}_{\ell m}(f)$ can be obtained by computing analytically the Fourier transform of $h_{\ell m}(t)$ and $k(t)$, using the \ac{SPA} method~\cite{Buonanno:2009zt}. Its expression is
\begin{equation}
\label{eq:phase_hm_22_FD}
\Psi_{\ell m}(f) \propto m\Psi_k(f/m),
\end{equation}
where $\Psi_k(f)$ is the phase of $\tilde{k}(f)$, the Fourier transform of $k(t)$. The function $\Psi_k(f)$ is the first function that I interpolate over the binary parameter space for the construction of the \ac{ROM}, but it only provides an approximate representation of the phase of $\tilde{h}_{\ell m}(f)$, and contains no information on the amplitude of the modes. For this reason, to accurately reconstruct the functions $\tilde{h}_{\ell m}(f)$, I also have to  interpolate over the binary parameter space the real and imaginary part of the complex functions $\tilde{h}^c_{\ell m}(f)$, defined as
\begin{equation}
\label{eq:cmodes}
\tilde{h}^c_{\ell m}(f) \equiv \tilde{h}_{\ell m}(f)e^{im\Psi_k(f/m)} = \tilde{A}_{\ell m}(f)e^{i\left(m\Psi_k(f/m) - \Psi_{\ell m}(f)\right)}.
\end{equation}
The functions $\Re{(\tilde{h}^c_{\ell m}(f))}$ and $\Im{(\tilde{h}^c_{\ell m}(f))}$ are not as difficult to interpolate as $\Re{(\tilde{h}_{\ell m}(f))}$ and $\Im{(\tilde{h}_{\ell m}(f))}$, because most of the oscillatory behaviour present in $\tilde{h}_{\ell m}(f)$ is removed in $\tilde{h}^c_{\ell m}(f)$ by virtue of Eq.~\eqref{eq:phase_hm_22_FD}. Having the interpolation over the binary parameter space of $\Psi_k(f)$, $\Re{(\tilde{h}^c_{\ell m}(f))}$ and $\Im{(\tilde{h}^c_{\ell m}(f))}$ allows to reconstruct $\tilde{h}_{\ell m}(f)$ simply by inverting Eq.~\eqref{eq:cmodes}. The interpolation of these functions over the binary parameter space has to be performed only on the $3D$ space of the mass ratio $q$, and the two $z$-component of the \ac{BH} spins, $\chi_{1\mathrm{z}}$ and $\chi_{2\mathrm{z}}$, since the dependence of the modes on all the other parameters is trivial. To produce the interpolations of $\Psi_k(f)$, $\Re{(\tilde{h}^c_{\ell m}(f))}$ and $\Im{(\tilde{h}^c_{\ell m}(f))}$, I first decompose them into their respective \ac{SVD} bases, and then I interpolate the projection coefficients using tensor-product spline interpolation. This is a standard technique already used in Refs.~\cite{Bohe:2016gbl,Purrer:2014fza,Purrer:2015tud}.

I test the accuracy of this \ac{ROM} against \texttt{SEOBNRv4HM} with metric given by the averaged unfaithfulness defined in Eq.~\eqref{eq:avg_unfaith_intro}. I find that this unfaithfulness between \texttt{SEOBNRv4HM\_ROM} and \texttt{SEOBNRv4HM} is on average $\mathcal{O}(0.001 \%)$, when computed for many different values of the masses and the \ac{BH}'s spins. Since the same unfaithfulness is $\mathcal{O}(0.1\%)$ when computed between \texttt{SEOBNRv4HM} and \ac{NR} waveforms (see Fig.~\ref{fig:skyaverageallnew}),  the modeling error added by the \ac{ROM} is negligible with respect to the error of the original model against the \ac{NR} simulations.

Finally, I evaluate the speed of the \ac{ROM} with respect to \texttt{SEOBNRv4HM}.
\robert{I find that generating \texttt{SEOBNRv4HM\_ROM} waveforms is $\sim 100$ times faster than generating \texttt{SEOBNRv4HM} waveforms in the total mass and mass ratio ranges $5 M_\odot \leq M \leq 200 M_\odot$ and $1\leq q \leq 50$, and for every value of the \ac{BH} spins}. \robert{\texttt{SEOBNRv4HM\_ROM} is therefore much more efficient to use } in data analysis applications. Also for the waveform model \texttt{SEOBNRv4PHM}, a procedure similar to the \ac{ROM} technique is being used to produce a faster version of the model~\cite{Gadre:2021aa}.


\section{Binary black-holes characterization using waveform models with higher-order modes}
\label{sec:PE}
In this section, I summarize the results I obtained when using the waveform models with \acp{HM}, outlined in the previous sections, to measure the parameters of \ac{BBH} systems. In Secs.~\ref{sec:synthetic_AS} and~\ref{sec:synthetic_PS}, I use synthetic \ac{GW} signals to test the accuracy of the waveform models, in the case for which the parameters of the binary are already known.
Then, in Secs.~\ref{sec:pe_GW170729} and~\ref{sec:pe_GW190412}, I extend the analysis to two \textit{real} \ac{GW} signals, detected by the LIGO and Virgo interferometers, \roberto{respectively during O2 and O3a}: GW170729~\cite{LIGOScientific:2018mvr} and GW190412~\cite{LIGOScientific:2020stg}. These \ac{GW} signals are particularly interesting because their sources lie in regions of the binary parameter space where the \acp{HM} are expected to be relevant in the \ac{GW} signal. \roberto{In addition to the signal GW190412, the waveform models outlined in the previous sections, were also used for the inference of the \ac{BBH} parameters of all the other \ac{GW} signals detected during O3a~\cite{Abbott:2020niy}. In particular, they were also used to analyze other signals for which the effect of \acp{HM} was expected to be important, namely GW190521~\cite{Abbott:2020tfl} and GW190814~\cite{Abbott:2020khf}. Here, I restrict my attention to the signals GW170729 and GW190412 because I directly analyzed them. }
\subsection{Synthetic signal I: the aligned-spins case}
\label{sec:synthetic_AS}

\roberto{In Chapter 4, Sec.~\ref{sec:PEsec}, } I test the accuracy of the waveform model \texttt{SEOBNRv4HM\_ROM}, when used to measure the parameters of a \ac{BBH} system from a synthetic \ac{GW} signal.
\roberto{For this study, I choose to consider a \ac{GW} signal measured by the network composed by the two LIGO detectors: LIGO Hanford, LIGO Livingston, and the Virgo detector. To construct the synthetic signal $d_\mathrm{syn}(t) = n_\mathrm{syn}(t) + h_\mathrm{syn}(t)$, it is necessary to specify the detector noise $n_\mathrm{syn}(t)$, and the waveform $h_\mathrm{syn}(t)$. For the noise of the three detectors, I choose to use the mean value $\langle n_\mathrm{syn}(t) \rangle = 0$. } 
This is a typical choice, when the purpose of the study is to test the accuracy of a waveform model. In fact, it allows to avoid biases in the parameter estimation due to a particular realization of the Gaussian detector noise, when one is interested in the biases due to the inaccuracy of the waveform model. \roberto{The noise also enters the parameter-estimation analysis through the \ac{PSD} in the likelihood function, see Eq.~\eqref{eq:LIGO_likelihood}. For this analysis, I use the \acp{PSD} of the  LIGO and Virgo detectors at design sensitivity~\cite{Barsotti:2018,TheVirgo:2014hva}. }

I generate $h_\mathrm{syn}(t)$, the waveform for the synthetic signal, using the \ac{NR} surrogate model \texttt{NRHybSur3dq8}~\cite{Varma:2019csw} introduced in Sec.~\ref{sec:HM_importance}. Typically these studies are performed using \ac{NR} waveforms as synthetic signals, but the waveforms generated with this model are indistinguishable from \ac{NR} waveforms \roberto{at the \ac{SNR} of this study, which I indicate below. }

I choose the parameters of the \ac{BBH} system for the synthetic signal to \textit{enhance the \acp{HM} contribution in the waveform}. In particular, for the mass ratio I choose the value $q = 8$, and I use a large total mass $M = 67.5 M_\odot$.  \roberto{Also to \textit{enhance the \acp{HM} contribution in the waveform}}, I set the inclination angle $\theta_\mathrm{JN}$ to $\pi/2$. I focus on \ac{BBH} systems with \ac{BH} spins aligned with the orbital angular momentum of the binary and, for this synthetic signal, I set their magnitude to $|\bm{\chi_1}| = 0.5$ and $|\bm{\chi_2}| = 0.3$. 
The network-\ac{SNR} of the synthetic signal\roberto{, defined as the the root sum squared of the \acp{SNR} in each detector}, is $21.8$. All the other parameters are less relevant for the discussion, and can be found in Chapter 4 Sec.~\ref{sec:PEsec}.

Finally, I use the waveform model \texttt{SEOBNRv4HM\_ROM} to perform the Bayesian parameter estimation on this synthetic signal, and measure the parameters of the \ac{BBH} system. As a comparison, I also use other three waveform models for the same analysis: \texttt{NRHybSur3dq8}~\cite{Varma:2019csw}, \texttt{IMRPhenomHM}~\cite{London:2017bcn} and \texttt{SEOBNRv4\_ROM}~\cite{Bohe:2016gbl}, which only includes the mode $(\ell, |m|) = (2,2)$. \roberto{In Table~\ref{tbl:wf_models_inj_1}, I summarize the waveform models used to analyze this synthetic signal. }

\begin{table}
\centering
	\begin{tabular}{lc}
		\hline
		\hline
		Model name & \ac{GW} modes \\
		\hline
		\texttt{SEOBNRv4\_ROM} & $(\ell, |m|) = (2,2)$  \\
		\texttt{SEOBNRv4HM\_ROM} & $(\ell, |m|) = (2,2),(2,1),(3,3),(4,4),(5,5)$  \\
		\hline
		\texttt{IMRPhenomHM} & $(\ell, |m|) = (2,2),(2,1),(3,3),(3,2),(4,4),(4,3)$  \\
		\hline
		\texttt{NRHybSur3dq8} & $\ell \leq 8$  \\
		\hline
		\hline
	\end{tabular}
	\caption{The waveform models used to analyze the synthetic signal described in Sec.~\ref{sec:synthetic_AS}. I also specify the \ac{GW} modes included in each waveform model.}
	\label{tbl:wf_models_inj_1}
\end{table}

The first binary parameters that are interesting to examine are the mass ratio $q$ and the effective spin~\cite{Damour:2001tu,Racine:2008qv,Ajith:2009bn,Santamaria:2010yb}
\begin{equation}
\label{eq:chieff}
\chi_\mathrm{eff} \equiv \frac{(m_1 \bm{\chi_1} + m_2 \bm{\chi_2})}{M} \cdot \bm{\hat{L}_\mathrm{N}} = \frac{1}{1+q}\left(q\,\bm{\chi_1} + \bm{\chi_2} \right) \cdot\bm{\hat{L}_\mathrm{N}},
\end{equation}
where \roberto{I remind that } $\bm{\hat{L}_\mathrm{N}}$ is the direction of the Newtonian angular momentum of the binary.
The mass ratio $q$ and the effective spin $\chi_\mathrm{eff}$ are typically well measured because \roberto{they determine $t_\mathrm{c}$, the time of coalescence of the binary from a given frequency, which the \ac{GW} detectors can measure quite well. In addition, $q$ and $\chi_\mathrm{eff}$ are degenerate because the phase of the inspiral waveform depends on them at similar \ac{PN} orders, respectively $1$\ac{PN} and $1.5$\ac{PN}}\footnote{In reality, the phase of the inspiral waveform at $1$\ac{PN} order is proportional to $\nu = q/M$, while at $1.5$\ac{PN} is proportional to $\beta \equiv 113/12(\chi_\mathrm{eff} - 76\nu/113 \chi_\mathrm{s})$ with $\chi_\mathrm{s} \equiv (\bm{\chi_1} + \bm{\chi_2})\cdot \bm{\hat{L}_\mathrm{N}}$. Nevertheless, in many applications the parameters $q$ and $\chi_\mathrm{eff}$ are used in substitution of $\nu$ and $\beta$ for simplicity.}.
In fact, it can be easily shown that the effect of a smaller mass ratio on the phase can be compensated by that of a larger effective spin~\cite{Cutler:1994ys,Poisson:1995ef,Baird:2012cu,Hannam:2013uu,Ohme:2013nsa}. As a consequence, these two parameters are expected to be correlated in the parameter estimation of \ac{BBH} systems.

\begin{figure}[hbt]
  \centering
  \includegraphics[width=0.7\textwidth]{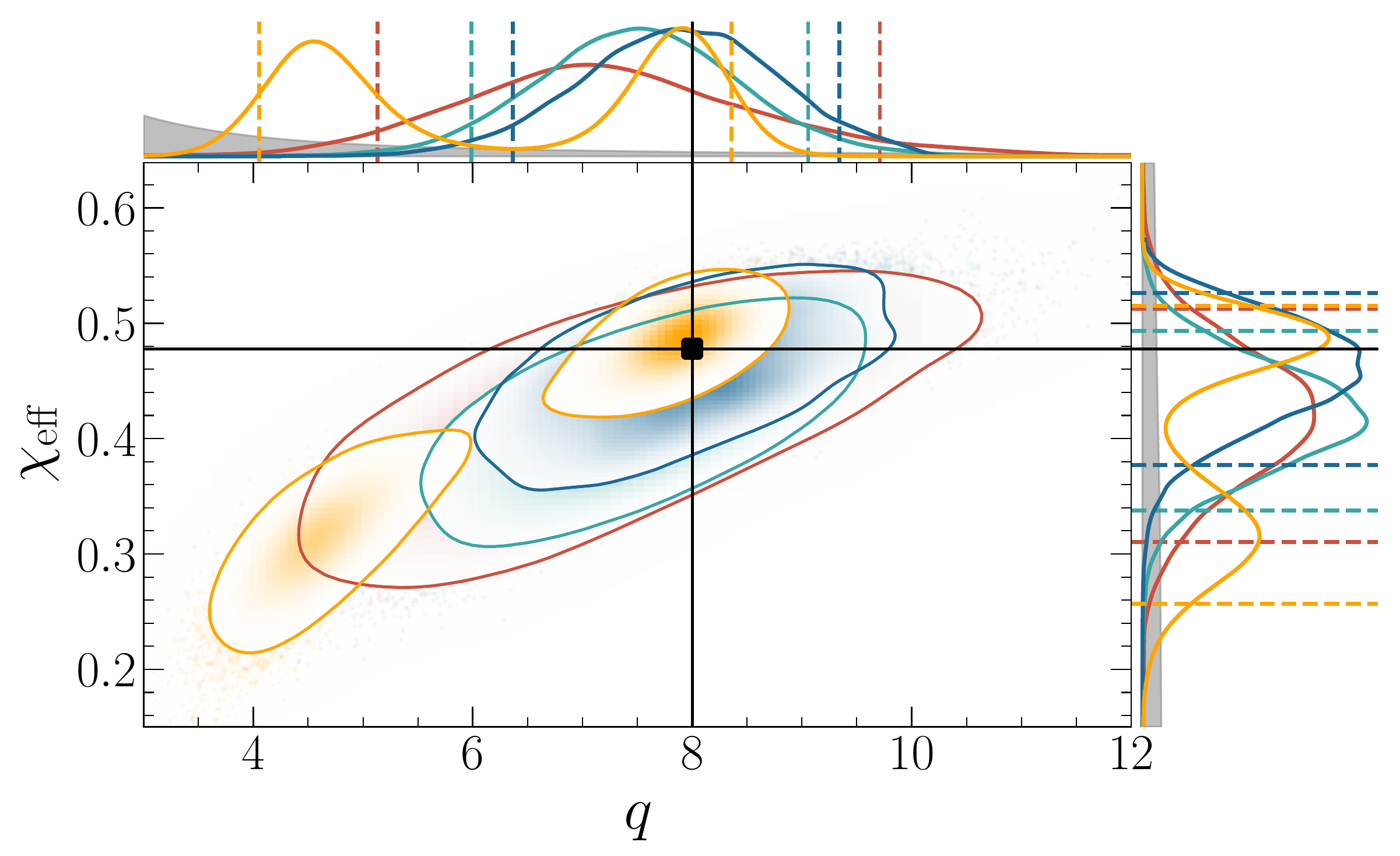}
\includegraphics[trim={8cm 0cm 8cm 0cm},clip,width=\textwidth]{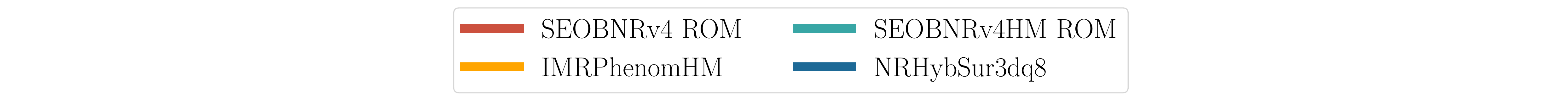}
	\caption{
  2D and 1D posterior distributions for the mass ratio $q$ and the effective spin $\chi_\mathrm{eff}$ measured from the synthetic BBH signal described in the text. In the 2D posteriors solid
    contours represent $90\%$ credible intervals and the black dot shows
    the value of the parameters used in the synthetic signal. In the 1D
    posteriors they are represented respectively by dashed lines and
    black solid lines. The gray shaded regions are the prior distributions. The parameter estimation is performed with the
    waveform models \texttt{SEOBNRv4\_ROM} (red), \texttt{SEOBNRv4HM\_ROM} (cyan), \texttt{NRHybSur3dq8} (blue) and \texttt{IMRPhenomHM} (orange).  Figure adapted from Ref.~\cite{Cotesta:2020qhw}.
  }
  \label{fig:PE_all_NRinj_q_chieff}
\end{figure}

In Fig.~\ref{fig:PE_all_NRinj_q_chieff}, I show, for each waveform model, the $1$D and $2$D posterior distributions for $q$ and $\chi_\mathrm{eff}$. The black dot in the plot represents the values used in the synthetic signal. 
All waveform models are able to measure, within the $90\%$ credible regions, the binary parameters used in the synthetic signal.  There are, however, some differences in the shape of the $90\%$ credible regions that are worth discussing.

First of all, in the measurement of $q$ and $\chi_\mathrm{eff}$ obtained with \texttt{SEOBNRv4HM\_ROM} (cyan curve) these two parameters are less correlated, when compared to those obtained with \texttt{SEOBNRv4\_ROM} (red curves). As a consequence, they are measured with more precision. This is expected, since the contribution of  \acp{HM} to the waveform increases when the binary system is more asymmetric, as discussed in Sec.~\ref{sec:HM_importance}. Therefore having \acp{HM} in the signal helps in measuring the mass ratio of the binary system and, consequently, in breaking its degeneracy with the effective spin.

The $2$D posterior distribution obtained with \texttt{SEOBNRv4HM\_ROM} (cyan curve) has a quite similar size to that computed with \texttt{NRHybSur3dq8} (blue curve), but it is less centered around the true values. This shift in the \texttt{SEOBNRv4HM\_ROM} posterior from the true values is a systematic bias due to the inaccuracy of the waveform model. In fact, in the case of \say{small} waveform systematic errors, the posterior distribution is expected to be shifted with respect to the true value, as discussed in Sec.~\ref{sec:source_characterization}.  Despite this systematic error, at the \ac{SNR} of this synthetic signal \texttt{SEOBNRv4HM\_ROM} is still able to measure the true parameters within the $90\%$ credible interval. 
\roberto{This result is consistent also with the expectation coming from  Lindblom's criterion, discussed in Sec.~\ref{sec:source_characterization}. In this case, the unfaithfulness between the \texttt{SEOBNRv4HM\_ROM} and the \texttt{NRHybSur3dq8} waveform is $0.3\%$. For this value of the unfaithfulness, Lindblom's criterium predicts a bias larger than the statistical uncertainty in at least one of \ac{BBH} parameters for values of the network-\ac{SNR} larger than $\sim22$, which is greater than $21.8$ used for this synthetic signal.  } The inaccuracy of the model \texttt{IMRPhenomHM} has instead a larger impact on the posterior distribution obtained with this model (orange curve). In fact, its shape is very different with respect to that obtained with the model \texttt{NRHybSur3dq8}, and it even features a bimodality. This is not unexpected since in \texttt{IMRPhenomHM} the \acp{HM} are modeled approximately without using any information from \ac{NR} waveforms. 
\roberto{Also this result is consistent with the expectation from Lindblom's criterium. In fact, the unfaithfulness between the \texttt{IMRPhenomHM} and the \texttt{NRHybSur3dq8} waveform is $3.2\%$, and, according to Lindblom's criterium, unbiased measurements of the parameters are only possible for network-\acp{SNR} $\lesssim 7$. }

Another two interesting binary parameters to study are, $\theta_\mathrm{JN}$, the angle between the total angular momentum $\bm{J}$ and the direction of the observer\footnote{In the case of \ac{BBH} systems with non-precessing spins the angles $\theta_\mathrm{JN}$ and $\iota$ coincide.}, and the luminosity distance $D_\mathrm{L}$. As in the case of $q$ and $\chi_\mathrm{eff}$, they are also correlated when only the dominant mode $(\ell, |m|) = (2,2)$ is included in the waveform. The reason for this correlation can be easily understood from Eq.~\eqref{eq:sph_harm_expansion_2} expressed for simplicity in the limit of $\theta_\mathrm{JN} \approx 0$. Under this approximation, the relevant modes with negative $m$ can be neglected, as ${}_{-2}Y_{\ell m}(\theta_\mathrm{JN} \rightarrow 0, \varphi_0) \rightarrow 0$. In this limit, and in the case that only the $(\ell, |m|) = (2,2)$ is included in the waveform, Eq.~\eqref{eq:sph_harm_expansion_2} reduces to
\begin{equation}
\label{eq:theta_jn_distance_deg}
h_+ - i h_\times \approx \frac{{}_{-2}Y_{2 2}(\theta_\mathrm{JN}, \varphi_0)}{D_\mathrm{L}} h_{22}(t; \bm{\lambda}).
\end{equation}
\roberto{It is important to highlight that ${}_{-2}Y_{2 2}(\theta_\mathrm{JN}, \varphi_0)$ is a complex function only because of the term $e^{2i\varphi_0}$, while the dependence on $\theta_\mathrm{JN}$ is a real function}. Eq.~\eqref{eq:theta_jn_distance_deg} implies that $h_+$ and $h_\times$ depend on $\theta_\mathrm{JN}$ and $D_\mathrm{L}$ only through the combination ${}_{-2}Y_{2 2}(\theta_\mathrm{JN}, \varphi_0)/D_\mathrm{L}$, hence they are correlated. \roberto{In addition,  $h_+$ and $h_\times$ depend on $\theta_\mathrm{JN}$ and $D_\mathrm{L}$ through a real function, therefore changing $\theta_\mathrm{JN}$ and $D_\mathrm{L}$ only effects their amplitude}. As a consequence, $\theta_\mathrm{JN}$ and $D_\mathrm{L}$ are particularly difficult to measure since LIGO and Virgo detectors are most sensitive to the phase of $h_+$ and $h_\times$. When at least one \ac{HM} is included in the waveform, for example the $(\ell, |m|) = (3,3)$ mode, Eq.~\eqref{eq:theta_jn_distance_deg} is modified to 
\begin{equation}
h_+ - i h_\times \approx \frac{1}{D_\mathrm{L}}\left({}_{-2}Y_{2 2}(\theta_\mathrm{JN}, \varphi_0)h_{22}(t; \bm{\lambda}) + {}_{-2}Y_{3 3}(\theta_\mathrm{JN}, \varphi_0)h_{33}(t; \bm{\lambda}) \right).
\end{equation}
In this case, there is not a simple correlation between $\theta_\mathrm{JN}$ and $D_\mathrm{L}$. In addition, the phases of $h_+$ and $h_\times$ are now functions of the phases of the individual modes $h_{22}$ and $h_{33}$, but also of ${}_{-2}Y_{2 2}(\theta_\mathrm{JN}, \varphi_0)$ and ${}_{-2}Y_{3 3}(\theta_\mathrm{JN}, \varphi_0)$. As a consequence, $\theta_\mathrm{JN}$ has an effect on the phases of $h_+$ and $h_\times$ and, for this reason, it can be better measured by LIGO and Virgo detectors.

\begin{figure}[hbt]
  \centering
  \includegraphics[width=0.7\textwidth]{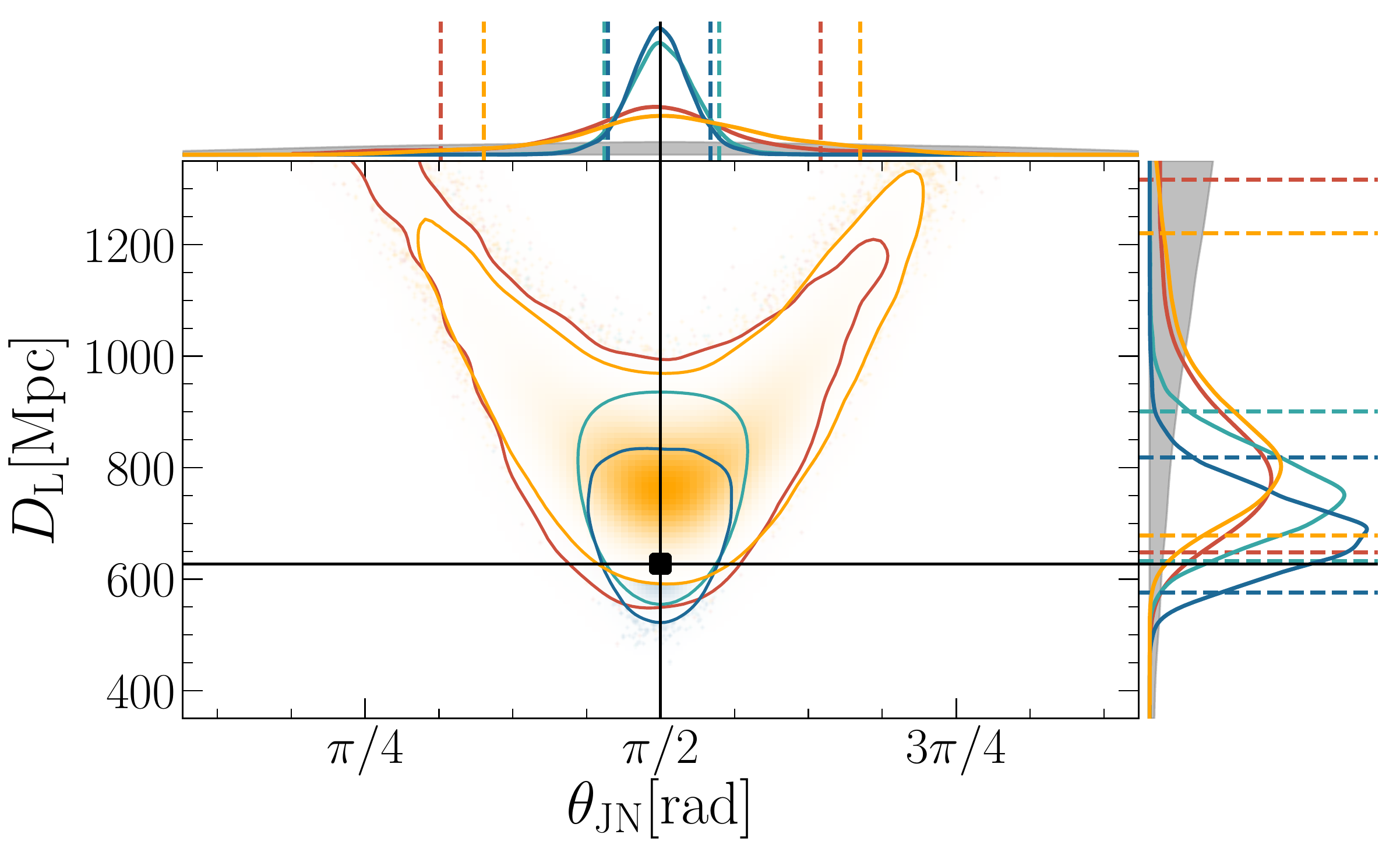}
\includegraphics[trim={8cm 0cm 8cm 0cm},clip,width=\textwidth]{event_legend_AS_injection.pdf}
	\caption{
  2D and 1D posterior distributions for the angle $\theta_\mathrm{JN}$ and the luminosity distance $D_\mathrm{L}$ measured from the synthetic BBH signal described in the text. In the 2D posteriors solid
    contours represent $90\%$ credible intervals and the black dot shows
    the value of the parameters used in the synthetic signal. In the 1D
    posteriors they are represented respectively by dashed lines and
    black solid lines. The gray shaded regions are the prior distributions. The parameter estimation is performed with the
    waveform models \texttt{SEOBNRv4\_ROM} (red), \texttt{SEOBNRv4HM\_ROM} (cyan), \texttt{NRHybSur3dq8} (blue) and \texttt{IMRPhenomHM} (orange).  Figure adapted from Ref.~\cite{Cotesta:2020qhw}.
  }
  \label{fig:PE_all_NRinj_theta_dL}
\end{figure}

These expectations about the parameters $\theta_\mathrm{JN}$ and $D_\mathrm{L}$ are confirmed by their measurements obtained from the synthetic signal, which I present in Fig.~\ref{fig:PE_all_NRinj_theta_dL}. Here, I show the $1$D and $2$D posterior distributions for these two parameters, when measured with the different waveform models listed above. The degeneracy between $\theta_\mathrm{JN}$ and $D_\mathrm{L}$ is present when using for the parameter estimation the waveform model \texttt{SEOBNRv4\_ROM} (red curve), but it is broken when the \acp{HM} are correctly included in the waveform, as in the case of the models \texttt{SEOBNRv4HM\_ROM} (cyan curve) and \texttt{NRHybSur3dq8} (blue curve). As a result, the measurement of both these parameters are more precise when obtained with one of these waveform models. As in the case of the $q$ and $\chi_\mathrm{eff}$, also here there is a shift between the $2$D posterior distribution obtained, with the model \texttt{SEOBNRv4HM\_ROM}, and the true values of the parameters. Also in this case, the explanation for this shift is a systematic bias due to the inaccuracy in the waveforms calculated with \texttt{SEOBNRv4HM\_ROM}. At the \ac{SNR} of this synthetic signal, the systematic bias is smaller than the statistical uncertainty and the true values are correctly measured within the $90\%$ credible intervals. On the contrary, the posterior distribution obtained with the waveform model \texttt{IMRPhenomHM} has a completely different shape with respect to those obtained with the other two models with \acp{HM}. The measurement of $\theta_\mathrm{JN}$ and $D_\mathrm{L}$, in this case, is similar to that obtained with the waveform model \texttt{SEOBNRv4\_ROM}, which does not include \acp{HM}. As before for $q$ and $\chi_\mathrm{eff}$, also this imprecise measurement of $\theta_\mathrm{JN}$ and $D_\mathrm{L}$ is likely due to the approximation used in the construction of the \acp{HM} in this waveform model.

This study confirms that including \acp{HM} in waveform models improves the measurement of the binary parameters of spinning \ac{BBH} systems, as also found in Refs.~\cite{Kumar:2018hml,Shaik:2019dym,Kalaghatgi:2019log}.
Most importantly, this analysis demonstrates that \texttt{SEOBNRv4HM\_ROM} can be used for parameter estimation yielding unbiased measurements, even for signals with moderately high \ac{SNR} and configurations where the effect of \acp{HM} in the waveform is large.

\FloatBarrier
\subsection{Synthetic signal II: the precessing-spins case}
\label{sec:synthetic_PS}

\roberto{In Chapter 3 Sec.~\ref{sec:peEOBNR}}, I extend the study described in the section above to \acp{BH} with precessing spins. For this purpose, I use \texttt{SEOBNRv4PHM} to measure the parameters of a \ac{BBH} system from a synthetic \ac{GW} signal. \roberto{Also in this case I use the three-detectors network, described in Sec.~\ref{sec:synthetic_AS}. In addition, for the construction of the synthetic signal I use, as before, the mean value of the noise $\langle n_\mathrm{syn}(t) \rangle = 0$. Also here, I use the \acp{PSD} of the noise of the Advanced LIGO and Advanced Virgo detectors at design sensitivity. }
For the waveform, I use the \ac{NR} waveform \texttt{SXS:BBH:0165}~\cite{Mroue:2013xna,Boyle:2019kee} having mass ratio $q = 6$ and initial spin components $\bm{\chi_1} = (-0.06,0.78,-0.47)$ and $\bm{\chi_2} = (0.08,-0.17,-0.23)$. The large mass ratio, and the in-plane spin components of this \ac{BBH} system, guarantee that the waveform features both a \textit{substantial \acp{HM} contribution} and \textit{strong precessional effects}. For the synthetic signal, I use a large total mass of the system, $M = 76 M_\odot$, and an inclination angle \roberto{close to edge-on, } $\theta_\mathrm{JN} \approx 1.3$, also to emphasize the \acp{HM} in the waveform. The network-\ac{SNR} of the synthetic signal is $21$. All the other parameters for the synthetic signal are not relevant for this discussion and are reported in Chapter 3 Sec.~\ref{sec:peEOBNR}.

\begin{figure}[hbt]
  \centering
  \includegraphics[width=0.7\textwidth]{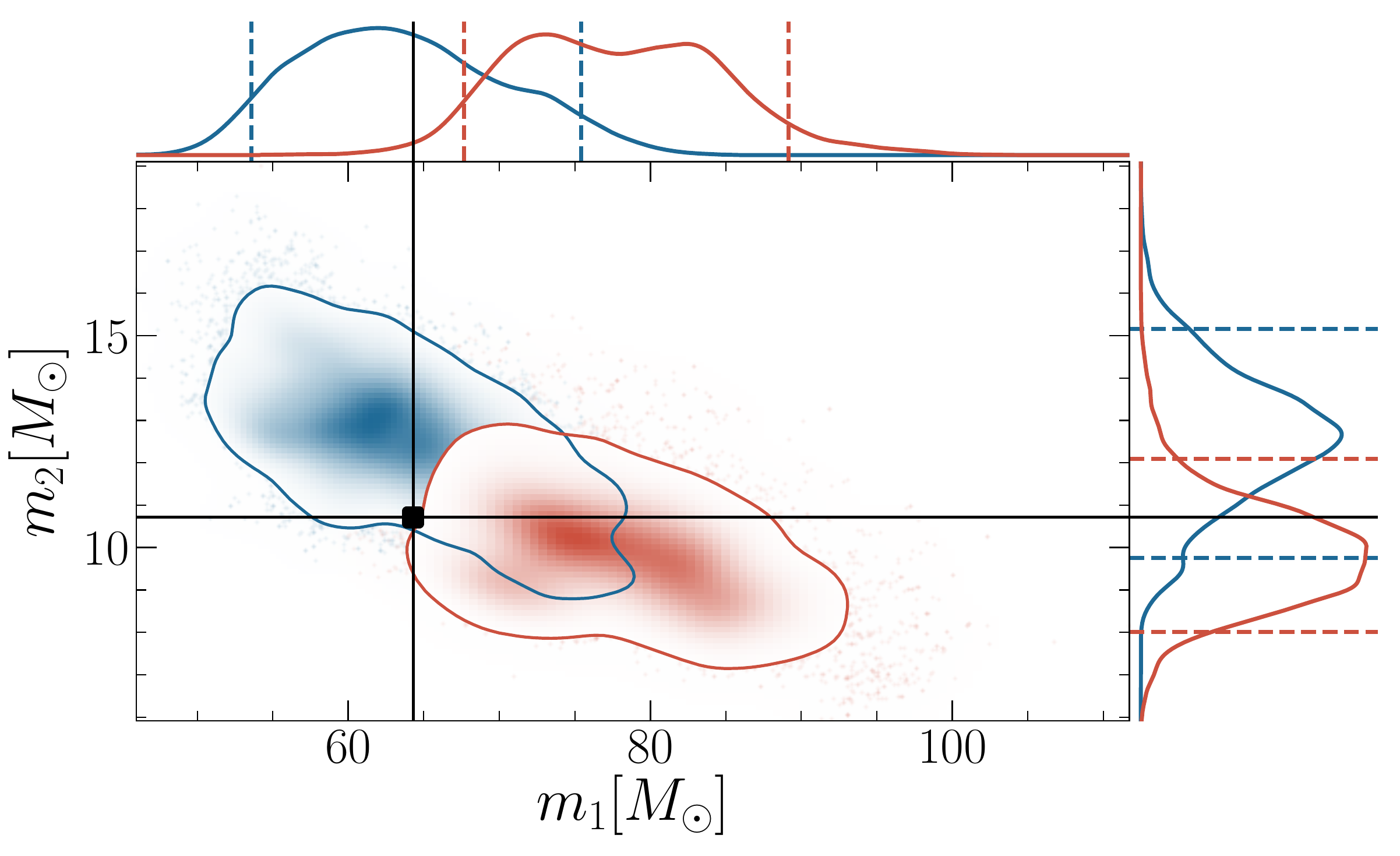}
  \includegraphics[clip, width=0.8\textwidth]{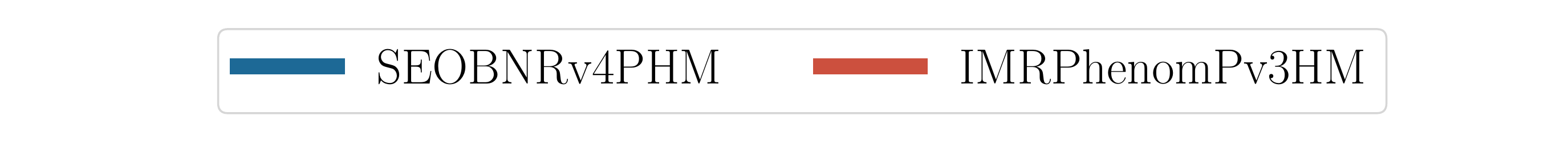}
  \caption{2D and 1D posterior distributions for the component masses in the source frame measured from the synthetic BBH signal described in the text. In the 2D posteriors solid
    contours represent $90\%$ credible intervals and the black dot shows
    the value of the parameters used in the synthetic signal. In the 1D
    posteriors they are represented respectively by dashed lines and
    black solid lines. The parameter estimation is performed with the
    waveform models \texttt{SEOBNRv4PHM} (blue) and \texttt{IMRPhenomPv3HM} (red). Figure taken from Ref.~\cite{Ossokine:2020kjp}}
  \label{fig:PE_q_6_masses}
\end{figure}

I perform the Bayesian parameter estimation on the signal using the waveform models \texttt{SEOBNRv4PHM} and \texttt{IMRPhenomPv3HM}~\cite{Khan:2019kot} for comparison.
In Fig.~\ref{fig:PE_q_6_masses}, I show the $2$D and $1$D posterior distributions for the component masses $m_1$ and $m_2$, obtained when using the two models for the analysis.
While with \texttt{SEOBNRv4PHM} the true value is correctly measured within the $90\%$ credible interval for both masses, this is not the case for \texttt{IMRPhenomPv3HM}. In fact, with this waveform model, the true value of $m_1$ is excluded from the $90\%$ credible interval. \roberto{This is not unexpected, as the unfaithfulness between the \texttt{IMRPhenomPv3HM} and the \ac{NR} waveform is $8.8\%$, while it is only $4.4\%$ when computed between the \texttt{SEOBNRv4PHM} and the \ac{NR} waveform. In this case, Lindblom's criterium predicts a bias larger than the statistical uncertainty in the measurement of at least one of the \ac{BBH} parameter for both waveform models. In fact, with the unfaithfulness of \texttt{IMRPhenomPv3HM} and \texttt{SEOBNRv4PHM}, measurements with bias smaller than the statistical uncertainty are predicted for network-\acp{SNR} values smaller than $\sim 6$ and $\sim 9$, respectively. I will show later that also \texttt{SEOBNRv4PHM} provides a measurement with bias larger than the statistical uncertainty in one of the binary parameters, as predicted by Lindblom's criterium. }

\begin{figure}[hbt]
  \centering
  \includegraphics[width=0.7\textwidth]{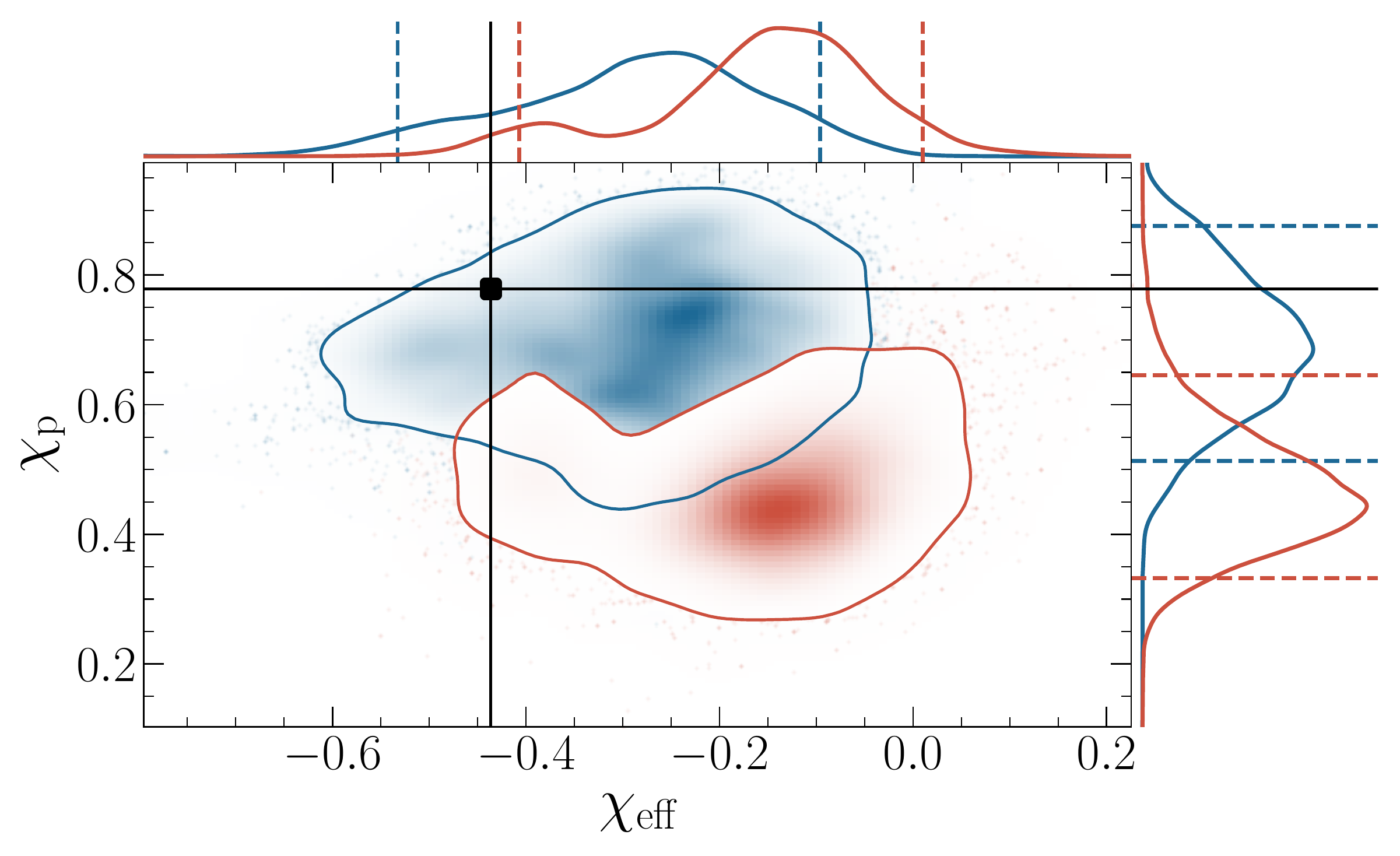}
  \includegraphics[clip, width=0.8\textwidth]{event_legend_posterior_samples_q6_NR_evolved.pdf}
  \caption{2D and 1D posterior distributions for the $\chi_{\mathrm{eff}}$ and the $\chi_{\mathrm{p}}$
    parameters measured from the synthetic BBH signal described in the text. In the 2D posteriors solid
    contours represent $90\%$ credible intervals and the black dot shows
    the value of the parameters used in the synthetic signal. In the 1D
    posteriors they are represented respectively by dashed lines and
    black solid lines. The parameter estimation is performed with the
    waveform models \texttt{SEOBNRv4PHM} (blue) and \texttt{IMRPhenomPv3HM} (red). Figure taken from Ref.~\cite{Ossokine:2020kjp}}
  \label{fig:PE_q_6_chip_chieff}
\end{figure}

Regarding the \ac{BH} spins, the parameter $\chi_\mathrm{eff}$, defined in Eq.~\eqref{eq:chieff}, ignores the spin components perpendicular to $\bm{\hat{L}_\mathrm{N}}$. In the case of precessing \ac{BBH} systems, these spin components are also interesting to measure and, a commonly used spin parameter to quantify their magnitude is the effective precession parameter~\cite{Schmidt:2014iyl}
\begin{equation}
\label{eq:chip}
\chi_\mathrm{p} = \max{\{|\bm{\chi_{1\perp}}|,k\,|\bm{\chi_{2\perp}} |\}},
\end{equation}
with $\bm{\chi_{i\perp}} = \bm{\chi_i} - (\bm{\chi_i}\cdot \bm{\hat{L}_\mathrm{N}}) \bm{\hat{L}_\mathrm{N}}$ and $k = q(4q+3)/(4+3q)$.
In Fig.~\ref{fig:PE_q_6_chip_chieff}, I show the $2$D and $1$D posterior distributions for $\chi_\mathrm{eff}$ and $\chi_\mathrm{p}$ obtained using \texttt{SEOBNRv4PHM} and \texttt{IMRPhenomPv3HM}. Also in this case \texttt{SEOBNRv4PHM}, is able to measure the values of these two spin parameters within the $90\%$ credible interval, while, with \texttt{IMRPhenomPv3HM}, the true values are excluded at this level.

\begin{figure}[hbt]
  \centering
  \includegraphics[width=0.7\textwidth]{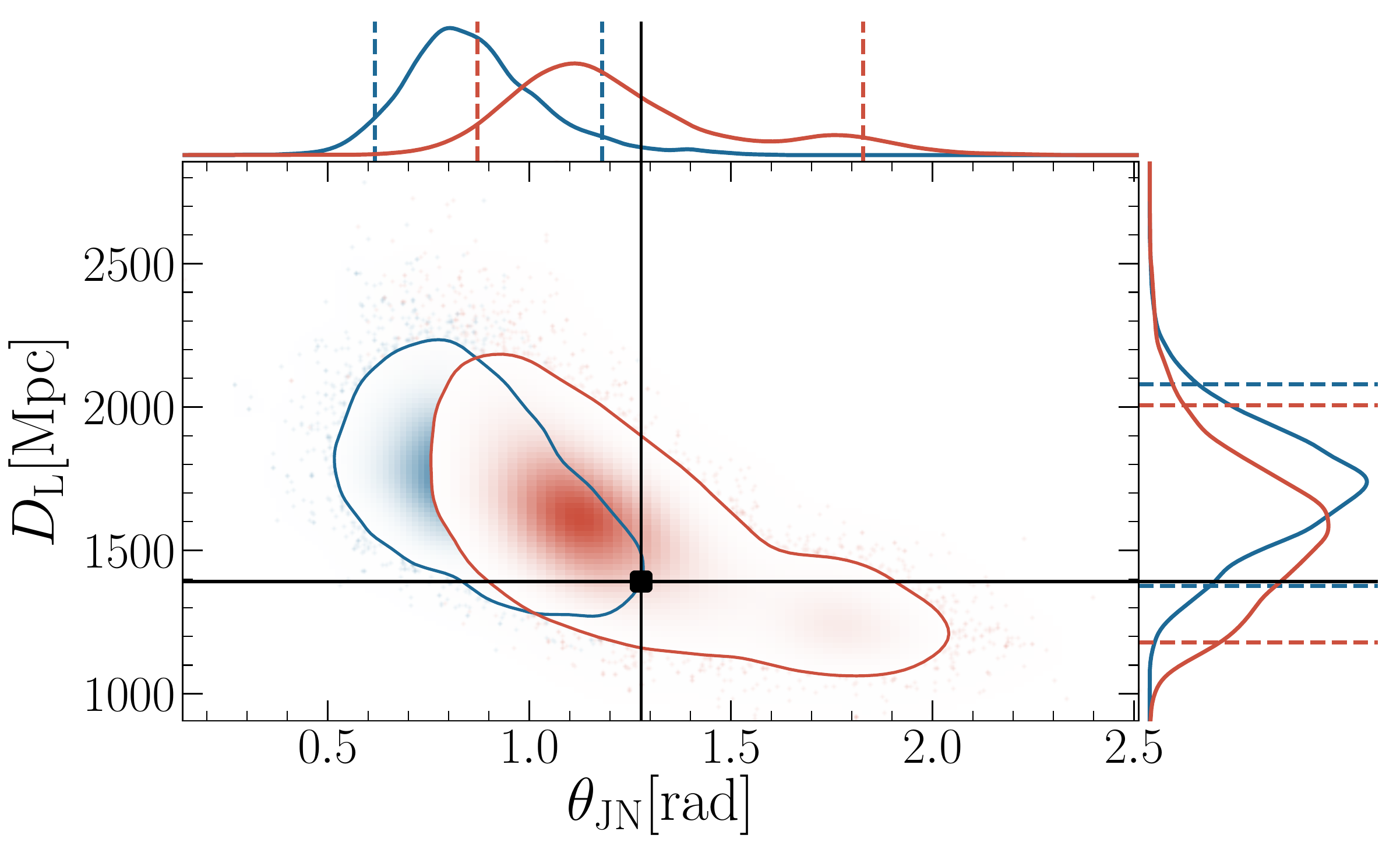}
  \includegraphics[clip, width=0.8\textwidth]{event_legend_posterior_samples_q6_NR_evolved.pdf}
  \caption{2D and 1D posterior distributions for the inclination angle $\theta_{\mathrm{JN}}$ and the luminosity distance $D_\mathrm{L}$ measured from the synthetic BBH signal described in the text. In the 2D posteriors solid
    contours represent $90\%$ credible intervals and the black dot shows
    the value of the parameters used in the synthetic signal. In the 1D
    posteriors they are represented respectively by dashed lines and
    black solid lines. The parameter estimation is performed with the
    waveform models \texttt{SEOBNRv4PHM} (blue) and \texttt{IMRPhenomPv3HM} (red). Figure taken from Ref.~\cite{Ossokine:2020kjp}}
  \label{fig:PE_q_6_thetajn_distance}
\end{figure}

Finally, in Fig.~\ref{fig:PE_q_6_thetajn_distance}, I show the $2$D and $1$D posterior distributions for the inclination angle $\theta_\mathrm{JN}$ and the luminosity distance $D_\mathrm{L}$, obtained with the two waveform models.
In this case, the inclination angle $\theta_\mathrm{JN}$ measured using the model \texttt{SEOBNRv4PHM} has a bias larger than the statistical uncertainty, as the true value of $\theta_\mathrm{JN}$ lies outside the $90\%$ credible interval of the posterior distribution. \roberto{This confirms the prediction made before, using Lindblom's criterium, that at least one of the \ac{BBH} parameter measured by \texttt{SEOBNRv4PHM} would have a bias larger than the statistical uncertainty. } On the contrary, \texttt{IMRPhenomPv3HM} provides a measurement of this parameter within the $90\%$ credible interval.
Both \texttt{SEOBNRv4PHM} and \texttt{IMRPhenomPv3HM} are able to recover the true value of the luminosity distance within the $90\%$ credible interval. 

This study demonstrates that, even for this \roberto{\ac{BBH} configuration with a large contribution of the \acp{HM} and strong precessional effects, } the model \texttt{SEOBNRv4PHM} yields an unbiased measurement for the most relevant binary parameters (masses, spins and luminosity distance) in the case of a signal \roberto{with the moderately high value of the \ac{SNR} of $21$}.

In Chapter 3 Sec.~\ref{sec:peEOBNR}, I also test the ability of \texttt{SEOBNRv4PHM} to provide unbiased measurements of the binary parameters, in the case of a synthetic signal with \textit{very large \ac{SNR}}, but for a \ac{BBH} system with \roberto{smaller mass ratio and \ac{BH}'s spin magnitudes. } Also in this case, the synthetic signal only includes the waveform and no detector noise.
For the signal, I use a waveform generated from the \ac{NR} surrogate model \texttt{NRSur7dq4}~\cite{Varma:2019csw}, instead of a \ac{NR} waveform, \roberto{because they are indistinguishable at the \ac{SNR} of this study, which I indicate below}.  The \ac{BBH} system emitting this waveform has mass ratio $q = 3$, total mass $M = 70 M_\odot$, initial spins $\bm{\chi}_1 = (0.3,0.0,0.5)$ and $\bm{\chi}_2 = (0.2,0.0,0.3)$ and inclination angle $\theta_\mathrm{JN} \approx 0.9$. The network-\ac{SNR} of the signal, in the same three-detector network used before, is $50$. The other binary parameters are not relevant for the discussion, and can be found in Chapter 3 Sec.~\ref{sec:peEOBNR}.

\begin{figure}[hbt]
  \centering
  \includegraphics[width=0.7\textwidth]{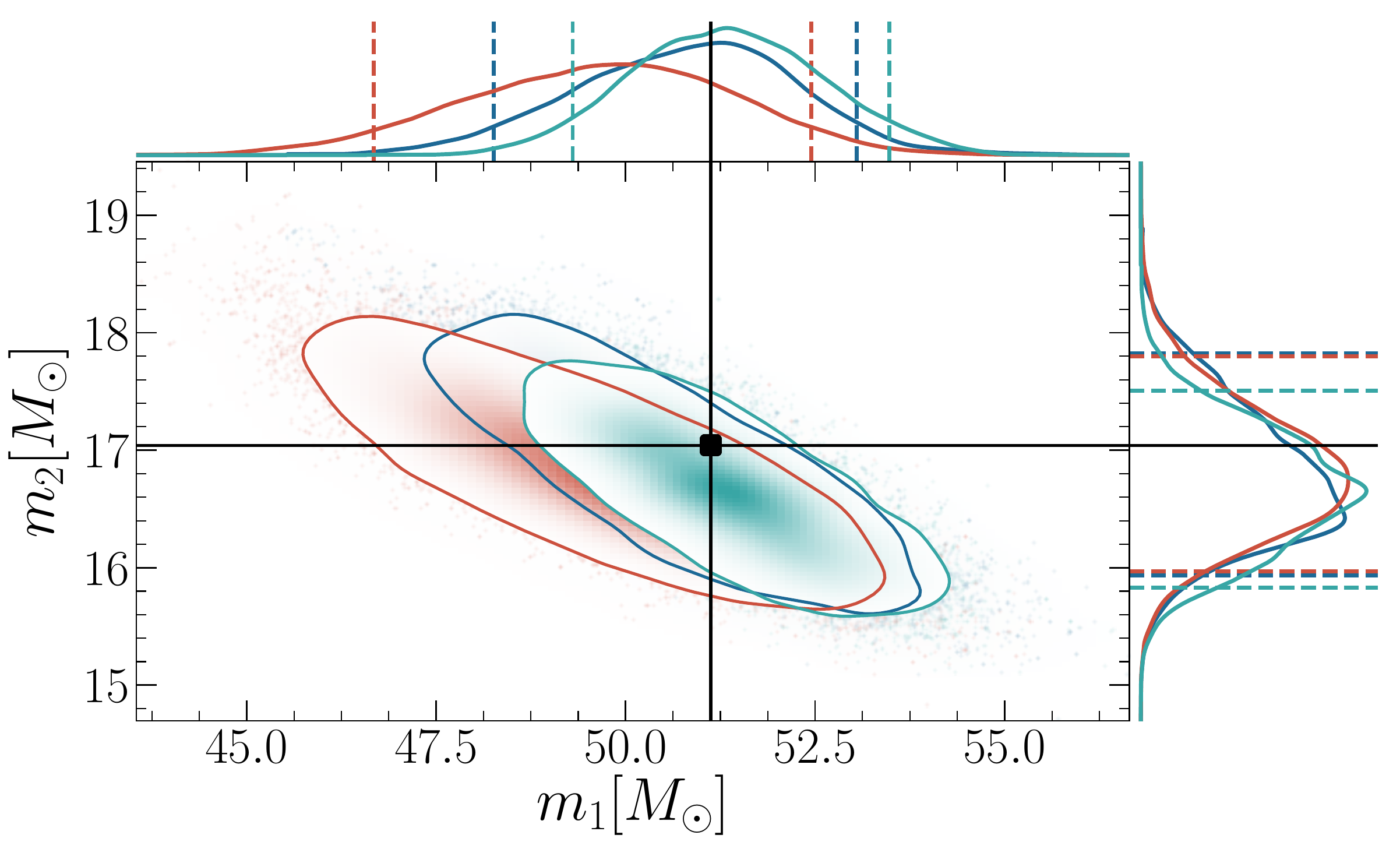}
  \includegraphics[trim={10cm 0cm 8cm 0cm},clip, width=\textwidth]{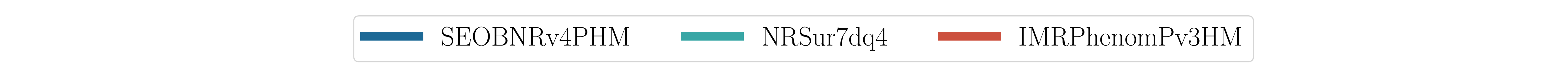}
  \caption{2D and 1D posterior distributions for the component masses in the source frame measured from the synthetic BBH signal described in the text. In the 2D posteriors solid
    contours represent $90\%$ credible intervals and the black dot shows
    the value of the parameters used in the synthetic signal. In the 1D
    posteriors they are represented respectively by dashed lines and
    black solid lines. The parameter estimation is performed with the
    waveform models \texttt{SEOBNRv4PHM} (blue), \texttt{IMRPhenomPv3HM} (red) and \texttt{NRSur7dq4} (cyan). Figure taken from Ref.~\cite{Ossokine:2020kjp}}
  \label{fig:PE_q_3_masses}
\end{figure}

\begin{figure}[hbt]
  \centering
  \includegraphics[width=0.7\textwidth]{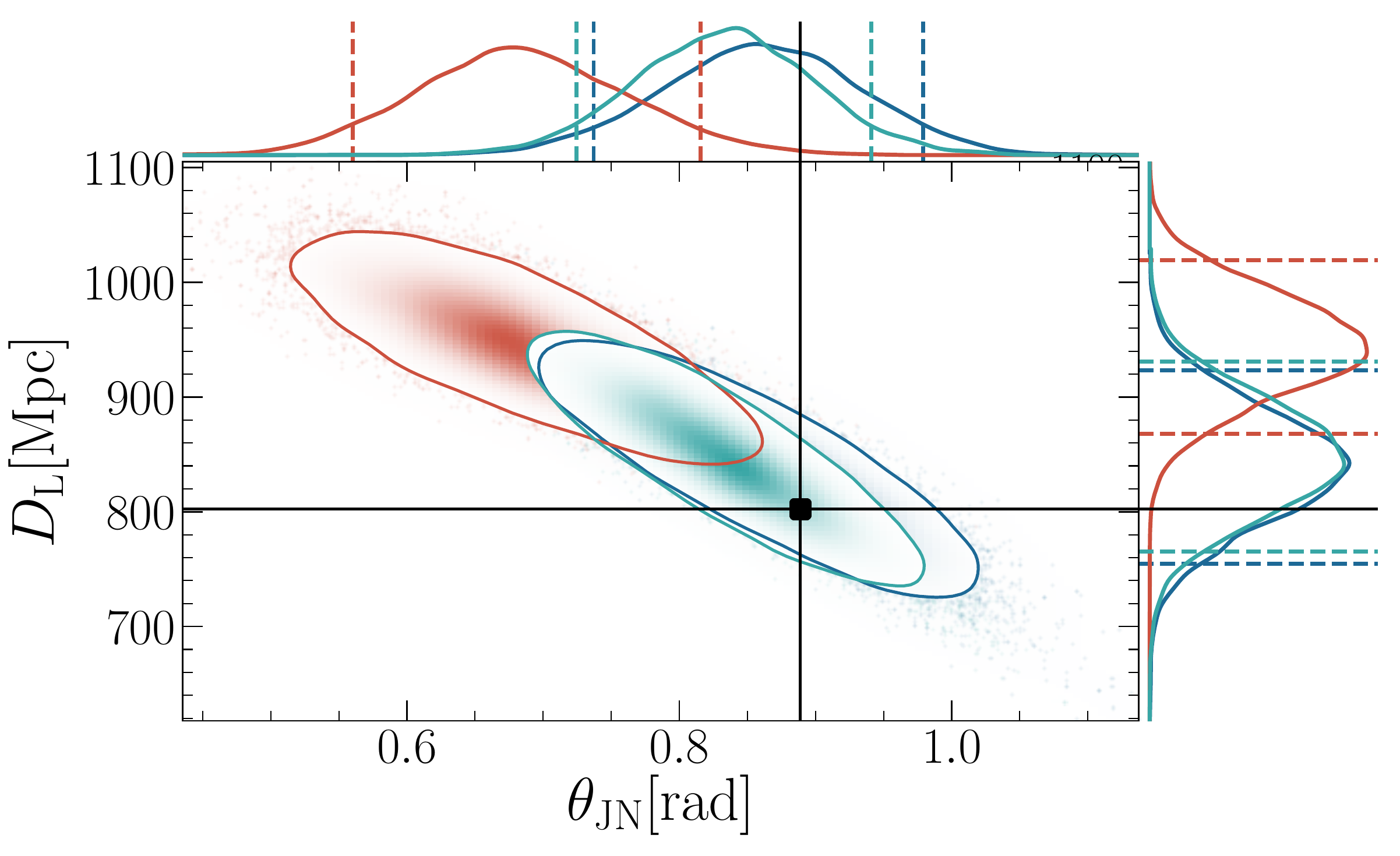}
  \includegraphics[trim={10cm 0cm 8cm 0cm},clip, width=\textwidth]{event_legend_posterior_samples_q3.pdf}
  \caption{2D and 1D posterior distributions for the inclination angle $\theta_{\mathrm{JN}}$ and the luminosity distance $D_\mathrm{L}$ measured from the synthetic BBH signal described in the text. In the 2D posteriors solid
    contours represent $90\%$ credible intervals and the black dot shows
    the value of the parameters used in the synthetic signal. In the 1D
    posteriors they are represented respectively by dashed lines and
    black solid lines. The parameter estimation is performed with the
    waveform models \texttt{SEOBNRv4PHM} (blue), \texttt{IMRPhenomPv3HM} (red) and \texttt{NRSur7dq4} (cyan). Figure taken from Ref.~\cite{Ossokine:2020kjp}}
  \label{fig:PE_q_3_thetajn_distance}
\end{figure}

I perform the parameter estimation on this signal using the waveform model \texttt{NRSur7dq4}, in addition to \texttt{SEOBNRv4PHM} and \texttt{IMRPhenomPv3HM}. In Table~\ref{tbl:wf_models_inj_2}, I summarize the waveform models used in this study and in the study above.
\begin{table}
\centering
	\begin{tabular}{lc}
		\hline
		\hline
		Model name & \ac{GW} modes in co-precessing frame \\
		\hline
		\texttt{SEOBNRv4PHM} & $(\ell, |m|) = (2,2),(2,1),(3,3),(4,4),(5,5)$  \\
		\hline
		\texttt{IMRPhenomPv3HM} & $(\ell, |m|) = (2,2),(2,1),(3,3),(3,2),(4,4),(4,3)$  \\
		\hline
		\texttt{NRSur7dq4} & $\ell \leq 8$  \\
		\hline
		\hline
	\end{tabular}
	\caption{The waveform models used to analyze the synthetic signals described in Sec.~\ref{sec:synthetic_PS}. I also specify the \ac{GW} modes included in each waveform model.}
	\label{tbl:wf_models_inj_2}
\end{table}
\roberto{In Fig.~\ref{fig:PE_q_3_masses}, I show the $2$D and $1$D posterior distributions for the masses $m_1$ and $m_2$. Even for this large \ac{SNR}, \texttt{SEOBNRv4PHM} (blue curve) is able to correctly measure the true values of the masses within the $90\%$ credible intervals}. In addition, the posterior distributions for the $m_1$ and $m_2$ obtained with this model agree very well with those obtained using the \ac{NR} surrogate model (cyan curve).
\roberto{The waveform model \texttt{IMRPhenomPv3HM} (red curve) is also able to recover the true values of $m_1$ and $m_2$ within the $90\%$ credible interval. Both waveform models are also able to measure with bias smaller than the statistical uncertainty the spin parameters $\chi_\mathrm{eff}$ and $\chi_\mathrm{p}$, see Chapter 3 Sec.~\ref{sec:peEOBNR} for the detailed analysis}.
\roberto{Finally, in Fig.~\ref{fig:PE_q_3_thetajn_distance}, I show the $2$D and $1$D posterior distributions for the inclination angle $\theta_\mathrm{JN}$ and the luminosity distance $D_\mathrm{L}$, as measured by the waveform models \texttt{SEOBNRv4PHM} (blue), \texttt{IMRPhenomPv3HM} (red) and \texttt{NRSur7dq4} (cyan). Also for these parameters \texttt{SEOBNRv4PHM} provides unbiased measurements, in excellent agreement with the \ac{NR} waveform model \texttt{NRSur7dq4}. Conversely, the parameters $\theta_\mathrm{JN}$ and $D_\mathrm{L}$ measured with \texttt{IMRPhenomPv3HM} have a bias larger than the statistical uncertainty, as their true values lie outside the $90\%$ credible intervals obtained with this model. This is in agreement with the expectation from Lindblom's criterium. In fact, the unfaithfulness between the \texttt{NRSur7dq4} and the \texttt{IMRPhenomPv3HM} waveform is $1\%$. For this value of the unfaithfulness, Lindblom's criterium predicts a bias larger than the statistical uncertainty in one of the \ac{BBH} parameter for network-\ac{SNR} values larger than $19$. Also for \texttt{SEOBNRv4PHM}, Lindblom's criterium predicts a bias larger than the statistical uncertainty in one of the measured \ac{BBH} parameters. In fact, the unfaithfulness between the \texttt{SEOBNRv4PHM} and the \texttt{NRSur7dq4} waveform is $0.2\%$, for which Lindblom's criterium predicts bias larger than the statistical uncertainty in at least one of the \ac{BBH} parameters, for network-\acp{SNR} larger than $\sim 42$. However, for \texttt{SEOBNRv4PHM} I find that none of the measured parameters have biases larger than the statistical uncertainty. This confirms that Lindblom's criterium is sometimes too conservative, as already observed in Ref.~\cite{Purrer:2019jcp}. }

In conclusion, this analysis shows that using \texttt{SEOBNRv4PHM} for parameter estimation will also be appropriate to analyze the properties of \acp{BBH} detected by upgraded version of the LIGO and Virgo detectors, when \ac{GW} signals with such a large \ac{SNR} are expected as a consequence of the improvements in detector sensitivity.

\FloatBarrier
\subsection{The LIGO-Virgo signal GW170729}
\label{sec:pe_GW170729}

\roberto{Chapter 5 is the publication in which } I use the waveform model \texttt{SEOBNRv4HM} (as its \ac{ROM} was not available at the time of this study) to analyze the signal GW170729~\cite{LIGOScientific:2018mvr} detected by the LIGO and Virgo interferometers.
This \textit{real} \ac{GW} signal is interesting to study using waveform models with \acp{HM}, because its source is likely the most massive one observed during the \roberto{the first and second (O1 and O2) } LIGO and Virgo observing runs, with a total mass of $\sim 85 M_\odot$.

\begin{figure}[hbt]
  \centering
  \includegraphics[width=0.7\textwidth]{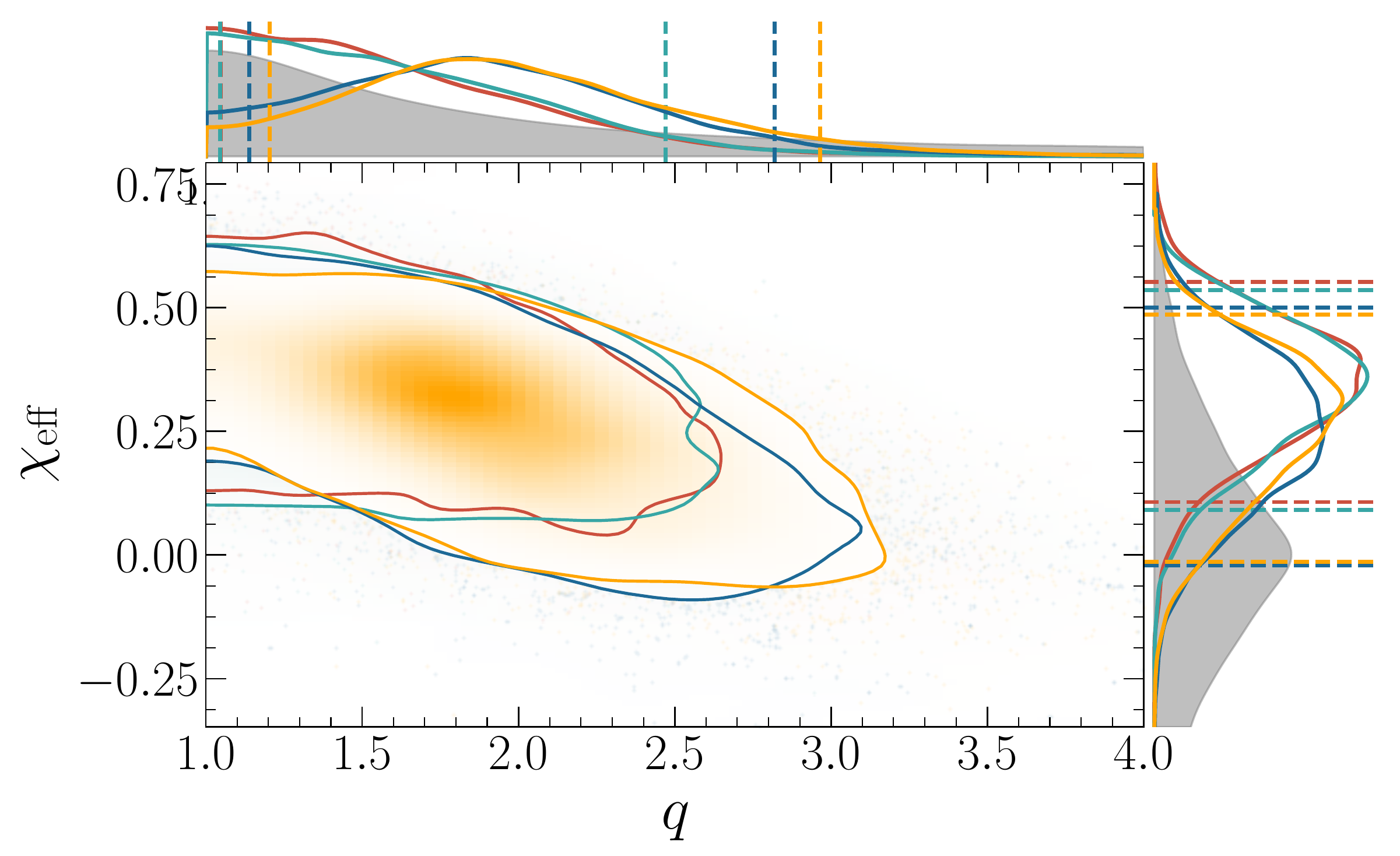}
\includegraphics[trim={8cm 0cm 8cm 0cm},clip,width=\textwidth]{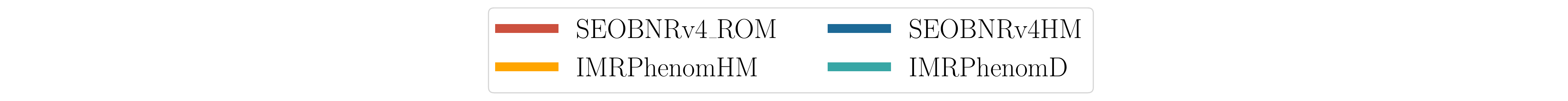}
	\caption{
  2D and 1D posterior distributions for the mass ratio $q$ and the effective spin $\chi_\mathrm{eff}$ measured from the signal GW170729. In the 2D posteriors solid
    contours represent $90\%$ credible intervals and the black dot shows
    the value of the parameters used in the synthetic signal. In the 1D
    posteriors they are represented respectively by dashed lines and
    black solid lines. The gray shaded regions are the prior distributions. The parameter estimation is performed with the
    waveform models \texttt{SEOBNRv4\_ROM} (red), \texttt{SEOBNRv4HM} (blue), \texttt{IMRPhenomD} (cyan) and \texttt{IMRPhenomHM} (orange).  Figure adapted from Ref.~\cite{Chatziioannou:2019dsz}.
  }
  \label{fig:PE_GW170729_q_chieff}
\end{figure}

In this real \ac{GW} signal, the binary parameters that are most affected, when waveform models with \acp{HM} are used for the parameter estimation, are the mass ratio $q$ and the effective spin $\chi_\mathrm{eff}$.
In Fig.~\ref{fig:PE_GW170729_q_chieff}, I show the $2$D and $1$D posterior distributions for these two parameters, obtained from the Bayesian inference on the signal performed using the waveform models \texttt{SEOBNRv4HM} and \texttt{SEOBNRv4\_ROM}. I also compare these measurements with those obtained by the waveform models \texttt{IMRPhenomHM} and \texttt{IMRPhenomD}, to test their robustness against eventual biases, due to the inaccuracy of the waveform models. See Table~\ref{tbl:wf_models_inj_GW170729} for the summary of the waveform models used for this study.
\begin{table}
\centering
	\begin{tabular}{lc}
		\hline
		\hline
		Model name & \ac{GW} modes \\
		\hline
		\texttt{SEOBNRv4\_ROM} & $(\ell, |m|) = (2,2)$  \\
		\texttt{SEOBNRv4HM\_ROM} & $(\ell, |m|) = (2,2),(2,1),(3,3),(4,4),(5,5)$  \\
		\hline
		\texttt{IMRPhenomD} & $(\ell, |m|) = (2,2)$  \\
		\texttt{IMRPhenomHM} & $(\ell, |m|) = (2,2),(2,1),(3,3),(3,2),(4,4),(4,3)$  \\
		\hline
		\hline
	\end{tabular}
	\caption{The waveform models used to analyze the real \ac{GW} signal GW170729. I also specify the \ac{GW} modes included in each waveform model.}
	\label{tbl:wf_models_inj_GW170729}
\end{table}
The mass ratio posterior distributions, obtained with the waveform models \texttt{SEOBNRv4\_ROM} and \texttt{IMRPhenomD}, which only include the mode $(\ell, |m|) = (2,2)$, (red and cyan curves) are approximately flat in the region $1 \lesssim q \lesssim 2$, and exclude $q\gtrsim 2.5$ with $95\%$ probability. On the contrary, the posterior distributions obtained with \texttt{SEOBNRv4HM} and \texttt{IMRPhenomHM} (blue and orange curves) are both peaked around $q \approx 2$ and, most importantly, they exclude with a larger confidence the hypothesis of a merger of \acp{BH} with the same masses. In fact, according to the measurements obtained with \texttt{SEOBNRv4HM} and \texttt{IMRPhenomHM}, there is a $40\%$ probability that the mass ratio of the system is $q\geq 2$, while this probability is only $20\%$ if one considers the measurements with \texttt{SEOBNRv4\_ROM} and \texttt{IMRPhenomD}.
Regarding the $\chi_\mathrm{eff}$ parameter, the measurements obtained with \texttt{SEOBNRv4\_ROM} and \texttt{IMRPhenomD} imply that $\chi_\mathrm{eff} > 0$ with a probability of $99\%$. Among the signals observed during \roberto{O1 and O2}, only for this signal and GW151226~\cite{Abbott:2016nmj} a negative value for $\chi_\mathrm{eff}$ is excluded with such large probability. When the waveform models with \acp{HM} are used for the parameter estimation, the $\chi_\mathrm{eff}$ posterior distribution shift to smaller values, and the probability of a positive $\chi_\mathrm{eff}$ is reduced to $94\%$.

\begin{figure}[hbt]
  \centering
  \includegraphics[width=0.7\textwidth]{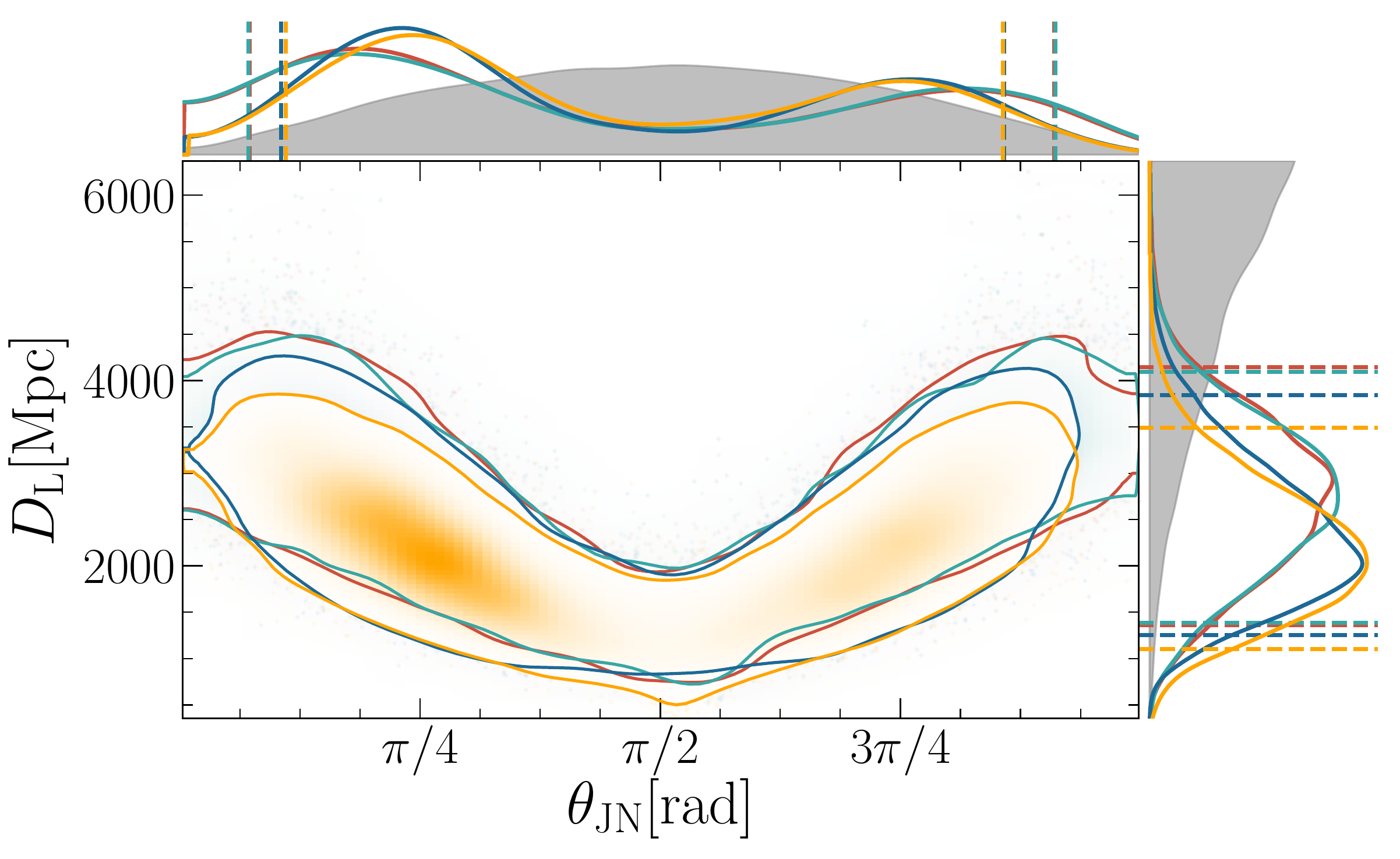}
\includegraphics[trim={8cm 0cm 8cm 0cm},clip,width=\textwidth]{event_legend_GW170729.pdf}
	\caption{
  2D and 1D posterior distributions for the inclination angle $\theta_\mathrm{JN}$ and the luminosity distance $D_\mathrm{L}$ measured from the signal GW170729. In the 2D posteriors solid
    contours represent $90\%$ credible intervals and the black dot shows
    the value of the parameters used in the synthetic signal. In the 1D
    posteriors they are represented respectively by dashed lines and
    black solid lines. The gray shaded regions are the prior distributions. The parameter estimation is performed with the
    waveform models \texttt{SEOBNRv4\_ROM} (red), \texttt{SEOBNRv4HM} (blue), \texttt{IMRPhenomD} (cyan) and \texttt{IMRPhenomHM} (orange).  Figure adapted from Ref.~\cite{Chatziioannou:2019dsz}.
  }
  \label{fig:PE_GW170729_thetajn_dist}
\end{figure}

Using waveform models with \acp{HM} has a smaller impact on the measurements of the inclination angle $\theta_\mathrm{JN}$ and the luminosity distance $D_\mathrm{L}$ of this system. In Fig.~\ref{fig:PE_GW170729_thetajn_dist}, I show the $2$D and $1$D posterior distributions of these two parameters, obtained with the waveform models described before. The only relevant difference is that both waveform models with \acp{HM} are able to exclude with larger confidence inclination angles $\theta_\mathrm{JN}$ close to $0$ and $\pi$. As consequence of the correlation between $\theta_\mathrm{JN}$ and $D_\mathrm{L}$, also larger luminosity distances are excluded with a greater confidence.

The study presented here assumes \acp{BH} spins aligned with the angular momentum of the binary. In Ref.~\cite{Khan:2019kot} the authors generalize this analysis by using the waveform model \texttt{IMRPhenomPv3HM}, which includes the effect of \acp{HM} and precessing \ac{BH} spins. The results of their analysis are in agreement with those discussed here. This is not unexpected, since they report a measured value of the spin parameter $\chi_\mathrm{p}$ consistent with $0$, as it should be in the case of \ac{BH} spins aligned with the angular momentum of the binary.

\FloatBarrier
\subsection{The LIGO-Virgo signal GW190412}
\label{sec:pe_GW190412}

In this section, I summarize the analysis I performed as member of the editorial team of the publication that reported the discovery of GW190412~\cite{LIGOScientific:2020stg} by the LIGO and Virgo detectors. 
This \textit{real} \ac{GW} signal is particularly interesting because its source is a \ac{BBH} system with a mass ratio $q$ measured precisely enough to \textit{exclude with large confidence the scenario of a merger between \acp{BH} with equal masses}. In fact, the mass ratio of this \ac{BBH} system lies in the region $3 \lesssim q \lesssim 5$, while most of the \acp{BBH} observed during \roberto{O2 and O3a } have mass ratios consistent with 1 (see Fig.5 in Ref.~\cite{LIGOScientific:2018mvr} and Fig.6 in Ref.~\cite{Abbott:2020niy}). 

In the following, I discuss the most interesting properties of this system, with particular emphasis on (i) the impact of the improved waveform models discussed in Sec.~\ref{sec:waveform_modeling_intro} on the precise measurement of these properties and (ii) potential systematic biases in these measurements due to the inaccuracy of the waveform models. \roberto{In fact, because of its interesting properties, this \ac{BBH} system was analyzed with a large number of waveform models. I summarize in Table~\ref{tbl:wf_models_inj_GW190412} the waveform models I consider here. }

\begin{table}
\centering
	\begin{tabular}{lcc}
		\hline
		\hline
		Model name & \ac{GW} modes in co-precessing frame & Precession \\
		\hline
		\texttt{SEOBNRv4\_ROM} & $(\ell, |m|) = (2,2)$ & $\times$  \\
		\texttt{SEOBNRv4HM\_ROM} & $(\ell, |m|) = (2,2),(2,1),(3,3),(4,4),(5,5)$ & $\times$  \\
		\texttt{SEOBNRv4PHM} & $(\ell, |m|) = (2,2),(2,1),(3,3),(4,4),(5,5)$ & $\checkmark$  \\
		\hline
		\texttt{IMRPhenomD} & $(\ell, |m|) = (2,2)$ & $\times$  \\
		\texttt{IMRPhenomHM} & $(\ell, |m|) = (2,2),(2,1),(3,3),(3,2),(4,4),(4,3)$ & $\times$  \\
		\texttt{IMRPhenomXHM} & $(\ell, |m|) = (2,2),(2,1),(3,3),(3,2),(4,4)$ & $\times$  \\
		\texttt{IMRPhenomPv3HM} & $(\ell, |m|) = (2,2),(2,1),(3,3),(3,2),(4,4),(4,3)$ & $\checkmark$  \\
		\texttt{IMRPhenomXPHM} & $(\ell, |m|) = (2,2),(2,1),(3,3),(3,2),(4,4)$ & $\checkmark$  \\
		\hline

		\hline
		\hline
	\end{tabular}
	\caption{The waveform models used to analyze the real \ac{GW} signal GW190412. I also specify the \ac{GW} modes included in each waveform model, and whether they include the effect of spin precession.}
	\label{tbl:wf_models_inj_GW190412}
\end{table}

\begin{table}
\center
\begin{tabular}{lc}
\hline\hline
 parameter
 & Measurement with \texttt{SEOBNRv4PHM} \\ \hline
$m_1 / M_\odot$ & $31.7^{+3.6}_{-3.5}$ \\
$m_2 / M_\odot$ & $8.0^{+0.9}_{-0.7}$ \\
$M / M_\odot$ & $39.7^{+3.0}_{-2.8}$ \\
$q$ & $4.00^{+0.76}_{-0.77}$ \\
\hline
$\chi_\mathrm{eff}$ & $0.28^{+0.06}_{-0.08}$  \\
$\chi_\mathrm{p}$ & $0.31^{+0.14}_{-0.15}$  \\
$\chi_1$ & $0.46^{+0.12}_{-0.15}$ \\
\hline
$D_\mathrm{L} / \textrm{Mpc}$ & $740^{120}_{-130}$ \\
$\theta_{JN}$ & $0.71^{+0.23}_{-0.21}$ \\
\hline\hline
\end{tabular}
\caption{Measured parameters for the \ac{GW} signal GW190412 and
their
90\% credible intervals, obtained using the waveform model \texttt{SEOBNRv4PHM}.}
\label{tab:PEresults_intro}
\end{table}

\begin{figure}[hbt]
  \centering
  \includegraphics[width=0.7\textwidth]{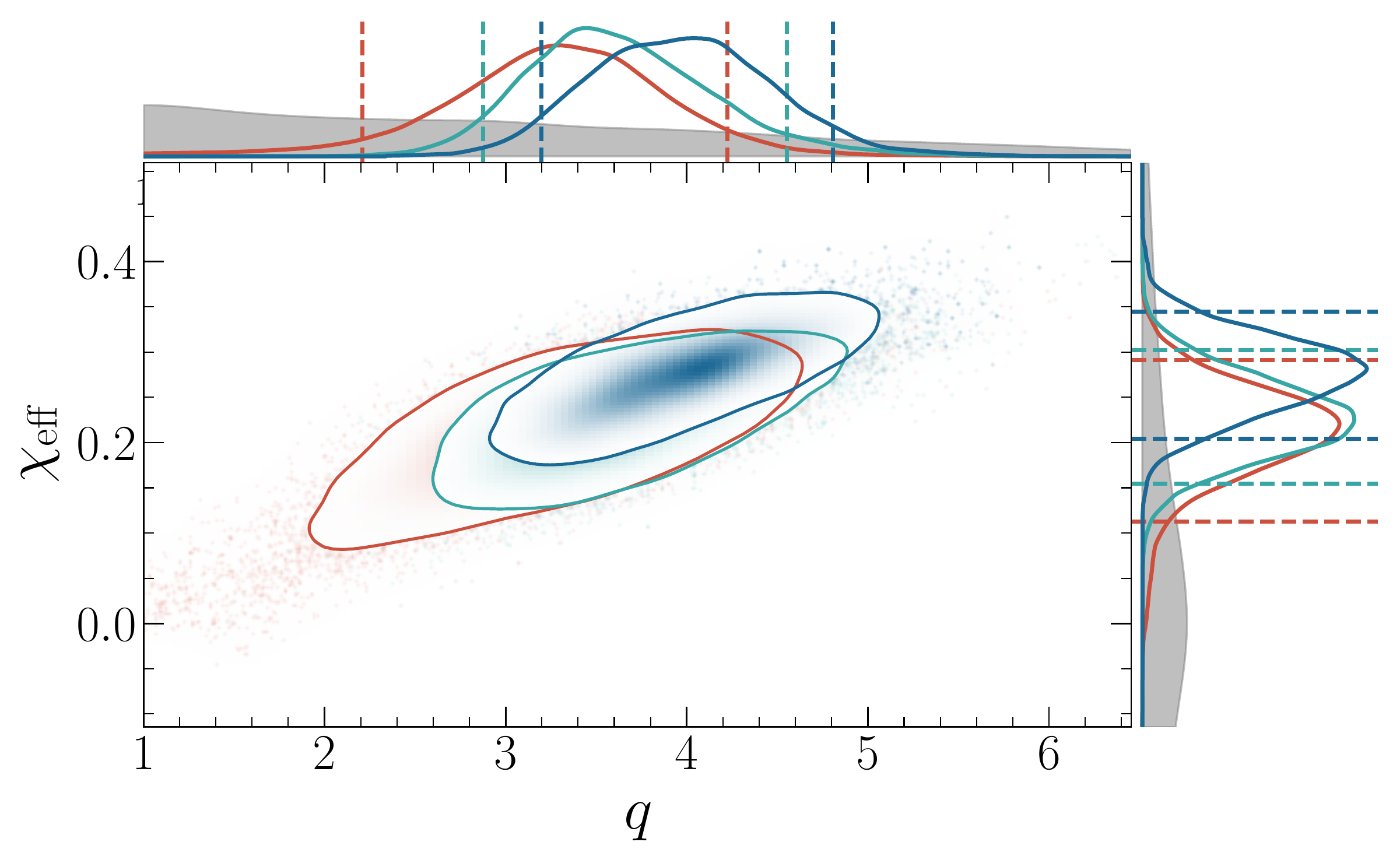}
  \includegraphics[trim={10cm 0cm 8cm 0cm}, clip, width=\textwidth]{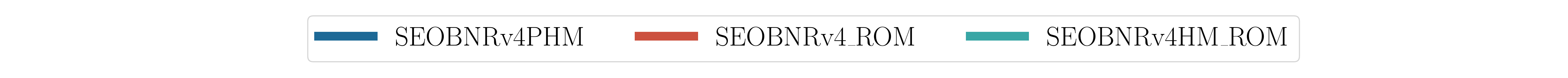}
  \caption{2D and 1D posterior distributions for the mass ratio $q$ and the effective spin $\chi_\mathrm{eff}$ of the signal GW190412. $90\%$ credible intervals are represented by solid contours in the 2D posteriors and dashed lines in the 1D posteriors. The gray shaded regions are the priors.  The parameter estimation is performed with the waveform models \texttt{SEOBNRv4PHM} (blue), \texttt{SEOBNRv4\_ROM} (red) and \texttt{SEOBNRv4HM\_ROM} (cyan).}
  \label{fig:PE_GW190412_q_chieff}
\end{figure}

\roberto{In Table~\ref{tab:PEresults_intro}, I summarize the measurement of the most interesting \ac{BBH} parameters obtained with \texttt{SEOBNRv4PHM}. In the following, I discuss in detail each of these measurements. }
I begin this discussion with the mass ratio measurement, which is the most interesting property of this \ac{BBH} system. As discussed in Sec.~\ref{sec:synthetic_AS}, the mass ratio measurement is expected to be degenerate with that of the effective spin, therefore it is beneficial to examine them together.
For this purpose, in Fig.~\ref{fig:PE_GW190412_q_chieff} I show 2D and 1D posterior distributions for the mass ratio $q$ and the effective spin $\chi_\mathrm{eff}$, obtained by performing Bayesian parameter estimation on the signal using increasingly sophisticated waveform models: \texttt{SEOBNRv4\_ROM} (no \acp{HM}, \acp{BH} spins aligned with the binary angular momentum), \texttt{SEOBNRv4HM\_ROM} (\acp{HM}, \acp{BH} spins aligned with the binary angular momentum) and \texttt{SEOBNRv4PHM} (\acp{HM}, generic \acp{BH} spins). 
The largest improvement in the precision of the measurement is obtained when moving from \texttt{SEOBNRv4\_ROM} to \texttt{SEOBNRv4HM\_ROM}. In fact, the size of the $90\%$ credible intervals of the 1D mass ratio posterior distribution obtained with \texttt{SEOBNRv4HM\_ROM} (cyan curve) is $\sim 40 \%$ smaller with respect to that obtained with \texttt{SEOBNRv4\_ROM} (red curve). This larger precision in the mass ratio measurement is due to the fact that, including the \acp{HM} in the waveform model, allows to partially break the degeneracy between $q$ and $\chi_\mathrm{eff}$. This is consistent with what I find in Sec.~\ref{sec:synthetic_AS} for the synthetic signal. Because of this broken degeneracy, also the $90\%$ credible interval of the 1D $\chi_\mathrm{eff}$ posterior is $\sim 26\%$ tighter when computed using the waveform model with \acp{HM}.  The precision of the measurements improves more modestly when going from \texttt{SEOBNRv4HM\_ROM} (cyan curve) to \texttt{SEOBNRv4PHM} (blue curve). In this case, the size of the $90\%$ credible interval of the 1D posterior for $q$ and $\chi_\mathrm{eff}$ decrease respectively by $\sim 23 \%$ and $\sim 14 \%$ compared to \texttt{SEOBNRv4HM\_ROM}. The most noticeable difference between the measurements made with these two waveform models, consists in a shift of the \texttt{SEOBNRv4PHM} posterior distribution towards smaller $q$ and larger $\chi_\mathrm{eff}$. This shift is likely due to a bias in the measurement with \texttt{SEOBNRv4HM\_ROM}, originating from the fact that this model neglects precessional effects, while the value of $\chi_\mathrm{p}$ measured with \texttt{SEOBNRv4PHM} is in the range $0.2 \lesssim \chi_\mathrm{p} \lesssim 0.5$, indicating a small evidence for precession.

\begin{figure}[hbt]
  \centering
  \includegraphics[width=0.7\textwidth]{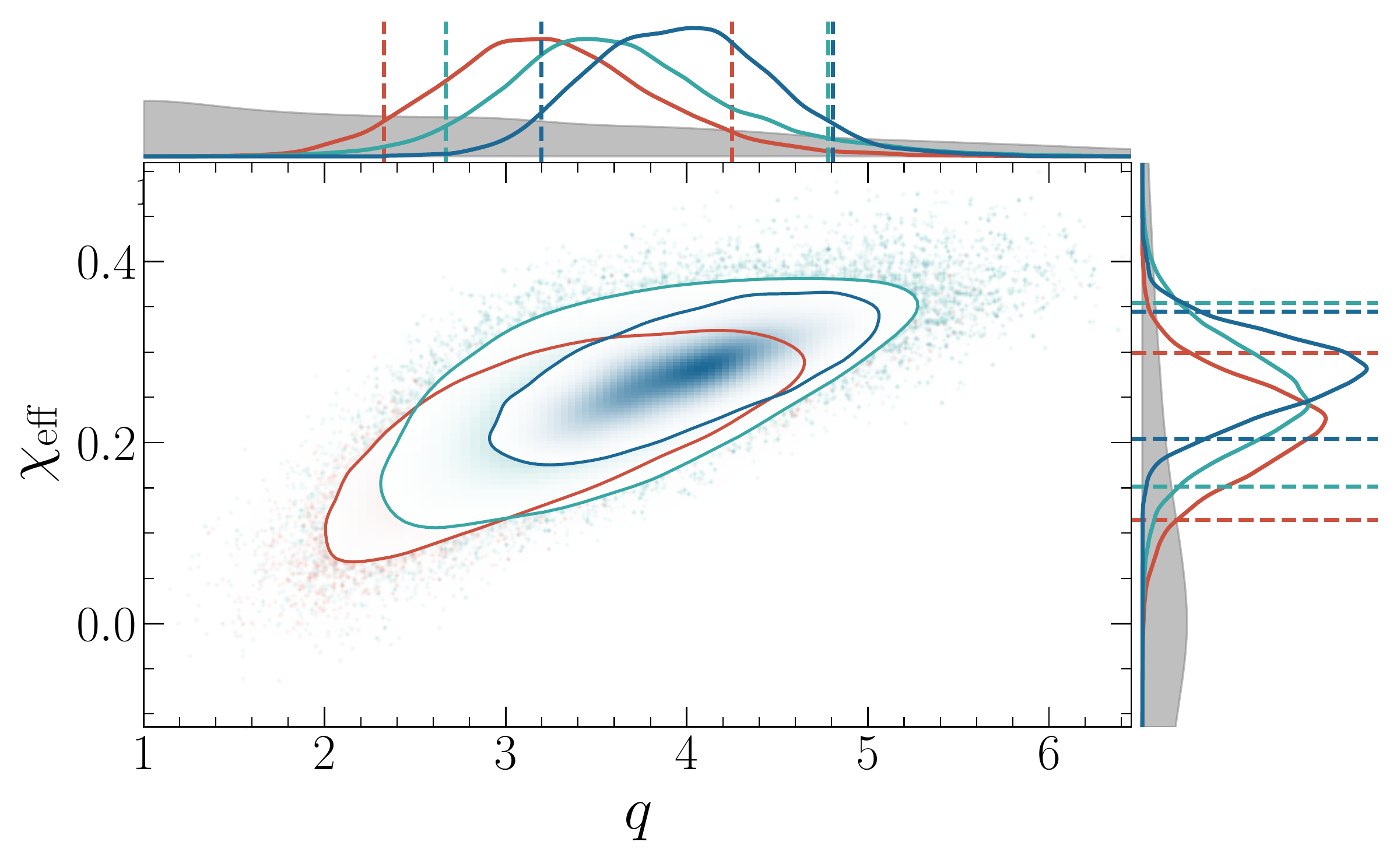}
  \includegraphics[trim={10cm 0cm 8cm 0cm}, clip, width=\textwidth]{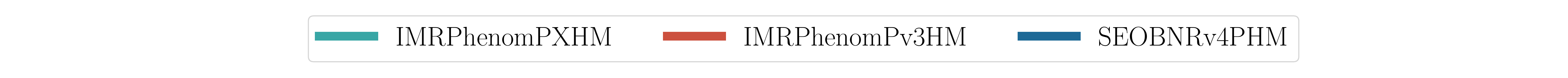}
  \caption{2D and 1D posterior distributions for the mass ratio $q$ and the effective spin $\chi_\mathrm{eff}$ of th signal GW190412. $90\%$ credible intervals are represented by solid contours in the 2D posteriors and dashed lines in the 1D posteriors. The gray shaded regions are the priors.  The parameter estimation is performed with the waveform models \texttt{SEOBNRv4PHM} (blue), \texttt{IMRPhenomPv3HM} (red) and \texttt{IMRPhenomXPHM} (cyan).}
  \label{fig:PE_GW190412_q_chieff_systematics_precessing}
\end{figure}

\begin{figure}[hbt]
  \centering
 \includegraphics[width=0.7\textwidth]{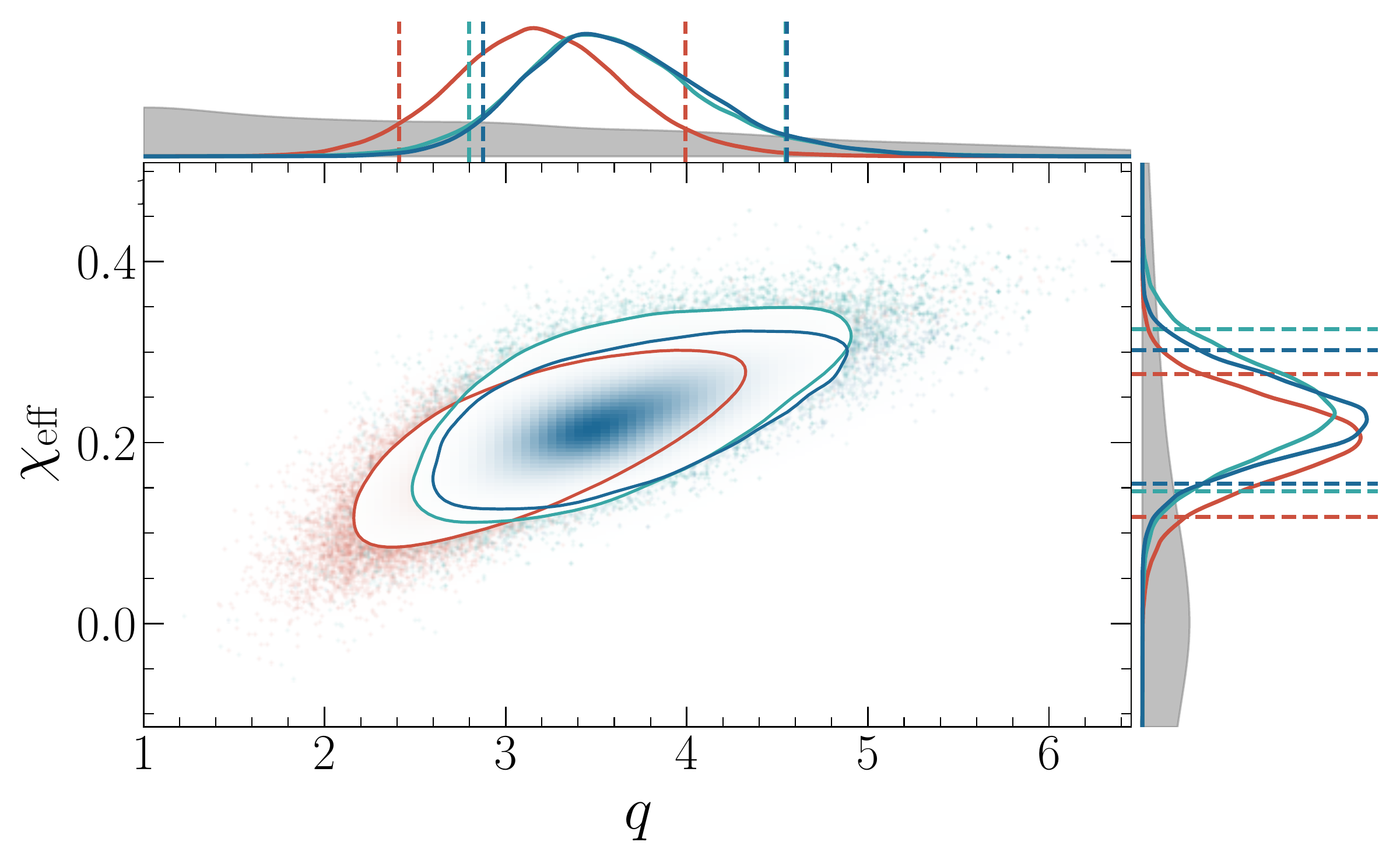}
  \includegraphics[trim={10cm 0cm 8cm 0cm}, clip, width=\textwidth]{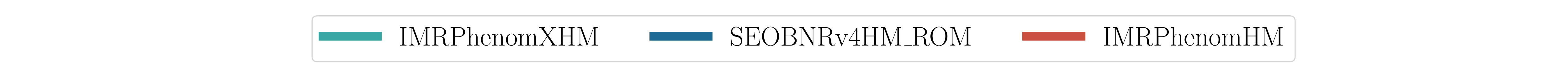}
 \caption{2D and 1D posterior distributions for the mass ratio $q$ and the effective spin $\chi_\mathrm{eff}$ of the signal GW190412. $90\%$ credible intervals are represented by solid contours in the 2D posteriors and dashed lines in the 1D posteriors. The gray shaded regions are the priors. The parameter estimation is performed with the waveform models \texttt{SEOBNRv4HM\_ROM} (blue), \texttt{IMRPhenomHM} (red) and \texttt{IMRPhenomXHM} (cyan).}
 \label{fig:PE_GW190412_q_chieff_systematics_aligned}
\end{figure}

The mass ratio and effective spin measured with the model \texttt{SEOBNRv4PHM} are respectively $q = 4^{+0.76}_{-0.77}$ and $\chi_\mathrm{eff} = 0.28^{+0.06}_{-0.08}$, and they are in a small tension with those measured using the waveform model \texttt{IMRPhenomPv3HM}, which gives $q = 3.22^{+0.95}_{-0.90}$ and $\chi_\mathrm{eff} = 0.22^{+0.08}_{-0.11}$. Since the two waveform models include the same physical effects (i.e. \acp{HM} and spin precession), this disagreement indicates a potential bias in the measurement obtained with one of the two models (or both of them) due to their inaccuracy in representing the true \ac{GR} waveforms. A potential source of this disagreement, as already discussed in Ref.~\cite{LIGOScientific:2020stg}, is the fact that the model \texttt{IMRPhenomPv3HM} includes \acp{HM} in the co-precessing frame that are not calibrated to \ac{NR} simulations during the merger and ringdown regime, differently from \texttt{SEOBNRv4PHM}. The authors in Ref.~\cite{Colleoni:2020tgc} partially shed light on this disagreement, by performing parameter estimation on this signal with their newly developed waveform model \texttt{IMRPhenomXPHM}~\cite{Pratten:2020ceb}, for which the \acp{HM} in the co-precessing frame are calibrated to \ac{NR} simulations during the merger and ringdown regimes, similarly to what is done in \texttt{SEOBNRv4PHM}\footnote{A similar study has been performed also in Ref.~\cite{Islam:2020reh}, by using the \ac{NR} surrogate model \texttt{NRSur7dq4} to analyze this signal. A direct comparison between their results and those discussed here is not possible, because the posterior distributions obtained as result of their analysis are not yet publicly available.}. In Fig.~\ref{fig:PE_GW190412_q_chieff_systematics_precessing}, I compare their measurement of $q$ and $\chi_\mathrm{eff}$ with those obtained with \texttt{SEOBNRv4PHM} and \texttt{IMRPhenomPv3HM}, by plotting the 2D and 1D posterior distributions for these parameters. From the plot, it is clear that the tension between the measurements is partially resolved when using \texttt{IMRPhenomXPHM} instead of \texttt{IMRPhenomPv3HM} for the comparison with the results obtained using \texttt{SEOBNRv4PHM}. 
To clarify the source of the residual difference between the measurements obtained using \texttt{IMRPhenomXPHM} and \texttt{SEOBNRv4PHM}, it is useful to repeat the same analysis assuming that the \acp{BH} have spins aligned with the orbital angular momentum of the binary. For this purpose, I use the \roberto{results obtained with the } waveform models \texttt{SEOBNRv4HM\_ROM}, \texttt{IMRPhenomXHM}\roberto{ (taken from Ref.~\cite{Colleoni:2020tgc}) } and \texttt{IMRPhenomHM}, which represent respectively \texttt{SEOBNRv4PHM}, \texttt{IMRPhenomXPHM} and \texttt{IMRPhenomPv3HM} under this assumption. The results of this analysis are summarized in Fig.~\ref{fig:PE_GW190412_q_chieff_systematics_aligned}\roberto{, where I show 2D and 1D posterior distributions for $q$ and $\chi_\mathrm{eff}$ obtained with \texttt{SEOBNRv4HM\_ROM}, \texttt{IMRPhenomXHM} and \texttt{IMRPhenomHM}. } In the non-precessing limit, the measurements obtained with \texttt{IMRPhenomXHM} and \texttt{SEOBNRv4HM\_ROM} agree much better than in the generic case of precessing \acp{BH}. This suggests that the difference in the measurements obtained with \texttt{SEOBNRv4PHM} and \texttt{IMRPhenomXPHM} originates from the different approaches used to describe spin precession in the two models.

\begin{figure}[hbt]
  \centering
  \includegraphics[width=0.7\textwidth]{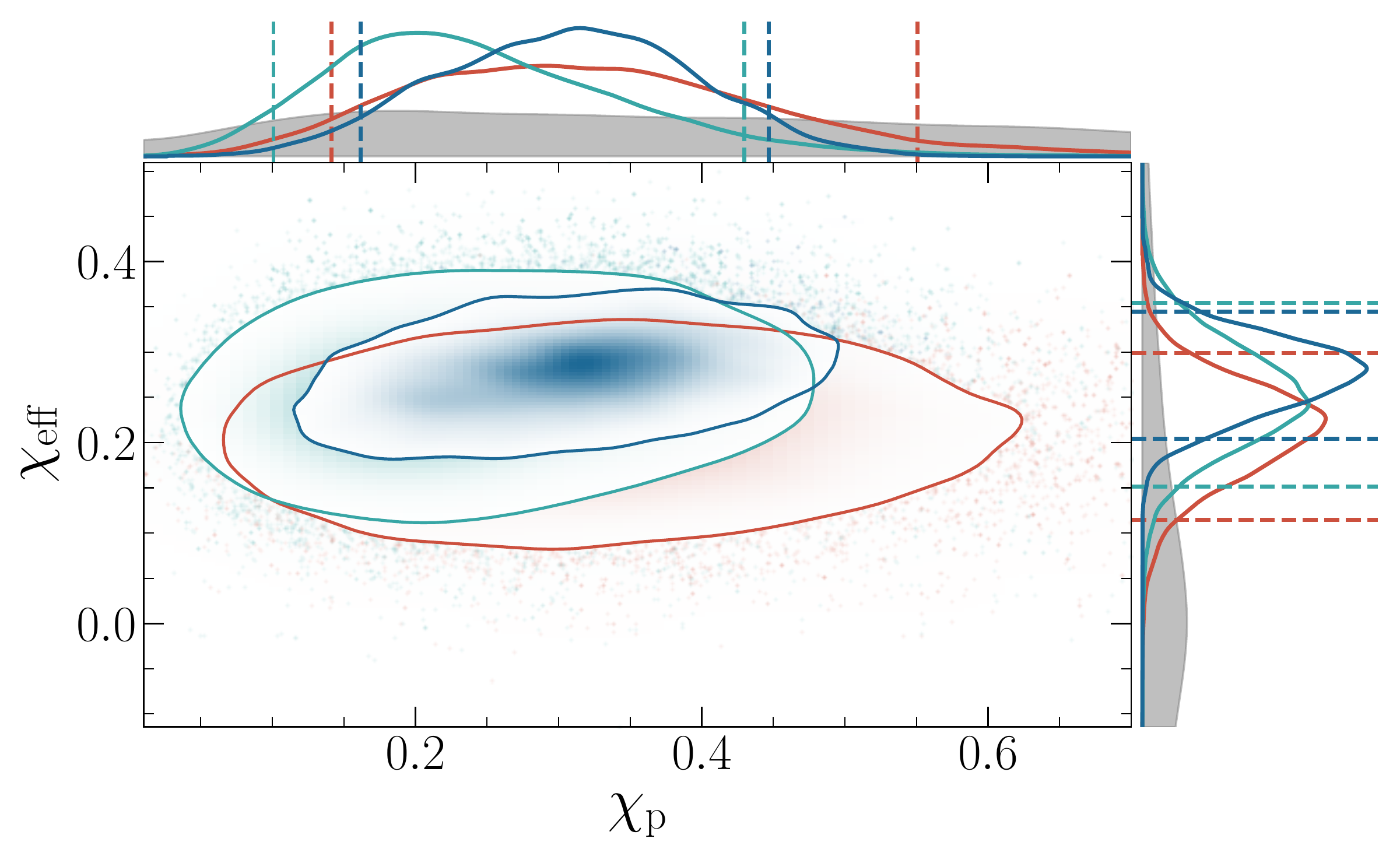}
  \includegraphics[trim={10cm 0cm 8cm 0cm}, clip, width=\textwidth]{event_legend_GW190412_systematics_precession}
  \caption{2D and 1D posterior distributions for $\chi_\mathrm{p}$ and the effective spin $\chi_\mathrm{eff}$ of the signal GW190412. $90\%$ credible intervals are represented by solid contours in the 2D posteriors and dashed lines in the 1D posteriors. The gray shaded regions are the priors. The parameter estimation is performed with the waveform models \texttt{SEOBNRv4PHM} (blue), \texttt{IMRPhenomPv3HM} (red) and \texttt{IMRPhenomXPHM} (cyan).}
  \label{fig:PE_GW190412_chieff_chip}
\end{figure}

Additional evidence for this can be found in the fact that the values of $\chi_\mathrm{p}$, measured by \texttt{SEOBNRv4PHM} and \texttt{IMRPhenomXPHM}, are also in small tension. To demonstrate this difference in the two measurements, I show, in Fig.~\ref{fig:PE_GW190412_chieff_chip}, the 2D and 1D posterior distributions for $\chi_\mathrm{p}$ and $\chi_\mathrm{eff}$, when measured with \texttt{SEOBNRv4PHM}, \texttt{IMRPhenomXPHM} and \texttt{IMRPhenomPv3HM}. It is clear that the posterior distribution obtained with \texttt{IMRPhenomXPHM} favours smaller values of $\chi_\mathrm{p}$ with respect to those obtained with \texttt{SEOBNRv4PHM}. The $\chi_\mathrm{p}$ posterior distribution measured using \texttt{IMRPhenomPv3HM} is broader with respect to the other two and it is in  agreement with \roberto{both of them}.
Despite these small differences, the values of $\chi_\mathrm{p}$ measured with \texttt{SEOBNRv4PHM}, \texttt{IMRPhenomPv3HM} and \texttt{IMRPhenomXPHM}, respectively $0.31_{-0.15}^{+0.14}$, $0.31_{-0.17}^{+0.24}$ and $0.23_{-0.13}^{+0.20}$, are in good agreement with each other.

The parameters $\chi_\mathrm{eff}$ and $\chi_\mathrm{p}$ are functions of the spins of the two \acp{BH}, $\bm{\chi_1}$ and $\bm{\chi_2}$, and the mass ratio $q$. When the mass ratio of the \ac{BBH} system is large, the contribution of $\bm{\chi_2}$ to $\chi_\mathrm{eff}$ and $\chi_\mathrm{p}$ is subdominant with respect to the contribution of $\bm{\chi_1}$, as it is clear from the definition of these two parameters in Eqs.~\eqref{eq:chieff} and~\eqref{eq:chip}, respectively. For this reason, despite the precise measurements of $\chi_\mathrm{eff}$ and $\chi_\mathrm{p}$, the \ac{BH} spin $\bm{\chi_2}$ remains unconstrained for this system. On the other side, while the orientation of $\bm{\chi_1}$ is not well constrained, its magnitude is one of the best measured among the \ac{BBH} systems detected by the LIGO and Virgo interferometers during O1, O2 and O3a~\cite{LIGOScientific:2018mvr, Abbott:2020niy}.
In particular, the value of the spin magnitude $\bm{\chi_1}$ obtained using the waveform models \texttt{SEOBNRv4PHM}, \texttt{IMRPhenomPv3HM} and \texttt{IMRPhenomXPHM} is respectively $0.46^{+0.12}_{-0.15}$, $0.41^{+0.22}_{-0.24}$ and $0.39^{+0.16}_{-0.17}$.

\begin{figure}[hbt]
  \centering
  \includegraphics[width=0.7\textwidth]{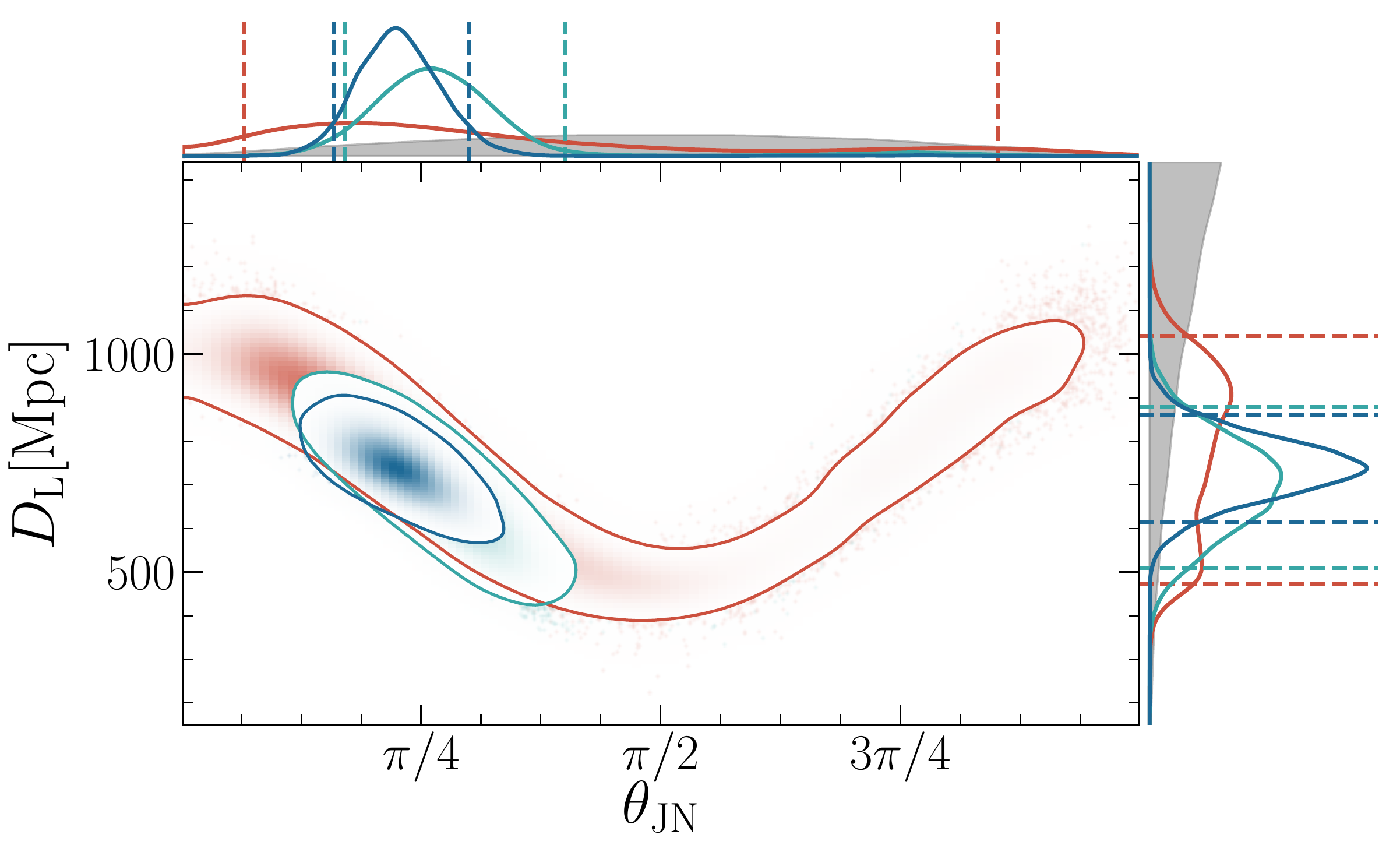}
  \includegraphics[trim={10cm 0cm 8cm 0cm}, clip, width=\textwidth]{event_legend_GW190412.pdf}
  \caption{2D and 1D posterior distributions for the angle $\theta_\mathrm{JN}$ and the luminosity distance $D_\mathrm{L}$ of the signal GW190412. $90\%$ credible intervals are represented by solid contours in the 2D posteriors and dashed lines in the 1D posteriors. The gray shaded regions are the priors. The parameter estimation is performed with the waveform models \texttt{SEOBNRv4PHM} (blue), \texttt{SEOBNRv4\_ROM} (red) and \texttt{SEOBNRv4HM\_ROM} (cyan).}
  \label{fig:PE_GW190412_distance_theta}
\end{figure}

The angle $\theta_\mathrm{JN}$, and the luminosity distance $D_\mathrm{L}$, are other two source parameters for which is interesting to study the improvement in their measurement, when using more sophisticated waveform models, as done before in the case of $q$ and $\chi_\mathrm{eff}$. For this purpose, in Fig.~\ref{fig:PE_GW190412_distance_theta}, I show 2D and 1D posterior distributions for these parameters obtained using the waveform models \texttt{SEOBNRv4\_ROM}, \texttt{SEOBNRv4HM\_ROM} and \texttt{SEOBNRv4PHM}. As already discussed in Sec. ~\ref{sec:synthetic_AS}, including the \acp{HM} in the waveform model partially breaks the degeneracy between $\theta_\mathrm{JN}$ and $D_\mathrm{L}$, allowing to measure them much more precisely. In fact, when going from \texttt{SEOBNRv4\_ROM} (red curve) to \texttt{SEOBNRv4HM\_ROM} (cyan curve), the size of the $90\%$ credible intervals of the posterior distribution for $\theta_\mathrm{JN}$ and $D_\mathrm{L}$ decrease, respectively by $\sim 70\%$ and $\sim 33\%$. Using \texttt{SEOBNRv4PHM} (blue curve) for the parameter estimation allows to further increase the precision of the measurement of these two parameters. In particular, the size of the $90\%$ credible interval of the posterior distribution for $\theta_\mathrm{JN}$ and $D_\mathrm{L}$ decrease respectively by $\sim 50\%$ and $\sim 33\%$, compared to those obtained using the model \texttt{SEOBNRv4HM\_ROM}.
The reason for this increased precision is that the precession of the orbital plane, caused by the in-plane \ac{BH} spin components, has a different imprint on the waveform when observed by different $\theta_\mathrm{JN}$ angles. This allows to put additional constraints on the angle $\theta_\mathrm{JN}$ and, consequently, on the luminosity distance $D_\mathrm{L}$. Differently from the case of the parameters $q$ and $\chi_\mathrm{eff}$, for $\theta_\mathrm{JN}$ and $D_\mathrm{L}$, the measurements obtained by \texttt{IMRPhenomPv3HM} and \texttt{IMRPhenomXPHM} are in very good agreement with those obtained when using \texttt{SEOBNRv4PHM}. This suggests that, for these two parameters, the systematic errors, due to the waveform model inaccuracies, are negligible with respect to the statistical uncertainty.

In the case of \roberto{GW190412}, waveform models that include the effect of \acp{HM} provide more precise measurements for all binary parameters, compared to waveform models that neglect this effect. This already suggests that \acp{HM} may be detectable in this signal. To make this concrete, it is useful to compute the Bayes factor between the hypothesis of the \ac{GW} signal including \acp{HM}, against the hypothesis of the signal only including the dominant mode $(\ell, |m|) = (2,2)$. This calculation can be performed assuming \ac{BH} spins aligned with the angular momentum of the binary system, hence computing the Bayes factor using the posterior distributions obtained with \texttt{SEOBNRv4\_ROM} and \texttt{SEOBNRv4HM\_ROM}, or in the case of generic spins, using the posterior distributions computed with the waveform models \texttt{SEOBNRv4P} and \texttt{SEOBNRv4PHM}. In both cases, the Bayes factor in favour of the hypothesis that this \ac{GW} signal includes \acp{HM} is larger than $10^3$, which is considered a strong evidence for this thesis. It is useful to recall that the Bayes factor does not account for the prior belief on the hypotheses. In this case, since \ac{GR} predicts the existence of \acp{HM}, and assuming that \ac{GR} is correct, the prior probability associated with the hypothesis of a \ac{GW} signal including only the mode $(\ell, |m|) = (2,2)$ would be $0$.

The improved measurement of the \ac{BBH} parameters of the source of  GW190412, obtained using the sophisticated waveform models described in this thesis, also had an impact on our understanding of the population of \acp{BBH}. An interesting property of the \ac{BBH} population is its mass-ratio distribution. In certain models for the \ac{BBH} population, this quantity is parametrized as a power law, $p(q|m_1) \propto q^{-\beta_\mathrm{q}}$~\cite{Kovetz:2016kpi,Fishbach:2017dwv,Talbot:2018cva}\footnote{In the literature on this subject, the mass-ratio definition is different compared to this thesis, and correspond to the definition $1/q$ adopted here. For this reason, in the literature, the mass-ratio distribution is parametrized as $p(q|m_1) \propto q^{\beta_\mathrm{q}}$ i.e. without the minus sign in the exponential. }. Since all the \acp{BBH} detected during O1 and O2 had a mass ratio consistent with $1$, the measured $\beta_\mathrm{q}$ indicated a preference for $\beta_\mathrm{q} > 0$~\cite{LIGOScientific:2018jsj}. \robert{However, its value was not precisely constrained by the \acp{BBH} detected during O1 and O2. When using only the \acp{BBH} detected during O1 and O2, the measured value of $\beta_\mathrm{q}$ was $0 \lesssim \beta_\mathrm{q} \lesssim 12$ at $90\%$ credible interval \cite{LIGOScientific:2018jsj}.  Including GW190412 in the \ac{BBH} population analysis, in addition to the signals detected during O1 and O2, allows to put a stronger upper bound on the value of $\beta_\mathrm{q}$, which is constrained to be $\beta_\mathrm{q} < 2.7$ at $90\%$ credible level, see Sec. VI of Ref.~\cite{LIGOScientific:2020stg} for more details}. The analysis of the \ac{BBH} population using all the signals in O3a (including GW190412), which were also analyzed with the waveform models described in this thesis, later confirmed the upper bound on $\beta_\mathrm{q}$ discussed before. See Sec. 5.1 in Ref.~\cite{Abbott:2020gyp} for the detailed discussion. The precise characterization of the \ac{BBH} population will allow us to understand the formation mechanism (or mechanisms) that is producing the observed population of \acp{BBH}.


  
\chapter{Enriching the Symphony of Gravitational Waves from Binary Black Holes by Tuning Higher Harmonics}
\chaptermark{}
\label{chap:two}

\hspace{\parindent}\textbf{Authors}\footnote{Originally published as Phys.Rev.D 98 (2018) 8, 084028.}: Roberto Cotesta, Alessandra Buonanno, Alejandro Boh\'{e}, Andrea Taracchini, Ian Hinder, Serguei Ossokine\\

\textbf{Abstract}:  For the first time, we construct an inspiral-merger-ringdown waveform model within the
  effective-one-body formalism for spinning, nonprecessing binary
  black holes that includes gravitational modes beyond the dominant 
$(\ell,|m|) = (2,2)$ mode, specifically $(\ell,|m|)=(2,1),(3,3),(4,4),(5,5)$. 
Our multipolar waveform model incorporates 
  recent (resummed) post-Newtoni\-an results for the inspiral and
  information from 157 numerical-relativity simulations, and 13
  waveforms from black-hole perturbation theory for the
  (plunge-)merger and ringdown. We quantify the improvement in accuracy when including 
higher-order modes by computing the faithfulness of the waveform model against the numerical-relativity 
waveforms used to construct the model. We define the faithfulness as the match maximized over 
time, phase of arrival, gravitational-wave polarization and sky position of the waveform model,
and averaged over binary orientation, gravitational-wave
  polarization and sky position of the numerical-relativity waveform. When the waveform
  model contains only the $(2,2)$ mode, we find that the averaged
  faithfulness to numerical-relativity waveforms containing all modes
  with $\ell \leq$ 5 ranges from 90\% to 99.9\% for binaries with
  total mass $20-200 M_\odot$ (using the Advanced LIGO's design noise
  curve). By contrast, when the $(2,1),(3,3),(4,4),(5,5)$ modes are also included in
  the model, the faithfulness improves to 99\% for all but four
 configurations in the numerical-relativity catalog, for which the
  faithfulness is greater than 98.5\%. Starting from the complete inspiral-merger-ringdown model, we develop also a (stand-alone) 
waveform model for the merger-ringdown signal, calibrated to numerical-relativity waveforms, which 
can be used to measure multiple quasi-normal modes.
The multipolar waveform model can be extended to include spin-precessional effects, and will be employed 
in upcoming observing runs of Advanced LIGO and Virgo.

\section{Introduction}
\label{sec:Intro}

The Advanced LIGO detectors \cite{TheLIGOScientific:2014jea} have reported, so far, 
the observation of five gravitational-wave (GW) signals from coalescing binary
black holes (BBHs): GW150914 \cite{Abbott:2016blz}, GW151226
\cite{Abbott:2016nmj}, GW170104 \cite{Abbott:2017vtc}, GW170608 \cite{Abbott:2017gyy}, GW170814
\cite{Abbott:2017oio} (observed also by the Virgo detector
\cite{TheVirgo:2014hva}), and one GW signal from a coalescing binary
neutron star (BNS) \cite{TheLIGOScientific:2017qsa}. 
The modeled search for GWs from binary systems and the extraction of binary parameters, 
such as the masses and spins, are based on the matched-filtering technique~\cite{Nitz:2017svb,Veitch:2014wba,Usman:2015kfa,Canton:2014ena,Cannon:2011vi,Cannon:2012zt}, 
which requires accurate knowledge of the waveform of the incoming signal. 
During the first two observing runs (O1 and O2), the Advanced LIGO and Virgo 
modeled-search pipelines employed, for binary total masses below $4 M_\odot$, templates~\cite{Sathyaprakash:1991mt} 
built within the post-Newtonian (PN) approach~\cite{Arun:2008kb,Buonanno:2009zt,Mishra:2016whh,Blanchet:2013haa}, 
and, for binary total masses in the range $4\mbox{--} 200 M_\odot$, templates developed using the effective-one-body (EOB) formalism calibrated 
to numerical-relativity (NR) simulations~\cite{Buonanno:1998gg,Buonanno:2000ef,Damour:2008qf,Damour:2009kr,Barausse:2009xi,Taracchini:2013rva,Purrer:2015tud,Bohe:2016gbl} (i.e. EOBNR waveforms). 
For parameter-estimation analyses~\cite{TheLIGOScientific:2016pea,TheLIGOScientific:2016wfe,TheLIGOScientific:2017qsa,Veitch:2014wba} and tests of General Relativity (GR) ~\cite{TheLIGOScientific:2016src}, PN~\cite{Arun:2008kb,Buonanno:2009zt,Mishra:2016whh}, 
EOBNR~\cite{Pan:2013rra,Taracchini:2013rva,Bohe:2016gbl,Babak:2016tgq} and 
also inspiral-merger-ringdown phenomenological (IMRPhenom) waveform models~\cite{Husa:2015iqa,Khan:2015jqa,Hannam:2013oca} were used.

The -2 spin-weighted spherical harmonics comprise a convenient basis into which one can decompose the two polarizations of GWs.
The spinning, nonprecessing EOBNR waveform model~\cite{Bohe:2016gbl}  employed in searches and parameter-estimation studies during the O2 run 
(henceforth, \texttt{SEOBNRv4} model),  
only used the dominant $(\ell,|m|) = (2,2)$ mode to build the gravitational polarizations.
This approximation was accurate enough for detecting and inferring astrophysical information of the sources observed during O2 (and also O1), as discussed in Refs.~\cite{Littenberg:2012uj,Brown:2012nn,Capano:2013raa,Harry:2017weg,Varma:2014jxa,Graff:2015bba,Varma:2016dnf,Bustillo:2016gid,Abbott:2016wiq}.

Because of the expected increase in sensitivity during the third observing run (O3), which is planned to start in the Fall of 2018, 
some GW signals are expected to have 
much larger signal-to-noise ratio (SNR) with respect to the past, and may lie in regions of parameter space so far unexplored (e.g., 
more massive and/or higher mass-ratio systems than observed in O1 and O2). This poses an excellent 
opportunity to improve our knowledge of astrophysical and gravitational properties of the sources, but it also requires more accurate 
waveform models to be able to take full advantage of the discovery and inference potential.
More accurate waveform models would be
useful, as well, from the detection point of view to further increase the effective volume reached by the search, in
particular for regions of the parameter space where the approximation of restricting to the 
(2,2) mode starts to degrade~\cite{Brown:2012nn,Capano:2013raa,Harry:2017weg}. Following these motivations, we build here an improved version of the 
\texttt{SEOBNRv4} waveform model that includes the modes $(\ell,|m|) = (2,1),(3,3),(4,4),(5,5)$ beyond the dominant $(2,2)$ mode (henceforth, 
\texttt{SEOBNRv4HM} model). 
Similar work was done for the nonspinning case for the EOBNR waveform model of Ref.~\cite{Pan:2011gk} (henceforth, \texttt{EOBNRv2HM} model), and for the nonspinning and spinning, nonprecessing \texttt{IMRPhenom} models in Refs.~\cite{Mehta:2017jpq, London:2017bcn}. 

In building the \texttt{SEOBNRv4HM} model we incorporate new informations from PN calculations~\cite{Marsatetal2017,Fujita:2012cm}, 
from NR simulations (produced with the (pseudo) Spectral Einstein code (\texttt{SpEC}) \cite{Chu:2015kft} of the Simulating
eXtreme Spacetimes (\texttt{SXS}) project and the \texttt{Einstein Toolkit} code \cite{Zilhao:2013hia,Loffler:2011ay}), and also from merger-ringdown 
waveforms computed in BH perturbation theory solving the Teukolsky equation~\cite{Barausse:2011kb,Taracchini:2014zpa}.  
The NR waveforms are described in Refs.~\cite{Mroue2013,Kumar:2016dhh,Chu:2015kft,Kumar:2015tha,Lovelace:2010ne,Scheel:2014ina,Mroue:2013xna,Bohe:2016gbl}, 
and summarized in Appendix~\ref{sec:NRcatalog}. They were also employed to build the \texttt{SEOBNRv4} waveform model in 
Ref.~\cite{Bohe:2016gbl} 
(see Sec.~\Romannum{3} therein). However, here, we do not use the BAM simulation 
\texttt{BAMq8s85s85} \cite{Bruegmann:2006at,Husa:2007hp}, because the higher-order modes are not available to us. Thus, for the same binary configuration, 
we produce a new NR simulation using the \texttt{Einstein Toolkit} code and extract higher-order modes (henceforth, \texttt{ET:AEI:0004}).

As by product of the \texttt{SEOBNRv4HM} model, we obtain a (stand-alone) merger-ringdown model~\cite{Baker:2008mj,Damour:2014yha,London:2014cma,Nagar:2016iwa,Bohe:2016gbl,London:2018gaq}, tuned to the NR and Teukolsky-equation waveforms, which can be employed to extract multiple quasi-normal modes from GW signals, and test General Relativity~\cite{Dreyer:2003bv,Berti:2005ys,Meidam:2014jpa,Yang:2017zxs}.

The paper is organized as follows. In
  Sec.~\ref{sec:motivations} we use the NR waveforms at our disposal
  to quantify the importance of higher harmonics in presence of
  spins. In Sec.~\ref{sec:faithfulness} we determine, taking also into account 
  the error in NR waveforms, which gravitational modes are
  crucial to achieve at least $\sim 99\%$ accuracy. In Sec.~\ref{sec:eob_formalism} we
  develop the multipolar EOB waveform model, and describe how to
  enhance its performance by including information from NR simulations
  and BH perturbation theory. We also highlight the construction and use 
of the multipolar (stand-alone) merger-ringdown model. 
In Sec.~\ref{sec:comparison} we compare
  the newly developed \texttt{SEOBNRv4HM} model to 157 NR waveforms. In
  Sec.~\ref{sec:concl} we summarize our main conclusions, and outline
  possible future work. Finally, in Appendices \ref{app:modes},
  \ref{app:NQCfits} and \ref{app:ringdownfits} we provide interested
  readers with explicit expressions of all quantities entering the
  higher-order modes of the \texttt{SEOBNRv4HM} model, and point out the 
presence of numerical artifacts in the (4,4) and (5,5) modes of some 
NR simulations. For convenience, we summarize in Appendix~\ref{sec:NRcatalog} the NR
  waveforms used in this paper. In Appendix~\ref{sec:EOBNRv2HM} we
  also compare the model \texttt{SEOBNRv4HM} with the nonspinning
  \texttt{EOBNRv2HM} waveform model, developed in
  2011~\cite{Pan:2011gk}. Finally in Appendix~\ref{sec:time_domain} we
  compare the \texttt{SEOBNRv4HM} model with an NR waveform in time
  domain.

In this paper we adopt the geometric units $G = c = 1$.

\section{Motivations to model higher-order modes for binary black holes}
\label{sec:motivations}

\begin{figure}[h]
  \centering
  \includegraphics[width=0.7\textwidth]{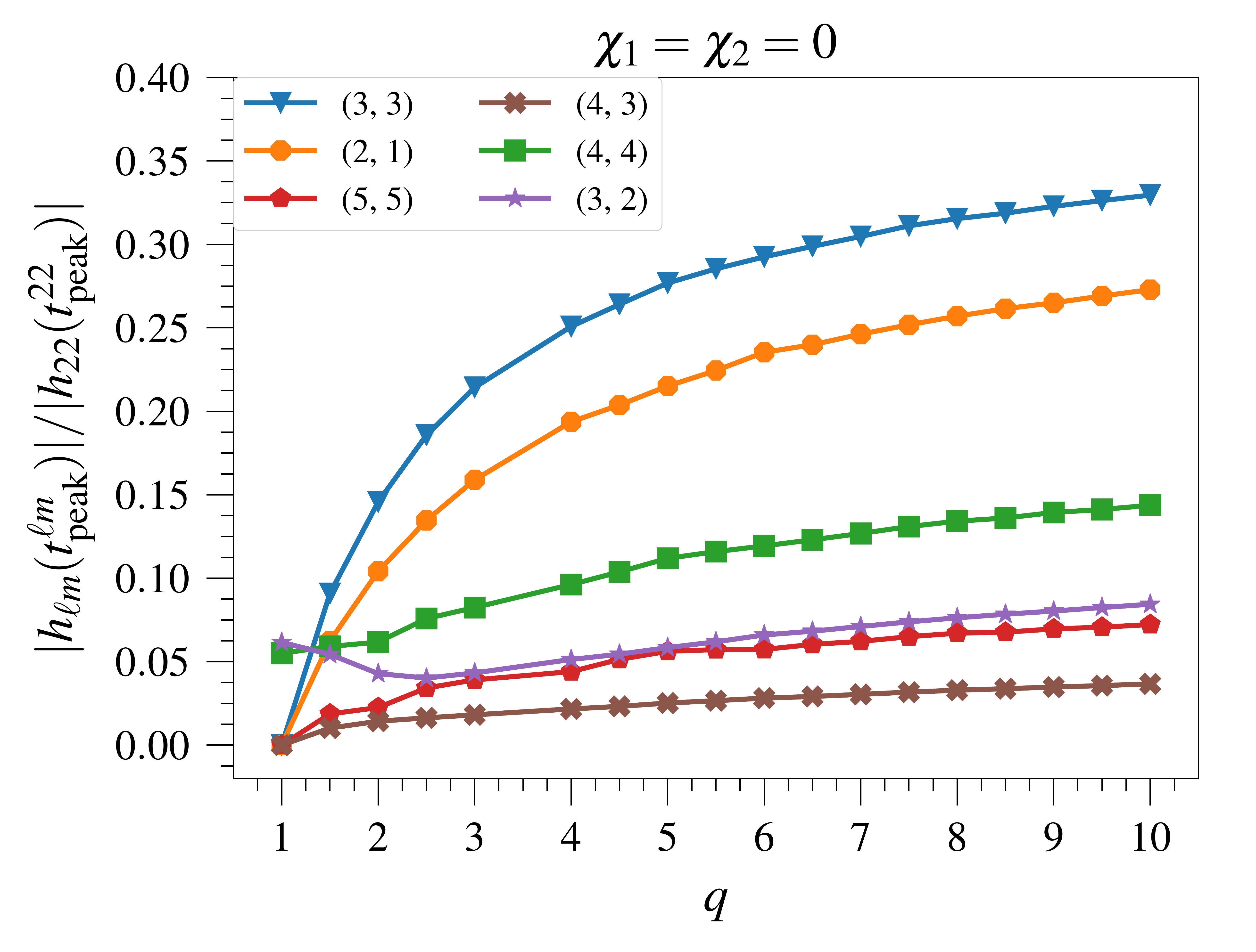}
\caption{Amplitude ratio between the $(\ell,m)$ mode and the dominant $(2,2)$ mode, both 
evaluated at their peak, as function of the mass ratio. 
We use only nonspinning NR waveforms. 
(Note that the markers represent the NR data, and we connect them by a line). We note that the importance of a given higher-order mode with respect to the dominant one is not controlled only by the amplitude ratio between the two, but also by the -2 spin-weighted spherical harmonic associated to the mode (see Eq.~(\ref{eq:spHarmDec})).}
\label{fig:rel_amp_hm_0spin}
\end{figure}

In this section we describe the spherical-mode decomposition of the gravitational polarizations and 
discuss the motivations for building an inspiral-merger-ringdown waveform model (\texttt{SEOBNRv4HM}) 
with higher harmonics for spinning BHs. 

Henceforth, we denote the binary's total mass with $M = m_1 + m_2$, and choose the body's masses $m_1$ and $m_2$ such that the mass ratio $q = m_1/m_2 \geq 1$. Since we consider only spinning, nonprecessing BHs (i.e., 
spins aligned or antialigned with the direction perpendicular to the orbital plane $\mathbf{\hat{L}})$, we only 
have one (dimensionless) spin parameter for each BH, $\chi_{1,2}$, defined as $\mathbf{S_{\mathrm{1,2}}}= 
\chi_{1,2} m_{1,2}^2 \mathbf{\hat{L}}$, where $\mathbf{S_{\mathrm{1,2}}}$ are the BH's spins ($-1 \leq \chi_{1,2} \leq 1$).

The observer-frame's gravitational polarizations read 
\begin{equation}
\label{eq:spHarmDec}
h_+(\iota,\varphi_0;t) - i \ h_x(\iota,\varphi_0;t) = \sum_{\ell=2}^\infty \sum_{m=-\ell}^\ell \tensor[_{-2}]{Y}{_{\ell m}}(\iota,\varphi_0) \ h_{\ell m}(t),
\end{equation}
where we denote with $\iota$ the inclination angle (computed with respect to the direction perpendicular to the 
orbital plane), $\varphi_0$ the azimuthal direction to the observer, and 
$\tensor[_{-2}]{Y}{_{\ell m}}(\iota,\varphi_0)$'s the -2 spin-weighted spherical harmonics. For spinning, nonprecessing BHs' 
we have $h_{\ell m} = (-1)^\ell h^*_{\ell -m}$. Thus, without loss of generality, we 
restrict the discussion to $(\ell,m)$ modes with $m>0$. 

\begin{figure}
\center
\includegraphics[width=0.7\textwidth]{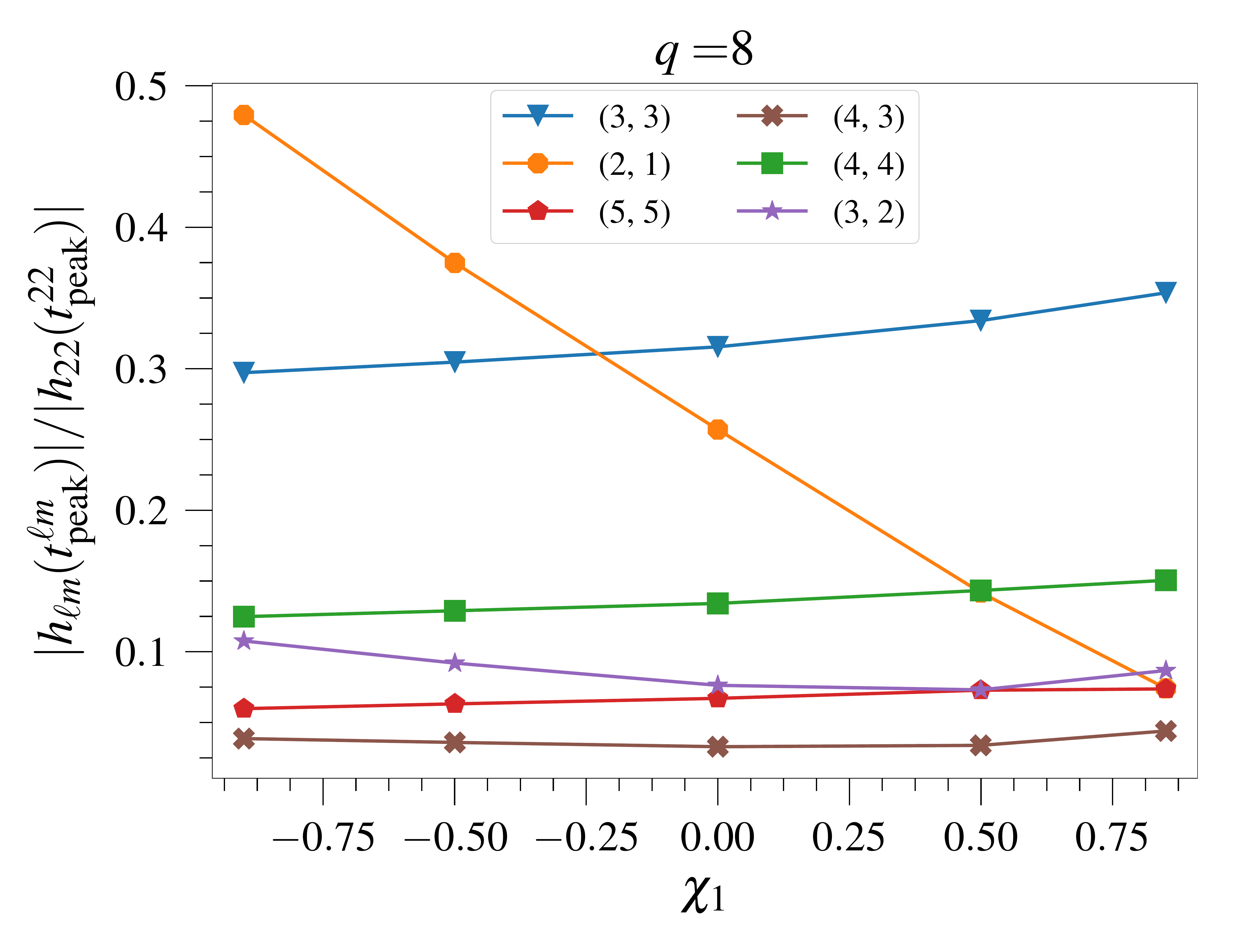} 
\\
\includegraphics[width=0.7\textwidth]{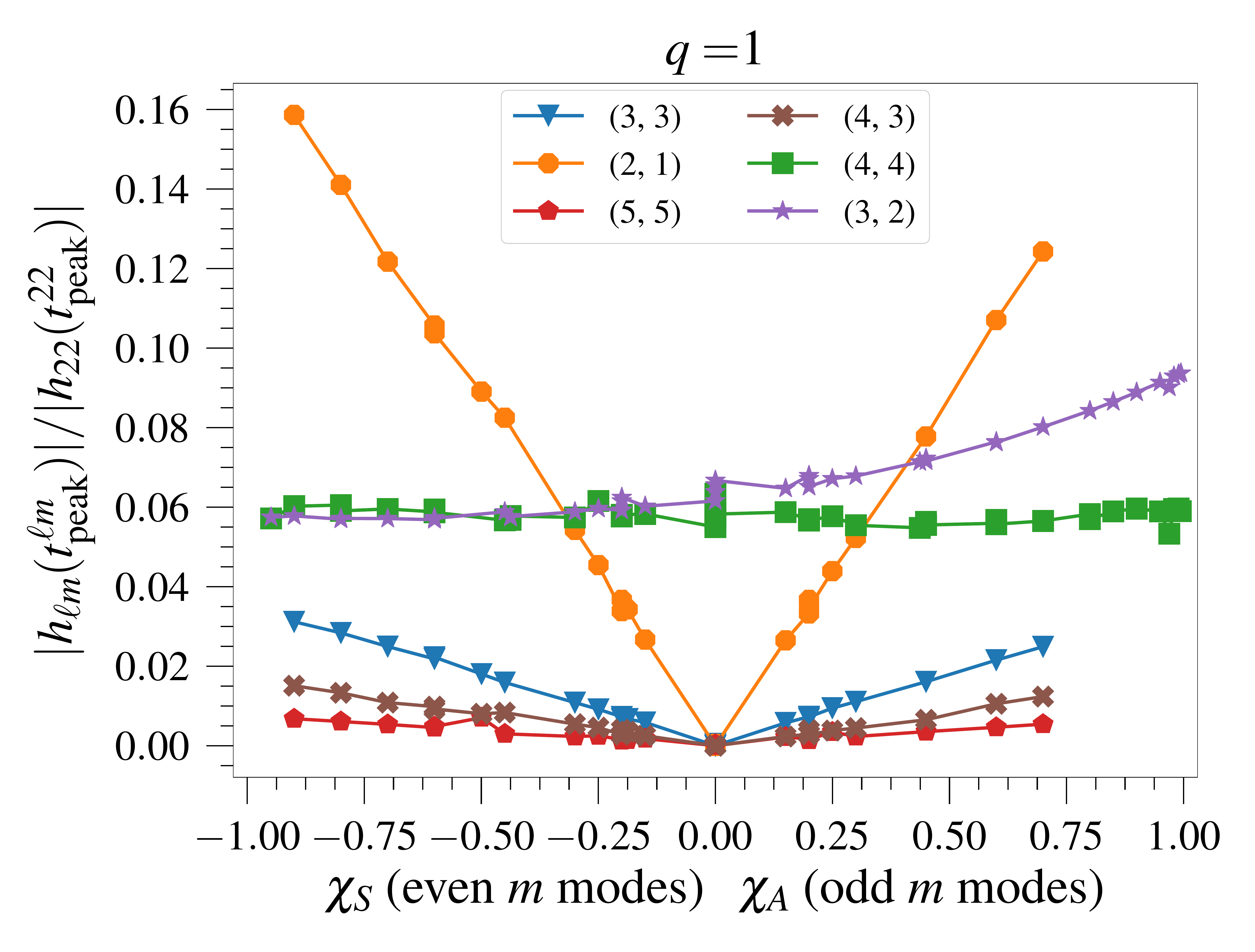} 
\caption{Amplitude ratio between the $(\ell,m)$ mode and the dominant $(2,2)$ mode, both 
evaluated at their peak. In the top (bottom) panel we plot these quantities for mass ratio $q = 8$  versus 
the spin of the heavier BH ($q=1$ versus $\chi_A = (\chi_1-\chi_2)/2$ for modes with odd $m$, and 
$\chi_S = (\chi_1 + \chi_2)/2$ for modes with even $m$). The markers represent the NR data, and we connect them by a line.}
\label{fig:rel_amp_hm_spin}
\end{figure}

As we shall discuss below, for face-on/face-off binary configurations, 
the dominant mode is the $(\ell, m) = (2,2)$ mode. For generic binary orientations the modes $(\ell, m) 
\neq (2,2)$ could be comparable to the $(2,2)$ mode. Nevertheless, we will loosely refer to $(\ell, m) \neq (2,2)$ as subdominat modes; 
sometime we also refer to them as higher-order modes or higher harmonics, even if they include the $(2,1)$ mode.

Several authors in the literature have investigated the impact of
  neglecting higher-order modes for detection and parameter
  estimation. From the detection perspective, Refs.~\cite{Brown:2012nn,
    Pekowsky:2012sr,Varma:2014jxa,Capano:2013raa} showed that neglecting 
  higher-order modes in nonspinning gravitational waveforms can cause a loss in detection volume bigger
  than $10 \%$ when the mass ratio $q \geq 4$ and total mass $M \geq
  100 M_\odot$. To overcome this issue, Ref.~\cite{Harry:2017weg} suggested 
  a new method to search for GW signals with templates that include higher modes,  
  increasing the search sensitivity up to a factor of 2 in volume 
for high mass-ratio, and high total-mass binaries. While those works consider only nonspinning systems, 
the authors of Ref.~\cite{Bustillo:2015qty} show that for spinning
  systems, the loss in detection volume due to neglecting 
  higher-order modes is smaller with respect to the nonspinning
  case. This happens because the spin parameters provide an additional degree
  of freedom that templates with only the dominant $(2,2)$ mode can employ
  to better match signals containing higher-order modes.

From the parameter-estimation perspective, as discussed in
Ref.~\cite{Varma:2014jxa}, for nonspinning systems with mass ratio $q
\geq 4$ and total masses $M \geq 150 M_\odot$ the systematic error due
to neglecting higher-order modes is larger than the $1\sigma$
statistical error for signals with signal-to-noise ratio (SNR) of
8. Signals with a larger SNR yield smaller statistical errors and, the
constraints discussed before become more stringent~\cite{Littenberg:2012uj}. Indeed even for equal-mass systems,
where the higher-order modes are expected to be negligible, if the
signal has an SNR of 48, the systematic error from neglecting 
higher-order modes can be bigger than the statistical error~\cite{Littenberg:2012uj}. 
(The SNRs above refer to Advanced LIGO’s “zero-detuned
high-power” design sensitivity curve~\cite{Shoemaker:2010}).

Here we briefly review known results, and highlight some features that will be exploited below when  
building the \texttt{SEOBNRv4HM} waveform model.

In Fig.~\ref{fig:rel_amp_hm_0spin} we show the ratio between the largest subdominant $(\ell,m)$ modes and the 
$(2,2)$ mode amplitudes, evaluated at their peak, $t_{\mathrm{peak}}^{\ell m}$ and $t_{\mathrm{peak}}^{22}$, respectively,  
as function of mass ratio for all the nonspinning waveforms in our NR catalog. We note that the well-known 
mode hierarchy $(\ell, m) = (2,2), (3,3),(2,1),(4,4),(3,2),(5,5),(4,3)$ changes when approaching the
equal-mass (equal-spin) limit (see, e.g., Ref.~\cite{Healy:2013jza}). 
Indeed, in this limit all modes with odd $m$ have to vanish  in order to enforce 
the binary's symmetry under rotation $\varphi_0 \rightarrow \varphi_0 + \pi$. Thus, when 
$\nu \rightarrow 1/4$ ($\chi_{1}=\chi_2$), the $(3,2)$ and $(4,4)$ modes become the 
most important subdominant modes.
In Fig.~\ref{fig:rel_amp_hm_spin} we show how the modes' hierarchy in 
the nonspinning case (see Fig.~\ref{fig:rel_amp_hm_0spin}) changes when BH's spins are included. 
In particular, in the left panel of Fig.~\ref{fig:rel_amp_hm_spin} we fix the mass ratio to $q = 8$ and
plot the relative amplitude of the modes as function of the spin of the more massive BH. Note that for $q = 8$ 
all NR waveforms in our catalog (with the exception of \texttt{ET:AEI:0004}, $q = 8, \, \chi_1 = \chi_2 = 0.85$)  
have the spin only on the more massive BH.  We see that the relative amplitude of 
the modes $(3,3),(4,4),(3,2),(5,5),(4,3)$ depends weakly on the spins, except for the $(2,1)$ mode. Indeed, 
for $\chi_1 \gtrsim 0.5$, the $(2,1)$ mode becomes smaller than the $(4,4)$ mode and
for $\chi_1 \gtrsim 0.75$ is as small as the modes $(3,2),(5,5)$. On the other side, 
for $\chi_1 \lesssim -0.25$ the mode $(2,1)$ is larger than the $(3,3)$
mode.  We find that for smaller mass ratios the effect of $\chi_2$ (i.e., the spin of the
lighter BH), becomes more important. In particular, for a fixed value
of $\chi_1$ the amplitude ratio $|h_{\ell m}(t_{\mathrm{peak}}^{\ell
  m})|/|h_{2 2}(t_{\mathrm{peak}}^{2 2})|$ for the modes $(3,3),(4,4),(5,5)$ 
decreases with increasing $\chi_2$, while the ratio increases for the modes $(2,1),(3,2),(4,3)$.

The special case of equal-mass systems, $q = 1$, is discussed in the right panel of 
Fig.~\ref{fig:rel_amp_hm_spin}. Here we show the
amplitude ratio between the $(\ell,m)$ mode and the dominant $(2,2)$ mode, 
both evaluated at their peak, as function of $\chi_A
= (\chi_1-\chi_2)/2$ for modes with odd $m$ and as function of
$\chi_S = (\chi_1 + \chi_2)/2$ for modes with even $m$. As discussed
before, the modes with odd $m$ vanish for equal-mass, equal-spins
configurations ($\chi_A = 0$) from symmetry arguments and, the
amplitude ratio grows proportionally to $|\chi_A|$ for these modes. In
particular, we note that in this case the $(2,1)$ mode
behaves differently from the other modes, undergoing a much more significant
growth in the amplitude ratio. Regarding the modes with even $m$, we notice 
that whereas the $(4,4)$ mode is nearly
constant as function of $\chi_S$ in the spin range considered, the 
$(3,2)$ mode increases as a function of $\chi_S$ in the same range.
The amplitude of the $(2,1)$ mode has a stronger dependence 
on the spins with respect to the other modes because in its PN expansion the spin term enters at a lower
relative order (see Eqs. (38a)--(38i) in Ref.~\cite{Pan:2010hz}). A similar spin-dependence 
was found in Ref.~\cite{Kamaretsos:2012bs} for the amplitudes ratio $(A_{\ell m}/A_{2 2})$ of the quasi-normal mode oscillations.

  Finally, it is worth emphasizing that in understanding the 
  relevance of subdominant modes for the observer, it is important
  to take into account the -2 spin-weighted spherical-harmonic factor
  $\tensor[_{-2}]{Y}{_{\ell m}}(\iota,\varphi_0)$ that enters
  Eq.~\eqref{eq:spHarmDec}, notably its dependence on the angles
  $(\iota,\varphi_0)$. Indeed, the -2 spin-weighted spherical harmonic
  associated to the dominant mode starts from a maximum in the face-on
  orientation ($\iota = 0$) and decreases to a minimum at edge-on
  ($\iota = \pi/2$). On the other hand, the spherical harmonics favour the
  higher-order modes with respect to the dominant one in orientations
  close to edge-on where ${ |\tensor[_{-2}]{Y}{_{\ell m}}(\iota
    \rightarrow \pi/2)|/|\tensor[_{-2}]{Y}{_{2 2}}(\iota \rightarrow
    \pi/2)| > 1}$.  Furthermore, a direct inspection of the harmonic factor shows that 
  the modes $(3,2),(4,3)$ are suppressed (i.e., ${ |\tensor[_{-2}]{Y}{_{\ell
        m}}(\iota)|/|\tensor[_{-2}]{Y}{_{2 2}}(\iota)| < 1}$) for a larger region
  in $\iota$ than for the modes $(3,3),(2,1),(4,4),(5,5)$. For this reason the contribution of the
  former to the gravitational polarizations is limited to a smaller number of orientations with respect
  to the latter.

\section{Selecting the most-important higher-order modes for modeling}
\label{sec:faithfulness}

\begin{figure}
\centering
\includegraphics[width=0.6\textwidth]{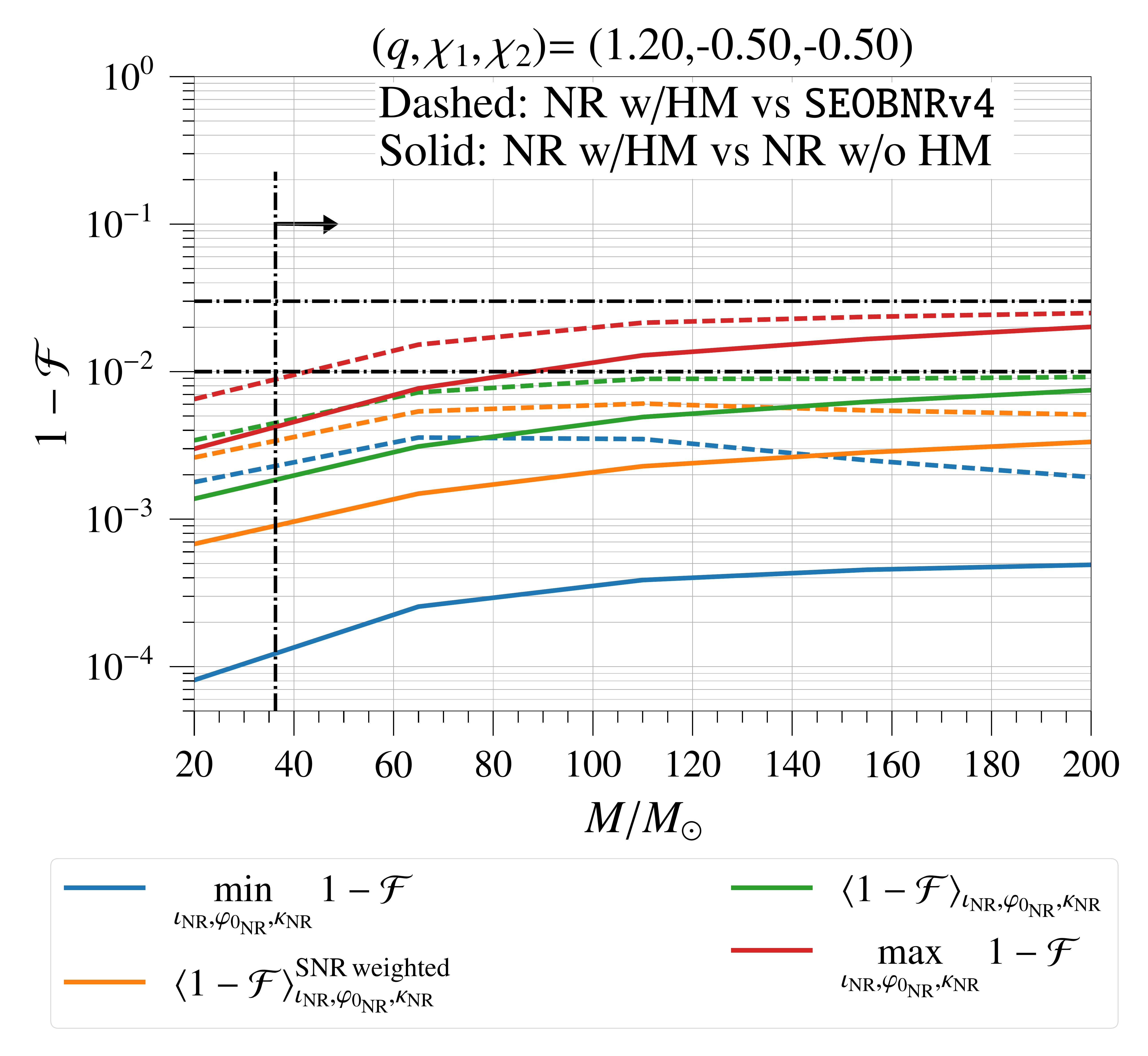} 
\includegraphics[width=0.6\textwidth]{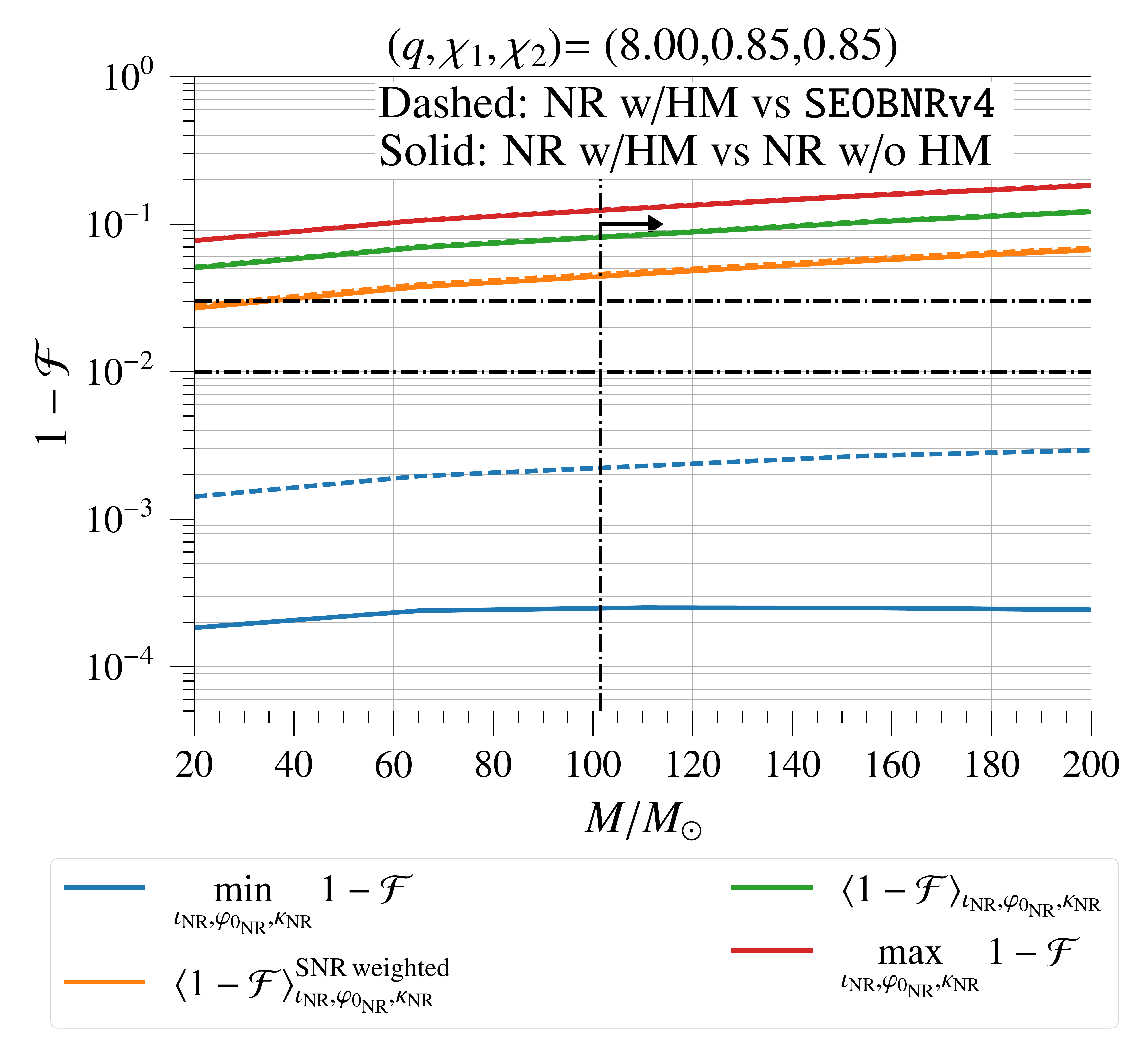} 
\caption{Unfaithfulness $(1-\mathcal{F})$ for the configurations $(q = 1.2, \,\chi_1 = -0.5,\,\chi_2=-0.5)$ (top panel) and $(q = 8, \,\chi_1 = 0.85,\,\chi_2= 0.85)$ (bottom panel)  in the mass range $20 M_\odot \leq M \leq 200 M_\odot$. In dashed the results for the \texttt{SEOBNRv4} model and in solid the results for the NR waveform containing only the dominant mode, both against the NR waveform with the modes $(\ell \leq 5, \, m \neq 0)$. The minimum of the unfaithfulness (blue curves) correspond to a face-on orientation.      We also show the unfaithfulness averaged over the three angles $\iota_{\textrm{NR}},{\phi_0}_{\textrm{NR}},\kappa_{\textrm{NR}}$ (green curves) and weighted by the cube of the SNR (orange curves). Finally the minimum of the unfaithfulness (red curves) which in practice correspond to edge-on and minimized over the other two angles. The vertical dotted-dashed black line is the smallest mass for which the $(\ell, m) = (2,1)$ mode is entirely in the Advanced LIGO band. The $(\ell, |m'|)$ mode is entirely in the Advanced LIGO band starting from a mass $m'$ times the mass associated with the $(\ell, m) = (2,1)$ mode.  The horizontal dotted-dashed black lines represent the values of $1\%$ and $3\%$ unfaithfulness.}
\label{fig:unfaith_mass}
\end{figure}

In this section we first introduce the faithfulness function as a tool to assess the closeness of two waveforms when higher-order modes 
are included. Then, we use it to estimate how many gravitational modes we need to model in order not to loose 
more than $10\%$ in event rates when rectricting to the binary's configurations in the NR catalog 
at our disposal. We also determine the loss in faithfulness of the NR waveforms due to numerical error.  

The GW signal measured from a spinning, nonprecessing and noneccentric 
BBH is characterized by 11 parameters, namely the masses of the two bodies
$m_{1}$ and $m_{2}$, the (constant) projection of the spins in the
direction perpendicular to the orbital plane, $\chi_{1}$ and $\chi_{2}$,
the angular position of the line of sight measured in the source's
frame $(\iota, \varphi_0)$ (see Eq. \eqref{eq:spHarmDec}), the sky
location of the source in the detector frame $(\theta, \phi)$, the
polarization angle $\psi$, the luminosity distance of the source
$D_{\mathrm{L}}$ and the time of arrival $t_{\mathrm{c}}$. 
The signal measured by the detector takes the form:
\begin{equation}
\label{eq:det_strain}
h \equiv  F_+(\theta,\phi,\psi) \ h_+(\iota,\varphi_0, D_{\mathrm{L}}, \boldsymbol{\xi},t_{\mathrm{c}};t) + F_\times(\theta,\phi,\psi)\ h_\times(\iota,\varphi_0, D_{\mathrm{L}}, \boldsymbol{\xi},t_\mathrm{c};t)\,,
\end{equation}
where for convenience we introduce $\boldsymbol{\xi} \equiv (m_{1}, m_{2}, \chi_{1}, \chi_{2})$.
The functions $F_+(\theta,\phi,\psi)$ and $F_\times(\theta,\phi,\psi)$ are the antenna patterns~\cite{Sathyaprakash:1991mt,Finn:1992xs}:
\begin{align}
F_+(\theta,\phi,\psi) &= \frac{1+ \cos^2(\theta)}{2} \ \cos(2\phi) \ \cos(2\psi)
 -\cos(\theta) \ \sin(2\phi)\ \sin(2\psi),\\ 
F_\times(\theta,\phi,\psi) &= \frac{1+ \cos^2(\theta)}{2} \ \cos(2\phi) \ \sin(2\psi)
+ \cos(\theta) \ \sin(2\phi)\ \cos(2\psi).
\end{align}
Equation \eqref{eq:det_strain} can be rewritten as:
\begin{align}
h \equiv & \mathcal{A}(\theta,\phi)\big[\cos\kappa(\theta,\phi,\psi) \ h_+(\iota, \varphi_0, D_{\mathrm{L}}, \boldsymbol{\xi}, t_{\mathrm{c}};t) \nonumber \\
&+ \sin\kappa(\theta,\phi,\psi) \ h_\times (\iota, \phi, D_{\mathrm{L}}, \boldsymbol{\xi},t_{\mathrm{c}};t) \big],
\end{align}
where $\kappa(\theta,\phi,\psi)$ is the \textit{effective polarization}~\cite{Capano:2013raa} defined in the region $[0, 2\pi)$ as:
\begin{equation}
\label{eq:effective_pol}
e^{i \kappa(\theta,\phi,\psi)} = \frac{F_+(\theta,\phi,\psi) + i F_\times(\theta,\phi,\psi)}{\sqrt{F_+^2(\theta,\phi,\psi) + F_\times^2(\theta,\phi,\psi)}},
\end{equation}
while $\mathcal{A}(\theta,\phi)$ reads:
\begin{equation}
\label{eq:amp_decomp}
\mathcal{A}(\theta,\phi) = \sqrt{F_+^2(\theta,\phi,\psi) + F_\times^2(\theta,\phi,\psi)}\,.
\end{equation}
We stress that $\mathcal{A}(\theta,\phi)$ does not depend on $\psi$ despite the fact $F_+$ and $F_\times$ depend on it.
Henceforth, to simplify the notation we suppress the dependence of $\kappa$ on $(\theta,\phi,\psi)$.  
Given a GW signal $h_{\mathrm{s}}$ and a template waveform $h_{\mathrm{t}}$, we define the faithfulness as~\cite{Capano:2013raa,Harry:2016ijz}
\begin{equation}
\label{eq:faith}
\mathcal{F}(\iota_{\textrm{s}},{\varphi_0}_{\textrm{s}},\kappa_{\textrm{s}}) \equiv  \max_{t_c, {\varphi_0}_{\mathrm{t}}, \kappa_{\textrm{t}}} \left[\left . \frac{ \left( h_{\mathrm{s}},\,h_{\mathrm{t}} \right)}{\sqrt{ \left( h_{\mathrm{s}},\,h_{\mathrm{s}} \right) \left( h_{\mathrm{t}},\,h_{\mathrm{t}} \right)}}\right \vert_{\substack{\iota_{\mathrm{s}} = \iota_{\mathrm{t}} \\\boldsymbol{\xi}_{\mathrm{s}} = \boldsymbol{\xi}_{\mathrm{t}}}} \right ],
\end{equation}
where parameters with the subscript ``s'' (``t'') refer to the signal (template) waveform. 
The inner product is defined as~\cite{Sathyaprakash:1991mt,Finn:1992xs}:
\begin{equation}
\left( a, b\right) \equiv 4\ \textrm{Re}\int_{f_\textrm{l}}^{f_\textrm{h}} df\,\frac{\tilde{a}(f) \ \tilde{b}^*(f)}{S_n(f)},
\end{equation}
where a tilde indicates the Fourier transform, a star the complex
conjugate and $S_n(f)$ is the one-sided power spectral density (PSD)
of the detector noise, and we employ the Advanced LIGO’s “zero-detuned
high-power” design sensitivity curve~\cite{Shoemaker:2010}.  The integral is evaluated between the
frequencies $f_{\textrm{l}} = 20 \mathrm{Hz}$ and $f_{\textrm{h}} = 3
\mathrm{kHz}$. 

The maximizations over $t_c$ and ${\varphi_0}_{\mathrm{t}}$ in Eq.~\eqref{eq:faith} are computed numerically,
  while the maximization over $\kappa_{\mathrm{t}}$ is done analytically
  following the procedure described in Ref. ~\cite{Capano:2013raa}
  (see Appendix A).  When $h_{\mathrm{t}}$ does not include
  higher-order modes, the maximization over the effective polarization
  $\kappa_{\mathrm{t}}$ in Eq. \eqref{eq:faith} becomes degenerate
  with the maximization over ${\varphi_0}_{\mathrm{t}}$ and we recover the usual 
definition of faithfulness.

The faithfulness given in Eq.~\eqref{eq:faith} depends on the signal parameters
$(\iota_{\textrm{s}},{\varphi_0}_{\textrm{s}},\kappa_{\textrm{s}})$. To understand 
how the faithfulness varies as function of those parameters, 
we introduce the minimum, maximum, average and average
weighted with the SNR unfaithfulness
$[1-\mathcal{F}(\iota_{\textrm{s}},{\varphi_0}_{\textrm{s}},\kappa_{\textrm{s}})]$
over these parameters, namely~\cite{Buonanno:2002fy,Capano:2013raa,Harry:2016ijz}:
\begin{align}
\min_{\iota_{\mathrm{s}},{\varphi_0}_{\mathrm{s}},\kappa_{\mathrm{s}}}(1 -\mathcal{F}) \equiv &  1 - \max_{\iota_{\mathrm{s}},{\varphi_0}_{\mathrm{s}},\kappa_{\mathrm{s}}}\mathcal{F}(\iota_{\textrm{s}},{\varphi_0}_{\textrm{s}},\kappa_{\textrm{s}}) \label{eq:min_unfaith}\,, \\
\max_{\iota_{\mathrm{s}},{\varphi_0}_{\mathrm{s}},\kappa_{\mathrm{s}}}(1 -\mathcal{F}) \equiv &  1 - \min_{\iota_{\mathrm{s}},{\varphi_0}_{\mathrm{s}},\kappa_{\mathrm{s}}}\mathcal{F}(\iota_{\textrm{s}},{\varphi_0}_{\textrm{s}},\kappa_{\textrm{s}}) \label{eq:max_unfaith}\,, 
\end{align}
\begin{equation}
\langle
1-\mathcal{F}\rangle_{\iota_{\mathrm{s}},{\varphi_0}_{\mathrm{s}},\kappa_{\mathrm{s}}} \equiv 1 - \frac{1}{8\pi^2}\int_{0}^{2\pi} d\kappa_{\mathrm{s}} \int_{-1}^{1} d(\cos\iota_s) \int_{0}^{2\pi} d{\varphi_0}_{\mathrm{s}} \ \mathcal{F}(\iota_{\textrm{s}},{\varphi_0}_{\textrm{s}},\kappa_{\textrm{s}})\,, \label{eq:avg_unfaith}
\end{equation}

\begin{align}
&\langle
1-\mathcal{F}\rangle_{\iota_{\mathrm{s}},{\varphi_0}_{\mathrm{s}},\kappa_{\mathrm{s}}}^{\mathrm{SNRweighted}} \equiv \nonumber \\
& \equiv 1 - \sqrt[3]{\frac{\int_{0}^{2\pi} d\kappa_ {\mathrm{s}} \int_{-1}^{1} d(\cos\iota_s) \int_{0}^{2\pi} d{\varphi_0}_{\mathrm{s}} \ \mathcal{F}^3(\iota_{\textrm{s}},{\varphi_0}_{\textrm{s}},\kappa_{\textrm{s}}) \ \mathrm{SNR}^3(\iota_{\textrm{s}},{\varphi_0}_{\textrm{s}},\kappa_{\textrm{s}})}{\int_{0}^{2\pi} d\kappa_{\mathrm{s}} \int_{-1}^{1} d(\cos\iota_s) \int_{0}^{2\pi} d{\varphi_0}_{\mathrm{s}} \ \mathrm{SNR}^3(\iota_{\textrm{s}},{\varphi_0}_{\textrm{s}},\kappa_{\textrm{s}})}} \label{eq:wavg_unfaith}\,, \\ \nonumber
\end{align}
where the $\mathrm{SNR}(\iota_{\textrm{s}},{\varphi_0}_{\textrm{s}},\theta_\textrm{s}, \phi_\textrm{s},\kappa_{\textrm{s}},{D_{\mathrm{L}}}_{\mathrm{s}},
\boldsymbol{\xi}_\mathrm{s},{t_c}_\mathrm{s})$ is defined as:
\begin{equation}
\mathrm{SNR}(\iota_{\textrm{s}},{\varphi_0}_{\textrm{s}},\theta_\textrm{s}, \phi_\textrm{s}, \kappa_{\textrm{s}},{D_{\mathrm{L}}}_{\mathrm{s}},\boldsymbol{\xi}_\mathrm{s},{t_c}_\mathrm{s}) \equiv \sqrt{\left(h_{\mathrm{s}},h_{\mathrm{s}}\right)}.
\end{equation}
We note that for the average unfaithfulness weighted with the SNR in Eq.~\eqref{eq:wavg_unfaith}, we drop in the SNR the 
explicit dependence 
on ${\cal A}(\theta,\phi)$ and $D_{\mathrm{L}}$, because they cancel out. It is important to highlight that the unfaithfulness weighted with the cube 
of the SNR is a conservative upper limit of the fraction of detection volume lost. Indeed, weighting the unfaithfulness with the SNR 
takes into account that, at a fixed distance, configurations closer to an edge-on orientation have a smaller SNR with respect to configurations closer to a face-on orientation, therefore they are less likely to be observed. The definitions of minimum, maximum and averaged unfaithfulness in Eqs.~\eqref{eq:min_unfaith}-\eqref{eq:avg_unfaith} are similar to those in Ref.~\cite{Babak:2016tgq}, with the difference that in the latter they  minimize, maximize and average also over the source orientation $\iota_\mathrm{s}$. The average weighted with the SNR in Eq.~\eqref{eq:wavg_unfaith} was introduced in Ref.~\cite{Buonanno:2002fy} and used for a similar purpose also in Ref.~\cite{Harry:2016ijz}.

In the following we shall show results  
where all the averages are computed assuming an isotropic distribution for the source orientation and sky position.

\begin{figure}[h]
\centering
\includegraphics[width=0.7\textwidth]{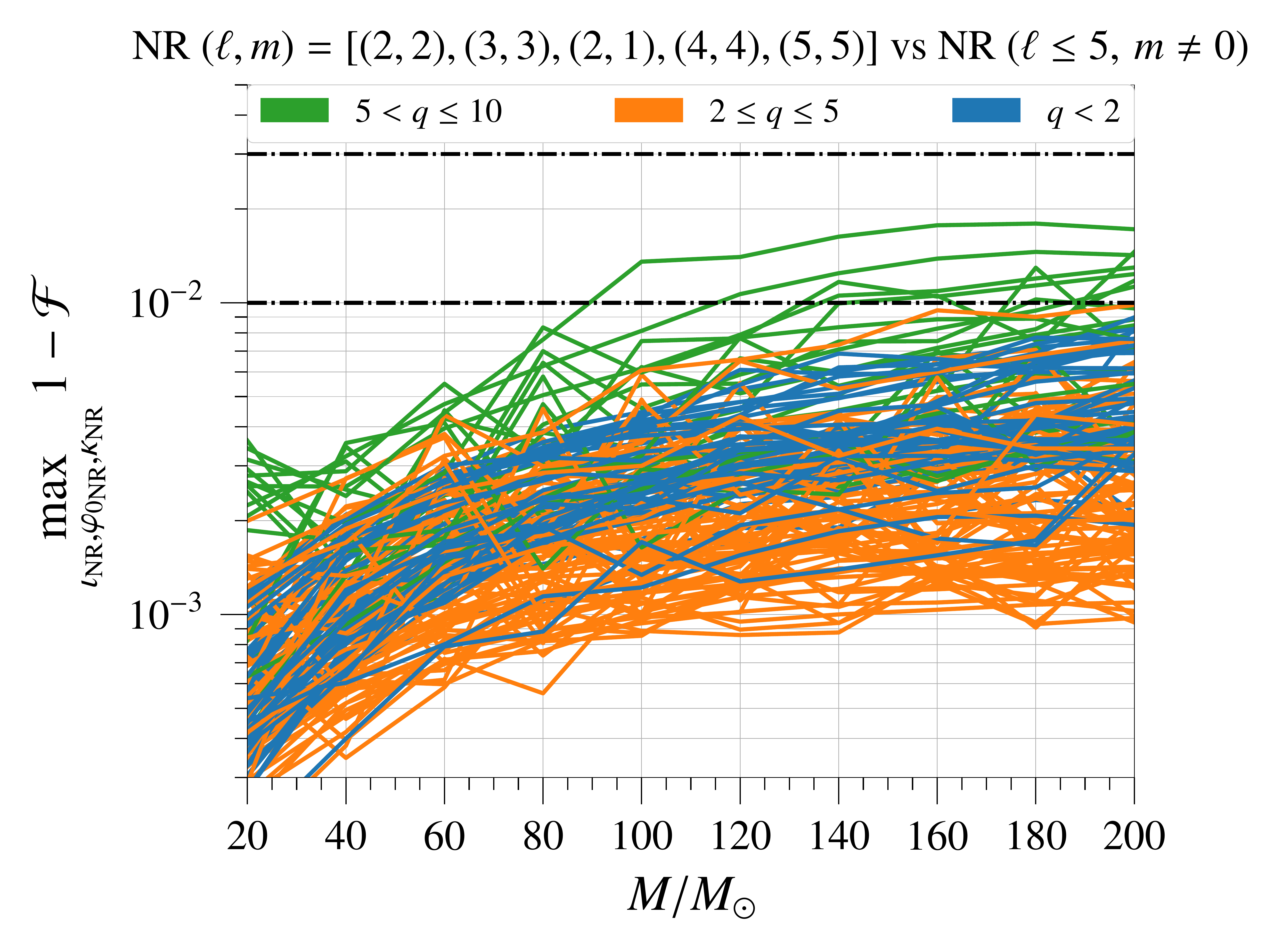}
\caption{Maximum of unfaithfulness $(1-\mathcal{F})$ over the three
  angles
  $(\iota_{\textrm{NR}},{\varphi_0}_{\textrm{NR}},\kappa_{\textrm{NR}})$
  as a function of the total mass, in the range $20 M_\odot \leq M
  \leq 200 M_\odot$ of the NR waveform with
  $(2,2),(2,1),(3,3),(4,4),(5,5)$ modes against NR waveform with
  $(\ell \leq 5, m \neq 0)$ modes. The maximum 
unfaithfulness is typically reached for edge-on orientations. The
  jaggedness of the curves is caused by the numerical noise present in
  higher-order modes that are less resolved in the NR simulations. We
  find that this feature is not present when these noisy modes are
  removed from the calculation of the faithfulness.}
\label{fig:unfaith_NR55vsNRmax}
\end{figure}

Using the aforementioned definitions (\ref{eq:min_unfaith})--(\ref{eq:wavg_unfaith}), we compute the unfaithfulness 
assuming that the signal is an NR waveform with modes $(\ell \leq 5, m \neq 0)$~\footnote{Since the
  nonoscillating $m=0$ modes are not well reproduced by NR
  simulations and their contribution is small, we do not include them
  in these calculations. We find that the contribution of the modes with $\ell \geq
  6$ is neglibigle.}, and the template is either an NR waveform 
or a \texttt{SEOBNRv4} waveform with only the $(2,2)$ mode. 

In the left panel of Fig.~\ref{fig:unfaith_mass}
we show results for the simulation \texttt{SXS:BBH:0610} having $q = 1.2,
\,\chi_1 = -0.5,\,\chi_2=-0.5$. Given the small mass ratio, 
we do not expect the higher-modes to play an important role. Indeed 
both the NR with only the dominant mode and the \texttt{SEOBNRv4} model have averaged
unfaithfulness $<1\%$ in the mass range $20 M_\odot \leq M \leq 200
M_\odot$. In both cases the unfaithfulness is maximum for an
edge-on orientation and is $<3\%$. Conversely the minima of the unfaithfulness occur for a
face-on configuration and they are always much smaller than $1\%$.
The situation is very different in the right panel of Fig.~\ref{fig:unfaith_mass} 
where we consider the simulation \texttt{ET:AEI:0004} that has 
larger mass ratio and spins: $q = 8, \,\chi_1 = \chi_2=0.85$. 
In this case the minima of the unfaithfulness correspond to a face-on orientation where the
higher-order modes are negligible and for this reason both NR with
only the dominant mode and the \texttt{SEOBNRv4} model have unfaithfulness 
smaller than $1\%$. By contrast, the results for the maximum of the
unfaithfulness correspond to an edge-on orientation and they are
equally large for the NR with only the dominant mode and for the 
\texttt{SEOBNRv4} model. They have unfaithfulness in the range $[10\%,20\%]$
for masses $20 M_\odot \leq M \leq 200 M_\odot$. In this case also the
averaged unfaithfulness are large, in the range $[5\%,15\%]$ and $[3\%,8\%]$ 
for the weighted averages.  

Thus, for this high mass-ratio configuration the
error from neglecting higher-order modes supersedes the modeling error
of the dominant mode when the orientation is far from face-on/face-off. This is not surprising because the \texttt{SEOBNRv4} waveform model was constructed requiring $1\%$ of maximum unfaithfulness against the NR waveforms when only the $(2,2)$ mode 
was included~\cite{Bohe:2016gbl}.

Only by properly including the largest subdominant modes can one hope to achieve an
unfaithfulness of the waveform model below $1\%$~\footnote{We notice that using a waveform model with 
unfaithfulness smaller than $3\%$ (or $1\%$ depending on the
features of the template bank) is a sufficient condition for a
template bank to have a loss in event rates due to modeling
error and discreteness of the template bank smaller than $10\%$
(e.g., see Ref.~\cite{Buonanno:2009zt})}. Which subdominnat modes should we include
to achieve such an accuracy? To address this question, we compute the faithfulness between
NR waveforms including the modes $(2,2),(2,1),(3,3),(4,4),(5,5)$ and 
NR waveforms including only the $(\ell \leq 5, m \neq 0)$ modes. We find that the unfaithfulness
averaged over the three angles $(\iota_{\textrm{NR}},{\varphi_0}_{\textrm{NR}},\kappa_{\textrm{NR}})$
ranges between $0.01\% \lessapprox (1-\mathcal{F}) \lessapprox 0.5\%$
for the total mass interval $20 M_\odot \leq M \leq 200
M_\odot$. Thus, we conclude that the modes $(2,2),(2,1),(3,3),(4,4),(5,5)$ 
are sufficient to model the full GW signal if we want to achieve 
an average unfaithfulness smaller than $1\%$. Furthermore, we note that these modes 
are not enough to ensure that the maximum of the unfaithfulness is smaller than
$1\%$. In fact, for some of the configurations with higher mass ratio,
the unfaithfulness is slightly larger than $1\%$ in the mass range $20
M_\odot \leq M \leq 200 M_\odot$, as it is clear from the plot in
Fig.~\ref{fig:unfaith_NR55vsNRmax}. The maximum unfaithfulness
decreases, almost reaching the requirement of being below $1\%$ for
all the waveforms in the catalog, if we add also the more subdominant
modes $(3,2),(4,3)$.  However, given that the overall improvement
in the maximum of unfaithfulness when including also the modes
$(3,2),(4,3)$ is small (of the order of a few $0.1 \%$)
with respect to the results obtained using only the $(2,2),(2,1),(3,3),(4,4),(5,5)$ 
modes, it is worth comparing this improvement with the estimation of the maximum of the unfaithfulness
due to the numerical error of the NR waveforms. The numerical errors
we consider are numerical truncation error
\cite{Hinder:2013oqa,Kumar:2015tha} and waveform extrapolation error
\cite{Hinder:2013oqa,Kumar:2015tha,Taylor:2013zia}. For our NR catalog, we estimate the
numerical truncation error computing the maximum of the unfaithfulness
between the same NR waveforms with the same modes (i.e., 
$(2,2),(2,1),(3,3),(4,4),(5,5)$), but with different resolutions, notably 
the highest (maximum) resolution and the second highest. The
waveform extrapolation error is estimated in the same way, but
employing different extrapolation orders (i.e., $N = 2$ and $N = 3$). 
We find that the contribution of each of these errors to the maximum of the unfaithfulness is in the
range $[0.1\%, 1\%]$ for the total mass interval $20 M_\odot \leq M
\leq 200 M_\odot$~\footnote{The unfaithfulness averaged over the three
  angles
  $(\iota_{\textrm{NR}},{\varphi_0}_{\textrm{NR}},\kappa_{\textrm{NR}})$ due
  to numerical errors is much smaller than $1\%$. The
  reason is that the main contribution to this average unfaithfulness
  is the numerical error of the dominant mode. The latter is 
  much smaller than $1\%$, as well. This conclusion is in agreement
  with Ref.~\cite{Kumar:2015tha} where the authors studied the
  numerical errors of the dominant mode for a subset of the waveforms in
  our NR catalog.}.
 
Since adding the modes $(3,2),(4,3)$ is a non trivial task because of the mode mixing between spherical and spheroidal harmonics~\cite{Buonanno:2006ui,Kelly:2012nd, Berti:2014fga,London:2014cma}, and considering that their contribution is at the same level of the numerical error of the NR waveforms, we decide not to include them in the \texttt{SEOBNRv4HM} model. The results of the maximum of the unfaithfulness due to the numerical errors suggest that in order to use NR waveforms to build an EOBNR model having maximum unfaithfulness against NR smaller than $1\%$ it would be necessary to have more accurate higher-order modes from NR simulations.

\section{Effective-one-body multipolar waveforms for nonprecessing binary black holes}
\label{sec:eob_formalism}

In this section we describe the main ingredients used to build the multipolar spinning, nonprecessing 
\texttt{SEOBNRv4HM} waveform model. We start briefly describing the dynamics in Sec.~\ref{subsec:dynamics}, and then 
focus on the structure of the gravitational modes in Sec. \ref{subsec:waveform}.

In the EOB formalism the real dynamics of two bodies with masses $m_{1,2}$ and
spins $\mathbf{S_{1,2}}$ is mapped into the effective dynamics of a
test particle with mass $\mu$ and spin $\mathbf{S_*}$ moving in a
deformed Kerr metric with mass $M = m_1 + m_2$ and spin
$\bm{S_{\textrm{Kerr}}}$ (for details see Ref.~\cite{Barausse:2011ys}). As discussed above, here we limit to nonprecessing 
spins $\mathbf{S_{1,2}}$ and introduce the dimensionless spin parameters $\chi_{1,2}$
defined as $\mathbf{S_\textrm{i}} = \chi_\textrm{i} m_\textrm{i}^2
\mathbf{\hat{L}}$, with $-1 \leq \chi_i \leq 1$.

\subsection{Effective-one-body dynamics}
\label{subsec:dynamics}

The EOB conservative orbital dynamics is obtained from the resummed EOB Hamiltonian through the 
energy mapping~\cite{Buonanno:1998gg}
\begin{equation}
H_{\textrm{EOB}} = M\sqrt{1+ 2\nu\left(\frac{H_{\textrm{eff}}}{\mu} -1 \right)},
\end{equation}
where $\mu=m_1m_2/(m_1+m_2)$ is the reduced mass of the BBH and $\nu = \mu/M$ is the
symmetric mass ratio. When spins are nonprecessing the motion is constrained to a plane. 
Thus, the dynamical variables entering the Hamiltionian are the orbital phase $\phi$~\footnote{Abusing notation, we indicate the orbital phase with $\phi$, which we use to denote the azimuthal angle describing the sky location of the source in the detector frame. It will be clear from the context which of the two angles we refer to.},  
the radial separation $r$ (normalized to M) and their conjugate momenta $p_\phi$ and $p_r$
(normalized to $\mu$). The explicit form of $H_{\textrm{eff}}$ that we adopt here 
was derived in Refs.~\cite{Barausse:2009xi,Barausse:2011ys}, based on the
linear-in-spin Hamiltonian for spinning test particles of
Ref.~\cite{Barausse:2009aa}. The radial potential entering the 00-component of the EOB 
deformed metric, which also enters the effective Hamiltonian 
$H_{\textrm{eff}}$, is explicitly given in Eqs. (2.2) and (2.3) in Ref.~\cite{Bohe:2016gbl}. 
The Hamiltonian $H_{\textrm{eff}}$ depends also on the calibration parameters ($K$, $d_{\textrm{SO}}$, $d_{\textrm{SS}}$
$\Delta_{\textrm{peak}}^{22}$), which were determined in Ref.~\cite{Bohe:2016gbl} by requiring agreement against 
a large set of NR simulations (see Eqs. (4.12)--(4.15) therein). Here, we adopt the same values for these calibration parameters.

The dissipative dynamics in the EOB formalism is described by the radiation-reaction force given in 
Eq. (2.9) in Ref.~\cite{Bohe:2016gbl}. We notice that in this paper we do not change the dissipative and conservative 
dynamics of the \texttt{SEOBNRv4} model, and that the \texttt{SEOBNRv4HM} waveform models share the same two-body dynamics of 
\texttt{SEOBNRv4}. Here, we improve the accuracy of the gravitational modes with $(\ell, m) \neq (2,2)$, and use them in 
the gravitational waveform, but we do not employ these more accurate version of the modes in the radiation-reaction force. 
Furthermore, we note that the gravitational modes with $(\ell, m) \neq (2,2)$
are present in the radiation-reaction force, but they do not include the NQCs corrections (see Eq.~\eqref{eq:NQC_corrections}). As discussed also in Ref.~\cite{Pan:2011gk}, the latter modify the amplitude of the already subdominant higher-order modes (see Fig.~\ref{fig:rel_amp_hm_0spin}) by $\sim 10\%$ close to merger, where the effect of the radiation reaction is not very important for the plunging dynamics.

\subsection{Effective-one-body  gravitational modes} 
\label{subsec:waveform}

As usual in the EOB formalism~\cite{Buonanno:2000ef}, the gravitational modes entering Eq.~(\ref{eq:spHarmDec}) are composed of 
two main parts: inspiral \& plunge, and merger \& ringdown. We can write the generic mode as:
\begin{align}
\label{eq:complete_mode}
h_{\ell m}(t) = \begin{cases}
h_{\ell m}^{\mathrm{insp-plunge}}(t), &t < t_{\textrm{match}}^{\ell m}\\
h_{\ell m}^{\mathrm{merger-RD}}(t), &t > t_{\textrm{match}}^{\ell m},\\
\end{cases}
\end{align}
where $t_{\textrm{match}}^{\ell m}$ is defined as:
\begin{align}
t_{\textrm{match}}^{\ell m} =\begin{cases} 
t_{\textrm{peak}}^{22}, &  (\ell, m) =(2,2),(3,3),(2,1),(4,4)\\
t_{\textrm{peak}}^{22} - 10 M, & (\ell, m) = (5,5),\\
\end{cases}
\label{eq:matchtime}
\end{align}
with $t_{\textrm{peak}}^{22}$ being the peak of the amplitude of the
$(2,2)$ mode. By construction the amplitude and phase
of $h_{\ell m}(t)$ are $C^1$ at $t = t_{\textrm{match}}^{\ell m}$.  In
the following we shall discuss in more detail how
these two parts of the gravitational modes are built and why we choose 
a different matching point for the mode $(5,5)$. We note again that the mode
$(2,2)$ in the \texttt{SEOBNRv4HM} model is the same as in the 
\texttt{SEOBNRv4} model, and for this reason below we focus on the higher-order modes 
$(3,3),(2,1),(4,4),(5,5)$.

\subsection{Effective-one-body  waveform modes: inspiral-plunge} 
\label{subsec:waveform_inspiral}

The inspiral-plunge EOB modes are expressed in the following multiplicative form:
\begin{equation}
\label{eq:hlm_NQC}
h_{\ell m}^{\textrm{insp-plunge}} = h_{\ell m}^\textrm{F} N_{\ell m},
\end{equation}
where $h_{\ell m}^\textrm{F}$ is the factorized form of the PN GW modes~\cite{Arun:2008kb,Buonanno:2012rv} for quasi-circular orbits, aimed  
at capturing strong-field effects, as discussed in the test-mass limit~\cite{Damour:2007xr,Damour:2008gu,Pan:2010hz}. 
The factor $N_{\ell m}$ in Eq.~(\ref{eq:hlm_NQC}) is the nonquasi-circular (NQC) term, which includes possible radial  
effects that are no longer negligible during the late inspiral and 
plunge, and that are not captured by the rest of the waveform. More explicitly, the factorized term reads: 
\begin{equation}
\label{eq:hlm_factorized}
h_{\ell m}^{\textrm{F}} = h_{\ell m}^{(N, \epsilon)}\, \hat{S}^{(\epsilon)}_{\textrm{eff}}\, T_{\ell m}\,f_{\ell m}\, e^{i \delta_{\ell m}}\,,
\end{equation}
where $\epsilon$ is the parity of the multipolar waveform, defined as
\begin{equation}
\epsilon = \begin{cases}
0, \,\, \ell + m\,\, \text{is even} \\
1, \,\, \ell + m\,\, \text{is odd}.
\end{cases}
\end{equation}
The Newtonian term $h_{\ell m}^{(N,\epsilon)}$ reads: 
\begin{equation}
\label{eq:Newtonian}
h_{\ell m}^{(N,\epsilon)} = \frac{M \nu}{\textrm{D}_L}\, n_{\ell m}^{(\epsilon)} \,\ c_{\ell + \epsilon}(\nu)\,\, V_{\phi}^{\ell}\,\, Y^{\ell - \epsilon, -m}\left(\frac{\pi}{2},\phi\right),
\end{equation}
where $\textrm{D}_L$ is the distance from the source, $Y^{\ell m}(\theta,\phi)$ are the scalar spherical harmonics and the expression of the functions $n_{\ell m}^{(\epsilon)}$ and $c_{\ell + \epsilon}(\nu)$ are given in Appendix \ref{app:modes}. 
The function $V_{\phi}^{\ell}$ is defined as:
\begin{equation}
V^{\ell}_{\phi} \equiv v_{\phi}^{(\ell +  \epsilon)} \equiv M\,\, \Omega \,\, r_{\Omega},
\end{equation}
where
\begin{equation}
r_{\Omega} = \left[ \frac{\partial H_{\textrm{EOB}}(r, \phi, p_r = 0, p_\phi)}{\partial p_{\phi}}\right]^{-\frac{2}{3}},
\end{equation}
$\Omega = d\phi/dt$ being the angular frequency. We also define $v_\Omega = (M \, \Omega)^{1/3}$. 
 The term $\hat{S}^{(\epsilon)}_{\textrm{eff}}$ in Eq.~\eqref{eq:hlm_factorized} is an effective source term:
\begin{equation}
\hat{S}^{(\epsilon)}_{\textrm{eff}} = \begin{cases}
H_{\textrm{eff}}(r,p_{r_*},p_\phi),\,\,\epsilon = 0\\
L_{\textrm{eff}} = p_\phi \, (M \, \Omega)^{\frac{1}{3}},\,\,\,\, \epsilon = 1.\\
\end{cases}
\end{equation}
The function $T_{\ell m}$ in \eqref{eq:hlm_factorized} is a resummation of the leading-order logarithms of tail effects:
\begin{align}
T_{\ell m} &= \frac{\Gamma (\ell +1 -2 \ i H_{\textrm{EOB}} \Omega)}{\Gamma (\ell +1)} \exp[\pi \ m \ \Omega \ H_{\textrm{EOB}}] \nonumber\\
& \times \exp[2 \ i \ m \ \Omega \ H_{\textrm{EOB}} \ \log(2 \ m \ \Omega \ r_0)], 
\end{align}
where $r_0 = 2M/\sqrt{e}$. 

The functions $f_{\ell m}$ and $e^{\delta_{\ell m}}$ in
  Eq.~\eqref{eq:hlm_factorized} contain terms such
that when expanding in PN order $h_{\ell m}^\mathrm{F}$ one recovers $h_{\ell m}^{\mathrm{PN}}$ (i.e., the PN expansion of
the $(\ell, m)$ mode up to the PN order at which $h_{\ell
  m}^{\mathrm{PN}}$ is known today). In the \texttt{SEOBNRv4HM} model the expression for $f_{\ell m}$ and $\delta_{\ell m}$ are mostly taken from the \texttt{SEOBNRv4} model \cite{Bohe:2016gbl} with the addition of some newly computed PN terms (for more details and explicit
expressions of $f_{\ell m}$ and $\delta_{\ell m}$ see Appendix
\ref{app:modes}). For the modes $(2,1)$ and $(5,5)$, $f_{\ell m}$ includes also the calibration term $c_{\ell m} \
v_\Omega^{\beta_{\ell m}}$, where $\beta_{\ell m}$ denotes the first-order term 
at which the PN series of $h_{\ell m}^{\mathrm{PN}}$ is not known today with its complete dependence on mass ratio and spins (see Eqs.~\eqref{eq:f_21}--\eqref{eq:f_55}). The calibration parameter $c_{\ell m}$ is evaluated
to satisfy the condition: 
\begin{align}
\left|h_{\ell m}^{F}(t_{\textrm{match}}^{\ell m})\right| &\equiv \left|h_{\ell m}^{(N, \epsilon)} \hat{S}^{(\epsilon)}_{\textrm{eff}} T_{\ell m} e^{\mathrm{i} \delta_{\ell m}} f_{\ell m}(c_{\ell m})\right| \bigg|_{t = t_{\textrm{match}}^{\ell m}}\,, \nonumber \\
&= \left|h_{\ell m}^\textrm{NR}(t_{\textrm{match}}^{\ell m})\right|, \qquad \mathrm{for}\,\, (\ell, m) = (2,1), \
(5,5),
\label{eq:cal_par}
\end{align}
where $\left|h_{\ell m}^\textrm{NR}(t_{\textrm{match}}^{\ell m})\right|$ is the amplitude of the NR modes at the matching point $t_{\textrm{match}}^{\ell m}$. The latter are given as fitting formulae for every point of the
parameter space $(\nu,\chi_1,\chi_2)$ in Appendix \ref{app:NQCfits}. We need to include the calibration parameter $c_{\ell m}$ for the modes $(\ell,m) = (2,1),(5,5)$ for reasons that we explain below in Sec.~\ref{subsec:waveform_zeros}.

Finally, the term $N_{\ell m}$ in Eq. \eqref{eq:hlm_NQC} is the NQC correction:
\begin{align}
\label{eq:NQC_corrections}
N_{\ell m} &= \left[1+ \frac{p_{r^*}^2}{(r\ \Omega)^2}\left(a_1^{h_{\ell m}} + \frac{a_2^{h_{\ell m}}}{r} + \frac{a_3^{h_{\ell m}}}{r^{3/2}} \right)\right] \nonumber \\ 
& \times \exp\left[i \left(b_1^{h_{\ell m}}\frac{p_{r^*}}{r \ \Omega} + b_2^{h_{\ell m}}\frac{p_{r^*}^3}{r \ \Omega} \right) \right], 
\end{align}
which is used to reproduce the shape of the NR modes close to the matching point $t_{\ell m}^{\textrm{match}}$. As done in the past~\cite{Taracchini:2013rva,Bohe:2016gbl}, the 5 constants $(a_1^{h_{\ell m}}$, $a_2^{h_{\ell m}}$, $a_3^{h_{\ell m}}$, $b_1^{h_{\ell m}}$, $b_2^{h_{\ell m}})$ are fixed by requiring that:
\begin{itemize}
\item The amplitude of the EOB modes is the same as that of the NR modes at the matching point $t_{\textrm{match}}^{\ell m}$:
\begin{equation}
\label{eq:NQC_condition_1}
\left| h_{\ell m}^{\textrm{insp-plunge}}(t_{\textrm{match}}^{\ell m}) \right| = \left|h_{\ell m}^\textrm{NR}(t_{\textrm{match}}^{\ell m})\right|;
\end{equation}
We notice that this condition is different from that in Eq.~\eqref{eq:cal_par} because it affects $h_{\ell m}^{\textrm{insp-plunge}}(t_{\textrm{match}}^{\ell m})$ and not $h_{\ell m}^\textrm{F}(t_{\textrm{match}}^{\ell m})$.
\item The first derivative of the amplitude of the EOB modes is the same as that of the NR modes at the matching point $t_{\textrm{match}}^{\ell m}$:
\begin{equation}
\label{eq:NQC_condition_2}
\left. \frac{d\left| h_{\ell m}^{\textrm{insp-plunge}}(t) \right|}{dt} \right|_{t =t_{\textrm{match}}^{\ell m}} = 
\left. \frac{d\left| h_{\ell m}^{\textrm{NR}}(t) \right|}{dt} \right|_{t =t_{\textrm{match}}^{\ell m}};
\end{equation}
\item The second derivative of the amplitude of the EOB modes is the same as that of the NR modes at the matching point $t_{\textrm{match}}^{\ell m}$:
\begin{equation}
\label{eq:NQC_condition_3}
\left. \frac{d^2\left| h_{\ell m}^{\textrm{insp-plunge}}(t) \right|}{dt^2} \right|_{t = t_{\textrm{match}}^{\ell m}} 
= \left. \frac{d^2\left| h_{\ell m}^{\textrm{NR}}(t) \right|}{dt^2} \right|_{t = t_{\textrm{match}}^{\ell m}};
\end{equation}
\item The frequency of the EOB modes is the same as that of the NR modes at the matching point $t_{\textrm{match}}^{\ell m}$:
\begin{equation}
\label{eq:NQC_condition_4}
\omega_{\ell m}^{\textrm{insp-plunge}}(t_{\textrm{match}}^{\ell m}) = \omega_{\ell m}^{\textrm{NR}}(t_{\textrm{match}}^{\ell m});
\end{equation}
\item The first derivative of the frequency of the EOB modes is the same as that of the NR modes at the matching point $t_{\textrm{match}}^{\ell m}$:
\begin{equation}
\label{eq:NQC_condition_5}
\left .\frac{d {\omega}_{\ell m}^{\textrm{insp-plunge}}(t)}{dt}\right|_{t = t_{\textrm{match}}^{\ell m}}  =
\left . \frac{d {\omega}_{\ell m}^{\textrm{NR}}(t)}{dt}\right|_{t = t_{\textrm{match}}^{\ell m}},
\end{equation}
\end{itemize}
where the RHS of Eqs. \eqref{eq:NQC_condition_1}--\eqref{eq:NQC_condition_5} (usually called ``input values''), are given as fitting formulae for every point of the
parameter space $(\nu,\chi_1,\chi_2)$ in Appendix \ref{app:NQCfits}. These fits are produced using the NR 
catalog and BH-perturbation-theory waveforms, as described in Appendix~\ref{sec:NRcatalog}. 

As we discuss in Appendices~\ref{app:NQCfits} and ~\ref{app:ringdownfits}, we find that for several binary configurations in the NR catalog, the numerical error is quite large for the mode $(5,5)$ close to merger. To minimize the impact of the numerical error on the fits of the input values,  
we are obliged to choose the  matching point for this mode earlier than for other modes, as indicated 
in Eq.~(\ref{eq:matchtime}).

\subsection{Minima in $(2,1)$, $(5,5)$-modes' amplitude and $c_{\ell m}$'s calibration parameters}
\label{subsec:waveform_zeros}

\begin{figure}[h]
\centering
\includegraphics[width=0.7\textwidth]{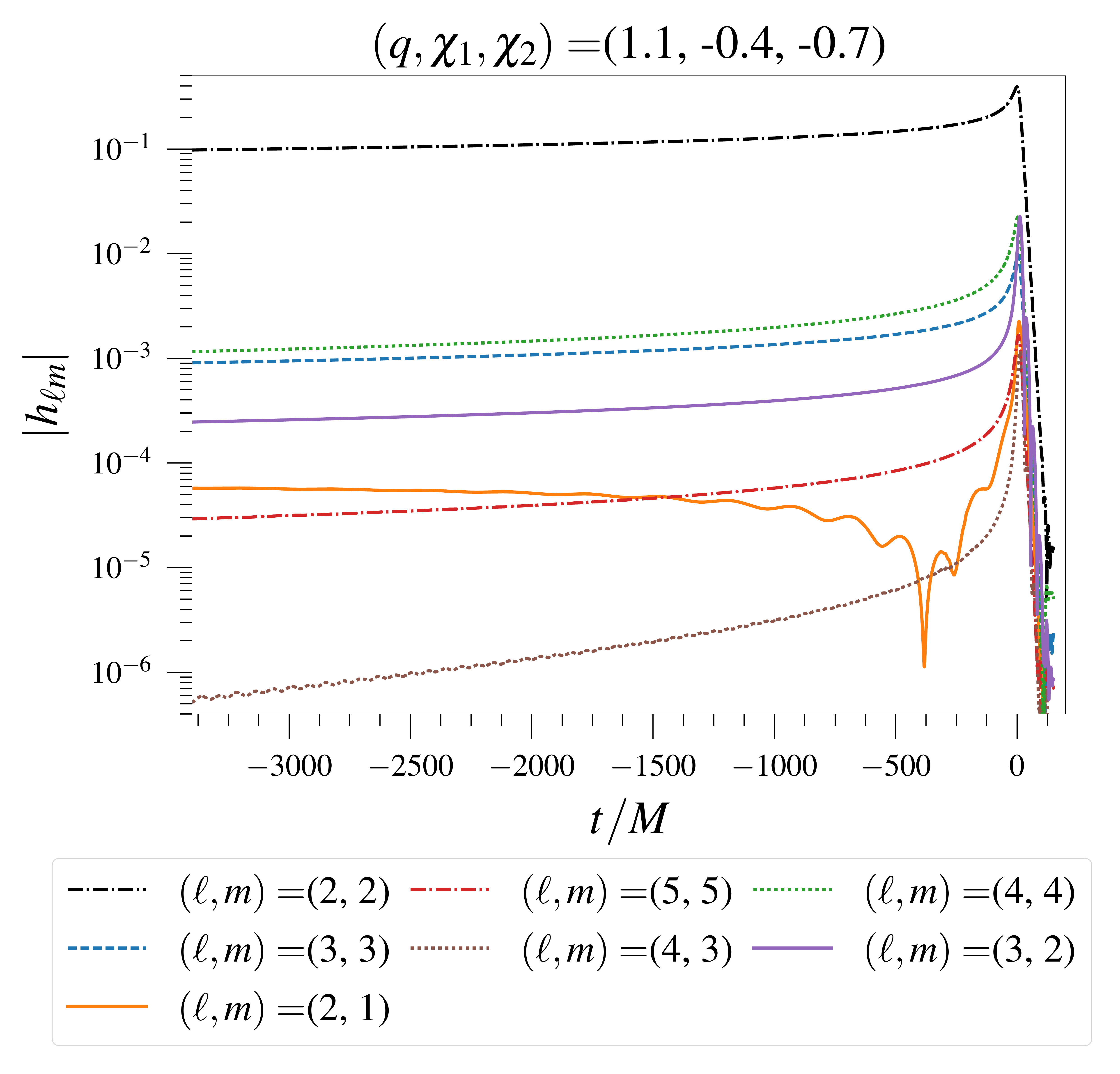}
\caption{Amplitude of the $(2,2),(2,1),(3,3),(4,4),(5,5)$,$(3,2),(4,3)$ modes versus time for the NR simulation 
\texttt{SXS:BBH:1377} with parameters $q = 1.1, \,\chi_1 = -0.4,\,\chi_2 = -0.7$. We produce such simulation to check if the 
analytical prediction that the $(2,1)$-mode's amplitude would have a non-monotonic behaviour toward merger holds. 
We choose as origin of time the peak of the $(2,2)$ mode.}
\label{fig:zero}
\end{figure}

We want now to come back to the motivation of introducing the $c_{\ell m}$'s calibration parameters in 
  Eq.~\eqref{eq:cal_par} for the modes $(2,1)$ and $(5,5)$. We note that those parameters are determined 
and included in the waveform  before applying the NQC conditions (\ref{eq:NQC_condition_1})--(\ref{eq:NQC_condition_5}). 
We introduce the $c_{\ell m}$'s to ``cure'' the behaviour of the modes $(2,1),(5,5)$ close to the matching point for a particular 
region of the parameter space. Indeed, we find that the factorized expression of the amplitude $\left|h_{\ell m}^\textrm{F}(t)\right|$ starts to  
decrease toward plunge and merger, approaching minimum values close to zero for $t \sim
  t_{\ell m}^{\textrm{match}}$ when the binary parameters have $q \sim 1$ and large $|\chi_A| = |(\chi_1-\chi_2)|/2$. 
Although the term $f_{\ell m}$ in Eq.~(\ref{eq:hlm_factorized}) is responsible of the zeros in the amplitude, we find 
that this unexpected behaviour is also present in the PN-expanded form of the mode, and persist in other 
mode resummations, like those suggested in Ref.~\cite{Pan:2010hz} (see Eq. 2 therein) and in Refs.~\cite{Nagar:2016ayt,Messina:2018ghh}.

Quite interestingly, in the case of the $(5,5)$ mode, we do not find such a non-monotonic behaviour toward merger 
in the NR simulations at our disposal, but we do find it for the $(2,1)$ mode in the same region of parameter 
space predicted by the analytical computation. In particular, we notice minima toward merger in \texttt{SXS:BBH:0612} with $(q = 1.6,\,
  \chi_1 = 0.5,\, \chi_2 = -0.5)$, \texttt{SXS:SXS:BBH:0614} $(q =
  2,\, \chi_1 = 0.75,\, \chi_2 = -0.5)$, \texttt{SXS:BBH:0254} $(q =
  2,\, \chi_1 = 0.6,\, \chi_2 = -0.6)$. We also produce a new NR simulation 
\texttt{SXS:BBH:1377} with $q = 1.1, \,\chi_1 = -0.4,\,\chi_2 = -0.7$ to check the 
presence of a minimum in the amplitude mode. Figure~\ref{fig:zero} shows indeed the presence of such a mimimum in 
the $(2,1)$ mode amplitude for  \texttt{SXS:BBH:1377}.

The minima (or zeros) of the $(2,1),(5,5)$ modes can sometime 
occur at times $t \sim t_{\ell m}^{\textrm{match}}$, that is close to the times where 
we impose the NQC conditions \eqref{eq:NQC_condition_1}--\eqref{eq:NQC_condition_5}. 
When that happens, the enforcement of such conditions yield a waveform which contains 
unwanted features\footnote{Since $|h_{\ell m}^{\mathrm{insp-plunge}}(t_{\mathrm{match}}^{\ell m})| \sim 0$, imposing the condition in Eq.~\eqref{eq:NQC_condition_1} with $\left|h_{\ell m}^\textrm{NR}(t_{\textrm{match}}^{\ell m})\right| \neq 0$ forces the function $|N_{\ell m}|$, hence the amplitude $|h_{\ell m}^{\mathrm{insp-plunge}}(t)|$, to assume unphysically large values for $t < t_{\mathrm{match}}^{\ell m}$.}. Considering that for the mode $(5,5)$ the mimima are absent in the NR simulations, 
thus they are likely an artefact of the analytical waveform, and that for the mode $(2,1)$ 
the minima are present only in the region of parameter space where the 
$(2,1)$ mode is much smaller than the other modes (i.e., when $q \sim 1$ and $|\chi_A| = |(\chi_1-\chi_2)|/2$ is large, 
see also Fig.~\ref{fig:zero}), we decide to remove the minima from the $(2,1)$ and $(5,5)$ EOB modes. We achieve 
this by introducing the calibration parameter $c_{\ell m}$, which enforces the condition that the EOB amplitude 
at $t_{\ell m}^{\textrm{match}}$ is equal to the NR amplitude (see Eq.~(\ref{eq:cal_par})). Note that the 
latter is imposed before the NQC conditions and removes the minima only when they appear for $t \sim t_{\mathrm{match}}^{\ell m}$. Modeling the minima in the $(2,1)$ modes could be considered 
in the future, when more accurate waveforms would be needed at higher SNRs.
  
Henceforth, we attempt to describe why the analytical modes (both in the PN and factorized form) present minima or zeros 
for the $(2,1)$ and $(5,5)$ cases when $q \sim 1$ and $|\chi_A| = |(\chi_1-\chi_2)|/2$ is large. 
Readers who might not be interested in this technical discussion, could skip the rest of this section 
and move to Sec.~\ref{subsec:waveform_merger}.

As discussed in Sec.~\ref{sec:motivations}, because of binary symmetry under rotation 
($\varphi_0 \rightarrow \varphi_0 + \pi$) the modes with odd $m$ vanish 
for equal-mass and equal-spins configurations. Thus, the nonspinning terms in those modes are proportional 
to $\delta m = (m_1 - m_2)/M$ while the spinning terms are an antisymmetric combination of $\delta m$, $\chi_A$ and $\chi_S =
(\chi_1+\chi_2)/2$ (e.g, $\chi_A$, $\chi_S \delta m$, $\chi_A^2 \delta m$), see for example Eqs.(38a)--(38i) in Ref.~\cite{Pan:2010hz}. In the limit $q\sim 1$ all 
the nonspinning and spinning terms proportional to $\delta m$ are suppressed, and the leading
spinning terms are proportional to $\chi_A$. For large values of $\chi_A$ and small values of $\delta m$ (very
unequal spins, almost equal mass) a cancellation between the leading-order spin 
correction and the dominant nonspinning PN term (which despite being of lower PN order 
is supressed by $\delta m$) can occur at some given frequency. The higher the difference in PN orders
between these two leading spinning and nonspinning contributions, the
higher the frequency at which the cancellation happens. For the
$(2,1)$ mode, there is only a half PN order difference between these terms (see Eq.~(38b) in Ref.~\cite{Pan:2010hz}), so the
cancellation arises at sufficiently low frequencies where this PN
analysis based on two leading terms can be reliable, and, indeed, we do
observe these minima in the NR simulations. In  Table~\ref{tab:zeros} we 
list the configurations in our NR catalog where the minimum happens and
its orbital frequency as measured in the NR simulation~\footnote{We estimate the orbital
  frequency in the NR simulation as half of the gravitational frequency of the $(2,2)$ mode.} 
and as predicted by PN modeling at 3PN order~\cite{Blanchet:2013haa,Buonanno:2012rv,Marsatetal2017}. As expected, the lower the
frequency, the more accurate the PN prediction. We note that the last row
shows results of a NR simulation that we specifically produce to confirm the presence of
the minimum in the mode (see also Fig.~\ref{fig:zero}). We note that for the binary's 
configuration listed in the first row of Table~\ref{tab:zeros},
the NR simulation shows a high-frequency minimum, which is not
reproduced by PN calculations, confirming that this analysis becomes less reliable
in the high-frequency regime.

Lastly, as already pointed out above, for the $(5,5)$
mode we do not observe any minimum in the NR simulations at our disposal. The most
likely explanation is that the cancellation of the leading terms
happens at frequencies high enough that the higher-order PN corrections
would change the result (i.e., they completely remove the minimum or push it
at frequency higher than the merger frequency).

\begin{table}[h]
\centering
 \begin{tabular}{|c|c| c| c| c| c|} 
 \hline
 NR name & $q$ & $\chi_1$ & $\chi_2$ & $M \Omega_{0}^{\mathrm{NR}}$ & $M \Omega_{0}^{\mathrm{PN}}$ \\ [0.5ex] 
 \hline
 \texttt{SXS:BBH:0254}&2 & 0.6 & -0.6 & 0.17 & n/a \\ 
 \hline
 \texttt{SXS:BBH:0614}&2 & 0.75 & -0.5 & 0.082 & 0.057 \\
 \hline
 \texttt{SXS:BBH:0612}&1.6 & 0.5 & -0.5 & 0.068 & 0.047 \\
 \hline
 \texttt{SXS:BBH:1377}&1.1 & -0.4 & -0.7 & 0.033 & 0.029 \\
\hline
\end{tabular}
\caption{For each NR simulation, binary's parameters and values of the orbital 
frequencies $M\Omega_{0}^{\mathrm{NR}}$ and $M\Omega_{0}^{\mathrm{PN}}$ 
at which the minimum of the $(2,1)$ mode occurs.}
\label{tab:zeros}
\end{table}

\subsection{Effective-one-body  waveform modes: merger-ringdown} 
\label{subsec:waveform_merger}

We build the merger-ringdown EOB waveforms following Refs.~\cite{Baker:2008mj,Damour:2014yha,Nagar:2016iwa,Bohe:2016gbl}, notably 
the implementation in Ref.~\cite{Bohe:2016gbl}. The merger-ringdown mode reads:
\begin{equation}
\label{eq:merger-RD_wave}
h_{\ell m}^{\textrm{merger-RD}}(t) = \nu \ \tilde{A}_{\ell m}(t)\ e^{i \tilde{\phi}_{\ell m}(t)} \ e^{-i \sigma_{\ell m 0}(t-t_{\textrm{match}}^{\ell m})},
\end{equation}
where $\sigma_{\ell m 0}$ is the (complex) frequency of the
  least-damped QNM of the final BH. We denote $\sigma_{\ell m}^\textrm{R} \equiv \Im (\sigma_{\ell m0}) < 0$ and 
$\sigma_{\ell m}^\textrm{I} \equiv -\Re (\sigma_{\ell m0})$. For each mode $(\ell,m)$, we employ the  
  frequency values tabulated in Refs.~\cite{Berti:2005ys,Berti:2009kk} as
  functions of the BH's mass and spin. We compute the remnant-BH's mass using the same fitting 
  formula in Ref.~\cite{Taracchini:2013rva}, which is based on the
  phenomenological formula in Ref.~\cite{Barausse:2012qz}, but
  we replace its equal-mass limit (see Eq. (11) in Ref.~\cite{Barausse:2012qz}) with the fit
  in Ref.~\cite{Hemberger:2013hsa} (see Eq. (9) of Ref.~\cite{Hemberger:2013hsa}). The remnant-BH's spin 
is computed using the spin formula in Ref.~\cite{Hofmann:2016yih} (see Eq. (7) in Ref.~\cite{Hofmann:2016yih}).

For the two functions $\tilde{A}_{\ell m}(t)$ and $\tilde{\phi}_{\ell m}(t)$, we use the ans\"atze~\cite{Bohe:2016gbl}:
\begin{equation}
\label{eq:ansatz_amp}
\tilde{A}_{\ell m}(t) = c_{1,c}^{\ell m} \tanh[c_{1,f}^{\ell m}\ (t-t_{\textrm{match}}^{\ell m}) \ +\ c_{2,f}^{\ell m}] \ + \ c_{2,c}^{\ell m},
\end{equation}
\begin{equation}
\label{eq:ansatz_phase}
\tilde{\phi}_{\ell m}(t) = \phi_{\textrm{match}}^{\ell m} - d_{1,c}^{\ell m} \log\left[\frac{1+d_{2,f}^{\ell m} e^{-d_{1,f}^{\ell m}(t-t_{\textrm{match}}^{\ell m})}}{1+d_{2,f}^{\ell m}}\right],
\end{equation}
where $ \phi_{\textrm{match}}^{\ell m}$ is the phase of the
inspiral-plunge mode $(\ell, m)$ at $t = t_{\textrm{match}}^{\ell m}$.
The coefficients $d_{1,c}^{\ell m}$ and $c_{i,c}^{\ell m}$ \footnote{The subscript ``c'' means ``constrained'' 
while ``f'' stands for ``free''.} with $i = 1,2$
are fixed by imposing that the functions $\tilde{A}_{\ell m}(t)$ and $\tilde{\phi}_{\ell m}(t)$ 
in Eq. \eqref{eq:complete_mode} are of class $C^1$ at $t = t_{\textrm{match}}^{\ell m}$. Those 
constraints allow us to express $c_{i,c}^{\ell m}$ in terms of $c_{1,f}^{\ell
    m}$,$\ c_{2,f}^{\ell m},\ \sigma^\textrm{R}_{\ell m}$, $\ |h_{\ell
    m}^{\textrm{insp-plunge}}(t_{\textrm{match}}^{\ell
    m})|$,$\ \partial_t|h_{\ell
    m}^{\textrm{insp-plunge}}(t_{\textrm{match}}^{\ell m})|$ as
\begin{align}   
\label{eq:continuity_amp} 
c_{1,c}^{\ell m} &= \frac{1}{c_{1,f}^{\ell
    m} \nu} \big[ \partial_t|h_{\ell
    m}^{\textrm{insp-plunge}}(t_{\textrm{match}}^{\ell m})| \nonumber \\
    &- \sigma^\textrm{R}_{\ell m} |h_{\ell
    m}^{\textrm{insp-plunge}}(t_{\textrm{match}}^{\ell
    m})|\big] \cosh^2{(c_{2,f}^{\ell m})}, \\
c_{2,c}^{\ell m} &= -\frac{ |h_{\ell
    m}^{\textrm{insp-plunge}}(t_{\textrm{match}}^{\ell
    m})|}{\nu} + \frac{1}{c_{1,f}^{\ell
    m} \nu} \big[ \partial_t|h_{\ell
    m}^{\textrm{insp-plunge}}(t_{\textrm{match}}^{\ell m})|  \nonumber \\
    &- \sigma^\textrm{R}_{\ell m} |h_{\ell
    m}^{\textrm{insp-plunge}}(t_{\textrm{match}}^{\ell
    m})|\big] \cosh{(c_{2,f}^{\ell m})}\sinh{(c_{2,f}^{\ell m})}, \\ \nonumber   
\end{align}
and 
$d_{1,c}^{\ell m}$ in terms of $d_{1,f}^{\ell m},\ d_{2,f}^{\ell m}, \sigma^\textrm{I}_{\ell
      m},\ \omega_{\ell
      m}^{\textrm{insp-plunge}}(t_{\textrm{match}}^{\ell m})$ as
\begin{align} 
\label{eq:continuity_phase}    
d_{1,c}^{\ell m} &= \left[\omega_{\ell m}^{\textrm{insp-plunge}}(t_{\textrm{match}}^{\ell m}) -  \sigma^\textrm{I}_{\ell
      m}\right]\frac{1+ d_{2,f}^{\ell m}}{d_{1,f}^{\ell m}d_{2,f}^{\ell m}}.
\end{align}
We emphasize again that the values of $|h_{\ell
    m}^{\textrm{insp-plunge}}(t_{\textrm{match}}^{\ell
    m})|$, $\partial_t|h_{\ell
    m}^{\textrm{insp-plunge}}(t_{\textrm{match}}^{\ell m})|$ and $\omega_{\ell
      m}^{\textrm{insp-plunge}}(t_{\textrm{match}}^{\ell m})$ are fixed by the NQCs conditions in Eqs.~\eqref{eq:NQC_condition_1} \eqref{eq:NQC_condition_2} \eqref{eq:NQC_condition_4} to be the same as the NR values $\left|h_{\ell m}^\textrm{NR}(t_{\textrm{match}}^{\ell m})\right|, \partial_t|h_{\ell m}^\textrm{NR}(t_{\textrm{match}}^{\ell m})|$ and $\omega_{\ell m}^{\textrm{insp-plunge}}(t_{\textrm{match}}^{\ell m})$ which are given in Appendix~\ref{app:NQCfits} as function of $\nu$ and a combination of the spins $\chi_1$ and $\chi_2$.     
Thus, we are left 
with only two free parameters in the amplitude $c_{i,f}^{\ell m}$ and 
in the phase $d_{i,f}^{\ell m}$. To obtain those 
parameters we first extract them applying a least-square fit in each point of the
parameter space $(\nu,\chi_1,\chi_2)$ for which we have NR and 
Teukolsky-equation--based waveforms. Then, we interpolate those values in the rest 
of the parameter space using polynomial fits in $\nu$ and a combination of $\chi_1$ and $\chi_2$, as given 
explicitly in Appendix~\ref{app:ringdownfits}.

Regarding the accuracy of our merger-ringdown model, for the modes (2,1) and (3,3) the average fractional difference in the amplitude between the model and the NR waveform is of the
  order of percent, while the average phase difference is $\lesssim 0.1$ radians. For the modes (4,4) and (5,5) we are unable to 
  determine a similar average error, because those modes are 
  affected by numerical error at merger and during ringdown, as we discuss in Appendix \ref{app:ringdownfits}. 
  We find that the average fractional difference in the amplitude (phase) between the
  model and the NR simulation can be in some cases on the order of $10\%$ ($\lesssim 0.3$ rad), but this can be comparable to the difference between NR waveforms at different 
extraction radius (see Fig.~\ref{fig:noisyNR} in Appendix~\ref{app:ringdownfits}). We notice that although the
  errors in those modes are not as small as those of the modes (2,1) and (3,3), they
  are still acceptable considering the relatively small amplitude of
  the modes (4,4) and (5,5) with respect to the (2,1) and (3,3).

In summary, given a binary configuration $(m_1, m_2, \chi_1,\chi_2)$, 
the merger-ringdown model that we have developed is uniquely
  determined by the following parameters $(m_1,m_2,\chi_1,\chi_2,t_{\textrm{match}}^{\ell m},
  \phi_{\textrm{match}}^{\ell m}, \sigma^\textrm{I}_{\ell m},
  \sigma^\textrm{R}_{\ell m})$, the latter being a function of the
  remnant-BH's mass and spin determined by the NR fits. It is possible to use
  this merger-ringdown model as a stand-alone model (i.e., independently from the 
inspiral-plunge part), if we also provide equations relating $\phi_{\textrm{match}}^{\ell m}$ (i.e., the phase of the mode $(\ell, m)$ at $t_{\textrm{match}}^{\ell m}$) with $\phi_{\textrm{match}}^{2 2}$.
Indeed even if a
  global time and phase shift is possible, the relations between the phases of different modes are fixed.
  The latter are given as a fit for every point of the parameter space $(\nu,
  \chi_1,\chi_2)$ in Appendix~\ref{app:fitphasediff}.  We note that in this stand-alone
  merger-ringdown model, one can also treat $\sigma^\textrm{I}_{\ell m}$
  and $\sigma^\textrm{R}_{\ell m}$ as free parameters (i.e., we do not 
  compute them from Refs.~\cite{Berti:2005ys,Berti:2009kk}). In this case the 
  merger-ringdown model is a function of
  $(m_1,m_2,\chi_1,\chi_2,t_{\textrm{match}}^{\ell m},
  \phi_{\textrm{match}}^{\ell m}, \sigma^\textrm{I}_{\ell m},
  \sigma^\textrm{R}_{\ell m}, M_{\textrm{final}})$ where
  $M_{\textrm{final}}$ is the remnant-BH's, which is used only
  to rescale $\sigma^\textrm{I}_{\ell m}$ and $\sigma^\textrm{R}_{\ell
    m}$.

\begin{figure}
\centering
\includegraphics[width=0.7\textwidth]{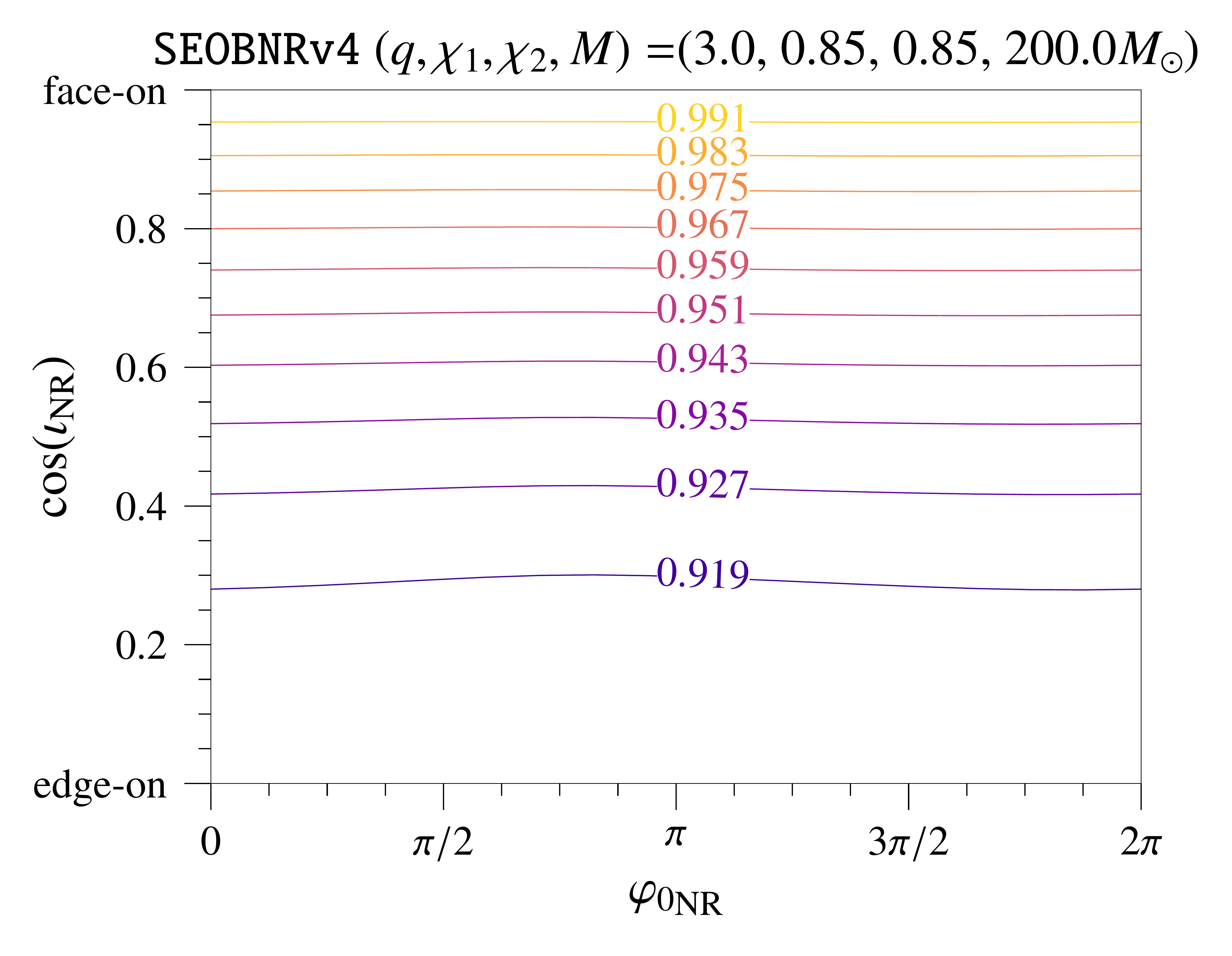} 
\includegraphics[width=0.7\textwidth]{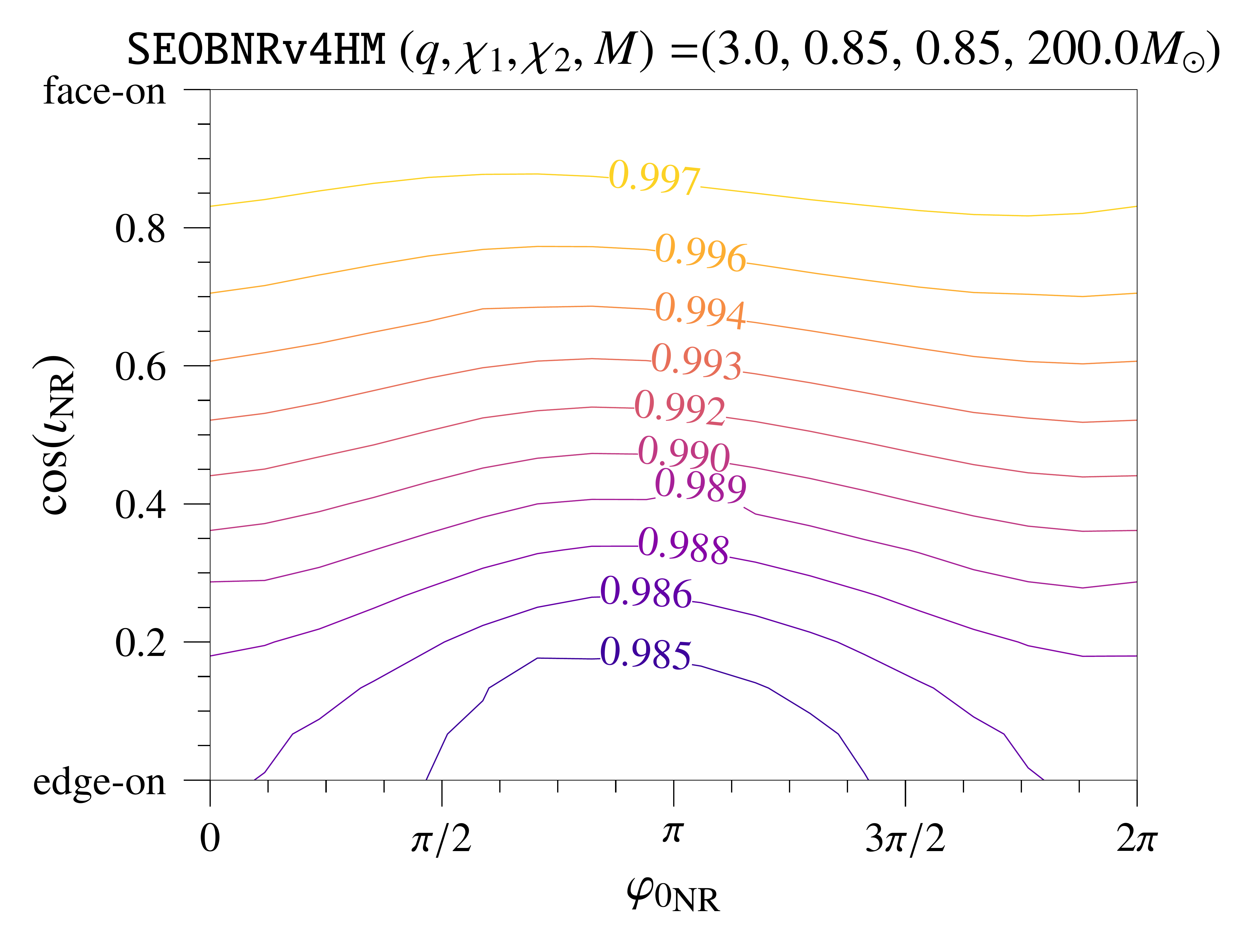} 
\caption{Faithfulness $\mathcal{F}(\cos(\iota_{\textrm{NR}}),{\varphi_0}_{\textrm{NR}},\kappa_{\textrm{NR}} = 0)$ for the configuration $(q = 3,\, M = 200 M_\odot,\, \chi_1 = 0.85, \, \chi_2 = 0.85)$: NR $(\ell \leq 5, \, m \neq 0)$ vs \texttt{SEOBNRv4} (left panel), NR $(\ell \leq 5, \, m \neq 0)$ vs \texttt{SEOBNRv4HM} (right panel). We plot the faithfulness for a fixed $\kappa_{\textrm{NR}}$ because we have noted that $\mathcal{F}(\iota_{\textrm{NR}},{\varphi_0}_{\textrm{NR}},\kappa_{\textrm{NR}})$ is mildly dependent on this variable.}
\label{fig:q3v4vsHMthetaphi}
\end{figure}

\section{Performance of the multipolar effective-one-body waveform model}
\label{sec:comparison}

We study the accuracy of the multipolar waveform model
\texttt{SEOBNRv4HM} by computing its faithfulness against waveforms
in the NR catalog at our disposal.  In Secs.~\ref{comparison:NR1} and
\ref{comparison:NR2}, we perform a detailed comparison against three
NR simulations, notably a moderate--mass-ratio configuration,
\texttt{SXS:BBH:0293} $(q = 3,\, \chi_1 = 0.85,\, \chi_2 = 0.85)$, and
two high--mass-ratio configurations, \texttt{SXS:BBH:0065} $(q = 8,\,
\chi_1 = 0.5,\, \chi_2 = 0)$ and \texttt{ET:AEI:0004} $(q = 8,\,
\chi_1 = 0.85,\, \chi_2 = 0.85)$. We also compare the results above
with those obtained when the (2,2)--waveform-model \texttt{SEOBNRv4}
is employed. Finally, in
Sec.~\ref{comparison:NRcatalog} we summarize the agreement of
the \texttt{SEOBRNv4HM} model against the entire NR catalog composed of 157
simulations.

\subsection{Moderate mass ratio: \texttt{SXS:BBH:0293}}
\label{comparison:NR1}

In the left panel of Fig.~\ref{fig:q3v4vsHMthetaphi} we show a contour
plot of the faithfulness
$\mathcal{F}(\cos(\iota_{\textrm{NR}}),{\varphi_0}_{\textrm{NR}},\kappa_{\textrm{NR}})\big|_{\kappa_{\textrm{NR}}
  = 0}$ between the NR waveform \texttt{SXS:BBH:0293} with modes
($\ell \leq 5, \,m \neq 0$), and the waveform generated with
\texttt{SEOBNRv4}, for a total mass of $M = 200 M_\odot$. In order to
reduce the dimensionality of the plot, we fix the value of
$\kappa_{\textrm{NR}}$. However, we find that the dependence of the faithfulness on this
variable is mild. We can see that the 
faithfulness depends mainly on the inclination angle
$\iota_{\textrm{NR}}$ and degrades when we move from a face-on
$\{\mathcal{F}(\cos(\iota_{\textrm{NR}}) = 0) \sim 99\%\}$ to an
edge-on orientation $\{\mathcal{F}(\cos(\iota_{\textrm{NR}}) = 1)
\sim 92\%\}$. This situation is different if we include the
higher-order modes in the model (i.e, $(3,3),(2,1),(4,4),(5,5)$), 
as can be seen in the right panel of Fig.~\ref{fig:q3v4vsHMthetaphi} 
where we use the \texttt{SEOBNRv4HM} waveform model.
In this case the faithfulness degrades much less if we go from a
face-on ($\mathcal{F} \sim 99.7\%$) to an edge-on ($\mathcal{F} \sim
98.5\%$) orientation. The small residual degradation is due to the
fact that the dominant mode is still better modeled than the
higher-order modes and for this reason for a face-on orientation
(where the signal is dominated by the dominant mode) the faithfulness
is larger than for an edge-on orientation where the higher-order modes
contribute the most. Another contribution to the residual degradation
in an edge-on orientation stems from the fact that in the
\texttt{SEOBNRv4HM} model we still miss some subdominant higher-order
modes, which instead we have included in the NR waveform.

\begin{figure}[htb]
\centering
\includegraphics[width=0.7\textwidth]{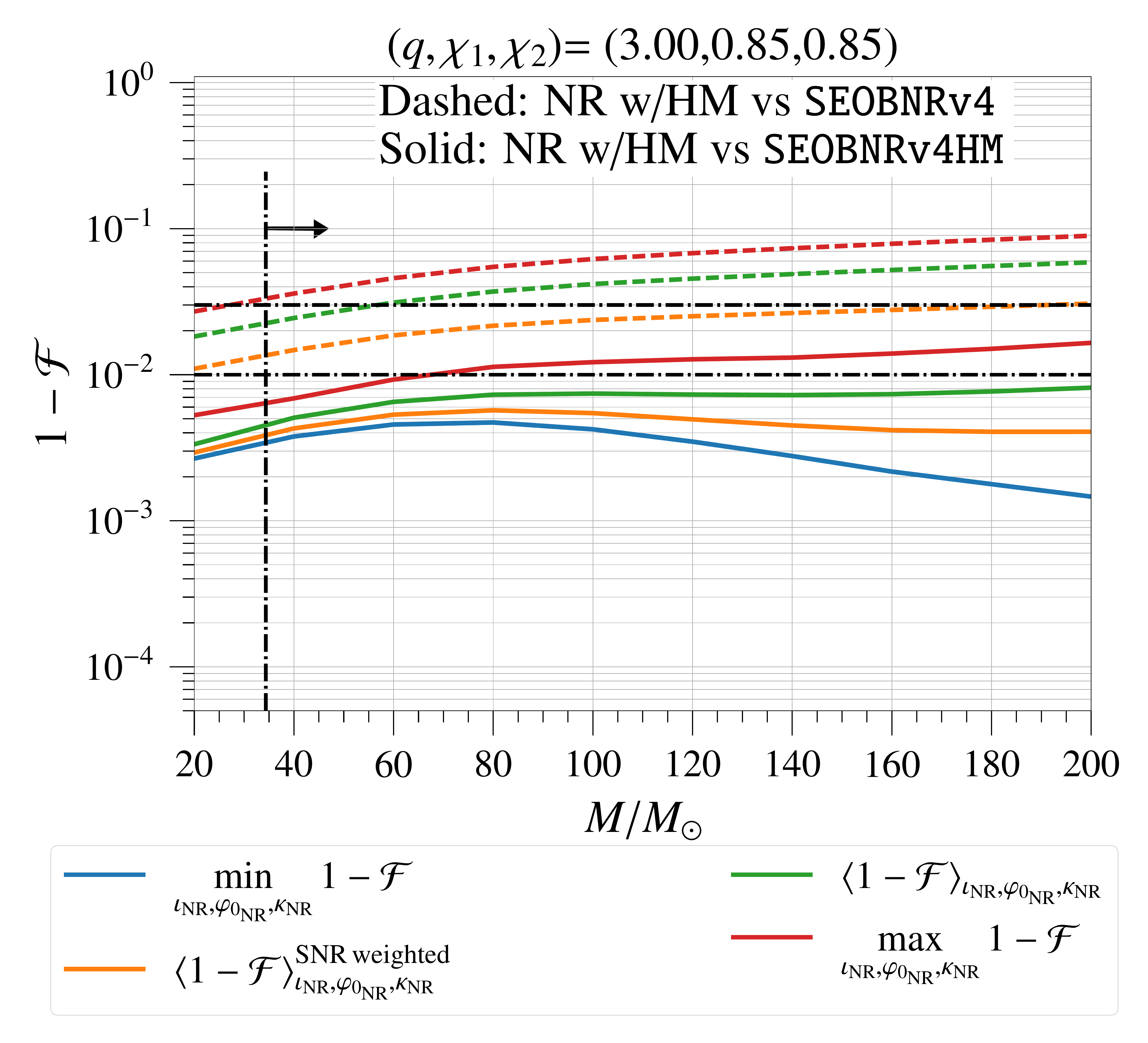}
\caption{Unfaithfulness $(1-\mathcal{F})$ for the
  configuration $(q = 3,\, \chi_1 = \chi_2 = 0.85)$ in the mass range
  $20 M_\odot \leq M \leq 200 M_\odot$. Dashed (plain) curves refer to results for
  \texttt{SEOBNRv4} (\texttt{SEOBNRv4HM}). The minima of the unfaithfulness for the two
  models (blue curves), lie on top of each other because they are
  reached for a face-on orientation, where the higher modes
  contribution is zero. The unfaithfulness averaged over the three
  angles
  $\iota_{\textrm{NR}},{\varphi_0}_{\textrm{NR}},\kappa_{\textrm{NR}}$ are
  obtained assuming an isotropic distribution for the source
  orientation, homogeneous distribution in GW polarization and isotropic distribution in sky position (green curves and orange curves for the average weighted
  with the SNR). The minimum of the unfaithfulness (red curves) in
  practice correspond to an edge-on orientation, minimized over the
  other two angles. The vertical dotted-dashed black line is the
  smallest mass at which the $(2,1)$ mode is entirely in
  the Advanced LIGO band. The $(\ell, m')$ mode is entirely in the Advanced LIGO band
  starting from a mass $m'$ times the mass associated with the $(2,1)$ mode.  
 The horizontal dotted-dashed black lines
  represent the values of $1\%$ and $3\%$ unfaithfulness.}
\label{fig:unfaith_mass_q3chi085}
\end{figure}

As done in Sec. \ref{sec:faithfulness} we summarize the results of the
faithfulness calculation in Fig.~\ref{fig:unfaith_mass_q3chi085}, where 
we show the minimum and maximum of the unfaithfulness over the
NR orientations, GW polarization and sky position, respectively indicated as $\min_{\iota_{\mathrm{NR}},{\varphi_0}_{\mathrm{NR}},\kappa_{\mathrm{NR}}}
(1 -\mathcal{F})$ (blue) and
$\max_{\iota_{\mathrm{NR}},{\varphi_0}_{\mathrm{NR}},\kappa_{\mathrm{NR}}}
(1 -\mathcal{F})$ (red); the average of the unfaithfulness
over these three angles $
\langle1-\mathcal{F}\rangle_{\iota_{\mathrm{NR}},{\varphi_0}_{\mathrm{NR}},\kappa_{\mathrm{NR}}}$
(green), and the average of the unfaithfulness weighted with the
cube of the SNR: $\langle1-
\mathcal{F}\rangle_{\iota_{\mathrm{NR}},{\varphi_0}_{\mathrm{NR}},\kappa_{\mathrm{NR}}}^{\mathrm{SNRweighted}}$
(orange). All the averages are computed assuming an isotropic distribution 
for the source orientation, homogeneous distribution in GW polarization and isotropic distribution in sky position.  
All these quantities are shown as a function of the total mass of the system. In the plots the plain curves are the
results of the unfaithfulness between the NR and \texttt{SEOBNRv4HM} waveforms, 
while dashed curves are the results of the unfaithfulness between NR
and \texttt{SEOBNRv4} waveforms. In this case, the maximum and the averaged
values of the unfaithfulness for the \texttt{SEOBNRv4} model are one order of
magnitude larger than the ones with the \texttt{SEOBNRv4HM} model. The minimum of
the unfaithfulness is the same for both models (blue curves lying on
top of each other) because it is reached for a face-on orientation, where
the contribution of the higher-order modes used for \texttt{SEOBNRv4HM} is zero. Indeed the -2 spin-weighted 
spherical harmonics associated to these higher-order modes 
go to zero for face-on orientations. We note also that in \texttt{SEOBNRv4}, as expected, the disagreement grows strongly with the
total mass of the system, because higher-order modes are more 
important toward merger and ringdown.

\subsection{High mass ratios: \texttt{SXS:BBH:0065} and \texttt{ET:AEI:0004}}
\label{comparison:NR2}

\begin{figure}
\centering
\includegraphics[width=0.7\textwidth]{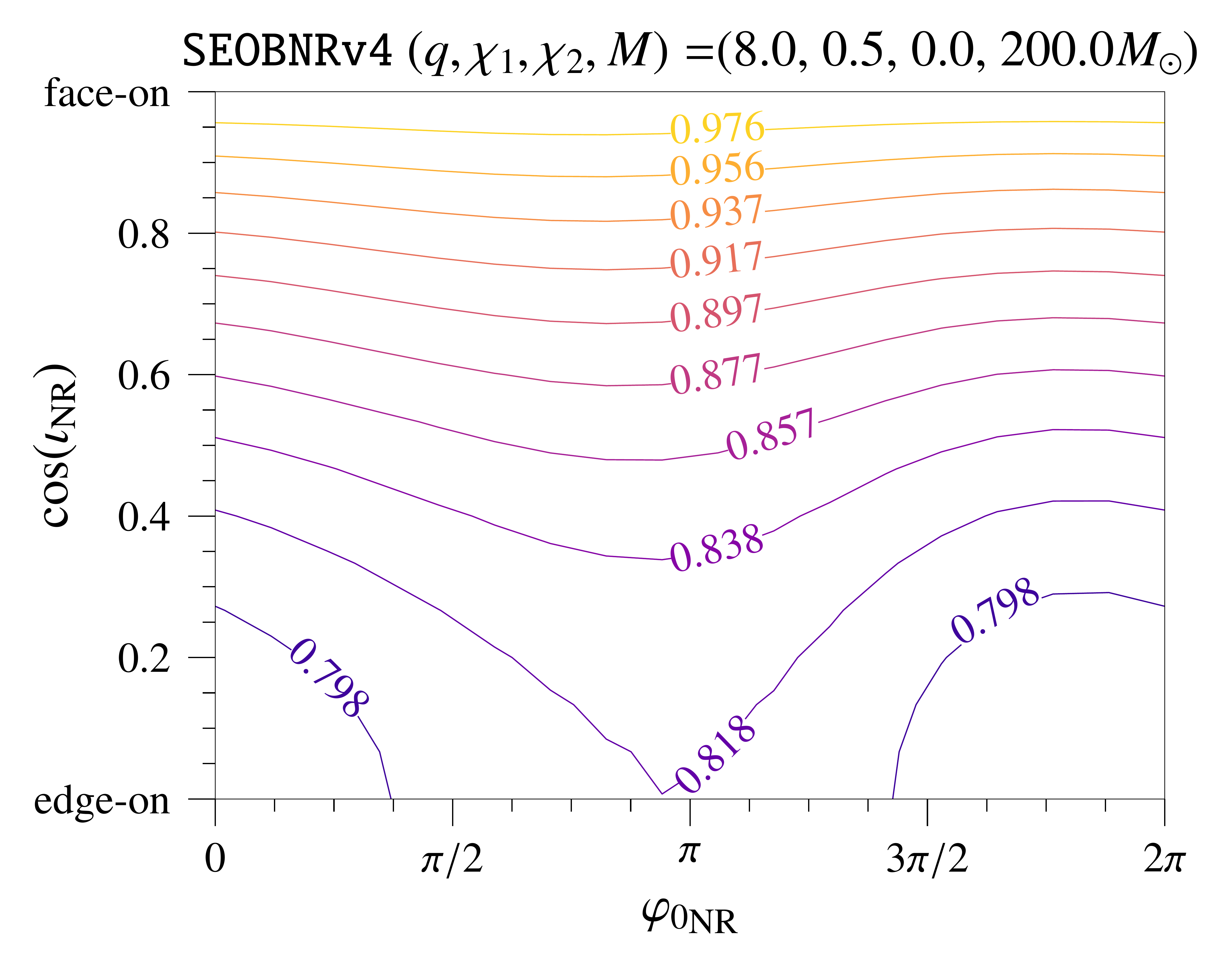} 
\\
\includegraphics[width=0.7\textwidth]{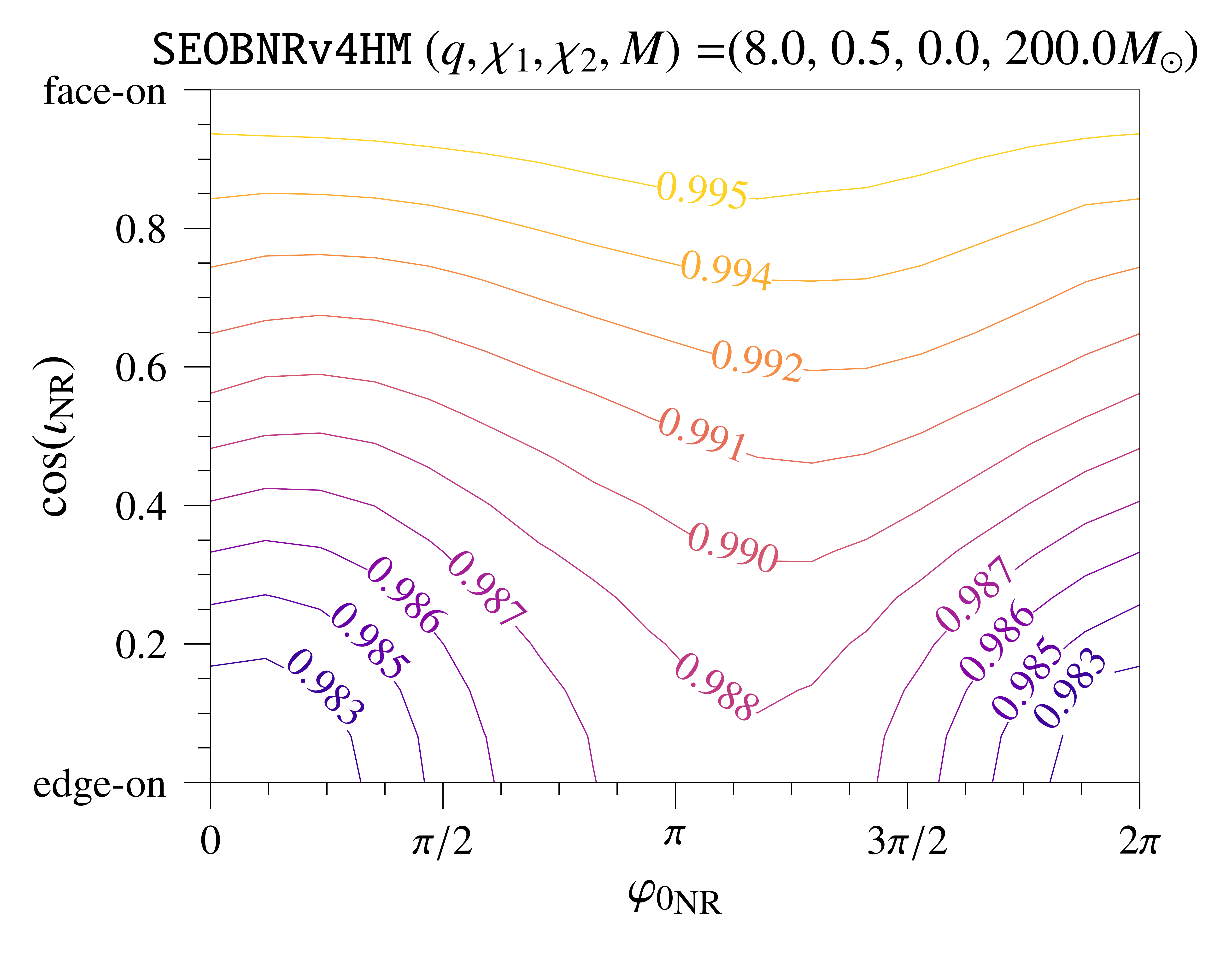} %
\caption{Faithfulness $\mathcal{F}(\cos(\iota_{\textrm{NR}}),{\varphi_0}_{\textrm{NR}},\kappa_{\textrm{NR}} = 0)$ for the configuration $(q = 8,\, M = 200 M_\odot,\, \chi_1 = 0.5, \, \chi_2 = 0)$: NR $(\ell \leq 5,\, m \neq 0)$ vs \texttt{SEOBNRv4} (left panel), NR $(\ell \leq 5, \, m \neq 0)$ vs \texttt{SEOBNRv4HM} (right panel).}
\label{fig:q8v4vsHMthetaphi}
\end{figure}

\begin{figure}
    \centering
      \includegraphics[width=0.7\textwidth]{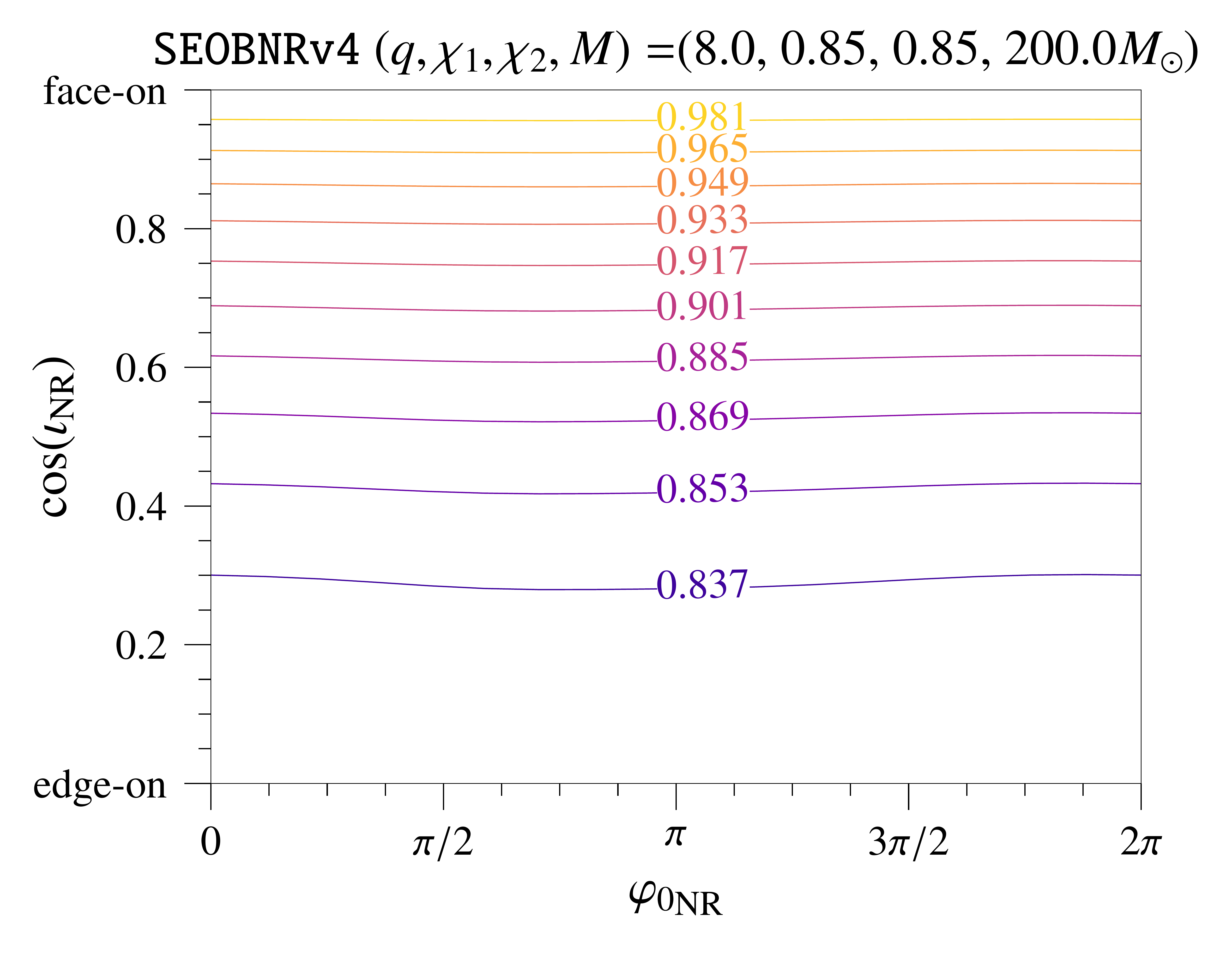} 
        \includegraphics[width=0.7\textwidth]{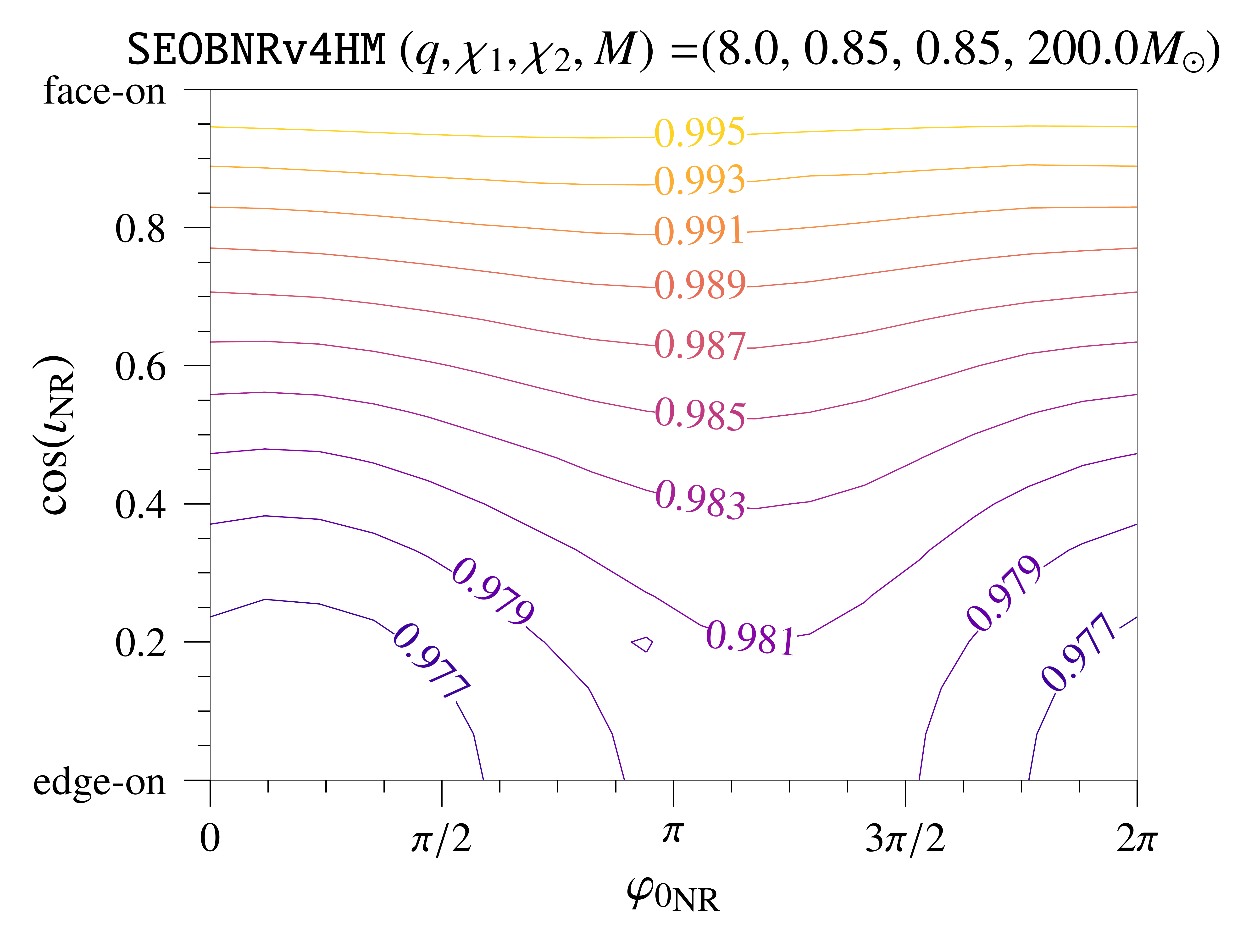} 
\caption{Faithfulness $\mathcal{F}(\cos(\iota_{\textrm{NR}}),{\varphi_0}_{\textrm{NR}},\kappa_{\textrm{NR}} = 0)$ for the configuration $(q = 8,\, M = 200 M_\odot,\, \chi_1 = 0.85, \, \chi_2 = 0.85)$: NR $(\ell \leq 5,\, m \neq 0)$ vs \texttt{SEOBNRv4} (left panel), NR $(\ell \leq 5,\, m \neq 0)$ vs \texttt{SEOBNRv4HM} (right panel).}
\label{fig:q8sv4vsHMthetaphi}
\end{figure}

More striking conclusions about the improvement of the waveform model due to the inclusion of higher-order modes can be drawn looking at the comparison with the two NR simulations \texttt{SXS:BBH:0065} and \texttt{ET:AEI:0004}, 
for which higher-order modes are expected to be more important,  
because of the higher mass ratio. For the first
configuration $(q = 8,\, M = 200 M_\odot,\, \chi_1 = 0.5, \, \chi_2 =
0)$ we see in Fig.~\ref{fig:q8v4vsHMthetaphi} that the
faithfulness between the NR $(\ell \leq 5, \, m\neq 0)$ and the
\texttt{SEOBNRv4} waveforms (left panel) degrades much faster than
before as a function of the inclination angle $\iota_{\textrm{NR}}$,
reaching $\mathcal{F} \lesssim 90\%$ already for values of
$\cos(\iota_{\textrm{NR}}) \sim 0.7$ ($\iota_{\textrm{NR}} \sim 45
\degree)$, being very large for the edge-on inclination
$\mathcal{F} \sim 80\%$. Similarly to what happens for the example discussed 
in Sec.~\ref{comparison:NR1}, the situation is much better if we include in the 
model the higher modes, as can be seen in Fig.~\ref{fig:q8v4vsHMthetaphi} (right panel). 
Now, the degradation as a function of $\iota_{\textrm{NR}}$ is much weaker
and for edge-on orientations the faithfulness reaches values close to
$\mathcal{F} \sim 98 \%$. Similar conclusions can be drawn by looking at Fig.~\ref{fig:q8sv4vsHMthetaphi}, whch refers 
to the simulation \texttt{ET:AEI:0004} $(q = 8,\, M = 200 M_\odot,\, \chi_1 = 0.85, \,
\chi_2 = 0.85)$. The only relevant difference with respect to the
aforementioned case is that in this case the faithfulness of
the \texttt{SEOBNRv4HM} waveform is a little bit smaller and it goes down to
$\mathcal{F} \sim 97.7\%$ in the edge-on orientations. At a fixed binary orientation, the faithfulness of the (2,2)--waveform-model \texttt{SEOBNRv4} against the NR waveform for the configuration $(q = 8,\, M = 200 M_\odot,\, \chi_1 = 0.85 = \chi_2 =
0.85)$ is always larger than that for the configuration  $(q = 8,\, M = 200 M_\odot,\, \chi_1 = 0.5, \, \chi_2 =
0)$. This can be explained considering that, as discussed in Sec.~\ref{sec:motivations}, for a fixed mass ratio the $(2,1)$ mode is increasingly suppressed when the spin of the heavier BH grows, while the other higher-order modes are mostly constant as a function of the spins. Since in the first case $\chi_1$, that is the spin of the heavier BH, is larger than in the second case, the $(2,1)$ mode is more suppressed in the first case than in the second one. For this reason the faithfulness with the \texttt{SEOBNRv4} model, including only the dominant mode, is higher for the first configuration.

\begin{figure}
    \centering
        \includegraphics[width=0.7\textwidth]{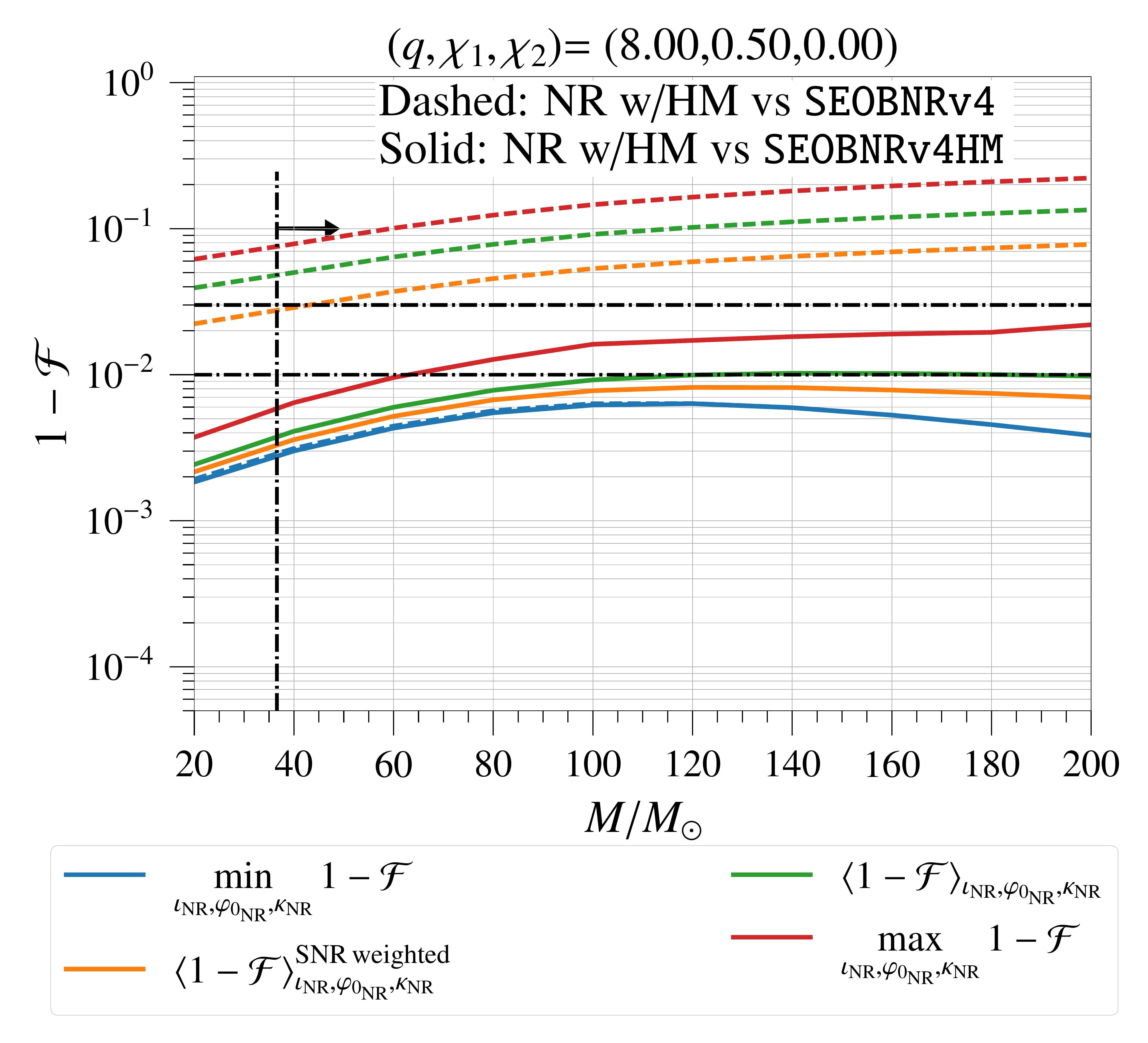} 
        \includegraphics[width=0.7\textwidth]{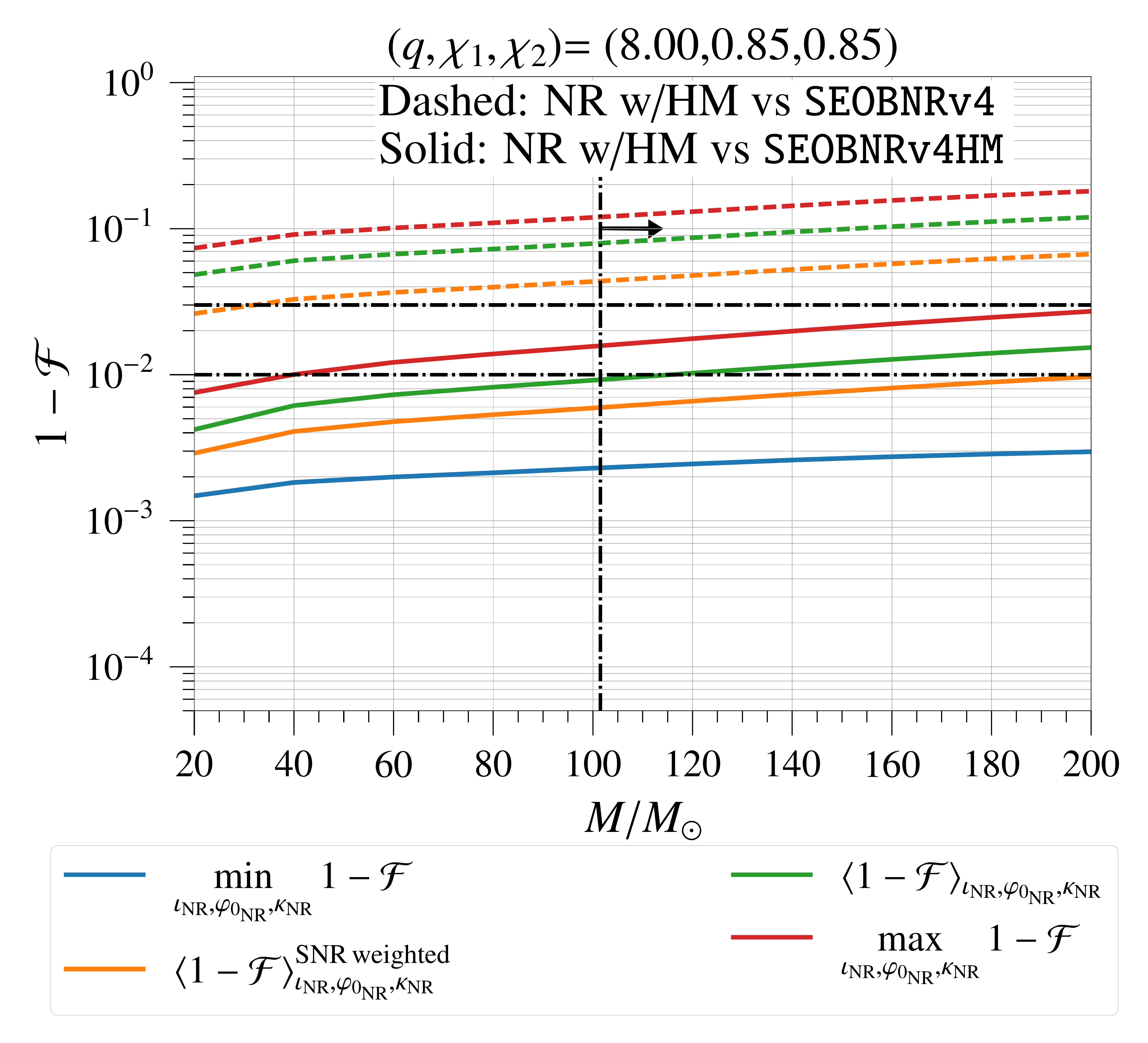} 
\caption{Unfaithfulness $(1-\mathcal{F})$ in the mass range $20 M_\odot \leq M \leq 200 M_\odot$ for the configuration $(q = 8,\, \chi_1 = 0.5,\, \chi_2 = 0)$ (left panel) and $(q = 8,\, \chi_1 = 0.85,\, \chi_2 = 0.85)$ (right panel). Plotted data as in Fig.~\ref{fig:unfaith_mass_q3chi085}}
\label{unfaith_mass_q8}
\end{figure}

As for the previous configuration, in Fig.~\ref{unfaith_mass_q8}, we
show the summary of the faithfulness results as maximum, minimum and
averages of the unfaithfulness, respectively for \texttt{SXS:BBH:0065}
(left panel) and \texttt{ET:AEI:0004} (right panel). For these binary
configurations, even if the maxima of the unfaithfulness have larger
values with respect to the case discussed in the previous section (
$\sim 2\%$ for \texttt{SXS:BBH:0065} and $\sim 2.7\%$ for
\texttt{ET:AEI:0004} at a total mass of $M = 200M_\odot$), we still
have acceptable values of the unfaithfulness averaged over the
orientations, sky position and polarizations: respectively $\sim 1\%$
and $\sim1.6\%$ for a total mass of $M = 200M_\odot$. This is a big
improvement with respect to the \texttt{SEOBNRv4} model, which gives
averaged values of the unfaithfulness larger than $10\%$ for both
configurations and the same total mass.  For the configuration
  with $q = 8,\, \chi_1 = 0.85 = \chi_2 = 0.85$, the unfaithfulness
  against the NR simulation was also computed for the multipolar
  waveform model developed in Ref.~\cite{London:2017bcn}, and found to 
  be around $\sim 5\%$ for $\iota_\mathrm{s} =
  \pi/2$, when averaging over the angles $\kappa_\mathrm{s}$ and
  ${\varphi_0}_s$ for a total mass $M = 100 M_\odot$. In our model the
  maximum of the unfaithfulness
  (i.e., $\max_{\iota_{\mathrm{s}},{\varphi_0}_{\mathrm{s}},\kappa_{\mathrm{s}}}(1
  -\mathcal{F})$) over the angles
  $\iota_{\mathrm{s}},{\varphi_0}_{\mathrm{s}}$ and
  $\kappa_{\mathrm{s}}$ is around $1.5\%$ at $M = 100 M_\odot$. The
  reason for the better accuracy of \texttt{SEOBNRv4HM} model with respect
  to the waveform model in Ref.~\cite{London:2017bcn} for this ``extreme''
  binary configuration might be due to the fact that the simple scaling argument used
  there to build the higher-order modes is not very accurate for
  high-mass ratio and high-spin binary systems. We leave to the future a 
direct, comprehensive comparison between the two waveform models.

As discussed in Sec.~\ref{sec:faithfulness}, an important quantity to assess the 
improvement that \texttt{SEOBNRv4HM} could yield for detecting BBHs 
is the average unfaithfulness weighted with the cube of the
SNR. For this quantity our model yields values of $\sim 0.7\%$ for
\texttt{SXS:BBH:0065} and $\sim 1\%$ for \texttt{ET:AEI:0004} at a
total mass of $M = 200M_\odot$ compared to values around $\sim 7\%$
returned by the \texttt{SEOBNRv4} model.

\subsection{Comparison with entire numerical-relativity catalog}
\label{comparison:NRcatalog}

Having studied in detail some particular configurations, we can now examine how the model works over the entire NR waveform catalog at our 
disposal. In Fig.~\ref{fig:skyaverageall} we plot the angle-averaged 
unfaithfulness as a function of the total mass of the system,
computed between the NR waveforms with modes $(\ell
\leq 5, m \neq 0)$ and the \texttt{SEOBNRv4} model (left panel), 
\texttt{SEOBNRv4HM} model (right panel). Comparing the two panels, we can see that \texttt{SEOBNRv4HM} 
yields unfaithfulnesses one order of magnitude
smaller than those of the \texttt{SEOBNRv4} model.  In the plots
different colors correspond to different ranges of mass ratios, and
from the left panel it is visible that in the case of the 
\texttt{SEOBNRv4} model, there is a clear hierarchy for which configurations
with higher mass ratios have also larger unfaithfulness. This effect
is removed in the \texttt{SEOBNRv4HM} model, as visible in the right panel of the
same figure. In general for all of NR simulations the averaged
unfaithfulness against \texttt{SEOBNRv4HM} is always smaller than $1 \%$ in the
mass range $20 M_\odot \leq M \leq 200 M_\odot$ with the exception of
few simulations for which the unfaithfulness reaches values $\leq 1.5\%$ for a total mass of $M = 200 M_\odot$: \texttt{SXS:BBH:0202} $(q = 7,\, \chi_1 = 0.6,\,
\chi_2 = 0)$, \texttt{ET:AEI:0004} $(q = 8,\, \chi_1 = 0.85,\,
\chi_2 = 0.85)$, \texttt{ET:AEI:0001} $(q = 5,\, \chi_1 = 0.8,\,
\chi_2 = 0)$ and \texttt{SXS:BBH:0061} $(q = 5,\, \chi_1 = 0.5,\,
\chi_2 = 0)$. These are the configurations in the NR catalog having the most extreme values of mass ratio and spins.
The results of this analysis does not change considerably if
we include in the NR waveforms only the modes used in the \texttt{SEOBNRv4HM} 
model, because, when looking at averaged unfaithfulness, the error is dominated by the imperfect modeling of the 
$(2,1), (3,3), (4,4),(5, 5)$ modes, and not by neglecting other 
subdominant higher modes, as discussed in Sec.~\ref{sec:faithfulness}.

\begin{figure}
\centering
\includegraphics[width=0.7\textwidth]{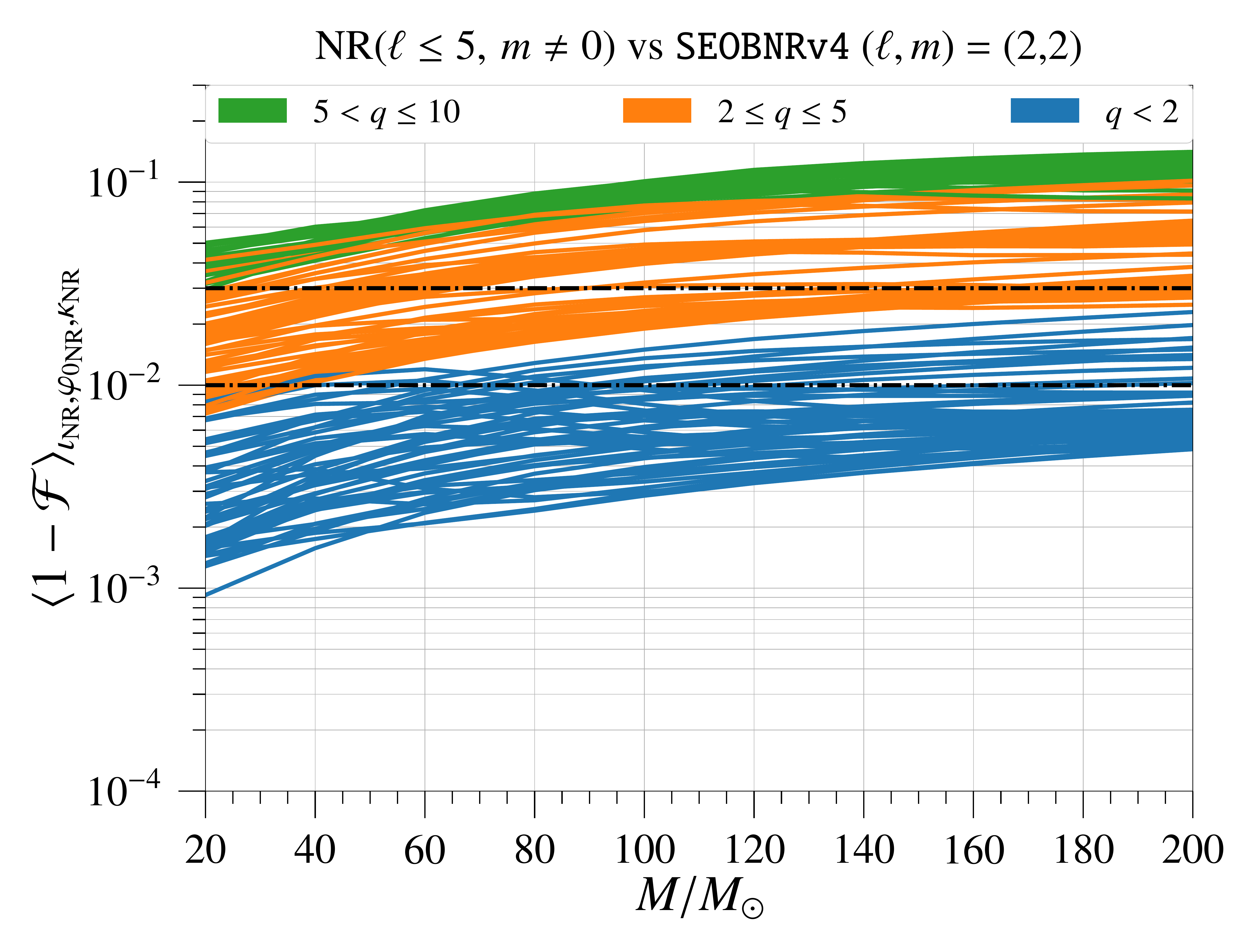} 
\\
\centering
\includegraphics[trim=-2cm 0 2cm 0,clip,width=0.8\textwidth]{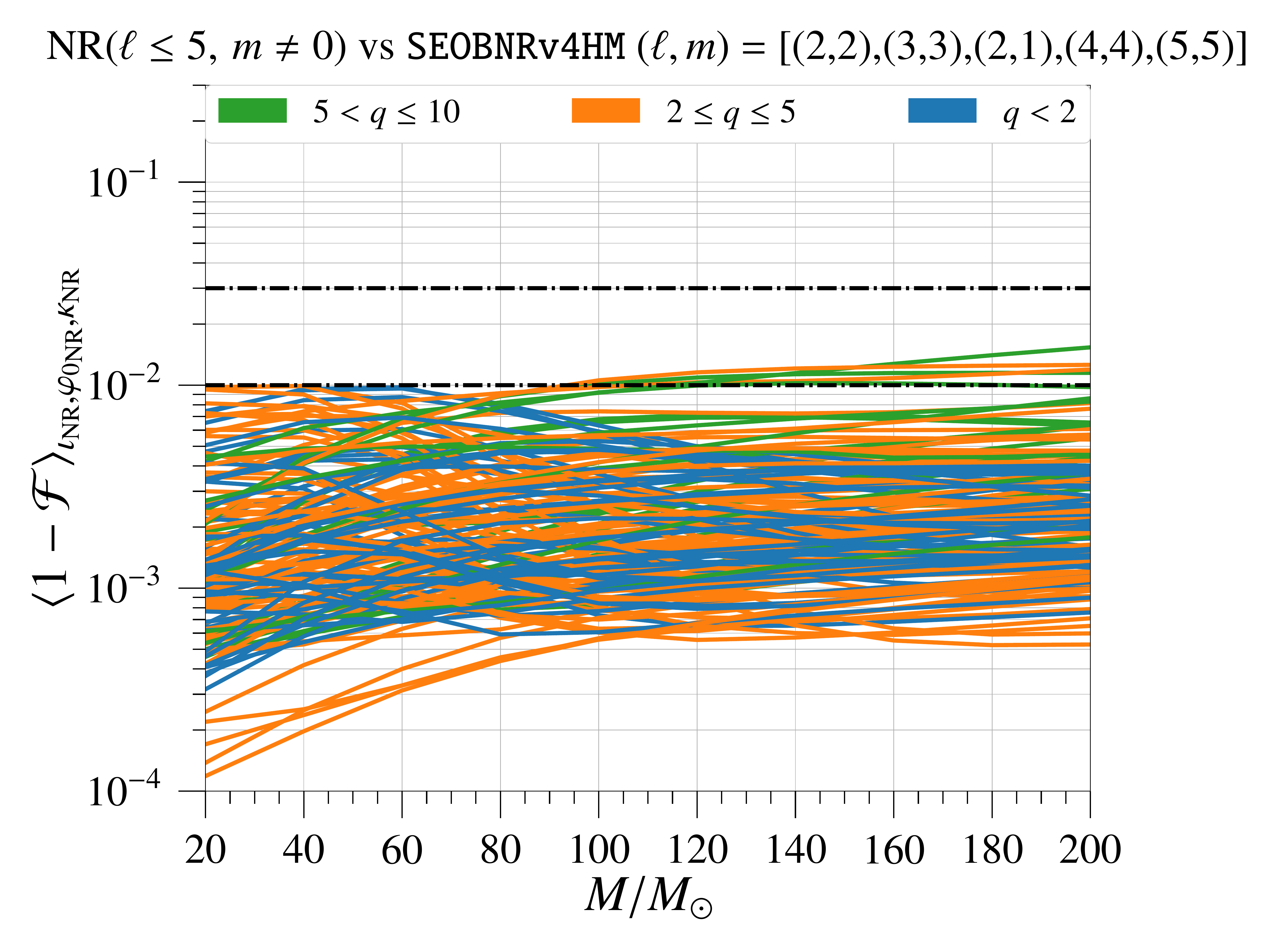} 
\caption{Unfaithfulness $(1-\mathcal{F})$ averaged over the three angles $(\iota_{\textrm{NR}},{\varphi_0}_{\textrm{NR}},\kappa_{\textrm{NR}})$ as a function of the total mass, in the range $20 M_\odot \leq M \leq 200 M_\odot$. Left panel NR $(\ell \leq 5, \, m \neq 0)$ vs \texttt{SEOBNRv4}, right panel NR $(\ell \leq 5,\, m \neq 0)$ vs \texttt{SEOBNRv4HM}. The horizontal dotted-dashed black lines represent the values of $1\%$ and $3\%$ unfaithfulness.}
\label{fig:skyaverageall}
\end{figure}

The comparison between the unfaithfulness averaged
over the three angles $(\iota_{\textrm{NR}},{\varphi_0}_{\textrm{NR}},\kappa_{\textrm{NR}})$ and
weighted by the cube of the SNR of two waveform models against NR waveforms displays similar features, with the only difference of having overall smaller values of the unfaithfulness 
(always $\leq 1\%$ for the \texttt{SEOBNRv4HM} model). This happens because weighting with the SNR 
favours orientations closer to face-on for which the best
modeled $(2,2)$ mode is dominant.

\begin{figure}
    \centering
        \includegraphics[width=0.7\textwidth]{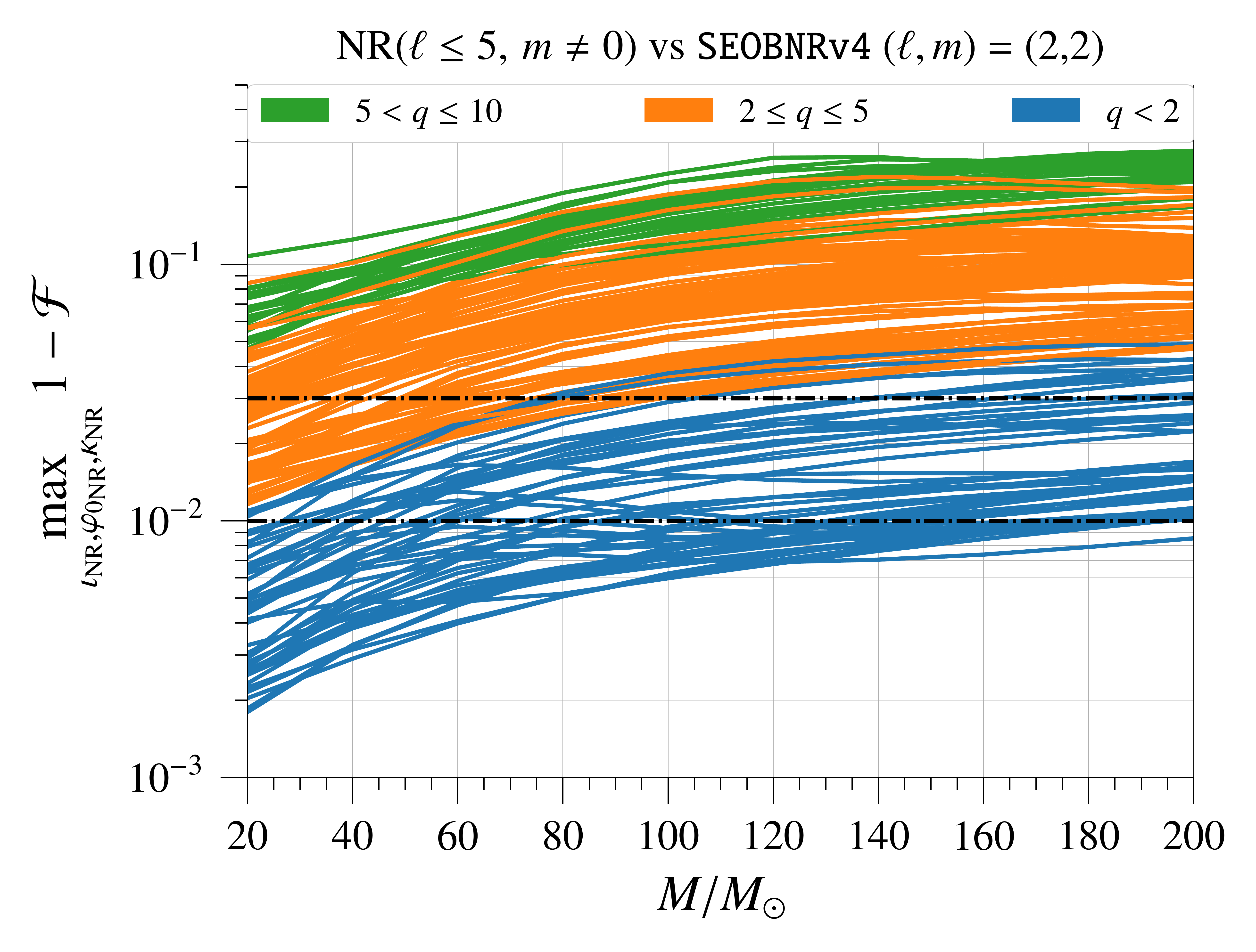} 
        \includegraphics[trim=-3cm 0 2cm 0,clip,width=0.8\textwidth]{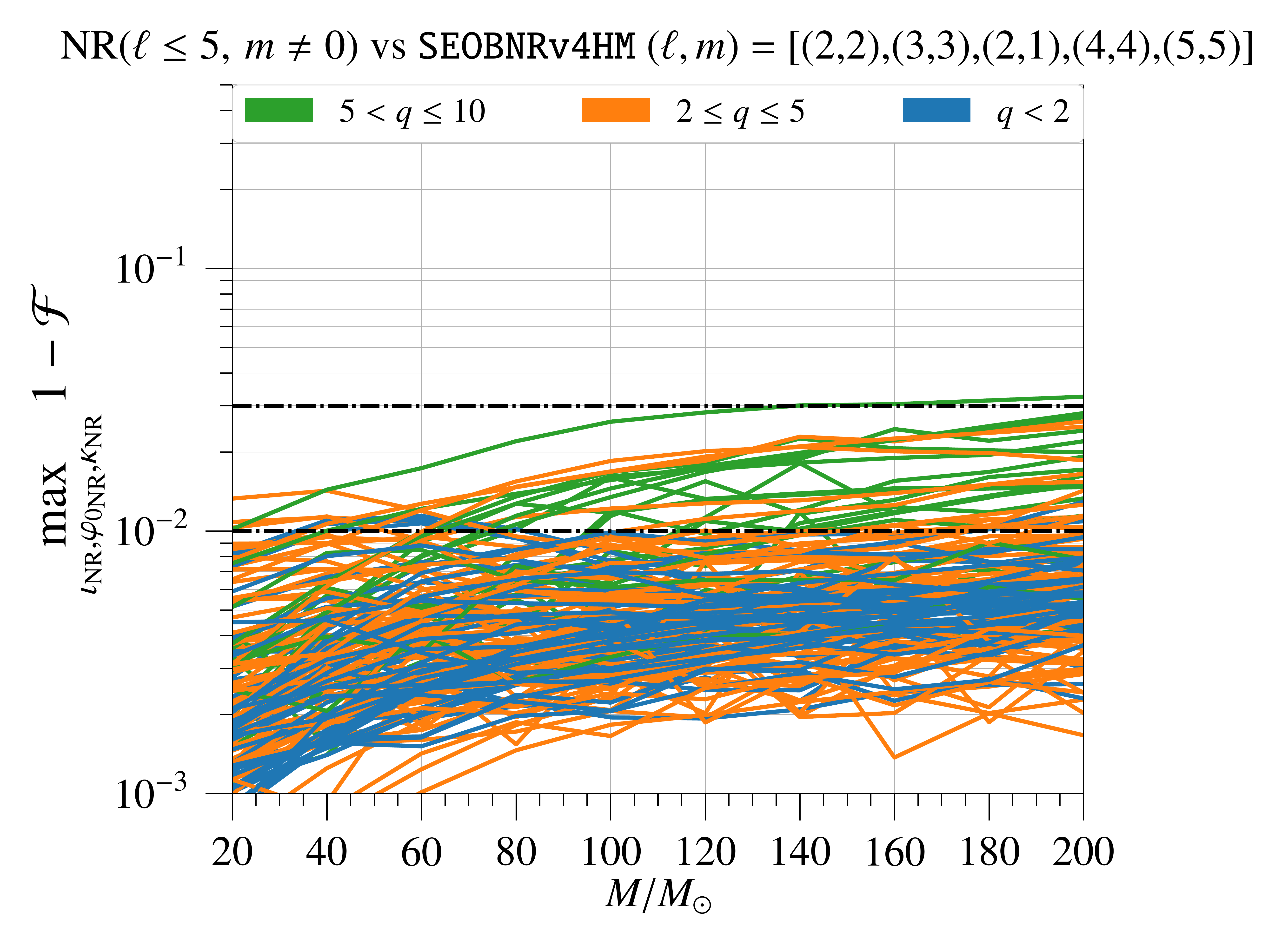} 
\caption{Maximum of unfaithfulness $(1-\mathcal{F})$ over the three angles $(\iota_{\textrm{NR}},{\varphi_0}_{\textrm{NR}},\kappa_{\textrm{NR}})$ as a function of the total mass, in the range $20 M_\odot \leq M \leq 200 M_\odot$. Left panel NR $(\ell \leq 5, \, m \neq 0)$ vs \texttt{SEOBNRv4}, right panel NR $(\ell \leq 5,\, m \neq 0)$ vs \texttt{SEOBNRv4HM}. The horizontal dotted-dashed black lines represent the values of $1\%$ and $3\%$ unfaithfulness. The jaggedness of the curves in the plot (right panel) is caused by the numerical noise present in the NR 
higher-order modes, which are not very well resolved. We find that this feature is not present when these noisy modes are removed from the calculation of the faithfulness.}
\label{fig:skyworstall}
\end{figure}

Finally, in the right panel of Fig.~\ref{fig:skyworstall} we show the maximum of the
unfaithfulness over the three angles $(\iota_{\textrm{NR}},{\varphi_0}_{\textrm{NR}},\kappa_{\textrm{NR}})$  
between the \texttt{SEOBNRv4HM} model and the NR waveforms with 
the modes $(\ell \leq 5, \, m\neq 0)$. In the left panel of the same figure we show
the same comparison but this time using the \texttt{SEOBNRv4} model. 
Here we see that the \texttt{SEOBNRv4HM} waveforms have unfaithfulness 
smaller than $3\%$ in the mass range considered for all the NR simulations 
with the exception of one case, namely \texttt{SXS:BBH:0621} $(q = 7,\, \chi_1 = -0.8,\, \chi_2 = 0)$ for which the unfaithfulness at $M = 200 M_\odot$ is $(1-\mathcal{F}) \sim 3.1\%$. 

In general, over the NR simulations of our catalog, the maximum of the
unfaithfulness is always smaller than $1\%$ in the total mass range
$20 M_\odot \leq M \leq 200 M_\odot$ for nonspinning configurations up
to mass ratio $q = 8$. Nonspinning cases with $q \geq 8$ and
configurations with high spins and mass ratios $q \geq 5$ have maximum
unfaithfulness in the range $1\% \leq (1-\mathcal{F}) \leq 3\%$. For
the former the unfaithfulness decreases to values smaller than $1\%$
when the comparison is done including only the modes $(2,2),(2,1),(3,3),(4,4),(5,5)$ 
in the NR waveforms (i.e., excluding smaller higher-order modes like $(3,2),(4,3)$). This is
not true for high-spin, high--mass-ratio configurations where the
unfaithfulness due to a nonperfect modeling dominates over that due
to neglecting smaller higher-order modes.  It is important to stress
that, as discussed in Sec.~\ref{sec:faithfulness}, the maximum
unfaithfulness due to the numerical error in the NR waveforms of our
catalog is in the range $[0.1 \%, 1\%]$. This means that 
when comparing the NR waveforms with the
\texttt{SEOBNRv4HM} model a fraction of
the maximum unfaithfulness as large as $1\%$ could be due to numerical error. Given that
maximum unfaithfulness are reached for edge-on configurations where
the higher-order modes are more relevant, NR waveforms with better
resolved higher-order modes would be needed in order to attempt to
build a model with maximum unfaithfulness smaller than $1\%$.

\section{Conclusions}
\label{sec:concl}

We have worked within the spinning EOB framework and have built a multipolar waveform
model for BBHs with nonprecessing spins that includes the higher-order modes
$(\ell, m) = (2,1),(3,3),(4,4),(5,5)$, besides the dominant $(2,2)$
mode. In order to improve the agreement with the NR results we
included recently computed PN corrections~\cite{Marsatetal2017,Faye:2014fra,Fujita:2012cm} 
in the resummed GW modes, and also used nonperturbative informations from
NR waveforms in the NQCs corrections of the higher-order modes, and in the 
calibration parameters $c_{\ell m}$'s (the latter only for the modes $(2,1),(5,5)$).  
We also extended to higher-order modes the phenomenological ansatz for 
the merger-ringdown signal that was originally proposed in 
Refs.~\cite{Baker:2008mj,Damour:2014yha,Nagar:2016iwa,Bohe:2016gbl} for the 
dominant $(2,2)$ mode.
 
We have found that the unfaithfulness averaged over orientations, polarizations and
sky positions between the \texttt{SEOBNR4HM} model and NR waveforms of the 
catalog at our disposal, is always smaller than $1\%$ with the exception of four configurations
for which the unfaithfulness is smaller than $1.5\%$. Moreover, the unfaithfulness are 
one order of magnitude smaller than those obtained with the \texttt{SEOBNRv4} model~\cite{Bohe:2016gbl}, 
which only contains the $(2,2)$ mode. The maximum unfaithfulness over orientations, polarizations and sky 
positions between \texttt{SEOBNR4HM} and NR waveforms is always smaller than $3\%$ with the exception of one configuration
for which the faithfulness is smaller than $3.1\%$. Also for the maximum unfaithfulness the results are 
one order of magnitude smaller than those obtained with the \texttt{SEOBNRv4} model~\cite{Bohe:2016gbl}. 
We have also found that, in the nonspinning limit, the \texttt{SEOBNRv4HM} model returns values of the unfaithfulness 
smaller than its (nonspinning) predecessor waveform model, that is \texttt{EOBNRv2HM}~\cite{Pan:2011gk} 
(see Appendix~\ref{sec:EOBNRv2HM}).

Other studies are needed to fully assess the accuracy of \texttt{SEOBNRv4HM} for GW astronomy. In
particular it will be important to understand if unfaithfulnesses below $1\%$ can affect the recovery of 
binary parameters, and if so which parameters will be mainly biased, for which SNR and in which region 
of the parameter space. In particular, we expect that the multipolar \texttt{SEOBNRv4HM} model will be more precise 
than the \texttt{SEOBNRv4} model for recovering the binary's inclination angle and the distance from the source. 
Indeed, those parameters are degenerate with each other when only the $(2,2)$ mode is present, and the inclusion 
of higher-order modes can help in disentagle them (e.g., see Ref.~\cite{OShaughnessy:2014shr}).  We postpone this kind of studies to the future because for
computational reasons, we would need to develop a reduced-order-model (ROM)~\cite{Purrer:2015tud} version of the 
\texttt{SEOBNRv4HM} model. Another important test for the future would be the comparison between 
\texttt{SEOBNRv4HM} model and other multipolar, inspiral-merger-ringdown in the literature, such as the \texttt{IMRPhenom} 
models proposed in Refs.~\cite{Mehta:2017jpq, London:2017bcn}. It will be relevant to compare those models especially 
outside the range of binary configurations where the NR waveforms are available, in order to identify if there are 
regions where the two models predict significantly different waveforms.

We also expect that the multipolar spinning, nonprecessing waveform model developed here 
will be a more accurate model to carry out parameterized tests of General Relativity~\cite{TheLIGOScientific:2016src} 
when BBHs with high mass-ratio, high total mass and in a non face-on orientation will be detected. Furthermore, the 
\texttt{SEOBNRv4HM} model can be employed to search for more than one gravitational quasi-normal mode in the ringdown 
portion of the signal, coherently with multiple detections~\cite{Dreyer:2003bv,Berti:2005ys,Meidam:2014jpa,Yang:2017zxs}. 
In fact, those studies can also be performed with our multipolar, stand-alone merger-ringdown model.

The \texttt{SEOBNRv4HM} waveform model employs the same conservative and dissipative dynamics of the
\texttt{SEOBNRv4} model, which was calibrated to NR simulations by requiring very good 
agreement with the NR $(2,2)$ GW mode. Further improvements of the 
\texttt{SEOBNRv4} waveform model could be achieved in the future by recalibrating the two-body dynamics.  
Such calibration would require the production of a new set of NR waveforms (with more accurate higher-order modes) 
in the region of high mass-ratios, say $q\geq 4$, and high spins, say $\chi_{1,2} \geq 0.6$ where few NR
simulations are currently available and where the disagreement between current analytical inspiral-merger-ringdown 
waveforms is the worst (e.g., see Figs.~ 5 and 6 in Ref.~\cite{Bohe:2016gbl}). Those NR waveforms would need to be sufficiently 
long to make the calibration procedure sufficiently robust (see Sec.VI, and Fig.~7 and 8 in Ref.~\cite{Bohe:2016gbl}).

In the near future our priority is to include the next largest modes in the \texttt{SEOBNRHM} model, notably the 
$(3,2),(4,3)$ modes. The work would need to take into account the mixing between spherical-harmonic and spheroidal harmonics during the merger-ringdown 
stage, as observed in Refs.~\cite{Buonanno:2006ui,Kelly:2012nd}, and investigated more recently in Refs.~\cite{Berti:2014fga,London:2014cma}. 
Insights might need to be gained also from merger-ringdown waveforms in the test-particle limit~\cite{Taracchini:2013wfa,Harms:2015ixa,Harms:2016ctx}. However, to develop a more accurate multipolar model, one would also need to reduce the numerical error in NR waveforms 
around merger and during ringdown, in particular for the modes (4,4) and (5,5). 
Another important and timely application of this work, is its extension to the spinning, precessing case, 
thus improving, the current \texttt{SEOBNRv3} model~\cite{Pan:2013rra,Babak:2016tgq,Abbott:2016izl}, which only contains the $(2,2)$ and $(2,1)$ modes.

\FloatBarrier
\section*{Acknowledgments}

It is a pleasure to thank Juan Calderon Bustillo, Ian Harry, Sylvain Marsat, Harald Pfeiffer, and Noah Sennett for helpful discussions.
Computational work for this manuscript was carried out on the computer clusters {\texttt{Vulcan}} 
and {\texttt{Minerva}} at the Max Planck Institute for Gravitational Physics in Potsdam.

\chapter{Multipolar Effective-One-Body Waveforms for Precessing Binary Black Holes: Construction and Validation}
\chaptermark{}
\label{chap:three}

\def\be{\begin{equation}}
\def\ee{\end{equation}}
\def\bea{\begin{eqnarray}}
\def\eea{\end{eqnarray}}
\newcommand{\bes}{\begin{subequations}}
\newcommand{\ees}{\end{subequations}}

\newcommand{\vS}{\mbox{\boldmath${S}$}}
\newcommand{\vR}{\mbox{\boldmath${r}$}}
\newcommand{\vP}{\mbox{\boldmath${p}$}}
\newcommand{\vL}{\mbox{\boldmath${L}$}}
\newcommand{\vN}{\mbox{\boldmath${N}$}}
\newcommand{\vLhat}{\mbox{\boldmath${\hat{L}}$}}
\newcommand{\vJhat}{\mbox{\boldmath${\hat{J}}$}}
\newcommand{\vJ}{\mbox{\boldmath${J}$}}
\newcommand{\vE}{\mbox{\boldmath${\hat{e}}$}}
\newcommand{\vZ}{\mbox{\boldmath${\hat{Z}}$}}
\newcommand{\vchi}{\mbox{\boldmath$\chi$}}

\newcommand{\aconfigs}{1523} 
\newcommand{\EOBbthree}{$94\%$}
\newcommand{\EOBbone}{$57\%$}
\newcommand{\Phenombthree}{$83\%$}
\newcommand{\Phenombone}{$20\%$}

\def\vct#1{{\bm{#1}}}

\newcommand{\doubleline}{\hline \hline}

\allowdisplaybreaks

\hspace{\parindent}\textbf{Authors}\footnote{Originally published as Phys.Rev.D 102 (2020) 4, 044055.}: Serguei Ossokine, Alessandra Buonanno, Sylvain Marsat, Rober\-to Cotesta, Stanislav Babak, Tim Dietrich, Roland Haas, Ian Hinder, Harald P. Pfeiffer, Michael P\"urrer, Charles J. Woodford, Michael Boyle, Lawrence E. Kidder, Mark  A. Scheel and B\'ela Szil\'agyi\\

\textbf{Abstract}: As gravitational-wave detectors become more sensitive and broaden their frequency bandwidth, 
we will access a greater variety of signals emitted by compact binary systems, shedding light on their 
astrophysical origin and environment. A key physical effect that can distinguish among different 
formation scenarios is the misalignment of the spins with the orbital angular momentum, causing the spins 
and the binary's orbital plane to precess. To accurately model such precessing signals, especially 
when masses and spins vary in the wide astrophysical range, it is crucial to include multipoles 
beyond the dominant quadrupole. Here, we develop the first \textit{multipolar} precessing waveform model 
in the effective-one-body (EOB) formalism for the entire coalescence stage (i.e., inspiral, 
merger and ringdown) of binary black holes: \verb+SEOBNRv4PHM+. In the nonprecessing limit, the model reduces to 
\verb+SEOBNRv4HM+, which was calibrated to numerical-relativity (NR) simulations, and 
waveforms from black-hole perturbation theory. 
We validate \verb+SEOBNRv4PHM+ by comparing it to the public catalog of 1405 precessing NR waveforms 
of the Simulating eXtreme Spacetimes (SXS) collaboration, and also to 118 SXS precessing NR waveforms, 
produced as part of this project, which span mass ratios 1-4 and (dimensionless) black-hole's spins 
up to 0.9. We stress that \verb+SEOBNRv4PHM+ is not calibrated to NR simulations in the 
precessing sector. We compute the unfaithfulness against 
the \aconfigs \ SXS precessing NR waveforms, and find that, for \EOBbthree\ (\EOBbone\ ) of the cases, the maximum value, in the 
total mass range $20\mbox{--}200 M_\odot$, is below $3\%$ ($1\%$). Those numbers change to \Phenombthree\ (\Phenombone\ ) 
when using the inspiral-merger-ringdown, multipolar, precessing phenomenological model \verb+IMRPhenomPv3HM+. 
We investigate the impact of such unfaithfulness values with two Bayesian, parameter-estimation studies on synthetic signals.
We also compute  the unfaithfulness between 
those waveform models as a function of the mass and spin parameters to identify in which part of the parameter 
space they differ the most. We validate them also against the multipolar, precessing NR surrogate model \verb+NRSur7dq4+, 
and find that the \verb+SEOBNRv4PHM+ model outperforms \verb+IMRPhenomPv3HM+.

\section{Introduction}
\label{sec:Intro}

Since the Laser Interferometer Gravitational wave Observatory (LIGO) 
detected a gravitational wave (GWs) from a binary--black-hole (BBH) 
in 2015~\cite{Abbott:2016blz}, multiple observations of GWs from BBHs have been made with 
the LIGO~\cite{TheLIGOScientific:2014jea} and Virgo~\cite{TheVirgo:2014hva} detectors~\cite{TheLIGOScientific:2016pea,LIGOScientific:2018mvr,Zackay:2019tzo,Venumadhav:2019lyq,Nitz:2019hdf,LIGOScientific:2020stg}.  
Two binary neutron star (BNSs) systems have been observed~\cite{TheLIGOScientific:2017qsa,Abbott:2020uma}, one of them both 
in gravitational and electromagnetic radiation~\cite{GBM:2017lvd,Monitor:2017mdv}, opening the exciting new chapter 
of multi-messenger GW astronomy. Mergers of compact-object binaries are expected to be detected at an
even higher rate with LIGO and Virgo ongoing and future, observing runs~\cite{Aasi:2013wya}, and with 
subsequent third-generation detectors on the ground, such as the Einstein Telescope and 
Cosmic Explorer, and the Laser Interferometer Space Antenna (LISA). In order to 
extract the maximum amount of astrophysical and cosmological information, the accurate modeling of 
GWs from binary systems is more critical than ever. Great progress has been made in this direction, 
both through the development of analytical methods to solve the two-body problem in General
Relativity (GR), and by ever-more expansive numerical-relativity (NR) simulations. 

One of the key areas of interest is to improve the modeling of systems
where the misalignment of the spins with the orbital angular momentum
causes the spins and the orbital plane to
precess~\cite{Apostolatos:1994mx}. Moreover, when the binary's
component masses are asymmetric, gravitational radiation is no longer
dominated by the quadrupole moment, and higher multipoles need to be
accurately modeled~\cite{Blanchet:2013haa}. Precession and higher
multipoles lead to very rich dynamics, which in turn is imprinted on
the GW signal (see
e.g.~\cite{Apostolatos:1994mx,Kidder:1995zr,Schnittman:2004vq,Kesden:2014sla,Gerosa:2015tea,Buonanno:2002fy,Buonanno:2005xu,Schmidt:2010it,OShaughnessy:2012iol,Pekowsky:2013ska,Pan:2013rra,Campanelli:2006fy,Lousto:2015uwa,Ossokine:2015vda,Lewis:2016lgx,Afle:2018slw}). Their
measurements will be able to shed light on the formation mechanism of
the observed systems, probe the astrophysical environment, break
degeneracy among parameters, allowing more accurate measurements of
cosmological parameters, masses and spins, and more sophisticated
tests of GR.

Faithful waveform models for precessing compact-object binaries have been developed within the 
effective-one-body (EOB) formalism~\cite{Taracchini:2013wfa,Pan:2013rra,Babak:2016tgq}, 
and the phenomenological approach~\cite{Hannam:2013oca,Khan:2015jqa,Khan:2018fmp,Pratten:2020fqn,Garcia-Quiros:2020qpx} through calibration 
to NR simulations. Recently, an inspiral-merger-ringdown phenomenological waveform model that tracks precession and 
includes higher modes was constructed in Ref.~\cite{Khan:2019kot} (henceforth, \verb+IMRPhenomPv3HM+)~\footnote{During the final preparation of this work, a new frequency-domain phenomenological model with precession and higher modes (\texttt{ IMRPhenomXPHM}~\cite{Pratten:2020ceb}), and a time-domain phenomenological precessing model with the  dominant mode (\texttt{ IMRPhenomTP}~\cite{Estelles:2020c}) were developed. We leave the comparison to these models for future work.} 
The model describes the six spin degrees of freedom in the inspiral phase, but not in the late-inspiral, merger and ringdown stages.  
In the co-precessing frame~\cite{Buonanno:2002fy,Schmidt:2010it,Boyle:2011gg,O'Shaughnessy:2011fx,Schmidt:2012rh},  
in which the BBH is viewed face-on at all times and the GW radiation resembles the 
nonprecessing one, it includes the modes $(\ell, m)= (2,\pm 2), (2,\pm 1), (3,\pm 3), (3,\pm 2), 
(4,\pm 4)$ and $(4,\pm 3)$. Here, building on the multipolar 
aligned-spin EOB waveform model of Ref.~\cite{Bohe:2016gbl,Cotesta:2018fcv}, 
which was calibrated to 157 NR simulations~\cite{Mroue:2013xna,Chu:2015kft}, and 13 waveforms from 
BH perturbation theory for the (plunge-)merger and ringdown~\cite{Barausse:2011kb}, we develop the first EOB waveform model that 
includes both spin-precession and higher modes (henceforth, \verb+SEOBNRv4PHM+). The model describes 
the six  spin degrees of freedom throughout the BBH coalescence. It differs from the one 
of Refs.~\cite{Pan:2013rra,Babak:2016tgq}, not only because it includes  in the co-precessing 
frame the $(3,\pm 3)$, $(4,\pm 4)$ and $(5,\pm 5)$ 
modes, beyond the $(2,\pm 2)$ and $(2,\pm  1)$ modes, but also because it uses an improved description 
of the two-body dynamics, having been calibrated~~\cite{Bohe:2016gbl} to a large set of NR waveforms~\cite{Mroue:2013xna}. We note that \verb+IMRPhenomPv3HM+ and \verb+SEOBNRv4PHM+ are not completely independent 
because the former is constructed fitting (in frequency domain) hybridized waveforms obtained by stitching together EOB and NR waveforms.  
We stress that both \verb+SEOBNRv4HM+ and  \verb+IMRPhenomPv3HM+ are not calibrated to NR simulations in the 
precessing sector. Finally, the surrogate approach, which  interpolates NR waveforms, has been used to construct several waveform models that include higher modes~\cite{Varma:2018mmi} and precession~\cite{Blackman:2017pcm}. In this paper, we  consider the state-of-the-art surrogate waveform model with full spin precession and higher modes~\cite{Varma:2019csw} (henceforth, \texttt{NRSur7dq4}), 
developed for binaries with mass ratios 1-4, (dimensionless) BH's spins up to $0.8$ and binary's 
total masses larger than $\sim 60 M_\odot$. It includes in the co-precessing frame all modes up to  $\ell =4$. 
Table~\ref{tbl:wf_models} summarizes the waveform models used in this paper. 

\begin{table}
		\begin{tabularx}{\textwidth}{lll}
		\hline\hline
		Model name & Modes in the co-precessing frame & Reference\\
		\hline
		\texttt{ SEOBNRv3P} &  $(2,\pm2)$, $(2,\pm1)$ & \cite{Pan:2013rra,Babak:2016tgq}\\
		\texttt{ SEOBNRv4P} & $(2,\pm2)$, $(2,\pm1)$ &  this work \\
		\texttt{ SEOBNRv4PHM} & $(2,\pm2)$, $(2,\pm1)$, $(3,\pm3)$, $(4,\pm4)$\\
		& $(5,\pm5)$ & this work \\
		\texttt{ IMRPhenomPv2} & $(2,\pm2)$ & \cite{Hannam:2013oca}\\
		\texttt{ IMRPhenomPv3} & $(2,\pm2)$  & \cite{Khan:2018fmp}\\
		\texttt{ IMRPhenomPv3HM} & $(2,\pm2)$, $(2,\pm1)$, $(3,\pm3)$, $(3,\pm2)$,\\
		& $(4,\pm 4)$, $(4,\pm3)$&\cite{Khan:2019kot} \\
		\texttt{ NRSur7dq4} & all with $\ell\leq4$ &\cite{Varma:2019csw} \\
	\hline\hline
	\end{tabularx}
	\caption{The waveform models used in this paper. We also specify which modes are included in the co-precessing frame}
	\label{tbl:wf_models}
\end{table}

The best tool at our disposal to validate waveform models built from approximate solutions of the 
Einstein equations, such as the ones obtained from post-Newtonian (PN) theory, BH perturbation theory 
and the EOB approach, is their comparison to NR waveforms. So far, NR simulations of  BBHs have been mostly limited to mass ratio $\leq 4$ and (dimensionless) spins $\leq 0.8$, and length   
of $15\mbox{--}20$ orbital cycles before merger~\cite{Jani:2016wkt,Healy:2017psd,Healy:2019jyf,Huerta:2019oxn,Boyle:2019kee} (however, see Ref.~\cite{Hinder:2018fsy}
where simulations with larger spins and mass ratios were obtained through a synergistic use of NR codes).
Here, to test our newly constructed EOB precessing waveform model, we enhance the 
NR parameter-space coverage, while maintaining a manageable computational cost, and perform $118$ 
new NR simulations with the pseudo spectral Einstein code (SpEC) of the Simulating eXtreme 
Spacetimes (SXS) collaboration.  The new NR simulations span BBHs with  
mass ratios  $1\mbox{--}4$, and dimensionless spins in the range $0.3\mbox{--}0.9$, and different 
spins' orientations.  To assess the accuracy of the different precessing waveform models, 
we compare them to the NR waveforms of the public SXS catalogue~\cite{Boyle:2019kee}, 
and to the new $118$ NR waveforms produced for this paper.

The paper is organized as follows. In Sec.~\ref{sec:NR} we discuss the new NR simulations of BBHs, 
and assess their numerical error. In Sec.~\ref{sec:multiEOB} we develop the multipolar EOB waveform 
model for spin-precessing BBHs, \verb+SEOBNRv4PHM+, and highlight the improvements with respect to the 
previous version~\cite{Pan:2013rra,Babak:2016tgq}, which was used in LIGO and Virgo 
inference analyses~\cite{Abbott:2016izl,Abbott:2017vtc,LIGOScientific:2018mvr}. 
In Sec.~\ref{sec:compEOBNR} we validate the accuracy of the multipolar precessing EOB model 
by comparing it to NR waveforms. We also compare the performance of \verb+SEOBNRv4PHM+ 
against the one of \verb+IMRPhenomPv3HM+, and study in which region of the parameter space 
those models differ the most from NR simulations, and also from each other. In Sec.~\ref{sec:peEOBNR} 
we use Bayesian analysis to explore the impact of the accuracy of the precessing waveform models 
when extracting astrophysical information and perform two synthetic NR injections in zero noise. 
In Sec.~\ref{sec:concl} we summarize our main conclusions and discuss future  work. Finally, 
in Appendix~\ref{sec:comparisonNRSurr} we compare the precessing waveform models to the 
NR surrogate \texttt{ NRSur7dq4}, and in Appendix~\ref{sec:NRparam} we list the parameters 
of the 118 NR simulations done for this paper.
 
In what follows, we use geometric units $G=1=c$ unless otherwise specified. 

\section{New numerical-relativity simulations of spinning, precessing binary black holes}
\label{sec:NR}

Henceforth, we denote with $m_{1,2}$ the two BH masses (with $m_1 \geq m_2$), $\vS_{1,2} \equiv
m_{1,2}^2\,\vchi_{1,2}$ their spins, $q = m_1/m_2$ the mass ratio, $M = m_1+m_2$ the 
binary's total mass, $\mu= m_1m_2/M$ the reduced mass, and $\nu = \mu/M$ the 
symmetric mass ratio. We indicate with $\vJ = \vL + \vS $ the total angular momentum, 
where $\vL$ and $\vS = \vS_1 +\vS_2$, are the orbital angular momentum and the 
total spin, respectively

\subsection{New $118$ precessing numerical-relativity waveforms}

\begin{figure}
\centering
\includegraphics[width=0.7\linewidth]{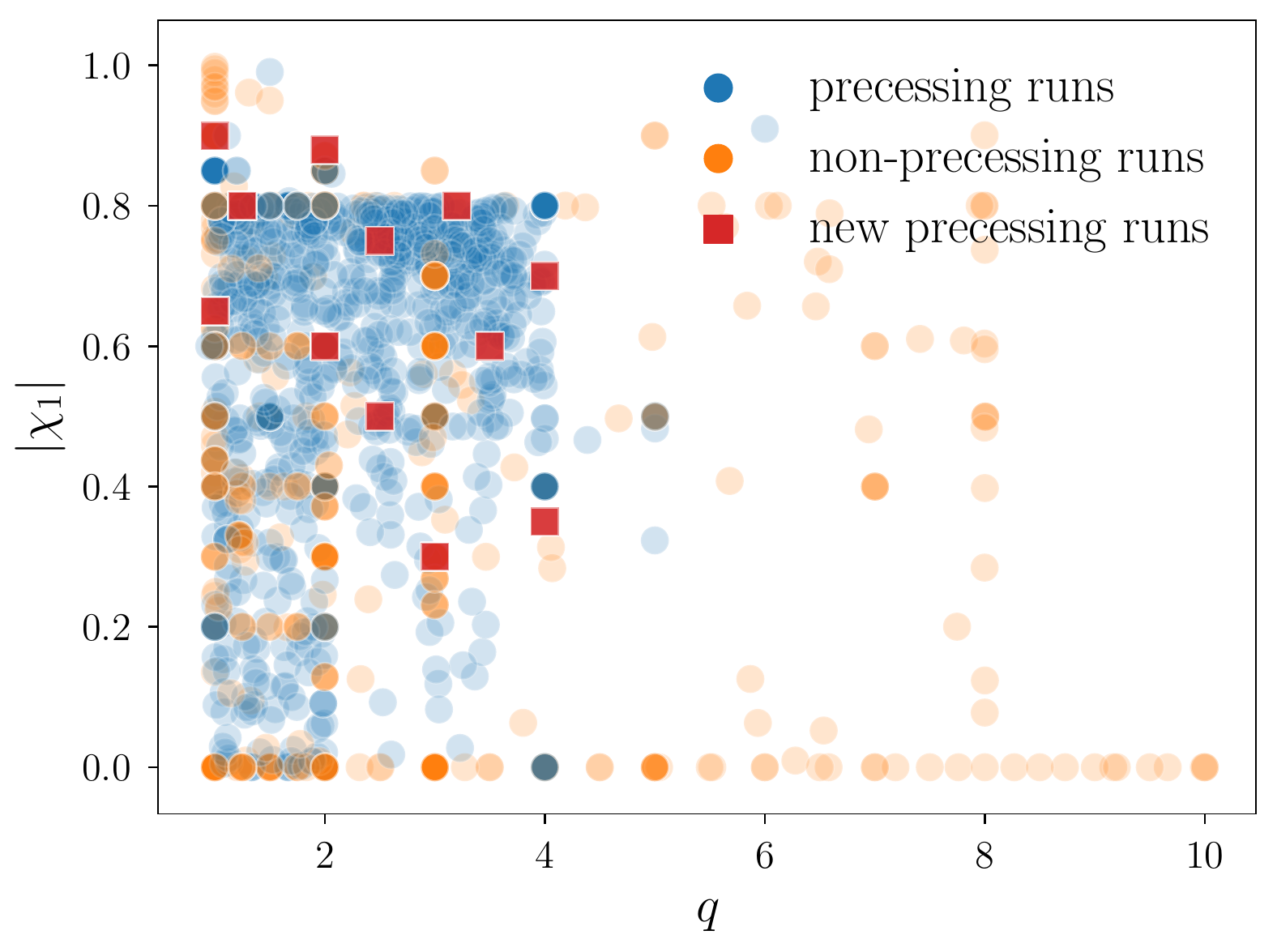}
\caption{Parameter space coverage in $q\mbox{--}\chi_{1}$ space for \texttt{ SpEC} waveforms. For runs
  from the \texttt{ SpEC} catalog~\cite{Boyle:2019kee} the opacity was changed so that runs with similar
  parameters are clearly visible. We indicate with squares precessing BBH runs performed as part of this paper. 
\label{fig:Param_space}}
\end{figure}

The spectral Einstein code (\texttt{SpEC}) \footnote{\url{www.black-holes.org}} of the Simulating eXtreme Spacetimes
(SXS) collaboration is a multi-domain collocation code designed for the solution 
of partial differential equations on domains with simple topologies.
It has been used extensively to study the mergers of compact-object
binaries composed of BH~\cite{Scheel:2014ina,Lovelace:2014twa,Szilagyi:2015rwa,Lovelace:2016uwp,Afle:2018slw,Boyle:2019kee} and NSs~\cite{Foucart:2015gaa,Haas:2016cop,Foucart:2018lhe,Vincent:2019kor}, including in theories 
beyond GR~\cite{Okounkova:2018abo,Okounkova:2019dfo,Okounkova:2019zjf,Okounkova:2020rqw}. 
SpEC employs a first-order symmetric-hyperbolic formulation of Einstein's equations~\cite{Lindblom:2005qh} in the damped harmonic 
gauge~\cite{Lindblom:2009tu,Szilagyi:2009qz}. Dynamically controlled
excision boundaries are used to treat spacetime singularities~\cite{Hemberger:2012jz,Scheel:2014ina} 
(see Ref.~\cite{Boyle:2019kee} for a recent, detailed overview).

Significant progress has been made in recent years by several NR groups to improve the coverage of the BBH parameter 
space~\cite{Jani:2016wkt,Healy:2017psd,Healy:2019jyf,Huerta:2019oxn,Boyle:2019kee,Hinder:2018fsy}, 
mainly motivated by the calibration of analytical waveform models and surrogate models 
used in LIGO and Virgo data analysis.  While large strides have been made for aligned-spin cases, the exploration of
precessing waveforms has been mostly limited to $q\leq 4, \chi_{1,2} \equiv 
|\vchi_{1,2}| \leq 0.8$, typically $15\mbox{--}20$ orbital cycles before merger, 
and a large region of parameter space remains to be explored. Simulations with high mass ratio ($q\ge4$) and high spin ($|\vchi_{1}|>0.5$) are challenging, primarily due to the need
to resolve the disparate length scales in the binary system, which increases the computational cost for a given level of accuracy. Furthermore, for high spin, the apparent horizons can be  dramatically smaller, which makes it more difficult to control the excision boundaries, further increasing the computational burden.

Here, we want to improve the parameter-space coverage of the SXS catalog~\cite{Boyle:2019kee}, while 
maintaining a manageable computational cost, thus we restrict to  simulations in 
the range of mass ratios $q=1\mbox{--}4$ 
and (dimensionless) spins $\chi_{1,2}=0.3\mbox{--}0.9$, with the spin magnitudes decreasing as the mass ratio
increases. In Fig.~\ref{fig:Param_space} we display, in the $q\mbox{--} \chi_1$ parameter space,  
the precessing and non-precessing waveforms from the published SXS catalogue~\cite{Boyle:2019kee}, 
and the new precessing waveforms produced as part of this work. 

We choose to start all the simulations at the same (angular) orbital frequency, $M\Omega_0 \approx 0.0157$, where the value 
is not exact as it was modified slightly during the eccentricity-reduction procedure in \texttt{ SpEC}~\cite{Buonanno:2010yk}. 
This corresponds to a physical GW starting frequency of $20$ Hz at $50 M_{\odot}$ and 
results in the number of orbits up to merger varying between $15$ and $30$  in 
our new catalog.

We parametrize the directions of the spins by three angles, the angles $\theta_{1,2}$ between the spins 
and the unit vector along the Newtonian orbital angular momentum, $\vLhat_\mathrm{N}$, and the angle
$\Delta \phi$ between the projections of the spins in the orbital plane. Explicitly, 
\begin{subequations}
\begin{align}
\theta_{i}  &=\arccos (\vchi_{i} \cdot \vLhat_\mathrm{N})\,, \\
\Delta \phi &= \arccos \left(\frac{\cos\theta_{12}-\cos\theta_1\cos\theta_2}{\sin\theta_1\sin\theta_2}\right)\,,
\end{align}
\end{subequations}
where $\cos\theta_{12}=\vchi_{1}\cdot \vchi_{2}$. We make the choice
that $\vchi_{1}$ lies in the $\vLhat_\mathrm{N}-\mathbf{n}$ plane, where $\mathbf{n}$  
is the unit vector along the binary's radial separation, at the start of the
simulation.  The angles are chosen to be $\theta_{i,0}\in \{60^{\circ},\theta_\mathrm{
max}\}$, and $\Delta\phi_0 \in \{0,90^{\circ}\}$. Here $\theta_{max}$ is the angle that maximizes the opening angle of $\vL_\mathrm{N}$ around the total angular momentum
$\vJ$ and is computed assuming that the two spins are co-linear, giving
\begin{equation}
\cos \theta_\mathrm{max} = -\frac{|\mathbf{S}|}{|\vL_\mathrm{N}|} =
-\frac{m_{1}^{2}\,\chi_{1}+m_{2}^{2}\,\chi_{2}}{|\vL_\mathrm{N}|}\,,
\end{equation}
with $|\vL_\mathrm{N}|=\mu M^{2/3}\Omega^{-1/3}$ for circular orbit, 
being $\Omega$ the orbital angular frequency. For each choice
of $\{q,\chi\}$ we choose 10 different configurations divided into two 
categories: i) $\chi_{1}=\chi_{2}=\chi,\theta_{i,0}\in\{60^{\circ},\theta_\mathrm{max}\},\ \Delta \phi_0\in\{0,90^{\circ}\}$ giving eight runs, and ii)
$\chi_{1} = \chi, \chi_{2}=0, \theta_{1,0} \in \{60^{\circ},\theta_\mathrm{
  max}\}$ giving two runs. The detailed parameters can be found in Appendix~\ref{sec:NRparam}.

These choices of the spin directions allow us to test the multipolar 
precessing model \verb+SEOBNRv4PHM+ in many different regimes, including where the effects of 
precession are maximized, and where spin-spin effects are significant or diminished.

\subsection{Unfaithfulness for spinning, precessing waveforms}
\label{sec:unfaith_sec}

The gravitational signal emitted by non-eccentric BBH systems 
and observed by a detector depends on 15 parameters: the component masses $m_1$ and
$m_2$ (or equivalently the mass ratio $q = m_1/m_2 \geq 1$ and
the total mass $M = m_1 + m_2$), the dimensionless spins
$\vchi_1(t)$ and $\vchi_2(t)$, the direction to observer from the
source described by the angles $(\iota,\varphi_0)$, the luminosity
distance $d_L$, the polarization $\psi$, the location in the sky
$(\theta,\phi)$ and the time of arrival $t_c$. The gravitational strain 
can be written as:
\begin{align}
\label{eq:det_strain}
h(t) \equiv & F_+(\theta,\phi,\psi) \ h_+(\iota,\varphi_0, d_L, \boldsymbol{\xi},t_{\mathrm{c}};t) \nonumber \\
&+ F_\times(\theta,\phi,\psi)\ h_\times(\iota,\varphi_0, d_L, \boldsymbol{\xi},t_\mathrm{c};t)\,,
\end{align}
where to simplify the notation we introduce the function $\boldsymbol{\xi} \equiv (q$, $M$, $\vchi_{1}(t)$, $\vchi_{2}(t))$.
The functions $F_+(\theta,\phi,\psi)$ and $F_\times(\theta,\phi,\psi)$ are the antenna 
patterns~\cite{Sathyaprakash:1991mt,Finn:1992xs}:
\begin{subequations}
\begin{align}
F_+(\theta,\phi,\psi) &= \frac{1+ \cos^2(\theta)}{2} \ \cos(2\phi) \ \cos(2\psi) -\cos(\theta) \ \sin(2\phi)\ \sin(2\psi),\\ 
F_\times(\theta,\phi,\psi) &= \frac{1+ \cos^2(\theta)}{2} \ \cos(2\phi) \ \sin(2\psi) + \cos(\theta) \ \sin(2\phi)\ \cos(2\psi).
\end{align}
\end{subequations}
Equation \eqref{eq:det_strain} can be rewritten as:
\begin{align}
\label{eq:det_strain_kappa}
h(t) \equiv & \mathcal{A}(\theta,\phi)\big[\cos\kappa(\theta,\phi,\psi) \ h_+(\iota, \varphi_0, d_L, \boldsymbol{\xi}, t_{\mathrm{c}};t) \nonumber \\
&+ \sin\kappa(\theta,\phi,\psi) \ h_\times (\iota, \varphi_0, d_L, \boldsymbol{\xi},t_{\mathrm{c}};t) \big],
\end{align}
where $\kappa(\theta,\phi,\psi)$ is the \textit{effective polarization}~\cite{Capano:2013raa} defined as:
\begin{equation}
e^{i \kappa(\theta,\phi,\psi)} = \frac{F_+(\theta,\phi,\psi) + i F_\times(\theta,\phi,\psi)}{\sqrt{F_+^2(\theta,\phi,\psi) + F_\times^2(\theta,\phi,\psi)}},
\end{equation}
which has support in the region $[0, 2\pi)$, while $\mathcal{A}(\theta,\phi)$ reads:
\begin{equation}
\mathcal{A}(\theta,\phi) = \sqrt{F_+^2(\theta,\phi,\psi) + F_\times^2(\theta,\phi,\psi)}\,.
\end{equation}
Henceforth, to ease the notation we suppress the dependence on $(\theta,\phi,\psi)$ in $\kappa$.

Let us introduce the inner product between two waveforms $a$ and $b$~\cite{Sathyaprakash:1991mt,Finn:1992xs}:
\begin{equation}
\left( a, b\right) \equiv 4\ \textrm{Re}\int_{f_\mathrm{ in}}^{f_\mathrm{max}} df\,\frac{\tilde{a}(f) \ \tilde{b}^*(f)}{S_n(f)},
\end{equation}
where a tilde indicates the Fourier transform, a star the complex
conjugate and $S_n(f)$ is the one-sided power spectral density (PSD)
of the detector noise. We employ as PSD the Advanced LIGO’s “zero-detuned
high-power” design sensitivity curve~\cite{Barsotti:2018}. Here we use $f_\mathrm{in} = 10 \mathrm{ Hz}$  
and $f_\mathrm{ max} = 2 \mathrm{ kHz}$, when both waveforms fill the band. For cases where this is 
not the case (e.g the NR waveforms) we set $f_\mathrm{ in}=1.05f_\mathrm{ start}$, where $f_\mathrm{ start}$ is the starting frequency of the waveform.

To assess the closeness between two waveforms $s$ (e.g., the signal) and $\tau$ (e.g., the template), 
as observed by a detector, we define the following \textit{faithfulness} function~\cite{Cotesta:2018fcv}:
\begin{equation}
\label{eq:faith}
\mathcal{F}(M_{\textrm{s}},\iota_{\textrm{s}},{\varphi_0}_{\textrm{s}},\kappa_{\textrm{s}}) \equiv  \max_{t_c, {\varphi_0}_{\tau}, \kappa_{\tau}} \left[\left . \frac{ \left( s,\,\tau \right)}{\sqrt{ \left( s,\,s \right) \left( \tau,\,\tau \right)}}\right \vert_{\substack{\iota_{\mathrm{s}} = \iota_{\tau} \\\boldsymbol{\xi}_{\mathrm{s}}(t_{\mathrm{s}} = t_{0_\mathrm{s}}) = \boldsymbol{\xi}_{\tau}(t_\tau = t_{0_\mathrm{\tau}})}} \right ].
\end{equation}
While in the equation above we set the inclination angle $\iota$ of signal and template waveforms to  be the same, the  coalescence time $t_c$
and the  angles ${\varphi_0}_{\tau}$ and  $ \kappa_{\tau}$ of  the template
waveform are adjusted to maximize  the faithfulness. This is a typical
choice motivated by the fact these quantities are not interesting from
an  astrophysical  perspective.  The   maximizations  over  $t_c$  and
${\varphi_0}_{\tau}$ are performed numerically, while the maximization
over   $\kappa_{\tau}$  is   done  analytically   following  the
procedure described  in Ref. ~\cite{Capano:2013raa} (see Appendix  A therein).

The condition $\boldsymbol{\xi}_{\mathrm{s}}(t_{\mathrm{s}} =
t_{0_\mathrm{s}}) = \boldsymbol{\xi}_{\tau}(t_\tau =
t_{0_\mathrm{\tau}})$ in Eq.~\eqref{eq:faith} enforces that the mass
ratio $q$, the total mass $M$ and the spins $\vchi_{1,2}$ of the
template waveform at $t = t_0$ (i.e., at the beginning of the
waveform) are set to have the same values of the ones in the signal
waveform at its $t_0$. When computing the faithfulness between NR
waveforms with different resolutions this condition is trivially
satisfied by the fact that they are generated using the same initial
data. In the case of the faithfulness between NR and any model from the   \verb+SEOBNR+  family, it
is first required to ensure that $t_0$ has the same physical meaning
for both waveforms. Ideally $t = t_{0_\mathrm{\tau}}$ in the \verb+SEOBNR+ 
waveform should be fixed by requesting that the frequency of the \verb+SEOBNR+
$(2,2)$ mode at $t_{0_\mathrm{\tau}}$ coincides with the NR (2,2) mode
frequency at $t_{0_\mathrm{s}}$. This is in practice not possible
because the NR (2,2) mode frequency may display small oscillations
caused by different effects --- for example the persistence of the
junk radiation, some residual orbital eccentricity or spin-spin
couplings~\cite{Buonanno:2010yk}.  Thus, the frequency of the \verb+SEOBNR+ 
$(2,2)$ mode at $t = t_{0_\mathrm{\tau}}$ is chosen to guarantee the
same time-domain length of the NR waveform. \footnote{The difference
  between the NR (2,2) mode frequency and the \texttt{ SEOBNRv4PHM} (2,2) frequency
  chosen at $t = t_0$ is never larger than $5\%$.}.
In practice, we
require that the peak of $\sum_{\ell, m} |h_{\ell m}|^2$, as elapsed
respectively from $t_{0_\mathrm{s}}$ and $t_{0_\mathrm{\tau}}$, occurs
at the same time in NR and \verb+SEOBNR+. 
For waveforms from the \verb+IMRPhenom+ family we adopt a different approach, and following 
the method outlined in Ref.~\cite{Khan:2018fmp}, also optimize the faithfulness numerically over the
reference frequency of the waveform.

The faithfulness defined in Eq.~\eqref{eq:faith} is  still a function of 4
parameters (i.e., $M_{\textrm{s}},\iota_{\textrm{s}},{\varphi_0}_{\textrm{s}},\kappa_{\textrm{s}}$), 
therefore it does not allow to describe the agreement between waveforms in a compact form. For  this  purpose  we define  the \textit{sky-and-polarization-averaged faithfulness}~\cite{Babak:2016tgq} as:
\begin{equation}
\overline{\mathcal{F}}(M_\mathrm{s}, \iota_\mathrm{s}) \equiv  \frac{1}{8\pi^2}\int_{0}^{2\pi} d\kappa_{\mathrm{s}} \int_{0}^{2\pi} d{\varphi_0}_{\mathrm{s}} \ \mathcal{F}(M_\mathrm{s},\iota_{\textrm{s}},{\varphi_0}_{\textrm{s}},\kappa_{\textrm{s}}).
\label{eq:avg_faith}
\end{equation}
Despite the apparent difference, the sky-and-polarization-averaged
faithfulness $\overline{\mathcal{F}}$ defined above is equivalent to the one given in Eqs.~(9) 
and (B15) of Ref.~\cite{Babak:2016tgq}. The definition in 
Eq.~\eqref{eq:avg_faith} is less computationally expensive because,
thanks to the parametrization of the waveforms in
Eq.~\eqref{eq:det_strain_kappa}, it allows one to write the sky-and-polarization-averaged 
faithfulness as a double integral instead of the
triple integral in Eq.~(B15) of Ref.~\cite{Babak:2016tgq}.
We also define the sky-and-polarization-averaged, signal-to-noise (SNR)-\textit{ weighted} faithfulness as:
\begin{equation}
\overline{\mathcal{F}}_{\mathrm{SNR}}(M_\mathrm{s},\iota_{\mathrm{s}}) \equiv \sqrt[3]{\frac{\int_{0}^{2\pi} d\kappa_ {\mathrm{s}} \int_{0}^{2\pi} d{\varphi_0}_{\mathrm{s}} \ \mathcal{F}^{3}(M_{\textrm{s}},\iota_{\textrm{s}},{\varphi_0}_{\textrm{s}},\kappa_{\textrm{s}}) \ \mathrm{SNR}^3(\iota_{\textrm{s}},{\varphi_0}_{\textrm{s}},\kappa_{\textrm{s}})}{\int_{0}^{2\pi} d\kappa_{\mathrm{s}} \int_{0}^{2\pi} d{\varphi_0}_{\mathrm{s}} \ \mathrm{SNR}^3(\iota_{\textrm{s}},{\varphi_0}_{\textrm{s}},\kappa_{\textrm{s}})}}.
\end{equation}
where the $\mathrm{SNR}(\iota_{\textrm{s}},{\varphi_0}_{\textrm{s}},\theta_\textrm{s}, \phi_\textrm{s},\kappa_{\textrm{s}},{D_{\mathrm{L}}}_{\mathrm{s}},
\boldsymbol{\xi}_\mathrm{s},{t_c}_\mathrm{s})$ is defined as:
\begin{equation}
\mathrm{SNR}(\iota_{\textrm{s}},{\varphi_0}_{\textrm{s}},\theta_\textrm{s}, \phi_\textrm{s}, \kappa_{\textrm{s}},{D_{\mathrm{L}}}_{\mathrm{s}},\boldsymbol{\xi}_\mathrm{s},{t_c}_\mathrm{s}) \equiv \sqrt{\left(h_{\mathrm{s}},h_{\mathrm{s}}\right)}.
\end{equation}
This is also an interesting metric because weighting the faithfulness
with the SNR takes into account that, at fixed distance, 
the SNR of the signal depends on its phase and on the effective
polarization (i.e., a combination of waveform polarization and
sky-position). Since the SNR scales with the luminosity distance, the
number of detectable sources scale with the $\mathrm{SNR}^3$,
therefore signals with a smaller SNR are less likely to be
observed. Finally, we define the unfaithfulness (or
mismatch) as
\begin{equation}
\overline{\mathcal M} = 1 -  \overline{\mathcal{F}}\,.
\label{mismatch}
\end{equation}

\subsection{Accuracy of new numerical-relativity waveforms}

To assess the accuracy of the new NR waveforms, we compute the
sky-and-polarization-averaged unfaithfulness defined in Eq.~(\ref{eq:avg_faith})
between the highest and second highest resolutions in the NR simulation. 

\begin{figure}
\centering
\includegraphics[width=0.7\linewidth]{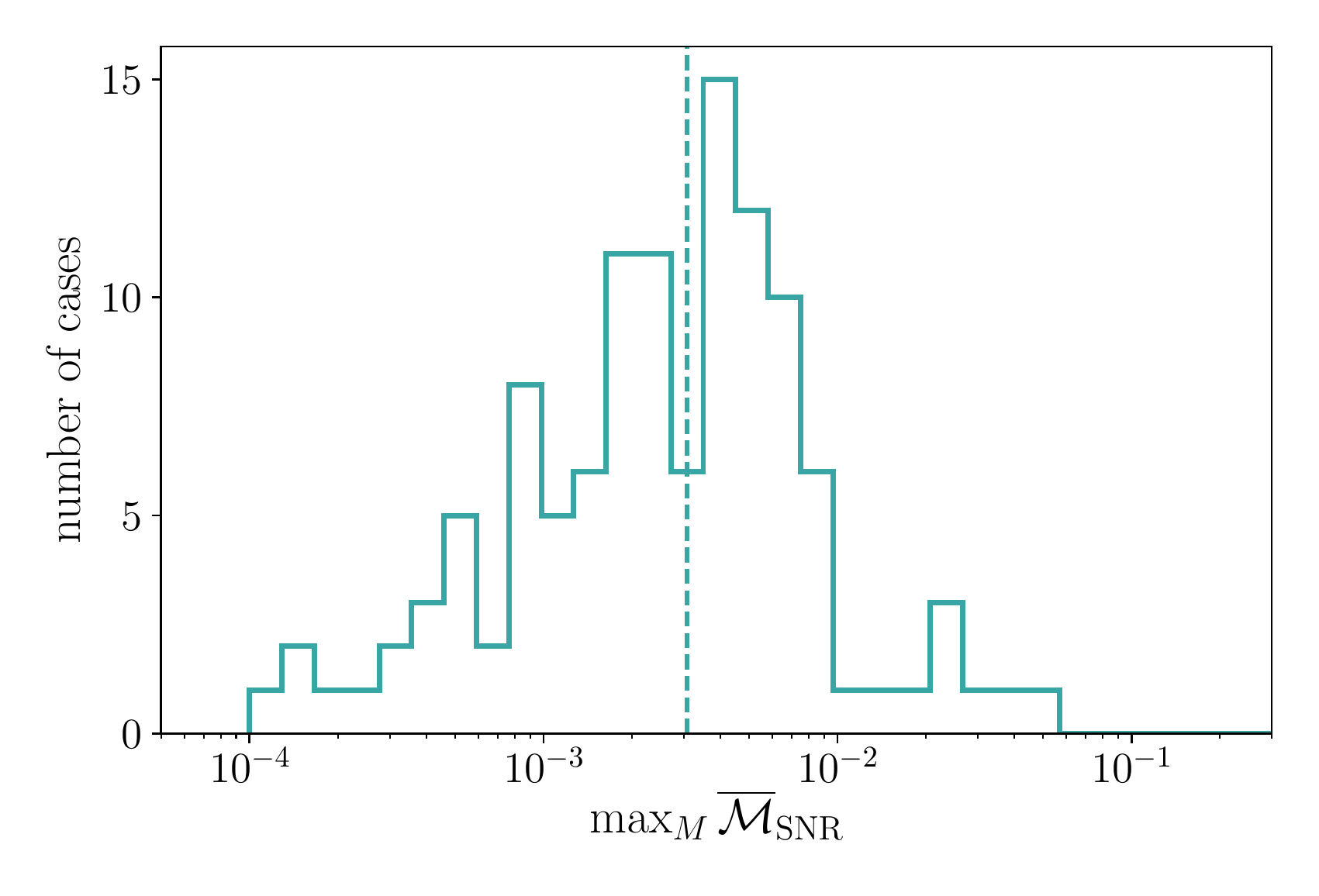}
 \caption{The sky-and-polarization-averaged unfaithfulness between 
the highest and second highest resolutions in the NR simulation 
maximized over the total mass for the new 
118 NR precessing waveforms. The inclination used is $\pi/3$. The vertical dashed line shows the median.}
  \label{fig:unfaith_NRNR}
\end{figure}

Figure~\ref{fig:unfaith_NRNR} shows a histogram of the unfaithfulness,
evaluated at $\iota_{s}={\pi}/{3}$ maximized over the total mass, between 20 and
200 $M_\odot$. It is apparent that the unfaithfulness is  below $1 \%$ for most
cases, but there are several cases with much higher unfaithfulness. 
This tail to high unfaithfulness has been observed previously, when evaluating 
the accuracy of SXS simulations in Ref.~\cite{Varma:2019csw}. Therein, it was 
established that, when the non-astrophysical junk radiation perturbs the parameters of
the simulation sufficiently, the different resolutions actually correspond to
different physical systems. Thus, taking the difference between adjacent resolutions 
is no longer an appropriate estimate of the error.

We also find that the largest unfaithfulness occurs when the difference in parameters is
largest, thus confirming that it is the difference in parameters that dominates
the unfaithfulness.

\subsection{Effect of mode asymmetries in numerical-relativity waveforms}
\label{sec:mode_asymm}

The gravitational polarizations at time $t$ and location $(\varphi_0,\iota)$ on the coordinate sphere from the binary can be decomposed in $-2$--spin-weighted spherical harmonics, as follows
\begin{equation}
h_{+}(\varphi_0,\iota;t) - i h_{\times}(\varphi_0,\iota;t) = \sum_{\ell =2} \sum_{m=-\ell}^{m=+\ell} {}_{-2} Y_{\ell m}(\varphi_0,\iota)\,h_{\ell m}(t) \,.
\end{equation}
For nonprecessing binaries, the invariance of the system under reflection across the orbital plane (taken to be the $x\mbox{--}y$ plane) implies 
$h_{\ell m}= (-1)^\ell h_{\ell -m}^*$. The latter is a very convenient relationship --- for example it renders unnecessary to model modes 
with negative values of $m$. However, this relationship is no longer satisfied for precessing binaries. 

As investigated in previous NR studies~\cite{Pekowsky:2013ska,Boyle:2014ioa}, we expect the asymmetries 
between opposite-$m$ modes to be small as compared to the dominant $(2,2)$-mode emission (at least during the
inspiral) in a co-rotating frame that maximizes emission in the $(2,\pm 2)$ modes, also known as the \textit{maximum-radiation 
frame}~\cite{Boyle:2011gg,Boyle:2013nka}. However, while the asymmetries are expected to be small during the inspiral, 
the difference in phase and amplitude between positive and negative $m$-modes might become non-negligible at merger. 

\begin{figure}
\centering
	\includegraphics[width=0.7\linewidth]{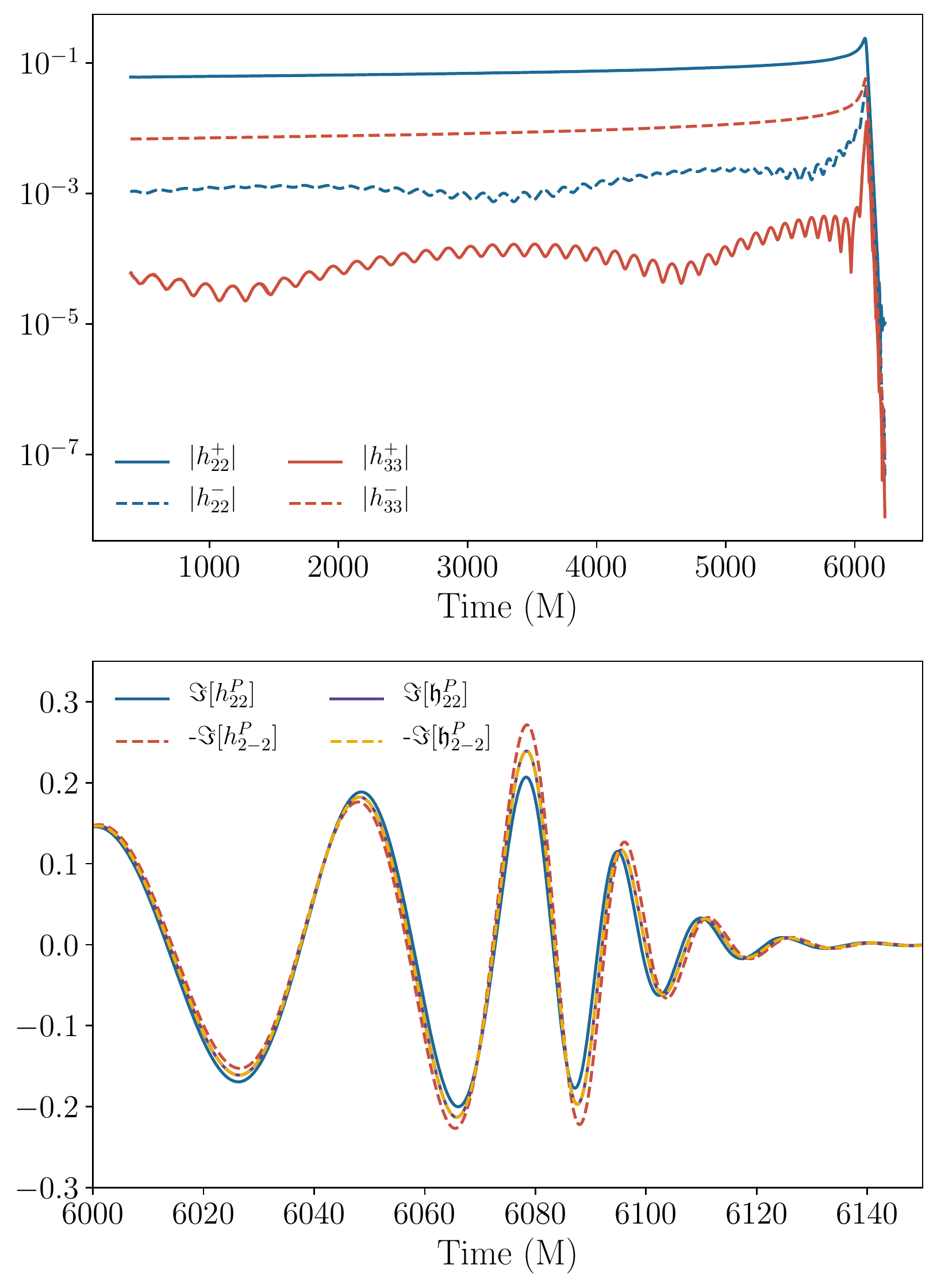}
        \caption{\emph{Top}: the behavior of $h_{\ell m}^{\pm}$ in the
          NR simulation {\texttt PrecBBH000078}. Note that especially during the inspiral,
          $|h_{22}^{+}|$ is much larger than $|h_{22}^{-}|$ while
          $|h_{33}^{-}|$ is much larger than $|h_{33}^{+}|$. \emph{Bottom}: an example of waveform symmetrization for the same NR
          case, shown in the co-precessing frame. The symmetrized
          waveform obeys the usual conjugation symmetry as expected,
          and represents a reasonable average to the behavior of the
          unsymmetrized modes.}
	\label{fig:example_symm_waveform}
\end{figure}

As we discuss in the next section, when building multipolar waveforms (\verb+SEOBNRv4PHM+) for precessing binaries by rotating 
modes from the co-precessing \cite{Buonanno:2002fy,Schmidt:2010it,Boyle:2011gg,O'Shaughnessy:2011fx,Schmidt:2012rh} 
to the inertial frame of the observer, we shall neglect the mode asymmetries. To quantify the error introduced by this assumption, 
we proceed as follows. We first take NR waveforms in the co-precessing frame and construct \emph{symmetrized} waveforms. Specifically, we 
consider the combination of waveforms in the co-precessing frame defined by (e.g., see Ref.~\cite{Varma:2019csw})
\begin{equation}
h_{\ell m}^{\pm} = \frac{h_{\ell m}^{P}\pm h_{\ell -m}^{P*}}{2}\,.
\end{equation}
Note that if the assumption of conjugate symmetry holds (i.e., if
$h^P_{\ell -m} = (-1)^{\ell}h^{P*}_{\ell m}$), then for even (odd)
$\ell$ modes, $h_{\ell m}^{+}$ ($h_{\ell m}^{-}$) is non-zero while
the other component vanishes. If the assumption does not hold, it is
still true that at given $\ell$, one of the components is much larger
than the other, as shown in top panel of Fig.~\ref{fig:example_symm_waveform}. 
Motivated by this, we define the symmetrized modes (for $m>0$) as~\cite{Varma:2019csw} 
\begin{equation}
\mathfrak{h}_{\ell m}^{P} =
\begin{cases}
h_{\ell m}^{+} & \text{if}\ \ell\ \text{is even}\,, \\
h_{\ell m}^{-} & \text{if}\ \ell\ \text{is odd}\,. 
\end{cases} 
\end{equation}
The other modes are constructed as $\mathfrak{h}_{\ell -m}^{P}=\mathfrak{h}_{\ell m}^{P*}$ for $m<0$, 
and we set $m=0$ modes to zero. The bottom panel of Fig.~\ref{fig:example_symm_waveform} shows an example 
of asymmetrized waveform for the case \verb|PrecBBH000078| of the SXS catalogue, in the
co-precessing frame. It is obvious that the asymmetry between the
modes has been removed and that the symmetrized waveform does indeed
represent a reasonable ``average'' between the original modes. The symmetrized waveforms in the
inertial frame are obtained by rotating the co-precessing frames modes back to the inertial frame.

\begin{figure}
\centering
	\includegraphics[width=0.7\linewidth]{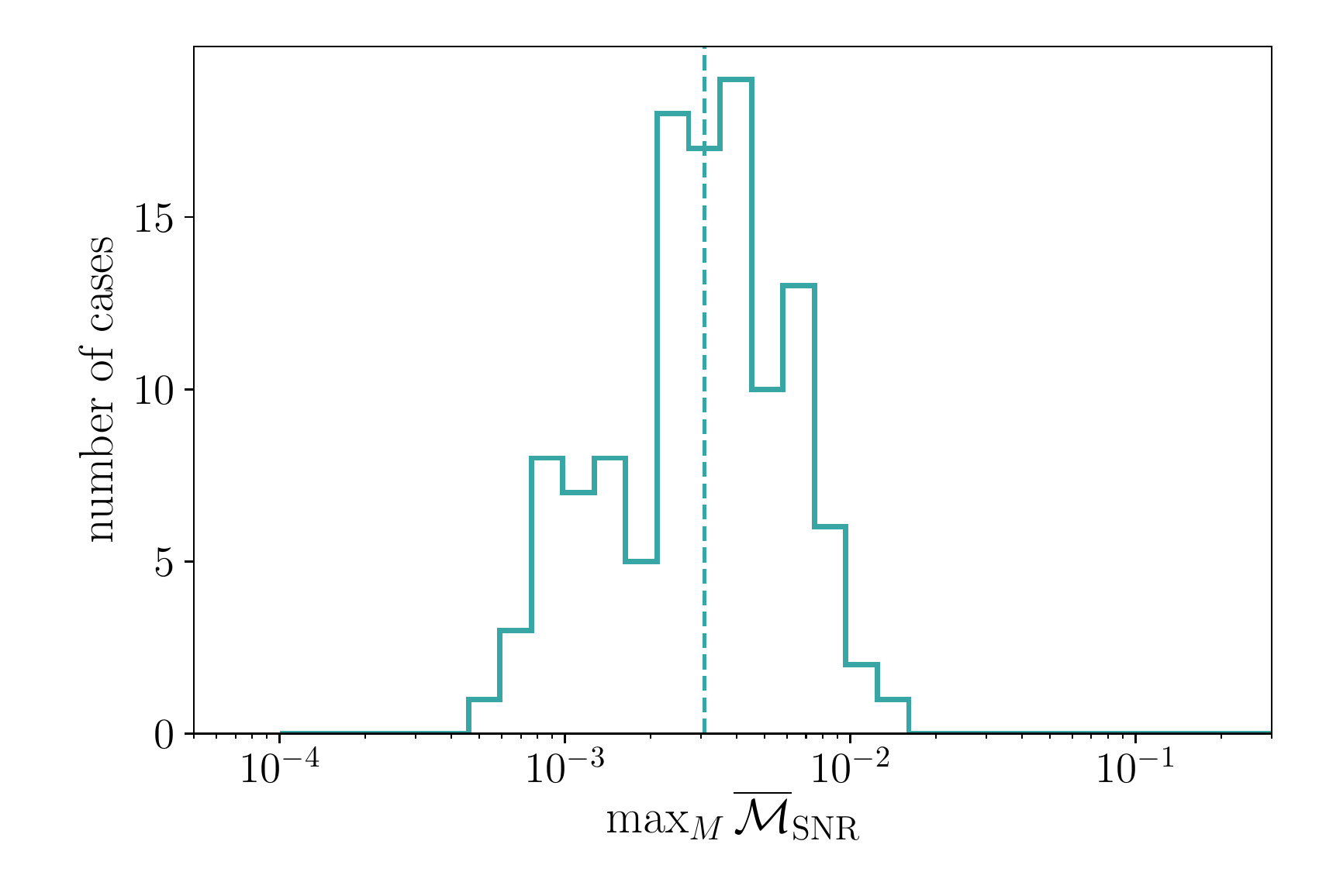}
	\caption{The sky-and-polarization-averaged unfaithfulness between NR and symmetrized 
NR waveforms, maximized over the total mass for the new 118 NR precessing waveforms. 
The inclination used is $\pi/3$. The vertical dashed line shows the median.}
	\label{fig:unfaith_NRNR_symm}
\end{figure}

In Fig.~\ref{fig:unfaith_NRNR_symm}, we show the sky-and-polarization 
averaged unfaithfulness between the NR waveforms and the
symmetrized waveforms described above, maximized over the total mass,
including all modes available in the NR simulation, that is up to $\ell=8$. 
For the vast majority of the cases, the unfaithfulness is $\sim0.5\%$, 
and all cases have unfaithfulness below $2\%$. This demonstrates that the effect of
neglecting the asymmetry is likely subdominant to other sources of
error such as the modeling of the waveform phasing, although the best 
way of quantifying the effect is to perform a Bayesian 
parameter-estimation study, which we leave to future work.  

\section{Multipolar EOB waveforms for spinning, precessing binary black holes}
\label{sec:multiEOB}

We briefly review the main ideas and building blocks of the EOB approach, and then describe 
an improved spinning, precessing EOBNR waveform model, which, for the first time, also contains 
multipole moments beyond the quadrupolar one. The model  is already
available in the LIGO Algorithm Library~\cite{LAL} under the name of \verb+SEOBNRv4PHM+.
We refer the reader to Refs.~\cite{Taracchini:2012ig,Taracchini:2013rva,Pan:2013rra,Babak:2016tgq,Cotesta:2018fcv}
for more details of the EOB framework and its most recent waveform models. 
Here we closely follow Ref.~\cite{Babak:2016tgq}, highlighting when needed differences with respect to the 
previous precessing waveform model developed in Ref.~\cite{Babak:2016tgq} (\verb+SEOBNRv3P+~\footnote{We note that 
whereas in LAL the name of this waveform approximant is \texttt{ SEOBNRv3}, here we 
add a ``P'' to indicate ``precession'', making the notation uniform with respect to 
the most recent developed model \texttt{ SEOBNRv4P}~\cite{Babak:2016tgq}.}). 

\subsection{Two-body dynamics}

The EOB formalism~\cite{Buonanno:1998gg, Buonanno:2000ef,Damour:2000we,Damour:2001tu} can describe
analytically the GW emission of the entire coalescence process, notably inspiral, merger and ringdown,  
and it can be made highly accurate by including information from NR. For the two-body conservative 
dynamics, the EOB approach relies on a Hamiltonian $H_{\textrm{EOB}}$, which is
constructed through: (i) the Hamiltonian $H_{\textrm{eff}}$
of a spinning particle of mass $\mu \equiv m_1 m_2/(m_1 + m_2)$ and
spin $\vS_* \equiv \vS_*(m_1,m_2,\vS_1,\vS_2)$ moving in an effective,
deformed Kerr spacetime of mass $M\equiv m_1 + m_2$ and spin
$\vS_{\textrm{Kerr}} \equiv \vS_1 +
\vS_2$~\cite{Barausse:2009aa,Barausse:2009xi,Barausse:2011ys}, and (ii) an
energy map between $H_{\textrm{eff}}$ and
$H_{\textrm{EOB}}$~\cite{Buonanno:1998gg}
\be
H_{\textrm{EOB}} \equiv
M\sqrt{1+2\nu\left(\frac{H_{\textrm{eff}}}{\mu} - 1\right)}-M\,,
\ee
where $\nu = \mu/M$ is the symmetric mass ratio. The deformation
of the effective Kerr metric is fixed by requiring that at any given PN order,
the PN-expanded Hamiltonian $H_{\textrm{EOB}}$ agrees with the PN Hamiltonian for
BBHs~\cite{Blanchet:2013haa}. In the EOB Hamiltonian used in this 
paper~\cite{Barausse:2009xi,Barausse:2011ys}, the spin-orbit (spin-spin) couplings are
included up to 3.5PN (2PN) order~\cite{Barausse:2009xi,Barausse:2011ys}, while 
the non-spinning dynamics is incorporated through 4PN order~\cite{Cotesta:2018fcv}. The dynamical variables
in the EOB model are the relative separation $\vR$ and its canonically conjugate momentum $\vP$, and the
spins $\vS_{1,2}$. The conservative EOB dynamics is completely general
and can naturally accommodate precession~\cite{Pan:2013rra,Babak:2016tgq} and 
eccentricity~\cite{Hinderer:2017jcs,Liu:2019jpg,Chiaramello:2020ehz}.

When BH spins have generic orientations, both the orbital plane and the
spins undergo precession about the total angular momentum of the
binary, defined as $\vJ \equiv \vL + \vS_1 + \vS_2$, where $\vL \equiv
\mu\, \vR \times \vP$. We also introduce the Newtonian orbital
angular momentum $\vL_N\equiv \mu\, \vR \times \dot{\vR}$, which at any
instant of time is perpendicular to the binary's orbital plane.
Black-hole spin precession is described by the following equations 
\be
\frac{\mathrm{d}\vS_{1,2}}{\mathrm{d}t} = \frac{\partial H_{\mathrm{EOB}}}{\partial \vS_{1,2}} \times \vS_{1,2}\,.
\ee

In the EOB approach, dissipative effects enter in the equations of motion through a nonconservative 
radiation-reaction force that is expressed in terms of the GW energy flux through the waveform 
multipole moments~\cite{Buonanno:2005xu,Damour:2007xr,Damour:2008gu,Pan:2010hz} as
\be
\label{RRforce}
\boldsymbol{\mathcal{F}} \equiv \frac{\Omega}{16\pi}
\frac{\vP}{|\vL|}\sum_{\ell=2}^{8}\sum_{m=-\ell}^{\ell} m^2\vert
d_{\textrm{L}} h_{\ell m}\vert^2\,,
\ee
where $\Omega \equiv |\vR \times \dot{\vR}|/|\vR|^2$ is the (angular) orbital frequency, $d_{\textrm L}$
is the luminosity distance of the BBH to the observer, and the
$h_{\ell m}$'s are the GW multipole modes. As discussed in 
Refs.~\cite{Cotesta:2018fcv,Bohe:2016gbl}, the $h_{\ell m}$ used in the 
energy flux are not the same as those used for building the 
gravitational polarizations in the inertial frame, since the latter  
include the nonquasi-circular corrections, which enforce that the 
SEOBNR waveforms at merger agree with the NR data, when available.

\begin{figure}
\centering
\includegraphics[angle=0,width=0.7\linewidth]{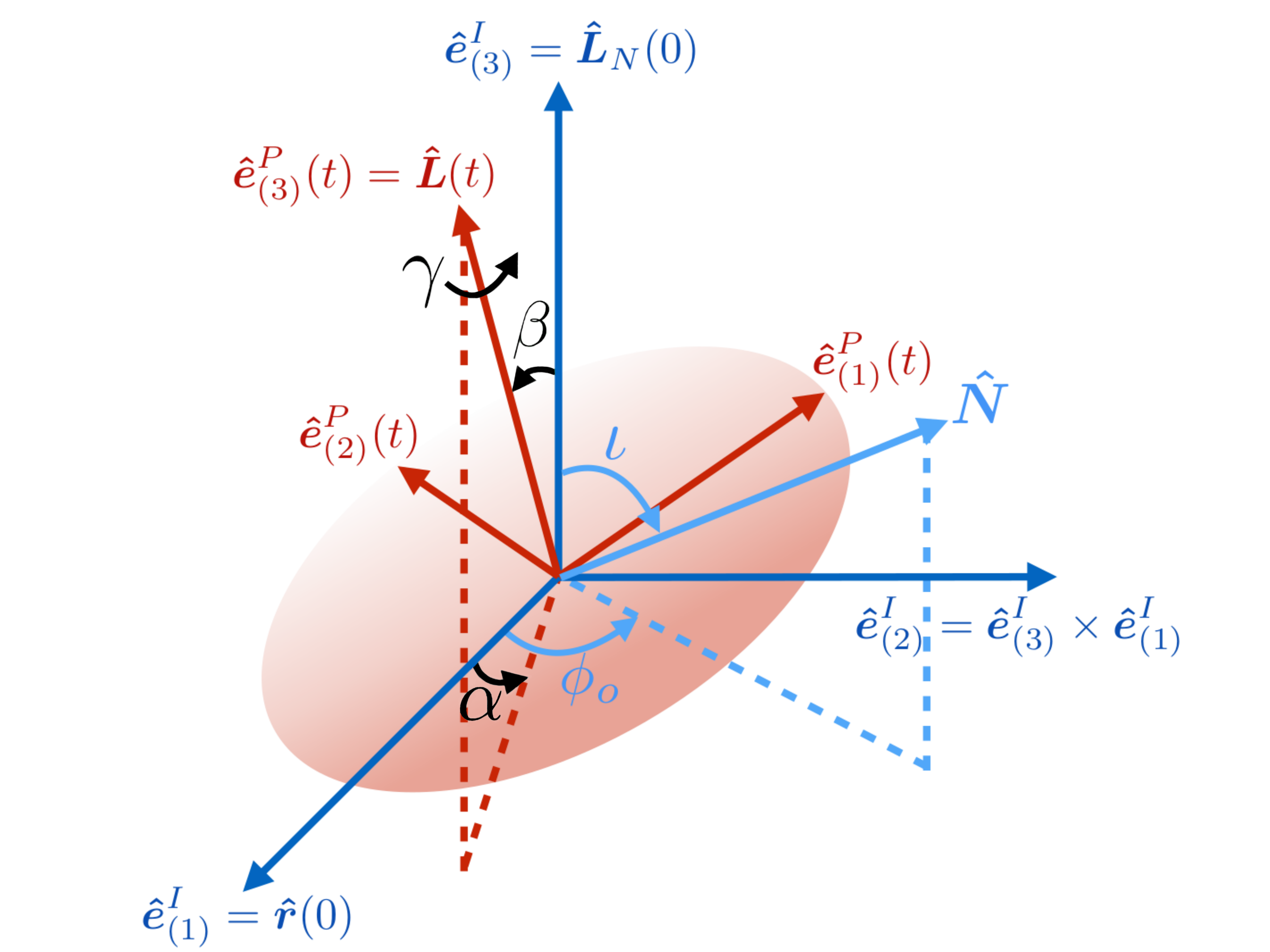}
\caption{Frames used in  the construction of the \texttt{ SEOBNRv4PHM} model: the observer's frame (blue), defined by the directions of the
  initial orbital angular momentum $\vLhat(0)$ and separation
  $\vR(0)$, and co-precessing frame (red), instantaneously aligned
  with $\vLhat(t)$ and described by the Euler angles $(\alpha, \beta,
  \gamma)$ (see text below for details). \label{F:Pframe}}
\end{figure}

\subsection{Inspiral-plunge waveforms}

For the inspiral-plunge waveform, the EOB approach uses a factorized, resummed 
version~\cite{Damour:2007xr,Damour:2008gu,Pan:2010hz,Cotesta:2018fcv} of the 
frequency-domain PN formulas of the modes~\cite{Arun:2008kb,Mishra:2016whh}.
As today, the factorized resummation has been developed only for quasicircular, 
nonprecessing BBHs~\cite{Damour:2008gu,Pan:2010hz}, and it has been shown to improve the accuracy of the PN expressions in the 
test-particle limit, where one can compare EOB predictions to numerical solutions of the
Regge-Wheeler-Zerilli and Teukolsky equations~\cite{Bernuzzi:2011aj,Barausse:2011kb,Taracchini:2013wfa,Harms:2015ixa}. 

The radiation-reaction force $\boldsymbol{\mathcal{F}}$ in Eq.~(\ref{RRforce}) 
depends on the amplitude of the individual GW modes $|h_{\ell m}|$, which, in the non-precessing
case, are functions of the constant aligned-spin magnitudes $\boldsymbol{\chi}_{1,2}\cdot \vLhat$.
In the precessing case, these modes depend on time, as  
$\boldsymbol{\chi}_{1,2}(t)\cdot \vLhat(t)$, and they depend on the generic, 
precessing orbital dynamics through the radial separation $r$ and orbital frequency $\Omega$, which carry 
modulations due to spin-spin couplings whenever precession is present. However, we stress that 
with this choice of the radiation-reaction force and waveform model, not all spin-precession effects are included, since 
the PN formulas of the modes~\cite{Arun:2008kb} also contain terms that depend 
on the in-plane spin components. 

For data-analysis purposes, we need to compute the GW polarizations in the inertial-frame
of the observer (or simply \textit{observer's frame}). We denote quantities 
in this frame with the superscript $I$. The observer's frame is defined  
by the triad $\{\vE^I_{(i)}\}$ ($i=1,2,3$), where
$\vE_{(1)}^I\equiv\boldsymbol{\hat{r}}(0)$, $\vE^I_{(3)} \equiv
\vLhat_N(0)$ and $\vE_{(2)}^I\equiv\vE_{(3)}^I \times \vE_{(1)}^I$. Moreover, in
this frame, the line of sight of the observer is parametrized as
$\boldsymbol{\hat{N}} \equiv
(\sin{\iota}\cos{\phi_o},\sin{\iota}\sin{\phi_o},\cos{\iota})$ (see
Fig.~\ref{F:Pframe}). We also introduce the observer's frame with the 
polarization basis $\{\vE_{(1)}^r,\vE_{(2)}^r\}$ such that $\vE_{(1)}^r \equiv
(\vE^I_{(3)}\times\boldsymbol{\hat{N}})/|\vE^I_{(3)}\times\boldsymbol{\hat{N}}|$
and $\vE_{(2)}^r \equiv \boldsymbol{\hat{N}} \times \vE_{(1)}^r$, 
which spans the plane orthogonal to $\boldsymbol{\hat{\vN}}$.

To compute the observer's-frame modes $h_{\ell m}^I$ during the
inspiral-plunge stage, it is convenient to introduce a non-inertial reference
frame that tracks the motion of the orbital plane, the so-called 
\textit{co-precessing frame} (superscript $P$), described 
by the triad $\{\vE^P_{(i)}\}$ ($i=1,2,3$). At each instant, its $z$-axis
is aligned with $\vLhat$: $\vE_{(3)}^P \equiv \vLhat(t)$~\footnote{Note that in 
Ref.~\cite{Babak:2016tgq}, the $z$-axis is aligned with $\vLhat_\mathrm{ N}$ instead of $\vLhat$.}. In this frame, 
the BBH is viewed face-on at all times, and the GW radiation looks very much
nonprecessing~\cite{Buonanno:2002fy,Schmidt:2010it,Boyle:2011gg,O'Shaughnessy:2011fx,Schmidt:2012rh}. The
other two axes lie in the orbital plane and are defined such as they minimize
precessional effects in the precessing-frame modes $h_{\ell
  m}^P$~\cite{Buonanno:2002fy,Boyle:2011gg}. After introducing the
vector $\boldsymbol{\Omega}_e \equiv \vLhat\times {\textrm
  d}\vLhat/{\textrm d}t$, we enforce the minimum-rotation condition 
by requiring that ${\textrm d}\vE_{(1),(2)}^P/{\textrm d}t =
\boldsymbol{\Omega}_e\times \vE_{(1),(2)}^P$ and $\vE_{(1),(2)}^P(0) =
\vE^I_{(1),(2)}$ (see also Fig.~\ref{F:Pframe}). As usual, we parametrize the
rotation from the precessing to the observer's frame through 
time-dependent Euler angles $(\alpha(t),\beta(t),\gamma(t))$, which 
we compute using Eqs.~(A4)--(A6) in Appendix A of Ref.~\cite{Babak:2016tgq}. We notice that 
the minimum-rotation condition can also be expressed through the following differential 
equation for $\gamma$: $\dot{\gamma} = -\dot{\alpha}\cos{\beta}$ with $\gamma(0)=-\alpha(0) =
\pi/2$.

We compute the precessing-frame inspiral-plunge modes just like we do for the GW flux, 
namely by evaluating the factorized, resummed nonprecessing multipolar waveforms 
along the EOB precessing dynamics, and employing the time-dependent spin projections
$\boldsymbol{\chi}_{1,2}(t)\cdot \vLhat(t)$. Finally, the observer's-frame inspiral-plunge 
modes are obtained by rotating the precessing-frame inspiral-plunge modes using 
Eq.~(A13) in Appendix A of Ref.~\cite{Babak:2016tgq}. 

Following Ref.~\cite{Cotesta:2018fcv}, where an EOBNR  
nonprecessing multipolar  waveform  model was developed (\verb+SEOBNRv4HM+), 
here we include in the precessing frame of the \verb+SEOBNRv4PHM+ model the $(2,\pm 2), (2,\pm 1), (3,\pm 3), (4,\pm 4)$ 
and $(5,\pm 5)$ modes, 
and make the assumption $h^P_{l-m} = (-1)^{l}h^{P*}_{lm}$.  As shown in Sec.~\ref{sec:mode_asymm}, we expect that inaccuracies 
due to neglecting mode asymmetries should remain mild, or at most at the level of other modeling errors.

\subsection{Merger-ringdown waveforms}

The description of a BBH as a system composed of two individual
objects is of course valid only up to the merger. After that point,
the EOB model builds the GW emission (ringdown stage) via a 
phenomenological model of the quasinormal modes (QNMs) of the remnant BH, 
which forms after the coalescence of the progenitors. 
The QNM frequencies and decay times are known (tabulated) functions of the mass
$M_f$ and spin $\vS_f \equiv M_f^2 \boldsymbol{\chi}_f$ of the remnant
BH~\cite{Berti:2005ys}. Since the QNMs are defined with respect to the direction 
of the final spin, the specific form of the ringdown signal, as a linear
combination of QNMs, is formally valid only in an inertial frame whose
$z$-axis is parallel to $\boldsymbol{\chi}_f$. 

A novel feature of the \verb+SEOBNRv4PHM+ waveform model presented here is that we attach 
the merger-ringdown waveform (notably each multipole mode $h_{\ell m}^\mathrm{ mergr-RD}$) 
directly in the co-precessing frame, instead of the observer's frame. As a consequence, 
we can employ here the merger-ringdown multipolar model developed for non-precessing BBHs 
(\verb+SEOBNRv4HM+) in Ref.~\cite{Cotesta:2018fcv} (see Sec.~IVE therein for details). 
By contrast, in the \verb+SEOBNRv3P+ waveform model~\cite{Babak:2016tgq}, the 
merger-ringdown waveform was built as a superposition of 
QNMs in an inertial frame aligned with the direction of the remnant spin.  
This construction was both more complicated to implement and more prone to numerical
instabilities.

To compute the waveform in the observer's frame, our approach requires 
a description of the co-precessing frame Euler angles $(\alpha, \beta, \gamma)$ that
extends beyond the merger. To prescribe this, we take advantage of
insights from NR simulations~\cite{OShaughnessy:2012iol}.  In
particular, it was shown that the co-precessing frame continues to
precess roughly around the direction of the final spin with a
precession frequency approximately equal to the differences between
the lowest overtone of the (2,2) and (2,1) QNM frequencies, while the
opening angle of the precession cone decreases somewhat at merger. We
find that this behavior is qualitatively correct for the NR waveforms used for 
comparison in this paper.

To keep our model generic for a wide range of mass ratios and spins,
we need an extension of the behavior noticed in Ref.~\cite{OShaughnessy:2012iol} 
to the retrograde case, where the remnant spin is negatively aligned with the orbital 
angular momentum at merger. Such configurations can occur for high mass-ratio binaries, when
the total angular momentum $\vJ$ is dominated by the spin of the
primary $\vS_{1}$ instead of the orbital angular momentum
$\vL$. This regime is not well explored by NR simulations, 
and includes in particular systems presenting transitional
precession~\cite{Apostolatos:1994mx}. In our model we keep imposing simple precession around the
direction of the remnant spin at a rate $\omega_\mathrm{ prec} \geq 0$,
but we distinguish two cases depending on the direction of the final
spin $\bm{\chi}_{f}$ (approximated by the total angular momentum
$\vJ = \vL+ \vS_{1}+ \vS_{2}$ at merger) relative to the final orbital
angular momentum $\vL_{f}$:
\begin{equation}\label{eq:postmergerprec}
  \dot{\alpha} = \omega_\mathrm{ prec}= \left\{\begin{aligned}
   &\omega^\mathrm{ QNM}_{22}(\chi_{f})-\omega^\mathrm{ QNM}_{21}(\chi_{f}) \quad \text{if} \quad \bm{\chi}_{f} \cdot \bm{L}_{f} > 0\\
   &\omega^\mathrm{ QNM}_{2-1}(\chi_{f})-\omega^\mathrm{ QNM}_{2-2}(\chi_{f}) \quad \text{if} \quad \bm{\chi}_{f} \cdot \bm{L}_{f} < 0
\end{aligned}\right.
\end{equation}
where $\chi_{f} = |\bm{\chi}_{f}|$, and the zero-overtone QNM frequencies for
negative $m$ are taken on the branch $\omega^\mathrm{ QNM}_{lm} > 0$ that
continuously extends the $m>0$, $\omega^\mathrm{ QNM}_{lm} > 0$
branch~\cite{Berti:2005ys} (the QNM refers to zero overtone). In both cases, $\dot{\alpha} \geq 0$. We
do not attempt to model the closing of the opening angle of the
precession cone and simply consider it to be constant during the post-merger phase,
$\beta = \mathrm{const}$. The third Euler angle $\gamma$ is then
constructed from the minimal rotation condition $\dot{\gamma} =
-\dot{\alpha} \cos\beta$. The integration constants are determined by
matching with the inspiral at merger. We find that the behavior
of  Eq.~\eqref{eq:postmergerprec} in the case $\bm{\chi}_{f} \cdot
\bm{L}_{f} < 0$ is qualitatively consistent with an NR simulation investigated 
by one of us~\cite{Ossokine:2020}. However, we stress that this
prescription for the retrograde case is much less tested than for the
prograde case.

\begin{figure}
\centering
    \includegraphics[width=0.7\linewidth]{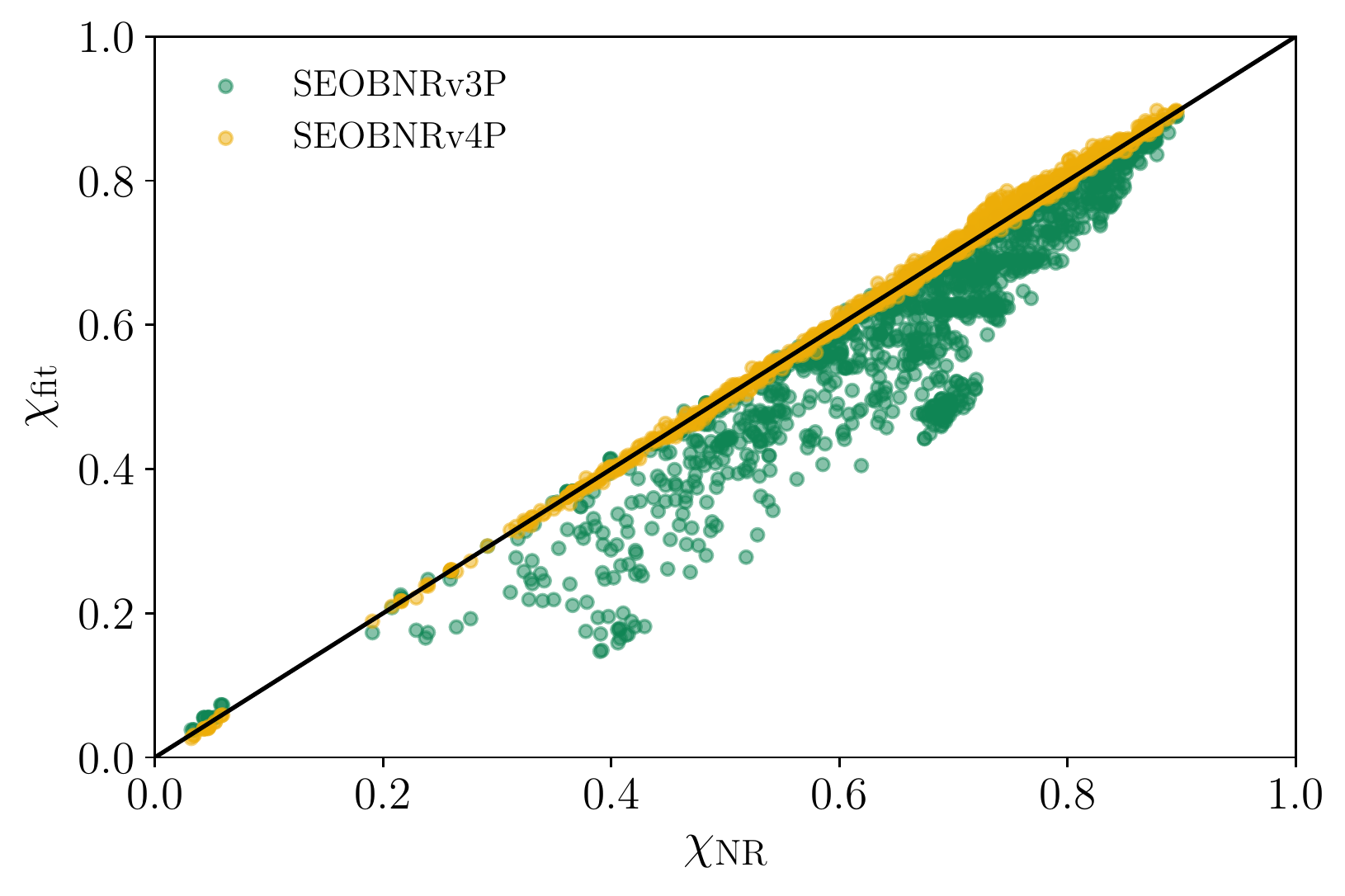}
    \caption{Comparison of the magnitude of the final spin  between the \texttt{ SEOBNRv3P} and
    \texttt{ SEOBNRv4P} models and NR results. For simplicity, the fits are evaluated 
using the NR data at the relaxed time. The black line is the identity. It is obvious that
    \texttt{ SEOBNRv4P} gives final-spin magnitudes much closer to the NR values.}
    \label{fig:final_spin}
\end{figure}

Furthermore, one crucial aspect of the above construction is the mapping from the
binary's component masses and spins to the final mass and spin, which is 
needed to compute the QNM frequencies of the merger remnant. Many groups have developed fitting formulae based
on a large number of NR simulations (e.g., see Ref.~\cite{Varma:2018aht} for an overview).
To improve the agreement of our EOB merger-ringdown model with NR, and to ensure agreement
in the aligned-spin limit with \verb+SEOBNRv4+~\cite{Bohe:2016gbl} and \verb+SEOBNRv4HM+~\cite{Cotesta:2018fcv},
we employ the fits from Hofmann et al.~\cite{Hofmann:2016yih}. In Fig.~\ref{fig:final_spin} we compare 
the performance of the fit used in the previous EOB precessing model \verb+SEOBNRv3P+~\cite{Pan:2013rra,Taracchini:2013rva,
Babak:2016tgq} to the fit from Hofmann et al. that we adopt for \verb+SEOBNRv4PHM+. It is clear that the new fit reproduces NR 
data much better. This in turn improves the correspondence between NR and EOB QNM frequencies.

For the final mass we employ the same fit as in previous EOB models, and we provide it here since it was not given 
explicitly anywhere before:

\bea
\frac{M_{f}}{M} &=& 1 - \left\{[1 - E_\mathrm{ ISCO}(a)] \nu + 16\nu^{2} \left [ 0.00258 \right. \right. \nonumber \\
&& \left. \left. - \frac{0.0773 }{\left [a\,(1 +1/q)^{2}/(1 + 1/q^{2}) - 1.6939 \right ]} \right . \right .
\nonumber \\
&& \left. \left. - \frac{1}{4}(1 - E_\mathrm{ ISCO}(a))\right ]\right\}\,,
\eea
where $a = \vLhat \cdot (\vchi_{1}+\vchi_{2}/q^{2})/(1+1/q)^{2}$, 
and $E_\mathrm{ ISCO}(a)$ is the binding energy of the Kerr spacetime at the innermost stable circular orbit~\cite{Bardeen:1972fi}.

Finally, for precessing binaries, the individual components of the spins vary
with time. Therefore, in applying the fitting formulae to obtain final mass and spin, one must make a crucial choice in
selecting the time during the inspiral stage at which the spin directions are evaluated. In fact, even if one considers 
a given physical configuration, evaluating the final spin formulae with spin directions from
different times yields different final spins and consequently different
waveforms. We choose to evaluate the spins at a time corresponding to the
separation of $r=10M$. This choice is guided by two considerations: by the
empirical finding of good agreement with NR (e.g., performing better than
using the time at which the inspiral-plunge waveform is attached to the merger-ringdown 
waveform~\cite{Cotesta:2018fcv}), and by the restriction that the waveform must
start at $r>10.5 M$ in order to have small initial eccentricity~\cite{Babak:2016tgq}. 
Thus, our choice ensures that a given physical configuration always
produces the same waveform regardless of the initial starting frequency.

\begin{figure*}
\centering
	\includegraphics[width=\linewidth]{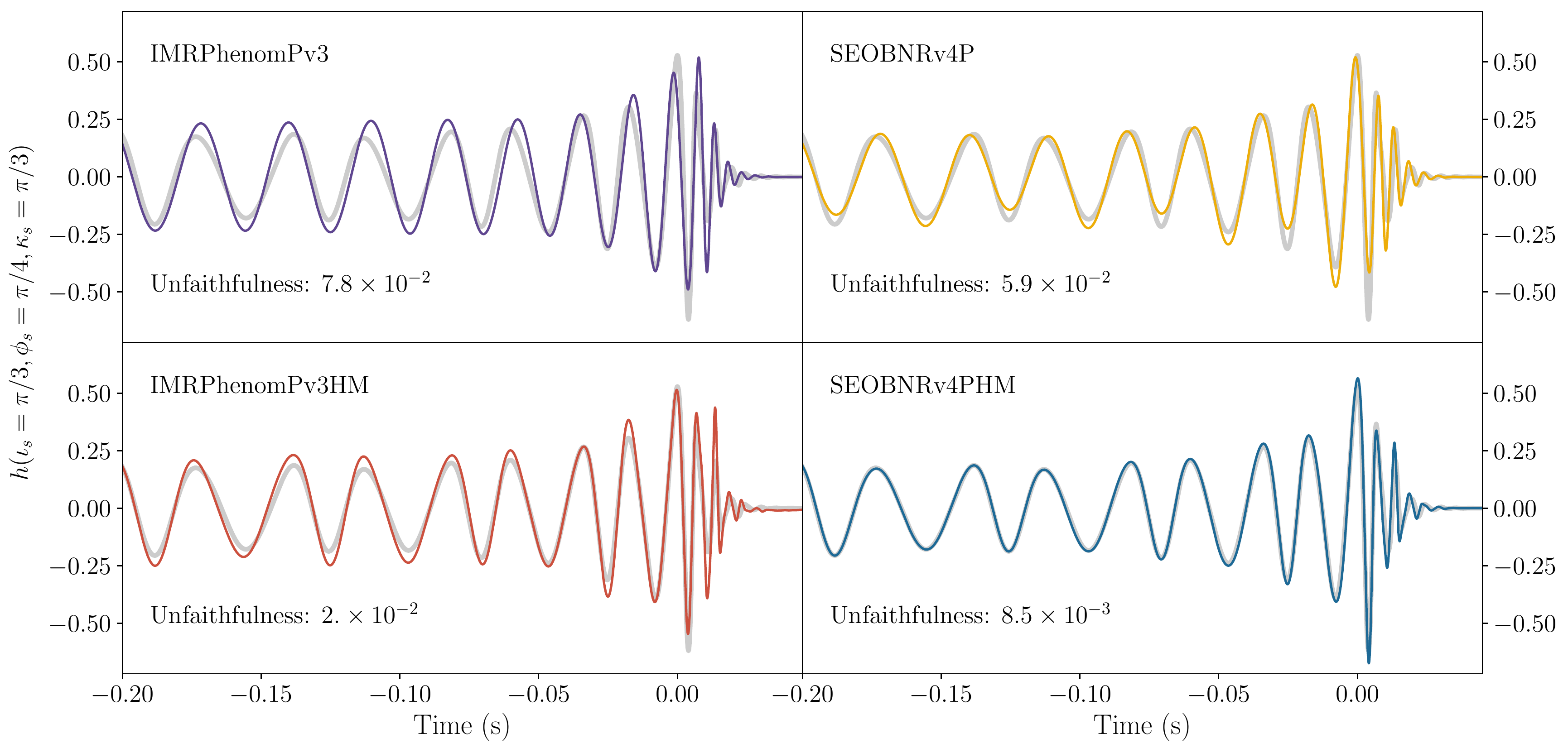}
	\caption{Time-domain comparison of state-of-the art waveform
          models to the NR waveform \texttt{ PrecBBH00078} with mass ratio
          $4$, BH's spins $0.7$ and total mass $M=70 M_\odot$. The
          source parameters are $\iota_{s}=\pi/3,\ \phi_{s}=\pi/4,\
          \kappa_{s}=\pi/4$. The NR waveform includes all modes up to
          and including $\ell=4$, and extends for 44 GW cycles before
          merger.  For models that include only $\ell=2$ modes, 
          the unfaithfulness are several
          percent 8\% for \texttt{ IMRPhenomPv3} and 6\% for \texttt{
            SEOBNRv4P}.  Meanwhile, adding the higher mode content
          drastically improves the agreement, with mismatches going
          down to 2\% for \texttt{ IMRPhenomPv3HM} and 1\% for \texttt{
            SEOBNRv4PHM}.  The agreement is particular good for \texttt{
            SEOBNRv4PHM}, which reproduces the higher mode features at
          merger and ringdown faithfully.}
	\label{fig:time_domain_waveform}
\end{figure*}

To obtain the inspiral-merger-ringdown modes in the inertial frame, $h^\mathrm{ I}_{\ell m}$, 
we rotate the inspiral-merger-ringdown modes $h^\mathrm{ P}_{\ell m}$ from the co-precessing  frame to 
the observer's frame using the rotation formulas and Euler angles in Appendix A of Ref.~\cite{Babak:2016tgq}. 
The inertial frame polarizations then read
\begin{equation}
h^\mathrm{ I}_{+}(\varphi_0,\iota;t) - i h^\mathrm{ I}_{\times}(\varphi_0,\iota;t) = \sum_{\ell, m} {}_{-2} Y_{\ell m}(\varphi_0,\iota)\,h^\mathrm{ I}_{\ell m}(t) \,.
\end{equation}

\subsection{On the fits of calibration parameters in presence of precession}

The \verb+SEOBNRv4PHM+ waveform model inherits the EOB Hamiltonian and GW energy flux from the aligned-spin model 
\verb+SEOBNRv4+~\cite{Bohe:2016gbl}, which features higher (yet unknown) PN-order terms in the dynamics 
calibrated to NR waveforms. These calibration parameters were denoted $K,d_\mathrm{ SO}$ and $d_\mathrm{ SS}$ in Ref.~\cite{Bohe:2016gbl}, 
and were fitted to NR and Teukolsky-equation--based waveforms
as polynomials in $\nu,\chi$ where $\chi \equiv {S_\mathrm{ Kerr}^{z}}/({1-2\nu}$) with
${\vS}_\mathrm{ Kerr}={\mathbf S}_{1}+{\mathbf S}_{2}$ the spin of the EOB background 
spacetime. In contrast to the \verb+SEOBNRv3P+ waveform model, which used the EOB Hamiltonian and GW energy flux 
from the aligned-spin model \verb+SEOBNRv2+\cite{Taracchini:2013rva}, the fits in 
Ref.~\cite{Bohe:2016gbl} include odd powers of $\chi$ and thus the sign of $\chi$ matters 
when the BHs precess. 

The most natural way to generalize these fits to the precessing case is to
project $\vS_\mathrm{ Kerr}$ onto the orbital angular momentum $\hat{\vL}$ 
 in the usual spirit of reducing precessing 
quantities to corresponding aligned-spin ones. To test the impact of this prescription, we compute 
the sky-and-polarization-averaged unfaithfulness with the set of 118 NR simulations
described in Sec.~\ref{sec:NR}, and find that while the majority of the cases have low 
unfaithfulness ($\sim 1$\%), there are a handful of cases where it is significant($\sim10$\%), 
with many of them having large in-plane spins. 

To eliminate the high mismatches, we introduce the \emph{augmented spin} that includes contribution of the in-plane spins: 
\begin{equation}
  \tilde{\chi}=\frac{{\vS}_\mathrm{ Kerr}\cdot {\vL}}{1-2\nu}+\alpha\frac{({\vS}_{1}^{\perp}+{\vS}_{2}^{\perp})\cdot 
{\vS}_\mathrm{ Kerr}}{|{\vS}_\mathrm{ Kerr}|(1-2\nu)}\,.
\label{eq:defn}
\end{equation}
Here ${\vS}_{i}^{\perp}\equiv {\vS}_{i}-({\vS}_{i}\cdot {\vL}) {\vL}$
and $\alpha$ is a \emph{positive} coefficient to be determined. Note that the
extra term in the definition of the augmented spin $\ge0$ for any combination
of the spins. We set $\tilde{\chi}=0$ when ${\vS}_\mathrm{ Kerr}=0$. Fixing \(\alpha={1}/{2}\) 
insures that the augmented spin obeys the Kerr bound. Using the augmented spin eliminates all mismatches 
above $6\%$, and thus greatly improves the agreement of the model with NR data. 

\section{Comparison of multipolar precessing models to numerical-relativity waveforms}
\label{sec:compEOBNR}

To assess the impact of the improvements incorporated in the \verb+SEOBNRv4PHM+ waveform model, we compare this model and other models publicly available 
in LAL (see Table~\ref{tbl:wf_models}) to the set of simulations described in Sec.~\ref{sec:NR}, as well as to all publicly available precessing \texttt{ SpEC} simulations~\footnote{The list of all
SXS simulations used can be found in \url{https://arxiv.org/src/1904.04831v2/anc/sxs_catalog.json}}.

We start by comparing in Fig.~\ref{fig:time_domain_waveform}, the precessing NR
waveform \verb+PrecBBH00078+ with mass ratio $4$, BH's spin magnitudes $0.7$, total mass $M=70 M_\odot$ 
and modes $\ell \leq 4$ from the new 118 SXS catalog (see Appendix~\ref{sec:NRparam}) to the precessing waveforms \verb+IMRPhenomPv3+ and \verb+SEOBNRv4P+ with modes $\ell = 2$ (upper panels), and to 
the precessing multipolar waveforms \verb+IMRPhenomPv3HM+ and \verb+SEOBNRv4PHM+
(lower panels). This NR waveform is the most ``extreme'' configuration from the
new set of waveforms and has about 44 GW cycles before merger, and the plot only shows 
the last $7$ cycles. More specifically, we plot the detector 
response function given in Eq.~(\ref{eq:det_strain_kappa}), but we leave out the overall constant amplitude. 
We indicate on the panels the unfaithfulness for the different cases. We note the improvement when including modes beyond the quadrupole. 
\verb+SEOBNRv4PHM+ agrees particularly well to this NR waveform, reproducing accurately the higher-mode features throughout merger and ringdown. 

\begin{figure}
\centering
  \includegraphics[width=0.8\linewidth]{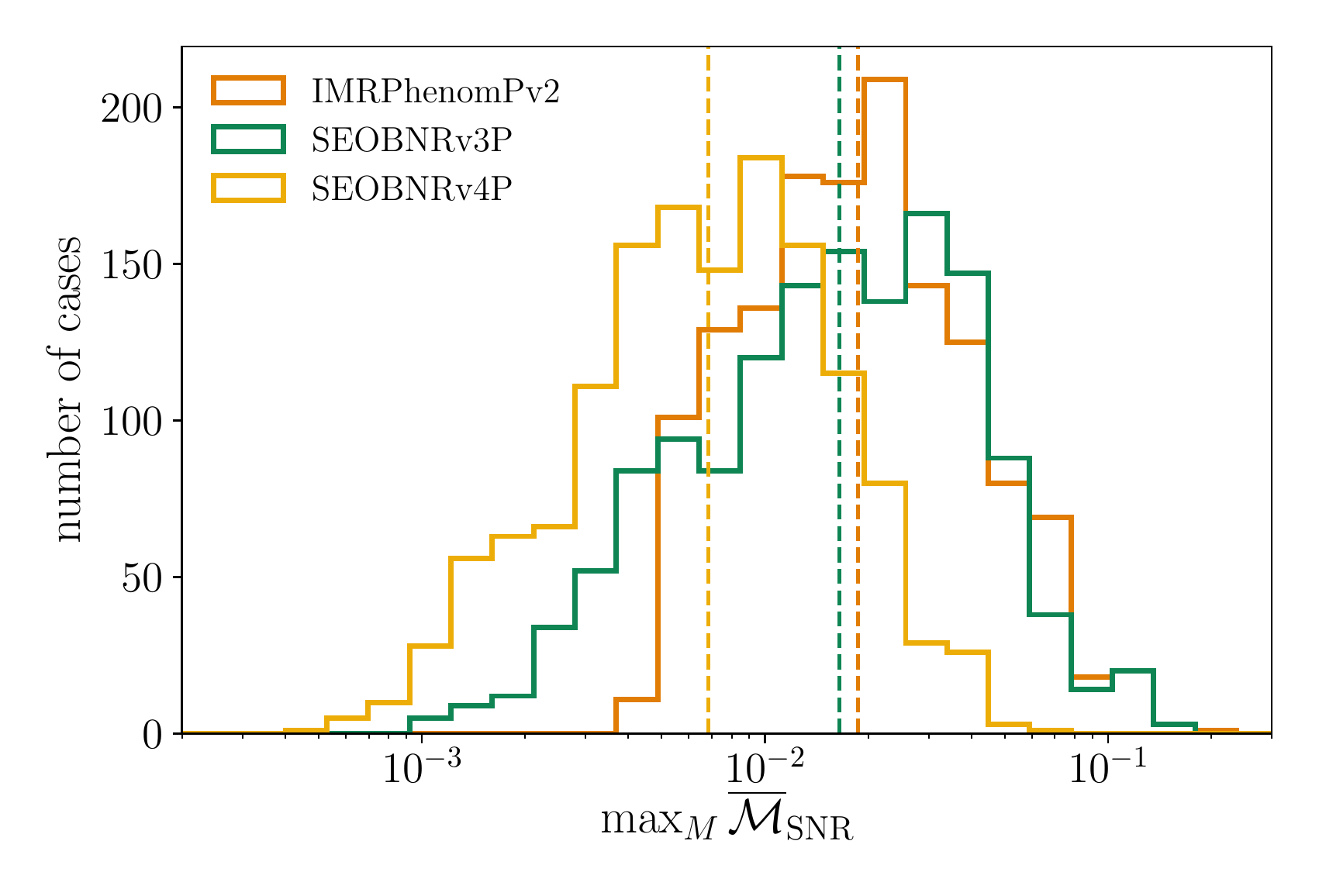}
  \caption{Sky-and-polarization averaged, SNR weighted unfaithfulness for
  an inclination $\iota=\pi/3$ between NR waveforms with $\ell =2$ and \texttt{ SEOBNRv4P}, and also 
\texttt{ SEOBNRv3P} and \texttt{ IMRPhenomPv2}, which were used in LIGO/Virgo publications.  The vertical dashed lines show the medians. 
It is evident the better performance of the newly developed precessing model \texttt{ SEOBNRv4P}.}
  \label{fig:all_approximants_ell2}
\end{figure}

\begin{figure}
\centering
  \includegraphics[width=0.8\linewidth]{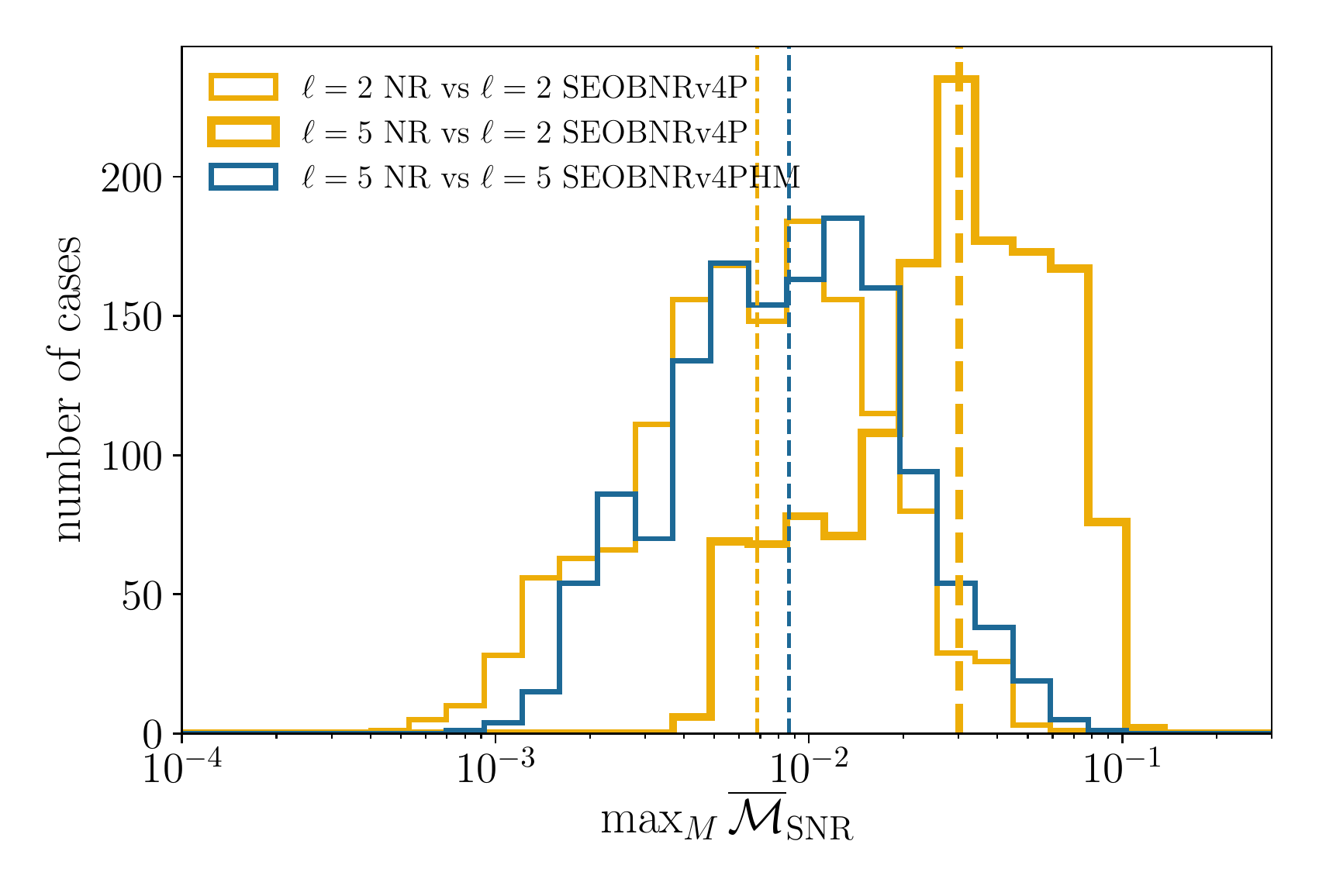}
  \caption{Sky-and-polarization averaged, SNR weighted unfaithfulness for
  an inclination $\iota=\pi/3$ between NR waveforms and \texttt{ SEOBNRv4PHM},
  including and omitting higher modes. The vertical dashed lines show the medians. Not including higher modes
  in the model results in high unfaithfulness. However, when they are 
included, the unfaithfulness between \texttt{ SEOBNRv4PHM} and NR is
  essentially at the same level as when only $\ell=2$ modes are compared (see Fig.~\ref{fig:all_approximants_ell2}).}
  \label{fig:higher_mode_effects}
\end{figure}

\begin{figure*}
\centering
	\includegraphics[width=\linewidth]{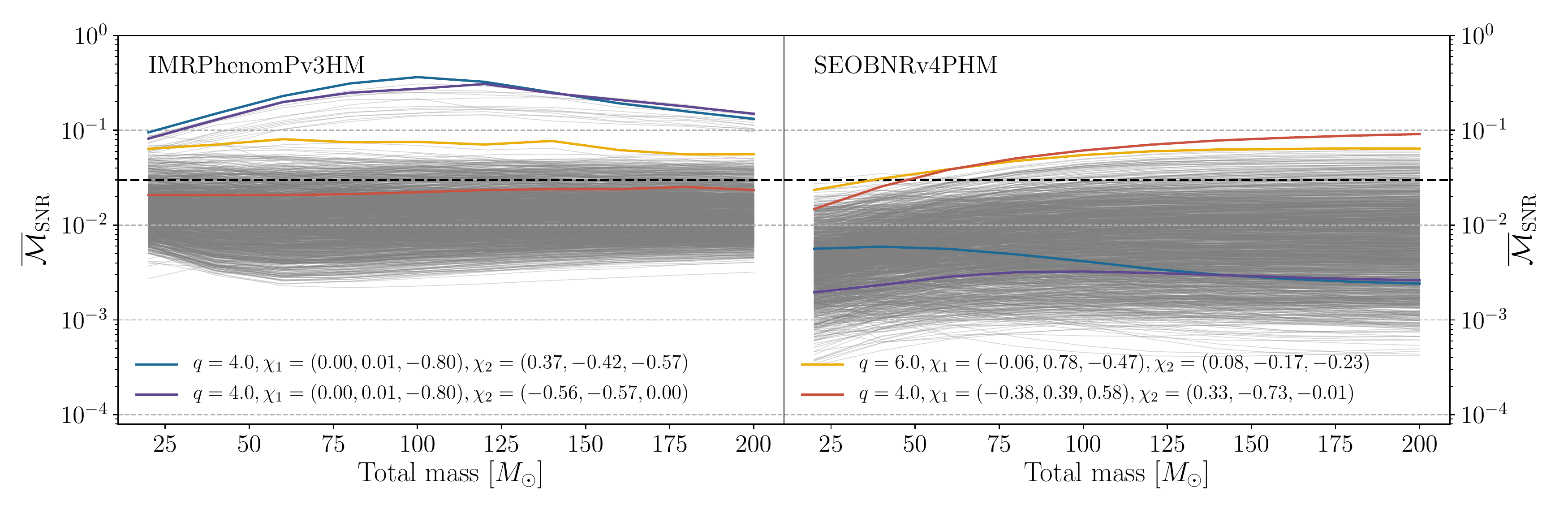}
	\caption{The sky-and-polarization averaged, SNR-weighted
          unfaithfulness as a function of binary's total mass for inclination
          $\iota=\pi/3$, between \texttt{ IMRPhenomPv3HM} and NR (left) and
          \texttt{ SEOBNRv4PHM} and NR (right) for 1404 quasi-circular
          precessing BBH simulations in the SXS public catalog. The
          colored lines highlight the cases with the worst maximum
          mismatches for both models. Note that for the majority of
          cases, both models have unfaithfulness below 5\%, but
          \texttt{ SEOBNRv4PHM} has no outliers beyond 10\% and many more cases
          at lower unfaithfulness.}
	 \label{fig:spaghetti_public}
\end{figure*}

We now turn to the public precessing SXS NR catalog of 1404 waveforms. First, to quantify the performance of the new precessing waveform model \verb+SEOBNRv4P+ 
with respect to previous precessing models used in LIGO and Virgo inference studies, we compute the unfaithfulness~\footnote{We always use the sky-and-polarization averaged, SNR-weighted faithfulness or 
unfaithfulness $\overline{\mathcal{M}}_\mathrm{ SNR}$ unless otherwise stated.} against the precessing NR catalog, including only the dominant
$\ell=2$ multipoles in the co-precessing frame. Figure~\ref{fig:all_approximants_ell2} shows the histograms of the largest
mismatches when the binary total mass varies in the range $[20,200]M_\odot$. Here, we also consider the
precessing waveform models used in the first GW Transient Catalog~\cite{LIGOScientific:2018mvr} of the 
LIGO and Virgo collaboration (i.e., \verb+SEOBNRv3P+ and \verb+IMRPhenomPv2+).
Two trends are apparent: firstly, \verb+SEOBNRv3P+ and \verb+IMRPhenomPv2+ distributions are
broadly consistent, with both models having mismatches which extend beyond $10\%$ , although \verb+SEOBNRv3+ has more cases at lower unfaithfulness; secondly, \verb+SEOBNRv4P+ 
has  a distribution which is shifted to much lower values of the unfaithfulness and does not include outliers with the largest
unfaithfulness below $7\%$. 

Next, we examine the importance of higher modes. To do so, we use \verb+SEOBNRv4PHM+ 
with and without the higher modes, while always including all modes up to
$\ell=5$ in the NR waveforms. As can be seen in Fig.~\ref{fig:higher_mode_effects},
if higher modes are omitted, the unfaithfulness can be very large, with a
significant number of cases having unfaithfulness $>7\%$, as has been seen in many past studies. 
On the other hand, once higher modes are included in the model, the distribution of mismatches becomes
much narrower, with all mismatches below $9\%$. Furthermore, the distribution now
closely resembles the distribution of mismatches when only $\ell=2$ modes were
included in the  NR waveforms. Thus, we see that higher modes play an important role
and are accurately captured by \verb+SEOBNRv4PHM+ waveform model.

Moreover, in Fig.~\ref{fig:spaghetti_public} we display, for a specific choice of the inclination, the 
unfaithfulness versus the binary's total mass between the public precessing SXS NR catalog and 
\verb+SEOBNRv4PHM+ and \verb+IMRPhenomPv3HM+. We highlight with curves in color the NR configurations having worst maximum  
mismatches for the two classes of approximants. For the majority of cases, both models have unfaithfulness below 5\%, but 
\verb+SEOBNRv4PHM+ has no outliers beyond 10\% and many more cases at lower unfaithfulness ($<2\times 10^{-3}$). We find that
the large values of unfaithfulness above $10\%$ for \verb+IMRPhenomPv3HM+ come from simulations with $q\gtrsim4$ and large anti-aligned primary spin,
i.e. $\chi_{1}^{z}=-0.8$.  An examination of the waveforms  in this region reveals that unphysical features develop
in the waveforms, with unusual oscillations both in amplitude and phase. For lower spin magnitudes these features are milder, and disappear
for spin magnitudes $\lesssim0.65$. These features are present also in \verb+IMRPhenomPv3+ and are thus connected to the
precession dynamics, a region already known to potentially pose a challenge when modeling the 
precession dynamics as suggested in Ref.~\cite{Chatziioannou:2017tdw}, and adopted in Ref.~\cite{Khan:2019kot}.

\begin{figure*}
\centering
	\includegraphics[width=\linewidth]{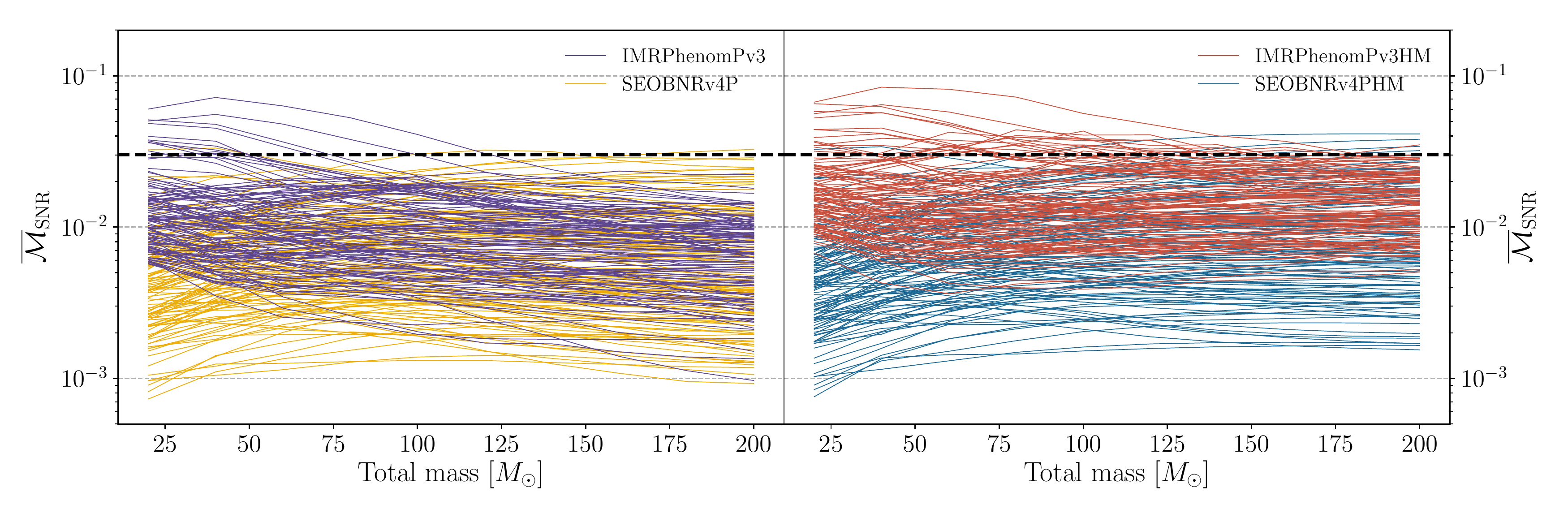}
	\caption{The sky-and-polarization averaged, SNR-weighted
		unfaithfulness as a function of binary's total mass for inclination
		$\iota=\pi/3$, between \texttt{ IMRPhenomPv3} and \texttt{ SEOBNRv4P} and NR (left),  and
		\texttt{ IMRPhenomPv3HM} and \texttt{ SEOBNRv4PHM} and NR (right) for the 118 SXS NR waveforms described in Appendix~\ref{sec:NRparam}. 
		The NR data has $\ell=2$ modes for the left panel, while all modes up to and including $\ell=4$ in the right panel. The unfaithfulness is low
	using both waveform families, however, \texttt{ SEOBNRv4P(HM)} has fewer cases above $3\%$, and the distribution is consistently shifted to lower values 
of unfaithfulness.}
	\label{fig:spaghetti_PrecBBH}
\end{figure*}

We now focus on the comparisons with the 118 SXS NR waveforms produced
in this paper. In Fig.~\ref{fig:spaghetti_PrecBBH} we show the
unfaithfulness for \verb+IMRPhenomPv3(HM)+ and \verb+SEOBNRv4P(HM)+ in
the left (right) panels. We compare waveforms without higher modes, to
NR data that has only the $\ell=2$ modes, and the other models to NR
data with $\ell\leq4$ modes. The performance of both waveform models
on this new NR data set is largely comparable to what was found for
the public catalog. Both families perform well on average, with most
cases having unfaithfulness below 2\% for models without higher modes
and 3\% for models with higher modes. However, for some configurations 
\texttt{ IMRPhenomPv3(HM)} reaches unfaithfulness values above $3\%$ for 
total masses below $125 M_\odot$. Once again, the overall
distribution is shifted to lower unfaithfulness values for
\verb+SEOBNRv4P(HM)+.

When studying the distribution of unfaithfulness for these 118 cases across parameter space, it is useful to introduce the widely used effective $\chi_\mathrm{ eff}$~\cite{Damour:2001tu,Racine:2008qv,Santamaria:2010yb} and precessing $\chi_{p}$~\cite{Schmidt:2014iyl} spins. These capture the leading order aligned-spin and precession effects respectively, and are defined as
\begin{subequations}
\begin{align}
\chi_\mathrm{ eff} &= \frac{(m_1\vchi_{1}+m_2\vchi_{2})}{m_1+m_2}\cdot\vLhat_N\,, \\
\chi_\mathrm{ p}  &= \frac{1}{B_{1}m_{1}^{2}}\max(B_{1}m_{1}^{2}\chi_{1\perp},B_{2}m_{2}^{2}\chi_{2\perp})\,,
\end{align}
\end{subequations}
where with $B_{1} = 2+3{m_{2}}/{m_{1}}$, $B_{2}=2+3{m_1}/{m_2}$ and we indicate with $\chi_{i\perp}$ the projection of the spins  
on the orbital plane. We find that the unfaithfulness shows 2 general trends. First, it tends to increase with increasing $\chi_\mathrm{ eff}$ and $\chi_\mathrm{ p}$. 
Secondly, that cases with positive $\chi_\mathrm{ eff}$ (i.e. aligned with Newtonian 
orbital angular momentum) tend to have larger unfaithfulness. This is likely driven 
by the fact that inspiral is longer for such cases and the binary merges at higher frequency. We do not find
any other significant trends based on spin directions.  It is interesting to note that the distribution of mismatches from the 118 cases is quite similar
to the distribution from the much larger public catalog. This suggests that the 118 cases do indeed explore many different regimes of precession. 

\begin{figure}
\centering
  \includegraphics[width=0.7\linewidth]{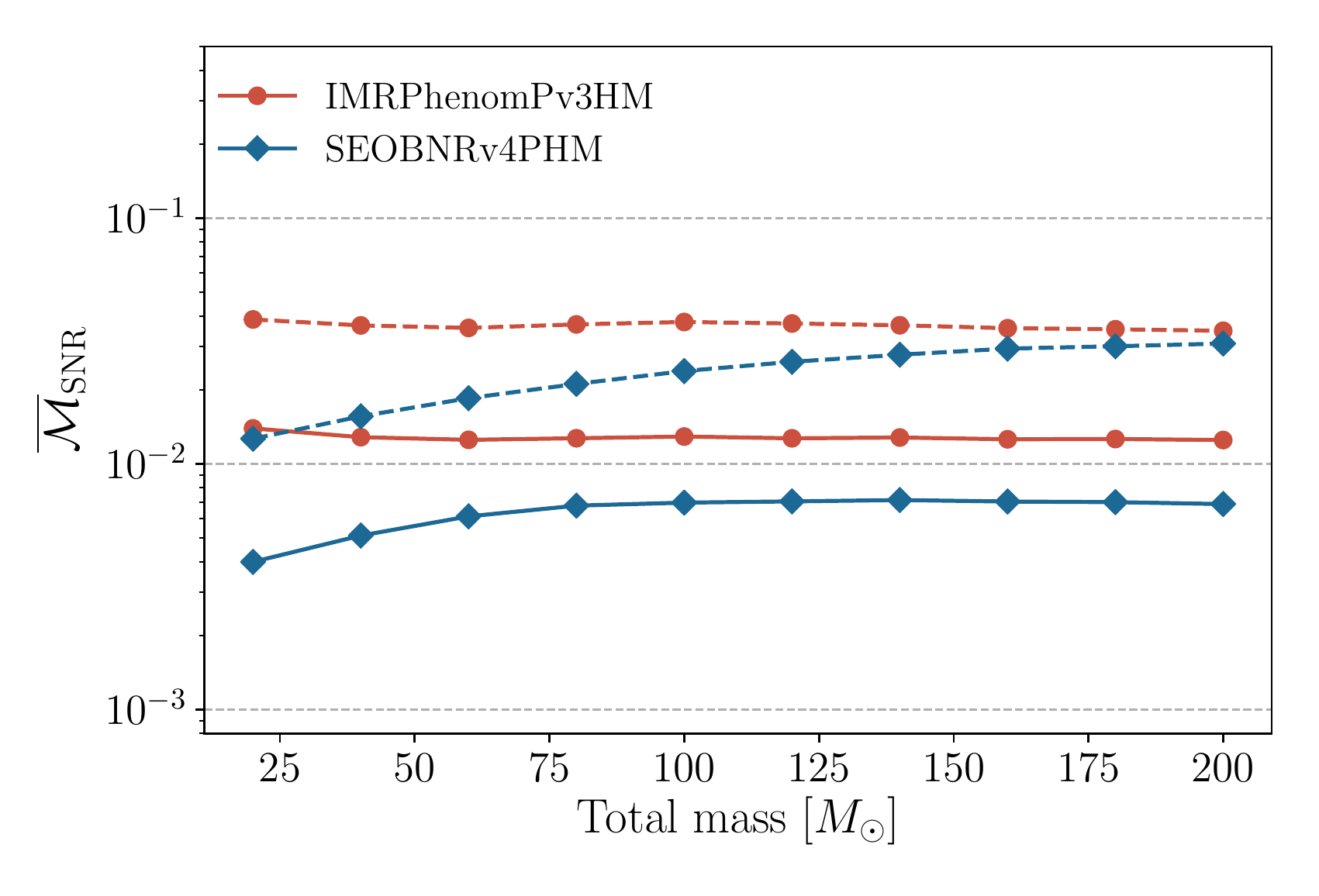}
  \caption{Summary of unfaithfulness as a function of the total
    mass, for all NR simulations considered as shown in Fig.~\ref{fig:spaghetti_public} and Fig~\ref{fig:spaghetti_PrecBBH}. The solid
    (dotted) line represents the median (95\%-percentile) of all 
    cases. For all total masses, we find that the median mismatch
    with \texttt{ SEOBNRv4PHM} is lower than 1\%, about a factor of 2
    lower than \texttt{ IMRPhenomPv3HM}. The 95th-percentile shows a
    stronger dependence on total mass for \texttt{ SEOBNRv4PHM}, with
    mismatches lower than \texttt{ IMRPhenomPv3HM} at low and medium total
    masses, becoming comparable at the highest total masses.}
\label{fig:all_runs_percentiles}
\end{figure}

\begin{figure}
\centering
	\includegraphics[width=0.8\linewidth]{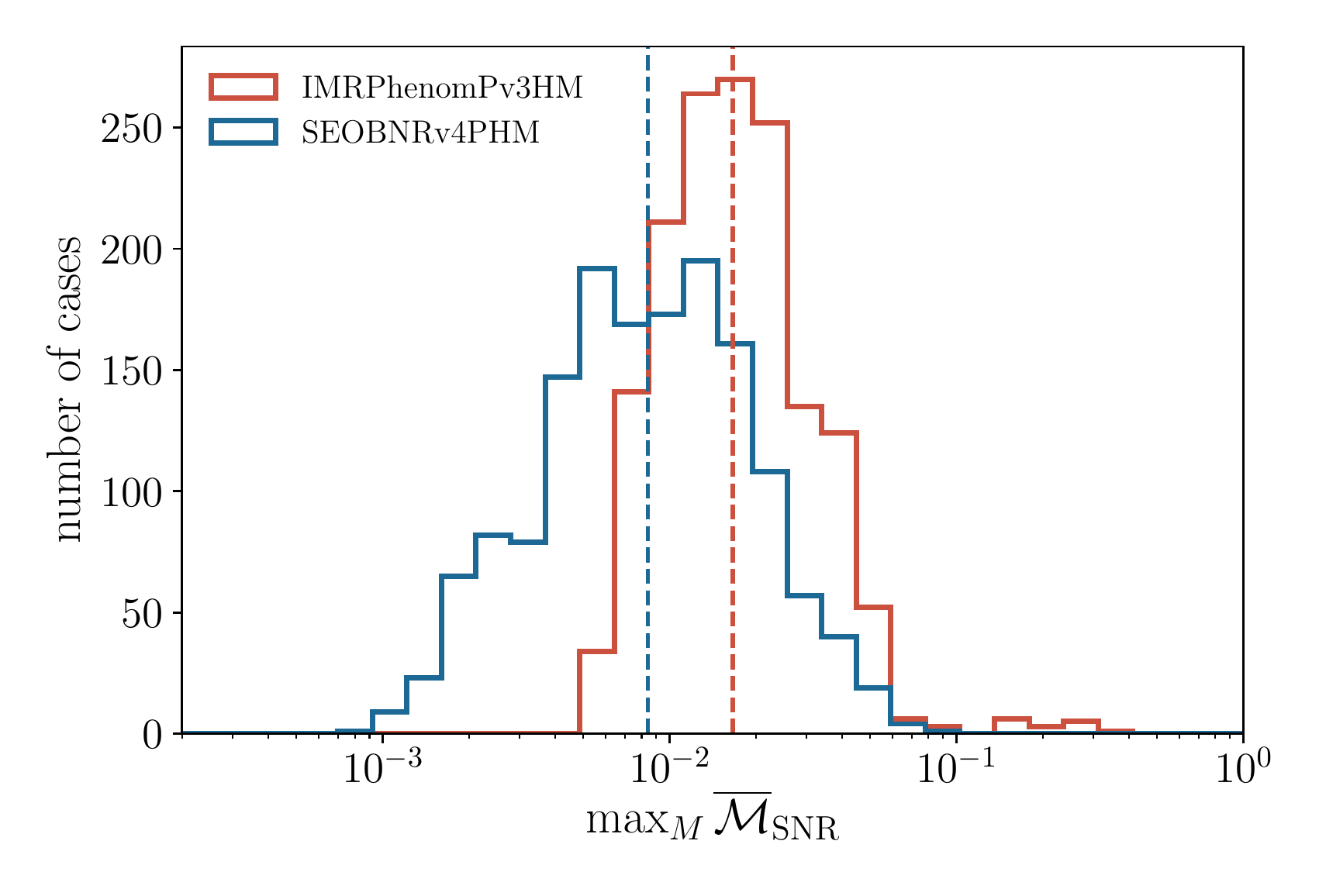}
	\caption{The highest unfaithfulness over total mass for all cases shown in Fig.~\ref{fig:all_runs_percentiles}. The median of unfaithfulness is 
		around 1\% for \texttt{ SEOBNRv4PHM} and 2\% for \texttt{ IMRPhenomPv3HM} (shown as dashed vertical lines). 
		Note that for \texttt{ SEOBNRv4PHM}, the worst unfaithfulness is below 10\% and the distribution is shifted to lower values.}
	\label{fig:all_runs_hist}
\end{figure}

To further quantify the results of the comparison between the precessing multipolar models \verb+SEOBNRv4PHM+ and \verb+IMRPhenomPv3HM+ and 
the NR waveforms, we show in Figs.~\ref{fig:all_runs_percentiles} and \ref{fig:all_runs_hist} the median and $95\%$-percentile of all cases,  
and the highest unfaithfulness as function of the total mass, respectively. These studies also demonstrate the better performance of 
\verb+SEOBNRv4PHM+ with respect to \verb+IMRPhenomPv3HM+. 

To summarize the performance against the entire SXS catalog (including the new 118 precessing waveforms) we find that 
for \verb+SEOBNRv4PHM+, out of a total of \aconfigs\ NR simulations we have considered, 864 cases (\EOBbone\ ) have a maximum unfaithfulness less than $1\%$,
and 1435 cases  (\EOBbthree\ ) have unfaithfulness less than $3\%$.  Meanwhile  for \verb+IMRPhenomPv3HM+ the numbers become 300 cases (\Phenombone\ ) below $1\%$, 1256 cases 
(\Phenombthree\ ) below $3\%$~\footnote{Due to technical details of the  \texttt{ IMRPhenomPv3HM} model, the total number of cases analyzed for this model is 1507 instead of 1523.}. 
The accuracy of the semi-analytical waveform models can be improved in the future by calibrating them to the precession sector of the SXS NR waveforms.

\begin{figure*}
\centering
	\includegraphics[width=\linewidth]{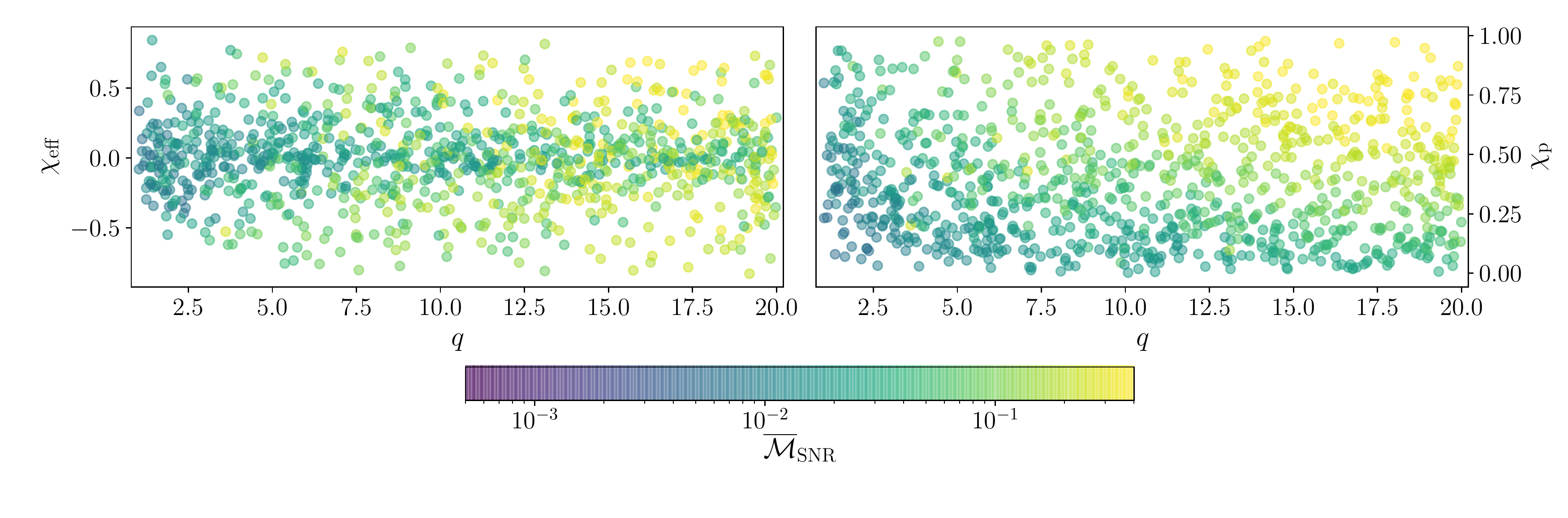}
	\caption{Sky-and polarization-averaged unfaithfulness between \texttt{ SEOBNRv4PHM} and \texttt{ IMRPhenomPv3HM} for 1000 random configurations.
	Notice that the unfaithfulness grows both with the mass ratio and the spin and can reach very large values for $q\approx20$ and high $\chi_{p}$. It's also
	clear that for cases with smaller spins the unfaithfulness remains much lower. }
	\label{fig:v4PHM_vs_Pv3HM}
\end{figure*}

An interesting question is to examine the behavior of the precessing models outside the region in which their underlying  aligned-spin waveforms were calibrated. To this effect we consider 1000 random cases
between mass ratios $q\in[1,20]$ and spin magnitudes $\chi_{1,2}\in[0,0.99]$ and compute $\overline{\mathcal{M}}_\mathrm{ SNR}$ between
\verb+SEOBNRv4PHM+ and \verb+IMRPhenomPv3HM+. Figure~\ref{fig:v4PHM_vs_Pv3HM} shows the dependence of the unfaithfulness on the binary parameters,
in particular the mass ratio, and the effective and precessing spins. We find that for mass ratios $q< 8$, $50\%$ of cases have unfaithfulness below 2\% and 90\%
have unfaithfulness below 10\%. The unfaithfulness grows very fast with mass ratio and spin, with the highest unfaithfulness occurring at the highest mass ratio and 
precessing spin.  This effect is enhanced due to the fact that we choose to start all the waveforms at the same frequency and for higher mass ratios, the number of cycles in
band grows as $1/\nu$ where $\nu$ is the symmetric mass ratio. These results demonstrate the importance of producing long NR simulations for large mass ratios 
and spins, which can be used to validate waveform models in this more extreme region of the parameter space. To design more accurate semi-analytical models 
in this particular region, it will be relevant to incorporate in the models the information from gravitational self-force~\cite{Damour:2009sm,Bini:2018ylh,Antonelli:2019fmq}, 
and also test how the choice of the underlying EOB Hamiltonians with spin effects~\cite{Rettegno:2019tzh,Khalil:2020mmr} affects the accuracy.

Finally, in Appendix~\ref{sec:comparisonNRSurr} we quantify the agreement of the precessing multipolar waveform models 
\verb+SEOBNRv4PHM+ and \verb+IMRPhenomPv3HM+ against the NR surrogate model \verb+NRSur7dq4+~\cite{Varma:2019csw}, which 
was built for binaries with mass ratios $1\mbox{--}4$, BH's spins up to $0.8$ and binary's 
total masses larger than $\sim 60 M_\odot$. We find that the unfaithfulness between the semi-analytic models and the NR surrogate largely mirrors the results of the 
comparison in \cref{fig:all_runs_percentiles,fig:all_runs_hist}. Notably, as it can be seen in Fig.~\ref{fig:models_vs_NRsurr}, the unfaithfulness is generally below 3\% for both waveform 
families, but \verb+SEOBNRv4PHM+ outperforms \verb+IMRPhenomPv3HM+ with the former having a median at $3.3 \times 10^{-3}$, while the latter is at $1.5 \times 10^{-2}$.

\section{Bayesian analysis with multipolar precessing waveform models}
\label{sec:peEOBNR}

We now study how the accuracy of the waveform model \verb+SEOBNRv4PHM+ (and also \verb+IMRPhenomPv3HM+), which we have 
quantified in the previous section through the unfaithfulness, affects parameter inference when synthetic signal injections are performed. 
To this end, we employ two mock BBH signals and do not add any detector noise to them 
(i.e., we work in zero noise), which is equivalent to average over many different noise realizations. This choice avoids 
arbitrary biases introduced by a random-noise realization, and it is reasonable 
since the purpose of this analysis is to estimate possible biases in the 
binary's parameters due to inaccuracies in waveform models. 

We generate the first precessing-BBH mock signal with the
\verb+NRSur7dq4+ model. It has mass ratio $q = 3$ and a total
source-frame mass of $M = 70 M_\odot$. The spins of the two BHs are
defined at a frequency of $20$ Hz, and have components $\vchi_1 =
(0.30, 0.00, 0.50)$ and $\vchi_2 = (0.20,0.00,0.30)$. The masses and
spins'' magnitudes ($0.58$ and $0.36$) of this injection are compatible
with those of BBH systems observed so far with LIGO and Virgo
detectors~\cite{TheLIGOScientific:2016pea,LIGOScientific:2018mvr,Zackay:2019tzo,Venumadhav:2019lyq,Nitz:2019hdf}.
Although the binary's parameters are not extreme, we choose the
inclination with respect to the line of sight of the BBH to be
 $\iota = \pi/3$, to emphasize the effect of higher
modes. The coalescence and polarization phase, respectively $\phi$ and
$\psi$, are chosen to be 1.2 rad and 0.7 rad. The sky-position is
defined by its right ascension of 0.33 rad and its declination of -0.6
rad at a GPS-time of 1249852257 s. Finally, the distance to the source
is set by requesting a network-SNR of $50$ in the three detectors
(LIGO Hanford, LIGO Livingston and Virgo) when using the Advanced LIGO
and Advanced Virgo PSD at design sensitivity~\cite{Barsotti:2018}. The
resulting distance is $800$ Mpc. The unfaithfulness against this
injection is $0.2\%$ and $1\%$ for \verb+SEOBNRv4PHM+ and
\verb+IMRPhenomPv3HM+, respectively. Although the value of
the network-SNR is large for this synthetic signal, it is not
excluded that the Advanced LIGO and Virgo detectors at design
sensitivity could detect such loud BBH. With this study we
want to test how our waveform model performs on a system with moderate
precessional effect when detected with a large SNR value, 
considering that it has an unfaithfulness of $0.2\%$.

For the second precessing-BBH mock signal, we use a binary with larger mass ratio 
and spin magnitude for the primary BH. We employ the NR waveform \verb+SXS:BBH:0165+ from 
the public SXS catalog having mass ratio $q = 6$, and we choose the source-frame total mass 
$M = 76 M_\odot$. The BH's spins, defined at a frequency of $20$ Hz, have values 
$\vchi_1 = (-0.06, 0.78, -0.47)$ and $\vchi_2 =(0.08,-0.17,-0.23)$. The BBH system in this simulation has strong 
spin-precession effects.  We highlight that this NR waveform is one of the worst cases in term of unfaithfulness 
against \verb+SEOBNRv4PHM+, as it is clear from Fig.~\ref{fig:spaghetti_public}. 
For this injection we choose the binary's inclination to be 
edge-on at $20$ Hz to strongly emphasize higher modes. All 
the other binary parameters are the same of the previous injection, 
with the exception of the luminosity distance, which 
in this case is set to be $1.2$ Gpc to obtain a
network-SNR of $21$. The NR waveform used for
this mock signal has unfaithfulness of $4.4\%$ for \verb+SEOBNRv4PHM+ and $8.8\%$ for \verb+IMRPhenomPv3HM+, 
thus higher than in the first injection. 

For the parameter-estimation study we use the function \texttt{pycbc\-\_generate\-\_hwinj} from the \texttt{PyCBC} software~\cite{alex_nitz_2020_3630601} to
prepare the mock signals, and we perform the Bayesian analysis with 
\texttt{parallel Bilby}~\cite{Smith:2019ucc}, a highly parallelized version of the
parameter-estimation software \texttt{Bilby}~\cite{Ashton:2018jfp}.  We choose a
uniform prior in component masses in the range $[5,150]
M_\odot$. Priors on the dimensionless spin magnitudes are uniform in
$[0,0.99]$, while for the spin directions we use prior isotropically
distributed on the unit sphere.  The priors on the other parameters
are the standard ones described in Appendix C.1 of 
Ref.~\cite{LIGOScientific:2018mvr}.

\begin{figure*}[hbt]
  \centering
  \includegraphics[width=0.45\textwidth]{m1m2_posterior_samples_q3_SNR_50.pdf}
  \includegraphics[width=0.45\textwidth]{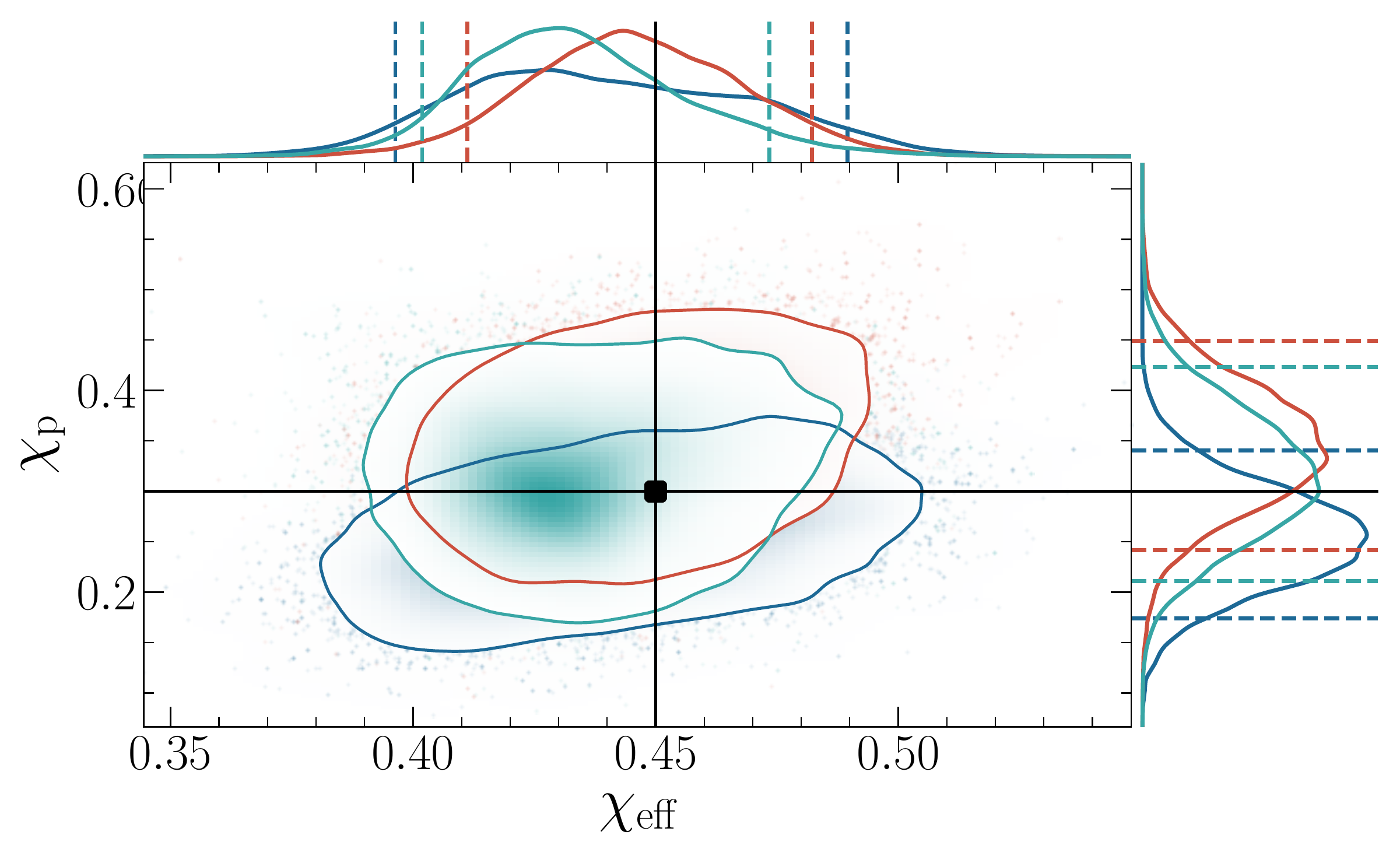}
  \includegraphics[width=0.45\textwidth]{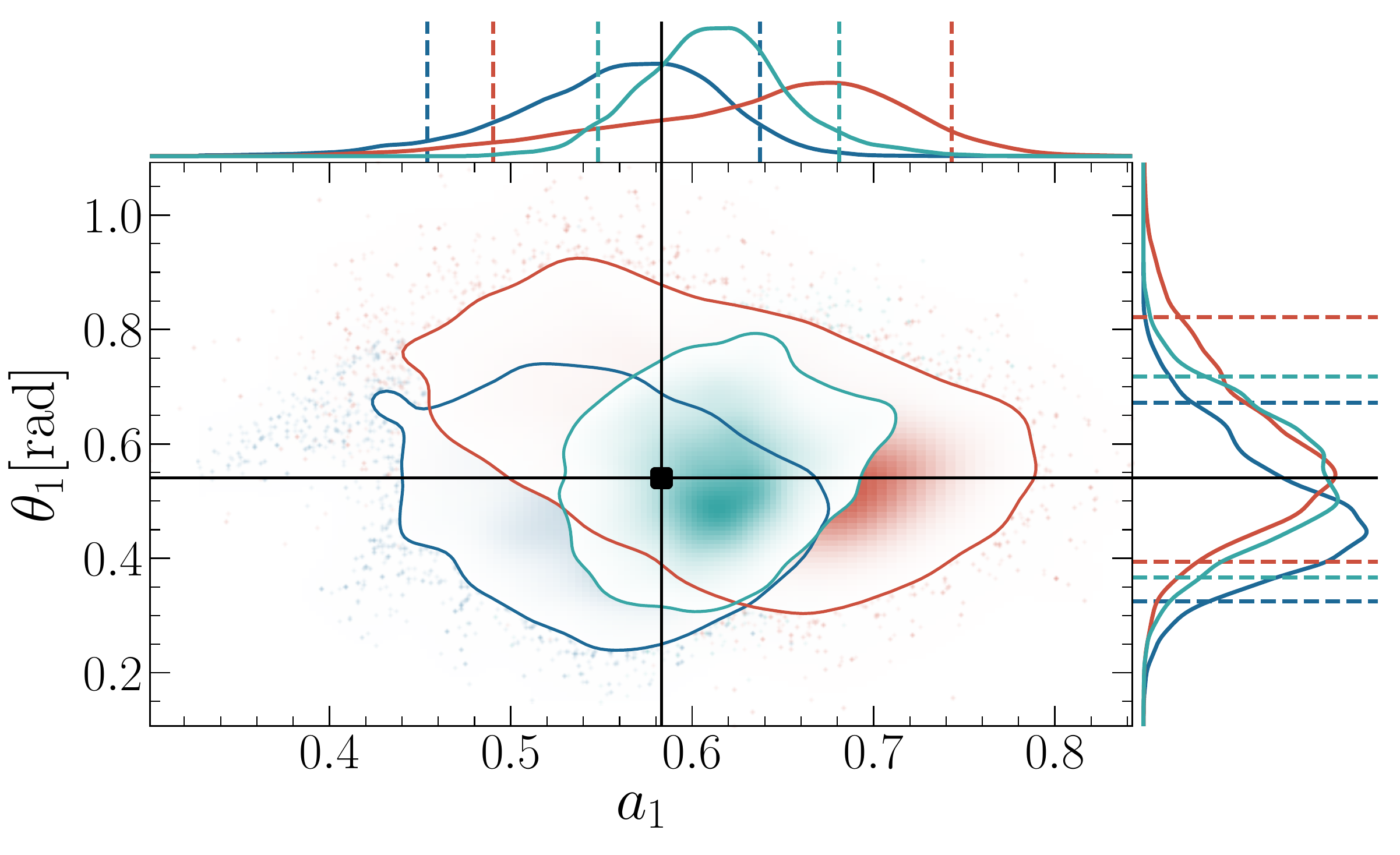}
  \includegraphics[width=0.45\textwidth]{thetaJNdistance_posterior_samples_q3_SNR_50.pdf}
  \includegraphics[clip, width=\textwidth]{event_legend_posterior_samples_q3.pdf}
  \caption{2D and 1D posterior distributions for some relevant
    parameters measured from the first synthetic BBH signal with mass ratio $q = 3$, total source-frame mass of $M = 70 M_\odot$,
spins of the two BHs $\vchi_1 = (0.30, 0.00, 0.50)$ and $\vchi_2 =
(0.20,0.00,0.30)$ defined at a frequency of $20$ Hz . The inclination with respect to the line of
sight of the BBH is $\iota = \pi/3$. The other parameters are specified in the text. The signal waveform is generated with the
    waveform model \texttt{NRSur7dq4}. In the 2D posteriors, solid
    contours represent $90\%$ credible intervals and black dots show
    the value of the parameter used in the synthetic signal. In the 1D
    posteriors they are represented respectively by dashed lines and
    black solid lines. The parameter estimation is performed with the
    waveform models \texttt{SEOBNRv4PHM} (blue), \texttt{NRSur7dq4}
    (cyan) and \texttt{IMRPhenomPv3HM} (red).  \emph{Top left:}
    component masses in the source frame, \emph{Top right:}
    $\chi_{\mathrm{eff}}$ and $\chi_{\mathrm{p}}$, \emph{Bottom left:}
    magnitude and tilt angle of the primary spin, \emph{Bottom right:}
    $\theta_{\mathrm{JN}}$ and luminosity distance.}
  \label{fig:PE_q_3}
\end{figure*}

We summarize in Fig.~\ref{fig:PE_q_3} the results of the parameter
estimation for the first mock signal for \verb+SEOBNRv4PHM+ (blue),
\verb+IMRPhenomPv3HM+ (red) and \verb+NRSur7dq4+ (cyan).  We report
the marginalized 2D and 1D posteriors for the component masses $m_1$
and $m_2$ in the source frame (top left), the effective spin
parameters $\chi_{\mathrm{eff}}$ and $\chi_\mathrm{p}$ (top right),
the spin magnitude of the more massive BH $a_1$ and its tilt angle
$\theta_1$ (bottom left) and finally the angle $\theta_{\mathrm{JN}}$
and the luminosity distance (bottom right). In the 2D posteriors, solid
contours represent $90\%$ credible intervals and black dots show the
value of the parameter used in the synthetic signal. In the 1D
posteriors, they are represented respectively by dashed lines and black
solid lines. As it is clear from Fig.~\ref{fig:PE_q_3}, when using the
waveform models \verb+SEOBNRv4PHM+ and \verb+NRSur7dq4+, all the
parameters of the synthetic signal are correctly measured within the
statistical uncertainty.  Moreover, the shape of the posterior
distributions obtained when using \verb+SEOBNRv4PHM+ are similar to
those recovered with \verb+NRSur7dq4+ (the model used to create the
synthetic signal). This means that the systematic error due to a non
perfect modeling of the waveforms is negligible in this case.

For the model \verb+IMRPhenomPv3HM+ while masses and spins
are correctly measured within the statistical uncertainty, the
luminosity distance $D_\mathrm{ L}$ and the angle $\theta_\mathrm{ JN}$ are
biased.  This is consistent with the prediction obtained using
Lindblom's criterion in Refs.~\cite{Flanagan:1997kp,Lindblom:2008cm,McWilliams:2010eq,Chatziioannou:2017tdw}~\footnote{The
  criterion says that a sufficient, but not necessary condition for
  two waveforms to become distinguishable is that the unfaithfulness $
  \geq (N_\mathrm{ intr}-1)/(2\mathrm{ SNR}^2)$, where $N_\mathrm{ intr}$ is the
  number of binary's intrinsic parameters, which we take to be 8 for a
  precessing-BBH system. Note, however, that in practice this factor
  can be much larger, see discussion in
  Ref.~\cite{Purrer:2019jcp}.}. In fact, according to this criterion,
an unfaithfulness of $1\%$ for \texttt{ IMRPhenomPv3HM} would be
sufficient to produce biased results at a network-SNR of $19$. Thus,
it is expected to observe biases when using \verb+IMRPhenomPv3HM+ at
the network-SNR of the injection, which is $50$. In the case of
\verb+SEOBNRv4PHM+ the unfaithfulness against the signal waveform is
$0.2\%$ and according to Lindblom's criterion we should also expect
biases for network-SNRs larger than $42$, but in practice we do not
observe them. We remind that Lindblom's criterion is only approximate
and it has been shown in Ref.~\cite{Purrer:2019jcp} to be too
conservative, therefore the lack of bias that we observe  is not
surprising. 

\begin{figure*}[hbt]
  \centering
  \includegraphics[width=0.45\textwidth]{m1m2_posterior_samples_q6_NR_evolved.pdf}
  \includegraphics[width=0.45\textwidth]{chieffchip_posterior_samples_q6_NR_evolved.pdf}
  \includegraphics[width=0.45\textwidth]{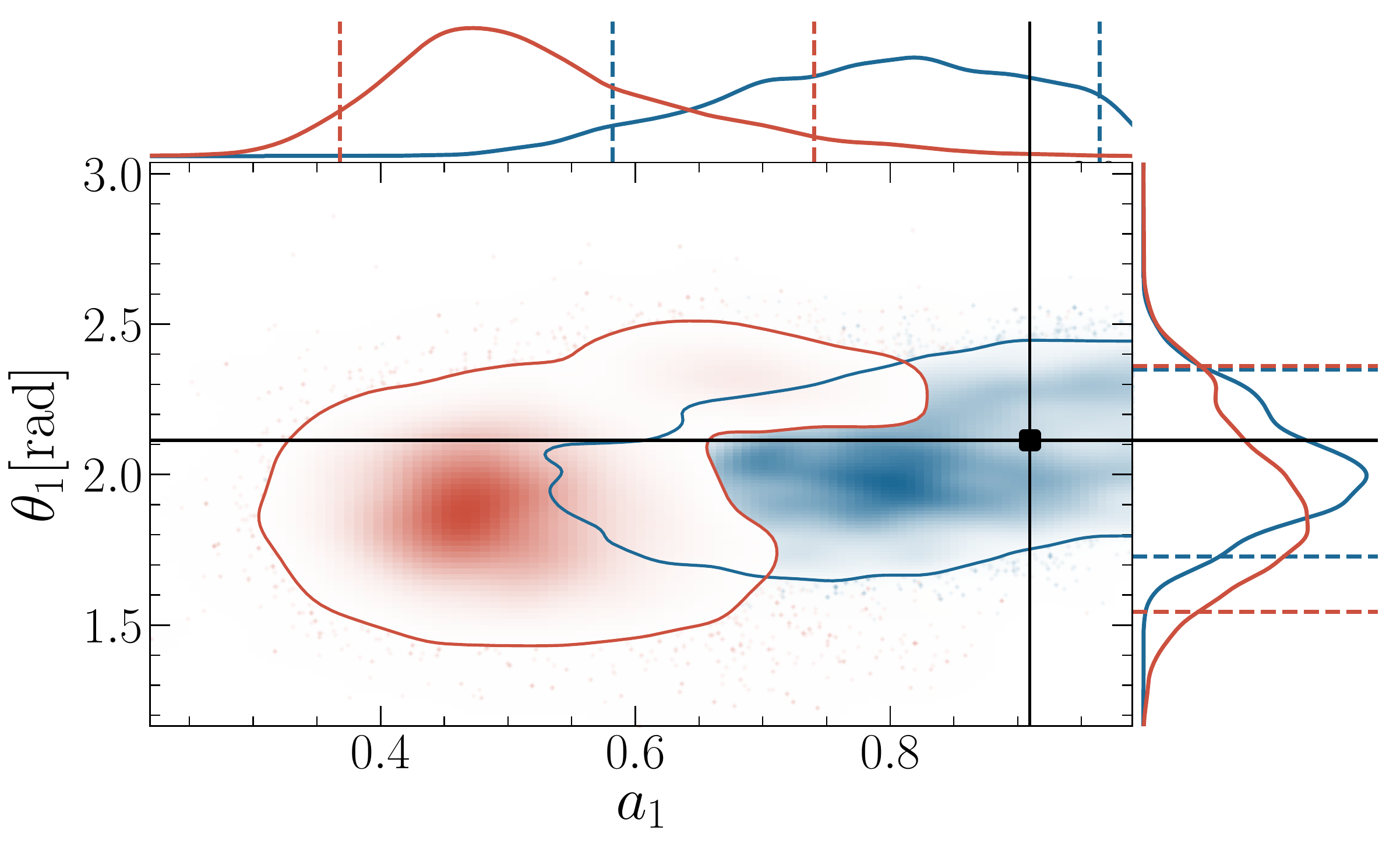}
  \includegraphics[width=0.45\textwidth]{thetaJNdistance_posterior_samples_q6_NR_evolved.pdf}
  \includegraphics[ clip, width=0.5\textwidth]{event_legend_posterior_samples_q6_NR_evolved.pdf}
  \caption{2D and 1D posterior distributions for some relevant
    parameters measured from the first synthetic BBH signal with mass ratio $q = 6$, total source-frame mass of $M = 76 M_\odot$,
spins of the two BHs $\vchi_1 = (-0.06, 0.78, -0.47)$ and $\vchi_2 =(0.08,-0.17,-0.23)$ defined at a frequency of $20$ Hz . The inclination with respect to the line of
sight of the BBH is edge-on, i.e., $\iota = \pi/2$. The other parameters are specified in the text. The signal waveform is generated using the NR waveform from the SXS public catalog \texttt{SXS:BBH:0165}. In the 2D posteriors solid
    contours represent $90\%$ credible intervals and black dots show
    the value of the parameter used in the synthetic signal. In the 1D
    posteriors they are represented respectively by dashed lines and
    black solid lines. The parameter estimation is performed with the
    waveform models \texttt{SEOBNRv4PHM} (blue) and \texttt{IMRPhenomPv3HM} (red).  \emph{Top left:}
    component masses in the source frame, \emph{Top right:}
    $\chi_{\mathrm{eff}}$ and $\chi_{\mathrm{p}}$, \emph{Bottom left:}
    magnitude and tilt angle of the primary spin, \emph{Bottom right:}
    $\theta_{\mathrm{JN}}$ and luminosity distance.}
  \label{fig:PE_q_6}
\end{figure*}

In Fig.~\ref{fig:PE_q_6} we summarize the results of 
the second mock-signal injection. The plots are the same as in Fig.~\ref{fig:PE_q_3} 
with the only exception that we do not have results for the \verb+NRSur7dq4+ 
model since it is not available in this region of the parameter space.  In
this case the unfaithfulness between \verb+SEOBNRv4PHM+
(\verb+IMRPhenomPv3HM+) and the NR waveform used for the mock signal
is $4.4\%$ ($8.8\%$). According to Lindblom's criterion, at the
network-SNR of this mock signal we should expect the bias due to
non-perfect waveform modeling to be dominant over the statistical
uncertainty for an unfaithfulness $\gtrsim
1\%$. Therefore we might expect some biases in inferring parameters 
for both models. Lindblom's criterion does not say which parameters 
are biased and by how much. The results in Fig.~\ref{fig:PE_q_6} clearly show 
that both models have biases in the measurement of some
parameters, but unfaithfulness of $4.4\%$ and $8.8\%$ induce different 
amount of biases and also on different set of parameters (intrinsic and extrinsic).

In particular for the component masses (top left panel of
Fig.~\ref{fig:PE_q_6}), the 2D posterior distribution obtained with
\verb+SEOBNRv4PHM+ barely include the value used for the mock signal
in the $90\%$ credible region. This measurement looks better when
focusing on the 1D posterior distributions for the individual masses
for which the injection values are well within the $90\%$ credible intervals. 
The situation is worst for the \verb+IMRPhenomPv3HM+ model, for which the 2D posterior
distribution barely excludes the injection value at 
$90\%$ credible level. In this case also the true value of $m_1$
is excluded from the $90\%$ credible interval of the marginalized 1D
posterior distribution. Furthermore, $\chi_\mathrm{ eff}$ and $\chi_\mathrm{ p}$ (top right
panel of Fig.~\ref{fig:PE_q_6}) are correctly measured with
\verb+SEOBNRv4PHM+ while the measurement with \verb+IMRPhenomPv3HM+
excludes the true value from the 2D $90\%$ credible region. From the
1D posterior distributions it is clear that the source of this inaccuracy
is the incorrect measurement of $\chi_\mathrm{ p}$, while $\chi_\mathrm{ eff}$
is correctly recovered within the $90\%$ credible interval. A similar
situation is observed in the measurement of $a_1$ the spin magnitude
of the heavier BH and $\theta_1$ its tilt angle (bottom left panel of
Fig.~\ref{fig:PE_q_6}). Also in this case \verb+SEOBNRv4PHM+ correctly
measures the parameters used in the mock signal, while
\verb+IMRPhenomPv3HM+ yields an incorrect measurement due to a bias in
the estimation of $a_1$. Finally, we focus on the measurement of the
angle $\theta_\mathrm{ JN}$ and the luminosity distance $D_L$ (bottom right
panel of Fig.~\ref{fig:PE_q_6}). In this case the value of these
parameters used in the synthetic signal is just slightly measured
within the $90\%$ credible region of the 2D posterior distribution
obtained with \verb+SEOBNRv4PHM+. As a consequence the luminosity
distance is also barely measured within the $90\%$ credible interval
from the marginalized 1D posterior distribution and the measured value
of $\theta_\mathrm{ JN}$ results outside the $90\%$ credible interval of
the 1D posterior distribution. The posterior distributions obtained
using \verb+IMRPhenomPv3HM+ are instead correctly measuring the
parameters of the mock signal. We can conclude that even with an
unfaithfulness of $4.4\%$ against the NR waveform used for the mock
signal the \verb+SEOBNRv4PHM+ model is able to correctly measure most
of the binary parameters, notably the intrinsic ones, such as masses 
and spins.

\section{Conclusions}
\label{sec:concl}

In this paper we have developed and validated the first inspiral-merger-ringdown precessing 
waveform model in the EOB approach, \verb+SEOBNRv4PHM+, that includes multipoles beyond the dominant quadrupole. 

Following previous precessing SEOBNR models~\cite{Pan:2013rra,Taracchini:2013rva,Babak:2016tgq}, 
we have built such a model twisting up the aligned-spin waveforms of \verb+SEOBNRv4HM+~\cite{Bohe:2016gbl,Cotesta:2018fcv} 
from the co-precessing~\cite{Buonanno:2002fy,Schmidt:2010it,Boyle:2011gg,O'Shaughnessy:2011fx,Schmidt:2012rh} 
to the inertial frame, through the EOB equations of motion for the spins and orbital angular momentum. 
With respect to the previous precessing SEOBNR model, \verb+SEOBNRv3P+~\cite{Babak:2016tgq}, 
which has been used in LIGO/Virgo data analysis~\cite{Abbott:2016izl,Abbott:2017vtc,LIGOScientific:2018mvr}, the new model 
(i) employs a more accurate aligned-spin two-body dynamics, since, in the non-precessing limit, it reduces to 
\verb+SEOBNRv4HM+, which was calibrated to 157 SXS NR simulations~\cite{Mroue:2013xna,Chu:2015kft}, and 13 waveforms~\cite{Barausse:2011kb} 
from BH perturbation theory,  (ii) includes in the co-precessing frame the modes $(2,\pm 2), (2,\pm 1), (3,\pm 3), (4,\pm 4)$ 
and $(5,\pm 5)$, instead of only $(2,\pm 2), (2,\pm 1)$, (iii) incorporates the merger-ringdown 
signal in the co-precessing frame instead of the inertial frame, (iv) describes the merger-ringdown stage through a phenomenological 
fit to NR waveforms~\cite{Bohe:2016gbl,Cotesta:2018fcv}, and (v) uses more accurate NR fits  for the final spin of the remnant BH.

\begin{figure*}
	\includegraphics[width=\linewidth]{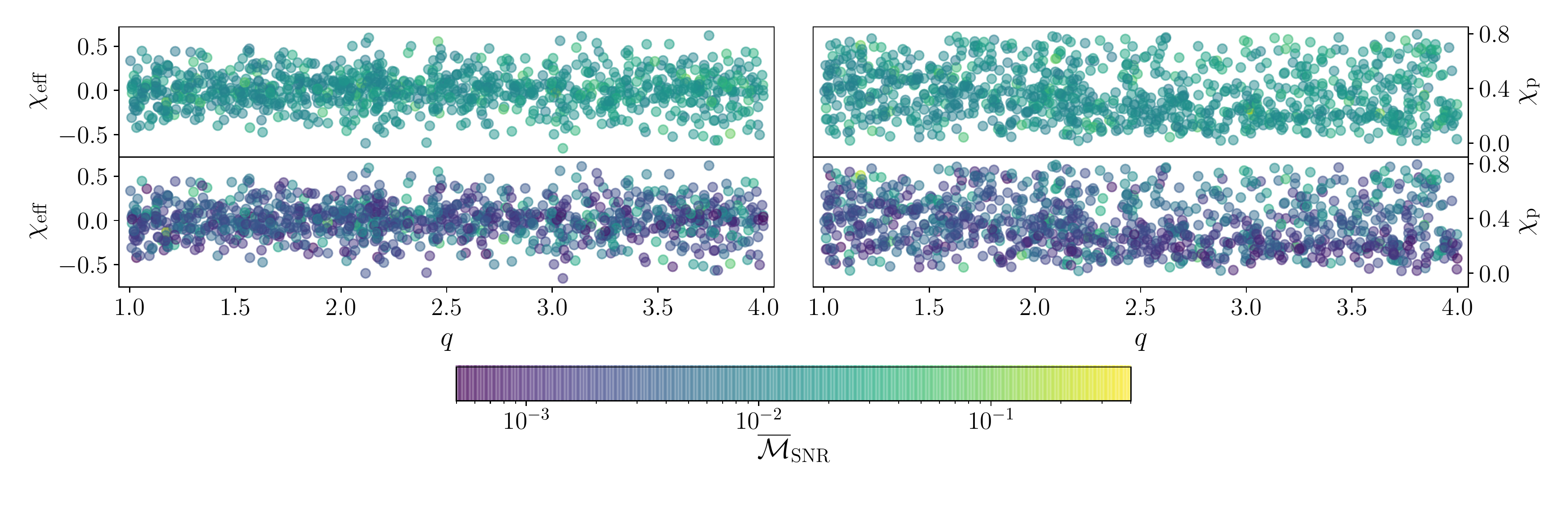}
	\caption{The maximum  sky-and-polarization averaged, SNR-weighted
		unfaithfulness as a function of binary's total mass for inclination
		$\iota=\pi/3$, between the models \texttt{IMRPhenomPv3HM} (\emph{top}) and \texttt{SEOBNRv4PHM} (\emph{bottom}),
		and the NR surrogate, cf. Fig.~\ref{fig:v4PHM_vs_Pv3HM}.
		The unfaithfulness is strongly dependent on the intrinsic parameters, especially the spins.
	}
	\label{fig:2d_models_vs_NRsurr}
\end{figure*}

\begin{figure*}
\centering
	\includegraphics[width=0.7\linewidth]{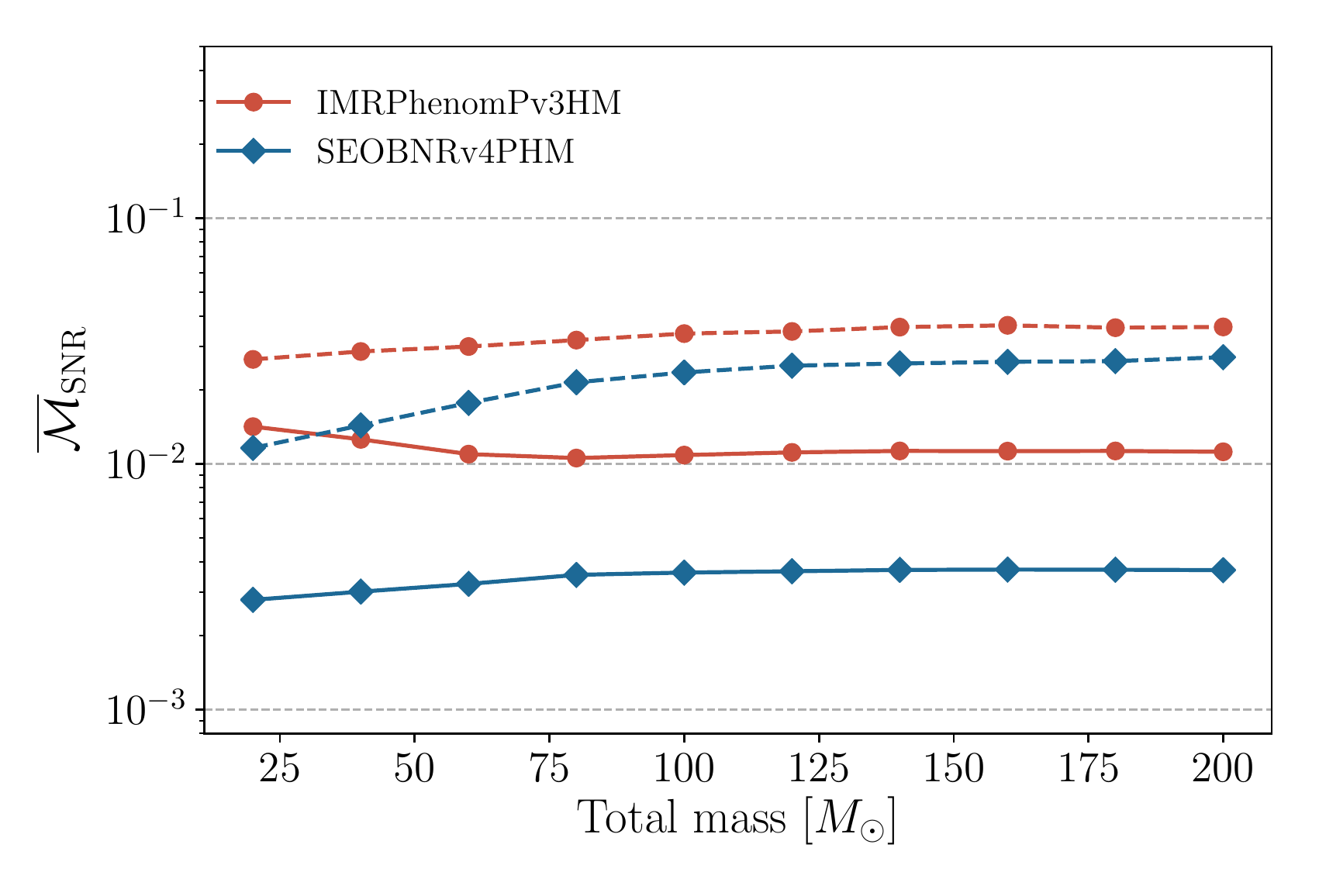}
		\includegraphics[width=0.7\linewidth]{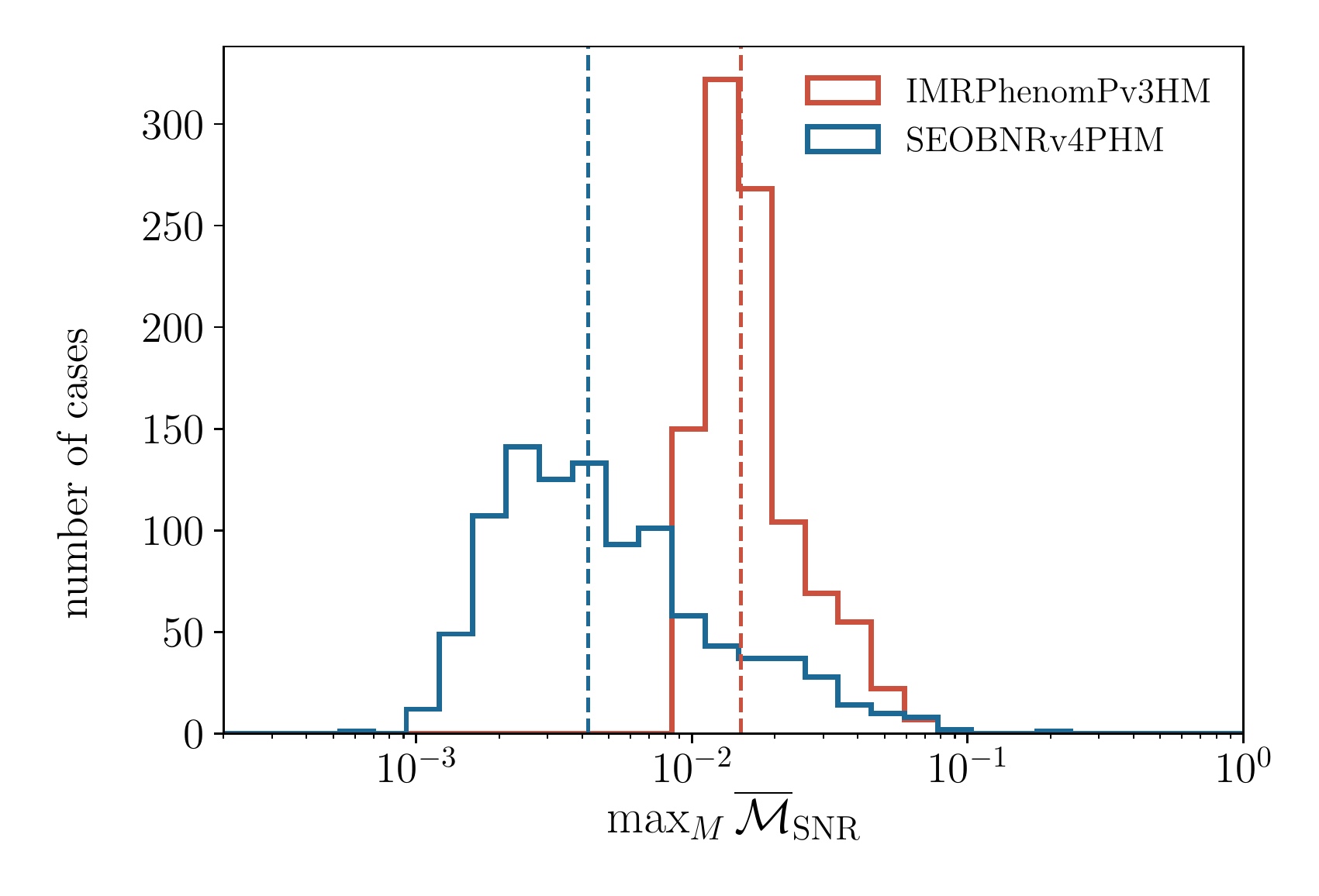}
	\caption{The summary of the sky-and-polarization averaged, SNR-weighted
		unfaithfulness as a function of binary's total mass for inclination
		$\iota=\pi/3$, among the \texttt{NRSur7dq4} model and the \texttt{IMRPhenomPv3HM} and \texttt{SEOBNRv4PHM} models. 
\emph{Left}: The solid (dashed) lines show the median (95th percentile) as a function of total mass,
		cf Fig.~\ref{fig:all_runs_percentiles}. \emph{Right}: maximum unfaithfulness over all total masses, cf. Fig.~\ref{fig:all_runs_hist}.
		The unfaithfulness is low 
		using both waveform families, however, \texttt{SEOBNRv4P(HM)} has lower median unfaithfulness by a factor of $\sim4.3$.}
	\label{fig:models_vs_NRsurr}
\end{figure*}

The improvement in accuracy between \verb+SEOBNRv4+ and
\verb+SEOBNRv3P+ (i.e., the models with only the $\ell = 2$ modes) is
evident from Fig.~\ref{fig:all_approximants_ell2}, where we have
compared those models to the public SXS catalog of 1405 precessing NR
waveforms, and the new 118 SXS NR waveforms produced for this
work. The impact of including higher modes in semi-analytical models
to achieve higher accuracy to multipolar NR waveforms is demonstrated
in
Fig.~\ref{fig:higher_mode_effects}. Figures~\ref{fig:spaghetti_public},
\ref{fig:spaghetti_PrecBBH}, \ref{fig:all_runs_percentiles} and
\ref{fig:v4PHM_vs_Pv3HM} quantify the comparison of the multipolar
precessing \verb+SEOBNRv4PHM+ and \verb+IMRPhenomPv3HM+ to all SXS NR
precessing waveforms at our disposal. We have found that for the
\verb+SEOBNRv4PHM+ model, \EOBbthree\ (\EOBbone\ ) of the cases have
maximum unfaithfulness value, in the total mass range $20\mbox{--}200
M_\odot$, below $3\%$ ($1\%$). Those numbers change to
\Phenombthree\ (\Phenombone\ ) when using the
\verb+IMRPhenomPv3HM+. We have found several cases with large
unfaithfulness ($>10\%$) for \texttt{ IMRPhenomPv3HM}, coming from a
region of parameter space with $q\gtrsim 4$ and large ($\simeq0.8$)
spins anti-aligned with the orbital angular momentum, which appear to
be connected to unphysical features in the underlying precession
model, and cause unusual oscillations in the waveform's amplitude and
phase. The better accuracy of \verb+SEOBNRv4PHM+ with respect to
\verb+IMRPhenomPv3HM+ is also confirmed by the comparisons with the NR
surrogate model \verb+NRSur7dq4+, as shown in
Fig.~\ref{fig:models_vs_NRsurr}. We have investigated in which region
of the parameter space the unfaithfulness against NR waveforms and
\texttt{ NRSur7dq4} lies, and have found, not surprisingly, that it occurs
where both mass ratios and spins are large (see
Fig.~\ref{fig:2d_models_vs_NRsurr}). When comparing \texttt{ SEOBNRv4PHM}
and \texttt{ IMRPhenomPv3HM} outside the region in which the aligned-spin
underlying model was calibrated, we have also found that the largest
differences reside when mass ratios are larger than 4 and spins larger
than 0.8 (see Fig.~\ref{fig:v4PHM_vs_Pv3HM}).  To improve the accuracy
of the models in those more challenging regions, we would need NR
simulations, but also more information from analytical methods, such
as the gravitational
self-force~\cite{Damour:2009sm,Bini:2018ylh,Antonelli:2019fmq}, and
resummed EOB Hamiltonians with
spins~\cite{Rettegno:2019tzh,Khalil:2020mmr}.

To quantify how the modeling inaccuracy, estimated by the unfaithfulness, impacts the inference of binary's parameters, we
have perfomed two parameter-estimation studies using Bayesian analysis. Working with
the Advanced LIGO and Virgo network at design sensitivity, we have injected in zero noise two precessing-BBH
mock signals with mass ratio 3 and 6, having SNR of 50 and 21, with inclination of  $\pi/3$ and $\pi/2$ with respect to the line of sight respectively,  and recovered them with \texttt{ SEOBNRv4PHM} and \texttt{ IMRPhenomPv3HM}.
The unfaithfulness values of those models against the synthetic signals considered (i.e., \texttt{ NRSurd7q4} and \texttt{ SXS:BBH:0165})
range from $0.2\%$ to $8.8\%$. The results are summarized in Figs.~\ref{fig:PE_q_3} and \ref{fig:PE_q_6}. Overall, we have found that Lindblom's criterion
\cite{Flanagan:1997kp,Lindblom:2008cm,McWilliams:2010eq,Chatziioannou:2017tdw,Purrer:2019jcp} is too conservative
and predicts visible biases at SNRs lower than what we have obtained through the Bayesian analysis. In particular, we 
have found, when doing inference with \texttt{ SEOBNRv4PHM}, that an unfaithfulness of $0.2\%$ may produce no biases up to SNR of 50, while 
an unfaithfulness of $2.2\%$ can produce biases only for some extrinsic parameters, such as distance and inclination, but not for binary's masses and 
spins at SNR of 21. A more comprehensive Bayesian study will be needed to quantify, in a more realistic manner, the modeling systematics of
\texttt{ SEOBNRv4PHM}, if this model were used during the fourth observation run of Avanced LIGO and Virgo in 2022 (i.e., the run at design sensitivity).

The newly produced 118 SXS NR waveforms extend the coverage of binary's parameter space, spanning  
mass ratios $q=1\mbox{--}4$, (dimensionless) spins $\chi_{1,2}=0.3\mbox{--}0.9$, and different orientations 
to maximize the number of precessional cycles. As we have emphasized, the waveform model \verb+SEOBNRv4HM+ 
is not calibrated to NR waveforms in the precessing sector, only the aligned-spin sector was calibrated 
in Refs.~\cite{Bohe:2016gbl,Cotesta:2018fcv}. Despite this, 
the accuracy of the model is very good, and it can be further improved in the future if we calibrate  
the model to the $1404$ plus $118$ SXS NR precessing waveforms at our disposal. This will be an important 
goal for the upcoming LIGO and Virgo O4 run in early 2022. Furthermore, \verb+SEOBNRv4HM+ assumes the following 
symmetry among modes $h_{\ell m}= (-1)^\ell h_{\ell -m}^*$ in the co-precessing frame, which however 
no longer holds in presence of precession. As discussed in Sec.~\ref{sec:mode_asymm}, forcing 
this assumption causes unfaithfulness on the order of a few percent. Thus, to achieve better accuracy, 
when calibrating the model to NR waveforms, the mode-symmetry would need to be relaxed.

Finally, \verb+SEOBNRv4HM+ uses PN-resumed factorized modes that were developed for aligned-spin BBHs~\cite{Damour:2008gu,Pan:2010hz}, 
thus they neglect the projection of the spins on the orbital plane. To obtain high-precision waveform models, it will 
be relevant to extend the factorized modes to precession. Considering the variety of GW signals 
that the improved sensitivity of LIGO and Virgo detectors is allowing to observe, it will also be important 
to include in the multipolar SEOBNR waveform models the more challenging $(3,2)$ and $(4,3)$ modes, which are characterized 
my mode mixing~\cite{Buonanno:2006ui,Berti:2014fga,Mehta:2019wxm}. Their contribution is no longer negligible for 
high total-mass and/or large mass-ratio binaries, especially if observed away from face-on (face-off). 

Lastly, being a time-domain waveform model generated by solving ordinary differential equations, \verb+SEOBNRv4HM+ is not a fast 
waveform model, especially for low total-mass binaries. To speed up the waveform generation, a reduced-order modeling version 
has been recently developed~\cite{Gadre:2020a}. Alternative methods that employ a fast evolution of the EOB Hamilton equations 
in the post-adiabatic approximation during the long inspiral phase have been suggested~\cite{Nagar:2018gnk}, and we are currently implementing 
them in the simpler nonprecessing limit in LAL.

\section*{Acknowledgments}
It is our pleasure to thank Andrew Matas for providing us with the scripts to make 
the parameter-estimation plots, and Sebastian Khan for useful
discussions on the faithfulness calculation.  We would also like to thank the SXS collaboration for
help and support with the \texttt{ SpEC} code in producing the new NR
simulations presented in this paper, and for making the large catalog
of BBH simulations publicly available. RH acknowledges support through
OAC-1550514.  The new 118 SXS NR simulations
were produced using the high-performance compute (HPC) cluster \texttt{
  Minerva} at the Max Planck Institute for Gravitational Physics in
Potsdam, on the \texttt{ Hydra} cluster at the Max Planck Society at the 
Garching facility, and on the \texttt{
  SciNet} cluster at the University of Toronto.  The data-analysis
studies were obtained with the HPC clusters \texttt{ Hypatia} and \texttt{
  Minerva} at the Max Planck Institute for Gravitational Physics. The
transformation and manipulation of waveforms were done using the \texttt{GWFrames} package~\cite{GWFrames,Boyle:2013nka}.

\chapter{Frequency domain reduced order model of aligned-spin effective-one-body waveforms with higher-order modes}
\chaptermark{}
\label{chap:four}

\hspace{\parindent}\textbf{Authors}\footnote{Originally published as Phys.Rev.D 101 (2020) 12, 124040.}: Roberto Cotesta, Sylvain Marsat, Michael P{\"u}rrer\\

\textbf{Abstract}:We present a frequency domain reduced order model (ROM) for the aligned-spin effective-one-body (EOB) model for binary black holes (BBHs) \texttt{SEOBNRv4HM} that includes the spherical harmonics modes $(\ell, |m|) = (2,1),(3,3),(4,4),(5,5)$ beyond the dominant $(\ell, |m|) = (2,2)$ mode. These higher modes are crucial to accurately represent the waveform emitted from asymmetric BBHs.
We discuss a decomposition of the waveform, extending other methods in the literature, that allows us to accurately and efficiently capture the morphology of higher mode waveforms.
We show that the ROM is very accurate with median (maximum) values of the unfaithfulness against \texttt{SEOBNRv4HM} lower than $0.001\% (0.03\%)$ for total masses in $[2.8,100] M_\odot$. For a total mass of $M = 300 M_\odot$ the median (maximum) value of the unfaithfulness increases up to $0.004\% (0.17\%)$. This is still at least an order of magnitude lower than the estimated accuracy of SEOBNRv4HM compared to numerical relativity simulations.
The ROM is two orders of magnitude faster in generating a waveform compared to \texttt{SEOBNRv4HM}. Data analysis applications typically require $\mathcal{O}(10^6-10^8)$ waveform evaluations for which SEOBNRv4HM is in general too slow. The ROM is therefore crucial to allow the SEOBNRv4HM waveform to be used in searches and Bayesian parameter inference.
We present a targeted parameter estimation study that shows the improvements in measuring binary parameters when using waveforms that includes higher modes and compare against three other waveform models.

\section{Introduction}
\label{sec:intro_v4HMROM}

In the past five years \ac{GW} observations~\cite{Abbott:2016blz,TheLIGOScientific:2016pea,LIGOScientific:2018mvr} have opened up a new window to the Universe. In the first two observing runs of the advanced LIGO~\cite{TheLIGOScientific:2014jea} and Virgo~\cite{TheVirgo:2014hva} detectors ten confident detections of \acp{BBH} and one \ac{BNS} were made~\cite{LIGOScientific:2018mvr} and tens of detection candidates~\cite{gracedb-O3-superevents} have been identified so far in the third observing run of this network, among them another confident detection of a binary neutron star system~\cite{Abbott:2020uma}.
Both the detection and inference of binary parameters for these compact binaries rely heavily on our knowledge of the gravitational waveform emitted in these coalescences as encapsulated in parametrized models of \acp{GW}. The construction of stochastic template banks and the Bayesian inference of binary parameters routinely require tens to hundreds of millions of waveform evaluations~\cite{Harry:2009ea,Manca:2009xw, Veitch:2014wba,Ashton:2018jfp}. At the same time the phasing of the \acp{GW} needs to be tracked to an accuracy better than a fraction of a wave cycle to avoid missing events or mis-measuring binary parameters. Therefore, waveform models need to be fast and accurate to extract the binary properties imprinted in the emitted \acp{GW}.

Inspiral-merger-ringdown models of \acp{GW} from coalescing \ac{BH} binaries have traditionally been constructed in the \ac{EOB}~\cite{Buonanno:1998gg,Buonanno:2000ef,Damour:2000we,Damour:2001tu,Damour:2008gu,Pan:2011gk,Taracchini:2012ig,Taracchini:2013rva,Pan:2013rra,Damour:2014sva,Nagar:2015xqa,Bohe:2016gbl,Babak:2016tgq,Cotesta:2018fcv,Nagar:2018zoe,Ossokine:2020kjp} or phenomenological~\cite{Ajith:2011,Santamaria:2010yb,Hannam:2013oca,Husa:2015iqa,Khan:2015jqa,Mehta:2017jpq,London:2017bcn,Khan:2018fmp,Khan:2019kot,Pratten:2020fqn} approaches, and, more recently, models for NR or EOB waveforms constructed with surrogate modeling techniques have come to prominence~\cite{Field:2013cfa,Purrer:2014fza,Purrer:2015tud,
Blackman:2015pia,Bohe:2016gbl,Blackman:2017pcm,Blackman:2017dfb,Varma:2018mmi,Lackey:2018zvw,Varma:2019csw}.
\ac{EOB} models incorporate physical descriptions of the inspiral, merger, and ringdown parts of \ac{BBH} coalescences. \ac{PN} solutions for the inspiral are re-summed and connected with an analytic description of the merger waveform which is tuned to data from \ac{NR} simulations~\cite{Pretorius:2005gq,Campanelli:2005dd,Baker:2005vv,
Bruegmann:2006at,Centrella:2010mx,Mroue:2013xna,Jani:2016wkt,Healy:2017psd,Boyle:2019kee}. \ac{EOB} models are posed as an initial value problem for a complicated system of \acp{ODE} describing the approximated\footnote{The full dynamics of a binary system is obtained by solving Einstein's equations which are \acp{PDE}.} dynamics of a compact binary. The emitted \acp{GW} are then computed from this orbital dynamics. \ac{EOB} models have provided accurate and generic descriptions of the \acp{GW} for the signals observed so far by LIGO and Virgo detectors. However, observations with third generation detector networks may require much more accurate waveform models~\cite{Purrer:2019jcp}.
The integration of the \acp{ODE} requires small time steps to obtain an accurate solution and especially for the long waveforms produced by low mass compact binaries can take on the order of hours, and thus be too slow for practical data analysis applications~\footnote{
  A faster method has been proposed, restricted to systems with spins aligned with the orbital angular momentum~\cite{Nagar:2018gnk}.
}.

Surrogate or reduced order modeling techniques~\cite{Field:2013cfa,Purrer:2014fza,Purrer:2015tud,Blackman:2015pia,Blackman:2017pcm,Blackman:2017dfb,Lackey:2018zvw,Doctor:2017csx,Setyawati:2019xzw} provide established methods for accelerating slow waveforms while retaining very high accuracy. These techniques have been successfully applied to \ac{EOB}~\cite{Field:2013cfa,Purrer:2014fza,Purrer:2015tud,Bohe:2016gbl,Lackey:2018zvw} and \ac{NR}~\cite{Blackman:2015pia,Blackman:2017pcm,Blackman:2017dfb,Varma:2018mmi,Varma:2019csw} waveforms. They work by decomposing waveforms from a training set in orthonormal bases on sparse grids in time or frequency, and interpolating or fitting the resulting waveform data pieces over the binary parameter space. The result is a smooth, accurate (as tested against an independent validation set), and fast to evaluate (compared to the original waveform data) \ac{GW} model. These surrogate models have proven invaluable for \ac{GW} data analysis.

In this paper we present a \ac{ROM} for \acp{GW} from coalescing binaries with spins aligned with the orbital angular momentum which include the most important higher harmonics of the waveform in addition to the dominant $(\ell. m) = (2, \pm 2)$ spherical harmonic mode, as described by the \texttt{SEOBNRv4HM} model~\cite{Cotesta:2018fcv}.
Higher harmonics in the expansion of the gravitational waveform become important for asymmetric and massive compact binaries~\cite{Brown:2012nn,Capano:2013raa,Harry:2017weg, Varma:2014jxa,Varma:2016dnf,Kalaghatgi:2019log}. The model we construct here, \texttt{SEOBNRv4HM\_ROM}, includes the $(\ell, |m|) = (2, 1), (3, 3), (4, 4), (5, 5)$ modes. We show that \texttt{SEOBNRv4HM\_ROM} has a mismatch less than $\mathcal{O}(0.1\%)$ with \texttt{SEOBNRv4HM} and that it accelerates waveform evaluation by a factor 100 -- 200. 

While \texttt{SEOBNRv4HM} and \texttt{SEOBNRv4HM\_ROM} include the contribution of higher harmonics in the waveform, they lack a description of spin-precession and eccentricity. In the \ac{EOB} waveform family the effect of spin-precession has been taken into account in Refs.~\cite{Babak:2016tgq,Pan:2013rra} and only recently in Ref.~\cite{Ossokine:2020kjp} for the case of waveforms with higher harmonics. A surrogate for the latter waveform model is currently under development (see  Ref.~\cite{Gadre:2020}). For the phenomenological and the numerical relativity surrogate families similar models including the effect of precession and higher harmonics have been described in Refs.~\cite{Blackman:2017dfb,Varma:2018mmi,Varma:2019csw,Khan:2019kot,Pratten:2020ceb}. Waveforms emitted from binary systems in an eccentric orbit have not been studied extensively. So far, only a few inspiral-merger-ringdown eccentric waveform models have been constructed for non-spinning~\cite{Huerta:2017kez,Hinder:2017sxy} and aligned-spin binaries~\cite{Cao:2017ndf,Chiaramello:2020ehz}.

This paper is organized as follows. In Sec.~\ref{sec:SEOBNRv4HM} we give a brief description of the time-domain \texttt{SEOBNRv4HM} model. In Sec.~\ref{sec:techniques} we discuss various techniques we use to build the ROM, notably waveform conditioning in Sec.~\ref{sec:preparation_and_decomposition_of_wf_data}.
We continue with a summary of the basis construction and decomposition in Sec.~\ref{sec:SVD}, tensor product interpolation in Sec.~\ref{sec:TPI}. Domain decomposition in frequency and in parameter space is discussed in Sec.~\ref{sec:patching-in-frequency} and Sec.~\ref{sec:patching_parameter_space}, respectively.
We summarize how we connect the ROM with \ac{PN} solutions at low frequency in Sec.~\ref{sec:hybridization}.
We present results in Sec. ~\ref{sec:results} where we demonstrate the accuracy of the ROM in Sec~\ref{sec:accuracy}, and its computational efficiency in Sec.~\ref{sec:computational_performance}. We showcase a parameter estimation application in Sec.~\ref{sec:PEsec}.
Finally we conclude in Sec.~\ref{sec:conclusion}.

\section{The \texttt{SEOBNRv4HM} model}
\label{sec:SEOBNRv4HM}

The gravitational wave signal emitted by a coalescing binary black hole is usually divided into three different regimes: \textit{inspiral}, \textit{merger} and \textit{ringdown}. During the inspiral the two black holes move at a relative speed $v$ that is small compared to the speed of light $c$, therefore the solution to the two body problem can be found using a perturbative expansion in the small parameter $v/c$, the so-called \ac{PN} expansion~\cite{Blanchet:2013haa}. At some point, during the evolution of the binary system, the parameter $v/c$ ceases to be small and the \ac{PN} expansion is not valid anymore. This happens roughly at the innermost stable circular orbit (ISCO) and demarcates the end of the inspiral and the beginning of the merger regime. The signal in this regime can only be computed using \ac{NR} simulations that solve Einstein's equations for a \ac{BBH} system, fully numerically. Finally, in the ringdown regime, the perturbed black hole formed after the merger of the binary emits gravitational waves at frequencies that can be computed within the black hole perturbation theory formalism~\cite{Berti:2009kk}.

The EOB formalism, first introduced by Buonanno and Damour in Refs.~\cite{Buonanno:1998gg,Buonanno:2000ef}, provides a natural framework to combine these three regimes and produce a complete waveform with inspiral, merger and ringdown. Within the EOB formalism the \ac{PN} conservative dynamics of a \ac{BBH} system during the inspiral is resummed in terms of the dynamics of a test particle with an effective mass and spin around a deformed Kerr metric. This improved conservative dynamics is combined with a resummed energy flux~\cite{Pan:2010hz,PhysRevD.79.064004,PhysRevD.76.064028} to produce an inspiral waveform that is close to \ac{NR} solutions. To improve the agreement with \ac{NR} waveforms the \ac{EOB} conservative dynamics is also \textit{calibrated} using information from \ac{NR} simulations~\cite{Taracchini:2012ig,Taracchini:2013rva}. In the EOB waveform the merger and ringdown part is built using a phenomenological fit produced using informations from \ac{NR} waveforms and black hole perturbation theory~\cite{Nagar:2016iwa,Bohe:2016gbl}. NR-tuned versions of EOB models are usually referred to as EOBNR.

In this paper we focus on the \texttt{SEOBNRv4HM}~\cite{Cotesta:2018fcv} model, an extension of \texttt{SEOBNRv4}~\cite{Bohe:2016gbl} that includes the modes $(\ell, |m|) = (2, 1), (3, 3), (4, 4), (5, 5)$ in the waveform  in addition to the $(\ell, |m|) = (2, 2)$ mode already present in  \texttt{SEOBNRv4}. This model assumes spins aligned or anti-aligned with the direction perpendicular to the orbital plane $\hat L_N$, and we define the dimensionless spin parameter for \ac{BH} $i$ as $\chi_i = \vec S_i \cdot \hat L_N / m_i^2$, where $\vec S_i$ are the BH’s spins and $m_i$ their masses. \texttt{SEOBNRv4HM} has been validated against several NR waveforms in the mass-ratio - aligned-spin parameter space in the region $q \equiv m_1/m_2 \in [1,10]$, $\chi_{1,2} \in [-1,1]$ yielding typical mismatches of $\mathcal{O}(\leq1\%)$ for total masses in the range $[20,200] M_\odot$. 

\section{Techniques for building the ROM}
\label{sec:techniques}

In this Section we describe the construction of our ROM, from the preparation of the waveforms to the reduced basis and interpolation techniques. We use techniques developed for previous ROM models~\cite{Purrer:2014fza,Purrer:2015tud,Bohe:2016gbl}, which we generalized to the higher-harmonics case. 

We start with a general discussion of how to prepare and decompose waveform data for higher mode waveforms in Sec.~\ref{sec:preparation_and_decomposition_of_wf_data}. In particular, we discuss time domain conditioning in Sec.~\ref{sec:TD-conditioning}, stationary phase approximation in Sec.~\ref{sec:SPA}, the orbital carrier phase in Sec.~\ref{sec:carrier}, the introduction of coorbital modes in Sec.~\ref{sec:coorbital_modes_and_waveform_building_blocks}, scaling of frequencies in Sec.~\ref{sec:scaling_of_frequencies}.
We summarize basis construction in Sec.~\ref{sec:SVD} and tensor product interpolation in Sec.~\ref{sec:TPI}.
We also explain how we decompose the model in both frequency range patches (see Sec.~\ref{sec:patching-in-frequency}) and parameter space patches (see Sec.~\ref{sec:patching_parameter_space}).
Hybridization with inspiral waveforms is discussed in Sec.~\ref{sec:hybridization}.

\subsection{Preparation and decomposition of waveform data}
\label{sec:preparation_and_decomposition_of_wf_data}

The waveform polarizations $h_{+}$, $h_{\times}$ are decomposed in spin-weighted spherical harmonics as
\begin{equation}
	h_{+} - i h_{\times} = \sum_{\ell \geq 2} \sum_{m=-\ell}^{\ell} {}_{-2}Y_{\ell,m} h_{\ell m} \,.
\end{equation}
The $h_{\ell m}$ are called the harmonics or simply the modes of the gravitational wave, with $h_{22}$ and $h_{2,-2}$ the dominant harmonics corresponding to quadrupolar radiation. These modes $h_{\ell m}$ are affected by convention choices: first, by the choice of polarization vectors defining $h_{+}$, $h_{\times}$, and secondly by the definition chosen for the source frame in which the waveform is described. For non-precessing systems, the $z$-axis of the source frame is taken to be the normal to the orbital plane, with a residual freedom in choosing the origin of phase. One can take two points of view for the definition of phase: either fix the definition of the source frame (for instance, imposing that the initial separation vector is along $x$) and call ``phase'' the azimuthal angle of the observer in the source frame, or fix the direction to the observer (for instance in the plane $(x,z)$) and call ``phase'' the binary's orbital phase at a given time. We can also consider the definition of the origin of time as part of the source frame definition.

During the inspiral, the individual harmonics obey a simple overall scaling with the orbital phase as
\begin{equation}
\label{eq:deforbitalphasing}
h_{\ell m} \propto \exp\left[ - i m \phi_\mathrm{orb}\right] 
\end{equation}
but this scaling does not apply post-merger where the modes are driven by their respective ringdown frequencies.

There are several challenges regarding the conditioning of higher-harmonics waveforms for the purpose of reduced order modelling. We recall that one relies on two kinds of interpolation here: one is the interpolation of waveform pieces along the tracking parameter, either time or frequency, used to compress data; the other is the interpolation across the parameter space (masses and spins) used in the internals of the ROM, either of waveform quantities directly (as in~\cite{Field:2013cfa,Blackman:2015pia,Blackman:2017pcm,Blackman:2017dfb,Lackey:2018zvw,Varma:2018mmi,Varma:2019csw} in the empirical interpolation formalism), or as in our case, of reduced basis projection coefficients~\cite{Purrer:2014fza,Purrer:2015tud,Bohe:2016gbl}. Both these interpolations require smoothness, and discrete jumps can cause significant (and non-local) errors.

As a result, zero-crossings in the subdominant harmonics $h_{\ell m}$ (as noticed in Refs.~\cite{Cotesta:2018fcv,Nagar:2020pcj}) cause difficulties for the usual amplitude/phase representation: if the envelope of a mode crosses zero with a positively defined amplitude, the phase jumps by $\pi$, a discontinuity that will harm the interpolation performed when reconstructing the waveform. Among other advantages, this is alleviated by the procedure used in~\cite{Blackman:2017pcm, Varma:2019csw} of modelling the waveform in a coorbital frame where the dominant phasing of Eq.~\eqref{eq:deforbitalphasing} has been scaled out, so that a more robust real/imaginary representation can be chosen instead; here we will use the same kind of coorbital quantities, but built in the Fourier domain.

The natural $2\pi$ degeneracy in phase also requires care when interpolating across parameter space. Discrete $2\pi$ phase jumps leave the waveforms themselves invariant, but can break the interpolation in-between waveforms. This issue is particularly relevant when dealing with numerical Fourier transforms of time-domain waveforms: when phase-unwrapping the output of the Discrete Fourier Transform starting from $f=0$, numerical noise causes the $2\pi$ interval to be essentially random. In~\cite{Purrer:2014fza,Purrer:2015tud,Bohe:2016gbl} a linear fit of the Fourier-domain phase was removed. Here we will keep time and phase alignment information throughout the conditioning procedure, so instead we will impose a given $2\pi$ range for the phase at a reference point, corresponding to the time of alignment.

Other difficulties are caused by the relative alignment of the different harmonics. Dividing the phase of the dominant $h_{22}$ mode by 2, whether in time or frequency domain, comes with an ambiguity of $\pi$ then propagated as $m \pi$ to the other modes. Such an ambiguity is not necessarily a problem if the phase alignment is done as a last step when generating the waveform (as is the case in the \texttt{IMRPhenomHM} model~\cite{London:2017bcn}); giving up the geometric interpretation of the source-frame definition, it is sufficient that a $[0,2\pi]$ range in the ``phase'' input by the user corresponds to a $[0,2\pi]$ range in geometric phase. It becomes a problem, however, when we need to interpolate across parameter space to build a ROM. In particular, when working from the Fourier domain waveform alone, we do not have access to the orbital phase (as read from trajectories) to lift these kind of degeneracies. Here we will make sure that the alignment is performed in the time domain before taking Fourier transforms, and we will further introduce an artificial carrier signal to have access to a proxy of the orbital phase in the Fourier domain.

We detail below our conditioning procedure, chosen to circumvent these issues.

\subsubsection{Time-domain conditioning}
\label{sec:TD-conditioning}

In building this ROM, we will carry along time and phase alignment information all the way to the final Fourier-domain waveforms. This is in contrast to previous Fourier-domain waveform models (\texttt{SEOBNRv4\_ROM}, \texttt{IMRPhenomD}) where the time and phase are adjusted after generating the waveform, as will be described below.

Individual harmonics are decomposed as an amplitude\footnote{In general, it would be preferable to consider $a_{\ell m}$ as a slowly-varying envelope rather than a positive amplitude, in particular allowing it to change sign, as we expect zero-crossings of certain subdominant modes like $h_{21}$.} and phase, following
\begin{equation}
\label{eq:hlmtd}
	h_{\ell m}(t) = a_{\ell m}(t) \exp \left[ -i \phi_{\ell m}(t)\right] \,,
\end{equation}
with the scaling
\begin{equation}
\label{eq:philmscaling}
	\phi_{\ell m} = m \phi_\mathrm{orb} + \Delta\phi_{\ell m} \,,
\end{equation}
In the early inspiral regime, for low frequencies, the phases $\Delta\phi_{\ell m}$ are approximately constant. We choose the same polarization convention as in~\cite{lrr-2006-4}, for which we have
\begin{align}
  \label{eq:delta_phi_constants}
	\Delta\phi_{22} &\rightarrow 0\,, \nonumber\\
	\Delta\phi_{21} &\rightarrow \frac{\pi}{2}\,, \nonumber\\
	\Delta\phi_{33} &\rightarrow -\frac{\pi}{2}\,, \nonumber\\
	\Delta\phi_{44} &\rightarrow \pi\,, \nonumber\\
	\Delta\phi_{55} &\rightarrow \frac{\pi}{2} \,,
\end{align}
in the low-frequency limit. When getting closer to merger, deviations from Eq.~\eqref{eq:delta_phi_constants} become more important. In the notations of the EOB factorized waveforms~\cite{Damour:2008gu,Pan:2011gk}, these phase deviations come from the phases $e^{i \delta_{\ell m}}$ and tail factors $T_{\ell m}$ (see Eqs.~(14) and~(21) in~\cite{Pan:2011gk}), and from non-quasicircular corrections (NQCs) close to merger (see Eq. (22) in~\cite{Pan:2011gk}).

 We choose the source frame convention for our model by imposing that its direction $x$ is along the separation vector between the two black holes $n(t_\mathrm{align})$, with an arbitrary time of alignment in the late inspiral $t_\mathrm{align} = -1000M$ (with $t = 0$ being defined as the amplitude peak of $h_{22}$). In practice, rather than using $n(t_\mathrm{align})$ we simply impose
\begin{equation}
	\phi_{22} (t_\mathrm{align}) = 0 \,,
\end{equation}
and we use the orbital phase $\phi_\mathrm{orb}$ as read from the EOB dynamics to resolve the $\pi$-ambiguity and impose $\phi_\mathrm{ orb} \simeq 0$ rather than $\phi_\mathrm{orb} \simeq \pi$. These alignment properties will be reproduced, up to small numerical errors, by the reconstructed ROM waveforms.

\subsubsection{Stationary phase approximation}
\label{sec:SPA}

As we will use it to guide our conditioning procedure, we recall here the Stationary phase approximation (SPA) for waveforms with higher harmonics. First, we introduce the Fourier transform for a time-domain signal $h$ as
\begin{equation}
	\tilde{h}(f) = \int dt \, e^{2i \pi f t} h(t) \,.
\end{equation}
Note the sign difference in the exponential with respect to the more usual definition (used in particular in LAL~\cite{LALSuiteGit}). This choice is made for convenience, as it will ensure that Fourier-domain modes $\tilde{h}_{\ell m}$ with $m>0$ and $m<0$ have support at positive and negative frequencies respectively. One can come back to the LAL Fourier convention by the simple map $f \leftrightarrow -f$, which for real signals $h(t) \in \mathbb{R}$ translates as $\tilde{h}_\mathrm{ LAL}(f) = \tilde{h}^{*}(f)$.

Let us first consider a generic signal with an amplitude and phase as $h(t) = a(t)e^{-i \phi(t)}$ and define $\omega \equiv \dot{\phi}$. The SPA is applicable under the assumptions $\left| \dot{a} / (a\omega)\right| \ll 1$, $\left| \dot{\omega} / \omega^{2}\right| \ll 1$ and $\left| (\dot{a} / a)^{2} / \dot{\omega}\right| \ll 1$, that are well verified in the inspiral. Defining a time-to-frequency correspondence $t(f)$ implicitly by
\be\label{eq:deftfSPA}
	\omega(t(f)) = 2\pi f \,,
\ee
the Fourier tranform of the signal is then $\tilde{h}_\mathrm{ SPA}(f) = A_\mathrm{SPA}(f) e^{-i \Psi_\mathrm{SPA}(f)}$ with
\begin{subequations}
\begin{align}
	A_\mathrm{ SPA}(f) &= a(t(f)) \sqrt{\frac{2 \pi}{\dot{\omega}(t(f))}} \,, \\
	\Psi_\mathrm{ SPA}(f) &= \phi(t(f)) - 2\pi f t(f) + \frac{\pi}{4} \,.
\end{align}
\end{subequations}

Applying this to the individual $h_{\ell m}$ modes~\eqref{eq:hlmtd}, treating the $\Delta\phi_{\ell m}$ as constants, defining $\omega_{\ell m} = \dot{\phi}_{\ell m} \simeq m \omega_\mathrm{ orb}$, each mode will have a separate time-to-frequency correspondence
\begin{equation}
	\omega_{\ell m} \left( t^{\ell m}(f) \right) = 2\pi f \,,
\end{equation}
and the phase
\begin{equation}
\label{eq:PsilmSPA}
	\Psi_{\ell m}^\mathrm{ SPA}(f) = m \phi_\mathrm{ orb}\left(t^{\ell m}(f) \right) - 2\pi f t^{\ell m}(f) + \Delta\phi_{\ell m} + \frac{\pi}{4} \,.
\end{equation}
This is the Fourier-domain equivalent to the time-domain relation~\eqref{eq:philmscaling}. We note useful relations between different mode numbers. The various $t^{\ell m}(f)$ functions are related by
\begin{equation}
	t^{\ell m}\left(\frac{mf}{2}\right) = t^{22}(f) \,,
\end{equation}
while the phases satisfy
\begin{equation}
\label{eq:PhilmRelationSPA}
	0 = 2 \Psi_{\ell m}\left( \frac{m f}{2} \right) - m \Psi_{22}\left( f\right) - 2 \Delta\phi_{\ell m} + m \Delta\phi_{22} - \left(2 - m\right)\frac{\pi}{4} \,.
\end{equation}
This last relation holds regardless of the time and phase alignment of the waveform: as a phase change $\delta \phi_{\ell m} = m\delta\phi$, or a time change $\delta\Psi_{\ell m} (f) = -2\pi f \delta t$ would both leave $2\Psi_{\ell m}(mf/2) - m \Psi_{22}(f)$ invariant. It is sensitive however to the quantities $\Delta\phi_{\ell m}$ (that we treat here as constants in the early inspiral), that depend on the choice of polarization convention.

Finally, we recall that we can build a time-to-frequency correspondence directly from the Fourier-domain waveform as
\begin{equation}
\label{eq:deftf}
	t(f) \equiv -\frac{1}{2\pi} \frac{d\Psi}{df}.
\end{equation}
Note that this definition of time is strictly speaking only accurate in the inspiral phase, where the SPA applies and the two definitions~\eqref{eq:deftf} and~\eqref{eq:deftfSPA} coincide. However, it can be used as a proxy for time everywhere, as we can evaluate~\eqref{eq:deftf} for any frequency\footnote{The converse is not true: since $t(f)$ is not monotonic at high frequencies, building an unambiguous mapping $f(t)$ is only possible in the inspiral.} $f$.

\subsubsection{Orbital carrier}
\label{sec:carrier}

In order to carry over information about the alignement of the respective mode from the time domain to the Fourier domain, we find it convenient to introduce a fictitious carrier signal $k(t)$, that evolves with the orbital phase instead of twice the orbital phase as the $h_{22}$ mode does.
\begin{equation}
\label{eq:defcarrier}
	k(t) \equiv a_{22}(t) \exp \left[-i \frac{\phi_{22}(t)}{2} \right] \,.
\end{equation}
The choice made here of keeping the same amplitude as the $h_{22}$ mode is quite arbitrary, but will ensure that it decays in the ringdown, giving us a smooth Fourier transform for this carrier. Note that this construction is artificial, as the carrier does not correspond to any physical signal.

As mentioned before, the carrier half-phase $\phi_{22}/2$ comes with a $\pi$-degeneracy. We can alleviate this by forcing the carrier phase to be within $\pi$ of the orbital phase, as read from the SEOB dynamics, at the time of alignment. This is, in fact, our main motivation for building this carrier in the time domain: it allows us to avoid the issues listed above, with all the conditioning being ultimately tied to the orbital phase, a quantity that is smooth across parameter space.

The Fourier transform of the carrier signal introduced in~\eqref{eq:defcarrier} is decomposed as an amplitude and phase as
\begin{equation}
\label{eq:carrierFD}
	\tilde{k}(f) = A_{k}(f) \exp\left[ -i \Psi_{k}(f)\right] \,,
\end{equation}
where $A_{k} = |\tilde{k}|$ will be discarded in the rest of the conditioning. When the SPA applies, we have approximately
\begin{equation}
\label{eq:PsikSPA}
	\Psi_{k}(f) \simeq \Psi_{k}^\mathrm{ SPA}(f) = \phi_\mathrm{ orb}(t^{k}(f)) - 2\pi f t^{k}(f) + \frac{\pi}{4}\,,
\end{equation}
with $t^{k}(f)$ defined like in~\eqref{eq:deftfSPA} as
\begin{equation}
	\omega_\mathrm{ orb}(t^{k}(f)) = 2\pi f \,.
\end{equation}

Since $\tilde{k}(f)$ is computed via an FFT, nothing forbids arbitrary jumps of $2\pi$ of the phases between different points in parameter space. We use the relation above to remove this $2\pi$-ambiguity in $\Psi_{k}$. At the frequency $f_\mathrm{ align}$ such that $t(f_\mathrm{ align}) = t_\mathrm{ align}$, we impose
\begin{equation}
	\left| \Psi_{k}(f_\mathrm{ align}) - \left(\phi_\mathrm{ orb}(t_\mathrm{ align}) - 2\pi f_\mathrm{ align} t_\mathrm{ align} + \frac{\pi}{4} \right)\right| < \pi \,.
\end{equation}
In this way, $\Psi_{k}$ is directly tied to $\phi_\mathrm{ orb}$ that is smooth in parameter space in our time-domain conditioning procedure.

We will factor out the Fourier domain phase of the carrier, build a ROM for the carrier separately, and then factor in the modelled carrier phase when reconstructing the waveform.

\subsubsection{Fourier-domain coorbital modes and waveform building blocks}
\label{sec:coorbital_modes_and_waveform_building_blocks}

Next, we build coorbital modes by scaling out the Fourier-domain phase of the carrier following
\begin{equation}
\label{eq:defhcoorb}
	\tilde{h}_{\ell m}^{c}(f) = \tilde{h}_{\ell m}(f) \exp\left[i m \Psi_{k}(f/m)\right] \,.
\end{equation}
These modes are built so as to factor out the main contribution to the phase of the Fourier-domain modes, to leave the coorbital modes $\tilde{h}_{\ell m}^{c}$ with an approximately constant phase in the inspiral regime. Namely, for the inspiral regime, where the SPA is valid, $t^{k}(f) = t^{\ell m}(mf)$ and applying~\eqref{eq:PsilmSPA} and \eqref{eq:PsikSPA} gives
\begin{equation}
\label{eq:PsilmfromPsik}
	\Psi_{\ell m}(f) \simeq m \Psi_{k}\left(\frac{f}{m}\right) + \Delta\phi_{\ell m} + (1 - m) \frac{\pi}{4} \,.
\end{equation}
Note that our Fourier-domain construction is approximate, and these ``coorbital'' quantities $\tilde{h}^{c}_{\ell m}$ do not correspond exactly to a coorbital frame defined in the time domain as in~\cite{Blackman:2017pcm, Varma:2019csw}.

We stress that these modes are not strictly coorbital, in the sense that they are not built from a time-domain coorbital frame built from the orbital phase. Indeed, the definition~\eqref{eq:defhcoorb} is rooted in the Fourier domain, and its physical meaning is unclear in the high-frequency range where the SPA does not apply anymore.

Thus, the basic building blocks for the ROM will be
\begin{itemize}
	\item $\Psi_{k} = - \mathrm{Arg}\left[ \tilde{k}\right]$, the Fourier-domain carrier phase;
	\item $\mathrm{Re}\left( \tilde{h}_{\ell m}^{c} \right)$, the real part of the coorbital modes;
	\item $\mathrm{Im}\left( \tilde{h}_{\ell m}^{c} \right)$, the imaginary part of the coorbital modes.
\end{itemize}
Conversely, to rebuild the modes $\tilde{h}_{\ell m}$ from these waveform pieces, it is enough to factor in the carrier phase as in~\eqref{eq:defhcoorb}.

\subsubsection{Scaling of frequencies using ringdown frequency}
\label{sec:scaling_of_frequencies}

One of the prerequisites of our ROM procedure is to represent the waveform on a common frequency grid. However, the frequency range covered varies with physical parameters, notably spin. In \texttt{SEOBNRv4\_ROM}, this was alleviated by extending waveforms to higher frequencies. Here, we choose to apply a scaling to the frequencies of the waveform building blocks, depending on the ringdown frequency. For every mode $(\ell,m)$ and the carrier we define
\begin{subequations}
\begin{align}
\label{eq:freq_scalings}
	y_{\ell m} &= \frac{2\pi}{\omega^\mathrm{ QNM}_{\ell m}} Mf \,, \\
	y_{k} &= \frac{4\pi}{\omega^\mathrm{ QNM}_{22}} Mf \,,
\end{align}
\end{subequations}
where $\omega^\mathrm{ QNM}_{\ell m}$ is the quasi-normal mode frequency, and varies for different waveforms as it depends on the spin of the remnant black hole. We will then use for all waveforms a common grid of this rescaled parameter $y$. Given this scaling, we have to carefully adjust the starting frequency of the waveforms of our training set so that the frequency range of the carrier phase $\Psi_{k}$ covers all modes after undoing the scaling. The maximal values of $y_{\ell m}$, $y_{k}$ where we cut the data are $(y^\mathrm{ \max}_{22}, y^\mathrm{ \max}_{21}, y^\mathrm{ \max}_{33}, y^\mathrm{ \max}_{44}, y^\mathrm{ \max}_{55}) = (1.7, 1.7, 1.55, 1.35, 1.25)$ and $y^\mathrm{ \max}_{k} = 2.5$. This technique is only used for building the high-frequency ROM (see Sec.~\ref{sec:patching_parameter_space}); for the low-frequency ROM, the ringdown frequency is irrelevant and the scaling would induce an additional cost in generating the waveforms of the training set.

\subsection{SVD decomposition}
\label{sec:SVD}

We decompose all waveform data pieces defined in Sec.~\ref{sec:coorbital_modes_and_waveform_building_blocks} into respective \ac{SVD} bases and subsequently interpolate the projection coefficients in each \ac{SVD} basis over the parameter space, as discussed in Sec.~\ref{sec:TPI}. This method follows earlier work in~\cite{Purrer:2014fza,Purrer:2015tud,Bohe:2016gbl}.

We start with a waveform data piece $X(f_i; \vec\theta)$, given on a discrete grid in frequency $\{f_i\}_{i=1}^m$, and on a regular grid of points $\vec\theta$ in the three-dimensional binary parameter space in mass-ratio 
$q$ 
and aligned \ac{BH} spins 
$\chi_i$, 
$(q, \chi_1, \chi_2)$. We flatten the parameter grid and arrange the data in matrix form $X_{ij} = X(f_i; \theta_j) \in \mathbb{R}^{m \times n}$, where
$n$ is the total number of input waveforms.

We then resample the data in a log-spaced frequency grid of $300$ points. The number of points used for resampling are based on previous studies (see Ref.~\cite{Bohe:2016gbl,Purrer:2015tud}).
We compute the \ac{SVD}~\cite{GolubVanLoan,Demmel} $X = V \Sigma U^T$ and obtain an orthonormal basis for the column space of the matrix $X$ in the first $r = \rank(X)$ columns of $V$. The \ac{SVD} provides us with a decomposition of the range space of $X$, $\range(X) =  \operatorname{span} \{ v_1, \dots, v_r \}$, where the $v_j$ are the \emph{left singular vectors} of $X$.

Given the basis $\mathcal{B}_X = V$, we expand the waveform data pieces $x_j(f_i)$ that make up the columns of $X$ in this basis and can write the expansion $x_j \approx \sum c_X(\theta_j) \cdot \mathcal{B}_X$ with projection coefficient matrix $c_X = \mathcal{B}^T_X \cdot X$.
To construct a waveform model we need to predict the coefficients $c_X$ at a desired parameter space point $\theta^*$. To do that we need to fit or interpolate $c_X$ over the parameter space. This is discussed in Sec.~\ref{sec:TPI}.

\subsection{Tensor-product spline interpolation}
\label{sec:TPI}

In the following we describe how we obtain projection coefficients at arbitrary parameter space points. In low dimensional spaces we can afford to use dense grids built from the Cartesian product of one-dimensional sets of points. We choose cubic splines as the univariate interpolants and obtain a \ac{TPI}~\cite{Purrer:2014fza,githubTPI,Setyawati:2019xzw} for a three-dimensional coefficient tensor $c_{ijk}$ which can be written as
\begin{equation}
  \label{eq:TPI}
  I_\otimes[c](q, \chi_1, \chi_2) = \sum_{ijk} c_{ijk}
    \left( \Psi_i \otimes \Psi_j \otimes \Psi_k \right)(q, \chi_1, \chi_2).
\end{equation}
Here, the $\Psi$ are B-spline basis functions~\cite{deBoor} of order 3 for the chosen one dimensional sets of parameter space points in each dimension. We use ``not-a-knot'' boundary conditions to avoid having to specify derivatives at the domain boundaries.
We built the model using \texttt{TPI}, a \texttt{Cython/C} package~\cite{githubTPI} to provide tensor product spline interpolation in \texttt{Python}, and later implemented the model in  LAL~\cite{LALSuiteGit}.

\subsection{Patching in geometric frequency}
\label{sec:patching-in-frequency}

Here we discuss dividing the waveform domain in geometric frequency into separate sub-domains,  where we build a separate ROMs. In Sec.~\ref{sec:patching_parameter_space} we will instead discuss how to tackle non-uniform resolution requirements over the binary parameter space.

In the early inspiral waveforms modes tend to be well approximated by the PN expansion and an accurate ROM can be built from a relatively small training set.
In contrast, the high geometric frequency part of the waveform modes encapsulates the late inspiral, merger and ringdown part of the signal which is more complex and non-linear, and this is also the regime where EOB waveforms are tuned against NR where they are available. Building an accurate ROM for the high geometric frequency part of the waveform modes consequently requires a higher density of training set waveforms.

Therefore, it is natural to treat the low and high geometric frequency part of the waveform separately, following the construction of previous ROMs~\cite{Purrer:2015tud}. This allows us to make the training set for the low geometric frequency part of the waveforms significantly smaller and reduce the computational cost of the training. Waveforms for low mass binaries are the most costly waveforms to generate. The cost is exacerbated due to the presence of higher modes with $|m|>2$, since they require a lower starting orbital frequency to cover the same gravitational wave frequency range as the dominant mode.

\begin{figure}[h]
        \centering
        \includegraphics[width=0.6\textwidth]{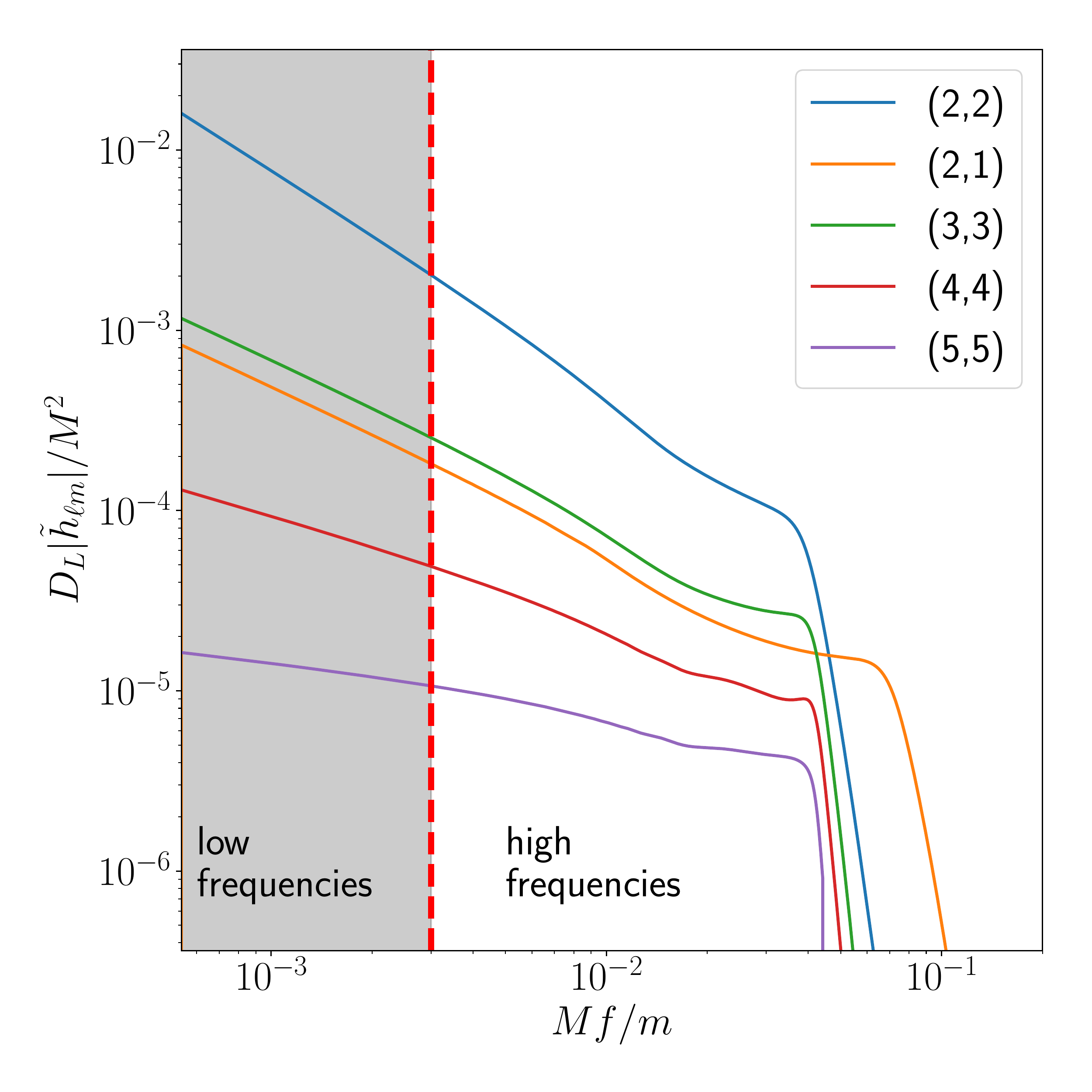}
	\caption{The complete ROM for each waveform mode is build by hybridizing a low and high frequency ROM. The $x$-axis shows the geometric frequency rescaled for each mode $(\ell,m)$ as $M f / m$, following the natural inspiral scaling of the frequency of the waveform modes with $m$.
  The low frequency sub-domain (black shaded region) starts at a geometric frequency of $Mf / m = 0.00025$, and
  transitions to the high frequency sub-domain at $Mf / m = 0.003$.
  }
  \label{fig:frequency-patches}
\end{figure}

In Fig.~\ref{fig:frequency-patches} we show the sub-division into low and high geometric frequency sub-domains. The low frequency sub-domain is connected with PN waveforms modes in the early inspiral,
as discussed in Sec.~\ref{sec:hybridization}. We generated the \texttt{SEOBNRv4HM} waveforms at a sufficiently low frequency (at $15$ Hz and a total mass of $5 M_\odot$ to allow for some tapering) to have the complete set of higher harmonics included in \texttt{SEOBNRv4HM} present at a frequency $Mf = 0.0005 * 5/2 \approx 0.0012$, where the low geometric frequency sub-domain starts. We choose the geometric transition frequency between the low and high frequency sub-domains to be $M f = 0.003*m$, using the natural inspiral scaling of the frequency of the waveform modes with $m$.. 
For the high frequency sub-domain we generated waveforms choosing the starting frequency as described in Sec.~\ref{sec:scaling_of_frequencies}, ensuring the generated waveforms after undoing the scaling~\eqref{eq:freq_scalings} will cover this transition frequency. The complete waveform modes are then generated by blending together the low and high frequency parts at the frequency using a variant of the Planck taper function described in Ref.~\cite{Hinder:2018fsy}.

\subsection{Hybridization with TaylorF2}
\label{sec:hybridization}

Here we describe how we carry out the hybridization of the \ac{ROM} waveform with the \texttt{TaylorF2} inspiral waveform.

After evaluating the \ac{ROM} waveform for all modes, we generate the \texttt{TaylorF2} amplitude and phase for the $(2,2)$ mode from the lowest frequency necessary to be able to start all inspiral modes at a user specified frequency. We blend the \texttt{TaylorF2} and \ac{ROM} amplitude and phase for the $(2,2)$ mode using the same Planck taper function used to connect high and low frequency ROM. We can obtain the higher mode \ac{PN} inspiral waveforms by rescaling the $(2,2)$ amplitude and phase.
For the phase we follow Eq.~\eqref{eq:PsilmfromPsik} 
and Eq.~\eqref{eq:delta_phi_constants} to compute the carrier phase from the \texttt{TaylorF2} (2,2) phase and rescale it to obtain the phase for each mode. We then align the inspiral phase with the \ac{ROM} phase for each mode and blend them together on a common frequency grid. For the amplitude we rescale the \texttt{TaylorF2} (2,2) amplitude according to the PN amplitudes given in Ref.~\cite{Mishra:2016whh} Eqs.(12a-12t).

\subsection{Patching in parameter space}
\label{sec:patching_parameter_space}

As already noted for the previous ROMs of EOBNR waveforms (see Refs.~\cite{Purrer:2014fza,Purrer:2015tud,Bohe:2016gbl}) model features often require more resolution in particular parts of the parameter space. However, regular grids do not allow for local refinement in $(q, \chi_1, \chi_2)$. Therefore, we partition the binary parameter space into subdomains on which resolution requirements can be satisfied with a particular regular grid choice.

\begin{figure}[h]
        \centering
        \includegraphics[width=0.5\textwidth]{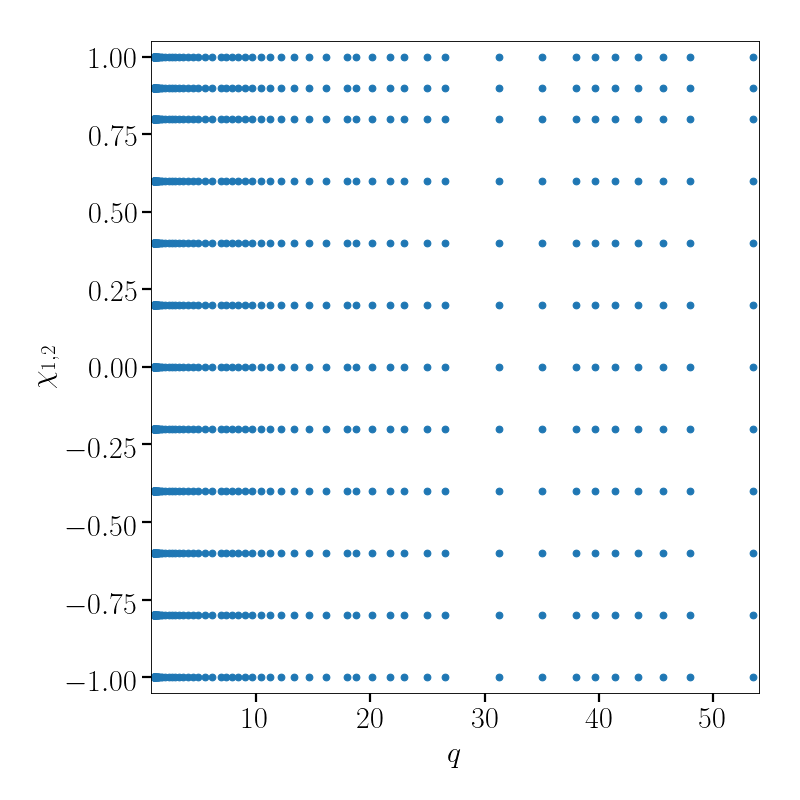}
	\caption{Location in parameter space $(q, \chi_1, \chi_2)$ of the waveforms used to build the inspiral ROM. For this ROM both spin components use the same grid.}
	\label{fig:LFGrid}
\end{figure}

The low frequency ROM does not need any special treatment and was built using waveforms placed on a Cartesian grid with 64 points in $q$ and 12 points in $\chi_{1,2}$ as shown in Fig.~\ref{fig:LFGrid}. Here, the 1D grids in $\chi_1$ and $\chi_2$ were chosen to be identical. The grids for $q$ and $\chi_{1,2}$ are the same as the ones used for \texttt{SEOBNRv4\_ROM} (see Sec.VII in Ref.~\cite{Bohe:2016gbl}), except that we limit the grid to $q = 50$.

On the other hand, as already noted in Refs.~\cite{Purrer:2015tud,Bohe:2016gbl}, modeling the non-linear merger and ringdown part of the waveform in the high geometric frequency ROM requires a higher resolution when approaching large positive values of the primary spin. Therefore, we build two different high frequency ROMs based on the value of the primary's spin, with one ROM having a finer grid in the region of high $\chi_1$.
The inclusion of higher modes in the \texttt{SEOBNRv4HM\_ROM} model also requires additional resolution near equal mass. The modes with odd $m$ vanish by symmetry on the line $q = 1$ and $\chi_1 = \chi_2$ and their behavior in the vicinity is non-trivial to model (see Refs.~\cite{Cotesta:2018fcv,Nagar:2020pcj}). Therefore, we build two different high geometric frequency ROMs in mass-ratio, one of which is covering the region $q \to 1$ with a finer grid. In total we then have four high frequency ROMs to correctly model the merger and ringdown part of the signal.

\begin{figure}[h]
        \centering
        \includegraphics[width=0.5\textwidth]{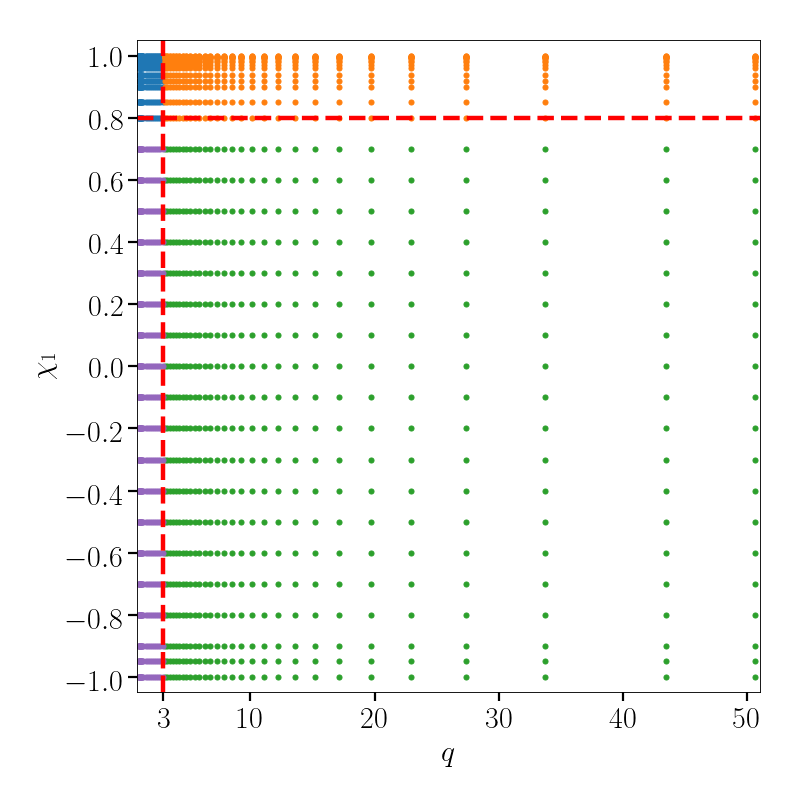}
	\caption{Location in the parameter space $(q, \chi_1)$ of the waveforms used to build the high-frequency ROM. The dashed red lines in the plot are the boundaries of the different patches, at $\chi_1 = 0.8$ and $q = 3$.
  }
\label{fig:HFGrid}

\end{figure}

The 2D projection of the grid in $(q, \chi_1)$ for these four ROMs is shown in Fig.~\ref{fig:HFGrid}. Since no special choice is made for the grid in $\chi_2$ we have omitted plotting the grid in this dimension. We set domain boundaries at $q = 3$ and $\chi_1 = 0.8$ for these four ROMs. In Table~\ref{tab:subdomains} we collect information on how the four patches are placed in parameter space and the number of gridpoints in each dimension.

\begin{table}
\centering
 \begin{tabular}{c|lclcl|c}
\hline
\hline
Patch    & \multicolumn{5}{c}{Intervals in $(q, \chi_1, \chi_2)$} & Points per interval\\
\hline
Patch 1  &  $[1,3]$  & $\cup$ & $[-1, 0.8]$ & $\cup$ & $[-1,1]$ & $24 \times 24 \times 24$\\
Patch 2  &  $[1,3]$  & $\cup$ & $(0.8, 1] $ & $\cup$ & $[-1,1]$ & $24 \times 17 \times 24$\\
Patch 3  &  $[3,50]$ & $\cup$ & $[-1, 0.8]$ & $\cup$ & $[-1,1]$ & $31 \times 24 \times 24$\\
Patch 4  &  $[3,50]$ & $\cup$ & $[0.8,1]  $ & $\cup$ & $[-1,1]$ & $30 \times 19 \times 24$\\
\hline
\hline
\end{tabular}
\caption{
  The grids for the high frequency \acp{ROM} on the four patches in parameter space shown in Fig.~\ref{fig:HFGrid}. The physical domain covered by each patch is defined by a Cartesian product of intervals in binary parameters 
  $(q, \chi_1, \chi_2)$. We also indicate the number of grid points in each parameter per patch.
}
\label{tab:subdomains}
\end{table}

\section{Results}
\label{sec:results}

In this section we discuss the accuracy and the increase in efficiency of \texttt{SEOBNRv4HM\_ROM} compared to \texttt{SEOBNRv4HM}.
Finally we also perform a parameter estimation study to demonstrate the potential of this model in data analysis applications.

\subsection{Accuracy of the model}
\label{sec:accuracy}

We start by defining the faithfulness function, used to assess the closeness between two waveforms when higher-order modes are included. We then use this faithfulness measure to determine how accurately the ROM reproduces \texttt{SEOBNRv4HM} waveforms.
\subsubsection{Definition of faithfulness}

A GW signal emitted by a spinning, non-precessing and non-eccentric
BBH is characterized by 11 parameters in the detector frame. These parameters are the BH masses $m_{1}$ and $m_{2}$, the (constant) projection
of the spins in the
direction perpendicular to the orbital plane $\chi_{1}$ and $\chi_{2}$, the angular position of the line of sight measured in the source's
frame $(\iota, \varphi_0)$,
the sky location of the source in the detector frame $(\theta, \phi)$, the luminosity distance $D_{\mathrm{L}}$,
the time of arrival $t_{\mathrm{c}}$ of the signal and finally the
polarization angle $\psi$.
The detector response can be written as:
\begin{equation}
\label{eq:det_strain}
h \equiv  F_+(\theta,\phi,\psi) \ h_+(\iota,\varphi_0, D_{\mathrm{L}}, \boldsymbol{\xi},t_{\mathrm{c}};t) 
+ F_\times(\theta,\phi,\psi)\ h_\times(\iota,\varphi_0, D_{\mathrm{L}}, \boldsymbol{\xi},t_\mathrm{c};t)\,,
\end{equation}
where masses and spins are combined in the vector  $\boldsymbol{\xi} \equiv (m_{1}, m_{2}, \chi_{1}, \chi_{2})$, and the functions $F_+(\theta,\phi,\psi)$ and $F_\times(\theta,\phi,\psi)$ are the antenna patterns~\cite{Sathyaprakash:1991mt,Finn:1992xs}.
This equation can be rewritten as:
\begin{align}
h = & \mathcal{A}(\theta,\phi)\big[\cos\kappa(\theta,\phi,\psi) \ h_+(\iota, \varphi_0, D_{\mathrm{L}}, \boldsymbol{\xi}, t_{\mathrm{c}};t) \nonumber \\
&+ \sin\kappa(\theta,\phi,\psi) \ h_\times (\iota, \phi, D_{\mathrm{L}}, \boldsymbol{\xi},t_{\mathrm{c}};t) \big],
\end{align}
with $\kappa(\theta,\phi,\psi)$ being the \textit{effective polarization}~\cite{Capano:2013raa} defined in the range $[0, 2\pi)$ as:
\begin{equation}
e^{i \kappa(\theta,\phi,\psi)} = \frac{F_+(\theta,\phi,\psi) + i F_\times(\theta,\phi,\psi)}{\sqrt{F_+^2(\theta,\phi,\psi) + F_\times^2(\theta,\phi,\psi)}},
\end{equation}
where the function $\mathcal{A}(\theta,\phi)$ is an overall amplitude and is defined as:
\begin{equation}
\mathcal{A}(\theta,\phi) = \sqrt{F_+^2(\theta,\phi,\psi) + F_\times^2(\theta,\phi,\psi)}\,.
\end{equation}
 Given a GW signal $h_{\mathrm{s}}$ (\texttt{SEOBNRv4HM} in our case) and a template waveform $h_{\mathrm{t}}$ (\texttt{SEOBNRv4HM\textunderscore ROM} in this context), we define the faithfulness (or match) as~\cite{Capano:2013raa,Harry:2016ijz,Cotesta:2018fcv}:
\begin{equation}
\label{eq:faith}
\mathcal{F}(\iota_{\textrm{s}},{\varphi_0}_{\textrm{s}},\kappa_{\textrm{s}}) \equiv  \max_{t_c, {\varphi_0}_{\mathrm{t}}, \kappa_{\textrm{t}}} \left[\left . \frac{ \left( h_{\mathrm{s}},\,h_{\mathrm{t}} \right)}{\sqrt{ \left( h_{\mathrm{s}},\,h_{\mathrm{s}} \right) \left( h_{\mathrm{t}},\,h_{\mathrm{t}} \right)}}\right \vert_{\substack{\iota_{\mathrm{s}} = \iota_{\mathrm{t}} \\\boldsymbol{\xi}_{\mathrm{s}} = \boldsymbol{\xi}_{\mathrm{t}}}} \right ],
\end{equation}
where parameters with the subscript ``s'' (``t'') refer to the signal (template) waveform. The expression above does not depend on $\mathcal{A}(\theta,\phi)$, therefore the only dependance on $(\theta,\phi,\psi)$ is encoded in $\kappa(\theta,\phi,\psi)$.
For the faithfulness calculation we optimize over the phases ${\varphi_0}_{\mathrm{t}}$ and $\kappa_{\textrm{t}}$ and the time of arrival $t_c$ because they are not interesting from an astrophysical perspective.
We use the standard definition of the inner product between two waveforms (see~\cite{Sathyaprakash:1991mt,Finn:1992xs}):
\begin{equation}
\left( a, b\right) \equiv 4\ \textrm{Re}\int_{f_\textrm{low}}^{f_\textrm{high}} df\,\frac{\tilde{a}(f) \ \tilde{b}^*(f)}{S_n(f)},
\end{equation}
where \char`\~ \, denotes the Fourier transform, * indicates the complex
conjugate and $S_n(f)$ is the one-sided power spectral density (PSD)
of the detector noise. For this computation we use the Advanced LIGO ``zero-detuned
high-power'' design sensitivity curve~\cite{Barsotti:2018}. The integral is computed between the
frequencies $f_{\textrm{low}} = 20 \mathrm{Hz}$ and $f_{\textrm{high}} = 3
\mathrm{kHz}$.
The same definition of faithfulness has been used to determine the agreement between \texttt{SEOBNRv4HM} and numerical relativity waveforms (see ~\cite{Cotesta:2018fcv})\footnote{In Ref.~\cite{Cotesta:2018fcv} we used an old version of the Advanced LIGO ``zero-detuned
high-power'' design sensitivity curve (in Ref.~\cite{Shoemaker:2010}). We have checked that the difference in the faithfulness calculation when using the new version of the sensitivity curve was negligible. For this reason here we report the calculations performed with the new curve described in Ref.~\cite{Barsotti:2018}.}.
Since the faithfulness given in Eq.~\eqref{eq:faith} depends on the signal parameters
$(\iota_{\textrm{s}},{\varphi_0}_{\textrm{s}},\kappa_{\textrm{s}})$, we will summarize the results using
the maximum and the average \textit{unfaithfulness} (or mismatch)
$[1-\mathcal{F}(\iota_{\textrm{s}},{\varphi_0}_{\textrm{s}},\kappa_{\textrm{s}})]$
over these parameters, namely~\cite{Buonanno:2002fy,Capano:2013raa,Harry:2016ijz}:

\begin{equation}
\mathcal{U}_{\mathrm{max}} \equiv \max_{\iota_{\mathrm{s}},{\varphi_0}_{\mathrm{s}},\kappa_{\mathrm{s}}}(1 -\mathcal{F}) \equiv  1 - \min_{\iota_{\mathrm{s}},{\varphi_0}_{\mathrm{s}},\kappa_{\mathrm{s}}}\mathcal{F}(\iota_{\textrm{s}},{\varphi_0}_{\textrm{s}},\kappa_{\textrm{s}}) \label{eq:max_unfaith}\,,
\end{equation}
\begin{equation}
\bar{\mathcal{U}} \equiv \langle
1-\mathcal{F}\rangle_{\iota_{\mathrm{s}},{\varphi_0}_{\mathrm{s}},\kappa_{\mathrm{s}}} \equiv  1 - \frac{1}{8\pi^2}\int_{0}^{2\pi} d\kappa_{\mathrm{s}} \int_{-1}^{1} d(\cos\iota_s) \int_{0}^{2\pi} d{\varphi_0}_{\mathrm{s}} \ \mathcal{F}(\iota_{\textrm{s}},{\varphi_0}_{\textrm{s}},\kappa_{\textrm{s}})\,. \label{eq:avg_unfaith}
\end{equation}

\subsubsection{Faithfulness against \texttt{SEOBNRv4HM}}

In order to avoid biases in data analysis applications when using the ROM instead of \texttt{SEOBNRv4HM},
it is important to verify that the additional modeling error introduced in the construction of the ROM is negligible compared to the inaccuracy of the \texttt{SEOBNRv4HM} waveforms with respect to the NR simulations.
Since the typical unfaithfulness between \texttt{SEOBNRv4HM} and NR waveforms is $\mathcal{O}(1\%)$ (see Figs. (11) and (12) in Ref.\cite{Cotesta:2018fcv}), it is therefore natural to require
the unfaithfulness between \texttt{SEOBNRv4HM} and \texttt{SEOBNRv4HM\textunderscore ROM} to be $\mathcal{O}(0.1\%)$ or less.
To that end we have generated 10000 \texttt{SEOBNRv4HM} waveforms with random (uniformly distributed) values of $(q, \chi_1, \chi_2)$ and computed their match against the same waveforms produced with \texttt{SEOBNRv4HM\_ROM}.

\begin{figure}[h]
        \centering
        \includegraphics[width=0.6\textwidth]{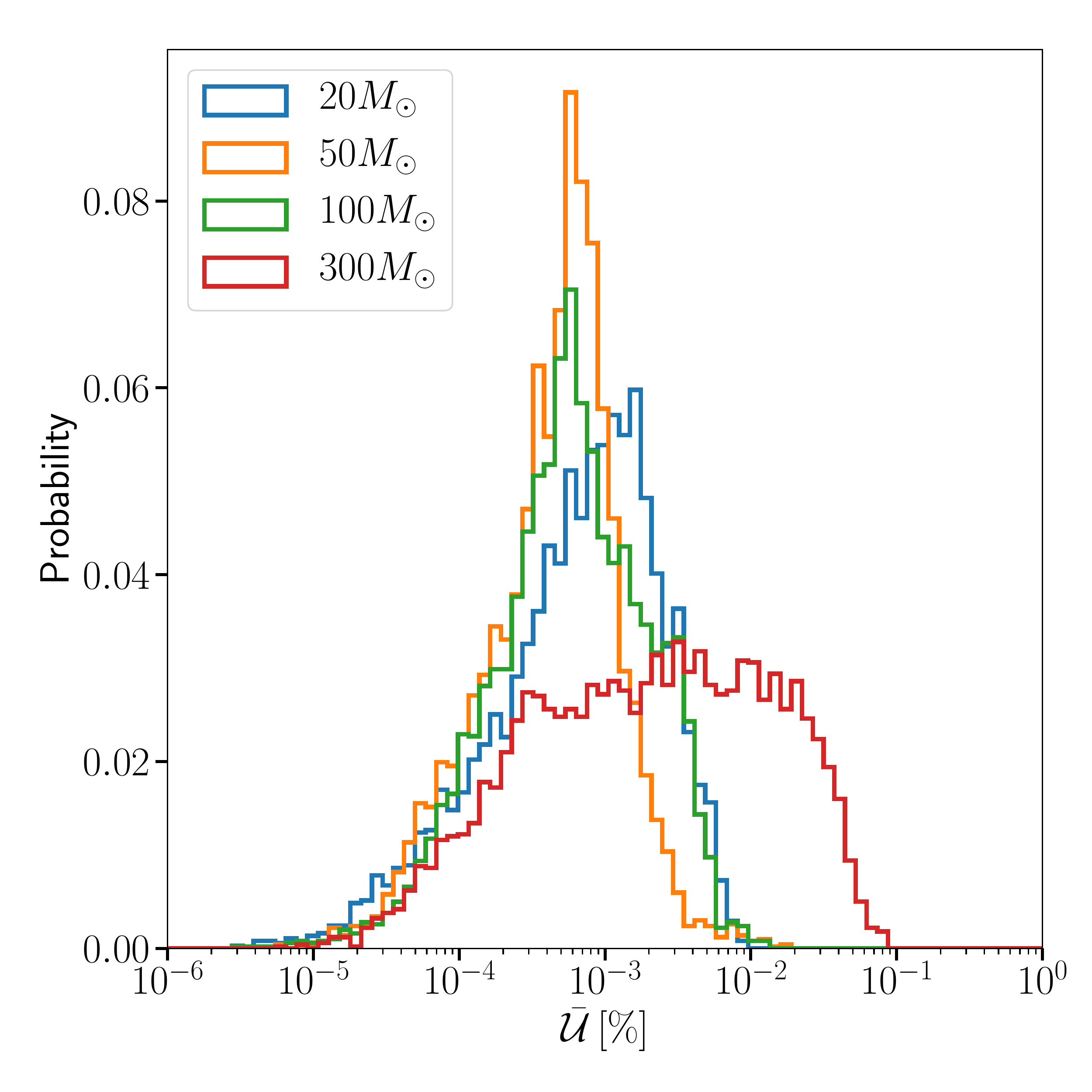}
	\caption{Histograms of the unfaithfulness $\bar{\mathcal{U}}$ (in percent) between \texttt{SEOBNRv4HM} and ROM waveforms for different values of the total mass.
  }
\label{fig:average_unfaith_M}
\end{figure}

We summarize these results in Fig.~\ref{fig:average_unfaith_M} where we show a histogram with the unfaithfulness $\bar{\mathcal{U}}$ computed between the ROM and \texttt{SEOBNRv4HM} waveforms for different values of the total mass. For each total mass we report in Tab.~\ref{tab:average_unfaith_M_summary_table} the median and maximum values of these unfaithfulness distributions.

\begin{table}
\centering
 \begin{tabular}{c|c|c|c|c}
\hline
\hline
Total mass $[M_\odot]$             & 20 & 50 & 100 & 300\\
\hline
$\underset{(q, \chi_1,\chi_2)}{\textrm{med}} \bar{\mathcal{U}} [\%]$  & 0.001 & 0.001 & 0.001 & 0.002   \\
\hline
$\underset{(q, \chi_1,\chi_2)}{\max} \bar{\mathcal{U}} [\%]$  & 0.01 & 0.02 & 0.01 & 0.08   \\
\hline
\hline
\end{tabular}
\caption{
  Median and maximum values of the $\bar{\mathcal{U}}$ distributions in
  Fig.~\ref{fig:average_unfaith_M} for different values of the total mass.
}
\label{tab:average_unfaith_M_summary_table}
\end{table}

The median of these mismatch distributions is weakly dependent on the total mass and it is always $\leq 0.002\%$ while their maximum value is always $\leq 0.08 \%$.
In Fig.~\ref{fig:skyaverageall} we display the distribution of mismatches shown in Fig.~\ref{fig:average_unfaith_M} as a function of $(q, \chi_1)$ and for different values of the total mass. The largest mismatches between the ROM and \texttt{SEOBNRv4HM} are obtained for $M = 300.0 M_\odot$ and large negative $\chi_1$, as it is clear from Fig.~\ref{fig:average_unfaith_M} and Fig.~\ref{fig:skyaverageall} (bottom right panel). ROM GW modes are generated up to a maximum frequency (in geometric units) that scales with the inverse of the total mass of the system. For large total masses the lack of signal above this maximum frequency is the main source of inaccuracy of the ROM. This maximum frequency for each mode is proportional to its least damped quasi-normal mode frequency as described in Eq.~\ref{eq:freq_scalings}. The mismatch is larger for large negative spins because the least damped quasi-normal mode frequency decreases in this region of the parameter space. We highlight that in this region the ROM waveforms still have mismatches $\lesssim 0.1\%$ against \texttt{SEOBNRv4HM} waveforms as demanded at the beginning of this section.
The results described above do not change substantially when considering the distribution of $\mathcal{U}_{\mathrm{max}}$ instead of $\bar{\mathcal{U}}$. In Tab.~\ref{tab:maximum_unfaith_M_summary_table} we report the median and maximum values of these distributions.

\begin{table}
\centering
 \begin{tabular}{c|c|c|c|c}
\hline
\hline
Total mass $[M_\odot]$             & 20 & 50 & 100 & 300\\
\hline
$\underset{(q, \chi_1,\chi_2)}{\textrm{med}}\mathcal{U}_{\mathrm{max}} [\%]$  & 0.001 & 0.001 & 0.001 & 0.004   \\
\hline
$\underset{(q, \chi_1,\chi_2)}{\max} \mathcal{U}_{\mathrm{max}}[\%]$  & 0.01 & 0.02 & 0.03 & 0.17   \\
\hline
\hline
\end{tabular}
\caption{
  Median and maximum values of the $\mathcal{U}_{\mathrm{max}}$ distributions for different values of the total mass.
}
\label{tab:maximum_unfaith_M_summary_table}
\end{table}

\begin{figure}
    \centering
    \begin{minipage}{0.5\textwidth}
        \centering
        \includegraphics[width=1.0\textwidth]{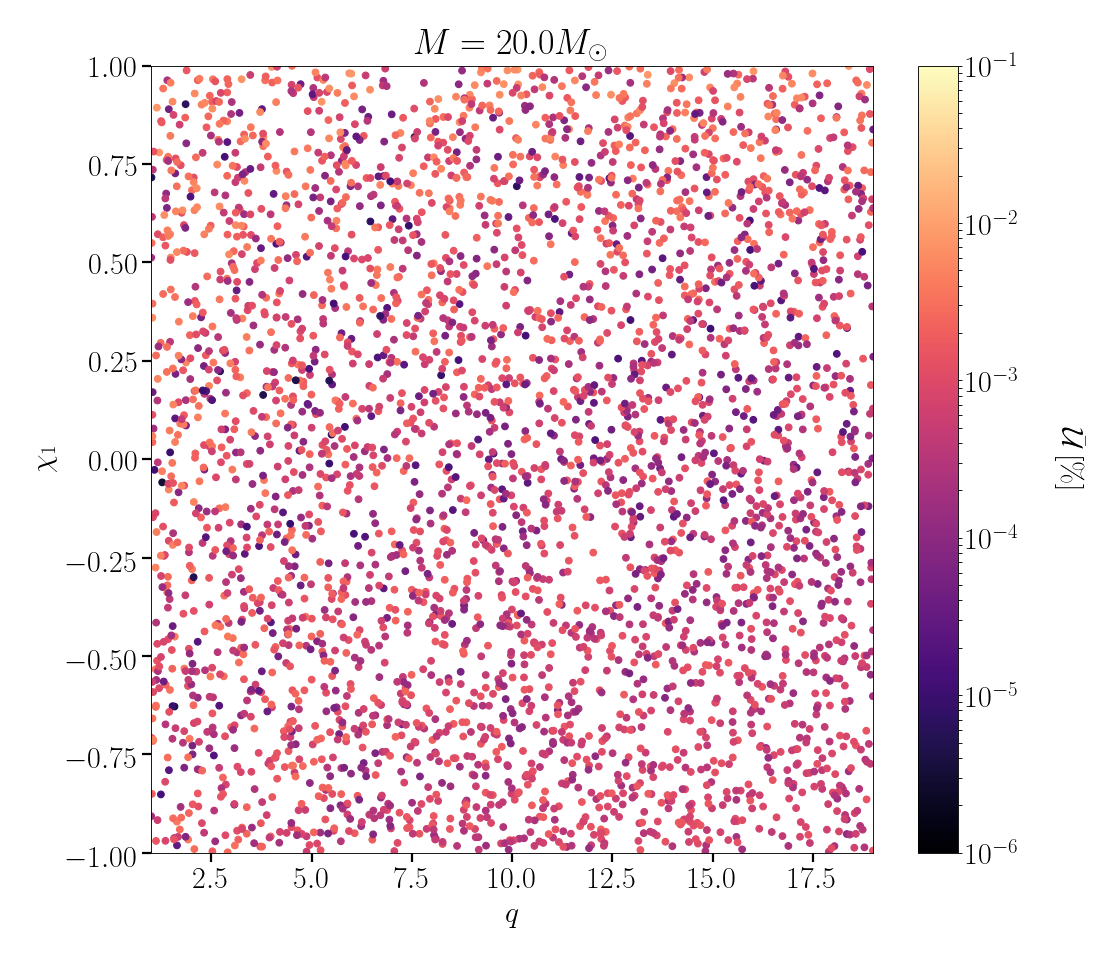}
    \end{minipage}\hfill
    \begin{minipage}{0.5\textwidth}
        \centering
        \includegraphics[width=1.0\textwidth]{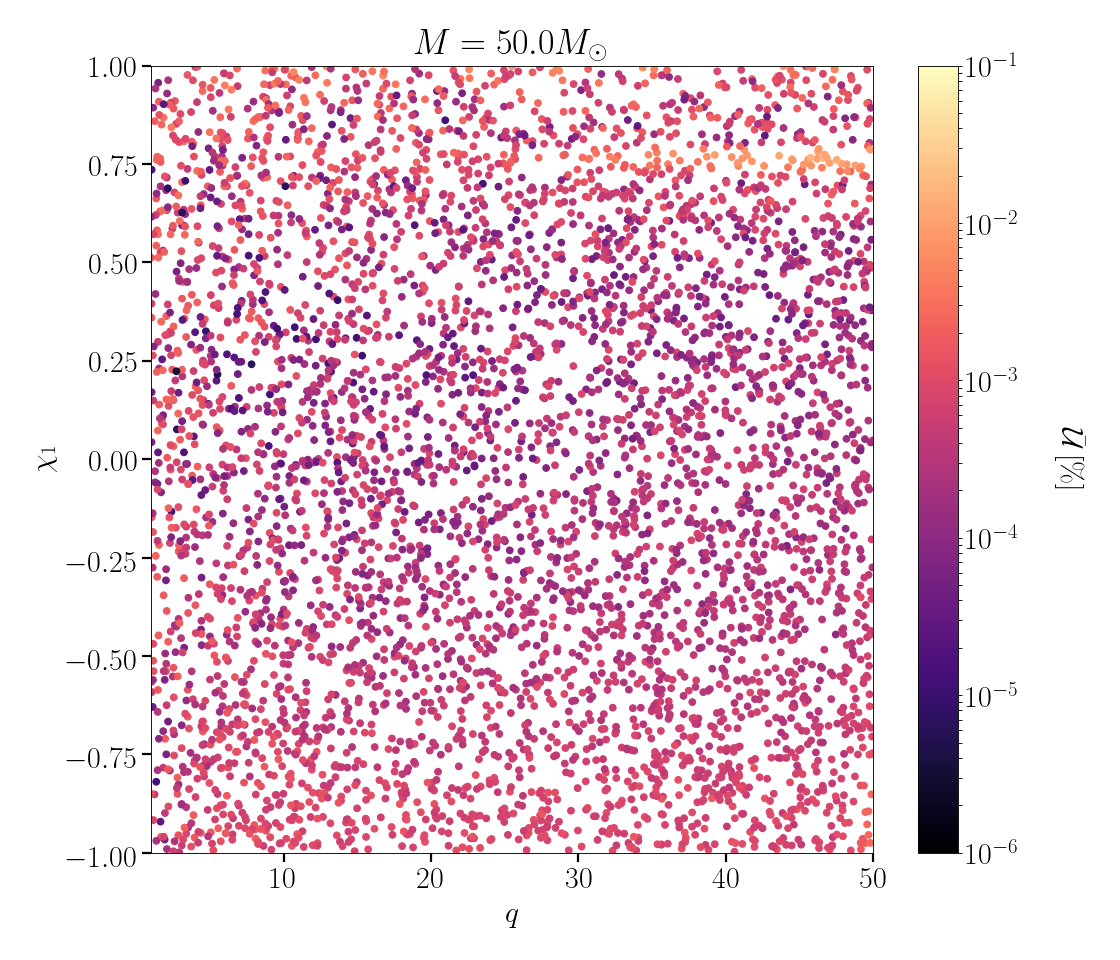}
    \end{minipage}
    \begin{minipage}{0.5\textwidth}
        \centering
        \includegraphics[width=1.0\textwidth]{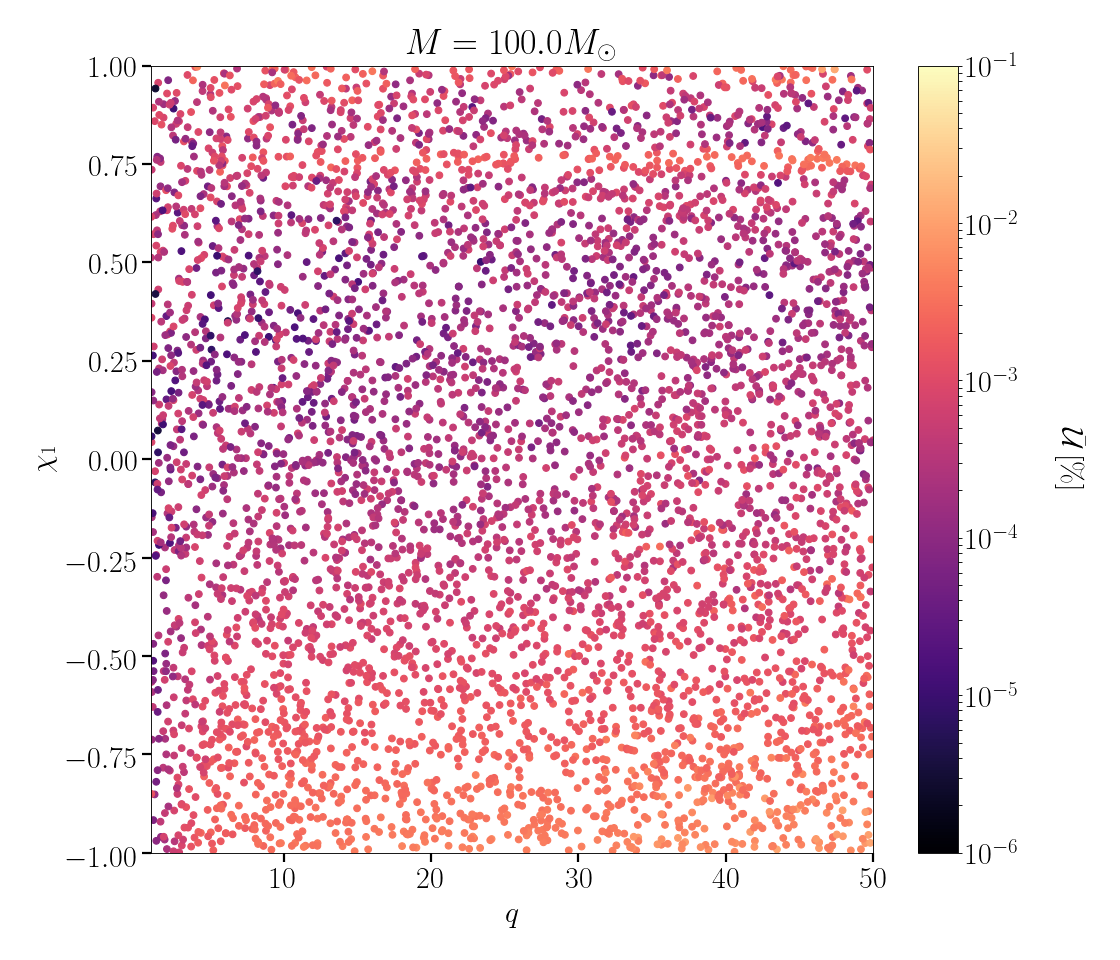}
    \end{minipage}\hfill
    \begin{minipage}{0.5\textwidth}
        \centering
        \includegraphics[width=1.0\textwidth]{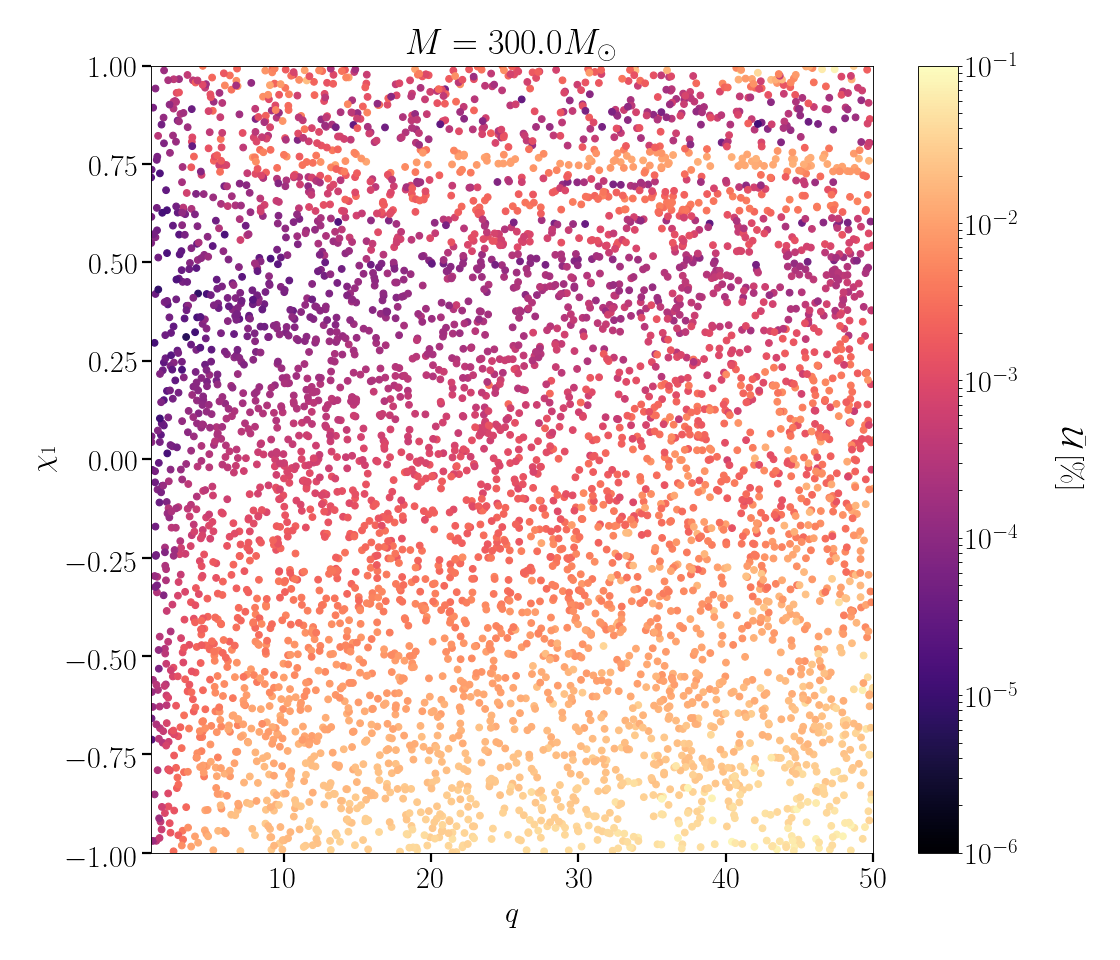}
    \end{minipage}
	\caption{Unfaithfulness $\bar{\mathcal{U}}$ between the ROM and \texttt{SEOBNRv4HM} as a function of $(q,\chi_1)$ and for different values of the total mass.  For $M = 20 M_\odot$ there are no data in the region $q > 19$ because for these system $m_2$ would have an unphysical subsolar mass.}
\label{fig:skyaverageall}
\end{figure}

\begin{figure}[htp]
        \centering
        \includegraphics[width=0.6\textwidth]{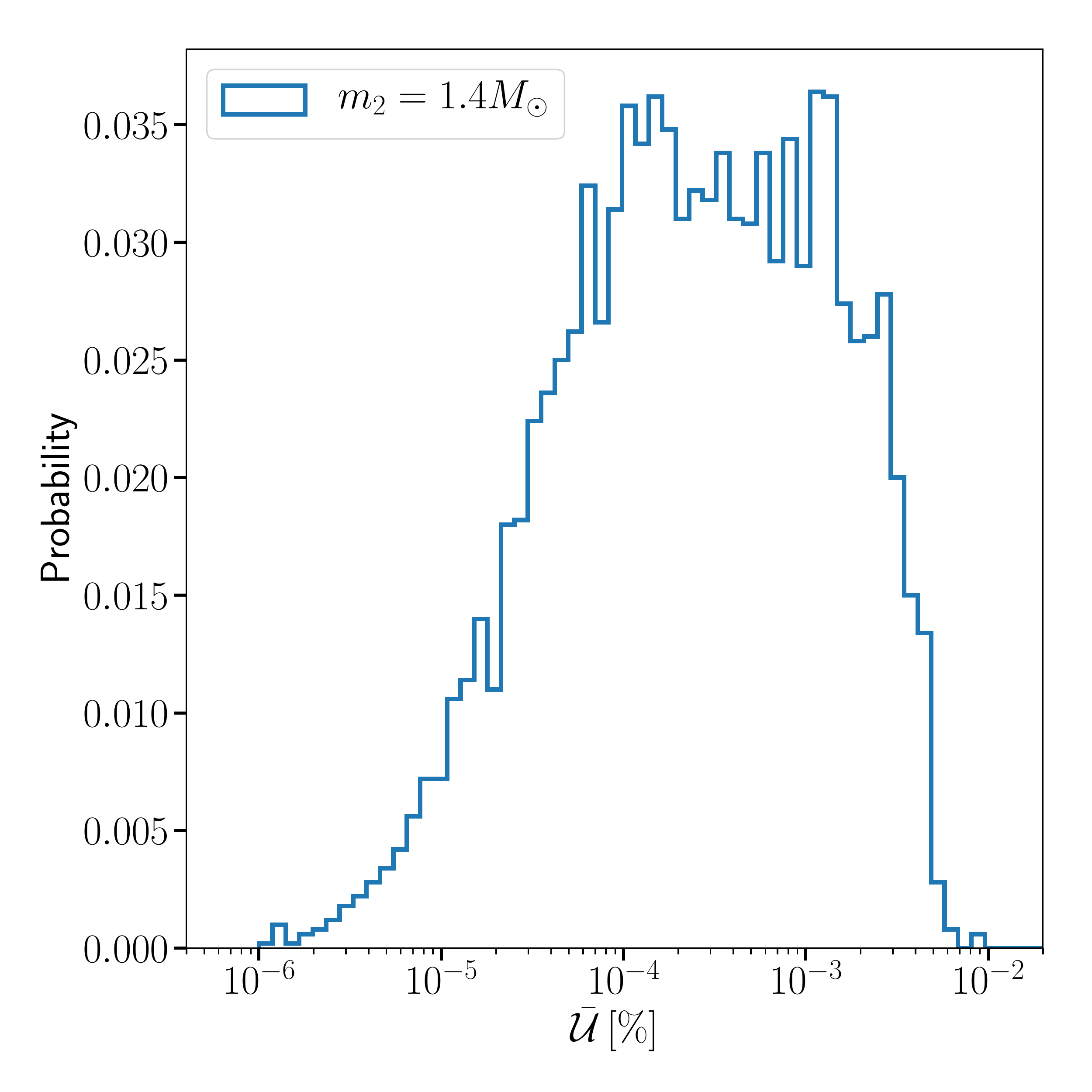}
	\caption{Histogram of the unfaithfulness $\bar{\mathcal{U}}$ between the ROM and \texttt{SEOBNRv4HM}. The \texttt{SEOBNRv4HM} waveforms used in the match calculations have $m_2$ fixed to $1.4 M_\odot$ and $m_1$ uniformly distributed in the range $1.4 M_\odot \leq m_1 \leq 18.6 M_\odot$.}
\label{fig:average_unfaith_m2}
\end{figure}

\begin{figure}[htp]
        \centering
        \includegraphics[width=0.6\textwidth]{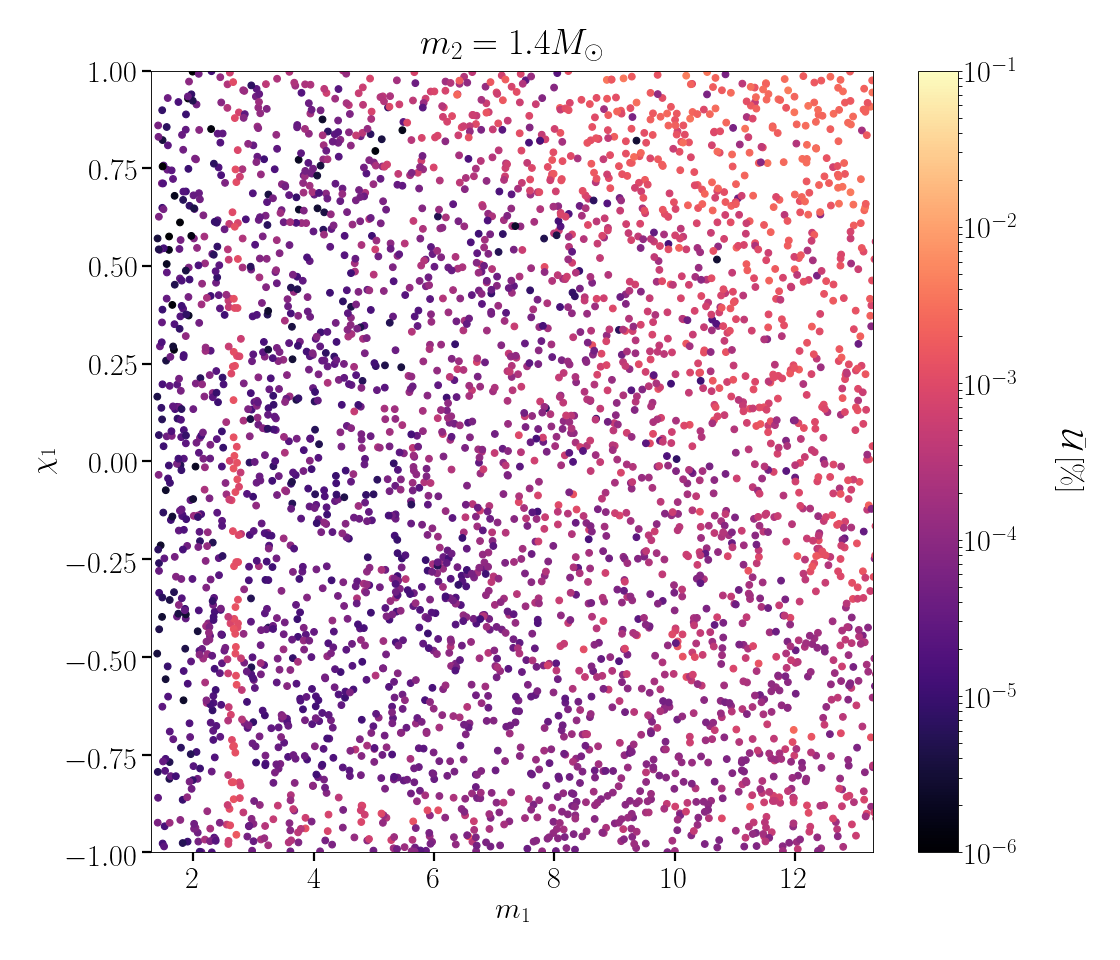}
	\caption{Unfaithfulness $\bar{\mathcal{U}}$ between the ROM and SEOBNRv4HM as a function of $(m_1,\chi_1)$ and with $m_2 = 1.4 M_\odot$.}
\label{fig:skyaverageallm2}
\end{figure}

For total masses $M \leq 20 M_\odot$ it is more convenient to summarize the results of the faithfulness calculations as an histogram with a fixed $m_2$ instead of the total mass.
In Fig.~\ref{fig:average_unfaith_m2} we show the $\bar{\mathcal{U}}$ distribution when fixing  $m_2 = 1.4 M_\odot$ and varying $m_1$ in the interval $1.4 M_\odot \leq m_1 \leq 18.6 M_\odot$ such that the total mass of the system is always $M \leq 20 M_\odot$.
The median of this distribution is $0.0003\%$ while its maximum is $0.01 \%$. In Fig.~\ref{fig:skyaverageallm2} we report the $\bar{\mathcal{U}}$ distribution in Fig.~\ref{fig:average_unfaith_m2} as a function of $(m_1,\chi_1)$. The accuracy of the ROM in this case degrades for large values of $m_1$ and large positive spins but it is still well within the requirements.  Also in this case the results are not very different when looking at the $\mathcal{U}_{\mathrm{max}}$ distribution for which the median is still $0.0003\%$ while the maximum increases to $0.02 \%$.

These analyses demonstrate that the modeling error introduced by the ROM is negligible with respect to the difference between \texttt{SEOBNRv4HM} and NR waveforms. For this reason the mismatch of the ROM against the NR waveforms is essentially the same as \texttt{SEOBNRv4HM} (see Figs.11-12 in Ref~\cite{Cotesta:2018fcv} and Fig.6 in Ref.~\cite{Varma:2018mmi}\footnote{The NR surrogate \texttt{NRHybSur3dq8} has a typical unfaithfulness against the NR simulations of $\mathcal{O}(10^{-3}\%)$, that is neglibigle with respect to the unfaithfulness between the NR simulations and the model \texttt{SEOBNRv4HM} (that is of $\mathcal{O}(1\%)$). Therefore in this case we can consider the \texttt{NRHybSur3dq8} waveform equivalent to an NR waveform. We make the same assumption in the parameter estimation study in Sec.~\ref{sec:PEsec}.}).

\subsection{Computational performance}
\label{sec:computational_performance}

In this section we discuss the computational performance of the ROM in terms of walltime for generating a waveform. We first compare the ROM to \texttt{SEOBNRv4HM} and then to other waveform models that include higher-order modes.

\subsubsection{Speedup with respect to \texttt{SEOBNRv4HM}}
\label{sec:speedup}

The speedup of the ROM with respect to \texttt{SEOBNRv4HM} is computed by the ratio of the walltimes of the two models for generating a frequency domain waveform at the same parameters. Since \texttt{SEOBNRv4HM} is a time domain model, we first generate the waveform in the time domain at a sample rate of 16384 Hz
and then compute its Fourier transform. The ROM waveform is already in the frequency domain and it is generated using the sampling interval set to $1/T$ where $T$ is the duration in seconds of the associated time domain waveform. The maximum frequency of the \texttt{SEOBNRv4HM\_ROM} waveform is set to 8192 Hz.

\begin{figure}[htp]
        \centering
        \includegraphics[width=0.6\textwidth]{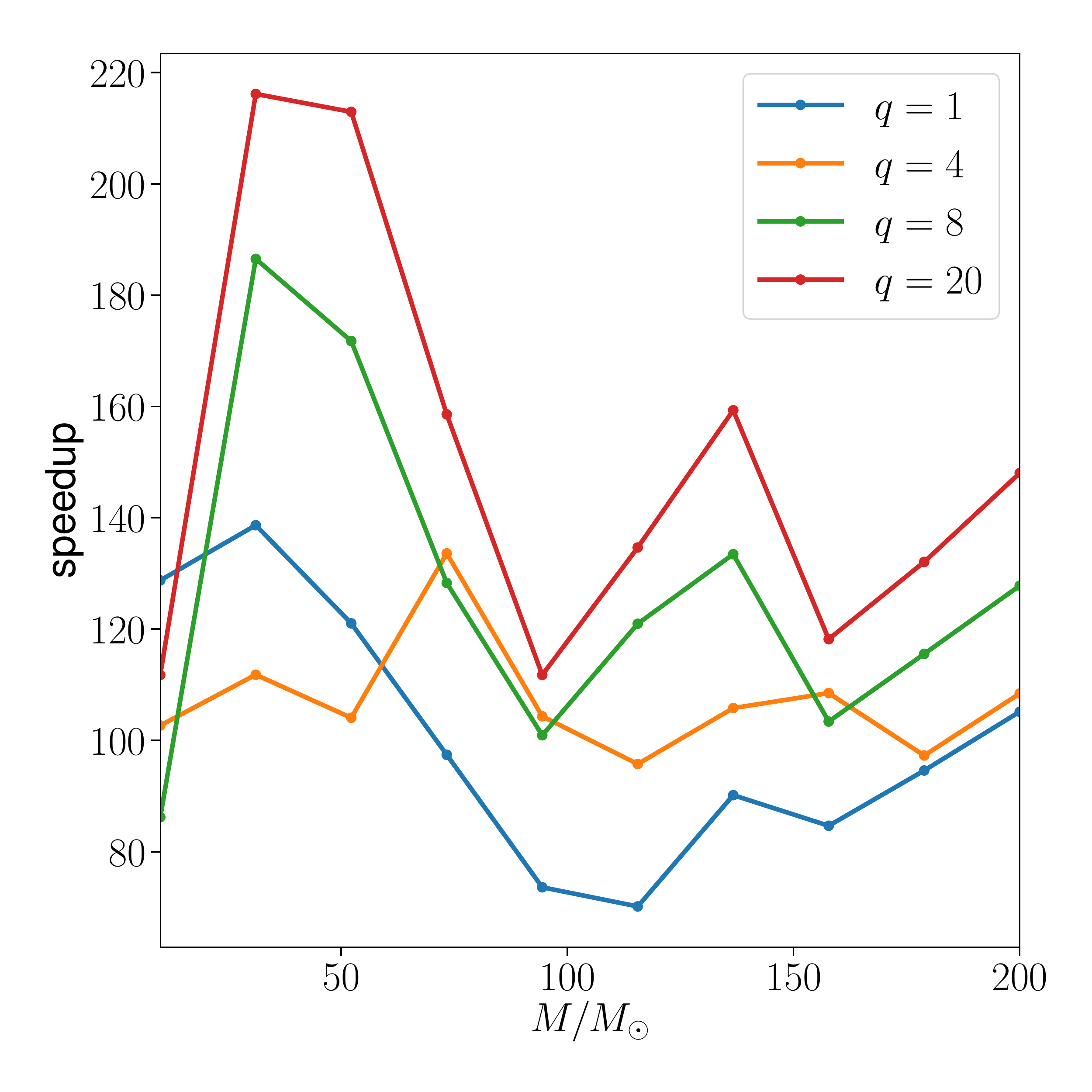}
	\caption{Speedup of waveform generation of the ROM with respect to \texttt{SEOBNRv4HM} as a function of the total mass and for different values of mass ratio.}
	\label{fig:speedup}
\end{figure}

In Fig.~\ref{fig:speedup} we show this speedup as a function of the total mass and for different values of the mass ratio. The speedup is of order $100$. It increases with mass ratio and decreases with total mass. The maximum speedup is found around a total mass of $50 M_\odot$.
Since the spins have only a limited effect on the waveform duration, the speedup depends only weakly on them.

\begin{figure}[htp]
         \centering
         \includegraphics[width=0.6\textwidth]{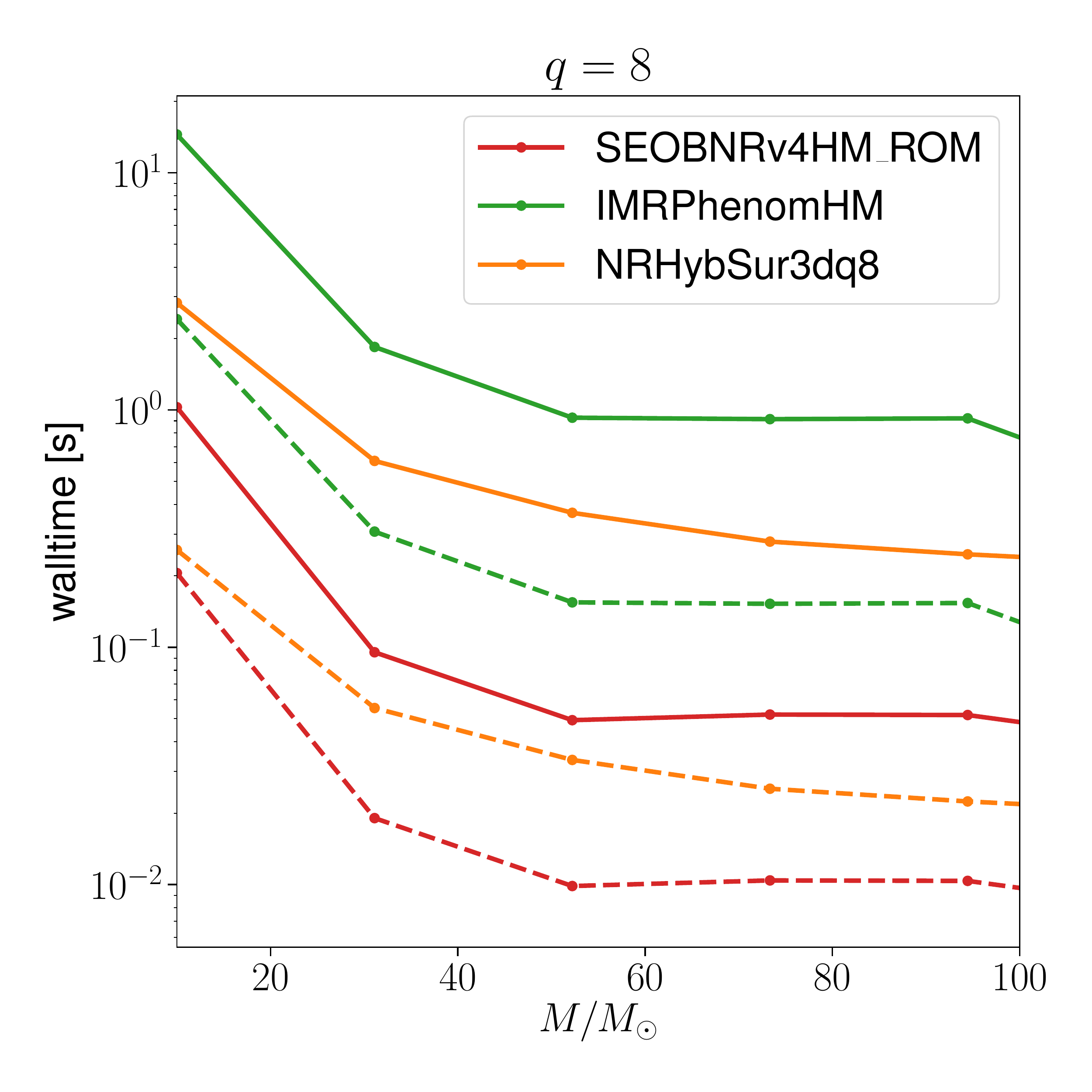}
   \caption{Walltime comparison between different spin-aligned waveform models with higher-order modes as a function of the total mass and for $q = 8$. The dashed lines indicate the walltime normalized by the number of modes included in the model respectively 5 fo \texttt{SEOBNRv4HM\_ROM}, 6 for \texttt{IMRPhenomHM} and 11 for \texttt{NRHybSur3dq8}.
   }
   \label{fig:walltimeq8}
\end{figure}

\subsubsection{Walltime comparison}
\label{sec:walltime}

We now perform a comparison of the walltime for generating a waveform between \texttt{SEOBNRv4HM\_ROM} and two waveform models that also include higher-order modes, namely \texttt{IMRPhenomHM}~\cite{London:2017bcn} and \texttt{NRHybSur3dq8}~\cite{Varma:2018mmi}.
As in Sec.~\ref{sec:speedup} we define walltime as the time to produce a frequency domain waveform at the same parameters. Since \texttt{NRHybSur3dq8} is a time domain model, we first generate the waveform in the time domain at a sample rate of 16384 Hz and then we compute its Fourier transform. For IMRPhenomHM and \texttt{SEOBNRv4HM\_ROM} the waveforms are generated in the frequency domain with a maximum frequency of 8192 Hz and a sampling interval of $1/T$ where $T$ is the duration in seconds of the associated time domain waveform.
The waveform models \texttt{SEOBNRv4HM\_ROM}, \texttt{IMRPhenomHM} and \texttt{NRHybSur3dq8} include a different numbers of modes in the waveform, respectively five $\left[(\ell, |m|) = (2,2),\allowbreak (2,1),\allowbreak (3,3),\allowbreak (4,4),\allowbreak (5,5)\right]$, six $[(\ell, |m|) = (2,2),\allowbreak (2,1),\allowbreak (3,3), \allowbreak (3,2), \allowbreak (4,4), \allowbreak (4,3)]$ and eleven $[(\ell,|m|) = (2,2), \allowbreak (2,1), \allowbreak (2,0), \allowbreak (3,3), \allowbreak (3,2), \allowbreak (3,1), \allowbreak (3,0), \allowbreak (4,4), \allowbreak (4,3),\allowbreak (4,2), \allowbreak (5,5)]$ modes. Since the total walltime is an increasing function of the number of modes, we also compute walltimes normalized by the number of modes to factor out this effect.
In Fig.~\ref{fig:walltimeq8} we show the walltime for generating a waveform with the different models as a function of the total mass for $q = 8$. \texttt{SEOBNRv4HM\_ROM} has walltimes of $\mathcal{O}(10)$ ms and is roughly 10 times faster than \texttt{IMRPhenomHM} or \texttt{NRHybSur3dq8}. When normalizing the walltime to the number of modes \texttt{SEOBNRv4HM\_ROM} is still about 10 times faster than \texttt{IMRPhenomHM}, but only $\sim 3$ times faster than \texttt{NRHybSur3dq8}.

\subsection{Parameter estimation study}
\label{sec:PEsec}

In this section we use the \texttt{SEOBNRv4HM\_ROM} model in a parameter estimation application.
For this purpose we create two mock signals (or injections) with the same binary parameters, using either \texttt{SEOBNRv4HM} or \texttt{NRHybSur3dq8} to generate the waveform. We then use \texttt{SEOBNRv4HM\_ROM}, \texttt{SEOBNRv4\_ROM}, and, as a comparison between waveform models for the second case, \texttt{IMRPhenomHM}~\cite{London:2017bcn}\footnote{A new version of the IMRPhenom waveform model with higher-order modes became only very recently available (see Refs.~\cite{Pratten:2020fqn,Garcia-Quiros:2020qpx}), therefore we have not been able to include it in our study. We defer comparisons with this model to future analysis.} and \texttt{NRHybSur3dq8} to compute posterior distributions from the mock signals. The analysis of the first mock signal will demonstrate the improvements in measuring binary parameters when using a model with higher harmonics with respect to a model including only the dominant $(\ell, |m|) = (2,2)$ mode.
The analysis of the second mock signal will give us a sense of possible biases due to modeling errors in the original \texttt{SEOBNRv4HM} model with respect to NR-surrogate  waveforms, which are close to NR simulations. In creating the mock signals we do not add detector noise. This choice is made to avoid additional uncertainty and bias introduced by a random noise realization. It is the natural choice given that the goal of this parameter estimation analysis is to check for possible biases due to inaccuracies in waveform models.

\subsubsection{Setup}

We choose parameters for the mock signals in order to emphasize the effect of higher-modes in the waveform.
Since the contribution of higher-order modes in the emitted \acp{GW} increases with the mass ratio, we use for the mock signals $q = 8$, the largest mass ratio available for the model \texttt{NRHybSur3dq8}. For the total mass we use $M = 67.5 M_\odot$ such that the values of the component masses $m_1 = 60 M_\odot$ and $m_2 = 7.5 M_\odot$ are consistent with the masses of BBH systems observed during O1 and O2 (see~\cite{LIGOScientific:2018mvr} and~\cite{LIGOScientific:2018jsj}). 
The waveform models are restricted to non-precessing spins and we pick $\chi_{1z} = 0.5$ and $\chi_{2z} = 0.3$. To maximize the effect of the higher-modes we inject the signal at edge-on inclination $(\iota = \pi/2)$ with respect to the observer. The coalescence phase $\phi_c$ is chosen to be 1.2 rad, while the polarization phase $\psi$ is set to 0.7 rad. The signal has been injected at gps-time 1249852257 s with a sky-position defined by its right ascension of 0.33 rad and its declination of -0.6 rad. Finally the distance of the mock signal is chosen by demanding a network-SNR of $21.8$ in the three detectors (LIGO Hanford, LIGO Livingston and Virgo) when using the Advanced LIGO and Advanced Virgo PSD at design sensitivity~\cite{Barsotti:2018}. The resulting distance is 627 Mpc.
We used \texttt{PyCBC}'s \texttt{pycbc\_generate\_hwinj}~\cite{alex_nitz_2020_3630601} to prepare the mock signal. To carry out Bayesian parameter estimation we used the Markov chain Monte Carlo code \texttt{LALInferenceMCMC}~\cite{Veitch:2014wba,LALInference}.
We choose a uniform prior in component masses in the range $[3,100] M_\odot$. Aligned component spins are assumed to be uniform in $[-1,1]$. The priors on the other parameters are the standard ones described in Appendix C.1 of Ref.~\cite{LIGOScientific:2018mvr}.

\subsubsection{Results}

\begin{figure*}[hbt]
  \centering
  \includegraphics[width=0.45\textwidth]{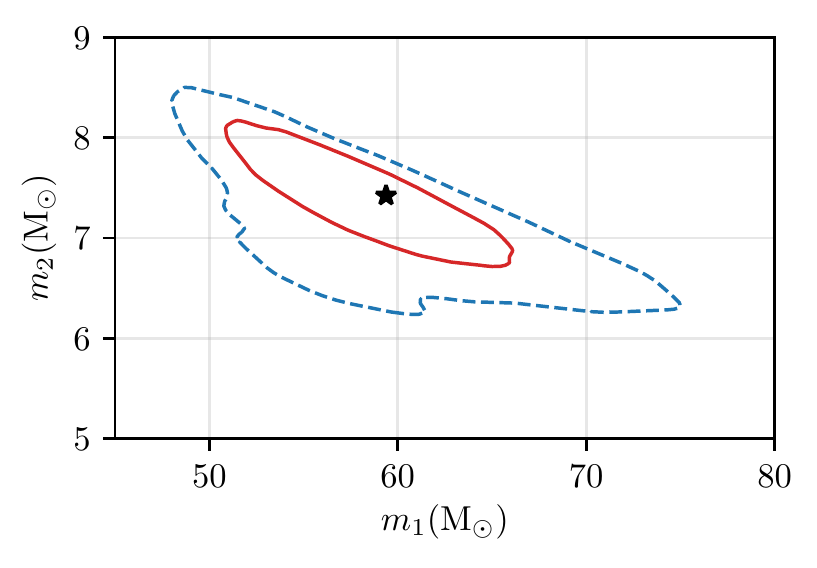}
  \includegraphics[width=0.45\textwidth]{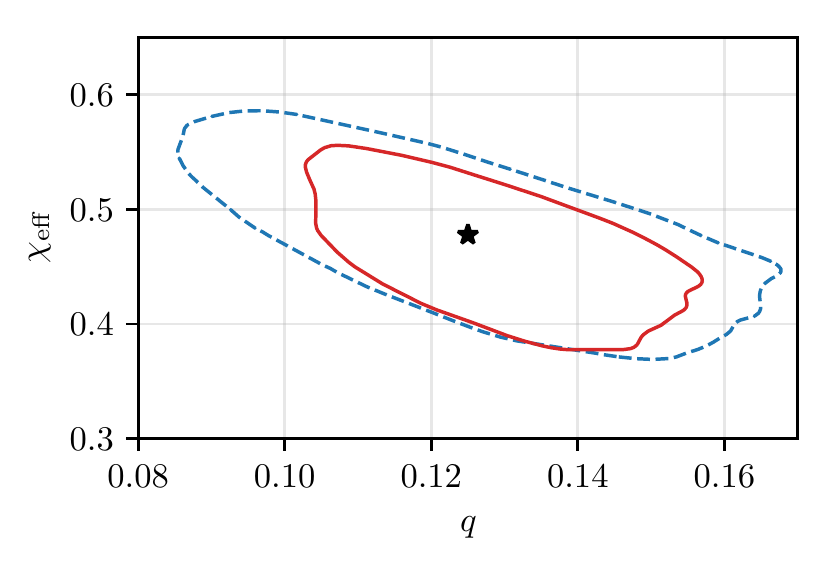}
  \includegraphics[width=0.45\textwidth]{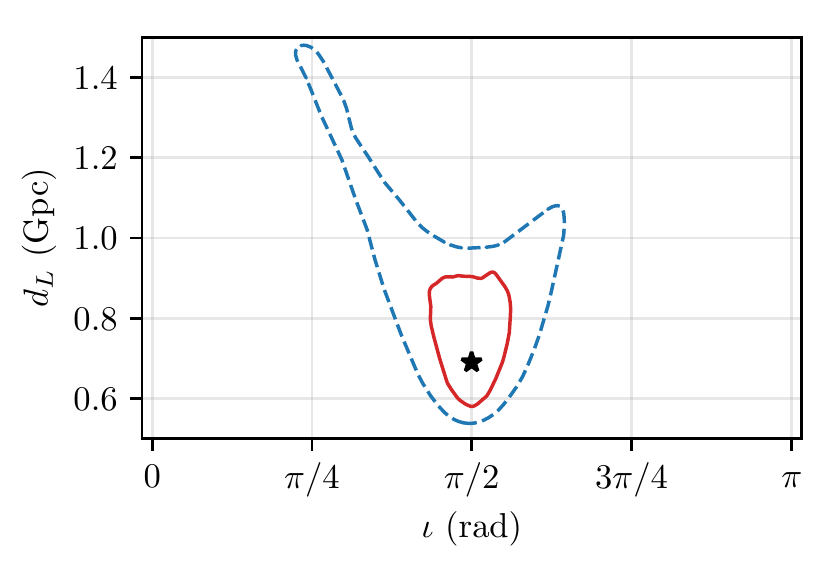}
  \includegraphics[width=0.45\textwidth]{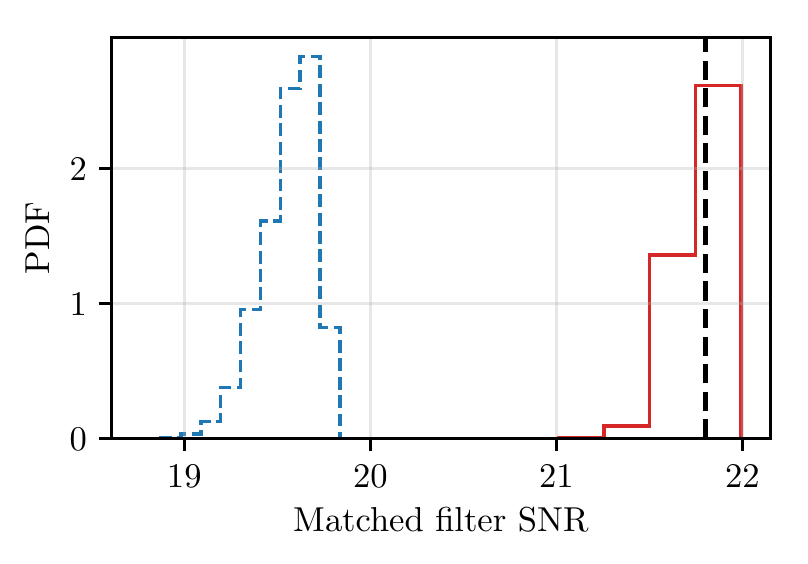}
  \includegraphics[trim={5.8cm 1cm 0 0}, clip, width=0.7\textwidth]{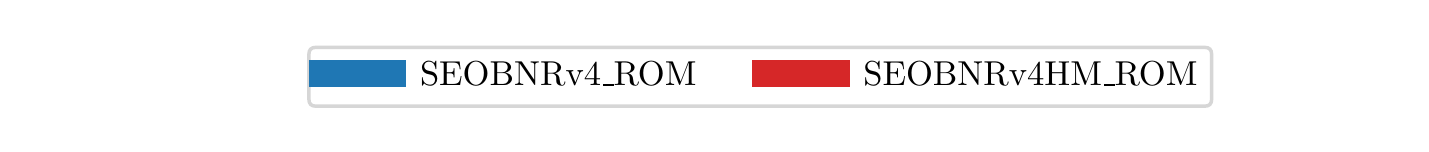}
	\caption{
  90\% credible regions and histograms of posterior distributions for a $q = 8$ \ac{BBH}. 
  The signal waveform is \texttt{SEOBNRv4HM\_ROM} and the stars represent binary parameters used for the signal. The mock signals are recovered with
  \texttt{SEOBNRv4\textunderscore ROM} and \texttt{SEOBNRv4HM\textunderscore ROM} waveform models.
 \emph{Top Left:} component masses in the source frame
  \emph{Top Right:} mass-ratio and effective aligned spin parameter.
  \emph{Bottom Left:} inclination angle and luminosity distance.
  \emph{Bottom Right:} matched filter \ac{SNR}
  }
  \label{fig:PE_all_EOBinj}
\end{figure*}

\begin{figure*}[hbt]
  \centering
  \includegraphics[width=0.45\textwidth]{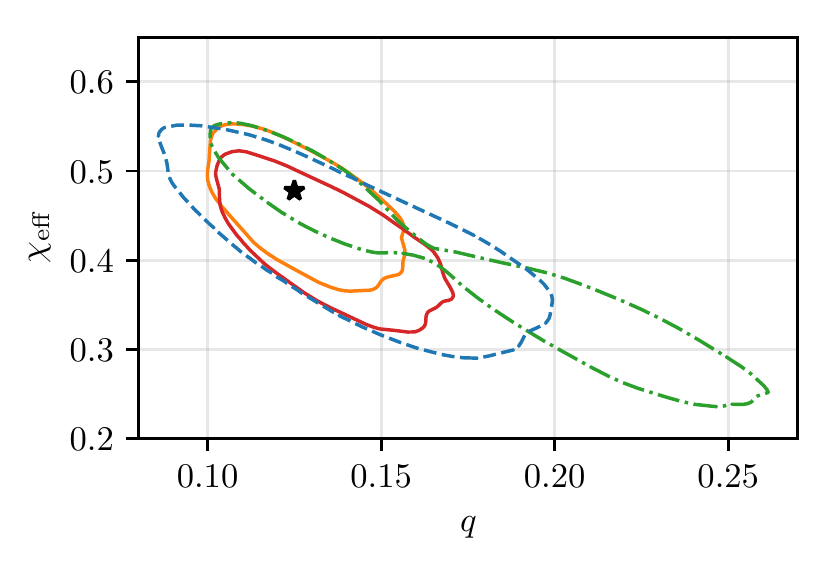}
  \includegraphics[width=0.45\textwidth]{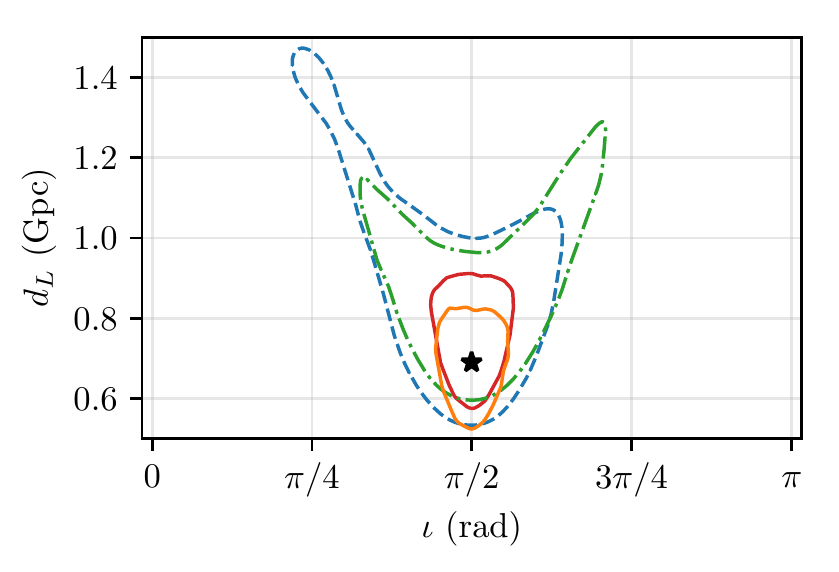}
  \includegraphics[trim={-1.0cm 1cm 2cm 0cm}, clip,width=0.9\textwidth]{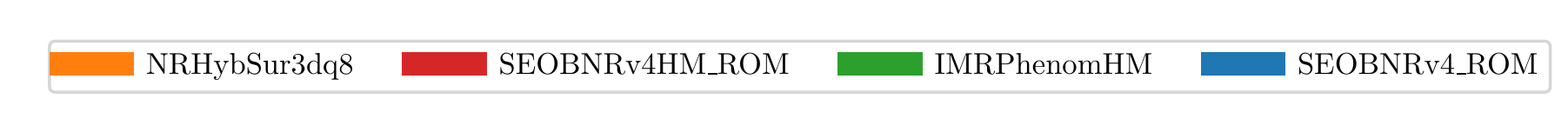}
	\caption{
  90\% credible regions and histograms of posterior distributions for a $q = 8$ \ac{BBH}. 
  The signal waveform is \texttt{NRHybSur3dq} and the stars represent binary parameters used for the signal. The mock signals are recovered with
  \texttt{SEOBNRv4\textunderscore ROM}, \texttt{SEOBNRv4HM\textunderscore ROM}, \texttt{IMRPhenomHM}, and  \texttt{NRHybSur3dq8} waveform models.
  \emph{Left:} mass-ratio and effective aligned spin parameter.
  \emph{Right:} inclination angle and luminosity distance.
  }
  \label{fig:PE_all_NRinj}
\end{figure*}

Let us first focus on the case in which the mock signal is generated with \texttt{SEOBNRv4HM}. In Fig.~\ref{fig:PE_all_EOBinj} we summarize the results of the parameter estimation analysis for some relevant binary parameters.
The top left panel shows the marginalized 2D posterior for the component source-frame masses, and the top right panel the marginalized 2D posterior for the mass ratio $q$ and the spin parameter $\chi_{\mathrm{eff}} = (m_1 \chi_1 + m_2 \chi_2) / (m_1 + m_2)$. In the bottom left panel we present the marginalized 2D posterior with inclination $\iota$ and luminosity distance $d_L$ and, finally, in the bottom right panel, we report the matched filter SNR. The star in the plots corresponds to the true value used for the mock signal, while the 2D contours of the posterior distributions represent $90\%$ credible regions. The waveform templates used to infer binary parameters are \texttt{SEOBNRv4\_ROM} (blue curve) and \texttt{SEOBNRv4HM\_ROM} (red curve). It is clear from the plots that all the  parameters reported in Fig.~\ref{fig:PE_all_EOBinj} are more precisely measured when using \texttt{SEOBNRv4HM\_ROM} instead of \texttt{SEOBNRv4\_ROM}. The posterior volume represents the degeneracy of the gravitational wave signal, and, in the absence of detector noise, this degeneracy is intrinsic to the waveforms. The inclusion of higher harmonics in \texttt{SEOBNRv4HM\_ROM} breaks the degeneracy between the parameters $q-\chi_{\mathrm{eff}}$ and $\iota-D_L$ and allows to measure them more precisely. These results are consistent with what was previously found in the literature~\cite{Graff:2015bba,Shaik:2019dym,Kalaghatgi:2019log}. As expected \texttt{SEOBNRv4HM\_ROM} also measures a larger matched filter SNR.

Let us now consider the case in which the mock signal is represented by \texttt{NRHybSur3dq8}. In Fig.~\ref{fig:PE_all_NRinj} we show the marginalized 2D posterior for the mass ratio $q$ and the spin parameter $\chi_{\mathrm{eff}}$ (left panel) and the marginalized 2D posterior with inclination $\iota$ and luminosity distance $d_L$ (right panel) as measured by the waveform models \texttt{SEOBNRv4HM\_ROM} (red curve), \texttt{SEOBNRv4\_ROM} (blue curve), \texttt{IMRPhenomHM} (green curve), and \texttt{NRHybSur3dq8} (orange curve). As before the star in the plots corresponds to the true value used for the mock signal, while the 2D contours of the posterior distributions represent $90\%$ credible regions. From the plots in Fig.~\ref{fig:PE_all_NRinj} it is clear that, as before, \texttt{SEOBNRv4HM\_ROM} recovers the binary parameters more precisely than \texttt{SEOBNRv4\_ROM}.
It is important to highlight that with \texttt{SEOBNRv4HM\_ROM} the binary parameters are recovered inside the $90\%$ credible regions. This means that for this quite asymmetric system at a moderately high SNR of $\sim 20$ the bias due to modeling errors in the original \texttt{SEOBNRv4HM} model compared to NR waveforms is negligible with respect to the statistical uncertainty.
In contrast, the marginal posterior distributions recovered for \texttt{IMRPhenomHM} are in general broader compared to the ones recovered by \texttt{SEOBNRv4HM\_ROM}, are notably bimodal in mass-ratio and effective spin, and extend a lot further along the line of $q$ - $\chi_\mathrm{eff}$ degeneracy. In distance and inclination the \texttt{IMRPhenomHM} posterior shows little improvement over \texttt{SEOBNRv4\_ROM} which does not include higher harmonics.
Finally, the marginal posteriors for \texttt{NRHybSur3dq8} are quite similar in size to those for \texttt{SEOBNRv4HM\_ROM}, but better centered around the true parameter values. This is as expected since the likelihood should peak at the true parameter values when the signal and template use the same waveform. The mock signal is also sufficiently loud for the posteriors to be likelihood- rather than prior-dominated, resulting in an unbiased parameter recovery.

This study shows that using \texttt{SEOBNRv4HM\_ROM} for parameter estimation yields unbiased measurements of the binary parameters at moderately high SNR even in a configuration where the effect of higher harmonics in the waveform is large. We defer a more comprehensive analysis to future studies.

\section{Conclusion}
\label{sec:conclusion}

In this paper we have presented a fast and accurate \ac{ROM} or surrogate model for the time domain \texttt{SEOBNRv4HM} \ac{EOB} waveform~\cite{Cotesta:2018fcv}. This model assumes spins aligned with the orbital angular momentum of the binary and includes the $(\ell, |m|) = (2, 1), (3, 3), (4, 4), (5, 5)$ spherical harmonic modes beyond the dominant $(\ell, |m|) = (2, 2)$ mode.

While the construction of this Fourier domain \ac{ROM} broadly follows previous work~\cite{Purrer:2014fza,Purrer:2015tud} we have introduced the following new features to accurately represent the higher harmonics and make the model more flexible (see Sec.~\ref{sec:techniques}).
While previous models used an amplitude / phase decomposition of the Fourier domain waveform, we here define a carrier signal (see Eq.~\eqref{eq:defcarrier}) based on the time domain orbital phase. Subsequently we extract the carrier phasing from each Fourier domain waveform mode (see Eq.~\eqref{eq:defhcoorb}). This essentially makes the phase of modes almost constant in the inspiral, defining what we here call ``coorbital modes''. This choice allows us to avoid zero-crossings in the subdominant harmonics which could spoil the smoothness of the training data and make accurate interpolation of the waveform data over parameter space very difficult.
We perform alignment in the time domain to keep track of the time of coalescence of the training set waveforms and this information is preserved in the \ac{ROM}.
We use here an alternative approach to dealing with the fact that the ringdown frequency varies over the parameter space, but waveform data needs to be given on a common frequency grid to build a \ac{ROM}. We rescale the geometric frequency parameter so that the ringdown is reached before a fixed termination frequency which demarcates the end of the frequency grid. We use the inverse rescaling during the evaluation of the \ac{ROM}.
We extend the \ac{ROM} to arbitrarily low frequencies by splicing it together with multipolar \ac{PN} waveforms. Therefore, it can in principle be used for arbitrarily light compact binary systems.
We decompose waveform input data in orthonormal bases using the \ac{SVD}, and build a model by constructing a tensor product spline over the 3-dimensional parameter space of mass-ratio and the two aligned spins of a binary. To increase model accuracy and efficiency we use domain decomposition in frequency and in parameter space.

In Sec.~\ref{sec:accuracy} we demonstrate that the \ac{ROM} has a very high faithfulness (or match) with \texttt{SEOBNRv4HM}. Maximizing over inclination, reference phase and effective polarization of the source waveform (see Eq.~\eqref{eq:max_unfaith}) the maximum mismatch over the remaining source parameters is below $0.03\%$ for binaries with a total mass below $100 M_\odot$ and below $0.2\%$ for binaries at $300 M_\odot$ (see Table~\ref{tab:maximum_unfaith_M_summary_table}). Even for this very conservative choice the mismatch is at least an order of magnitude lower than the unfaithfulness of \texttt{SEOBNRv4HM} against \ac{NR} simulations. Therefore the additional modeling error introduced in building the \ac{ROM} is strongly subdominant and the \ac{ROM} very accurately represents the \texttt{SEOBNRv4HM} waveform model.
In Sec.~\ref{sec:computational_performance} we show that our \ac{ROM} accelerates waveform evaluation by a factor 100 -- 200 compared to \texttt{SEOBNRv4HM} and favorably compares against other higher mode waveform models for \ac{BBH} systems, being about an order of magnitude faster.
We showcase in Sec.~\ref{sec:PEsec} (see Figs.~\ref{fig:PE_all_EOBinj}, and~\ref{fig:PE_all_NRinj}) that our \ac{ROM} can recover component masses and spins, and especially distance and inclination angle for a quite asymmetric and spinning \ac{BBH} with increased precision compared to the \texttt{SEOBNRv4\_ROM} waveform which only models the dominant mode. In addition we show that the \ac{ROM} accurately recovers binary parameters, irrespective of whether the source is represented by a \texttt{SEOBNRv4HM} or a \texttt{NRHybSur3dq8} waveform.
Our \ac{ROM} gives a significantly more accurate parameter recovery compared to the phenomenological \texttt{IMRPhenomHM} waveform and is close to the \texttt{NRHybSur3dq} NR-surrogate model, while being more versatile and covering a significantly larger parameter space.

This \ac{ROM} should prove a very useful tool for \ac{GW} data analysis to describe systems
where the contribution of higher harmonics is important in terms of additional signal-to-noise-ratio and discriminating power for detection and parameter inference.
We stress that the ROM is very fast and reproduces the \texttt{SEOBNRv4HM} model with a great accuracy over the widest range in parameter space of all inspiral-merger-ringdown higher mode models available to date, from mass-ratio 1 to 1:50 where aligned spins can take values in the full range allowed for Kerr \acp{BH}, up to extremal spins.

\section*{Acknowledgments}
The authors would like to thank Stas Babak, Alessandra Buonanno, Nils Fischer, Bhooshan Gadre, Jonathan Gair, Cecilio Garcia-Quiros, Steffen Grunewald, Ian Harry, Serguei Ossokine, Harald Pfeiffer and Vijay Varma for useful discussions.
The authors acknowledge usage of AEI's Hypatia computer cluster.

This is LIGO Document Number LIGO-P2000106.

\chapter{On the properties of the massive binary black hole merger GW170729}
\label{chap:five}
\chaptermark{}

\textbf{Authors}:\footnote{Originally published in Phys.Rev.D 100 (2019) 10, 104015.} Katerina Chatziioannou, Roberto Cotesta, Sudarshan Ghonge, Jacob Lange, Ken K.-Y. Ng, Juan Calderon Bustillo, James Clark, Carl-Johan Haster, Sebastian Khan, Michael Purrer, Vivien Raymond, Salvatore Vitale, Nousha Afshari, Stanislav Babak, Kevin Barkett, Jonathan Blackman, Alejandro Bohe, Michael Boyle, Alessandra Buonanno, Manuela Campanelli, Gregorio Carullo, Tony Chu, Eric Flynn, Heather Fong, Alyssa Garcia, Matthew Giesler, Maria Haney, Mark Hannam, Ian Harry, James Healy, Daniel Hemberger, Ian Hinder, Karan Jani, Bhavesh Khamersa, Lawrence E. Kidder, Prayush Kumar, Pablo Laguna, Carlos O. Lousto, Geoffrey Lovelace, Tyson B. Littenberg, Lionel London, Margaret Millhouse, Laura K. Nuttall, Frank Ohme, Richard O’Shaughnessy, Serguei Ossokine, Francesco Pannarale, Patricia Schmidt, Harald P. Pfeiffer, Mark A. Scheel, Lijing Shao, Deirdre Shoemaker, Bela Szilagyi, Andrea Taracchini, Saul A. Teukolsky and Yosef Zlochower.

\textbf{Abstract}:We present a detailed investigation into the properties of GW170729, the gravitational wave with the most massive and distant source confirmed to date. 
We employ an extensive set of waveform models, including new improved models that incorporate the effect of higher-order waveform modes which are particularly important for massive systems.
We find no indication of spin-precession, but the inclusion of higher-order modes in the models results in an improved estimate for the mass ratio of $(0.3-0.8)$ at the 90\% credible level. Our updated measurement excludes equal masses at that level.
We also find that models with higher-order modes lead to the data being more consistent with a smaller effective spin, with the probability that the effective spin is greater than zero being reduced from $99\%$ to $94\%$.
The 90\% credible interval for the effective spin parameter is now $(-0.01-0.50)$.
Additionally, the recovered signal-to-noise ratio increases by $\sim0.3$ units compared to analyses without higher-order modes;
the overall Bayes Factor in favor of the presence of higher-order modes in the data is 5.1:1.
We study the effect of common spin priors on the derived spin and mass measurements, and observe small shifts in the spins, while the masses remain unaffected.  
We argue that our conclusions are robust against systematic errors in the waveform models. 
We also compare the above waveform-based analysis which employs compact-binary waveform models to a more flexible wavelet- and chirplet-based analysis. 
We find consistency between the two, with overlaps of $\sim 0.9$, typical of what is expected from simulations of signals similar to GW170729, confirming that the data are well-described by the existing waveform models. 
Finally, we study the possibility that the primary component of GW170729 was the remnant of a past merger of two black holes and find this scenario
to be indistinguishable from the standard formation scenario.

\section{Introduction}

GW170729 was observed on July 29, 2017 by the Advanced LIGO~\cite{TheLIGOScientific:2014jea} twin detectors. 
Its detection was announced in~\cite{LIGOScientific:2018mvr} as part of GWTC-1, the gravitational-wave (GW) transient catalog of compact binary coalescences (CBCs)~\cite{GWOSC:GWTC}. 
As reported in~\cite{LIGOScientific:2018mvr}, GW170729 was emitted during the coalescence of two stellar-mass black holes (BH). 
It was observed during the offline analysis of the detection pipelines GstLAL~\cite{Messick:2016aqy} and PyCBC~\cite{Usman:2015kfa,Nitz_2017}, that search for signals from CBC events, as well as cWB~\cite{Klimenko:2015ypf}, a pipeline tuned to search for transient signals whose frequency increases with time. 

A number of reasons make GW170729 unique among the binary BHs (BBHs) presented in GWTC-1.
At a measured source-frame total mass of $\sim 85 M_{\odot}$ and a distance of $\sim 3$Gpc (median values), it is likely the most massive and distant BBH. 
Additionally, it is one of only two GW events that show evidence for nonzero spins with an effective spin of $(0.11-0.58)$ at the 90\% credible level~\cite{LIGOScientific:2018mvr}. 
Finally, it is the only event for which the more flexible, non-CBC-specific cWB search returns a lower false alarm rate than the CBC-specific GstLAL and PyCBC template-based searches. 
In spite of these, Ref.~\cite{LIGOScientific:2018jsj} concludes that GW170729 is consistent with the population of the other BBH detections
\footnote{An additional candidate claimed in~\cite{Zackay:2019tzo}, if confirmed, would also correspond to a binary with a non zero effective spin.}.

The fact that GW170729 is the most massive BBH found so far makes it a good candidate to observe the effects of higher-order waveform modes. 
The GW emission from BBHs can be described as a superposition of GW modes $h_\lm$ as $h\equiv h_+ - i h_{\times}=\sum_\lm    Y_\lm^{-2}(\iota,\varphi)h_\lm(t,\vec{\theta})$ \cite{Thorne:1980ru}. 
Here  $h_+$ and $h_\times$ are the two GW polarizations, $Y_\lm^{-2}$ denote spin-2 weighted spherical harmonics \cite{sharm}, which depend on the location $(\iota,\varphi)$ of the observer around the binary, while the modes $h_\lm$ depend on the masses and spins of the binary, denoted by $\vec{\theta}$. 
During most of the inspiral stage, $h$ is dominated by the quadrupole modes, $(\lm)=(2,\pm 2)$. 
The rest, known as higher-order modes, grow in strength during the merger and ringdown stages, their impact being larger for highly asymmetric and nearly edge-on binaries \cite{Pan:2011gk,Varma:2014jxa,Bustillo:2016gid,Pekowsky:2012sr,Varma:2016dnf,Bustillo:2015qty,Graff:2015bba}. 
Finally, for more massive BBH systems the inspiral emission moves out of the sensitive band of advanced detectors, while sensitivity to the merger-ringdown increases and so does the impact of higher-order modes \cite{VanDenBroeck:2006ar,Bustillo:2015qty,Graff:2015bba}.

Standard detection and parameter estimation of BBH events is usually performed using GW templates without the higher-order mode content of the signals~\cite{2018arXiv181205121M,DalCanton:2017ala,TheLIGOScientific:2016pea,LIGOScientific:2018mvr}. 
Reference~\cite{LIGOScientific:2018mvr} studied the fact that the event is recovered with higher significance by the flexible unmodeled, but less sensitive, search than the template-based searches~\cite{CalderonBustillo:2017skv}. 
By performing injections of signals without higher-order modes or spin-precession, it was argued that the difference in the measured significances is in fact not unlikely. 
It was shown that $\sim 4\%$ of the injected signals were recovered with a higher significance from cWB than PyCBC. 

At the same time, the presence of strong higher-order modes in the GW signal can potentially lead to biased parameter estimation if they are omitted in the waveform templates \cite{Littenberg:2012uj,Varma:2016dnf,Varma:2016dnf,Bustillo:2015qty,Graff:2015bba}. 
So far all reported events are consistent with nearly equal-mass, face-on BBHs, a fact that has prevented such biases, as shown in \cite{Abbott:2016wiq,Abbott:2016apu,Kumar:2018hml} for the case of events observed during the first observation run of Advanced LIGO. 
Even in this case, usage of models with higher-order modes can improve the accuracy of parameter estimation \cite{Littenberg:2012uj,Graff:2015bba,London:2017bcn,CalderonBustillo:2018zuq}. 
Consequently, a reanalysis of GW150914 and GW170104 events using models with higher-order modes obtains modestly tighter parameter constraints with respect to previous analyses \cite{Kumar:2018hml}.

In this paper we present a detailed investigation into the properties of GW170729. 
We carry out a parameter estimation analysis similar to the one in~\cite{LIGOScientific:2018mvr} in order to study the effect of spin-precession, higher-order modes, and spin priors on inferences drawn about GW170729. 
We make use of a more extended set of CBC waveform models belonging to three distinct waveform model families: phenomenological~\cite{Ajith:2007qp,Ajith:2009bn,Santamaria:2010yb,Husa:2015iqa,Khan:2015jqa,London:2017bcn,Hannam:2013oca}, effective-one-body~\cite{Taracchini:2013rva,Bohe:2016gbl,Pan:2013rra,Cotesta:2018fcv,Pan:2011gk,Babak:2016tgq}, and numerical relativity~\cite{2013PhRvL.111x1104M,2016CQGra..33t4001J,2017CQGra..34v4001H,2015PhRvL.115l1102B,2017PhRvD..95j4023B,2017PhRvD..96b4058B}. 
This set includes two new improved spin-aligned waveform models that include the effect of higher-order waveform modes.
We gauge the importance of a physical phenomenon, namely spin-precession and higher-order modes, by comparing posterior densities for various source parameters obtained through analyses using waveform models with and without that physical phenomenon included.

While we find no indication of spin-precession, higher-order modes have a distinct impact on the posterior density for various source parameters. 
We find that CBC waveform models that include higher-order modes result in posterior distributions for the mass ratio of the binary that are shifted away from unity, resulting in more support for unequal masses than originally concluded in ~\cite{LIGOScientific:2018mvr}. 
In particular, we find a highest probability density (HPD) interval of the mass ratio of $(0.3-0.8)$ at the 90\% credible level, while the corresponding upper limit of the HPD interval on the mass ratio without higher-order modes is $\sim 0.96$.
This improved measurement, obtained using waveform models that include more physical effects, shows that GW170729 is not consistent with the merger of two equal-mass BHs at the $90\%$ level.
At the same time, models with higher-order modes lead to marginally less support for positive effective spin $\chi_\mathrm{{eff}}$~\footnote{The effective spin parameter is defined as the sum of the mass-weighted projections of the component spins along the orbital angular momentum and it is conserved to at least the second post-Newtonian order~\cite{Racine:2008qv}.} and binary orientations where the orbital angular momentum points along or away from the line of sight. In particular, we find that the probability that the effective spin is positive is reduced from $99\%$ when higher-order modes are omitted to $94\%$ when they are included in the waveform models.
We obtain consistent results when we use various CBC waveform models from different waveform families to describe the data. 
We thus argue that our conclusions are robust against systematic errors in the waveform models.

As the source-frame mass of the more massive BH is close to the proposed mass upper limit due to pair instability\footnote{It has been suggested that pulsational pair instability supernovae will result in no BH remnants with masses above $\sim 50 M_{\odot}$ as the remnant is disrupted during the explosion, e.g.~\cite{Woosley_2017,Marchant:2018kun}.}~\cite{2002ApJ...567..532H} and the posterior for the binary mass ratio favors unequal masses, we further investigate the possibility of second generation (2g) merger~\cite{davide2g,Fishbach:2017dwv}.
In a 2g merger scenario, the primary BH is the remnant of an earlier BBH merger. 
As such, it is expected to be more massive than its companion in GW170729 resulting in unequal binary masses, and to have a relatively large spin magnitude.
We contrast this scenario to a first generation (1g) merger scenario which favors comparable component masses.
We reanalyze the data using two different priors tailored to 1g and 2g mergers and calculate the Bayes Factor (BF) of the 2g versus the 1g hypotheses.
We find a BF of 4.7:1(1.4:1) in favor of the 2g scenario when using waveforms with (without) higher-order modes. This value favors the 2g model, but not decisively so, in agreement with the results of~\cite{LIGOScientific:2018jsj}. This question has been earlier and independently addressed by Kimball et al.~\cite{Kimball:2019mfs} and our results are in agreement with their results.

Finally, we compare the signal reconstruction obtained with CBC waveform models to a morphology-independent reconstruction~\cite{Cornish:2014kda}.
We quantify the consistency between the CBC and the generic reconstruction by computing the noise-weighted overlap between the two. 
We find broad consistency between the two with overlap values typical of what is expected for this mass range and signal strength~\cite{TheLIGOScientific:2016uux}. 
This result is only minimally affected by the inclusion of higher-order modes. 

Posterior samples from all our analyses are available in~\cite{170729SamplesRelease}.
The rest of the paper presents our analysis and conclusions in detail. 
Section~\ref{analysis} describes the analysis we carry out including the CBC waveform models, the priors, and the generic analysis. 
Section~\ref{CBCresults} presents results derived under the CBC waveform models and posterior densities for the source parameters. 
Section~\ref{BWresults} gives the results of the generic analysis and how they compare to the CBC-specific analysis. 
Finally, Sec.~\ref{conclusions} presents our main conclusions.

\section{Analysis}
\label{analysis}

In this section we describe the details of the analysis we perform including the data, waveform models, and inference approaches we use. 
Our analysis follows closely and builds off of the work originally presented in~\cite{LIGOScientific:2018mvr}. 
We employ two complementary approaches: one is based on waveform models constructed specifically to describe compact binary coalescences, while the other uses a more flexible waveform model that can capture unexpected signal morphology. 
We describe both analyses in the following.

\subsection{Data and setup} 

We use the publicly available LIGO and VIRGO strain data for GW170729 from the Gravitational Wave Open Science Center~\cite{GWOSC,Vallisneri:2014vxa}. 
The LIGO strain data have been post-processed to subtract several sources of instrumental noise~\cite{Driggers:2018gii,Davis:2018yrz} and calibrated as described in~\cite{LIGOScientific:2018mvr}. 
In particular we analyze $4$s of strain data centered at the GW170729 trigger time.
The analysis covers a bandwidth from $f_\mathrm{ low}=20$~Hz with the upper frequency cutoff set to $f_\mathrm{ high}=1024$~Hz for waveform models without higher-order modes and $f_\mathrm{ high}=2048$~Hz when higher-order modes are included. 
For masses typical of GW170729 (a detector-frame total mass of 120$M_{\odot}$ and mass ratio of 0.5) this upper frequency cutoff ensures that the analysis includes up to at least the $\ell=5$ ringdown harmonic of a compact binary merger, the highest frequency mode available in the waveform models we use. 

We assume that the noise in the three detectors is Gaussian and stationary. 
The power spectral density (PSD) of the noise is obtained from the same $4$s of on-source data with the technique described in~\cite{Littenberg:2014oda}. 
Specifically, a model consisting of a cubic spline and a number of Lorenzians is used to obtain posterior samples for the PSD from which a median PSD value is computed separately for each frequency bin. 
This median PSD is used in the estimation of the likelihood function in \texttt{ LALInference}, \texttt{ RIFT}, and \texttt{ BayesWave} and we use the same PSD as~\cite{LIGOScientific:2018mvr} that is publicly available in~\cite{GW170729PSD}.

\subsection{ CBC waveform models} 
\begin{landscape}
\begin{table*}[]
\centering
  \begin{tabular}{c|ccccccccc}
    \hline \hline
Waveform Model         &&& Spin Dynamics &&& Modes ($\ell,|m|$) &&&  Algorithm   \\ \hline
\texttt{ IMRPhenomPv2}~\cite{Hannam:2013oca}   &&& Precessing &&& (2,2) &&& \texttt{ LALInference}\\
\texttt{ IMRPhenomD}~\cite{Husa:2015iqa,Khan:2015jqa}       &&& Aligned &&& (2,2) &&&  \texttt{ LALInference}\\
\texttt{ IMRPhenomHM}~\cite{London:2017bcn}    &&& Aligned &&& (2,2),(2,1),(3,3),(3,2),(4,4),(4,3) &&& \texttt{ LALInference}\\
\texttt{ SEOBNRv3}~\cite{2014PhRvD..89f1502T,2014PhRvD..89h4006P}           &&& Precessing &&& (2,2) (2,1)  &&&   \texttt{ LALInference}\\
\texttt{ SEOBNRv4}~\cite{Bohe:2016gbl}           &&& Aligned &&&(2,2) &&& \texttt{ LALInference}\\
\texttt{ SEOBNRv4HM}~\cite{Cotesta:2018fcv}     &&& Aligned &&& (2,2), (2,1), (3,3), (4,4), (5,5)  &&& \texttt{ LALInference}\\
NR    HM ~\cite{2013PhRvL.111x1104M,2016CQGra..33t4001J,2017CQGra..34v4001H}            &&& Aligned &&&   ($\ell \leq 4$,$|m|\leq\ell$) &&& \texttt{ RIFT}\\
NR/\NRSur{} HM ~\cite{2015PhRvL.115l1102B,2017PhRvD..95j4023B,2017PhRvD..96b4058B} &&& Aligned &&&  ($\ell \leq 4$,$|m|\leq\ell$) &&& \texttt{ RIFT}\\
NR  ~\cite{2013PhRvL.111x1104M,2016CQGra..33t4001J,2017CQGra..34v4001H}           &&& Aligned &&&  ($\ell=2$,$|m|\leq\ell$) &&&\texttt{ RIFT}\\
NR/\NRSur{} ~\cite{2015PhRvL.115l1102B,2017PhRvD..95j4023B,2017PhRvD..96b4058B} &&& Aligned &&&  ($\ell=2$,$|m|\leq\ell$)  &&& \texttt{ RIFT}\\
\hline
Wavelets~\cite{Cornish:2014kda}               &&& Flexible &&&  Flexible &&&\texttt{ BayesWave}\\
Chirplets~\cite{Millhouse:2018dgi}           &&& Flexible &&&  Flexible  &&& \texttt{ BayesWave}\\
\end{tabular}
\caption{List of waveform models we use to model the GW signal. 
The second and third columns indicate the spin dynamics and higher-order content $(\ell, |m|)$  of each model in the coprecessing frame respectively. 
The fourth column gives the algorithm we use with each model. 
The bottom horizontal line separates the CBC-specific models and the morphology-independent models.
}
\label{table:pe-table}
\end{table*}
\end{landscape}

The top half of Table~\ref{table:pe-table} lists the CBC waveform models we use; these models describe the inspiral, merger, and ringdown signal from the coalescence of two BHs as predicted by General Relativity (GR). 
All CBC waveforms we use can be divided into three main families: (i) phenomenological models (\texttt{ IMRPhenom}), effective-one-body models (\texttt{ SEOBNR}), and numerical relativity (NR). 
The first family is based on results of post-Newtonian theory~\cite{Blanchet:2014zz} to compute the inspiral phase and a phenomenological approach to describe the merger, aided by calibration to EOB-NR hybrid waveforms~\cite{Hannam:2013oca,Husa:2015iqa,Khan:2015jqa}.
The second family uses the effective-one-body approach~\cite{Buonanno:1998gg,Buonanno:2000ef}, which is based on a resummation of post-Newtonian results to describe the inspiral, and uses calibration to NR simulations for the late-inspiral and merger~\cite{Zilhao:2013hia,Loffler:2011ay,Lovelace:2010ne,2013PhRvL.111x1104M}. 
Both families describe the ringdown employing results of BH perturbation theory~\cite{Barausse:2011kb,Taracchini:2014zpa}. 
The NR waveforms are obtained by solving the full non-linear Einstein equations and are subject only to numerical errors~\cite{ 2010RvMP...82.3069C}.

Besides the waveform family, models also differ on whether they include the effect of spin-precession~\cite{Apostolatos:1994mx,Kidder:1995zr} and higher-order modes, as indicated in Table~\ref{table:pe-table}. 
From the \texttt{ IMRPhenom} and \texttt{ SEOBNR} families we use one aligned-spin and one spin-precessing model without higher-order modes as well as a spin-aligned model with higher-order modes. None of the models from these two waveform families currently include both precessing spins and higher-order modes.
In particular, from the \texttt{ IMRPhenom} family, we use: the spin-precessing \IMRP{}~\cite{Hannam:2013oca} and the spin-aligned \IMRPD{}~\cite{Husa:2015iqa,Khan:2015jqa} models. 
From the \texttt{ SEOBNR} family we use: the spin-precessing \SEOBP{}\footnote{\SEOBP{} includes the modes $(\ell, |m|) = (2,2), (2,1)$ in the coprecessing frame, the coordinate system for which the $z$-axis is instantaneously aligned with the Newtonian angular momentum, see~\cite{Buonanno:2002fy}.}~\cite{2014PhRvD..89f1502T,2014PhRvD..89h4006P} and the spin-aligned \SEOBA{}\footnote{In particular, we use the reduced order model implementation of  \SEOBA{} which is computationally less expensive~\cite{Purrer:2014fza,Purrer:2015tud}.}~\cite{Bohe:2016gbl} models. 
As far as higher-order modes are concerned, we use \IMRPHM{}~\cite{London:2017bcn}, a spin-aligned model of the \texttt{ IMRPhenom} family that includes the $(\ell, |m|) = [(2,2),(2,1),(3,3),(3,2),(4,4),(4,3)]$ higher-order modes and \SEOBHM{}~\cite{Cotesta:2018fcv}, a spin-aligned model of the \texttt{ SEOBNR} family that includes the $(\ell, |m|) = [(2,1),(3,3),(4,4),(5,5)]$ higher-order modes. 
Posterior samples obtained with \IMRP{} and \SEOBP{} have already been made publicly available by the LIGO-Virgo Collaborations and we use them directly~\cite{GW170729Samples}.

The NR simulations we use include a total of 763 spin-aligned and 625 spin-precessing simulations~\cite{2013PhRvL.111x1104M,2016CQGra..33t4001J,2017CQGra..34v4001H}.
We optionally augment the list of NR simulations with waveforms computed using \NRSur{}~\cite{2015PhRvL.115l1102B,2017PhRvD..95j4023B,2017PhRvD..96b4058B}, a surrogate model directly based on NR.
The surrogate model we use is valid for mass ratio $0.5\le q\le1$ and dimensionless spin magnitude $\chi\le0.8$; however, the analysis performed results in a full posterior due to the inclusion of the NR waveforms that cover the remaining region. For both NR-related analyses, we include results with higher-order modes $(\ell\leq4, |m|\leq\ell)$ and with only the $(\ell=2, |m|\leq\ell)$ modes. 
When using the NR simulations we also assume that the spins remain aligned to the orbital angular momentum (no spin-precession). 

\subsection{Priors} 

Our analysis employs the following priors for the source parameters. 
The detector-frame component masses $m_1, m_2$ are assumed to be uniform between $10M_{\odot}$ and $200M_{\odot}$ with $m_1>m_2$,
while the mass ratio $q\equiv m_2/m_1$ is restricted to be above 0.125.
The sky location and orientation of the binary, as well as the directions of the component spins are uniform on the unit sphere. 
The distance is uniform in volume with a maximum cut off of $7$Gpc, while the time and phase of arrival are uniform. 
We have verified that the mass and distance prior ranges encompass the entire region where the posterior distribution has non-negligible support.

For the magnitude of the dimensionless component spins $\chi_i, \,i\in \{1,2\},$ we make different choices in order to investigate how this affects the posterior. 
The first prior is uniform-in-$\chi$ up to 0.99 for both spin-aligned and spin-precessing waveform models. 
The second prior is uniform-in-$\chi_z$, where $\chi_z$ is the spin projection along the axis perpendicular to the orbital plane, with the restriction that the spin magnitude is below 1. 
For spin-precessing model, the in-plane $\chi_x$ and $\chi_y$ components are also uniform; in that case this prior is sometimes referred to as `volumetric' prior as it corresponds to the spin vector being uniformly distributed within the unit sphere.

Finally, we also use priors targeted toward CBCs where the primary component is the product of a past merger. We study two cases.
The 1g case uses a spin prior that is uniform-in-$\chi$ for both component spins and a mass ratio prior that favors equal masses
\begin{align}
p_{1g}(q) \propto \left\{ 1 + \exp{\left[ -k\left( q-q_0 \right) \right]} \right\}^{-1},
\end{align}
where $k=20$ and $q_0=0.8$. This prior choice was motivated in~\cite{salvoprior}.
In the case of a 2g merger, the primary is expected to be more massive than the secondary binary component. We use a Gaussian mass ratio prior with a mean of $0.5$ and a standard deviation of $0.2$, motivated by Fig. 2 of~\cite{davide2g}.
In the 2g case, the primary is also expected to be spinning more rapidly.
We therefore use a prior where $\chi_1$ is distributed according to a Gaussian centered at $0.7$ with a width of $0.1$~\cite{davide2g,Fishbach:2017dwv}.
The priors of secondary spin magnitude and both spin directions are uniform.

\subsection{\texttt{ LALInference} and \texttt{ RIFT}} 

Given a waveform model and a set of prior choices, we compute the joint multidimensional posterior distribution of the source parameters. 
For fast-to-evaluate waveform models we use the publicly available software library \texttt{ LALInference}~\cite{Veitch:2014wba,lalinference_o2} to directly sample the posterior distribution. 
This approach computes the likelihood exactly at various points of the parameter space, but in order to obtain enough independent samples, millions of likelihood evaluation are required. 
This is prohibitive for models that are slow to evaluate, such as NR. 

In these cases we use \texttt{ RIFT}~\cite{2015PhRvD..92b3002P,2018arXiv180510457L}. 
\texttt{ RIFT}'s three-stage algorithm first evaluates the likelihood on a dense grid; then approximates it via interpolation; and then uses Monte Carlo integration to produce the full posterior distribution. 
The number of grid points used for this particular NR-only analysis is 63,000, and the number of added surrogate points for the NR/\NRSur{} grid was
 40,000; this brought the total number of points for the NR/\NRSur{} to 103,000. For context, \texttt{ RIFT} in general calculates the marginalized likelihood on thousands grid points in parallel. For each marginalized likelihood, which has fixed intrinsic parameters, we evaluate the likelihood at $\approx10^6$ different extrinsic parameters. 
  Even though the number of evaluations are orders of magnitude larger than for \texttt{ LALInference}, the overall wallclock time is considerably lower because the likelihood evaluations are faster and done in parallel, see \cite{2019arXiv190204934W} for details. However, due to grid limitations (discreteness and limited range), \texttt{ RIFT} does not sample both extrinsic and intrinsic parameters jointly from the full posterior distribution. Instead it marginalizes over all extrinsic parameters to calculate the likelihood and posterior for just the intrinsic parameters.  

All results obtained using \texttt{ LALInference} marginalize over the same detector calibration amplitude and phase uncertainty as in~\cite{LIGOScientific:2018mvr} and publicly available in~\cite{GW170729CalEnv} using the method described in~\cite{TheLIGOScientific:2016wfe,SplineCalMarg-T1400682}. 
All \texttt{ RIFT} results assume perfect calibration; this choice was shown to not affect the intrinsic binary parameters~\cite{Vitale:2011wu}.

\subsection{\texttt{ BayesWave}} 

Finally, we also use a minimal-assumptions analysis that does not make use of CBC-specific waveform models. We use \texttt{ BayesWave}~\cite{Cornish:2014kda}, a publicly-available algorithm~\cite{bayeswave} that does not explicitly assume that the signal is a CBC\footnote{While \texttt{ BayesWave} does not assume an explicit signal morphology, it does assume that the signal is elliptically polarized, that it propagates at the speed of light, and that there is no phase decoherence during the propagation.}, and instead models it through a linear combination of basis functions, either sine Gaussian (known as Morlet Gabor) wavelets, or ``chirplets''~\cite{Millhouse:2018dgi}, as listed in the bottom half of Table~\ref{table:pe-table}. 
The latter are sine Gaussians modified with a linearly evolving frequency. \texttt{ BayesWave} relies on a transdimensional sampler~\cite{10.1093/biomet/82.4.711} to explore the multidimensional posterior of the parameters of the wavelets/chirplets (frequency, time, phase, amplitude, quality factor, and possibly the frequency derivative) as well as the number of wavelets/chirplets in the linear combination. 

We then compare the signal reconstruction obtained with the morphology-independent models of \texttt{ BayesWave} and with CBC waveform models. 
Broad agreement between the wavelet reconstruction and \texttt{ IMRPhenomPv2} was established in~\cite{LIGOScientific:2018mvr} and we here perform the same test for waveform models that include higher-order modes. 
We also quantify the level of consistency through the detector network overlap \cite{Apostolatos:1995pj}, defined as 
\begin{equation}
{\cal{O}}_{N} \equiv \frac{(h_1,h_2)_{N}}{\sqrt{(h_1,h_1)_{N}(h_2,h_2)_{N}}}
\end{equation}
where $(h_1,h_2)_{N}$ denotes the inner product over the network defined by

\begin{equation}
(h_1,h_2)_{N} = \sum_{i}^{n} (h_1^i, h_2^i)
\end{equation}
where $i$ sums over all the detectors in the network, and $(h_1^i,h_2^i)$ is the inner product in an individual detector defined by

\begin{equation}
(h_1^i,h_2^i)\equiv 4 \Re \int_{0}^{\infty} \frac{\tilde{h}_1^i(f)\tilde{h}^{i*}_2(f)}{S_n^i(f)}df.
\end{equation}
In the above, $\tilde{h}_1^i(f)$ denotes a signal reconstruction sample computed with CBC models and $\tilde{h}_2^i(f)$ is a reconstruction sample computed with \texttt{ BayesWave}. Finally, $S_n^i(f)$ denotes the PSD of the detector. The superscipt $i$ denotes the quantities as they appear in the the $i^\mathrm{{th}}$ detector.

\section{CBC-model-based analysis}
\label{CBCresults}

In this section we present results on inference with CBC waveform models. 
We study how the posteriors for the various source parameters are affected by the inclusion of spin-precession and higher-order modes in the waveforms. 
We also study the effect of spin priors and waveform systematics on the validity of our conclusions. 

\subsection{Higher-order modes}
\label{2HM}

The importance of higher-order modes on a GW signal observed in the detectors depends on both extrinsic parameters, such as the inclination of the binary, and intrinsic parameters, such as the mass ratio. 
In general, signals from edge-on and asymmetric binaries include more power in higher-order modes. 
To study the effect of higher-order modes on GW170729 we analyze the data with waveform models both with and without higher-order modes. 
Table~\ref{table:par-table} gives the median and 90\% symmetric credible interval and/or HPD (highest probability density interval) for various source parameters from multiple waveform models  and the two spin priors.

\begin{landscape}
\begin{table*}[]
\centering
  \begin{tabular}{cccccccccccccc}
    \hline \hline
Parameter   &$m_{1} (M_{\odot})$ & $m_{2} (M_{\odot})$ & $M (M_{\odot})$ &$q$ &$\chi_\mathrm{{eff}}$ & $\chi_\mathrm{{p}}$ & SNR & $D_L(\text{Gpc})$ & $\vert \cos{\theta_{JN}}\vert$\\ \hline
\texttt{ IMRPhenomPv2 ($\chi$)} & $51.0^{+13.9}_{-12.4}$& $31.9^{+9.3}_{-9.6}$& $83.7^{+13.0}_{-12.0}$& $0.62^{+0.36}_{-0.23}$& $0.35^{+0.22}_{-0.23}$& $0.42^{+0.34}_{-0.29}$& $10.7^{+0.4}_{-0.4}$& $2.85^{+1.31}_{-1.28}$& $0.83^{+0.17}_{-0.40}$ \\
\texttt{ IMRPhenomPv2 ($\chi_z$)} & $52.0^{+15.9}_{-11.6}$& $33.2^{+9.9}_{-9.7}$& $85.8^{+12.4}_{-12.4}$& $0.64^{+0.34}_{-0.25}$& $0.41^{+0.21}_{-0.21}$& $0.58^{+0.29}_{-0.29}$& $10.7^{+0.4}_{-0.4}$& $2.96^{+1.30}_{-1.38}$& $0.84^{+0.16}_{-0.42}$ \\
\texttt{ IMRPhenomD ($\chi$)} & $50.5^{+13.5}_{-11.2}$& $32.5^{+10.0}_{-8.8}$& $82.8^{+13.4}_{-12.6}$& $0.65^{+0.32}_{-0.23}$& $0.34^{+0.21}_{-0.23}$& -& $10.7^{+0.4}_{-0.4}$& $2.75^{+1.26}_{-1.41}$& $0.80^{+0.20}_{-0.46}$ \\
\texttt{ IMRPhenomD ($\chi_z$)} & $50.8^{+12.8}_{-11.5}$& $34.5^{+9.7}_{-9.2}$& $85.5^{+13.9}_{-11.4}$& $0.68^{+0.29}_{-0.25}$& $0.42^{+0.19}_{-0.21}$& -& $10.8^{+0.4}_{-0.3}$& $2.90^{+1.32}_{-1.48}$& $0.79^{+0.21}_{-0.46}$ \\
\texttt{ IMRPhenomHM} ($\chi$) & $57.0^{+12.6}_{-12.0}$& $29.5^{+9.2}_{-9.0}$& $86.0^{+12.2}_{-12.0}$& $0.52^{+0.26}_{-0.21}$& $0.27^{+0.23}_{-0.27}$& -& $11.1^{+0.4}_{-0.4}$& $2.15^{+1.19}_{-1.15}$& $0.70^{+0.30}_{-0.42}$ \\
\texttt{ IMRPhenomHM}  ($\chi_z$) & $58.1^{+12.5}_{-13.4}$& $32.0^{+9.5}_{-8.9}$& $89.7^{+13.4}_{-12.9}$& $0.55^{+0.26}_{-0.21}$& $0.36^{+0.21}_{-0.23}$& -& $11.1^{+0.4}_{-0.4}$& $2.30^{+1.24}_{-1.22}$& $0.68^{+0.32}_{-0.46}$ \\
\texttt{ SEOBNRv3 ($\chi$)} & $49.5^{+13.2}_{-10.8}$& $35.3^{+8.9}_{-8.6}$& $85.0^{+13.8}_{-12.8}$& $0.72^{+0.28}_{-0.22}$& $0.38^{+0.23}_{-0.22}$& $0.45^{+0.32}_{-0.31}$& $10.8^{+0.4}_{-0.4}$& $2.86^{+1.38}_{-1.38}$& $0.78^{+0.22}_{-0.47}$ \\
\texttt{ SEOBNRv4 ($\chi$)} & $50.8^{+12.4}_{-11.8}$& $33.4^{+9.6}_{-9.8}$& $83.9^{+13.7}_{-13.3}$& $0.66^{+0.30}_{-0.25}$& $0.35^{+0.20}_{-0.24}$& -& $10.8^{+0.4}_{-0.3}$& $2.77^{+1.30}_{-1.45}$& $0.79^{+0.21}_{-0.46}$ \\
\texttt{ SEOBNRv4 ($\chi_z$)} & $51.7^{+12.4}_{-12.6}$& $35.2^{+9.4}_{-9.6}$& $86.4^{+14.9}_{-12.7}$& $0.68^{+0.29}_{-0.24}$& $0.42^{+0.21}_{-0.26}$& -& $10.8^{+0.4}_{-0.3}$& $2.97^{+1.46}_{-1.43}$& $0.79^{+0.21}_{-0.47}$ \\
\texttt{ SEOBNRv4HM}  ($\chi$) & $55.2^{+10.2}_{-12.2}$& $29.8^{+9.9}_{-9.5}$& $84.6^{+12.5}_{-11.3}$& $0.54^{+0.31}_{-0.20}$& $0.25^{+0.26}_{-0.26}$& - & $11.0^{+0.4}_{-0.4}$& $2.30^{+1.36}_{-1.17}$ & $0.72^{+0.28}_{-0.39}$ \\
\texttt{ SEOBNRv4HM}  ($\chi_z$) & $54.8^{+11.1}_{-12.1}$& $32.8^{+9.8}_{-9.2}$& $87.5^{+12.2}_{-11.7}$& $0.60^{+0.32}_{-0.21}$& $0.34^{+0.26}_{-0.24}$& - & $11.0^{+0.4}_{-0.4}$& $2.66^{+1.38}_{-1.31}$ & $0.74^{+0.26}_{-0.41}$ \\
\end{tabular}
\caption{Parameters of GW170729 obtained with various waveform models and two spin priors, uniform-in-$\chi$ (labelled $\chi$) and uniform-in-$\chi_z$ (labelled $\chi_z$). 
We quote median values and 90\% credible intervals for the source-frame primary mass, the source-frame secondary mass, the source-frame total mass, the effective spin $\chi_\mathrm{{eff}}$, and the effective precession parameter $\chi_\mathrm{{p}}$~\cite{Schmidt:2014iyl}. 
For the mass ratio we quote the median value and the 90\% HPD. 
All masses are given in the source frame assuming the cosmological parameters of~\cite{Ade:2015xua} to convert luminosity distance to redshift.
The effective precession parameter $\chi_\mathrm{{p}}$ is absent in the spin-aligned models.
}
\label{table:par-table}
\end{table*}
\end{landscape}

\begin{figure*}[]
\centering
\includegraphics[width=0.7\textwidth,clip=true]{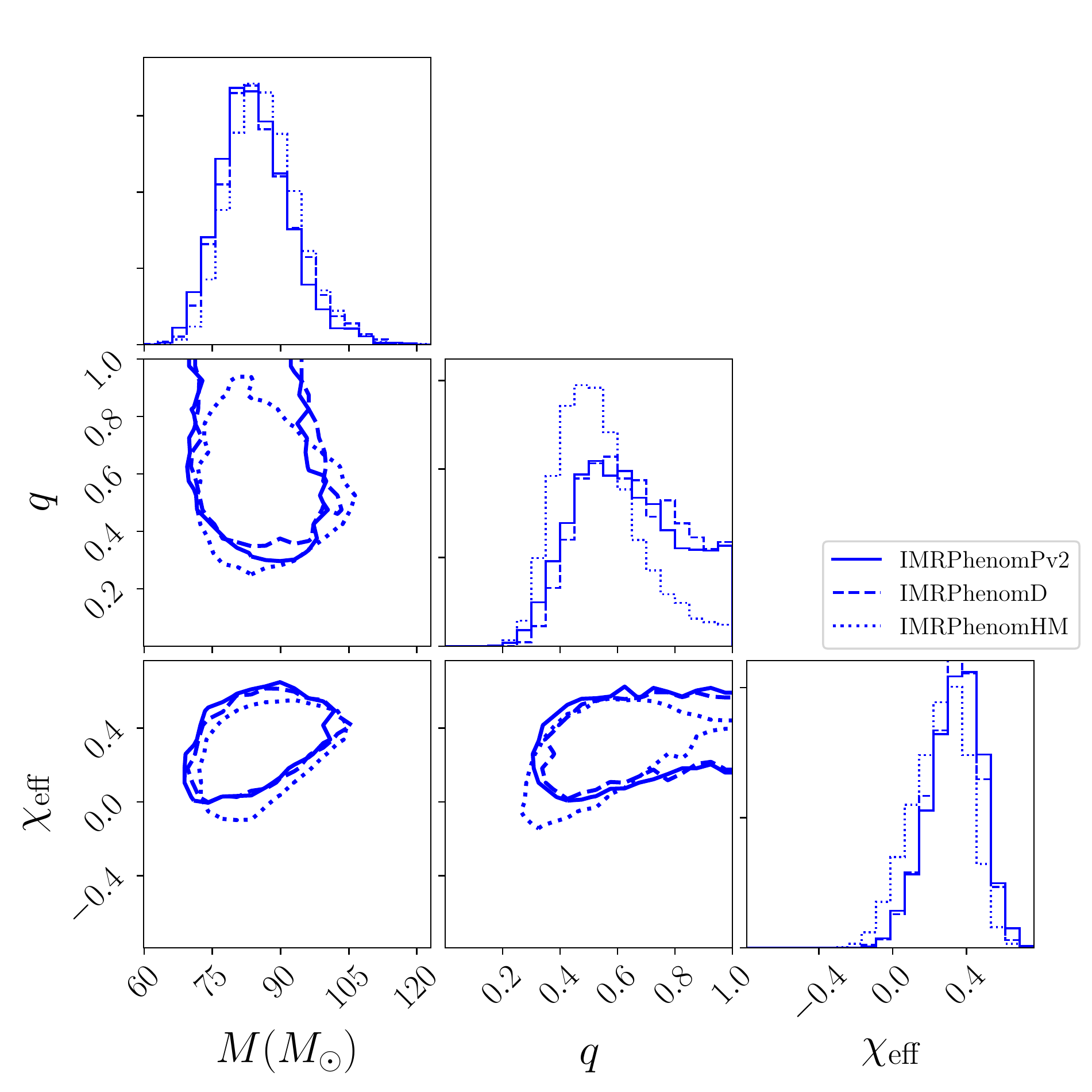}
\includegraphics[width=0.7\textwidth,clip=true]{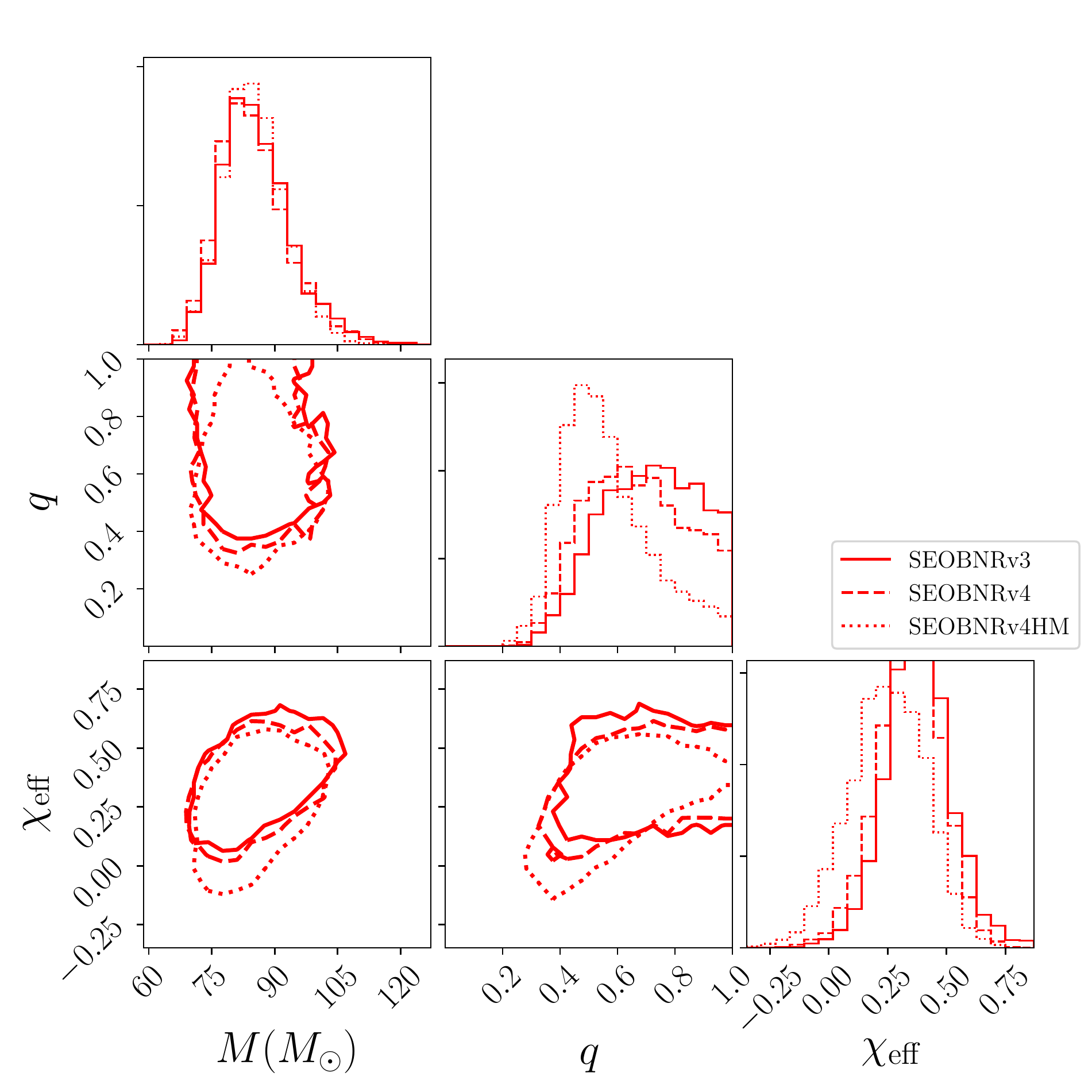}
\caption{Corner plots for the posterior densities of the total binary mass in the source frame, the mass ratio, and the effective spin parameter for the different waveform families. 
We show results obtained with CBC waveforms of the \texttt{ IMRPhenom} family (left) and the \texttt{ SEOBNR} family (right).
Results from both waveform families suggest that the inclusion of higher-order modes in the models results in evidence for more unequal-mass binaries and less support for nonzero effective spin. 
The total mass of the binary remains unchanged.}
\label{fig:intrinsic_IMREOB}
\end{figure*}

We start by discussing the binary intrinsic parameters, in particular masses and spins. 
In Fig.~\ref{fig:intrinsic_IMREOB} we show multi-dimensional corner plots for the posterior densities of the source-frame total mass $M$, the mass ratio $q$ and the effective spin parameter $\chi_\mathrm{{eff}}$ with waveform models of the \texttt{ IMRPhenom} (left) and the \texttt{ SEOBNR} family (right). 
For these figures and all the figures of this subsection we show results with the uniform-in-$\chi$ spin prior. 
Within each family, the spin-precessing model is given with a solid line, the spin-aligned model with a dashed line and the spin-aligned model with higher-order modes with a dotted line. 

In all cases we find that the inclusion of higher-order modes does not have a large effect on the total mass measurement. 
The mass ratio of the system and the effective spin posteriors are both shifted. 
In particular we find that waveforms with higher-order modes consistently provide more support for unequal-mass systems and smaller effective spins. 
For the mass ratio we find that the 90\% HPD interval is $(0.31-0.78) [(0.34-0.85)]$ with higher-order modes and $(0.42-0.97) [(0.41-0.96)]$ without them when using the spin-aligned \texttt{ IMRPhenomHM} [\texttt{ SEOBNRv4HM}] and \texttt{ IMRPhenomD} [\texttt{ SEOBNRv4}] models respectively. We conclude that GW170729 is not consistent with an equal-mass merger at the 90\% level at least. 

The combination of unchanged total mass but lower mass ratio means that the primary mass of GW170729 is inferred to be larger than previously measured. 
Reference~\cite{LIGOScientific:2018jsj} studied the population of the 10 detected BBHs and concluded that no more than 1\% of BHs in BBHs are expected to be above $45 M_{\odot}$.
We find that our updated primary mass measurement is not at odds with this conclusion. 
In particular, we find that the probability that $m_1$ is lower than $45 M_{\odot}$ is 17\% using \texttt{ IMRPhenomPv2} and reduced to 6\% with \texttt{ IMRPhenomHM}.
Given that we have detected 10 BBHs, it is not unlikely that the true $m_1$ of one of them is at the sixth posterior percentile.
A more detailed population analysis in needed to quantify this statement, but this is beyond the scope of this paper.

We also find that higher-order modes result in less support for a positive effective spin in GW170729. 
Reference~\cite{LIGOScientific:2018mvr} reported that for GW170729 $\chi_\mathrm{{eff}} \sim (0.11- 0.58)$ at the 90\% credible level using combined posterior samples between \texttt{ IMRPhenomPv2} and  \texttt{ SEOBNRv3}. 
Interestingly, this credible interval does not include zero, suggesting that at the 90\% level GW170729 has a nonzero effective spin.
The inclusion of higher-order modes slightly changes this picture as we now find that the corresponding 90\% credible intervals no longer exclude zero: $\chi_\mathrm{{eff}} \sim(-0.01-0.49)$ with \IMRPHM{} and $\chi_\mathrm{{eff}} \sim(-0.02-0.50)$ with \SEOBHM{}. 
The effective spin parameter is still probably positive with the probability of $\chi_\mathrm{{eff}}>0$ being $94\%$ with higher-order modes, which is slightly reduced from the corresponding probability of $99\%$ when higher-order modes are not taken into account.
Overall higher-order modes cause the 95\% lower limit for $\chi_\mathrm{{eff}}$ to shift by $-0.10$ for the \texttt{ IMRPhenom} family,  $-0.13$ for the \texttt{ SEOBNR} family.

\begin{figure*}[]
\centering
\includegraphics[width=0.7\textwidth,clip=true]{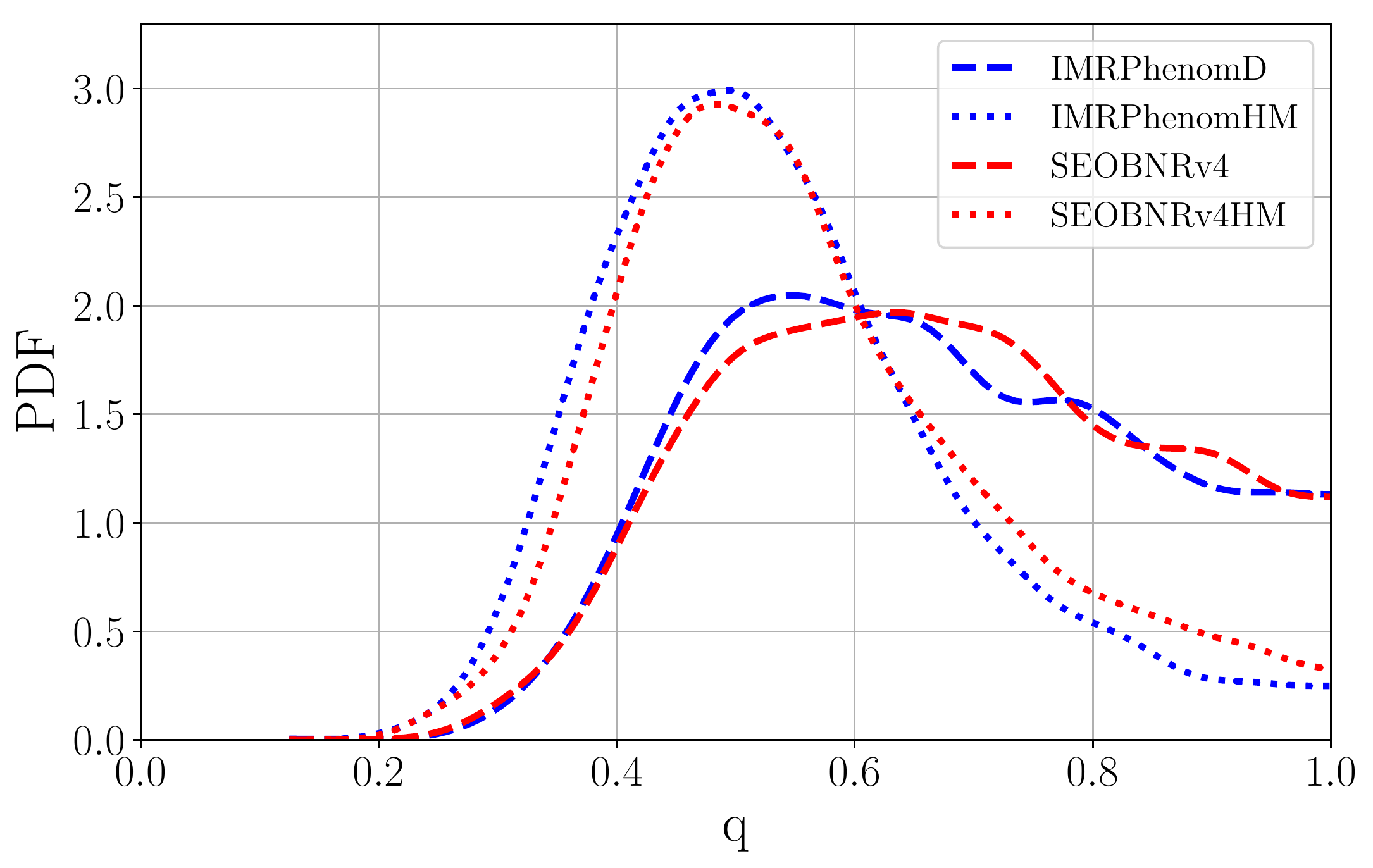}
\includegraphics[width=0.7\textwidth,clip=true]{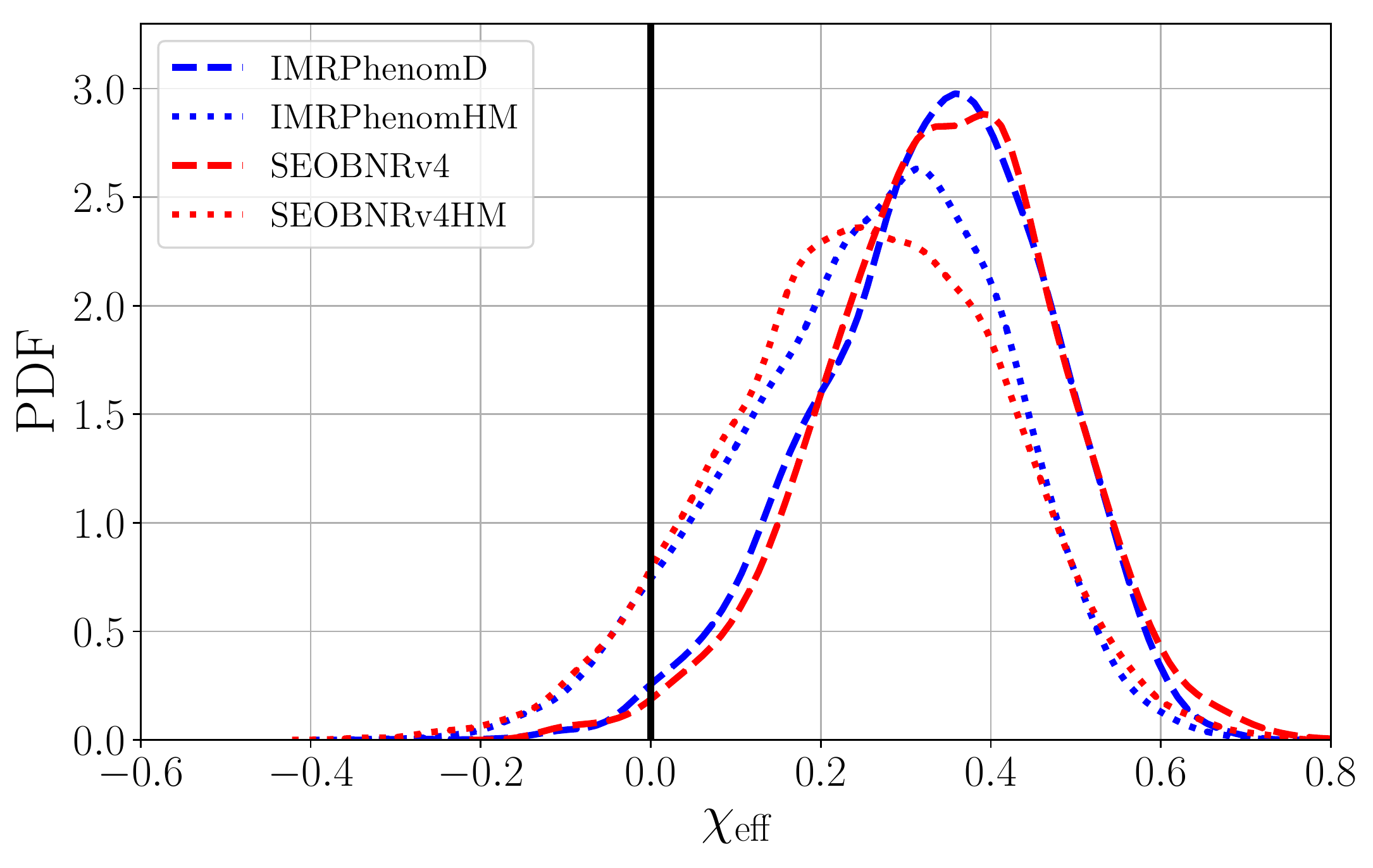}
\caption{Effect of waveform systematics. 
We show the mass ratio posterior (left) and the effective spin posterior (right) computed with different CBC waveform models that include higher-order modes (dotted lines) and models that do not include higher-order modes (dashed). 
Small differences between the posteriors from different waveform approximants are present, but these differences are much smaller than the effect of higher-order modes for both parameters. }
\label{fig:systematics}
\end{figure*}

Despite the broad consistency between results obtained with different waveform families, it is still possible that our results are partly affected by systematic uncertainties in the waveform models.
To address this in Fig.~\ref{fig:systematics} we plot again the mass ratio (left) and effective spin (right) posterior densities for the 
\texttt{ IMRPhenom} and \texttt{ SEOBNR} waveform models with (dashed lines) and without (dotted lines) higher-order modes. 
We find very small differences between both all the dotted and all the dashed lines and in particular between the new waveform
models \texttt{ IMRPhenomHM} and \texttt{ SEOBNRv4HM} that include higher-order modes.
More importantly, we find a clear separation between the dotted lines, i.e. the posteriors that include higher-order modes, and the dashed lines, i.e. the posteriors that do not include higher-order modes. 

\begin{figure}[]
\centering
\includegraphics[width=0.6\textwidth,clip=true]{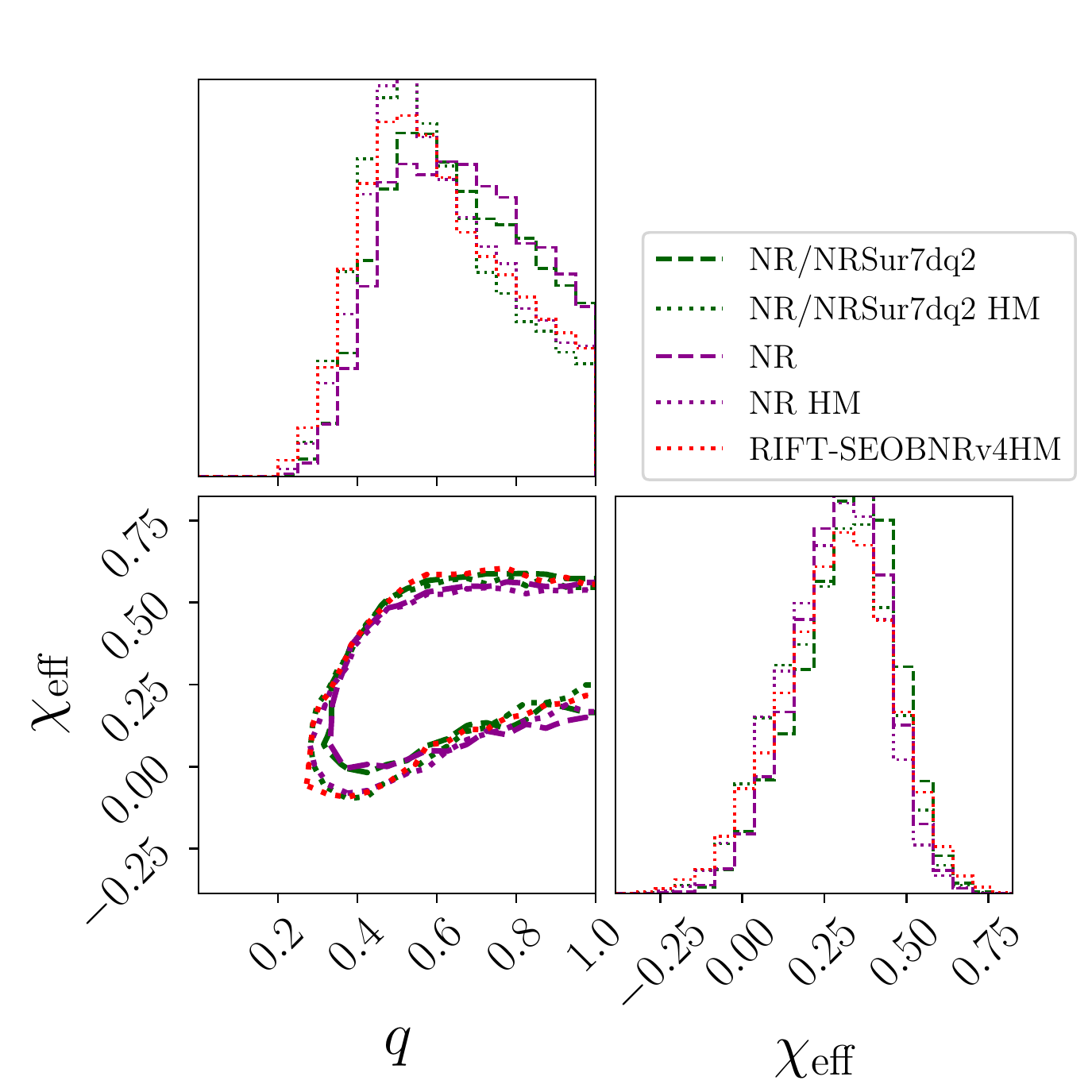}
\caption{Corner plots for the posterior densities of the mass ratio and the effective spin parameter for the NR waveform family with (magenta lines) and without (green lines) \NRSur{} and \texttt{ SEOBNRv4HM} (red line) computed with \texttt{ RIFT}. 
As before, we use dashed (dotted) lines for posteriors with (without) higher order modes. 
We observe excellent agreement between NR and NR/\NRSur{}, confirming the high accuracy of the NR surrogate model.
Additionally, we find very good agreement between the NR analysis and \texttt{ SEOBNRv4HM} when both are used with the
same inference code, \texttt{ RIFT}. 
}
\label{fig:intrinsic_NR}
\end{figure}

As a further test, in Fig.~\ref{fig:intrinsic_NR} we compare posteriors for  $q$ and $\chi_\mathrm{{eff}}$ computed 
with \texttt{ SEOBNRv4HM} (red line)  with NR waveforms (magenta lines), as well as with NR/\NRSur{} (green lines).
Compared to Fig.~\ref{fig:intrinsic_IMREOB} we omit the total mass posterior as \texttt{ RIFT} did not compute source-frame quantities.
In order to perform a fair comparison, all posteriors have been computed with \texttt{ RIFT}, while for technical reasons we cannot
use \texttt{ IMRPhenomHM} with \texttt{ RIFT}. We find excellent agreement between NR with higher-order modes, NR/\NRSur{} with higher-order modes, and \texttt{ SEOBNRv4HM}. This shows that \texttt{ SEOBNRv4HM} is as accurate as NR waveforms in 
describing GW170729. Moreover, the agreement between \texttt{ SEOBNRv4HM} and \texttt{ IMRPhenomHM} in Fig.~\ref{fig:systematics}
suggests that the latter \texttt{ IMRPhenom} waveform is also highly accurate for the event studied here.
While these posteriors are broadly consistent with those obtained in Fig.~\ref{fig:systematics}, we find a disagreement in results obtained 
with \texttt{ LALInference} and \texttt{ RIFT} with the same waveform model for the mass ratio at the $7\%$ level. 
The nature of this difference and further numerical estimates are described in Appendix~\ref{RIFTtroubles}.
Overall, Figures~\ref{fig:systematics} and~\ref{fig:intrinsic_NR} suggest that despite minor differences between the waveform models considered here, our main conclusions are robust.

\begin{figure}[]
\centering
\includegraphics[width=0.7\textwidth,clip=true]{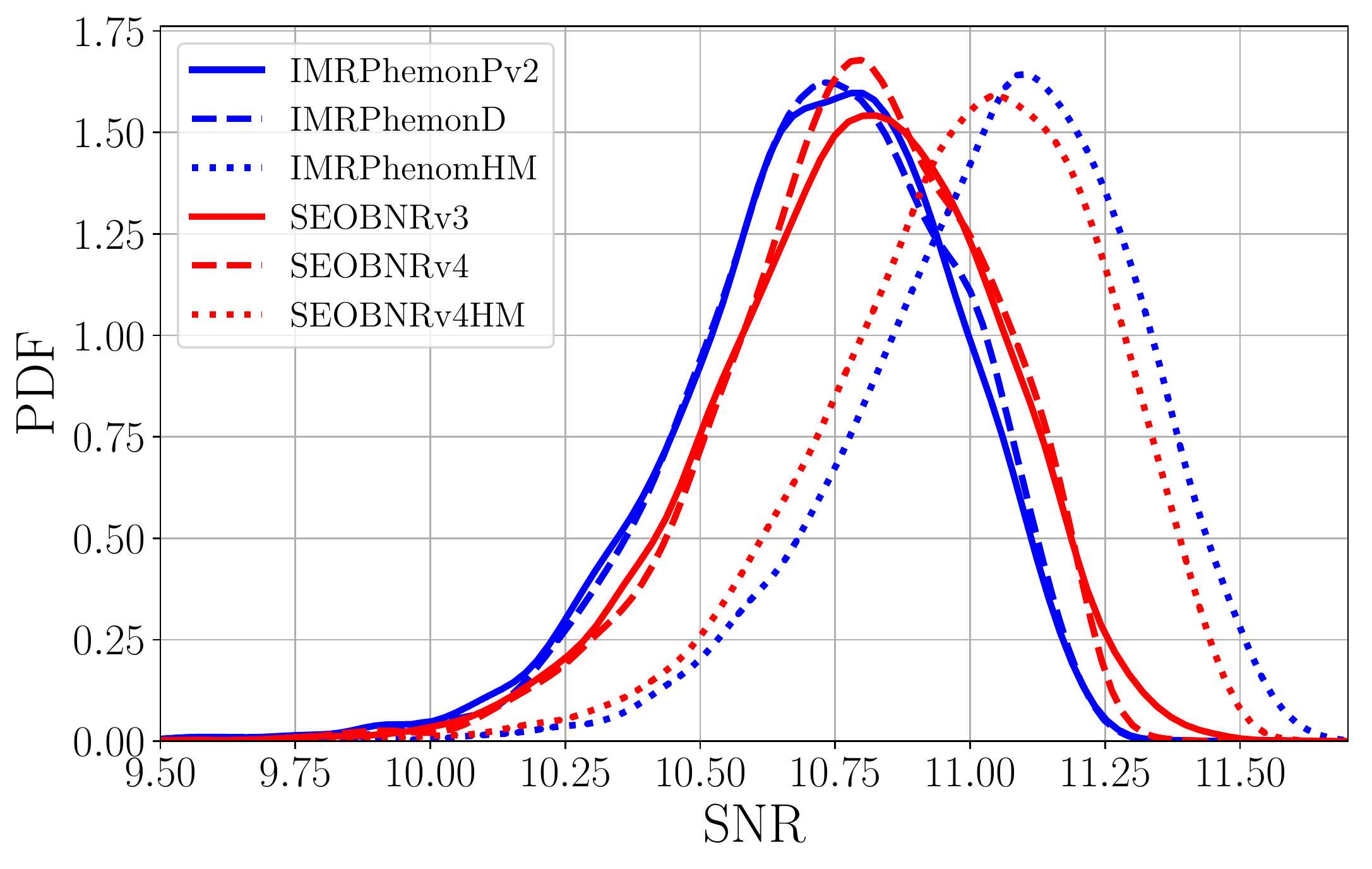}\\
\includegraphics[width=0.7\textwidth,clip=true]{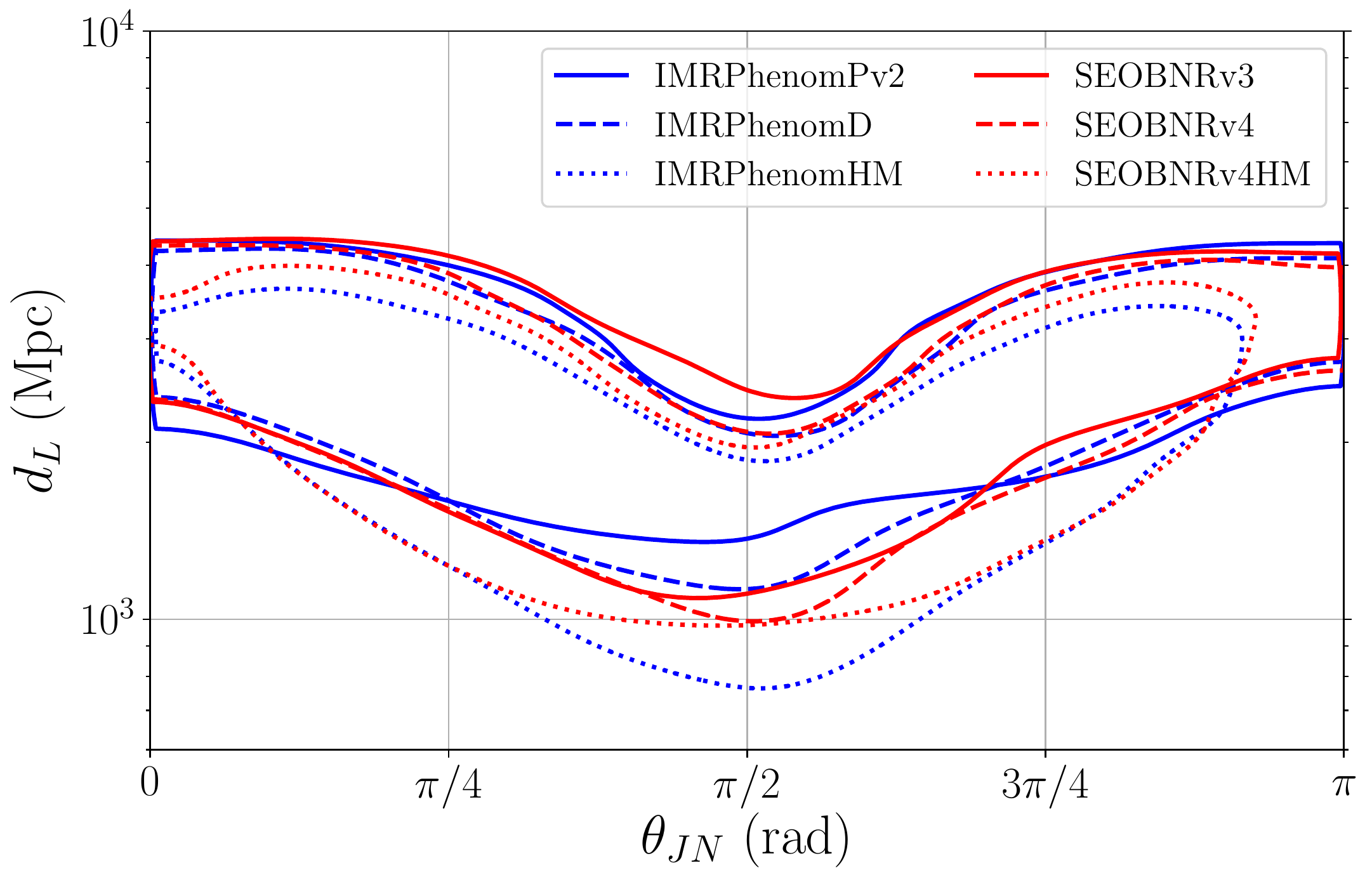}
\caption{Posterior density for the matched-filter SNR (top) and distance-inclination (bottom) for various waveform models. 
We find that both within the \texttt{ IMRPhenom} and the \texttt{ SEOBNR} waveform families, the waveform model that includes higher-order modes returns the highest value of matched-filter SNR. 
This higher SNR translates to lower distance and inclination angles closer to 90 degrees.
}
\label{fig:extrinsic}
\end{figure}

We now turn to the binary extrinsic parameters. 
Figure~\ref{fig:extrinsic} shows the matched-filter network signal-to-noise ratio (SNR) posterior density on the top and the two-dimensional posterior density for the luminosity distance and the inclination on the bottom. 
The network SNR is defined as the square root of the squared sums of the matched-filter SNR in each interferometer, calculated as $\rho = (h,d)/\sqrt{(h,h)}$, where $d$ is the data and $h$ the signal model.
The inclination is defined as the angle between the \emph{total} angular momentum vector of the binary, whose direction we treat as fixed,  and the line of sight. 
We present results for waveform models of the \texttt{ IMRPhenom} (blue) and the \texttt{ SEOBNR} (red) families. 

The SNR depends on both the intrinsic loudness of the data and the agreement between the signal and the template $h$. 
Since the data are common for all analyses, a larger SNR indicates a better agreement/overlap between the data and the template. 
While we find that within both families spin-precession has a minimal impact on the SNR, waveforms with higher-order modes report slightly larger SNR values, see the top panel of Fig.~\ref{fig:extrinsic}. 
This suggests that their inclusion leads to a marginally better fit of the data. 
The data, though, include both the GW170729 signal and a random realization of Gaussian noise, so a better fit of the data does not necessarily imply that waveforms with higher-order modes recover a larger fraction of the GW signal.

We quantify the impact of higher-order modes on GW170729 by computing the Bayes Factor in favor of \texttt{ IMRPhenomHM} compared to \texttt{ IMRPhenomD}. 
We find a BF of 5.1:1. While it favors the model with HM, this BF is consistent with the fact that the HM waveforms are able to extract marginally more SNR from the data\footnote{The BF is related to the SNR through $\log \mathrm{BF} \propto 1/2\, \mathrm{SNR}^2$. Therefore an SNR increase of $\sim 0.2$ compared to $\sim 10.8$ (see Table~\ref{table:par-table}) would result in a BF of $\sim 8$. This suggests that the measured BF of $5$ is consistent with the SNR increase due to HM.}.
Moreover, we emphasize that the BF is not the same as the odds ratio in favor of higher-order modes, which quantifies our degree of belief that higher-order modes are present in the signal. 
The odds ratio is the BF times the prior odds in favor of the presence of higher-order modes. 
The latter is formally infinite within GR, as the theory of gravity unequivocally predicts that higher-order modes are present in all CBC signals. 
The BF presented here only quantifies if higher-order modes are a necessary feature of the models in order to describe the data, and not whether we believe that they exist in general.

Regarding the bottom panel of Fig.~\ref{fig:extrinsic}, we find that waveforms with higher-order modes result in less support for face-on/off binary orientations. 
This observation, coupled to the fact that we see more support for unequal masses and lower spins, see Fig.~\ref{fig:intrinsic_IMREOB}, suggests that higher-order modes lead to more support for sources that are intrinsically of lower amplitude\footnote{We have also verified this by computing the posterior of the intrinsic loudness (defined as the product of the SNR and the distance) with and without higher-order modes. 
We find that higher-order modes lead to larger probability for intrinsically quieter sources than the  quadrupole templates.}. 
This in turn leads to a posterior distribution for the luminosity distance that is shifted to lower values, as compared to analyses without higher-order modes, as also seen in the bottom panel of Fig~\ref{fig:extrinsic}.

Overall, we find that the inclusion of higher-order modes induces small but noticeable shifts in the parameter posteriors. 
Specifically, the matched-filter SNR increases, the mass ratio posterior obtains more support for unequal masses, and the effective spin parameter is more consistent with lower values; parameter measurements are given in Table~\ref{table:par-table}. 
The general consistency between the waveform model families we study here shows that our conclusions are robust against waveform systematics.

\subsection{Spin prior}

\begin{figure*}[]
\centering
\includegraphics[width=0.7\textwidth,clip=true]{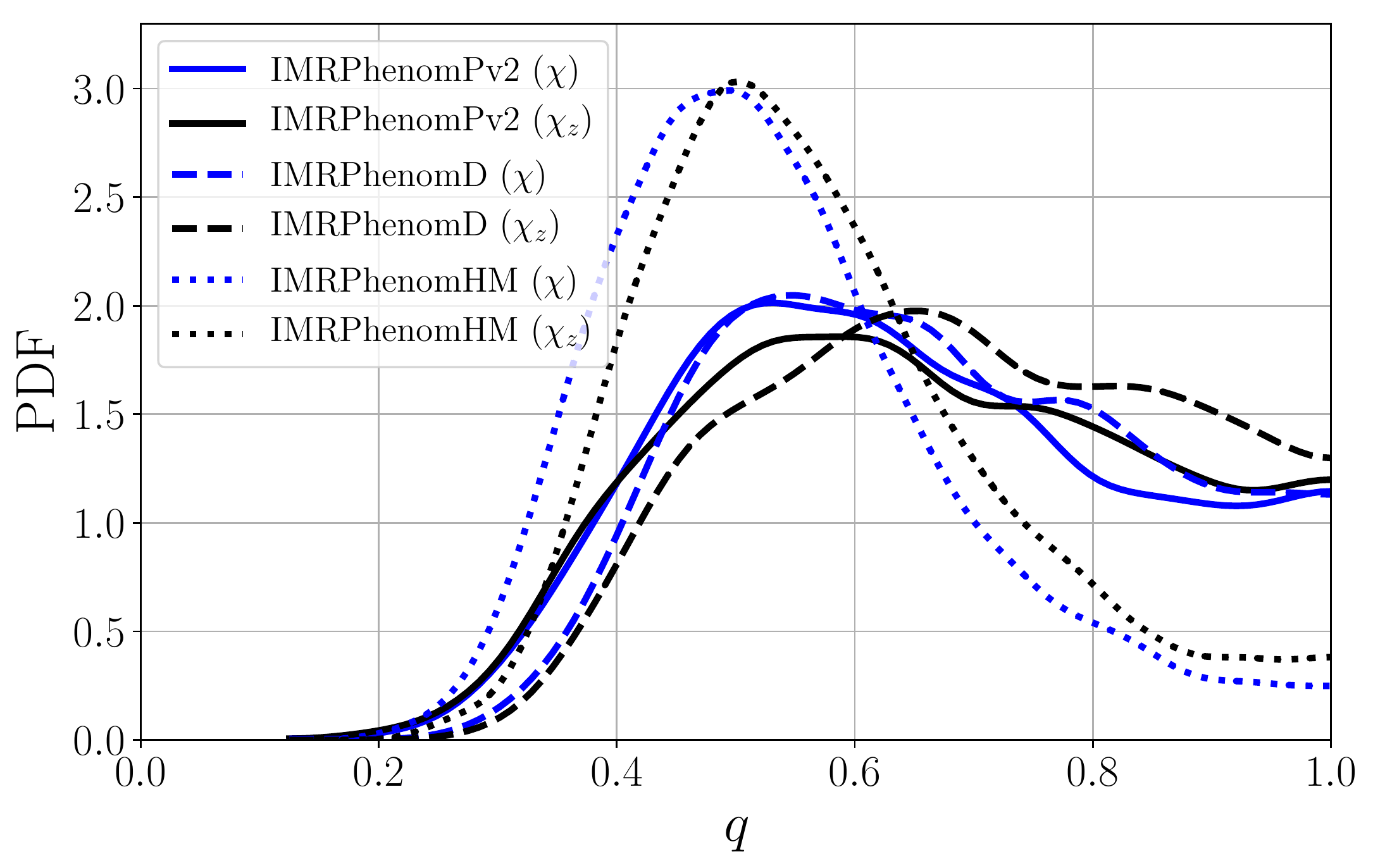}
\includegraphics[width=0.7\textwidth,clip=true]{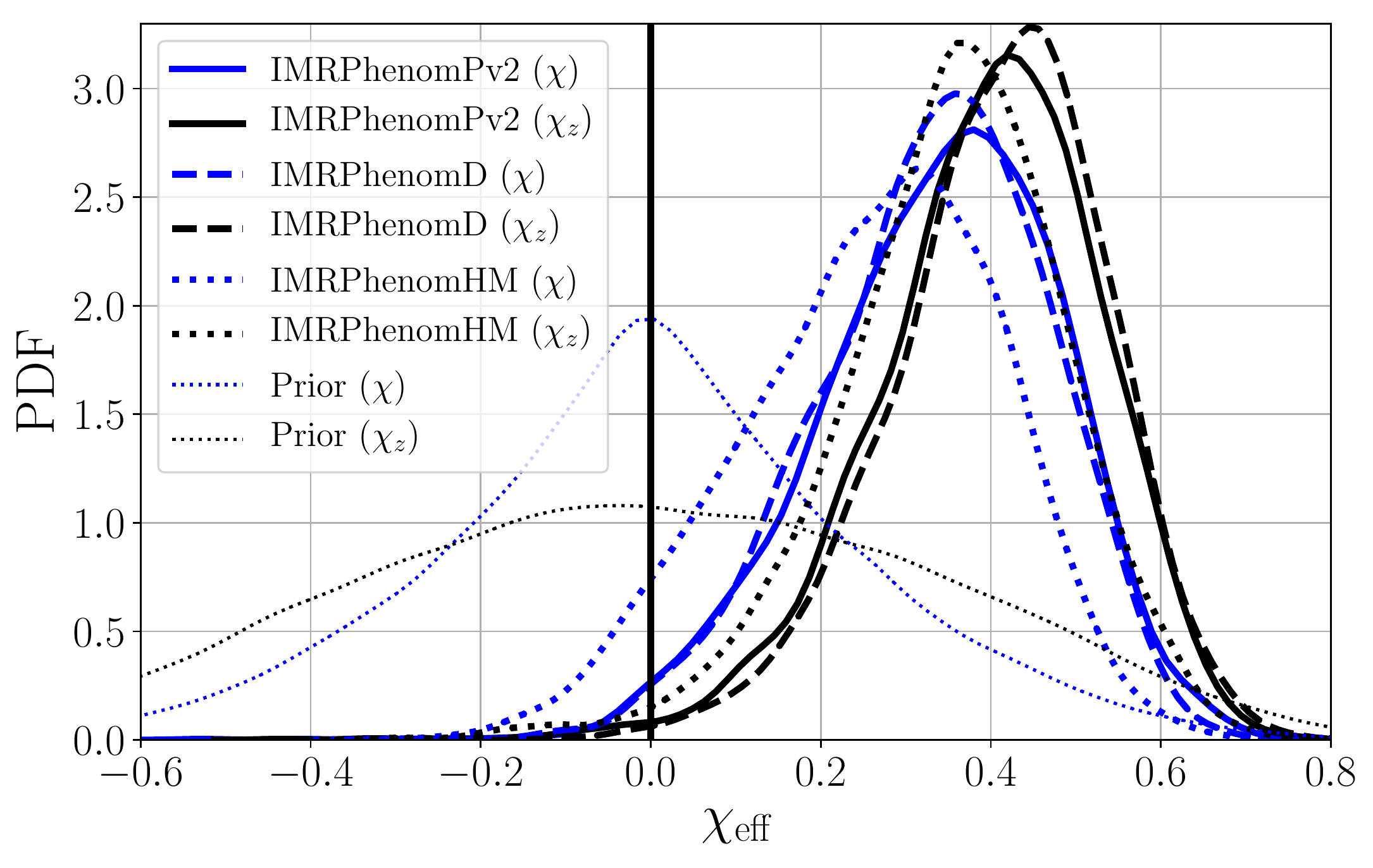}\\
\caption{Effect of spin prior on the mass ratio and effective spin of GW170729. 
As expected, we find that the mass ratio is minimally affected by the spin prior. 
The effective spin, however, is shifted to larger values with the uniform-in-$\chi_z$ prior, resulting in increased evidence for nonzero spins. 
We show results obtained with \texttt{ IMRPhenom} waveform models, but obtain similar results with \texttt{ SEOBNR} and NR waveform models as well.}
\label{fig:prior}
\end{figure*}

Besides the CBC waveform models, posterior measurements are also affected by prior choices, in particular the spin prior~\cite{salvoprior}. 
To test the effect of the spin prior, we reanalyze the data this time assuming a uniform-in-$\chi_z$ prior, where $\chi_z$ is the spin component perpendicular to the orbital plane. 
The results are presented in Fig.~\ref{fig:prior} for the mass ratio (left) and the effective spin (right) and for waveforms of the \texttt{ IMRPhenom} family. 
We have verified that we obtain qualitatively similar results when using \texttt{ SEOBNR} and NR waveforms models. 
Due to computational constraints we have only checked results with \texttt{ SEOBNRv3} and the uniform-in-$\chi$ prior, as computed in~\cite{LIGOScientific:2018mvr}.

We find that the spin prior has a minimal effect on the mass ratio posterior. 
This is expected as the correlation between mass ratio and effective spin is mostly present in the inspiral phase of a CBC. 
High-mass systems, such as GW170729 are instead dominated by the merger and ringdown in the LIGO sensitive frequency band. 
In this case little correlation exists between mass ratio and effective spin~\cite{Ng:2018neg}, and changing the spin prior doesnt affect the mass ratio posterior.
The effective spin parameter, on the contrary, is directly affected by the choice of the spin prior, and clear differences are visible. 
The uniform-in-$\chi_z$ prior favors larger spin magnitudes than the uniform-in-$\chi$ prior. 
As a result, the effective spin posterior is shifted to larger values. 
The median and 90\% credible interval for the effective spin is $0.41^{+0.21}_{-0.21}$ under the uniform-in-$\chi_z$ prior and $0.35^{+0.22}_{-0.23}$ under the uniform-in-$\chi$ prior using the \texttt{ IMRPhenomPv2} waveform model. 
Additional spin measurements for other waveform models are presented in Table~\ref{table:par-table}.

\subsection{Second generation merger}

\begin{figure*}[]
\centering
\includegraphics[width=0.7\textwidth,clip=true]{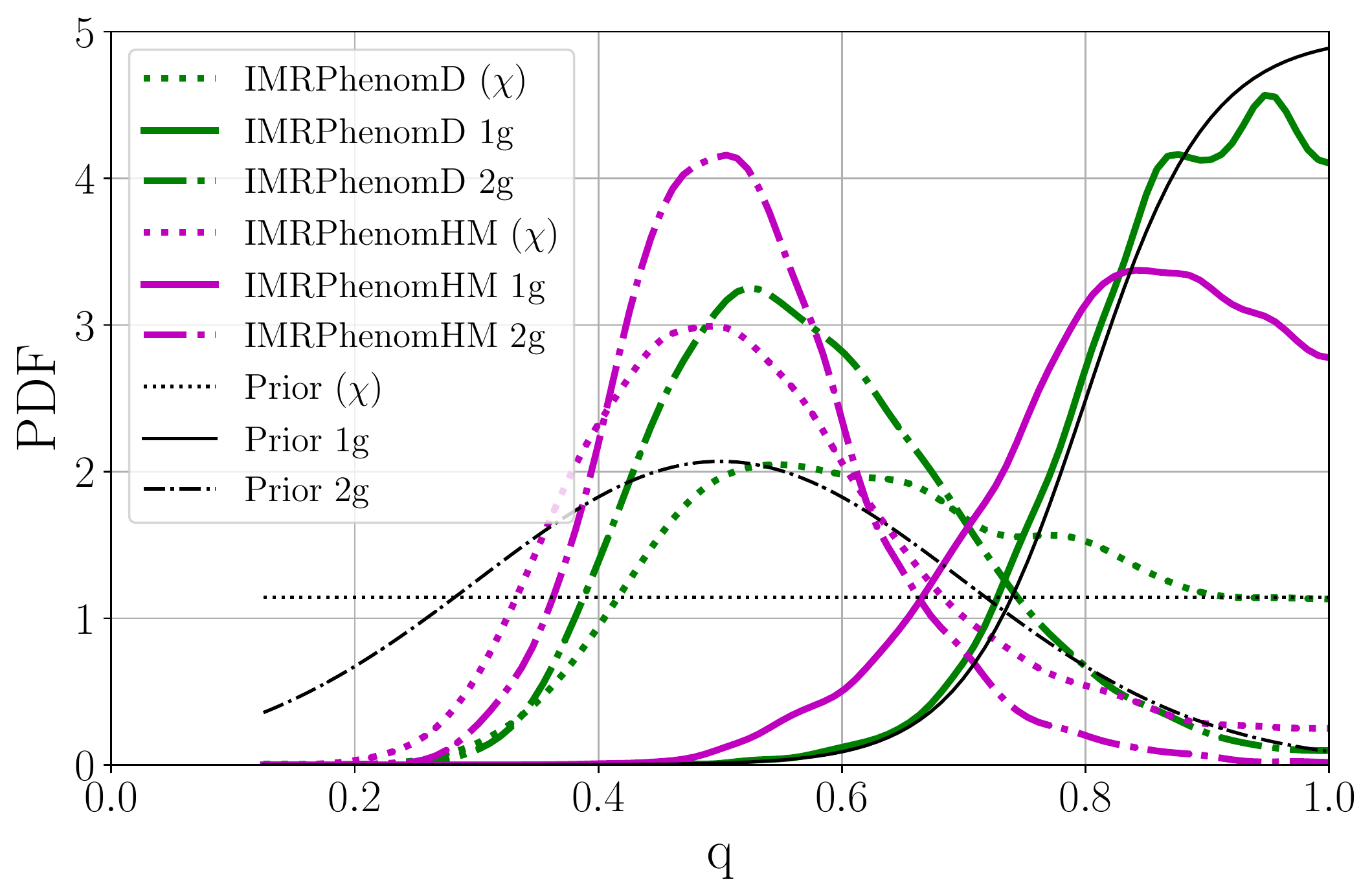}
\includegraphics[width=0.7\textwidth,clip=true]{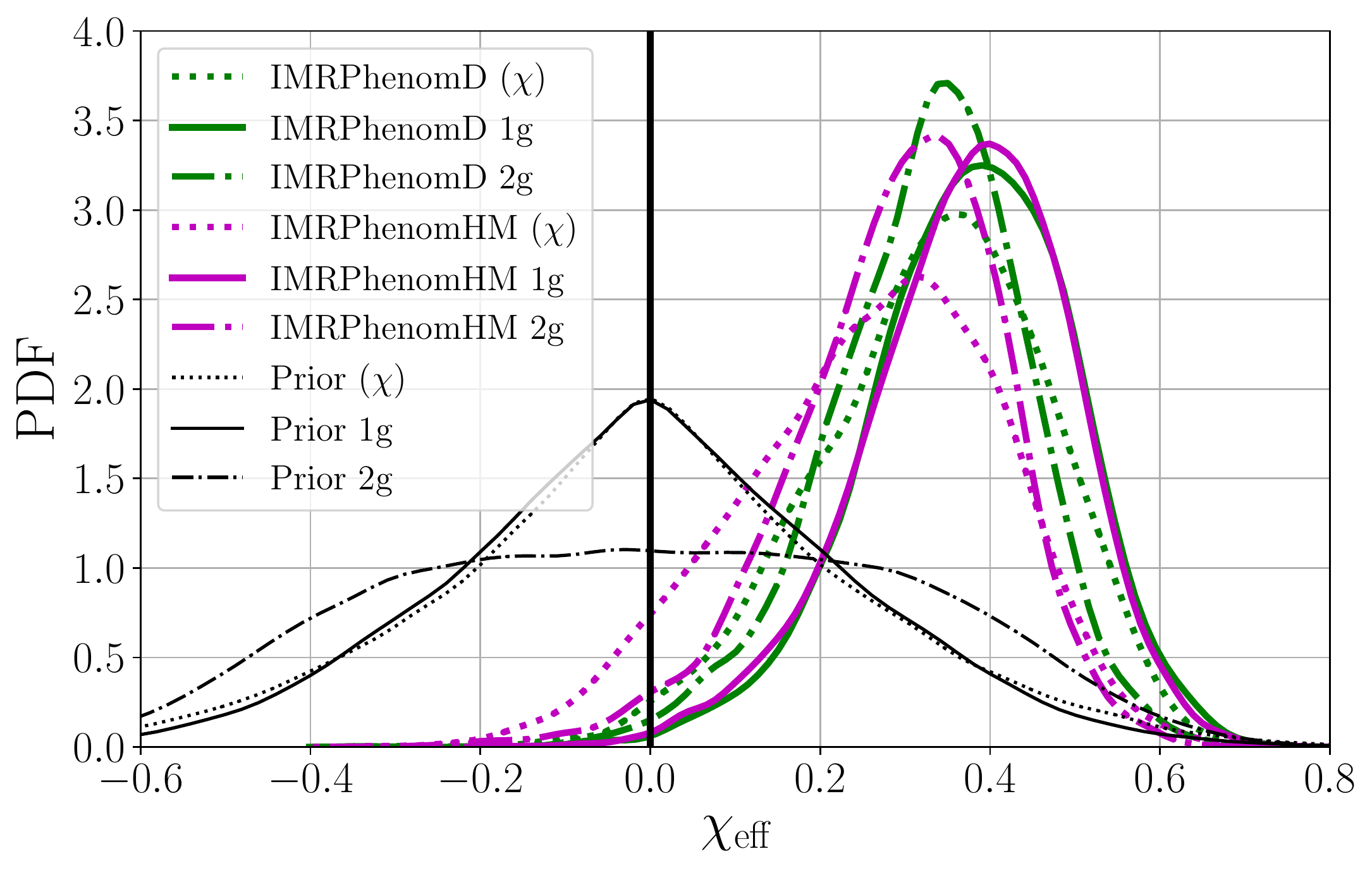}
\caption{Posterior density for the mass ratio (left) and the effective spin (right) under the 1g prior (solid lines) and the 2g prior (dashed lines) for various waveforms. 
The black lines show the prior density distributions for each parameter. 
We also show the uniform-in-$\chi$ posterior and prior from Fig.~\ref{fig:prior} for comparison.
We find that the 1g-2g prior has a minimal effect on the effective spin parameter, but affects the mass ratio considerably. 
All 2g runs have strong support for unequal masses, as expected from the prior.}
\label{fig:2G}
\end{figure*}

Finally, we study the possibility of a 2g merger. 
In that scenario, GW170729 is created in a dense environment such as a nuclear or a globular cluster and its primary mass is the product of a previous merger of two BHs~\cite{davide2g,Fishbach:2017dwv,DiCarlo:2019pmf}. 
In that case, the mass ratio of the system is expected to be closer to 2:1 (as we find when we use waveform models with HM) and the spin magnitude of the primary should be close to 0.7 (the typical spin of the remnant BH after the merger of two equal-mass, 
nonspinning BHs)~\cite{Gonzalez:2006md,Berti:2007fi,Buonanno:2007sv,Campanelli:2006gf,Baker:2003ds}
\footnote{Though a 2g merger scenario provides a simple way to produce a highly spinning BH, 
such systems could also be produced in certain astrophysical scenarios, e.g~\cite{Bouffanais:2019nrw}.}.

We repeat our analysis with two more priors tailored to the cases of a 1g and a 2g merger scenario. 
Table~\ref{table:par-table-2g} gives the median and 90\% symmetric credible interval and/or HPD for various source parameters obtained with the 1g and 2g priors.
Figure~\ref{fig:2G} shows the effect of this 2g prior on the mass ratio (left) and the effective spin (right) of the binary when using \texttt{ IMRPhenomD} and \texttt{ IMRPhenomHM}.  
We have verified that we obtain similar results with \texttt{ SEOBNR} and NR waveform models.
As expected from the priors we have selected, 2g runs show strong support for unequal masses. 
This support is even more evident for waveforms with higher-order modes, as anticipated from the results of the previous section. 
The effective spin parameter is similar, with 1g runs showing more support for nonzero binary components.

To further quantify the prior effect, we calculate the Kullback-Leibler (KL) divergence of posterior against prior~\cite{kullback1951}.
We find a KL divergence for the mass ratio in the 1g case that is $\sim10$ times smaller than the KL divergence in 2g or default prior scenarios, implying that we do not extract much information by applying the 1g mass ratio prior.
For the effective spin, we find that the KL divergence in both priors are comparable as expected because the prior on the effective spin does not change as drastically as the prior on the mass ratio.

We also compute the 2g-vs-1g BF.
We find 4.7:1 (1.4:1) for waveforms with (without) higher-order modes. We conclude that there is not enough support for the hypothesis that the GW170729 primary needs to be the product of a past merger in order to explain the data's properties.
Note that both 2g and 1g models have the same number of parameters, but different distributions in mass ratio and primary spin. The resulting BFs are therefore only affected by how well each model fits the data.

\begin{landscape}
\begin{table*}[]
\centering
  \begin{tabular}{cccccccccccccc}
    \hline \hline
Parameter   &$m_{1} (M_{\odot})$ & $m_{2} (M_{\odot})$ & $M (M_{\odot})$ &$q$ &$\chi_\mathrm{{eff}}$ & $\chi_\mathrm{{p}}$ & SNR & $D_L(\text{Gpc})$ & $\vert \cos{\theta_{JN}}\vert$\\ \hline
\texttt{ IMRPhenomPv2 (1g)} & $43.9^{+8.3}_{-6.7}$& $38.0^{+7.0}_{-6.3}$& $82.1^{+12.9}_{-11.6}$& $0.88^{+0.12}_{-0.14}$& $0.39^{+0.20}_{-0.22}$& $0.47^{+0.31}_{-0.31}$& $10.7^{+0.4}_{-0.4}$& $3.08^{+1.17}_{-1.47}$& $0.83^{+0.17}_{-0.42}$\\ 
\texttt{ IMRPhenomPv2 (2g)} & $55.1^{+11.3}_{-12.6}$& $30.2^{+8.0}_{-7.2}$& $85.5^{+12.1}_{-11.2}$& $0.55^{+0.23}_{-0.20}$& $0.37^{+0.18}_{-0.19}$& $0.48^{+0.25}_{-0.29}$& $10.7^{+0.4}_{-0.4}$& $2.83^{+1.19}_{-1.31}$& $0.83^{+0.17}_{-0.40}$\\
\texttt{ IMRPhenomD (1g)} & $43.8^{+7.9}_{-7.2}$& $38.2^{+6.9}_{-6.4}$& $81.9^{+13.9}_{-11.2}$& $0.88^{+0.12}_{-0.13}$& $0.38^{+0.18}_{-0.20}$& -& $10.8^{+0.4}_{-0.4}$& $2.99^{+1.26}_{-1.43}$& $0.80^{+0.20}_{-0.44}$\\
\texttt{ IMRPhenomD (2g)} & $53.7^{+11.0}_{-11.6}$& $30.2^{+7.9}_{-7.0}$& $84.0^{+13.3}_{-11.8}$& $0.56^{+0.21}_{-0.19}$& $0.34^{+0.17}_{-0.20}$& -& $10.8^{+0.4}_{-0.3}$& $2.69^{+1.23}_{-1.31}$& $0.79^{+0.21}_{-0.44}$\\
\texttt{ IMRPhenomHM (1g)} & $45.5^{+10.0}_{-7.9}$& $37.9^{+6.7}_{-6.4}$& $83.6^{+13.3}_{-11.7}$& $0.84^{+0.16}_{-0.16}$& $0.38^{+0.19}_{-0.20}$& -& $10.9^{+0.4}_{-0.5}$& $2.83^{+1.17}_{-1.41}$& $0.80^{+0.20}_{-0.38}$\\
\texttt{ IMRPhenomHM (2g)} & $58.2^{+10.5}_{-10.1}$& $29.7^{+7.1}_{-7.4}$& $87.6^{+12.7}_{-11.2}$& $0.51^{+0.17}_{-0.16}$& $0.30^{+0.18}_{-0.21}$& -& $11.1^{+0.4}_{-0.4}$& $2.18^{+1.18}_{-1.05}$& $0.69^{+0.31}_{-0.43}$\\
\texttt{ SEOBNRv4 (1g)} & $44.3^{+8.5}_{-7.2}$& $38.7^{+7.0}_{-6.9}$& $83.0^{+14.2}_{-11.6}$& $0.88^{+0.12}_{-0.14}$& $0.39^{+0.21}_{-0.20}$& -& $10.8^{+0.4}_{-0.4}$& $2.98^{+1.34}_{-1.50}$& $0.79^{+0.21}_{-0.45}$\\
\texttt{ SEOBNRv4 (2g)} & $53.9^{+11.5}_{-11.1}$& $31.2^{+8.3}_{-7.5}$& $85.1^{+14.3}_{-11.4}$& $0.57^{+0.21}_{-0.20}$& $0.34^{+0.19}_{-0.20}$& -& $10.8^{+0.4}_{-0.4}$& $2.68^{+1.28}_{-1.36}$& $0.78^{+0.22}_{-0.46}$\\
\texttt{ SEOBNRv4HM (1g)} & $45.1^{+8.4}_{-7.7}$& $38.5^{+7.8}_{-6.3}$& $83.7^{+14.3}_{-12.0}$& $0.87^{+0.13}_{-0.14}$& $0.38^{+0.20}_{-0.19}$& -& $10.9^{+0.4}_{-0.4}$& $2.94^{+1.35}_{-1.43}$& $0.78^{+0.22}_{-0.47}$\\
\texttt{ SEOBNRv4HM (2g)} & $56.6^{+9.3}_{-9.9}$& $30.4^{+6.2}_{-7.7}$& $86.8^{+11.5}_{-11.0}$& $0.54^{+0.16}_{-0.16}$& $0.29^{+0.19}_{-0.18}$& -& $11.1^{+0.4}_{-0.4}$& $2.43^{+1.21}_{-1.04}$& $0.72^{+0.28}_{-0.42}$\\
\end{tabular}
\caption{Parameters of GW170729 obtained with various waveform models and the 1g and 2g priors. 
We quote median values and 90\% credible intervals for the primary mass, the secondary mass, the total mass, the effective spin $\chi_\mathrm{{eff}}$, and the effective precession parameter $\chi_\mathrm{{p}}$. 
For the mass ratio we quote the median value and the 90\% HPD. 
All masses are given in the source frame. 
The effective precession parameter $\chi_\mathrm{{p}}$ is absent in the spin-aligned models.
}
\label{table:par-table-2g}
\end{table*}
\end{landscape}

\section{Morphology-Independent analysis}
\label{BWresults}

\begin{figure*}[]
\centering
\includegraphics[width=0.7\textwidth,clip=true]{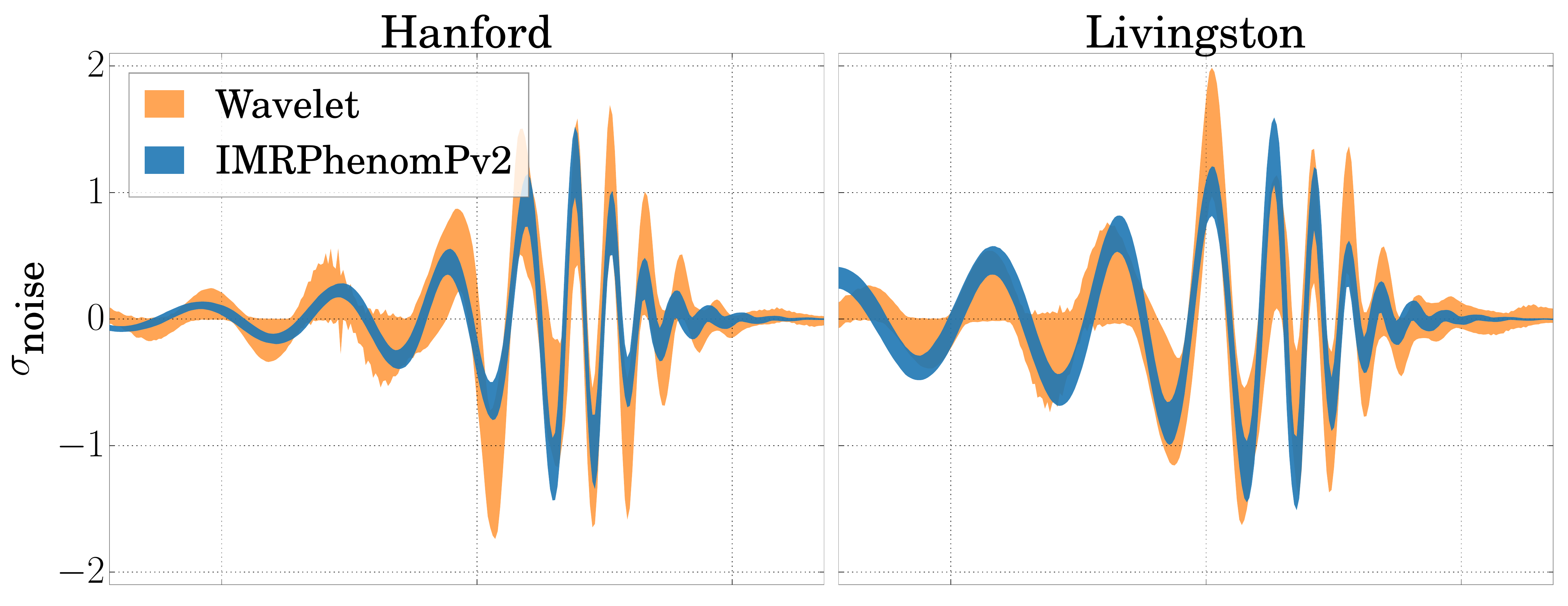}
\includegraphics[width=0.7\textwidth,clip=true]{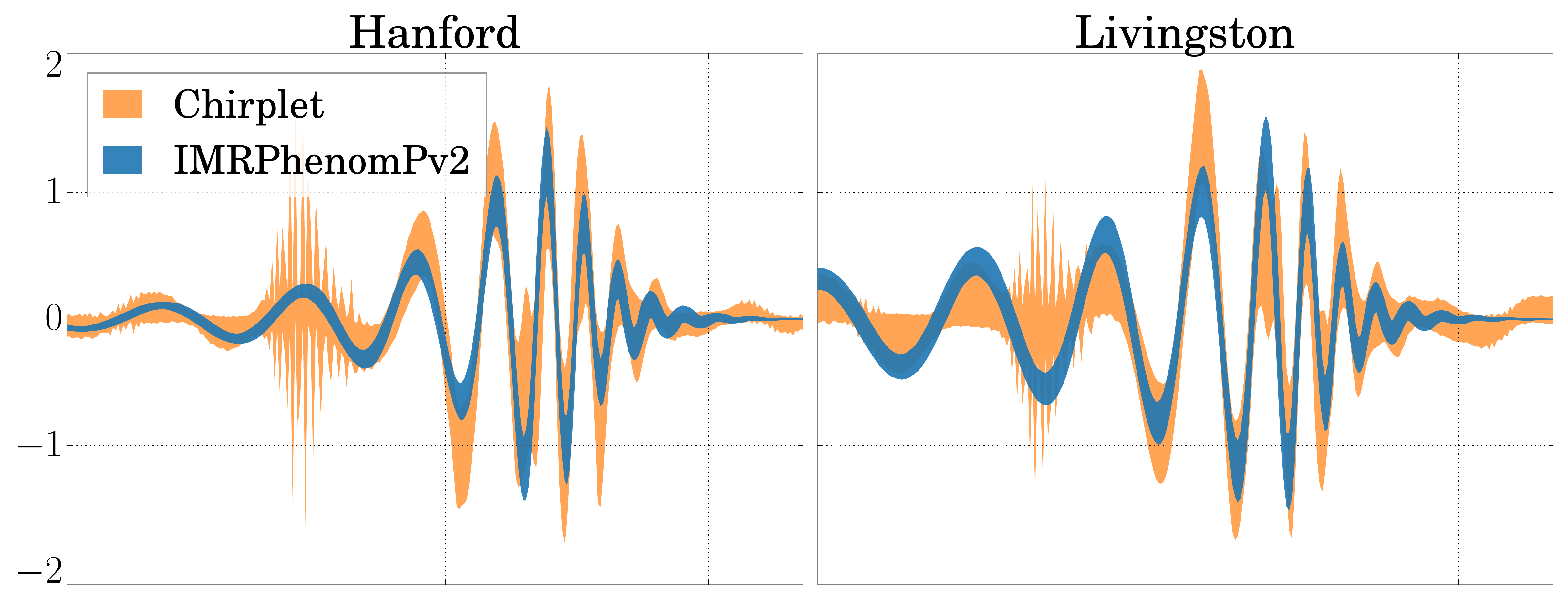}
\includegraphics[width=0.7\textwidth,clip=true]{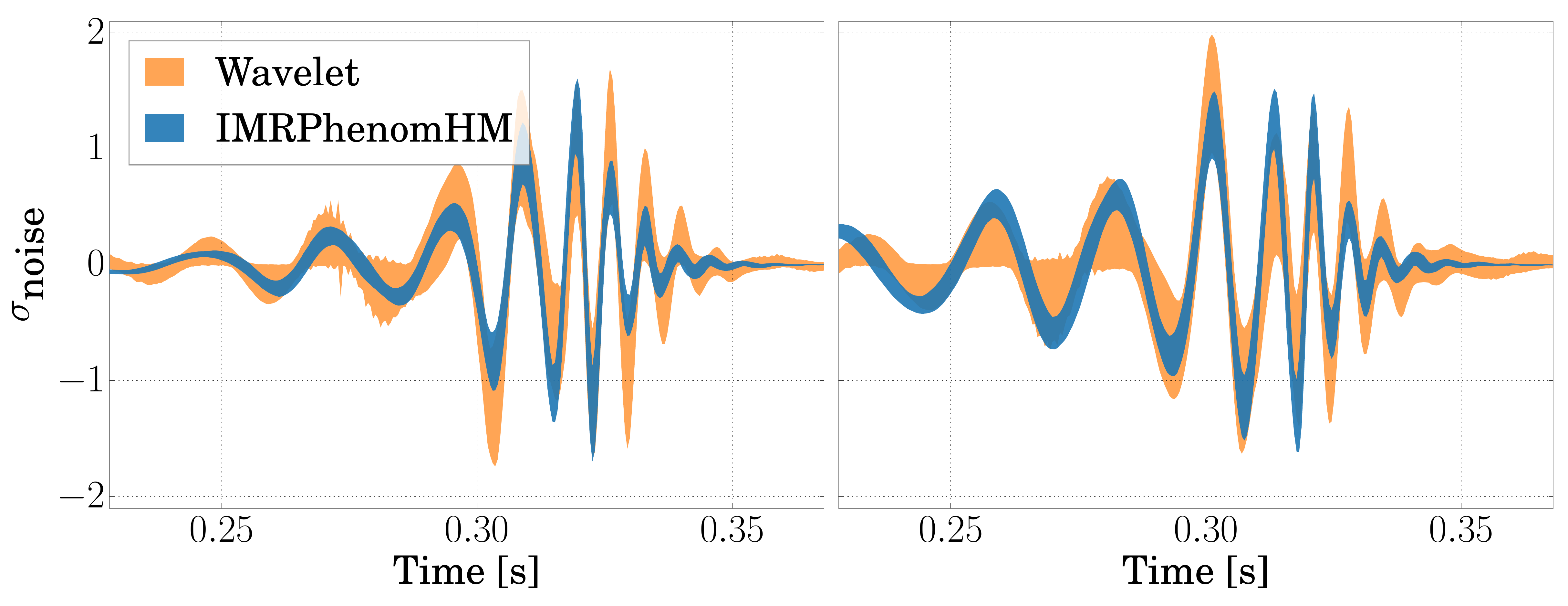}
\includegraphics[width=0.7\textwidth,clip=true]{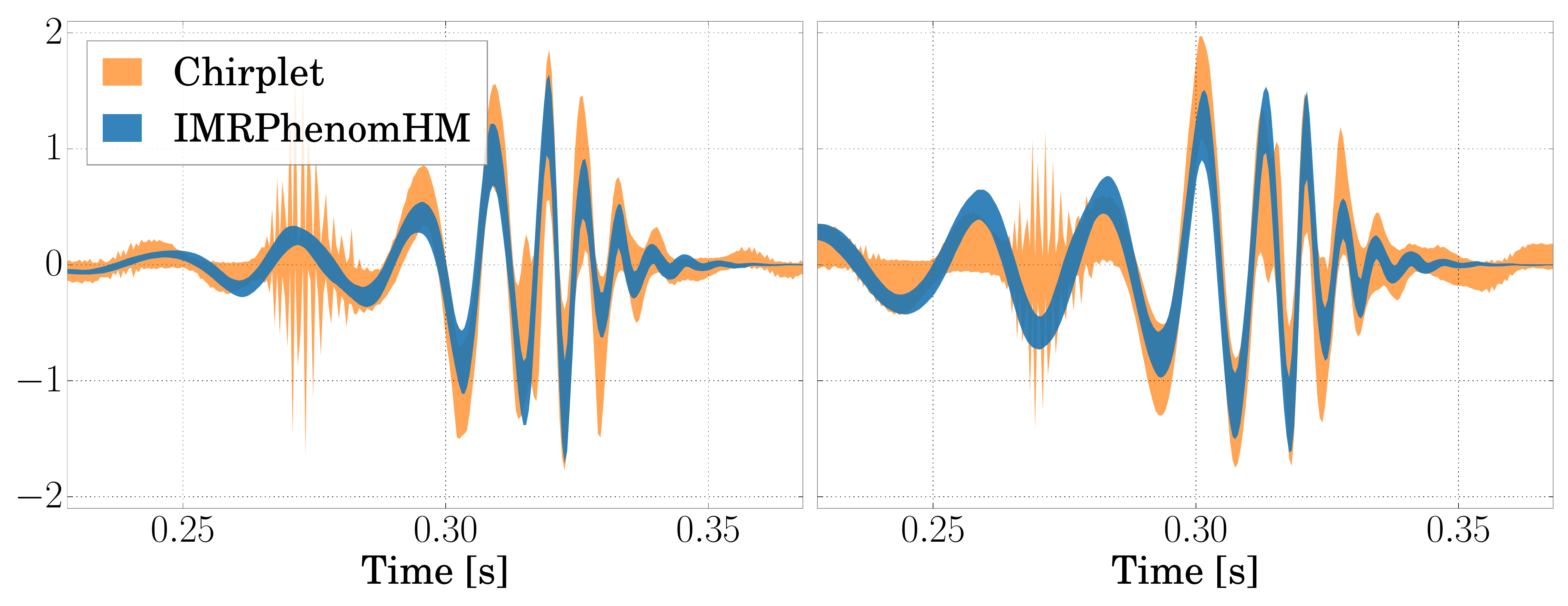}
\caption{Signal reconstruction comparison for GW170729. 
We plot the 90\% credible interval of the whitened strain data as a function of time for each of the LIGO detectors computed with CBC waveform models (blue) and \texttt{ BayesWave} (orange). 
The top plots show \texttt{ IMRPhenomPv2} while the bottom plots show \texttt{ IMRPhenomHM}. The left plots use the wavelet model of \texttt{ BayesWave}, while the right plots use the chirplet model. The $\textrm{x}$ axis represents the time in seconds from the nearest integer GPS time before the event. The $\textrm{y}$ axis represents the strain amplitude whitened using a filter which is the inverse Amplitude Spectral Density (ASD) of the noise in the detector. The units are in multiples of the standard deviation of the noise.
The generic signal reconstruction is consistent with the CBC signal reconstruction both when the latter includes higher-order modes and when it does not.}
\label{fig:bayeswave}
\end{figure*}

The studies presented in the previous section relied on specific waveform models for the signal emitted during a CBC as predicted by GR. 
We here follow a more generic approach and analyze GW170729 in a morphology-independent way that does not explicitly assume it is a CBC described by the currently available waveform models. 
We use \texttt{ BayesWave} to reconstruct the signal and then compare this reconstruction to the one obtained with CBC models. 
The comparison is shown in Fig.~\ref{fig:bayeswave}, where we plot the whitened strain as a function of time. 
At each time, the shaded region denotes the 90\% credible interval of the reconstruction using CBC waveform models (blue) and \texttt{ BayesWave} (orange). 
The left panel is obtained with the wavelets, while the right panel is obtained with chirplets. 
The top row is made with the CBC waveform model \texttt{ IMRPhenomPv2} which includes the effect of spin precession, but not higher-order modes. 
The bottom row uses \texttt{ IMRPhenomHM} which assumes that the spins remain aligned with the orbital angular momentum, but includes higher-order modes.

As discussed in~\cite{LIGOScientific:2018mvr}, \texttt{ BayesWave} can sometimes reconstruct features that are not present in the CBC reconstructions as can be seen in the right hand side panels of Fig.~\ref{fig:bayeswave}, for example around t=0.27s on the right panel. 
Unlike CBC model waveforms, wavelet-based models are not limited to a physically motivated waveform morphology. 
As a result, the \texttt{ BayesWave} sampler can sometimes pick up random coherence between nearby noise samples. 
However, these outlying wavelets do not point to any potential additional features in the waveform. 
In fact, they are absent in the 50\% credible intervals of the reconstruction, implying that they have a low significance. 
Similar outliers were observed in \texttt{ BayesWave} analyses applied to simulated signals added to real data.

We find broad agreement between the CBC reconstruction and the \texttt{ BayesWave} reconstruction in all cases.
In particular, the 90\% credible intervals obtained with the two methods overlap for all waveform models and \texttt{ BayesWave} basis functions. 
The agreement suggests that the omission of higher-order modes does not degrade the reconstruction enough to leave a coherent residual detectable 
by \texttt{ BayesWave}. 
This conclusion is in agreement with the results of the previous section, as well as the corresponding reconstruction plot in~\cite{LIGOScientific:2018mvr}.

\begin{figure*}[]
\centering
\includegraphics[width=0.7\textwidth,clip=true]{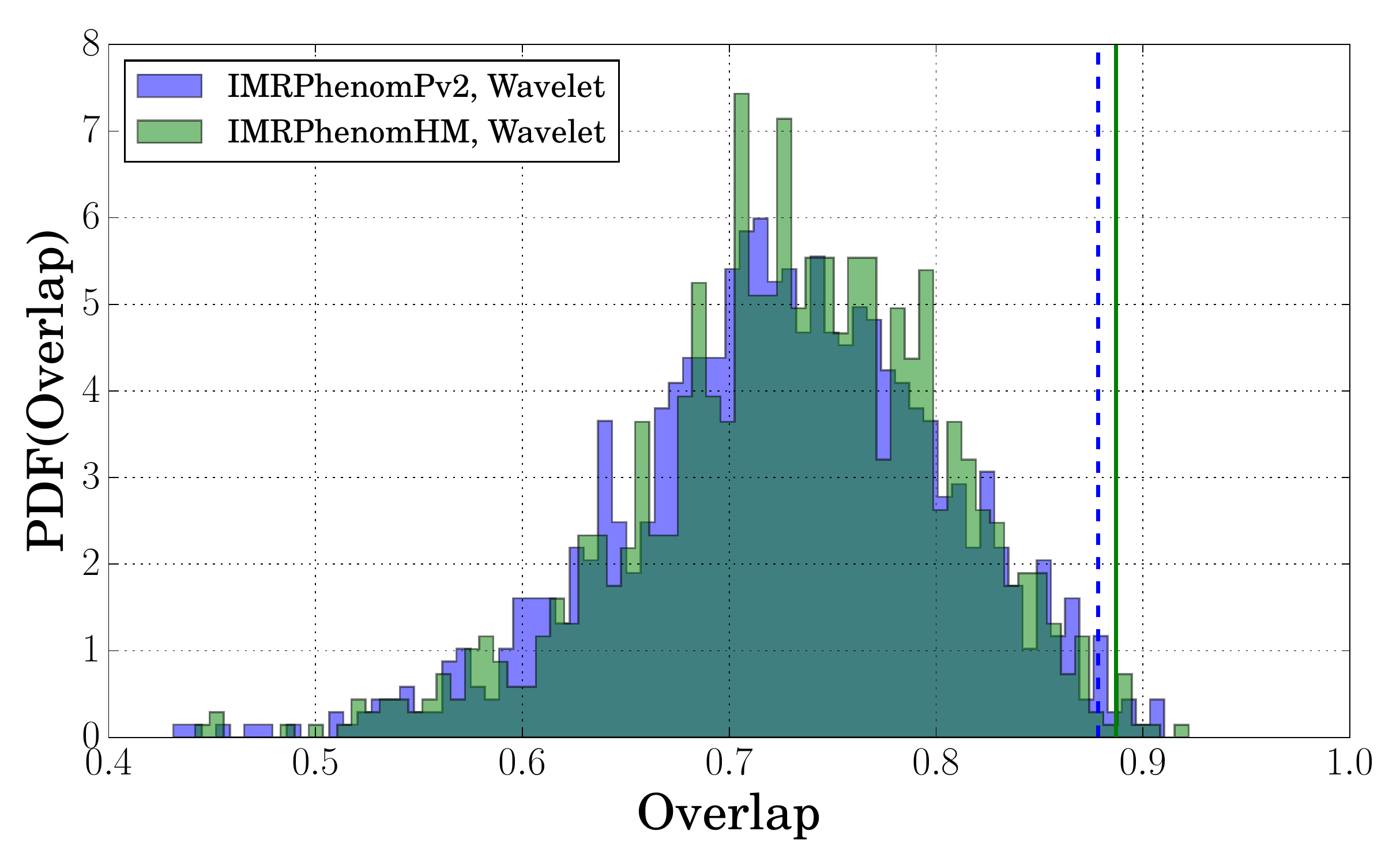}
\includegraphics[width=0.7\textwidth,clip=true]{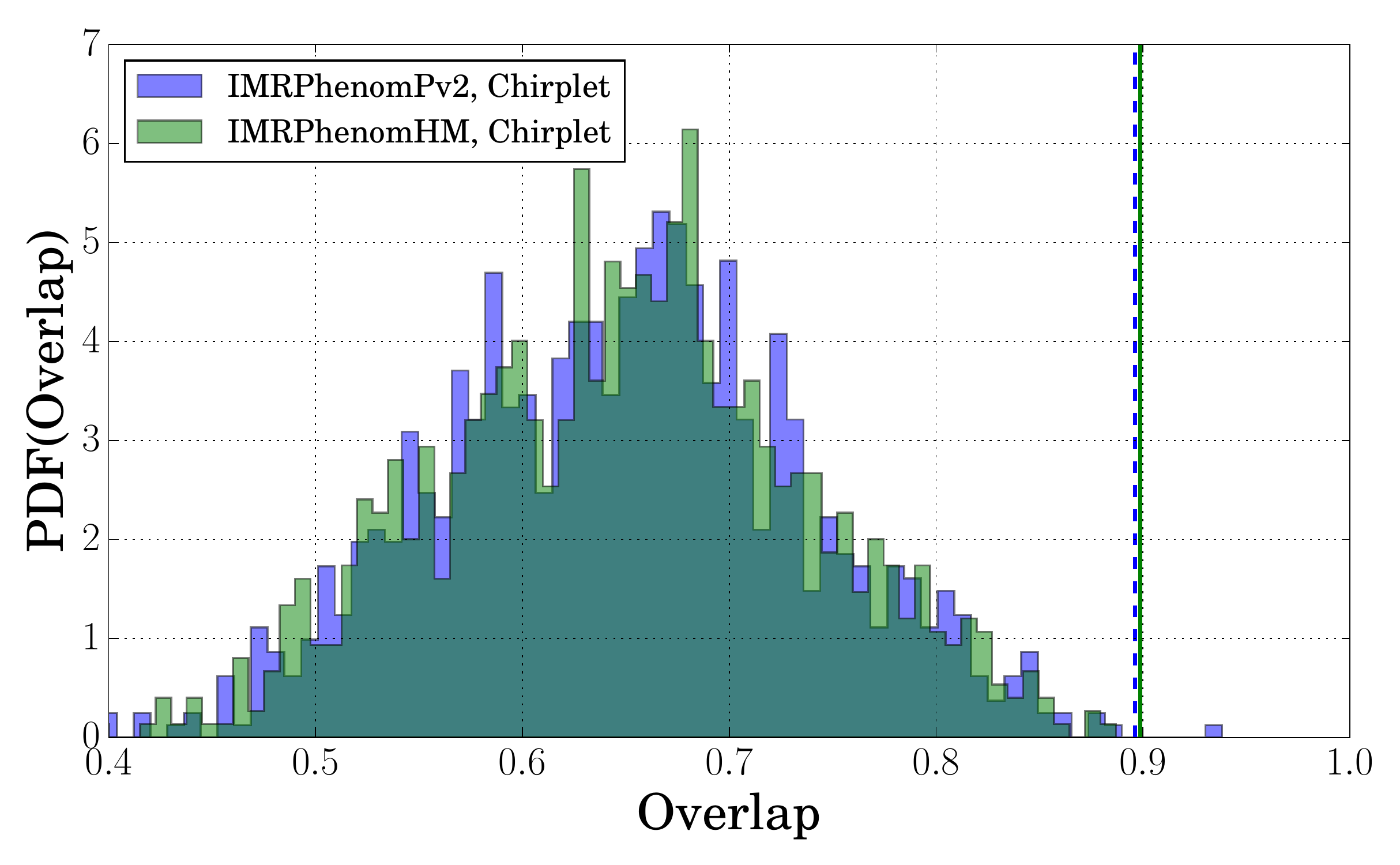}
\caption{Overlap between the \texttt{ BayesWave} and the CBC reconstruction for GW170729. We plot the overlap histogram between 1000 random \texttt{ BayesWave} waveform samples and the ML CBC waveform obtained with \texttt{ IMRPhenomPv2} (blue) and \texttt{ IMRPhenomHM} (green). 
The left panel uses \texttt{ BayesWave}'s wavelet model, while the right panel uses the chirplet model. 
The solid and dashed vertical lines represent the overlap of the MBW reconstruction with the ML CBC waveform, with the blue dashed and solid green lines representing the overlaps computed using \texttt{ IMRPhenomPv2} and \texttt{ IMRPhenomHM} respectively. 
As is described in the text, we expect these overlaps to be higher than those computed using individual \texttt{ BayesWave} samples.}
\label{fig:overlap}
\end{figure*}

In order to make this statement quantitative, we draw 1000 samples from the \texttt{ BayesWave} posterior and compute their overlap with the maximum likelihood (ML) CBC waveform from the analyses using \texttt{ IMRPhenomPv2} and \texttt{ IMRPhenomHM}.
The posterior distribution for the overlap is given in Fig.~\ref{fig:overlap} for wavelets (left) and chirplets (right). 
We find overlaps typically between $0.6-0.9$. This large spread in the overlaps is a result of the inherent flexibility in wavelet-based analyses causing a large variance in the reconstructed signal morphology. 
Therefore, unlike CBC waveform samples which are motivated by a physical theory, a single \texttt{ BayesWave} sample is not constrained by physical reconstruction considerations, other than propagation at the speed of light. 
Instead, the median \texttt{ BayesWave} waveform (MBW), defined as being the median across the sample waveforms at each time or frequency step, represents a collective estimate across samples, assumed to represent ``the wisdom of the crowd". 
The MBW is a stable estimate of the reconstruction as it is relatively immune to the stochastic fluctuations in the variable dimensional sampler. 
Each of the four vertical lines in Fig.~\ref{fig:overlap} represent the overlap values between the MBW and the ML CBC waveform. 
They are summarized in Table~\ref{table:overlaps} where we also show results obtained using \texttt{ SEOBNR} waveform models. 
The ML CBC waveform has in general a higher agreement with the MBW waveform than with each of the individual samples.

\begin{table}[]
\centering
  \begin{tabular}{l |c c}
    \hline 
       &Wavelets&Chirplets\\ \hline 
       \texttt{ IMRPhenomPv2}&$0.88$&$0.90$\\
       \texttt{ IMRPhenomD}&$0.87$&$0.89$\\
       \texttt{ IMRPhenomHM}&$0.89$&$0.90$\\
       \texttt{ SEOBNRv4}&$0.87$&$0.90$\\       
       \texttt{ SEOBNRv4HM}&$0.88$&$0.89$
   \end{tabular}
   \caption{Overlaps between the median \texttt{ BayesWave} reconstruction and the maximum likelihood CBC waveform with different waveform models.}
   \label{table:overlaps}
\end{table}

We find that waveforms both with and without higher-order modes achieve large overlaps with the MBW reconstruction, in the range of $0.87-0.9$, consistent with expectations. 
In fact, Ref.~\cite{TheLIGOScientific:2016uux} showed that for masses and SNRs typical of GW170729 the expected overlap between simulated signals and their median reconstructions is in the $0.85-0.9$ range at the 1-$\sigma$ level, similar to what we obtain here. 
The small remaining disagreement between the ML CBC reconstruction and the MBW reconstruction is due to the fact that, unlike modeled analyses, \texttt{ BayesWave} is only sensitive to excess signal that stands out and above the detector noise. 
This means that it is less sensitive than CBC analyses in the lower frequencies, $<40$ Hz.
We find that if we increase the low frequency cut off in the overlap calculation to $40$ Hz, the overlaps improve by $\sim 0.07$ for each pair of \texttt{ BayesWave} and CBC waveforms.

Besides the good reconstruction, we find that waveforms with higher-order modes lead to similar overlaps with the \texttt{ BayesWave} reconstruction to waveforms without higher-order modes. 
We perform the Kolmogorov--Smirnov test for the two overlap distributions for each panel of Fig.~\ref{fig:overlap} and find only 0.048 and 0.017, which implies that both \texttt{ IMRPhenomPv2} and \texttt{ IMRPhenomHM} reconstruct the data comparably well. This confirms that GW170729 is consistent with a CBC and that the higher-order modes are not strong enough to lead to a degradation of the signal reconstruction if neglected.

\section{Conclusions}
\label{conclusions}

We analyze the publicly available strain data for GW170729, the highest mass and most distant confirmed GW detection by the LIGO and Virgo detectors. 
In particular we investigate the effect of higher-order modes and spin priors on the inference of the source parameters. 
We find that higher-order modes leave small but noticeable effects, while spin priors affect the spin measurements as anticipated.

We find that the inclusion of higher-order modes in the models leads to changes in the estimates for the mass ratio, the effective spin, and the SNR. 
Our updated parameter measurements imply decreased support for equal binary component masses and nonzero effective spin. 
In particular we conclude that the mass ratio is $(0.3-0.8)$ at the 90\% credible level, a value that excludes equal masses. 
We also find that the 90\% credible interval for the effective spin parameter has changed from $(0.11- 0.58)$ as reported in~\cite{LIGOScientific:2018mvr} to $(-0.01-0.50)$, which now marginally includes zero. 
The effective spin parameter still has a 94\% probability of being positive.

Consistent with these findings, we compute the BF in favor of the presence of higher-order modes, and find it to be 5.1:1. 
Moreover, their omission does not dramatically change the measured parameters, which would happen if they were strong \cite{Varma:2016dnf,Bustillo:2015qty,Graff:2015bba}. 
This conclusion is also consistent with the fact that current matched-filter searches for CBCs have a reduced efficiency toward signals with strong higher harmonics \cite{Harry:2017weg,Capano:2013raa} and this event was indeed observed in both the GstLAL and PyCBC searches.

We argue that the observed changes in parameter measurements are not due to systematic errors in the CBC waveform models. 
We compare results obtained with different waveform models with and without higher-order modes, 
including waveforms computed with NR. 
That leads us to believe that both the increased support for unequal masses and the decreased support for nonzero spin are robust conclusions.

We emphasize that the fact that the evidence for higher-order modes is weak does not contradict the fact that waveforms with higher-order modes lead
to improved parameter measurements. In fact, accurate modeling of relevant physical effects can improve parameter measurements.
This can be because said physical effect is present even if it is too weak to unequivocally detect, or because it helps exclude regions of the parameter space for which that effect
would be larger than what we observe. Similar shifts in the posteriors (though in the opposite direction for the distance and inclination)
where in fact observed in~\cite{Kumar:2018hml} when reanalyzing GW150914 with waveforms that include higher-order modes.

We augment the analysis using CBC waveform models with a morphology-independent analysis using \texttt{ BayesWave}. 
We find broad agreement between the CBC analysis and the generic analyses regardless of whether the CBC model uses higher-order modes or not. 
We quantify this conclusion in terms of the overlap between the CBC and the generic reconstruction, which we find to be $\sim 0.9$, consistent with expectations for signals of this mass and SNR~\cite{TheLIGOScientific:2016uux}.

Posterior samples from all analyses are available at~\cite{170729SamplesRelease}.

\section*{Acknowledgements}
 
We thank Christopher Berry, Thomas Dent, Tristano DiGirolamo, Bhooshan Gadre, Roland Haas, Nathan Johnson-McDaniel, Riccardo Sturani, and Aaron Zimmerman for helpful comments and suggestions.
This research has made use of data, software and/or web tools obtained from the Gravitational Wave Open Science Center (https://www.gw-openscience.org), a service of LIGO Laboratory, the LIGO Scientific Collaboration and the Virgo Collaboration. LIGO is funded by the U.S. National Science Foundation. Virgo is funded by the French Centre National de Recherche Scientifique (CNRS), the Italian Istituto Nazionale della Fisica Nucleare (INFN) and the Dutch Nikhef, with contributions by Polish and Hungarian institutes.
The Flatiron Institute is supported by the Simons Foundation.
JCB acknowledges support from Australian Research Council Discovery Project DP180103155.
CJH, KKYN and SV acknowledges the support of the National Science Foundation and the LIGO Laboratory.
LIGO was constructed by the California Institute of Technology and Massachusetts Institute of 
Technology with funding from the National Science Foundation and operates under cooperative agreement PHY-1764464.
MH was supported by Science and Technology Facilities Council (STFC) grant ST/L000962/1 and European Research Council Consolidator Grant 647839.
Parts of this research were conducted by the Australian Research Council Centre of Excellence for Gravitational Wave Discovery (OzGrav), through project number CE170100004.
ROS and JL are supported by National Science Foundation (NSF) PHY-1707965 and PHY-1607520.
PS acknowledges support from the NWO Veni grant no. 680-47-460 and the Science and Technology Facilities Council grant ST/N000633/1.
MH acknowledges support from the Swiss National Science Foundation (SNSF) Grant IZCOZ0\_177057.
LS was supported by the Young Elite Scientists Sponsorship Program by the China Association for Science and Technology (2018QNRC001), and partially supported by the National Science Foundation of China (11721303), and XDB23010200.
The GT authors gratefully acknowledge the NSF for financial support from Grants No. PHY 1806580, PHY 1809572, and TG-PHY120016. Computational resources were provided by XSEDE and the Georgia Tech Cygnus Cluster.
The RIT authors gratefully acknowledge the NSF for financial support from Grants No. PHY-1607520,
No. PHY-1707946, No. ACI-1550436, No. AST-1516150,
No. ACI-1516125, No. PHY-1726215.  This work used the Extreme
Science and Engineering Discovery Environment (XSEDE) [allocation
  TG-PHY060027N], which is supported by NSF grant No. ACI-1548562.
Computational resources were also provided by the NewHorizons, BlueSky
Clusters, and Green Prairies at the Rochester Institute of Technology,
which were supported by NSF grants No.\ PHY-0722703, No.\ DMS-0820923,
No.\ AST-1028087, No.\ PHY-1229173, and No.\ PHY-1726215.
The SXS authors at Caltech acknowledge the Sherman Fairchild Foundation, and NSF grants
PHY-1708212 and PHY-1708213.
GL, NA, and AG are supported by NSF PHY-1606522, PHY-1654359, PHY-1654359. Computations were done using the orca cluster supported in part by PHY-1429873 and by Cal State Fullerton. AG is also supported in part by Nancy Goodhue-McWilliams.
The authors are grateful for computational resources provided by the LIGO Laboratory and supported by National Science Foundation Grants PHY-0757058 and PHY-0823459. 
Some of the computational work for this manuscript was also carried out on the computer cluster \texttt{ Vulcan} at the Max Planck Institute for Gravitational Physics in Potsdam.  
Plots have been made with \texttt{ matplotlib}~\cite{Hunter:2007} and \texttt{ corner}~\cite{corner}.

\chapter{Conclusions and future work}
\label{chap:six}
The work I have presented in this thesis aimed at improving the models for the \ac{GW} signals emitted by coalescing \acp{BBH} with spins. The accuracy of these waveform models is crucial, because they are key components for both the detection and the source characterization of \ac{GW} signals. For this reason, and considering that \ac{GW} detectors are costantly improving their sensitivity, it is important that also the accuracy of waveform models progresses in tandem.

This thesis focuses on the effect of \acp{HM} in the waveforms emitted by coalescing \acp{BBH} with spins. Their contribution is especially relevant when the total mass of the system is large, there is a big difference between the value of the masses of the two \acp{BH}, or the binary system is observed from a direction orthogonal to its \roberto{orbital } angular momentum \roberto{(edge-on)}.

In Chapters 2 and 3, I developed, within the \ac{EOB} formalism, inspiral-merger-ringdown waveform models for spinning \acp{BBH} which include \acp{HM}. The waveform model described in Chapter 2, \texttt{SEOBNRv4HM}, is limited to \acp{BH} with spins aligned with the \roberto{orbital } angular momentum of the binary, and served as a foundation for \texttt{SEOBNRv4PHM}, its generalization to \acp{BH} with generic spin orientations, described in Chapter 3.
I tested the accuracy of the waveforms generated with these models by computing the unfaithfulness against about $1700$ \ac{NR} simulations ($\sim 200$ for the first model and $\sim 1500$ for the second one). In the total mass range $20-200\,M_\odot$, \texttt{SEOBNRv4HM} yields average unfaithfulness values smaller than $1\%$ for almost all cases, with few exceptions for which the value is slightly above $1\%$. \texttt{SEOBNRv4PHM} is also very accurate and, for $94\%$ ($57\%$) of the cases, its unfaithfulness against \ac{NR} waveforms is below $3\%$ ($1\%$) in the same mass range, \roberto{even without being calibrated to \ac{NR} waveforms in the precessing sector. } The great accuracy reached by these models is also confirmed by the parameter estimation studies I performed, using them to analyze synthetic \ac{GW} signals generated with \ac{NR} waveforms. Both models managed to measure the most relevant binary parameters, such as masses and spins, with negligible bias compared to the statistical uncertainty, in binary systems with asymmetric masses, strong precessional effects (in the case of \texttt{SEOBNRv4PHM}) and moderately large values of the \ac{SNR}. \texttt{SEOBNRv4PHM} was even able to recover with trivial bias the parameters of a synthetic \ac{BBH} system for a less extreme binary configuration, but at an \ac{SNR} of $50$.

The number of observed \ac{GW} signals steadily increased in the past years \roberto{up to 46 confirmed detections of \acp{BBH}, } and it is expected to rise even more in the next decade, as consequence of the upgrades in the \ac{GW} detectors. Consequently, waveform models have to be not only accurate, but also fast to analyze the increasing number of detected signals on reasonable timescales. For this reason, in Chapter 3 I developed \texttt{SEOBNRv4HM\_ROM}, a \ac{ROM} version of \texttt{SEOBNRv4HM}, which accelerate the generation of the waveforms by two orders of magnitude, with negligible loss in the accuracy.

The models I developed in this thesis have been already used in the analysis of \ac{GW} signals detected by the LIGO and Virgo interferometers. In Chapter 4, I presented an investigation on the \ac{GW} signal GW170729 performed with waveform models with \acp{HM}, including \texttt{SEOBNRv4HM}. This signal was particularly interesting to study with these waveform models, because it was likely originating from the most massive \ac{BBH} system observed during O2. When using \texttt{SEOBNRv4HM} to analyze this signal, I obtained improved measurements of the mass ratio and the effective-spin parameter. In particular, the mass ratio measurement allowed to exclude with large confidence that the signal was emitted from a \ac{BBH} with similar masses. In fact, the study with \texttt{SEOBNRv4HM} revealed that with $40\%$ probability the signal was emitted by a \ac{BBH} where the mass of the most massive \ac{BH} in the binary was larger than the mass of the other \ac{BH} by more than a factor of two. This probability was only about $20\%$ when the properties of this signal were measured with waveform models without \acp{HM}. 

As a member of the LIGO Scientific Collaboration, I led the parameter estimation  analysis of the \ac{GW} signal GW190412~\cite{LIGOScientific:2020stg} observed during O3a. This signal was interesting to study for the asymmetric masses of its source. In Chapter 1, I summarize the analysis I performed on this system with the waveform models described in the thesis. This investigation confirmed, with large confidence, that the signal was emitted by a \ac{BBH} with very asymmetric masses. In fact, the median value of $q$ for this system was between $3$ and $5$, and $q \geq 2$ with a probability of $99\%$. This precise measurement of the mass ratio allowed to obtain also a tight constraint on the dimensionless spin magnitude of the most massive \ac{BH} in the binary, which value is between $0.22$ and $0.60$ at $90 \%$ probability. The luminosity distance was also very precisely measured in this system because the effect of \acp{HM} and spin precession were present in the waveform models used for the analysis. In fact, the bounds on the luminosity distance was $\sim 60\%$ tighter compared to those obtained by waveform models without these effects. A precise measurement of the luminosity distance can be very useful when trying to measure the Hubble constant $H_0$ using \ac{GW} signals~\cite{Schutz:1986gp,Abbott:2019yzh,Soares-Santos:2019irc,Fishbach:2018gjp,Abbott:2017xzu}, and employing waveform models with \acp{HM} has been proven to be useful for this purpose~\cite{Borhanian:2020vyr}.

Furthermore, the waveform models described in this thesis have been also used to analyze all the \ac{GW} signals emitted by \acp{BBH} systems detected during the O3a~\cite{Abbott:2020niy,Abbott:2020tfl,Abbott:2020jks}. Employing these waveform models for this purpose is very important, because their precise measurement of the binary parameters allows a better understanding of the observed population of \acp{BBH}, and their formation mechanism. \roberto{In particular, using these accurate waveform models allowed us to determine more precisely the mass ratio distribution of the observed \ac{BBH} population~\cite{Abbott:2020mjq}. } In addition, these waveform models were also employed in the study of the signal GW190814~\cite{Abbott:2020khf}. In this case, the nature of the lighter compact object in the binary is unclear. In fact, its measured mass seems to be too large for a neutron star and too light for a \ac{BH}. The peculiar nature of the source of this signal was only revealed thanks to the precise measurement of the masses obtained with the waveform models described in this thesis. \roberto{Finally, these improved waveform models were also used as baseline for some tests of \ac{GR}, performed using \acp{BBH} as laboratories, on the \ac{GW} signals detected during O3a~\cite{Abbott:2020jks}. }

There are at least two orthogonal future directions that the work of this thesis could follow. The first one involves the development of even better waveform models. A possible path for improvements is to overcome the limitation of the current models to \ac{BBH} systems in quasi-circular orbit. While there are good theoretical arguments in favour of the hypothesis that the majority of these systems feature a negligible eccentricity close to the merger, using a waveform models that allows for \acp{BBH} in generic eccentric orbits would be useful to experimentally test this hypothesis. The same is true for the \ac{BH} electric charge. 

The other possible direction for the future, is to take advantage of the improved accuracy of the waveform models developed in this thesis for some practical applications. For example, one could exploit the more precise determination of the luminosity distance, obtained with waveform models with \acp{HM}, in the context of the measurement of the Hubble constant $H_0$ in binary systems composed by a neutron star and a \ac{BH}. In fact, in this case the difference between the masses of the two objects is expected to be large, and the contribution of the \acp{HM} should therefore be important. \roberto{I am currently undertaking this study. }

\begin{appendices}

\chapter{Explicit expressions of higher-order \textit{factorized} modes}
\label{app:modes}

Here we list expressions needed to build the $h_{\ell m}$'s of the \texttt{SEOBNRv4HM} model.

The functions $n_{\ell m}^{(\epsilon)}$ and $c_{\ell + \epsilon}(\nu)$ used in Eq.~\eqref{eq:Newtonian} are defined as (see Ref.~\cite{Taracchini:2012ig}):
\begin{align}
n_{\ell m}^{(0)} &= (i m)^\ell \frac{8\pi}{(2\ell + 1)!!} \sqrt{\frac{(\ell + 1)(\ell + 2)}{\ell (\ell -1)}},\\
n_{\ell m}^{(0)} &= -(i m)^\ell \frac{16\pi i}{(2\ell +1)!!}\sqrt{\frac{(2\ell + 1)(\ell + 2)(\ell^2 - m^2)}{(2\ell - 1)(\ell + 1)\ell (\ell -1)}},
\end{align}
and
\begin{equation}
\label{eq:clm}
c_{\ell + \epsilon}(\nu) = \left(\frac{1}{2} - \frac{1}{2}\sqrt{1-4\nu}\right)^{\ell + \epsilon -1} + (-1)^{\ell + \epsilon}\left(\frac{1}{2} + \frac{1}{2}\sqrt{1-4\nu} \right)^{\ell + \epsilon -1}.
\end{equation}
We define also the function
\begin{equation}
\text{eulerlog}\left(m,v_{\Omega }\right)\equiv \gamma + \log(2mv_{\Omega}),
\end{equation}
which is used in the expression of the factorized modes. Here $\gamma$ is the Euler constant.

The quantity $f_{\ell m}$ in Eq. (\ref{eq:hlm_factorized}) is:
\begin{equation}
f_{\ell m} = \begin{cases}
\rho_{\ell m}^\ell, \qquad \qquad \quad \,\,\, \ell \,\, \text{is even}, \\
(\rho_{\ell m}^{\mathrm{NS}})^\ell + f_{\ell m}^{\mathrm{S}}, \qquad \ell \,\, \text{is odd}.
\end{cases}
\end{equation}
The functions $\rho_{\ell m}$, $\rho_{\ell m}^{\mathrm{NS}}$, $ f_{\ell
  m}^{\mathrm{S}}$ are defined below; the superscript ``NS'' stands for nonspinning,  
and the superscript ``S'' indicates spinning. Below, we also list the phase terms $\delta_{\ell
  m}$. 

The quantities $f_{\ell m}$ and $\delta_{\ell m}$ for the \texttt{SEOBNRv4HM} model are mostly taken from the \texttt{SEOBNRv4} 
model in Ref. \cite{Bohe:2016gbl} with the additions of several new terms:
\begin{itemize}
\item 3PN nonspinning terms in $\rho_{33}^{\mathrm{NS}}$ from Ref. \cite{Faye:2014fra};
\item 5PN test-mass, nonspinning terms in $\rho_{33}^{\mathrm{NS}}$ from Ref. \cite{Fujita:2012cm};
\item 5PN test-mass, nonspinning terms in $\rho_{21}^{\mathrm{NS}}$ from Ref. \cite{Fujita:2012cm};
\item 2PN and 2.5PN spinning terms in $\rho_{44}$ from Ref.~\cite{Marsatetal2017}; 
\item 3PN, 4PN and 5PN test-mass, nonspinning terms in $\rho_{55}^{\mathrm{NS}}$ from Ref. \cite{Fujita:2012cm};
\item 2PN, 2.5PN and 3PN spinning terms in $f_{33}^{\mathrm{S}}$ from Ref.~\cite{Marsatetal2017};
\item 2PN, 2.5PN and 3PN spinning terms in $f_{21}^{\mathrm{S}}$ from Ref.~\cite{Marsatetal2017}; 
\item 1.5PN and 2PN spinning terms in $f_{55}^{\mathrm{S}}$ from Ref.~\cite{Marsatetal2017};
\item  3PN and 4.5PN test-mass, nonspinning terms in $\delta_{55}$ from Ref. \cite{Fujita:2012cm}.
\end{itemize}
Furthermore, we find that resummations of the $f_{\ell m}$ function
for the $(3,3), (2,1),(4,4),(5,5)$ modes of the kind proposed in
Refs.~\cite{Nagar:2016ayt,Messina:2018ghh} (see Eq. (47) and Eq. (48)
in the latter) do not always improve the agreement with the NR
waveforms of our catalog. For this reason we decide not to implement
those resummations when building the \texttt{SEOBNRv4HM}
model. It is worth to mention that whereas in our model the resummed
  expressions are computed as a function of $v_\Omega =
  (M\Omega)^{1/3}$, in Refs.~\cite{Nagar:2016ayt,Messina:2018ghh} they
  are expressed as a function of $v_\phi$ defined in Eq.~(69) of
  Ref.~\cite{Damour:2014sva}. While the two variables are very similar
  at low frequency, they can  differ toward merger where the aforementioned resummation 
may be more effective.
\begin{landscape}
\begin{align}
\rho_{33}^{\mathrm{NS}} =  &1+\left(-\frac{7}{6}+\frac{2 \nu }{3}\right) v_{\Omega
   }^2+\left(-\frac{6719}{3960}-\frac{1861 \nu }{990}+\frac{149 \nu ^2}{330}\right)
   v_{\Omega
   }^4 \nonumber \\ 
   &+\left[\frac{3203101567}{227026800}+\left(-\frac{129509}{25740}+\frac{41 \pi
   ^2}{192}\right) \nu -\frac{274621 \nu ^2}{154440}+\frac{12011 \nu
   ^3}{46332}-\frac{26}{7} \text{eulerlog}\left(3,v_{\Omega }\right)\right] v_{\Omega
   }^6 \nonumber \\
   &+\left(-\frac{57566572157}{8562153600}+\frac{13}{3}
   \text{eulerlog}\left(3,v_{\Omega }\right)\right) v_{\Omega
   }^8+\left(-\frac{903823148417327}{30566888352000}+\frac{87347
   \text{eulerlog}\left(3,v_{\Omega }\right)}{13860}\right) v_{\Omega }^{10}\,, \\
 \rho_{21}^{\mathrm{NS}} =&1+\left(-\frac{59}{56}+\frac{23 \nu }{84}\right) v_{\Omega
   }^2+\left(-\frac{47009}{56448}-\frac{10993 \nu }{14112}+\frac{617 \nu
   ^2}{4704}\right) v_{\Omega }^4+\left(\frac{7613184941}{2607897600}-\frac{107}{105}
   \text{eulerlog}\left(1,v_{\Omega }\right)\right) v_{\Omega
   }^6 \nonumber \\
   &+\left(-\frac{1168617463883}{911303737344}+\frac{6313
   \text{eulerlog}\left(1,v_{\Omega }\right)}{5880}\right) v_{\Omega
   }^8+\frac{\left(-63735873771463+14061362165760 \text{eulerlog}\left(1,v_{\Omega
   }\right)\right) v_{\Omega }^{10}}{16569158860800}\,, \\
  \rho_{44} =& 1+\left(\frac{1614-5870 \nu +2625 \nu ^2}{1320 (-1+3 \nu )}\right) v_{\Omega }^2+ \Bigg[\left(\frac{2}{3 (-1+3 \nu )}-\frac{41 \nu }{15 (-1+3 \nu )}+\frac{14 \nu ^2}{5 (-1+3 \nu )}\right) \chi
   _S \nonumber \\
   &+\delta m \left(\frac{2}{3 (-1+3 \nu )}-\frac{13 \nu }{5 (-1+3 \nu )}\right) \chi _A\Bigg] v_{\Omega }^3 +\Bigg[-\frac{14210377}{8808800 (1-3 \nu )^2}+\frac{32485357 \nu
   }{4404400 (1-3 \nu )^2}-\frac{1401149 \nu ^2}{1415700 (1-3 \nu )^2} \nonumber \\
   &-\frac{801565 \nu ^3}{37752 (1-3 \nu )^2}+\frac{3976393 \nu ^4}{1006720 (1-3 \nu )^2}+\frac{\chi _A^2}{2}-2 \nu  \chi _A^2+\delta m \chi _A \chi _S+\frac{\chi _S^2}{2}\Bigg] v_{\Omega }^4 \nonumber \\
   &+ \Bigg[\left(-\frac{69}{55
   (1-3 \nu )^2}+\frac{16571 \nu }{1650 (1-3 \nu )^2}-\frac{2673 \nu ^2}{100 (1-3 \nu )^2}+\frac{8539 \nu ^3}{440 (1-3 \nu )^2}+\frac{591 \nu ^4}{44 (1-3 \nu )^2}\right) \chi _S\nonumber \\
   &+\delta m
   \left(-\frac{69}{55 (1-3 \nu )^2}+\frac{10679 \nu }{1650 (1-3 \nu )^2}-\frac{1933 \nu ^2}{220 (1-3 \nu )^2}+\frac{597 \nu ^3}{440 (1-3 \nu )^2}\right) \chi
   _A\Bigg] v_{\Omega }^5 \nonumber \\
   &+\left(\frac{16600939332793}{1098809712000}-\frac{12568 \text{eulerlog}\left(4,v_{\Omega }\right)}{3465}\right) v_{\Omega
   }^6 +\left(-\frac{172066910136202271}{19426955708160000}+\frac{845198 \text{eulerlog}\left(4,v_{\Omega }\right)}{190575}\right) v_{\Omega
   }^8\nonumber \\
   &+\left(-\frac{17154485653213713419357}{568432724020761600000}+\frac{22324502267 \text{eulerlog}\left(4,v_{\Omega }\right)}{3815311500}\right) v_{\Omega }^{10}\,, \\
   \rho_{55}^{\mathrm{NS}} =& 1+\left(\frac{487}{390 (-1+2 \nu )}-\frac{649 \nu }{195 (-1+2 \nu )}+\frac{256 \nu ^2}{195 (-1+2 \nu )}\right) v_{\Omega }^2-\frac{3353747 v_{\Omega
   }^4}{2129400} \nonumber \\
   &+\left(\frac{190606537999247}{11957879934000}-\frac{1546}{429} \text{eulerlog}\left(5,v_{\Omega }\right)\right) v_{\Omega
   }^6+\left(-\frac{1213641959949291437}{118143853747920000}+\frac{376451 \text{eulerlog}\left(5,v_{\Omega }\right)}{83655}\right) v_{\Omega
   }^8\nonumber \\
   &+\left(-\frac{150082616449726042201261}{4837990810977324000000}+\frac{2592446431 \text{eulerlog}\left(5,v_{\Omega }\right)}{456756300}\right) v_{\Omega }^{10}\,, \\
f_{33}^{\mathrm{S}} = &\left[\left(-2+\frac{19 \nu }{2}\right) \frac{\chi _A}{\delta m}+\left(-2+\frac{5 \nu }{2}\right) \chi _S\right] v_{\Omega }^3 + \left[\left(\frac{3}{2}-6 \nu\right)\chi
   _A^2+(3-12 \nu ) \frac{\chi _A}{\delta m} \chi _S+\frac{3 \chi
   _S^2}{2}\right] v_{\Omega }^4 \nonumber \\
   &+
   \left[\left(\frac{2}{3}-\frac{593 \nu }{60}+\frac{407 \nu ^2}{30}\right)\frac{\chi
   _A}{\delta m}+\left(\frac{2}{3}+\frac{11 \nu }{20}+\frac{241 \nu
   ^2}{30}\right) \chi _S\right]v_{\Omega }^5 \nonumber \\
   &+ \left[\left(-\frac{7}{4} + \frac{11\nu}{2} -12\nu^2 \right) \chi _A^2+\left(-\frac{7}{2}-\nu +44 \nu
   ^2\right)\frac{\chi _A}{\delta m} \chi _S+\left(-\frac{7}{4}-\frac{27 \nu }{2}+6
   \nu ^2\right) \chi _S^2\right]v_{\Omega }^6 \nonumber \\
   &+ \left[\left(-\frac{81}{20}+\frac{7339 \nu }{540}\right)
   \frac{\chi _A}{\delta m}+\left(-\frac{81}{20}+\frac{593 \nu }{108}\right) \chi
   _S\right] i (H_{\mathrm{EOB}} \Omega) ^2\,, \\
   \label{eq:f_21}
   f_{21}^{\mathrm{S}} =& \left(-\frac{3}{2} \chi _S-\frac{3 \chi _A}{2 \delta m}\right)v_{\Omega } + \left[\left(\frac{61}{12}+\frac{79 \nu }{84}\right) \chi
   _S+\left(\frac{61}{12}+\frac{131 \nu }{84}\right)\frac{\chi _A}{\delta m}\right]v_{\Omega }^3+ \left[(-3-2 \nu ) \chi _A^2+\left(-3+\frac{\nu }{2}\right) \chi
   _S^2+\left(-6+\frac{21 \nu }{2}\right)\chi _S \frac{\chi _A }{\delta m}\right]v_{\Omega }^4\nonumber \\
   &+ \bigg\{\left(\frac{3}{4 \delta m}-\frac{3 \nu }{\delta m}\right) \chi
   _A^3+\left[-\frac{81}{16}+\frac{1709 \nu }{1008}+\frac{613 \nu ^2}{1008}+\left(\frac{9}{4}-3 \nu \right) \chi _A^2\right] \chi _S + \nonumber \\
   &+\frac{3 \chi _S^3}{4}+\left[-\frac{81}{16}-\frac{703
   \nu ^2}{112}+\frac{8797 \nu }{1008}+\left(\frac{9}{4}-6 \nu \right) \chi _S^2\right]\frac{\chi _A }{\delta m}\bigg\}v_{\Omega }^5 \nonumber \\
   &+ \left[\left(\frac{4163}{252}-\frac{9287 \nu }{1008}-\frac{85 \nu
   ^2}{112}\right) \chi _A^2+\left(\frac{4163}{252}-\frac{2633 \nu }{1008}+\frac{461 \nu ^2}{1008}\right) \chi _S^2+\left(\frac{4163}{126}-\frac{1636 \nu }{21}+\frac{1088 \nu ^2}{63}\right)\chi _S
   \frac{\chi _A }{\delta m}\right]v_{\Omega }^6+\mathbf{c_{21}} v_{\Omega }^7\,,\\
   \label{eq:f_55}
   f_{55}^{\mathrm{S}} = & \left[\left(-\frac{70 \nu }{3 (-1+2 \nu )}+\frac{110 \nu ^2}{3 (-1+2 \nu )}+\frac{10}{3 (-1+2 \nu )}\right) \frac{\chi _A}{\delta m}+\left(\frac{10}{3
   (-1+2 \nu )}-\frac{10 \nu }{-1+2 \nu }+\frac{10 \nu ^2}{-1+2 \nu }\right) \chi _S\right]v_{\Omega }^3\nonumber \\
&+ \left[\frac{5}{2}\delta m^2 \chi _A^2+5 \delta m \chi _A \chi _S+\frac{5
   \chi _S^2}{2}\right]v_{\Omega }^4+\mathbf{c_{55}} v_{\Omega }^5\,.
\end{align}
\end{landscape}
\begin{align}
\delta_{33} =& \frac{13}{10}(H_{\mathrm{EOB}}\Omega) + \frac{39\pi}{7}(H_{\mathrm{EOB}}\Omega)^2 + \left(-\frac{227827}{3000} + \frac{78\pi^2}{7}\right)(H_{\mathrm{EOB}}\Omega)^3 + \nonumber \\
& - \frac{80897\nu}{2430} v_{\Omega}^5\,, \\
\delta_{21} =& \frac{2}{3} (\Omega  H_{\mathrm{EOB}})+\frac{107}{105} \pi  (\Omega H_{\mathrm{EOB}})^2+\left(-\frac{272}{81}+\frac{214 \pi ^2}{315}\right) \Omega ^3
   H_{\text{EOB}}^3 -\frac{493}{42} \nu  v_{\Omega}^{5}\,, \\
\delta_{44} =& \frac{(112+219 \nu )}{120 (1-3 \nu )}(\Omega  H_{\mathrm{EOB}})+\frac{25136 \pi}{3465}(\Omega  H_{\mathrm{EOB}})^2 | \nonumber \\
& +\left(\frac{201088}{10395}\pi^2 - \frac{55144}{375} \right) (\Omega  H_{\mathrm{EOB}})^3\,, \\
\delta_{55} =& \frac{(96875+857528 \nu )}{131250 (1-2 \nu )}(\Omega  H_{\mathrm{EOB}}) + \frac{3865\pi}{429}(\Omega  H_{\mathrm{EOB}})^2\nonumber \\
& + \frac{-7686949127 + 954500400\pi^2}{31783752}(\Omega  H_{\mathrm{EOB}})^3\,.
\end{align}
We notice that $f_{33}^S$ is a complex quantity because it contains an  
imaginary term recently computed in PN theory~\cite{Marsatetal2017}
\begin{equation}
i\delta_{33}^\mathrm{S} \equiv \left[\left(-\frac{81}{20}+\frac{7339 \nu }{540}\right)
   \frac{\chi _A}{\delta m}+\left(-\frac{81}{20}+\frac{593 \nu }{108}\right) \chi
   _S\right] i (H_{\mathrm{EOB}} \Omega) ^2,
\end{equation}
where with the superscript ``S'' we indicate the spin
  dependence. The term proportional to $\chi_A/\delta m$ seems to diverge when $\delta m \rightarrow 0$, but this divergence is apparent 
 because, as it happens for all the functions $f_{\ell
    m}^{\mathrm{S}}$, it is removed by the factor $\delta
  m$ that appears in the function $c_{\ell +\epsilon}(\nu)$ (see
  Eq.\eqref{eq:clm}) at Newtonian order (see
  Eq.\eqref{eq:Newtonian}). If one includes the term
  $\delta_{33}^\mathrm{S}$ in the resummation with the complex
  exponential, one obtains the expression $e^{i(\delta_{33}+
    \delta_{33}^\mathrm{S})}$ which is not well-behaved in the limit
  $\delta m \rightarrow 0$.  For this reason we do not include this
  new PN term in the resummation $f_{3 3}e^{i(\delta_{33}+
    \delta_{33}^\mathrm{S})}$, but, instead, we compute the latter quantity
  excluding this term (i.e., $f_{3 3}e^{i\delta_{33}}$) and we then add
  the new complex term to the real amplitude $f_{3 3}$. We can do so because 
  $e^{i\delta_{3 3}} i\delta_{33}^\mathrm{S} = i\delta_{33}^\mathrm{S}
  + \mathcal{O}(\Omega^3)$, where the latter is a PN correction at higher order with respect to the order at which 
we currently know PN terms. 

We remember also that the modes $(2,1),(5,5)$ contain the calibration 
parameters $c_{21}$ and $c_{55}$ computed imposing the condition in Eq.~\eqref{eq:cal_par}.

\chapter{Fits of nonquasi-circular input values}
\label{app:NQCfits}

We build the fits of the nonquasi-circular (NQC) input values using NR
waveforms with the highest level of resolution available and the
extrapolation order $N = 2$. Depending on the
  mode, the fits use a different number of NR waveforms, because for
  some binary configurations the large numerical error prevents us to
  use some NR modes. For each mode, in order to choose which NR
  simulations to use for the fits, we first remove all the NR
  simulations showing clearly unphysical features (e.g., strong
  oscillations in the post-merger stage that are not consistent among waveforms 
at different resolution and extrapolation order). For the modes (3,3) and (2,1) all the NR
  waveforms pass this selection, while for the modes (4,4) and (5,5)
  we remove respectively 10 and 42 NR simulations. 
For each NQC input value (i.e., amplitude and its first and second derivative, and
  frequency and its first derivative) 
we weight the value extracted by
  a given NR simulation with the inverse of the NR error. The latter
  is estimated as $\sqrt{(\delta^{\mathrm{NQC}}_{\mathrm{res}})^2 +
    (\delta^{\mathrm{NQC}}_{\mathrm{extr}}})^2$, where
  $\delta^{\mathrm{NQC}}_{\mathrm{res}}$ is the difference between the
  NQC input values extracted from the NR waveform with the same
  extrapolation order ($N = 2$) and different resolutions (i.e., the
  highest and second highest
  resolution). The quantity $\delta^{\mathrm{NQC}}_{\mathrm{extr}}$ is instead the
  difference between the NQC input values extracted from the NR
  waveform with the same resolution level (the highest) and different
  extrapolation order (i.e., $N = 2$ and $N = 3$).

We find it convenient to define a few variables that enter the fits below:
\begin{align}
\chi_{33} &= \chi_S\delta m + \chi_A\,,\\
\chi_{21A} &= \frac{\chi_S}{1-1.3\nu}\delta m + \chi_A\,,\\
\chi_{44A} &= (1-5\nu)\chi_S+\chi_A\delta_m\,,\\
\chi_{21D} &= \frac{\chi_S}{1-2\nu}\delta m + \chi_A\,,\\
\chi_{44D} &= (1-7\nu)\chi_S+\chi_A\delta_m\,, \\
\chi&= \chi_S+\chi_A\frac{\delta m}{1-2\nu}.\\ \nonumber
\end{align}
We notice that the variables $\chi_{33}, \,\chi_{21A}\,,\chi_{21D}$ are by definition
zero in the equal-mass, equal-spin limit. They are used for the fits
of the amplitude (and its derivative) to guarantee that in this limit 
the modes with  $m$ odd vanish, since they have to satisfy the 
symmetry under rotation $\varphi_0 \rightarrow \varphi_0 + \pi$.
\subsection{Amplitude's fits}
\begin{landscape}
\begin{align}
\frac{|h_{3 3}^{\textrm{NR}}(t^{3 3}_{\textrm{match}})|}{\nu} = &|(0.101092 - 0.470410 \nu+1.073546 \nu ^2) \chi
   _{33} \nonumber \\
   &+\delta m (0.563658 - 0.054609 \nu+2.309370 \nu
   ^2+0.029813 \chi _{33}^2 - 0.0968810 \nu \ \chi_{33}^2)|\,, \\
\frac{|h_{2 1}^{\textrm{NR}}(t^{2 1}_{\textrm{match}})|}{\nu} = &|\delta m (-0.428179 +0.113789 \nu -0.773677 \nu ^2  -0.0101951 \chi_{21A}+0.0470041
   \chi_{21A}^2 -0.0932613 \chi_{21A}^2 \nu) \nonumber \\
   &+\chi_{21A} (0.292567 -0.197103 \nu)+ \delta m 0.0168769 \chi_{21A}^3|\,, \\
\frac{|h_{4 4}^{\textrm{NR}}(t^{4 4}_{\textrm{match}})|}{\nu} = &0.264658\, +0.0675842 \chi _{44 A}+0.029251 \chi _{44 A}^2+ (-0.565825-0.866746 \chi _{44 A}+0.00523419 \chi _{44 A}^2)\nu \nonumber \\
&+ (-2.50083+6.88077 \chi _{44 A}-1.02347 \chi _{44
   A}^2)\nu ^2+ (7.69745\, -16.5515 \chi _{44 A})\nu ^3\,, \\
\frac{|h_{5 5}^{\textrm{NR}}(t^{5 5}_{\textrm{match}})|}{\nu} =&|0.128621\delta m -0.474201 \delta m \, \nu +1.0833  \delta m \, \nu^2 + 0.0322784 \chi_{33} -0.134511 \chi_{33} \nu +0.0990202 \chi_{33}\nu^2|
\end{align}
\end{landscape}
\subsection{Amplitude--first-derivative's fits}
\begin{landscape}
\begin{align}
\frac{1}{\nu}\left. \frac{d\left| h_{33}^{\textrm{NR}}(t) \right|}{dt} \right|_{t = t^{3 3}_{\textrm{match}}} = &\delta m (-0.00309944+0.0100765 \nu) \chi_{33}^2\nonumber \\
&+
   0.00163096\sqrt{\delta m^2 (8.81166\, +104.478 \nu)+\delta m
   (-5.35204+49.6862 \nu) \chi_{33}+ \chi_{33}^2}\,, \\
   \frac{1}{\nu}\left. \frac{d\left| h_{21}^{\textrm{NR}}(t) \right|}{dt} \right|_{t = t^{2 1}_{\textrm{match}}} = & \delta m (0.00714753\, -0.0356440 \nu ) +0.00801714  |
   -\delta m (0.787561\, +1.61127 \nu +11.30606 \nu ^2)+\chi _{21
   D}| \nonumber \\
   & +\delta m (-0.00877851+0.0305467 \nu ) \chi _{21 D}\,, \\
\frac{1}{\nu}\left. \frac{d\left| h_{44}^{\textrm{NR}}(t) \right|}{dt} \right|_{t = t^{4 4}_{\textrm{match}}} = &0.00434759\, -0.00146122 \chi _{44 D}-0.00242805 \chi _{44 D}^2+ (0.0233207\,
   -0.0224068 \chi _{44 D}+0.0114271 \chi _{44 D}^2)\nu\nonumber \\
   &+
   (-0.460545+0.433527 \chi _{44 D})\nu ^2+ (1.27963\, -1.24001 \chi
   _{44 D})\nu ^3\,, \\
   \frac{1}{\nu}\left. \frac{d\left| h_{55}^{\textrm{NR}}(t) \right|}{dt} \right|_{t = t^{5 5}_{\textrm{match}}} = &\delta m (-0.0083898+0.0467835 \nu )+\delta m (-0.00136056+0.00430271
   \nu ) \chi _{33} \nonumber \\
   &+\delta m (-0.00114121+0.00185904 \nu ) \chi
   _{33}^2+0.000294422 |\delta m (37.1113\, -157.799 \nu )+\chi
   _{33}|\,.
\end{align}
\end{landscape}
\subsection{Amplitude--second-derivative's fits}
\begin{landscape}
\begin{align}
\frac{1}{\nu}\left. \frac{d^2\left| h_{33}^{\textrm{NR}}(t) \right|}{dt^2} \right|_{t = t^{3 3}_{\textrm{match}}} = &\delta m (0.000960569\, -0.000190807 \nu)
   \chi_{33}\nonumber \\
   &- 0.000156238 | \delta m(4.67666 + 79.2019 \nu -1097.41 \nu^2 + 6512.96 \nu^3 -13263.4 \nu^4) + \chi_{33}|\,, \\
   \frac{1}{\nu}\left. \frac{d^2\left| h_{21}^{\textrm{NR}}(t) \right|}{dt^2} \right|_{t = t^{2 1}_{\textrm{match}}} = & 0.000371322 \delta m
    -| \delta m
   (-0.000365087-0.00305417 \nu ) +\delta m (-0.000630623 
   -0.000868048 \nu \nonumber \\
   &+0.0223062 \nu ^2) \chi _{21 D}^2 +0.000340243  \chi _{21
   D}^3+0.000283985  \delta m \, \chi _{21 D}|\,, \\
      \frac{1}{\nu}\left. \frac{d^2\left| h_{44}^{\textrm{NR}}(t) \right|}{dt^2} \right|_{t = t^{4 4}_{\textrm{match}}} = & -0.000301723+0.000321595 \chi +  (0.00628305\, +0.00115988 \chi)\nu\nonumber \\
   &+ (-0.0814352-0.0138195 \chi)\nu ^2+ (0.226849\, +0.0327575 \chi)\nu ^3\,, \\
       \frac{1}{\nu}\left. \frac{d^2\left| h_{55}^{\textrm{NR}}(t) \right|}{dt^2} \right|_{t = t^{5 5}_{\textrm{match}}}  = &\delta m (0.000127272\, +0.000321167 \nu )+\delta m
   (-0.0000662168+0.000328855 \nu ) \chi _{33} \nonumber \\
   &+(-0.0000582462+0.000139443 \nu ) \chi
   _{33}^2.
\end{align}
\end{landscape}
\subsection{Frequency and frequency-derivative fits}
\begin{align}
\omega_{33}^{\textrm{NR}}(t^{3 3}_{\textrm{match}}) = &0.397395\, +0.164193 \chi+0.163553 \chi^2+0.0614016 \chi^3 + \nonumber \\
&+
   (0.699506\, -0.362674 \chi-0.977547 \chi^2)\nu\nonumber \\
   &+
   (-0.345533+0.319523 \chi+1.93342 \chi^2)\nu^2\,, \\
   \omega_{21}^{\textrm{NR}}(t^{2 1}_{\textrm{match}}) = & 0.174319\, +0.0535087 \chi +0.0302288 \chi ^2+\nonumber\\
&+ (0.193894\, -0.184602 \chi
   -0.112222 \chi ^2)\nu\nonumber \\
   &+ (0.167006\, +0.218731 \chi )\nu ^2\,, \\
  \omega_{44}^{\textrm{NR}}(t^{4 4}_{\textrm{match}}) = & 0.538936\, +0.166352 \chi +0.207539 \chi ^2+0.152681 \chi ^3 \nonumber \\
  &+ \left(0.76174+0.00958786 \chi -1.3023 \chi
   ^2-0.556275 \chi ^3\right)\nu + \nonumber\\
&   + \left(0.967515\,
   -0.220593 \chi +2.6781 \chi ^2\right)\nu ^2\nonumber \\
   &- 4.89538 \nu^3 \,,\\
   \omega_{55}^{\textrm{NR}}(t^{5 5}_{\textrm{match}})   = &0.643755\, +0.223155 \chi +0.295689 \chi ^2+0.173278 \chi ^3\nonumber \\
   &+ 
   \left(-0.470178-0.392901 \chi -2.26534 \chi ^2-0.5513 \chi ^3\right)\nu +\nonumber \\
&   +\left(2.31148\, +0.882934 \chi +5.8176 \chi ^2\right)\nu ^2\,.
\end{align}
\begin{align}
\dot{\omega}_{33}^{\textrm{NR}}(t^{3 3}_{\textrm{match}}) = &0.0103372\, -0.00530678 \chi ^2-0.00508793 \chi ^3\nonumber \\
&+ \left(0.0277356\, +0.0188642
   \chi+0.0217545 \chi^2+0.0178548 \chi^3\right)\nu +\nonumber \\ &+(0.0180842\,
   -0.0820427 \chi)\nu^2, \\  
   \dot{\omega}_{21}^{\textrm{NR}}(t^{2 1}_{\textrm{match}}) = &0.00709874\, -0.00177519 \chi -0.00356273 \chi ^2-0.0019021 \chi ^3\nonumber \\&
   + 
   (0.0248168\, +0.00424406 \chi +0.0147181 \chi ^2)\nu + \nonumber \\ &+
   (-0.050429-0.0319965 \chi )\nu ^2\,, \\
   \dot{\omega}_{44}^{\textrm{NR}}(t^{4 4}_{\textrm{match}}) =&0.0139979\, -0.00511782 \chi -0.00738743 \chi ^2+\nonumber \\&+ \left(0.0528489\, +0.016323 \chi
   +0.0253907 \chi ^2\right)\nu  \nonumber \\
   &+ (-0.0652999+0.0578289 \chi )\nu ^2\,,\\
   \dot{\omega}_{55}^{\textrm{NR}}(t^{5 5}_{\textrm{match}})   =&0.0176343\, -0.000249257 \chi -0.0092404 \chi ^2-0.00790783 \chi ^3\nonumber \\ 
   &+\left(-0.13660+0.0561378 \chi +0.164063 \chi ^2+0.0773623 \chi ^3\right)\nu + \nonumber \\&
   +\left(0.987589\, -0.313921 \chi -0.592615 \chi ^2\right)\nu ^2 \nonumber \\
   &-1.694335\nu^3\,.
\end{align}

\chapter{Fits for amplitude and phase of merger-ringdown model}
\label{app:ringdownfits}

\begin{figure}[h]
  \centering
  \includegraphics[width=0.7\textwidth]{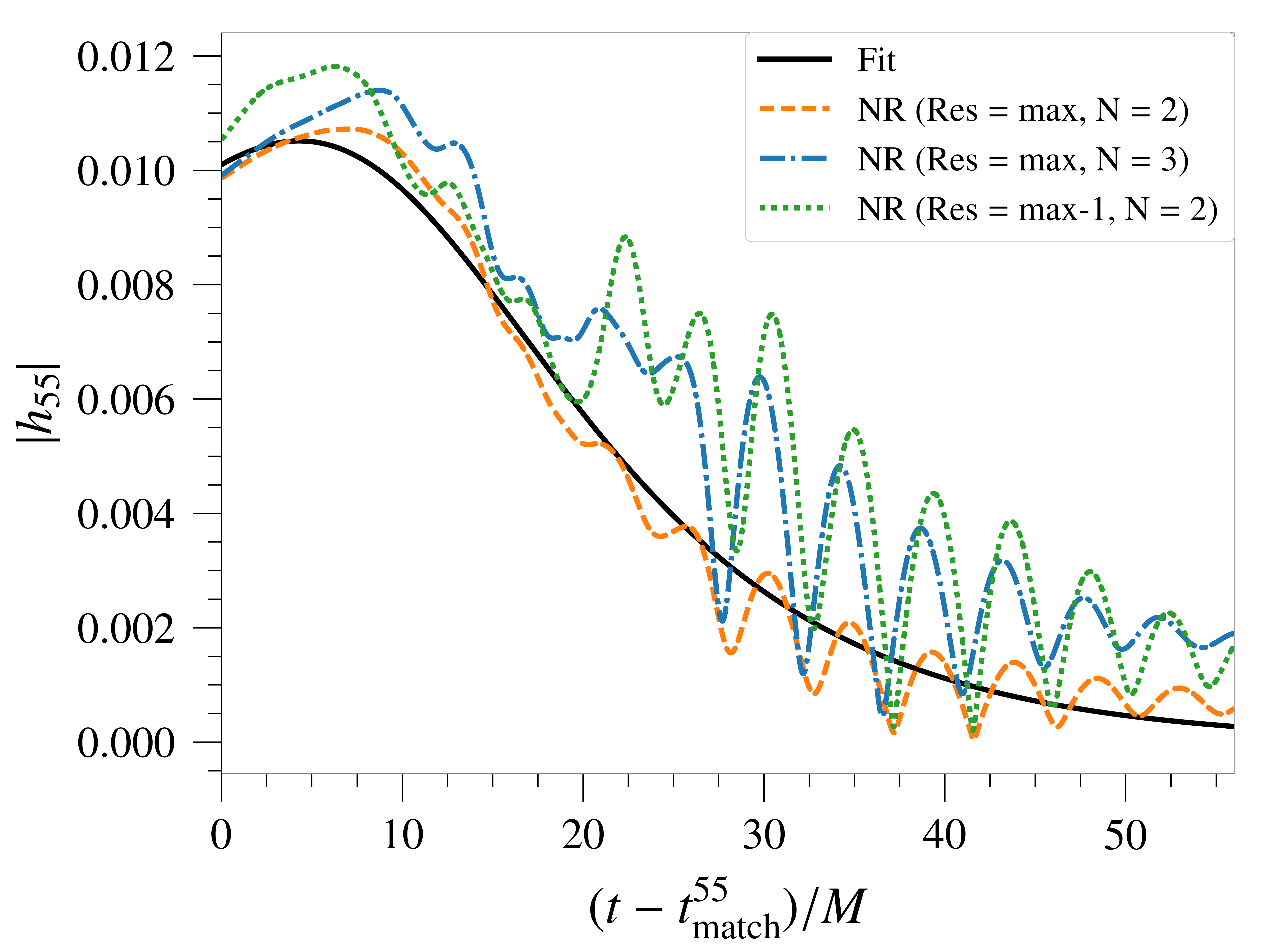}
\caption{Amplitudes of the (5,5) NR mode of the simulation \texttt{SXS:BBH:0065} $(q = 8, \chi_1 = 0.5, \chi_2 = 0)$ for extraction order $N = 2$ and highest resolution (dashed orange), extraction order $N = 3$ and highest resolution (dotted-dashed blue), extraction order $N = 2$ and second highest resolution (dotted green). In solid black we show the result of the fit of the merger-ringdown signal used in the \texttt{SEOBNRv4HM} model.}
\label{fig:noisyNR}
\end{figure}

For these fits we apply the same selection of the NR waveforms discussed for the fits of the input values for the NQC. In particular, in performing the fits for the amplitude (phase) of the merger-ringdown signal, we weigh the contribution of the values extracted from every NR waveform with the same weight used for the NQC input value of the amplitude (frequency). It should be noted that  in some cases, especially in the ringdown, the NR error in the (4,4) and (5,5) modes limits our ability to accurately model this part of the waveform (see Fig.~\ref{fig:noisyNR}).

\begin{align}
c_{1,f}^{33} = &0.0763873\, +0.254345 \nu -1.08927 \nu ^2-0.0309934 \chi\nonumber \\
& +0.251688 \nu  \chi -0.798091
   \nu ^2 \chi\,, \\
c_{2,f}^{33} = & -0.832529+2.76799 \nu -7.02815 \nu ^2-0.59888 \chi \nonumber \\
&+5.90437 \nu  \chi -18.2326 \nu ^2
   \chi\,, \\
c_{1,f}^{21} = &  0.0778033\, +0.24091 \nu -0.745633 \nu ^2-0.0507064 \chi\nonumber \\
& +0.385826 \nu  \chi -0.969553
   \nu ^2 \chi\,, \\
   c_{2,f}^{21} = & -1.24519+6.1342 \nu -14.6725 \nu ^2-1.19579 \chi \nonumber \\
   &+15.667 \nu  \chi -44.4198 \nu ^2 \chi\,, \\
   c_{1,f}^{44} = &-0.0639271+0.345195 \nu -1.76435 \nu ^2-0.0364617 \chi \nonumber \\
   &+1.27774 \nu  \chi -14.8253 \nu
   ^2 \chi +40.6714 \nu ^3 \chi\,, \\
   c_{2,f}^{44} = &0.781328\, -5.1869 \nu +14.0264 \nu ^2+0.809471 \chi\nonumber \\
   & -5.38343 \nu  \chi +0.105163 \nu ^2
   \chi +46.9784 \nu ^3 \chi\,,\\
c_{1,f}^{55} =&-0.0670461-0.247549 \nu +0.758804 \nu ^2+0.0219059 \chi \nonumber \\
   & -0.0943771 \nu  \chi +0.435777
   \nu ^2 \chi\,, \\
c_{2,f}^{55} =&1.67634\, -5.60456 \nu +16.7513 \nu ^2+0.49257 \chi\nonumber \\
   & -6.2091 \nu  \chi +16.7785 \nu ^2
   \chi\,.
\end{align}
\begin{align}
d_{1,f}^{33} = & 0.110853\, +0.99998 \nu -3.39833 \nu ^2+0.0189591 \chi \nonumber \\
& -0.72915 \nu  \chi +2.5192 \nu ^2
   \chi\,,\\
   d_{2,f}^{33} = & 2.78252\, -7.84474 \nu +27.181 \nu ^2+2.87968 \chi\nonumber \\
   & -34.767 \nu  \chi +127.139 \nu ^2
   \chi\,,\\
   d_{1,f}^{21} = &0.156014\, +0.0233469 \nu +0.153266 \nu ^2+0.1022 \chi\nonumber \\
   & -0.943531 \nu  \chi +1.79791 \nu
   ^2 \chi\,, \\
   d_{2,f}^{21} = &2.78863\, -0.814541 \nu +5.54934 \nu ^2+4.2929 \chi\nonumber \\
   & -15.938 \nu  \chi +12.6498 \nu ^2
   \chi\,, \\
   d_{1,f}^{44} = &0.11499\, +1.61265 \nu -6.2559 \nu ^2+0.00838952 \chi \nonumber \\
   & -0.806998 \nu  \chi +7.59565 \nu
   ^2 \chi -19.3237 \nu ^3 \chi\,, \\
   d_{2,f}^{44} = & 3.11182\, +15.8853 \nu -79.6493 \nu ^2+5.39934 \chi \nonumber \\
   & -87.9242 \nu  \chi +657.716 \nu ^2
   \chi -1555.3 \nu ^3 \chi\,,\\
    d_{1,f}^{55} = & 0.164654\, -0.191845 \nu +0.333284 \nu ^2-0.0265748 \chi\nonumber \\
    & -0.0551962 \nu  \chi +0.319427 \nu
   ^2 \chi\,, \\
    d_{2,f}^{55} = & 11.1024\, -58.6058 \nu +176.606 \nu ^2+6.01511 \chi\nonumber \\
    & -81.6803 \nu  \chi +266.473 \nu ^2 \chi\,.
\end{align}

\chapter{Fits for the phase difference between higher-order modes and (2,2) mode at the matching point $t_{\textrm{match}}^{\ell m}$}
\label{app:fitphasediff}

The relations between $\phi_{\textrm{match}}^{\ell m}$ (i.e., the phase of the $(\ell, m)$ modes computed at $t_{\textrm{match}}^{\ell m}$) and $\phi_{\textrm{match}}^{2 2}$ are
\begin{align}
\Delta\phi_{\textrm{match}}^{3 3} &\equiv \phi_{\textrm{match}}^{3 3} - \frac{3}{2}(\phi_{\textrm{match}}^{2 2} - \pi)  \pmod{\pi}\,, \\
\Delta\phi_{\textrm{match}}^{2 1} &\equiv \phi_{\textrm{match}}^{2 1} - \frac{1}{2}(\phi_{\textrm{match}}^{2 2} - \pi)  \pmod{\pi}\,,\\
\Delta\phi_{\textrm{match}}^{4 4} &\equiv \phi_{\textrm{match}}^{4 4} - (2\phi_{\textrm{match}}^{2 2} - \pi) \pmod{2\pi}\,,\\
\Delta\phi_{\textrm{match}}^{5 5} &\equiv \phi_{\textrm{match}}^{5 5} - \frac{1}{2}(5\phi_{\textrm{match}}^{2 2} - \pi) \pmod{\pi}, 
\end{align}
where the RHS is the scaling of the phase at leading PN order, and the LHS is the deviation from the latter, 
computed at $t_{\textrm{match}}^{\ell m}$. The term $\Delta\phi_{\textrm{match}}^{\ell m}$ is extracted from 
each NR and Teukolsky--equation-based waveforms in our catalog and then fitted as a function of $(\nu, \chi)$. We find
\begin{align}
\Delta\phi_{\textrm{match}}^{3 3} =&3.20275\, -1.47295 \sqrt{\delta m}+1.21021 \delta m-0.203442 \chi \nonumber \\
& +\delta m^2 (-0.0284949-0.217949 \chi ) \chi \pmod{\pi}\,,\\
\Delta\phi_{\textrm{match}}^{2 1} = & 2.28855\, +0.200895 \delta m-0.0403123 \chi \nonumber \\
&+\delta m^2 \left(-0.0331133-0.0424056 \chi -0.0244154 \chi ^2\right) \nonumber\\
&\pmod{\pi}\,, \\
\Delta\phi_{\textrm{match}}^{4 4} = & 5.89306\, +\nu ^2 (-36.7321-21.9229 \chi ) \nonumber \\
&-0.499652 \chi -0.292006 \chi ^2\nonumber \\
&+\nu ^3 (160.102\, +67.0793 \chi )\nonumber \\
&+\nu  \left(2.48143\, +3.26618 \chi +1.38065 \chi ^2\right) \pmod{2\pi}\,, \\
\Delta\phi_{\textrm{match}}^{5 5} =& 3.61933\, -1.52671 \delta m-0.172907 \chi \nonumber \\
& +\delta m^2 \left(0.72564\, -0.44462 \chi -0.528597 \chi ^2\right)\nonumber \\
& \pmod{\pi} \,.
\end{align}
The error on the phase of each mode caused by the fit of $\Delta\phi_{\textrm{match}}^{\ell m}$ is on average of the order of 0.05 rad.

\chapter{Fits for time difference between modes' amplitude peaks}
\label{app:timeshiftfits}

As originally observed in Refs.~\cite{Barausse:2011kb,Pan:2011gk}, gravitational modes peak 
at different times ($t_{\mathrm{peak}}^{\ell m}$) with respect to 
the dominant $(2,2)$ mode. Using the NR catalog at our disposal, we fit 
the times shifts $\Delta t_{\ell m} \equiv t_{\mathrm{peak}}^{\ell m} -
t_{\mathrm{peak}}^{2 2}$ as function of $\nu$ and $\chi_{\mathrm{eff}} = (m_1 \chi_1 +
m_2 \chi_2)/M$. We find
\begin{align}
\Delta t_{33} =& 4.20646\, +4.215 \chi_{\mathrm{eff}}+2.12487 \chi_{\mathrm{eff}}^2 \nonumber \\
&+(-10.9615+5.20758 a) \nu +(53.3674\, -65.0849 a) \nu ^2\,, \\
\Delta t_{21} =& 12.892\, +1.14433 \chi_{\mathrm{eff}}+1.12146 \chi_{\mathrm{eff}}^2 \nonumber \\
&+\left(-61.1508-96.0301\chi_{\mathrm{eff}} -85.4386 \chi_{\mathrm{eff}}^2\right) \nu \nonumber \\
&+\left(144.497\, +366.374 \chi_{\mathrm{eff}}+322.06 \chi_{\mathrm{eff}}^2\right) \nu ^2\,, \\
\Delta t_{44} =&  7.49641\, +6.7245 \chi_{\mathrm{eff}} +3.11618  \chi_{\mathrm{eff}}^2\nonumber \\
&+\left(-48.5578-78.8077 \chi_{\mathrm{eff}} -92.1608  \chi_{\mathrm{eff}}^2\right) \nu \nonumber \\
& +\left(91.483\, +231.917 \chi_{\mathrm{eff}} +388.074  \chi_{\mathrm{eff}}^2\right) \nu ^2 \,,\\
\Delta t_{55} =& 10.031\, +5.80884 \chi_{\mathrm{eff}}+(-103.252-75.8935 \chi_{\mathrm{eff}}) \nu \nonumber \\
& +(366.57\, +282.552 \chi_{\mathrm{eff}}) \nu ^2\,.
\end{align}
\begin{figure}[h]
  \centering
  \includegraphics[width=0.7\textwidth]{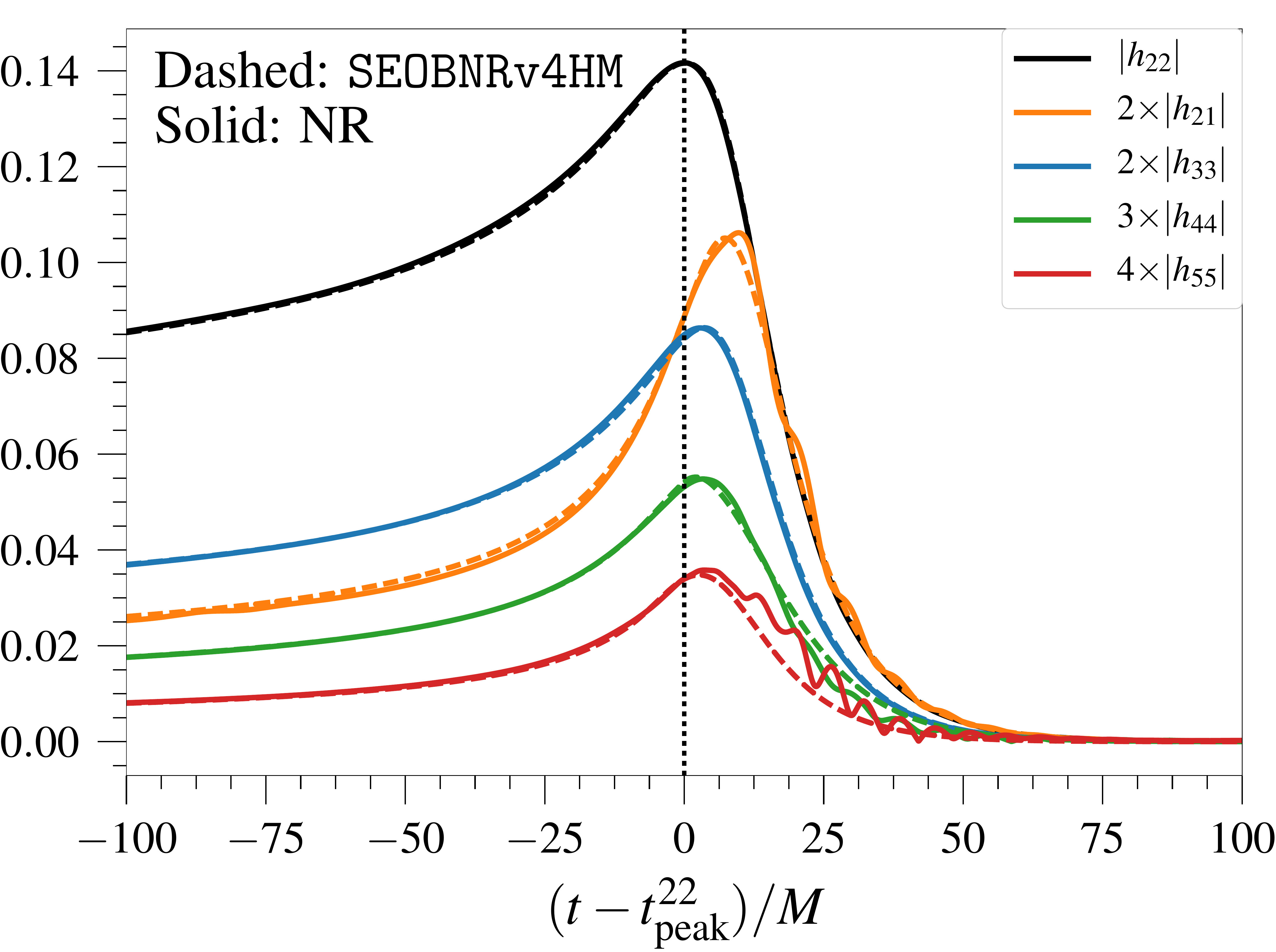}
\caption{Amplitudes of different modes for the \texttt{SEOBNRv4HM} (dashed) and NR (solid) waveforms 
with $(q = 8,\, \chi_1 = -0.5,\, \chi_2 = 0)$ (\texttt{SXS:BBH:0064}) versus time. The time origin corresponds to 
the $(2,2)$ mode's peak.}
\label{fig:timeshift}
\end{figure}

The above expressions could be employed in building phenomenological
  models for the ringdown signal when multipole modes are
  present~\cite{London:2014cma}. We notice that these fits are not
  used for building \texttt{SEOBNRv4HM} waveforms, whose merger-ringdown model 
is constructed through Eqs.~(\ref{eq:merger-RD_wave})--(\ref{eq:ansatz_phase}), 
starting from $t_{\mathrm{match}}^{\ell m}$ in Eq.~(\ref{eq:matchtime}). The 
merger-ringdown \texttt{SEOBNRv4HM} waveforms reproduce the time shifts $\Delta t_{\ell m}$ 
between the NR modes' amplitude peaks by construction, as it can be seen in Fig.~\ref{fig:timeshift} 
for a particular binary configuration.
  
We emphasize that while in the \texttt{EOBNRv2HM} model~\cite{Pan:2011gk} the 
merger-ringdown attachment was done at each modes' peak time, in
  \texttt{SEOBNRv4HM} we do it at the $(2,2)$ mode's peak for all modes 
except the $(5,5)$ mode. We make this change here because 
  typically $\Delta t_{\ell m} = t_{\mathrm{peak}}^{\ell m} - t_{\mathrm{peak}}^{2 2} > 0$, and 
at these late times we find that for some binary configurations either the
EOB dynamics becomes unreliable or the error in the NR waveforms is too large and prevents 
us to accurately extract the input values for the NQC conditions (i.e., Eqs. \eqref{eq:NQC_condition_1}
--\eqref{eq:NQC_condition_5}).

\chapter{Numerical-relativity catalog}
\label{sec:NRcatalog}

In the tables below we list the binary configurations of the NR
simulations used to build and test the \texttt{SEOBNRv4HM} waveform 
model. The NR waveforms were produced with the (pseudo) Spectral Einstein code (\texttt{SpEC}) 
of the Simulating eXtreme Spacetimes (\texttt{SXS}) project and the \texttt{Einstein Toolkit} (ET) 
code. In particular, we list the mass ratio $q$, the dimensionless
spins $\chi_{1,2}$, the eccentricity $e$, the initial frequency
$\omega_{22}$ of the dominant $(\ell, m) = (2,2)$ mode and the number
of orbits $N_{\mathrm{orb}}$ up to the waveform peak.

\begin{figure}[h]
  \centering
  \includegraphics[width=0.7\textwidth]{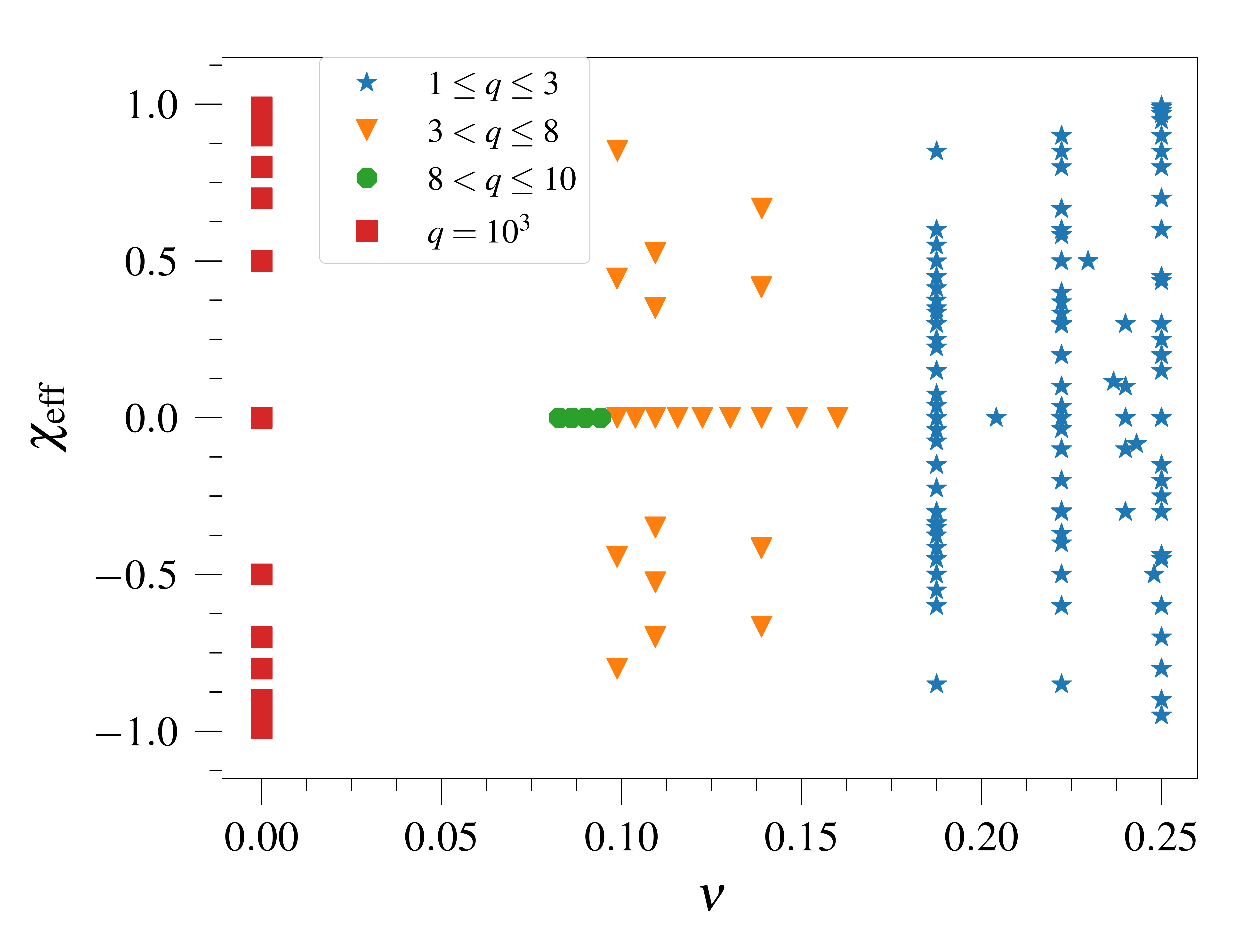}
\caption{2D projection of the 3D parameter space of the NR and BH-perturbation-theory waveforms used to build the \texttt{SEOBNRv4HM} model. 
The $x$-axis is $\nu$ and the $y$-axis is the effective spin $\chi_{\mathrm{eff}} = (\chi_1 m_1 + \chi_2 m_2)/M$. In the legend we 
highlight four different regions of coverage, as discussed in the text.}
\label{fig:NRcatalog}
\end{figure}

In Fig.~\ref{fig:NRcatalog} we show the coverage of NR and BH-perturbation-theory 
waveforms when projected on the binary's parameters $\nu$ and $\chi_\mathrm{eff}=(\chi_1 m_1 + \chi_2 m_2)/M$. 
We highlight four regions. In the first region $1 \leq q \leq 3$ there is a large number of configurations 
with both BHs carrying spin. The spins magnitude are as high as $\chi_{1,2}
= 0.99$ in the equal-mass limit, while they are limited to $\chi_{1,2}
= 0.85$ for $q = 3$. The second region is between $3 < q \leq 8$, and 
most of the simulations have spins only on the heavier BH. The values
of the spin of the heavier BH span in the region $-0.8 \leq \chi_1
\leq 0.85$. The third region is between $8 < q \leq 10$ and it includes
only nonspinning waveforms.  Finally, the fourth region covers 
13 waveforms computed solving the Teukolsky equation in the framework of BH
pertubation theory~\cite{Barausse:2011kb,Taracchini:2014zpa}. They
have $q = 10^{3}$ and dimensionless spins values in the range $-0.99
\leq \chi \leq 0.99$.

\newcommand{\numoriginal}{38}
\newcommand{\numpublicnew}{9}
\newcommand{\numchu}{94}

\subsection{SXS and ET waveform produced for testing \texttt{SEOBNRv4} (Ref.~\cite{Bohe:2016gbl})}
\label{sec:original}
\begingroup
 \begin{longtable}{ccccccc}
  \doubleline
  
ID & $q$ & $\chi_1$ & $\chi_2$ & $e$ & $M\omega_{22}$ & $N_\mathrm{orb}$\\
\hline
SXS:BBH:0610 & $1.2$ & $-0.50$ & $-0.50$ & $7.4\times 10^{-5}$ & $0.01872$ & $12.1$ \\
SXS:BBH:0611 & $1.4$ & $-0.50$ & $+0.50$ & $6.0\times 10^{-4}$ & $0.02033$ & $12.5$ \\
SXS:BBH:0612 & $1.6$ & $+0.50$ & $-0.50$ & $3.7\times 10^{-4}$ & $0.02156$ & $12.8$ \\
SXS:BBH:0613 & $1.8$ & $+0.50$ & $+0.50$ & $1.8\times 10^{-4}$ & $0.02383$ & $13.1$ \\
SXS:BBH:0614 & $2.0$ & $+0.75$ & $-0.50$ & $6.7\times 10^{-4}$ & $0.02355$ & $13.1$ \\
SXS:BBH:0615 & $2.0$ & $+0.75$ & $+0.00$ & $7.0\times 10^{-4}$ & $0.02401$ & $13.3$ \\
SXS:BBH:0616 & $2.0$ & $+0.75$ & $+0.50$ & $8.0\times 10^{-4}$ & $0.02475$ & $13.3$ \\
SXS:BBH:0617 & $2.0$ & $+0.50$ & $+0.75$ & $7.8\times 10^{-4}$ & $0.02342$ & $13.1$ \\
SXS:BBH:0618 & $2.0$ & $+0.80$ & $+0.80$ & $5.9\times 10^{-4}$ & $0.02578$ & $13.4$ \\
SXS:BBH:0620 & $5.0$ & $-0.80$ & $+0.00$ & $3.4\times 10^{-3}$ & $0.02527$ & $8.2$ \\
SXS:BBH:0621 & $7.0$ & $-0.80$ & $+0.00$ & $3.2\times 10^{-3}$ & $0.02784$ & $7.1$ \\
SXS:BBH:0619 & $2.0$ & $+0.90$ & $+0.90$ & $2.9\times 10^{-4}$ & $0.02520$ & $13.5$ \\
ET:AEI:0001 & $5.0$ & $+0.80$ & $+0.00$ & $9.2\times 10^{-4}$ & $0.03077$ & $10.5$ \\
ET:AEI:0002 & $7.0$ & $+0.80$ & $+0.00$ & $6.1\times 10^{-4}$ & $0.03503$ & $10.4$ \\
ET:AEI:0004 & $8.0$ & $+0.85$ & $+0.85$ & $3.0\times 10^{-3}$ & $0.04368$ & $7.4$ \\

  \doubleline
\end{longtable}
\label{tab:original}
\endgroup

\subsection{SXS waveforms from Ref.~\cite{Mroue:2013xna}}
\label{sec:original}
\begingroup
 \begin{longtable}{ccccccc}
  \doubleline
  ID & $q$ & $\chi_1$ & $\chi_2$ & $e$ & $M\omega_{22}$ & $N_\mathrm{orb}$\\
\hline
SXS:BBH:0004 & $1.0$ & $-0.50$ & $+0.00$ & $3.7\times 10^{-4}$ & $0.01151$ & $30.2$ \\
SXS:BBH:0005 & $1.0$ & $+0.50$ & $+0.00$ & $2.5\times 10^{-4}$ & $0.01227$ & $30.2$ \\
SXS:BBH:0007 & $1.5$ & $+0.00$ & $+0.00$ & $4.2\times 10^{-4}$ & $0.01229$ & $29.1$ \\
SXS:BBH:0013 & $1.5$ & $+0.50$ & $+0.00$ & $1.4\times 10^{-4}$ & $0.01444$ & $23.8$ \\
SXS:BBH:0016 & $1.5$ & $-0.50$ & $+0.00$ & $4.2\times 10^{-4}$ & $0.01149$ & $30.7$ \\
SXS:BBH:0019 & $1.5$ & $-0.50$ & $+0.50$ & $7.6\times 10^{-5}$ & $0.01460$ & $20.4$ \\
SXS:BBH:0025 & $1.5$ & $+0.50$ & $-0.50$ & $7.6\times 10^{-5}$ & $0.01456$ & $22.4$ \\
SXS:BBH:0030 & $3.0$ & $+0.00$ & $+0.00$ & $2.0\times 10^{-3}$ & $0.01775$ & $18.2$ \\
SXS:BBH:0036 & $3.0$ & $-0.50$ & $+0.00$ & $5.1\times 10^{-4}$ & $0.01226$ & $31.7$ \\
SXS:BBH:0045 & $3.0$ & $+0.50$ & $-0.50$ & $6.4\times 10^{-4}$ & $0.01748$ & $21.0$ \\
SXS:BBH:0046 & $3.0$ & $-0.50$ & $-0.50$ & $2.6\times 10^{-4}$ & $0.01771$ & $14.4$ \\
SXS:BBH:0047 & $3.0$ & $+0.50$ & $+0.50$ & $4.7\times 10^{-4}$ & $0.01743$ & $22.7$ \\
SXS:BBH:0056 & $5.0$ & $+0.00$ & $+0.00$ & $4.9\times 10^{-4}$ & $0.01589$ & $28.8$ \\
SXS:BBH:0060 & $5.0$ & $-0.50$ & $+0.00$ & $3.4\times 10^{-3}$ & $0.01608$ & $23.2$ \\
SXS:BBH:0061 & $5.0$ & $+0.50$ & $+0.00$ & $4.2\times 10^{-3}$ & $0.01578$ & $34.5$ \\
SXS:BBH:0063 & $8.0$ & $+0.00$ & $+0.00$ & $2.8\times 10^{-4}$ & $0.01938$ & $25.8$ \\
SXS:BBH:0064 & $8.0$ & $-0.50$ & $+0.00$ & $4.9\times 10^{-4}$ & $0.01968$ & $19.2$ \\
SXS:BBH:0065 & $8.0$ & $+0.50$ & $+0.00$ & $3.7\times 10^{-3}$ & $0.01887$ & $34.0$ \\
SXS:BBH:0148 & $1.0$ & $-0.44$ & $-0.44$ & $2.0\times 10^{-5}$ & $0.01634$ & $15.5$ \\
SXS:BBH:0149 & $1.0$ & $-0.20$ & $-0.20$ & $1.8\times 10^{-4}$ & $0.01614$ & $17.1$ \\
SXS:BBH:0150 & $1.0$ & $+0.20$ & $+0.20$ & $2.9\times 10^{-4}$ & $0.01591$ & $19.8$ \\
SXS:BBH:0151 & $1.0$ & $-0.60$ & $-0.60$ & $2.5\times 10^{-4}$ & $0.01575$ & $14.5$ \\
SXS:BBH:0152 & $1.0$ & $+0.60$ & $+0.60$ & $4.3\times 10^{-4}$ & $0.01553$ & $22.6$ \\
SXS:BBH:0153 & $1.0$ & $+0.85$ & $+0.85$ & $8.3\times 10^{-4}$ & $0.01539$ & $24.5$ \\
SXS:BBH:0154 & $1.0$ & $-0.80$ & $-0.80$ & $3.3\times 10^{-4}$ & $0.01605$ & $13.2$ \\
SXS:BBH:0155 & $1.0$ & $+0.80$ & $+0.80$ & $4.7\times 10^{-4}$ & $0.01543$ & $24.1$ \\
SXS:BBH:0156 & $1.0$ & $-0.95$ & $-0.95$ & $5.4\times 10^{-4}$ & $0.01643$ & $12.4$ \\
SXS:BBH:0157 & $1.0$ & $+0.95$ & $+0.95$ & $1.4\times 10^{-4}$ & $0.01535$ & $25.2$ \\
SXS:BBH:0158 & $1.0$ & $+0.97$ & $+0.97$ & $7.9\times 10^{-4}$ & $0.01565$ & $25.3$ \\
SXS:BBH:0159 & $1.0$ & $-0.90$ & $-0.90$ & $5.6\times 10^{-4}$ & $0.01588$ & $12.7$ \\
SXS:BBH:0160 & $1.0$ & $+0.90$ & $+0.90$ & $4.2\times 10^{-4}$ & $0.01538$ & $24.8$ \\
SXS:BBH:0166 & $6.0$ & $+0.00$ & $+0.00$ & $4.4\times 10^{-5}$ & $0.01940$ & $21.6$ \\
SXS:BBH:0167 & $4.0$ & $+0.00$ & $+0.00$ & $9.9\times 10^{-5}$ & $0.02054$ & $15.6$ \\
SXS:BBH:0169 & $2.0$ & $+0.00$ & $+0.00$ & $1.2\times 10^{-4}$ & $0.01799$ & $15.7$ \\
SXS:BBH:0170 & $1.0$ & $+0.44$ & $+0.44$ & $1.3\times 10^{-4}$ & $0.00842$ & $15.5$ \\
SXS:BBH:0172 & $1.0$ & $+0.98$ & $+0.98$ & $7.8\times 10^{-4}$ & $0.01540$ & $25.4$ \\
SXS:BBH:0174 & $3.0$ & $+0.50$ & $+0.00$ & $2.9\times 10^{-4}$ & $0.01337$ & $35.5$ \\
SXS:BBH:0180 & $1.0$ & $+0.00$ & $+0.00$ & $5.1\times 10^{-5}$ & $0.01227$ & $28.2$ \\

  \doubleline
\end{longtable}
\label{tab:original}
\endgroup

\subsection{SXS waveforms from Ref.~\cite{Kumar:2015tha,Lovelace:2010ne,Scheel:2014ina}}
\label{sec:publicnew}
\begingroup
\begin{longtable}{ccccccc}
  \doubleline
  ID & $q$ & $\chi_1$ & $\chi_2$ & $e$ & $M\omega_{22}$ & $N_\mathrm{orb}$\\
\hline
SXS:BBH:0177 & $1.0$ & $+0.99$ & $+0.99$ & $1.3\times 10^{-3}$ & $0.01543$ & $25.4$ \\
SXS:BBH:0178 & $1.0$ & $+0.99$ & $+0.99$ & $8.6\times 10^{-4}$ & $0.01570$ & $25.4$ \\
SXS:BBH:0202 & $7.0$ & $+0.60$ & $+0.00$ & $9.0\times 10^{-5}$ & $0.01324$ & $62.1$ \\
SXS:BBH:0203 & $7.0$ & $+0.40$ & $+0.00$ & $1.4\times 10^{-5}$ & $0.01322$ & $58.5$ \\
SXS:BBH:0204 & $7.0$ & $+0.40$ & $+0.00$ & $1.7\times 10^{-4}$ & $0.01044$ & $88.4$ \\
SXS:BBH:0205 & $7.0$ & $-0.40$ & $+0.00$ & $7.0\times 10^{-5}$ & $0.01325$ & $44.9$ \\
SXS:BBH:0206 & $7.0$ & $-0.40$ & $+0.00$ & $1.6\times 10^{-4}$ & $0.01037$ & $73.2$ \\
SXS:BBH:0207 & $7.0$ & $-0.60$ & $+0.00$ & $1.7\times 10^{-4}$ & $0.01423$ & $36.1$ \\
SXS:BBH:0306 & $1.3$ & $+0.96$ & $-0.90$ & $1.5\times 10^{-3}$ & $0.02098$ & $12.6$ \\

  \doubleline
\end{longtable}
\endgroup

\subsection{SXS waveforms from Ref.~\cite{Chu:2015kft}}
\label{sec:chu}
\begingroup
\begin{longtable}{ccccccc}
  \doubleline
  ID & $q$ & $\chi_1$ & $\chi_2$ & $e$ & $M\omega_{22}$ & $N_\mathrm{orb}$\\
\hline
\endfirsthead
ID & $q$ & $\chi_1$ & $\chi_2$ & $e$ & $M\omega_{22}$ & $N_\mathrm{orb}$\\
\hline
\endhead
SXS:BBH:0290 & $3.0$ & $+0.60$ & $+0.40$ & $9.0\times 10^{-5}$ & $0.01758$ & $24.2$ \\
SXS:BBH:0291 & $3.0$ & $+0.60$ & $+0.60$ & $5.0\times 10^{-5}$ & $0.01764$ & $24.5$ \\
SXS:BBH:0289 & $3.0$ & $+0.60$ & $+0.00$ & $2.3\times 10^{-4}$ & $0.01711$ & $23.8$ \\
SXS:BBH:0285 & $3.0$ & $+0.40$ & $+0.60$ & $1.6\times 10^{-4}$ & $0.01732$ & $23.8$ \\
SXS:BBH:0261 & $3.0$ & $-0.73$ & $+0.85$ & $1.0\times 10^{-4}$ & $0.01490$ & $21.5$ \\
SXS:BBH:0293 & $3.0$ & $+0.85$ & $+0.85$ & $9.0\times 10^{-5}$ & $0.01813$ & $25.6$ \\
SXS:BBH:0280 & $3.0$ & $+0.27$ & $+0.85$ & $9.7\times 10^{-5}$ & $0.01707$ & $23.6$ \\
SXS:BBH:0257 & $2.0$ & $+0.85$ & $+0.85$ & $1.1\times 10^{-4}$ & $0.01633$ & $24.8$ \\
SXS:BBH:0279 & $3.0$ & $+0.23$ & $-0.85$ & $6.0\times 10^{-5}$ & $0.01629$ & $22.6$ \\
SXS:BBH:0274 & $3.0$ & $-0.23$ & $+0.85$ & $1.6\times 10^{-4}$ & $0.01603$ & $22.4$ \\
SXS:BBH:0258 & $2.0$ & $+0.87$ & $-0.85$ & $1.8\times 10^{-4}$ & $0.01612$ & $22.8$ \\
SXS:BBH:0248 & $2.0$ & $+0.13$ & $+0.85$ & $7.0\times 10^{-5}$ & $0.01552$ & $23.2$ \\
SXS:BBH:0232 & $1.0$ & $+0.90$ & $+0.50$ & $2.8\times 10^{-4}$ & $0.01558$ & $23.9$ \\
SXS:BBH:0229 & $1.0$ & $+0.65$ & $+0.25$ & $3.1\times 10^{-4}$ & $0.01488$ & $23.1$ \\
SXS:BBH:0231 & $1.0$ & $+0.90$ & $+0.00$ & $1.0\times 10^{-4}$ & $0.01487$ & $23.1$ \\
SXS:BBH:0239 & $2.0$ & $-0.37$ & $+0.85$ & $9.1\times 10^{-5}$ & $0.01478$ & $22.2$ \\
SXS:BBH:0252 & $2.0$ & $+0.37$ & $-0.85$ & $3.8\times 10^{-4}$ & $0.01488$ & $22.5$ \\
SXS:BBH:0219 & $1.0$ & $-0.50$ & $+0.90$ & $3.3\times 10^{-4}$ & $0.01484$ & $22.4$ \\
SXS:BBH:0211 & $1.0$ & $-0.90$ & $+0.90$ & $2.6\times 10^{-4}$ & $0.01411$ & $22.3$ \\
SXS:BBH:0233 & $2.0$ & $-0.87$ & $+0.85$ & $6.0\times 10^{-5}$ & $0.01423$ & $22.0$ \\
SXS:BBH:0243 & $2.0$ & $-0.13$ & $-0.85$ & $1.8\times 10^{-4}$ & $0.01378$ & $23.3$ \\
SXS:BBH:0214 & $1.0$ & $-0.62$ & $-0.25$ & $1.9\times 10^{-4}$ & $0.01264$ & $24.4$ \\
SXS:BBH:0209 & $1.0$ & $-0.90$ & $-0.50$ & $1.7\times 10^{-4}$ & $0.01137$ & $27.0$ \\
SXS:BBH:0226 & $1.0$ & $+0.50$ & $-0.90$ & $2.4\times 10^{-4}$ & $0.01340$ & $22.9$ \\
SXS:BBH:0286 & $3.0$ & $+0.50$ & $+0.50$ & $8.0\times 10^{-5}$ & $0.01693$ & $24.1$ \\
SXS:BBH:0253 & $2.0$ & $+0.50$ & $+0.50$ & $6.7\times 10^{-5}$ & $0.01397$ & $28.8$ \\
SXS:BBH:0267 & $3.0$ & $-0.50$ & $-0.50$ & $5.6\times 10^{-5}$ & $0.01410$ & $23.4$ \\
SXS:BBH:0218 & $1.0$ & $-0.50$ & $+0.50$ & $7.8\times 10^{-5}$ & $0.01217$ & $29.1$ \\
SXS:BBH:0238 & $2.0$ & $-0.50$ & $-0.50$ & $6.9\times 10^{-5}$ & $0.01126$ & $32.0$ \\
SXS:BBH:0288 & $3.0$ & $+0.60$ & $-0.40$ & $1.9\times 10^{-4}$ & $0.01729$ & $23.5$ \\
SXS:BBH:0287 & $3.0$ & $+0.60$ & $-0.60$ & $7.0\times 10^{-5}$ & $0.01684$ & $23.5$ \\
SXS:BBH:0283 & $3.0$ & $+0.30$ & $+0.30$ & $7.6\times 10^{-5}$ & $0.01646$ & $23.5$ \\
SXS:BBH:0282 & $3.0$ & $+0.30$ & $+0.00$ & $7.5\times 10^{-5}$ & $0.01629$ & $23.3$ \\
SXS:BBH:0281 & $3.0$ & $+0.30$ & $-0.30$ & $6.7\times 10^{-5}$ & $0.01618$ & $23.2$ \\
SXS:BBH:0277 & $3.0$ & $+0.00$ & $+0.30$ & $7.0\times 10^{-5}$ & $0.01595$ & $22.9$ \\
SXS:BBH:0284 & $3.0$ & $+0.40$ & $-0.60$ & $1.5\times 10^{-4}$ & $0.01656$ & $22.8$ \\
SXS:BBH:0278 & $3.0$ & $+0.00$ & $+0.60$ & $2.1\times 10^{-4}$ & $0.01623$ & $22.8$ \\
SXS:BBH:0256 & $2.0$ & $+0.60$ & $+0.60$ & $7.8\times 10^{-5}$ & $0.01598$ & $23.9$ \\
SXS:BBH:0230 & $1.0$ & $+0.80$ & $+0.80$ & $1.3\times 10^{-4}$ & $0.01542$ & $24.2$ \\
SXS:BBH:0255 & $2.0$ & $+0.60$ & $+0.00$ & $4.0\times 10^{-5}$ & $0.01580$ & $23.3$ \\
SXS:BBH:0276 & $3.0$ & $+0.00$ & $-0.30$ & $6.7\times 10^{-5}$ & $0.01559$ & $23.0$ \\
SXS:BBH:0251 & $2.0$ & $+0.30$ & $+0.30$ & $7.5\times 10^{-5}$ & $0.01514$ & $23.5$ \\
SXS:BBH:0250 & $2.0$ & $+0.30$ & $+0.00$ & $7.5\times 10^{-5}$ & $0.01503$ & $23.2$ \\
SXS:BBH:0271 & $3.0$ & $-0.30$ & $+0.00$ & $6.3\times 10^{-5}$ & $0.01508$ & $22.5$ \\
SXS:BBH:0249 & $2.0$ & $+0.30$ & $-0.30$ & $7.2\times 10^{-5}$ & $0.01478$ & $23.2$ \\
SXS:BBH:0275 & $3.0$ & $+0.00$ & $-0.60$ & $1.2\times 10^{-4}$ & $0.01569$ & $22.6$ \\
SXS:BBH:0254 & $2.0$ & $+0.60$ & $-0.60$ & $6.0\times 10^{-5}$ & $0.01541$ & $22.9$ \\
SXS:BBH:0269 & $3.0$ & $-0.40$ & $+0.60$ & $1.2\times 10^{-4}$ & $0.01563$ & $22.3$ \\
SXS:BBH:0225 & $1.0$ & $+0.40$ & $+0.80$ & $3.5\times 10^{-4}$ & $0.01536$ & $23.5$ \\
SXS:BBH:0270 & $3.0$ & $-0.30$ & $-0.30$ & $6.2\times 10^{-5}$ & $0.01482$ & $22.8$ \\
SXS:BBH:0245 & $2.0$ & $+0.00$ & $-0.30$ & $6.8\times 10^{-5}$ & $0.01441$ & $23.0$ \\
SXS:BBH:0242 & $2.0$ & $-0.30$ & $+0.30$ & $6.7\times 10^{-5}$ & $0.01417$ & $23.1$ \\
SXS:BBH:0223 & $1.0$ & $+0.30$ & $+0.00$ & $6.7\times 10^{-5}$ & $0.01402$ & $23.3$ \\
SXS:BBH:0241 & $2.0$ & $-0.30$ & $+0.00$ & $6.6\times 10^{-5}$ & $0.01394$ & $23.1$ \\
SXS:BBH:0240 & $2.0$ & $-0.30$ & $-0.30$ & $6.4\times 10^{-5}$ & $0.01359$ & $23.5$ \\
SXS:BBH:0222 & $1.0$ & $-0.30$ & $+0.00$ & $7.4\times 10^{-5}$ & $0.01324$ & $23.6$ \\
SXS:BBH:0228 & $1.0$ & $+0.60$ & $+0.60$ & $3.2\times 10^{-4}$ & $0.01543$ & $23.5$ \\
SXS:BBH:0247 & $2.0$ & $+0.00$ & $+0.60$ & $1.0\times 10^{-4}$ & $0.01530$ & $22.6$ \\
SXS:BBH:0263 & $3.0$ & $-0.60$ & $+0.60$ & $1.9\times 10^{-4}$ & $0.01526$ & $22.0$ \\
SXS:BBH:0266 & $3.0$ & $-0.60$ & $+0.40$ & $1.8\times 10^{-4}$ & $0.01488$ & $22.0$ \\
SXS:BBH:0227 & $1.0$ & $+0.60$ & $+0.00$ & $3.1\times 10^{-4}$ & $0.01452$ & $23.1$ \\
SXS:BBH:0221 & $1.0$ & $-0.40$ & $+0.80$ & $2.7\times 10^{-4}$ & $0.01440$ & $22.7$ \\
SXS:BBH:0237 & $2.0$ & $-0.60$ & $+0.60$ & $6.1\times 10^{-5}$ & $0.01433$ & $22.6$ \\
SXS:BBH:0244 & $2.0$ & $+0.00$ & $-0.60$ & $7.5\times 10^{-5}$ & $0.01422$ & $23.2$ \\
SXS:BBH:0217 & $1.0$ & $-0.60$ & $+0.60$ & $1.5\times 10^{-4}$ & $0.01421$ & $22.7$ \\
SXS:BBH:0215 & $1.0$ & $-0.60$ & $-0.60$ & $1.8\times 10^{-4}$ & $0.01189$ & $25.8$ \\
SXS:BBH:0262 & $3.0$ & $-0.60$ & $+0.00$ & $2.0\times 10^{-4}$ & $0.01473$ & $22.5$ \\
SXS:BBH:0213 & $1.0$ & $-0.80$ & $+0.80$ & $1.4\times 10^{-4}$ & $0.01435$ & $22.3$ \\
SXS:BBH:0265 & $3.0$ & $-0.60$ & $-0.40$ & $9.0\times 10^{-5}$ & $0.01422$ & $23.4$ \\
SXS:BBH:0264 & $3.0$ & $-0.60$ & $-0.60$ & $2.8\times 10^{-4}$ & $0.01410$ & $23.4$ \\
SXS:BBH:0224 & $1.0$ & $+0.40$ & $-0.80$ & $2.5\times 10^{-4}$ & $0.01361$ & $22.9$ \\
SXS:BBH:0236 & $2.0$ & $-0.60$ & $+0.00$ & $1.2\times 10^{-4}$ & $0.01361$ & $23.4$ \\
SXS:BBH:0216 & $1.0$ & $-0.60$ & $+0.00$ & $2.6\times 10^{-4}$ & $0.01300$ & $23.6$ \\
SXS:BBH:0235 & $2.0$ & $-0.60$ & $-0.60$ & $6.1\times 10^{-5}$ & $0.01274$ & $25.1$ \\
SXS:BBH:0220 & $1.0$ & $-0.40$ & $-0.80$ & $1.0\times 10^{-4}$ & $0.01195$ & $25.7$ \\
SXS:BBH:0212 & $1.0$ & $-0.80$ & $-0.80$ & $2.4\times 10^{-4}$ & $0.01087$ & $28.6$ \\
SXS:BBH:0303 & $10.0$ & $+0.00$ & $+0.00$ & $5.1\times 10^{-5}$ & $0.02395$ & $19.3$ \\
SXS:BBH:0300 & $8.5$ & $+0.00$ & $+0.00$ & $5.7\times 10^{-5}$ & $0.02311$ & $18.7$ \\
SXS:BBH:0299 & $7.5$ & $+0.00$ & $+0.00$ & $5.9\times 10^{-5}$ & $0.02152$ & $20.1$ \\
SXS:BBH:0298 & $7.0$ & $+0.00$ & $+0.00$ & $6.1\times 10^{-5}$ & $0.02130$ & $19.7$ \\
SXS:BBH:0297 & $6.5$ & $+0.00$ & $+0.00$ & $6.4\times 10^{-5}$ & $0.02082$ & $19.7$ \\
SXS:BBH:0296 & $5.5$ & $+0.00$ & $+0.00$ & $5.2\times 10^{-5}$ & $0.01668$ & $27.9$ \\
SXS:BBH:0295 & $4.5$ & $+0.00$ & $+0.00$ & $5.2\times 10^{-5}$ & $0.01577$ & $27.8$ \\
SXS:BBH:0259 & $2.5$ & $+0.00$ & $+0.00$ & $5.9\times 10^{-5}$ & $0.01346$ & $28.6$ \\
SXS:BBH:0292 & $3.0$ & $+0.73$ & $-0.85$ & $1.8\times 10^{-4}$ & $0.01749$ & $23.9$ \\
SXS:BBH:0268 & $3.0$ & $-0.40$ & $-0.60$ & $1.7\times 10^{-4}$ & $0.01473$ & $22.9$ \\
SXS:BBH:0234 & $2.0$ & $-0.85$ & $-0.85$ & $1.4\times 10^{-4}$ & $0.01147$ & $27.8$ \\
SXS:BBH:0273 & $3.0$ & $-0.27$ & $-0.85$ & $2.0\times 10^{-4}$ & $0.01487$ & $22.9$ \\
SXS:BBH:0210 & $1.0$ & $-0.90$ & $+0.00$ & $1.8\times 10^{-4}$ & $0.01248$ & $24.3$ \\
SXS:BBH:0260 & $3.0$ & $-0.85$ & $-0.85$ & $3.5\times 10^{-4}$ & $0.01285$ & $25.8$ \\
SXS:BBH:0302 & $9.5$ & $+0.00$ & $+0.00$ & $6.0\times 10^{-5}$ & $0.02366$ & $19.1$ \\
SXS:BBH:0301 & $9.0$ & $+0.00$ & $+0.00$ & $5.5\times 10^{-5}$ & $0.02338$ & $18.9$ \\
SXS:BBH:0272 & $3.0$ & $-0.30$ & $+0.30$ & $6.4\times 10^{-5}$ & $0.01521$ & $22.7$ \\
SXS:BBH:0246 & $2.0$ & $+0.00$ & $+0.30$ & $7.2\times 10^{-5}$ & $0.01514$ & $22.9$ \\

  \doubleline
\end{longtable}
\endgroup

\chapter{Comparing the nonspinning \texttt{SEOBNRv4HM} and \texttt{EOBNRv2HM} models}
\label{sec:EOBNRv2HM}

\begin{figure}[htb]
\centering
\includegraphics[width=0.7\textwidth]{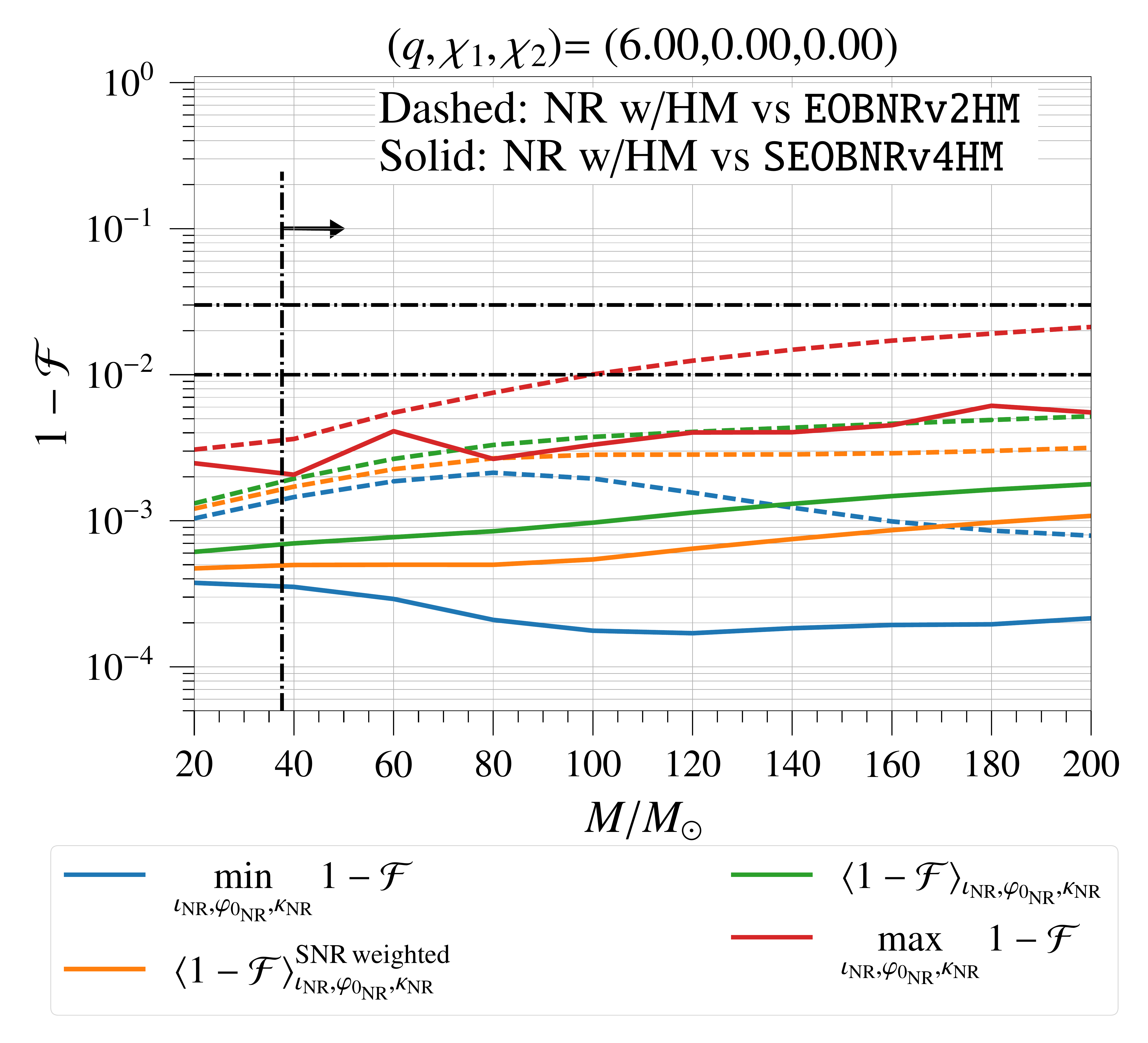}
\caption{Unfaithfulness $(1-\mathcal{F})$ in the mass range $20 M_\odot \leq M \leq 200 M_\odot$ for the configuration $(q = 6,\, \chi_1 = \chi_2 = 0)$. Dashed (plain) curves refer to results for \texttt{EOBNRv2HM} (\texttt{SEOBNRv4HM}). Plotted data as in Fig.~\ref{fig:unfaith_mass_q3chi085}.}
\label{fig:unfaith_mass_q6}
\end{figure}

Here we compare the nonspinning limit of \texttt{SEOBNRv4HM} to its predecessor,  
the \texttt{EOBNRv2HM} model developed in 2011~\cite{Pan:2011gk}, which is available 
in the LIGO Algorithm Library (LAL) and it has been used in Refs.~\cite{Capano:2013raa,Graff:2015bba,Harry:2016ijz}
to assess the importance of higher-order modes in Advanced LIGO searches and parameter estimation. The model \texttt{EOBNRv2HM} was also used 
to search for intermediate binary black holes~\cite{Aasi:2014iwa,Aasi:2014bqj,Aasi:2012rja,Virgo:2012aa}). The \texttt{EOBNRv2HM} model
includes the same higher-order modes as \texttt{SEOBNRv4HM}, that is 
$(2,2),(2,1),(3,3),(4,4),(5,5)$. Given that the
\texttt{EOBNRv2HM} model was calibrated against NR waveforms up to mass
ratio $q = 6$, we decide to compare first the two models for a configuration
with this mass ratio (\texttt{SXS:BBH:0166}). In
Fig.~\ref{fig:unfaith_mass_q6} we show the unfaithfulness results 
for maximum, minimum, average and SNR-weighted average with respect to the
angles $\iota_{\textrm{NR}},{\varphi_0}_{\textrm{NR}},\kappa_{\textrm{NR}}$ 
of the models against NR waveforms with the
modes $(\ell \leq 5,\, m\neq 0)$. The unfaithfulness is shown as a
function of total mass. The dashed (solid) lines
represent the results for \texttt{EOBNRv2HM} (\texttt{SEOBNRv4HM}).
The minimum of the unfaithfulness, reached for a face-on orientation,
is different for the two models and it is smaller for the
\texttt{SEOBNRv4HM} model. Since, for a face-on orientation, all the
higher-order modes included in the two models are exactly zero because of
the spherical harmonics, this difference is only due to a better
modeling of the dominant $(\ell, m) = (2,2)$ mode. This difference is
very small and both models yield a minimum of the unfaithfulness
much smaller than $1\%$ in the total mass range $20 M_\odot \leq M
\leq 200 M_\odot$. The most important quantity to compare is the
maximum of the unfaithfulness which is reached for an edge-on
orientation, where the higher-order modes are more relevant. Also in
this case the \texttt{SEOBNRv4HM} model has a lower unfaithfulness
against the NR waveform with respect to the \texttt{EOBNRv2HM}
model. In particular at a total mass of $M = 200 M_\odot$
\texttt{EOBNRv2HM} returns a maximum unfaithfulness $(1-\mathcal{F})
\sim 2\%$, while the \texttt{SEOBNRv4HM} model only $(1-\mathcal{F})
\sim 0.6\%$. This means that also the higher-order modes are better
modeled in \texttt{SEOBNRv4HM} with respect to \texttt{EOBNRv2HM}. 

We find that the model \texttt{SEOBNRv4HM} returns smaller values of the
unfaithfulness against the NR waveforms than the \texttt{EOBNRv2HM}
model for every nonspinning configuration in our NR catalog with $q \leq
6$. A comparison between the two models for mass ratio higher than $q
= 6$ is unfair because \texttt{EOBNRv2HM} is not calibrated in this
region. However it is worth mentioning that for the numerical
simulation with the largest mass ratio at our disposal $(q = 10)$ the
average unfaithfulness of \texttt{EOBNRv2HM} is larger than that of
\texttt{SEOBNRv4HM}, but still smaller than $1\%$ in the mass range
considered. For this configuration the value of the maximum of the
unfaithfulness is $(1-\mathcal{F}) \sim 3.5\%$ for \texttt{EOBNRv2HM}
at $M = 200 M_\odot$, while is $(1-\mathcal{F}) \sim 2\%$ for
\texttt{SEOBNRv4HM}.

\chapter{Comparing \texttt{SEOBNRv4HM} and numerical-relativity waveforms in time domain}
\label{sec:time_domain}

\begin{figure*}
  \centering
\includegraphics[scale=0.13]{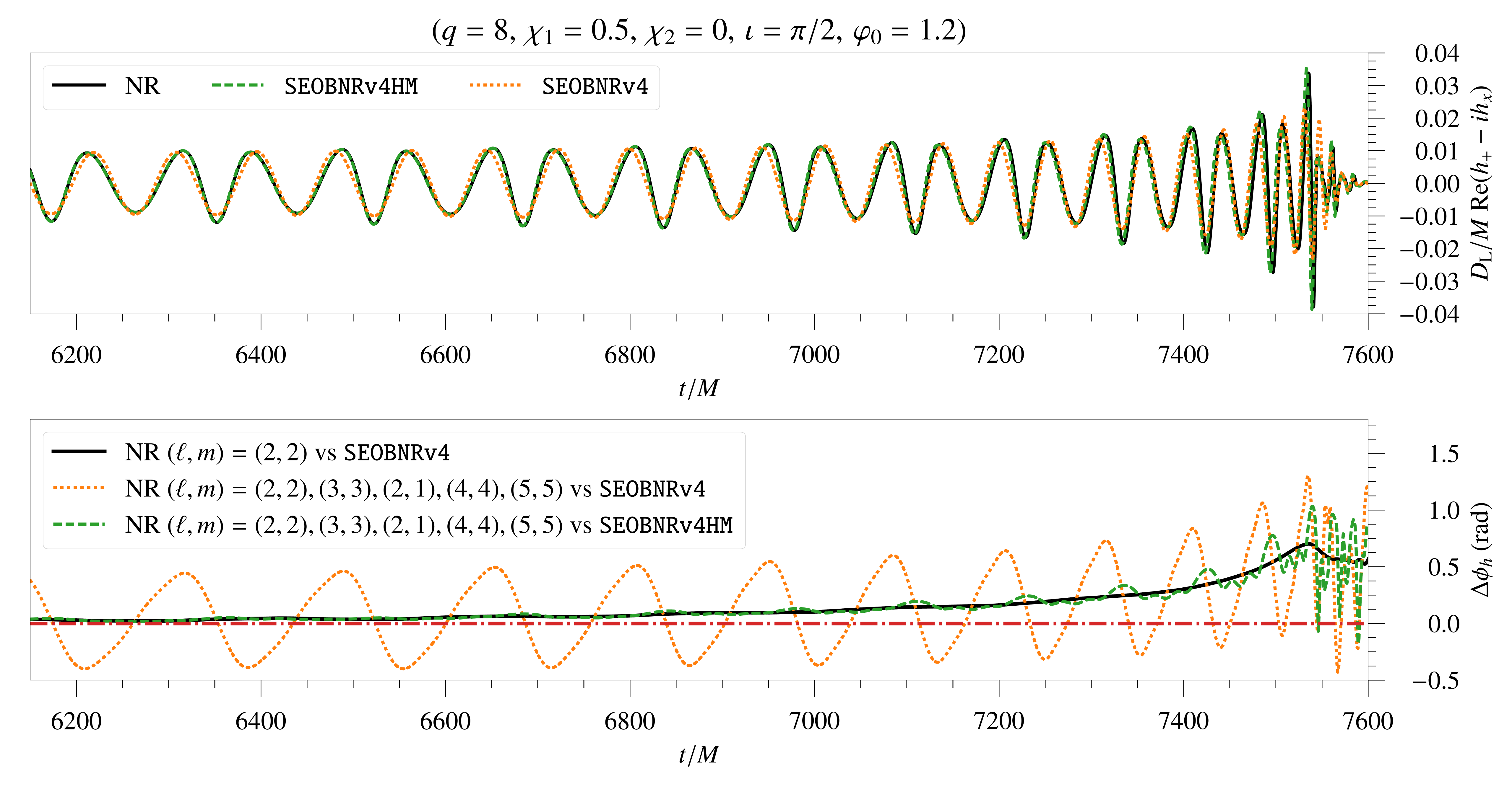}
\caption{Comparison between NR (solid black), \texttt{SEOBNRv4HM} (dashed green) and \texttt{SEOBNRv4} (dotted yellow) waveforms in an edge-on orientation $(\iota = \pi/2, \varphi_0 = 1.2)$ for the NR simulation \texttt{SXS:BBH:0065} $(q = 8,\, \chi_1 = 0.5,\, \chi_2 = 0)$. In the top panel is plotted the real part of the observer-frame's gravitational strain $h_+(\iota,\varphi_0;t) - i \ h_x(\iota,\varphi_0;t)$, while in the bottom panel the dephasing with the NR waveform $\Delta\phi_h$.The dotted-dashed red horizontal line in the bottom panel indicates zero dephasing with the NR waveform. Both \texttt{SEOBNRv4} and \texttt{SEOBNRv4HM} waveforms are phase aligned and time shifted at low frequency using as alignment window $t_{ini} = 1000 M$ and $t_{fin} = 3000 M$.}
\label{fig:q8v4vsHMwaveedgeon}
\end{figure*}

The improvement in waveform modeling obtained by including higher-order modes, 
can also be seen from a direct comparison of NR waveforms to \texttt{SEOBNRv4} and \texttt{SEOBNRv4HM} 
waveforms in time domain. We present this comparison in Fig.~\ref{fig:q8v4vsHMwaveedgeon} for the simulation
\texttt{SXS:BBH:0065}. We show the NR waveform with $(2,2),(2,1),(3,3),(4,4),(5,5)$ modes (solid black), the \texttt{SEOBNRv4HM} 
(dashed green) and \texttt{SEOBNRv4} (dotted yellow) waveforms in an edge-on 
orientation. The effect of neglecting higher-order modes results in an
oscillatory phase difference (dotted yellow curve of the bottom panel in Fig.~\ref{fig:q8v4vsHMwaveedgeon}) around the mean dephasing due to
the dominant $(2,2)$ mode (solid black curve of the same
panel). These oscillations in the dephasing are almost totally removed
up to merger when we include higher-order modes (dashed green of the
bottom panel in Fig.~\ref{fig:q8v4vsHMwaveedgeon}) where now the phase
difference with the NR waveform is dominated again by the discrepancy of the
$(2,2)$ mode. The residual oscillations of the dashed green curve
around the dephasing of the dominant $(2,2)$ mode is due
to the superposition of the different dephasing of the various
higher-order modes. The effect of the inclusion of higher-order modes
can be seen also in the amplitude of the waveform, in particular in
the last five cycle of the waveform there is an evident amplitude
difference between \texttt{SEOBNRv4} and NR waveforms, which is not present 
when the \texttt{SEOBNRv4HM} waveform is used.

\FloatBarrier


\chapter{Comparison of multipolar precessing models to numerical-relativity surrogate waveforms}
\label{sec:comparisonNRSurr}

In this appendix we compare directly \verb+SEOBNRv4PHM+ and \verb+IMRPhenomPv3HM+ to
the NR surrogate model \verb+NRSur7dq4+.  We choose a starting frequency corresponding to 20 Hz at 70 $M_{\odot}$ (this is  essentially the limit of the length for NR
surrogate waveforms). We generate 1000 random configurations, uniform
in mass ratio $q\in[1,4]$ and in spin magnitudes $\in[0,0.8]$,
and with random directions uniform on the unit sphere. The left panel  of  Fig.~\ref{fig:models_vs_NRsurr}
shows the summary of the unfaithfulness as a function of total mass for all the cases
considered, for \verb+IMRPhenomPv3HM+ and \verb+SEOBNRv4PHM+.  We see that the 
median and 95th percentile values for both models are close to the values in Fig.~\ref{fig:all_runs_percentiles}, with \verb+SEOBNRv4PHM+
having a median unfaithfulness below 1\% and \verb+IMRPhenomPv3HM+ about a factor of 3 larger. The right panel of Fig.~\ref{fig:models_vs_NRsurr} 
shows the maximum unfaithfulness distribution and the same trends are also observed. \verb+SEOBNRv4PHM+ outperforms 
\verb+IMRPhenomPv3HM+, with the median of the former being 4 times smaller than the one of the latter. Finally, to gain further insight 
into the behavior of the waveform models across the parameter space, we show in Fig.~\ref{fig:2d_models_vs_NRsurr} the maximum unfaithfulness 
as a function of mass ratio and the effective spin.

\clearpage

\chapter{Parameters of the new 118 NR simulations}
\label{sec:NRparam}

\begin{landscape}
\begin{longtable}{llllll}
            ID &     $q$ &                $\chi_{1}$ &                $\chi_{2}$ &  $M\Omega$ &  \# of  orbits \\
\hline
 PrecBBH000001 &  1.2499 &   (-0.272, -0.628, 0.414) &    (-0.212, -0.653, 0.41) &    0.01632 &        21 \\
 PrecBBH000002 &  1.2500 &    (-0.629, 0.202, 0.451) &    (-0.13, -0.708, 0.348) &    0.01645 &        20 \\
 PrecBBH000003 &  1.2499 &     (0.68, -0.104, 0.408) &    (0.71, -0.153, -0.335) &    0.01616 &        19 \\
 PrecBBH000004 &  1.2501 &    (0.309, -0.593, 0.439) &    (0.611, 0.325, -0.401) &    0.01627 &        18 \\
 PrecBBH000005 &  1.2500 &   (0.269, -0.684, -0.317) &       (0.393, -0.57, 0.4) &    0.01626 &        18 \\
 PrecBBH000006 &  1.2500 &   (0.561, -0.488, -0.293) &      (0.37, 0.611, 0.361) &    0.01623 &        18 \\
 PrecBBH000007 &  1.2499 &   (-0.671, 0.287, -0.328) &   (-0.694, 0.205, -0.339) &    0.01651 &        16 \\
 PrecBBH000008 &  1.2501 &     (-0.7, 0.269, -0.277) &  (-0.133, -0.669, -0.418) &    0.01653 &        16 \\
 PrecBBH000009 &  2.4998 &     (0.279, 0.579, 0.387) &     (0.138, 0.631, 0.381) &    0.01604 &        24 \\
 PrecBBH000010 &  2.5000 &     (-0.577, 0.26, 0.403) &   (-0.021, -0.679, 0.317) &    0.01633 &        24 \\
 PrecBBH000011 &  2.4999 &     (-0.604, 0.23, 0.381) &   (-0.608, 0.096, -0.428) &    0.01631 &        23 \\
 PrecBBH000012 &  2.4998 &    (-0.587, 0.238, 0.402) &   (-0.014, -0.576, -0.48) &    0.01630 &        23 \\
 PrecBBH000013 &  2.4998 &   (-0.531, 0.349, -0.399) &    (-0.65, -0.043, 0.371) &    0.01636 &        19 \\
 PrecBBH000014 &  2.4998 &   (-0.554, 0.332, -0.382) &    (0.012, -0.683, 0.309) &    0.01638 &        19 \\
 PrecBBH000015 &  2.4998 &    (0.052, 0.633, -0.399) &    (-0.096, 0.62, -0.411) &    0.01605 &        18 \\
 PrecBBH000016 &  2.4997 &    (0.615, 0.166, -0.396) &   (-0.326, 0.497, -0.457) &    0.01606 &        18 \\
 PrecBBH000017 &  3.4997 &     (0.421, 0.298, 0.306) &     (0.301, 0.417, 0.309) &    0.01598 &        27 \\
 PrecBBH000018 &  3.4992 &     (0.464, 0.218, 0.312) &    (-0.348, 0.402, 0.277) &    0.01599 &        27 \\
 PrecBBH000019 &  3.4996 &     (0.242, 0.455, 0.307) &    (0.127, 0.471, -0.349) &    0.01598 &        26 \\
 PrecBBH000020 &  3.4999 &     (0.514, -0.006, 0.31) &   (-0.139, 0.451, -0.371) &    0.01602 &        26 \\
 PrecBBH000021 &  3.4993 &     (-0.4, 0.297, -0.335) &   (-0.518, -0.054, 0.298) &    0.01631 &        22 \\
 PrecBBH000022 &  3.4995 &     (0.464, 0.18, -0.335) &    (-0.358, 0.395, 0.275) &    0.01605 &        22 \\
 PrecBBH000023 &  3.4991 &   (0.414, -0.273, -0.338) &    (0.472, -0.09, -0.358) &    0.01606 &        21 \\
 PrecBBH000024 &  3.4997 &   (0.256, -0.431, -0.329) &    (0.225, 0.401, -0.385) &    0.01609 &        21 \\
 PrecBBH000025 &  1.2501 &    (-0.661, 0.193, 0.407) &          (0.0, -0.0, 0.0) &    0.01645 &        19 \\
 PrecBBH000026 &  1.2501 &    (-0.466, -0.618, -0.2) &          (0.0, -0.0, 0.0) &    0.01638 &        17 \\
 PrecBBH000027 &  2.4999 &     (0.099, 0.637, 0.383) &          (0.0, -0.0, 0.0) &    0.01601 &        23 \\
 PrecBBH000028 &  2.5003 &    (0.557, 0.357, -0.354) &          (0.0, 0.0, -0.0) &    0.01609 &        19 \\
 PrecBBH000029 &  3.5006 &    (0.458, -0.242, 0.302) &           (0.0, 0.0, 0.0) &    0.01603 &        27 \\
 PrecBBH000030 &  3.4996 &   (-0.397, -0.32, -0.316) &          (0.0, -0.0, 0.0) &    0.01619 &        22 \\
 PrecBBH000031 &  1.0001 &    (-0.752, 0.179, 0.461) &         (-0.0, -0.0, 0.0) &    0.01646 &        19 \\
 PrecBBH000032 &  1.0002 &   (-0.836, 0.259, -0.206) &         (-0.0, -0.0, 0.0) &    0.01649 &        17 \\
 PrecBBH000033 &  2.0000 &    (-0.709, 0.269, 0.445) &         (-0.0, -0.0, 0.0) &    0.01638 &        22 \\
 PrecBBH000034 &  2.0004 &    (0.027, 0.793, -0.379) &         (-0.0, 0.0, -0.0) &    0.01605 &        18 \\
 PrecBBH000035 &  3.2002 &     (0.681, 0.112, 0.405) &           (0.0, 0.0, 0.0) &    0.01599 &        26 \\
 PrecBBH000036 &  3.1995 &    (0.162, 0.597, -0.507) &         (-0.0, -0.0, 0.0) &    0.01600 &        20 \\
 PrecBBH000037 &  3.9994 &    (0.596, -0.106, 0.352) &        (-0.0, -0.0, -0.0) &    0.01598 &        29 \\
 PrecBBH000038 &  4.0003 &  (-0.146, -0.481, -0.487) &         (-0.0, 0.0, -0.0) &    0.01613 &        22 \\
 PrecBBH000039 &  1.0000 &    (-0.542, 0.137, 0.332) &        (-0.0, -0.0, -0.0) &    0.01646 &        19 \\
 PrecBBH000040 &  1.0000 &   (-0.614, 0.183, -0.108) &        (-0.0, -0.0, -0.0) &    0.01649 &        17 \\
 PrecBBH000041 &  2.0001 &     (-0.48, 0.195, 0.303) &          (0.0, -0.0, 0.0) &    0.01639 &        21 \\
 PrecBBH000042 &  2.0003 &   (-0.509, 0.261, -0.181) &         (0.0, -0.0, -0.0) &    0.01644 &        19 \\
 PrecBBH000043 &  2.5002 &     (0.349, 0.252, 0.254) &          (0.0, 0.0, -0.0) &    0.01606 &        22 \\
 PrecBBH000044 &  2.5000 &     (0.456, 0.13, -0.158) &        (-0.0, -0.0, -0.0) &    0.01607 &        20 \\
 PrecBBH000045 &  3.9999 &    (-0.265, 0.146, 0.176) &          (0.0, 0.0, -0.0) &    0.01621 &        27 \\
 PrecBBH000046 &  4.0003 &    (0.25, -0.213, -0.122) &        (-0.0, -0.0, -0.0) &    0.01603 &        25 \\
 PrecBBH000047 &  2.9997 &     (0.249, 0.072, 0.152) &         (0.0, -0.0, -0.0) &    0.01604 &        23 \\
 PrecBBH000048 &  3.0000 &   (0.228, -0.183, -0.067) &        (-0.0, -0.0, -0.0) &    0.01610 &        22 \\
 PrecBBH000050 &  1.0001 &    (-0.709, 0.187, 0.522) &     (-0.18, -0.79, 0.391) &    0.01644 &        21 \\
 PrecBBH000051 &  1.0000 &    (-0.768, 0.118, 0.453) &   (-0.747, 0.299, -0.402) &    0.01646 &        18 \\
 PrecBBH000053 &  1.0000 &   (-0.747, 0.299, -0.402) &    (-0.768, 0.117, 0.453) &    0.01646 &        18 \\
 PrecBBH000054 &  1.0001 &    (-0.79, 0.265, -0.339) &   (-0.161, -0.801, 0.377) &    0.01648 &        18 \\
 PrecBBH000055 &  1.0000 &    (-0.748, 0.286, -0.41) &    (-0.748, 0.286, -0.41) &    0.01651 &        16 \\
 PrecBBH000056 &  1.0001 &   (-0.791, 0.266, -0.335) &   (-0.207, -0.71, -0.514) &    0.01655 &        15 \\
 PrecBBH000057 &  1.9997 &    (-0.715, 0.242, 0.452) &    (-0.757, 0.023, 0.448) &    0.01634 &        23 \\
 PrecBBH000058 &  2.0000 &    (-0.681, 0.276, 0.484) &   (-0.061, -0.797, 0.368) &    0.01636 &        23 \\
 PrecBBH000059 &  1.9997 &    (-0.725, 0.222, 0.447) &    (-0.706, 0.16, -0.499) &    0.01634 &        21 \\
 PrecBBH000060 &  2.0001 &    (-0.695, 0.242, 0.482) &  (-0.059, -0.655, -0.584) &    0.01636 &        21 \\
 PrecBBH000061 &  1.9998 &    (0.674, 0.294, -0.483) &     (0.529, 0.539, 0.451) &    0.01611 &        18 \\
 PrecBBH000062 &  1.9999 &  (-0.441, -0.618, -0.444) &    (0.762, -0.218, 0.381) &    0.01632 &        18 \\
 PrecBBH000063 &  1.9998 &   (-0.628, 0.392, -0.475) &     (-0.7, 0.137, -0.514) &    0.01643 &        16 \\
 PrecBBH000064 &  2.0000 &   (-0.188, 0.727, -0.458) &    (-0.608, -0.3, -0.561) &    0.01604 &        16 \\
 PrecBBH000065 &  3.1997 &    (-0.633, 0.268, 0.409) &   (-0.689, -0.046, 0.403) &    0.01622 &        27 \\
 PrecBBH000066 &  3.1998 &    (-0.611, 0.292, 0.426) &    (0.012, -0.728, 0.331) &    0.01623 &        27 \\
 PrecBBH000067 &  3.1996 &     (0.606, 0.327, 0.408) &    (0.436, 0.335, -0.581) &    0.01598 &        26 \\
 PrecBBH000068 &  3.1998 &     (-0.624, 0.27, 0.421) &   (0.018, -0.487, -0.634) &    0.01623 &        25 \\
 PrecBBH000069 &  3.1995 &   (-0.444, 0.373, -0.551) &   (-0.692, -0.085, 0.391) &    0.01634 &        20 \\
 PrecBBH000070 &  3.1992 &    (-0.51, -0.29, -0.544) &     (0.632, -0.342, 0.35) &    0.01627 &        20 \\
 PrecBBH000071 &  3.1991 &      (0.4, 0.409, -0.559) &    (0.228, 0.504, -0.577) &    0.01611 &        18 \\
 PrecBBH000072 &  3.1994 &    (0.245, 0.527, -0.549) &    (-0.51, 0.053, -0.613) &    0.01600 &        18 \\
 PrecBBH000073 &  4.0002 &     (0.604, 0.004, 0.354) &     (0.559, 0.214, 0.363) &    0.01597 &        30 \\
 PrecBBH000074 &  3.9992 &   (-0.004, -0.595, 0.369) &     (0.573, 0.241, 0.322) &    0.01607 &        30 \\
 PrecBBH000075 &  4.0000 &    (-0.549, 0.252, 0.354) &   (-0.441, 0.048, -0.542) &    0.01616 &        28 \\
 PrecBBH000076 &  3.9993 &    (-0.538, 0.262, 0.363) &   (0.034, -0.402, -0.572) &    0.01618 &        28 \\
 PrecBBH000077 &  4.0003 &   (-0.361, 0.309, -0.513) &     (-0.6, -0.101, 0.345) &    0.01623 &        22 \\
 PrecBBH000078 &  4.0001 &    (0.466, 0.089, -0.515) &    (-0.366, 0.503, 0.321) &    0.01604 &        22 \\
 PrecBBH000079 &  4.0003 &   (-0.435, 0.179, -0.518) &   (-0.416, -0.118, -0.55) &    0.01624 &        21 \\
 PrecBBH000080 &  4.0000 &    (0.139, 0.456, -0.513) &   (-0.422, -0.03, -0.557) &    0.01599 &        21 \\
 PrecBBH000081 &  1.0000 &     (-0.545, 0.12, 0.333) &     (-0.545, 0.12, 0.333) &    0.01643 &        20 \\
 PrecBBH000082 &  1.0000 &    (-0.519, 0.141, 0.365) &   (-0.132, -0.565, 0.293) &    0.01645 &        20 \\
 PrecBBH000083 &  1.0000 &     (-0.549, 0.107, 0.33) &   (-0.581, 0.198, -0.213) &    0.01646 &        18 \\
 PrecBBH000084 &  1.0000 &     (-0.52, 0.126, 0.369) &  (-0.163, -0.574, -0.258) &    0.01648 &        18 \\
 PrecBBH000085 &  1.0000 &   (-0.582, 0.197, -0.213) &      (-0.55, 0.106, 0.33) &    0.01646 &        18 \\
 PrecBBH000086 &  1.0000 &   (-0.596, 0.188, -0.176) &    (-0.123, -0.57, 0.286) &    0.01648 &        18 \\
 PrecBBH000087 &  1.0000 &   (-0.582, 0.192, -0.215) &   (-0.582, 0.192, -0.215) &    0.01650 &        17 \\
 PrecBBH000088 &  1.0000 &     (-0.6, 0.181, -0.172) &  (-0.151, -0.573, -0.266) &    0.01651 &        16 \\
 PrecBBH000089 &  1.9999 &    (0.513, -0.058, 0.305) &     (0.511, 0.046, 0.311) &    0.01615 &        22 \\
 PrecBBH000090 &  2.0003 &    (-0.467, 0.197, 0.322) &   (-0.039, -0.537, 0.264) &    0.01638 &        22 \\
 PrecBBH000091 &  2.0001 &     (0.278, 0.433, 0.308) &      (0.238, 0.5, -0.231) &    0.01604 &        21 \\
 PrecBBH000092 &  1.9999 &     (-0.47, 0.186, 0.323) &  (-0.041, -0.533, -0.273) &    0.01638 &        21 \\
 PrecBBH000093 &  1.9999 &   (-0.495, 0.257, -0.221) &    (-0.518, 0.005, 0.302) &    0.01639 &        19 \\
 PrecBBH000094 &  1.9999 &   (0.553, -0.092, -0.214) &    (-0.063, 0.531, 0.272) &    0.01612 &        19 \\
 PrecBBH000095 &  1.9999 &   (-0.494, 0.258, -0.221) &   (-0.544, 0.074, -0.242) &    0.01641 &        18 \\
 PrecBBH000096 &  1.9998 &  (-0.532, -0.185, -0.206) &   (0.332, -0.419, -0.273) &    0.01637 &        18 \\
 PrecBBH000097 &  2.4999 &      (0.003, 0.43, 0.255) &    (-0.085, 0.423, 0.252) &    0.01603 &        23 \\
 PrecBBH000098 &  2.5000 &      (0.075, 0.421, 0.26) &    (-0.435, 0.003, 0.246) &    0.01604 &        23 \\
 PrecBBH000099 &  2.5000 &      (0.128, 0.41, 0.256) &    (0.057, 0.461, -0.184) &    0.01601 &        22 \\
 PrecBBH000100 &  2.4999 &     (-0.06, 0.424, 0.259) &  (-0.435, -0.135, -0.206) &    0.01601 &        22 \\
 PrecBBH000101 &  2.5002 &   (0.195, -0.426, -0.176) &     (0.318, -0.294, 0.25) &    0.01616 &        20 \\
 PrecBBH000102 &  2.5001 &   (0.468, -0.022, -0.175) &    (-0.121, 0.428, 0.229) &    0.01610 &        20 \\
 PrecBBH000103 &  2.5002 &   (0.456, -0.095, -0.182) &    (0.459, 0.038, -0.194) &    0.01610 &        20 \\
 PrecBBH000104 &  2.5000 &    (0.293, 0.366, -0.173) &    (-0.41, 0.199, -0.206) &    0.01610 &        19 \\
 PrecBBH000105 &  4.0011 &    (-0.302, 0.006, 0.177) &    (-0.256, -0.156, 0.18) &    0.01616 &        28 \\
 PrecBBH000106 &  4.0001 &     (0.251, 0.165, 0.179) &    (-0.236, 0.197, 0.168) &    0.01596 &        28 \\
 PrecBBH000107 &  4.0008 &     (0.252, 0.166, 0.177) &    (0.198, 0.253, -0.138) &    0.01598 &        27 \\
 PrecBBH000108 &  3.9994 &     (0.038, -0.298, 0.18) &      (0.267, 0.17, -0.15) &    0.01608 &        27 \\
 PrecBBH000109 &  4.0002 &    (0.171, 0.277, -0.129) &     (0.048, 0.298, 0.177) &    0.01592 &        25 \\
 PrecBBH000110 &  4.0005 &  (-0.122, -0.303, -0.125) &     (0.305, 0.038, 0.168) &    0.01611 &        25 \\
 PrecBBH000111 &  3.9999 &     (0.278, 0.168, -0.13) &    (0.193, 0.257, -0.138) &    0.01597 &        24 \\
 PrecBBH000112 &  4.0001 &     (0.249, 0.21, -0.128) &   (-0.274, 0.165, -0.143) &    0.01598 &        24 \\
 PrecBBH000113 &  3.0002 &    (-0.232, 0.115, 0.152) &   (-0.257, -0.004, 0.154) &    0.01627 &        24 \\
 PrecBBH000114 &  3.0001 &    (0.001, -0.257, 0.154) &     (0.243, 0.096, 0.148) &    0.01614 &        24 \\
 PrecBBH000115 &  3.0000 &     (0.123, 0.227, 0.153) &    (0.069, 0.281, -0.078) &    0.01601 &        23 \\
 PrecBBH000116 &  2.9997 &     (0.236, 0.102, 0.154) &   (-0.179, 0.226, -0.083) &    0.01615 &        23 \\
 PrecBBH000117 &  2.9999 &   (-0.252, 0.144, -0.074) &   (-0.258, -0.009, 0.153) &    0.01629 &        22 \\
 PrecBBH000118 &  3.0001 &   (0.163, -0.242, -0.071) &      (0.14, 0.222, 0.144) &    0.01611 &        22 \\
 PrecBBH000119 &  3.0002 &     (0.018, 0.29, -0.073) &   (-0.055, 0.285, -0.076) &    0.01600 &        22 \\
 PrecBBH000120 &  3.0001 &   (-0.253, 0.144, -0.071) &    (0.011, -0.286, -0.09) &    0.01631 &        22 \\
\hline
\caption{The parameters of the runs in the new precessing catalog. Note that all the parameters are provided at the relaxed time and in the LAL source frame\cite{Schmidt:2017btt}.}

\end{longtable}

\end{landscape}


\chapter{Differences between \texttt{ LALInference} and \texttt{ RIFT}}
\label{RIFTtroubles}

\begin{table}[]
\centering
  \begin{tabular}{cccccc}
    \hline \hline
Parameter    &$q$ &$\chi_{\mathrm{eff}}$ \\ 
\hline
\texttt{ SEOBNRv4HM}  ($\chi$) &$0.59^{+0.34}_{-0.24}$& $0.29^{+0.25}_{-0.30}$ \\
NR/\NRSur{}   ($\chi$) & $0.64^{+0.32}_{-0.24}$& $0.33^{+0.22}_{-0.26}$ \\
NR/\NRSur{}   ($\chi_z$) & $0.67^{+0.30}_{-0.24}$& $0.40^{+0.22}_{-0.24}$ \\
NR/\NRSur{} HM  ($\chi$) & $0.58^{+0.34}_{-0.24}$& $0.29^{+0.23}_{-0.28}$ \\
NR/\NRSur{} HM  ($\chi_z$) & $0.62^{+0.32}_{-0.23}$& $0.37^{+0.25}_{-0.24}$\\
NR/\NRSur{}   (1g) &  $0.91^{+0.08}_{-0.12}$& $0.38^{+0.20}_{-0.25}$ \\
NR/\NRSur{}   (2g)&$0.58^{+0.23}_{-0.20}$& $0.34^{+0.19}_{-0.22}$ \\
NR/\NRSur{} HM  (1g) & $0.91^{+0.09}_{-0.12}$& $0.38^{+0.21}_{-0.23}$\\
NR/\NRSur{} HM  (2g) &$0.55^{+0.21}_{-0.19}$& $0.31^{+0.21}_{-0.23}$ \\
\end{tabular}
\caption{Estimates for the parameters of GW170729 obtained with \texttt{ RIFT} using various priors and waveform models. 
We quote median values and 90\% credible intervals for the the effective spin and HPD for the mass ratio. 
We follow similar notation as Tables~\ref{table:par-table} and~\ref{table:par-table-2g}.
}
\label{table:par-table_RIFT}
\end{table}

In Sec.~\ref{2HM} we discuss how higher-order modes affect the posterior distributions for the various source properties of GW170729.
We argue that waveform systematics are small since results with \texttt{ IMRPhenomHM} and \texttt{ SEOBNRv4HM} agree well with each other,
and the latter agrees well with NR waveform models. The investigation of waveform systematics, though, also reveals that there is a residual
small disagreement between results with \texttt{ SEOBNRv4HM} obtained with \texttt{ LALInference} and \texttt{ RIFT}. We have performed extensive 
investigations into the nature of this disagreement and have been ultimately unable to pinpoint its origin.

\texttt{ LALInference} and \texttt{ RIFT} are independently-implemented codes with differences in data-handling, likelihood estimation, algorithm, etc.
Despite these differences, in this work we have found good agreement between results obtained by the two algorithms for waveform approximants
that do not include higher-order modes, see Tables~\ref{table:par-table} and~\ref{table:par-table_RIFT}. However, for waveforms with higher-order modes and in particular the 
direct comparison using \texttt{ SEOBNRv4HM}, we find that the two codes produce results that differ for the mass ratio at the $7\%$ level. We also 
find that the two codes produce consistent results for the effective spin and the detector-frame total mass of GW170729, though
we are unable to check the source-frame total mass which \texttt{ RIFT} did not compute. See Table~\ref{table:par-table_RIFT} for more estimates.

We performed a number of reanalyses of the data in order to test the effects of various differences between the two algorithms.
On the \texttt{ RIFT} side these tests include: the NR grid, 
the specific choice of fitting coordinates, the noise PSD calculation, the data handling, the sampling rate, 
the lower frequency cut-off, the Monte-Carlo integration, the likelihood evaluation, the summation of higher-order modes to get the waveform, and the time window of the analysis. 
More technical details about these tests are presented in~\cite{Lange:2018}.
We also performed \texttt{ LALInference} runs ignoring the detector calibration uncertainty.
We found that none of these tests could account for the shift in the mass ratio posteriors.

Given that and the long history of testing and usage of \texttt{ LALInference}, in this paper we also follow previous studies by the LIGO/Virgo 
Collaborations, for example~\cite{Abbott:2018wiz,LIGOScientific:2018mvr}, and use \texttt{ LALInference}
 for our main results. We do note, though, that \texttt{ RIFT} results are qualitatively consistent and quantitatively close 
 to \texttt{ LALInference} and the discrepancy is only noticeable when higher-order modes are taken into account. 
 The small residual disagreement will be the focus of future investigations.

\end{appendices}

\backmatter
\pagenumbering{Roman}
\setcounter{page}{11}

\appendix

\bibliographystyle{hunsrt} 
\bibliography{literature}

\begin{thebibliography}{100}

\bibitem{Bohe:2016gbl}
Alejandro Bohé et~al.
\newblock {Improved effective-one-body model of spinning, nonprecessing binary
  black holes for the era of gravitational-wave astrophysics with advanced
  detectors}.
\newblock {\em Phys. Rev.}, D95(4):044028, 2017, 1611.03703.

\bibitem{Mroue:2013xna}
Abdul~H. Mroue et~al.
\newblock {Catalog of 174 Binary Black Hole Simulations for Gravitational Wave
  Astronomy}.
\newblock {\em Phys. Rev. Lett.}, 111(24):241104, 2013, 1304.6077.

\bibitem{Kumar:2015tha}
Prayush Kumar, Kevin Barkett, Swetha Bhagwat, Nousha Afshari, Duncan~A. Brown,
  Geoffrey Lovelace, Mark~A. Scheel, and Béla Szilágyi.
\newblock {Accuracy and precision of gravitational-wave models of inspiraling
  neutron star-black hole binaries with spin: Comparison with matter-free
  numerical relativity in the low-frequency regime}.
\newblock {\em Phys. Rev.}, D92(10):102001, 2015, 1507.00103.

\bibitem{Lovelace:2010ne}
Geoffrey Lovelace, Mark.A. Scheel, and Bela Szilagyi.
\newblock {Simulating merging binary black holes with nearly extremal spins}.
\newblock {\em Phys. Rev.}, D83:024010, 2011, 1010.2777.

\bibitem{Scheel:2014ina}
Mark~A. Scheel, Matthew Giesler, Daniel~A. Hemberger, Geoffrey Lovelace, Kevin
  Kuper, Michael Boyle, B.~Szilágyi, and Lawrence~E. Kidder.
\newblock {Improved methods for simulating nearly extremal binary black holes}.
\newblock {\em Class. Quant. Grav.}, 32(10):105009, 2015, 1412.1803.

\bibitem{Chu:2015kft}
Tony Chu, Heather Fong, Prayush Kumar, Harald~P. Pfeiffer, Michael Boyle,
  Daniel~A. Hemberger, Lawrence~E. Kidder, Mark~A. Scheel, and Bela Szilagyi.
\newblock {On the accuracy and precision of numerical waveforms: Effect of
  waveform extraction methodology}.
\newblock {\em Class. Quant. Grav.}, 33(16):165001, 2016, 1512.06800.

\bibitem{Einstein:1915by}
Albert Einstein.
\newblock {On the General Theory of Relativity}.
\newblock {\em Sitzungsber. Preuss. Akad. Wiss. Berlin (Math. Phys. )},
  1915:778--786, 1915.
\newblock [Addendum: Sitzungsber.Preuss.Akad.Wiss.Berlin (Math.Phys.) 1915,
  799--801 (1915)].

\bibitem{Einstein:1915ca}
Albert Einstein.
\newblock {The Field Equations of Gravitation}.
\newblock {\em Sitzungsber. Preuss. Akad. Wiss. Berlin (Math. Phys. )},
  1915:844--847, 1915.

\bibitem{VerrierThorieDM}
Le~Verrier and Urbain~Jean Joseph.
\newblock Th{\'e}orie du mouvement de mercure.

\bibitem{Will:2014kxa}
Clifford~M. Will.
\newblock {The Confrontation between General Relativity and Experiment}.
\newblock {\em Living Rev. Rel.}, 17:4, 2014, 1403.7377.

\bibitem{Schwarzschild:1916ae}
Karl Schwarzschild.
\newblock {On the gravitational field of a sphere of incompressible fluid
  according to Einstein's theory}.
\newblock {\em Sitzungsber. Preuss. Akad. Wiss. Berlin (Math. Phys. )},
  1916:424--434, 1916, physics/9912033.

\bibitem{Schwarzschild:1916uq}
Karl Schwarzschild.
\newblock {On the gravitational field of a mass point according to Einstein's
  theory}.
\newblock {\em Sitzungsber. Preuss. Akad. Wiss. Berlin (Math. Phys. )},
  1916:189--196, 1916, physics/9905030.

\bibitem{reissner_h_1916_1447315}
H.~Reissner.
\newblock {Über die Eigengravitation des elektrischen Feldes nach der
  Einsteinschen Theorie}, January 1916.

\bibitem{1918KNAB...20.1238N}
G.~{Nordstr{\"o}m}.
\newblock {On the Energy of the Gravitation field in Einstein's Theory}.
\newblock {\em Koninklijke Nederlandse Akademie van Wetenschappen Proceedings
  Series B Physical Sciences}, 20:1238--1245, January 1918.

\bibitem{Kerr:1963ud}
Roy~P. Kerr.
\newblock {Gravitational field of a spinning mass as an example of
  algebraically special metrics}.
\newblock {\em Phys. Rev. Lett.}, 11:237--238, 1963.

\bibitem{Newman:1965my}
E~T. Newman, R.~Couch, K.~Chinnapared, A.~Exton, A.~Prakash, and R.~Torrence.
\newblock {Metric of a Rotating, Charged Mass}.
\newblock {\em J. Math. Phys.}, 6:918--919, 1965.

\bibitem{Israel:1967wq}
Werner Israel.
\newblock {Event horizons in static vacuum space-times}.
\newblock {\em Phys. Rev.}, 164:1776--1779, 1967.

\bibitem{Israel:1967za}
Werner Israel.
\newblock {Event horizons in static electrovac space-times}.
\newblock {\em Commun. Math. Phys.}, 8:245--260, 1968.

\bibitem{Carter:1971zc}
B.~Carter.
\newblock {Axisymmetric Black Hole Has Only Two Degrees of Freedom}.
\newblock {\em Phys. Rev. Lett.}, 26:331--333, 1971.

\bibitem{Schodel:2003gy}
R.~Schodel, T.~Ott, R.~Genzel, A.~Eckart, N.~Mouawad, and T.~Alexander.
\newblock {Stellar dynamics in the central arcsecond of our Galaxy}.
\newblock {\em Astrophys. J.}, 596:1015--1034, 2003, astro-ph/0306214.

\bibitem{Ghez:2003qj}
A.M. Ghez, S.~Salim, Seth~D. Hornstein, A.~Tanner, M.~Morris, E.E. Becklin, and
  G.~Duchene.
\newblock {Stellar orbits around the galactic center black hole}.
\newblock {\em Astrophys. J.}, 620:744--757, 2005, astro-ph/0306130.

\bibitem{Barth:2003qu}
Aaron~J. Barth, Paul Martini, Charles~H. Nelson, and Luis~C. Ho.
\newblock {Iron emission in the z = 6.4 quasar SDSS J114816.64+525150.3}.
\newblock {\em Astrophys. J. Lett.}, 594:L95--L98, 2003, astro-ph/0308005.

\bibitem{Willott:2005pj}
Chris~J. Willott, Will~J. Percival, Ross~J. McLure, David Crampton, John~B.
  Hutchings, Matt~J. Jarvis, Marcin Sawicki, and Luc Simard.
\newblock {Imaging of SDSS z \ensuremath{>} 6 quasar fields: Gravitational
  lensing, companion galaxies and the host dark matter halos}.
\newblock {\em Astrophys. J.}, 626:657--665, 2005, astro-ph/0503202.

\bibitem{1972Natur.235..271B}
C.~T. {Bolton}.
\newblock {Identification of Cygnus X-1 with HDE 226868}.
\newblock {\em Nature}, 235(5336):271--273, February 1972.

\bibitem{Webster:1972bsw}
B.~Louise Webster and Paul Murdin.
\newblock {Cygnus X-1-a Spectroscopic Binary with a Heavy Companion ?}
\newblock {\em Nature}, 235:37--38, 1972.

\bibitem{doi:10.1146/annurev.astro.44.051905.092532}
Ronald~A. Remillard and Jeffrey~E. McClintock.
\newblock X-ray properties of black-hole binaries.
\newblock {\em Annual Review of Astronomy and Astrophysics}, 44(1):49--92,
  2006, https://doi.org/10.1146/annurev.astro.44.051905.092532.

\bibitem{Abbott:2016blz}
B.~P. Abbott et~al.
\newblock {Observation of Gravitational Waves from a Binary Black Hole Merger}.
\newblock {\em Phys. Rev. Lett.}, 116(6):061102, 2016, 1602.03837.

\bibitem{Abbott:2016nmj}
B.~P. Abbott et~al.
\newblock {GW151226: Observation of Gravitational Waves from a 22-Solar-Mass
  Binary Black Hole Coalescence}.
\newblock {\em Phys. Rev. Lett.}, 116(24):241103, 2016, 1606.04855.

\bibitem{Abbott:2017vtc}
Benjamin~P. Abbott et~al.
\newblock {GW170104: Observation of a 50-Solar-Mass Binary Black Hole
  Coalescence at Redshift 0.2}.
\newblock {\em Phys. Rev. Lett.}, 118(22):221101, 2017, 1706.01812.
\newblock [Erratum: Phys. Rev. Lett.121,no.12,129901(2018)].

\bibitem{Abbott:2017gyy}
B.~. P.~. Abbott et~al.
\newblock {GW170608: Observation of a 19-solar-mass Binary Black Hole
  Coalescence}.
\newblock {\em Astrophys. J.}, 851(2):L35, 2017, 1711.05578.

\bibitem{Abbott:2017oio}
Benjamin~P. Abbott et~al.
\newblock {GW170814: A Three-Detector Observation of Gravitational Waves from a
  Binary Black Hole Coalescence}.
\newblock {\em Phys. Rev. Lett.}, 119(14):141101, 2017, 1709.09660.

\bibitem{LIGOScientific:2018mvr}
B.~P. Abbott et~al.
\newblock {GWTC-1: A Gravitational-Wave Transient Catalog of Compact Binary
  Mergers Observed by LIGO and Virgo during the First and Second Observing
  Runs}.
\newblock {\em Phys. Rev.}, X9(3):031040, 2019, 1811.12907.

\bibitem{Abbott:2020tfl}
R.~Abbott et~al.
\newblock {GW190521: A Binary Black Hole Merger with a Total Mass of $150
  M_{\odot}$}.
\newblock {\em Phys. Rev. Lett.}, 125(10):101102, 2020, 2009.01075.

\bibitem{LIGOScientific:2020stg}
R.~Abbott et~al.
\newblock {GW190412: Observation of a Binary-Black-Hole Coalescence with
  Asymmetric Masses}.
\newblock {\em Phys. Rev. D}, 102(4):043015, 2020, 2004.08342.

\bibitem{Abbott:2020khf}
R.~Abbott et~al.
\newblock {GW190814: Gravitational Waves from the Coalescence of a 23 Solar
  Mass Black Hole with a 2.6 Solar Mass Compact Object}.
\newblock {\em Astrophys. J. Lett.}, 896(2):L44, 2020, 2006.12611.

\bibitem{Abbott:2020niy}
R.~Abbott et~al.
\newblock {GWTC-2: Compact Binary Coalescences Observed by LIGO and Virgo
  During the First Half of the Third Observing Run}.
\newblock 10 2020, 2010.14527.

\bibitem{2010A&ARv..18..279V}
Marta {Volonteri}.
\newblock {Formation of supermassive black holes}.
\newblock {\em Astron. Astrophys. Rev.}, 18(3):279--315, July 2010, 1003.4404.

\bibitem{Oppenheimer:1939ue}
J.R. Oppenheimer and H.~Snyder.
\newblock {On Continued gravitational contraction}.
\newblock {\em Phys. Rev.}, 56:455--459, 1939.

\bibitem{Penrose:1969pc}
R.~Penrose.
\newblock {Gravitational collapse: The role of general relativity}.
\newblock {\em Riv. Nuovo Cim.}, 1:252--276, 1969.

\bibitem{May:1966zz}
Michael~M. May and Richard~H. White.
\newblock {Hydrodynamic Calculations of General-Relativistic Collapse}.
\newblock {\em Phys. Rev.}, 141:1232--1241, 1966.

\bibitem{1967SvA....10..602Z}
Ya.~B. {Zel'dovich} and I.~D. {Novikov}.
\newblock {The Hypothesis of Cores Retarded during Expansion and the Hot
  Cosmological Model}.
\newblock {\em Soviet Ast.}, 10:602, February 1967.

\bibitem{Hawking:1971ei}
Stephen Hawking.
\newblock {Gravitationally collapsed objects of very low mass}.
\newblock {\em Mon. Not. Roy. Astron. Soc.}, 152:75, 1971.

\bibitem{1975ApJ...201....1C}
B.~J. {Carr}.
\newblock {The primordial black hole mass spectrum.}
\newblock {\em Astrophysical Journal}, 201:1--19, October 1975.

\bibitem{Sasaki:2018dmp}
Misao Sasaki, Teruaki Suyama, Takahiro Tanaka, and Shuichiro Yokoyama.
\newblock {Primordial black holes in gravitational wave astronomy}.
\newblock {\em Class. Quant. Grav.}, 35(6):063001, 2018, 1801.05235.

\bibitem{Gibbons:1975kk}
G.W. Gibbons.
\newblock {Vacuum Polarization and the Spontaneous Loss of Charge by Black
  Holes}.
\newblock {\em Commun. Math. Phys.}, 44:245--264, 1975.

\bibitem{Eardley:1975kp}
D.M. Eardley and W.H. Press.
\newblock {Astrophysical processes near black holes}.
\newblock {\em Ann. Rev. Astron. Astrophys.}, 13:381--422, 1975.

\bibitem{Einstein:1916cc}
Albert Einstein.
\newblock {Approximative Integration of the Field Equations of Gravitation}.
\newblock {\em Sitzungsber. Preuss. Akad. Wiss. Berlin (Math. Phys. )},
  1916:688--696, 1916.

\bibitem{1918SPAW.......154E}
Albert {Einstein}.
\newblock {{\"U}ber Gravitationswellen}.
\newblock {\em Sitzungsberichte der K{\"o}niglich Preu{\ss}ischen Akademie der
  Wissenschaften (Berlin}, pages 154--167, January 1918.

\bibitem{Hulse:1974eb}
R.A. Hulse and J.H. Taylor.
\newblock {Discovery of a pulsar in a binary system}.
\newblock {\em Astrophys. J. Lett.}, 195:L51--L53, 1975.

\bibitem{maggiore2008gravitational}
M.~Maggiore.
\newblock {\em Gravitational Waves: Volume 1: Theory and Experiments}.
\newblock Gravitational Waves. OUP Oxford, 2008.

\bibitem{1971ApJ...166..175I}
James~R. {Ipser}.
\newblock {Gravitational Radiation from Slowly Rotating, Fully Relativistic
  Stars}.
\newblock {\em Astrophysical Journal}, 166:175, May 1971.

\bibitem{Lasky:2015uia}
Paul~D. Lasky.
\newblock {Gravitational Waves from Neutron Stars: A Review}.
\newblock {\em Publ. Astron. Soc. Austral.}, 32:e034, 2015, 1508.06643.

\bibitem{Punturo:2010zz}
M.~Punturo et~al.
\newblock {The Einstein Telescope: A third-generation gravitational wave
  observatory}.
\newblock {\em Class. Quant. Grav.}, 27:194002, 2010.

\bibitem{Reitze:2019iox}
David Reitze et~al.
\newblock {Cosmic Explorer: The U.S. Contribution to Gravitational-Wave
  Astronomy beyond LIGO}.
\newblock {\em Bull. Am. Astron. Soc.}, 51:035, 7 2019, 1907.04833.

\bibitem{TheLIGOScientific:2014jea}
J.~Aasi et~al.
\newblock {Advanced LIGO}.
\newblock {\em Class. Quant. Grav.}, 32:074001, 2015, 1411.4547.

\bibitem{TheVirgo:2014hva}
F.~Acernese et~al.
\newblock {Advanced Virgo: a second-generation interferometric gravitational
  wave detector}.
\newblock {\em Class. Quant. Grav.}, 32(2):024001, 2015, 1408.3978.

\bibitem{Postnov:2014tza}
Konstantin~A. Postnov and Lev~R. Yungelson.
\newblock {The Evolution of Compact Binary Star Systems}.
\newblock {\em Living Rev. Rel.}, 17:3, 2014, 1403.4754.

\bibitem{Benacquista:2006ika}
Matthew~J. Benacquista.
\newblock {Relativistic Binaries in Globular Clusters}.
\newblock {\em Living Rev. Rel.}, 9:2, 2006.

\bibitem{Samsing:2013kua}
Johan Samsing, Morgan MacLeod, and Enrico Ramirez-Ruiz.
\newblock {The Formation of Eccentric Compact Binary Inspirals and the Role of
  Gravitational Wave Emission in Binary-Single Stellar Encounters}.
\newblock {\em Astrophys. J.}, 784:71, 2014, 1308.2964.

\bibitem{Samsing:2017xmd}
Johan Samsing.
\newblock {Eccentric Black Hole Mergers Forming in Globular Clusters}.
\newblock {\em Phys. Rev. D}, 97(10):103014, 2018, 1711.07452.

\bibitem{Flanagan:2005yc}
Eanna~E. Flanagan and Scott~A. Hughes.
\newblock {The Basics of gravitational wave theory}.
\newblock {\em New J. Phys.}, 7:204, 2005, gr-qc/0501041.

\bibitem{Blanchet:2013haa}
Luc Blanchet.
\newblock {Gravitational Radiation from Post-Newtonian Sources and Inspiralling
  Compact Binaries}.
\newblock {\em Living Rev. Rel.}, 17:2, 2014, 1310.1528.

\bibitem{1976ApJ...210..764W}
R.~V. {Wagoner} and C.~M. {Will}.
\newblock {Post-Newtonian gravitational radiation from orbiting point masses.}
\newblock {\em Astrophysical Journal}, 210:764--775, December 1976.

\bibitem{Kidder:1992fr}
Lawrence~E. Kidder, Clifford~M. Will, and Alan~G. Wiseman.
\newblock {Spin effects in the inspiral of coalescing compact binaries}.
\newblock {\em Phys. Rev. D}, 47(10):4183--4187, 1993, gr-qc/9211025.

\bibitem{Apostolatos:1994mx}
Theocharis~A. Apostolatos, Curt Cutler, Gerald~J. Sussman, and Kip~S. Thorne.
\newblock {Spin induced orbital precession and its modulation of the
  gravitational wave forms from merging binaries}.
\newblock {\em Phys. Rev.}, D49:6274--6297, 1994.

\bibitem{Racine:2008qv}
Etienne Racine.
\newblock {Analysis of spin precession in binary black hole systems including
  quadrupole-monopole interaction}.
\newblock {\em Phys.\ Rev.\ D}, 78:044021, 2008, 0803.1820.

\bibitem{Kidder:1995zr}
Lawrence~E. Kidder.
\newblock {Coalescing binary systems of compact objects to postNewtonian 5/2
  order. 5. Spin effects}.
\newblock {\em Phys. Rev.}, D52:821--847, 1995, gr-qc/9506022.

\bibitem{Damour:2014jta}
Thibault Damour, Piotr Jaranowski, and Gerhard Sch\"afer.
\newblock {Nonlocal-in-time action for the fourth post-Newtonian conservative
  dynamics of two-body systems}.
\newblock {\em Phys. Rev. D}, 89(6):064058, 2014, 1401.4548.

\bibitem{Bernard:2016wrg}
Laura Bernard, Luc Blanchet, Alejandro Boh\'e, Guillaume Faye, and Sylvain
  Marsat.
\newblock {Energy and periastron advance of compact binaries on circular orbits
  at the fourth post-Newtonian order}.
\newblock {\em Phys. Rev. D}, 95(4):044026, 2017, 1610.07934.

\bibitem{Bernard:2017ktp}
Laura Bernard, Luc Blanchet, Guillaume Faye, and Tanguy Marchand.
\newblock {Center-of-Mass Equations of Motion and Conserved Integrals of
  Compact Binary Systems at the Fourth Post-Newtonian Order}.
\newblock {\em Phys. Rev. D}, 97(4):044037, 2018, 1711.00283.

\bibitem{Foffa:2019rdf}
Stefano Foffa and Riccardo Sturani.
\newblock {Conservative dynamics of binary systems to fourth Post-Newtonian
  order in the EFT approach I: Regularized Lagrangian}.
\newblock {\em Phys. Rev. D}, 100(2):024047, 2019, 1903.05113.

\bibitem{Foffa:2019yfl}
Stefano Foffa, Rafael~A. Porto, Ira Rothstein, and Riccardo Sturani.
\newblock {Conservative dynamics of binary systems to fourth Post-Newtonian
  order in the EFT approach II: Renormalized Lagrangian}.
\newblock {\em Phys. Rev. D}, 100(2):024048, 2019, 1903.05118.

\bibitem{Hartung:2011te}
Johannes Hartung and Jan Steinhoff.
\newblock {Next-to-next-to-leading order post-Newtonian spin-orbit Hamiltonian
  for self-gravitating binaries}.
\newblock {\em Ann. Phys. (Berlin)}, 523:783--790, 2011, 1104.3079.

\bibitem{Hartung:2013dza}
Johannes Hartung, Jan Steinhoff, and Gerhard Schafer.
\newblock {Next-to-next-to-leading order post-Newtonian linear-in-spin binary
  Hamiltonians}.
\newblock {\em Annalen Phys.}, 525:359--394, 2013, 1302.6723.

\bibitem{Marsat:2012fn}
Sylvain Marsat, Alejandro Bohe, Guillaume Faye, and Luc Blanchet.
\newblock {Next-to-next-to-leading order spin-orbit effects in the equations of
  motion of compact binary systems}.
\newblock {\em Class. Quant. Grav.}, 30:055007, 2013, 1210.4143.

\bibitem{Bohe:2012mr}
Alejandro Bohe, Sylvain Marsat, Guillaume Faye, and Luc Blanchet.
\newblock {Next-to-next-to-leading order spin-orbit effects in the near-zone
  metric and precession equations of compact binaries}.
\newblock {\em Class. Quant. Grav.}, 30:075017, 2013, 1212.5520.

\bibitem{Levi:2015uxa}
Michele Levi and Jan Steinhoff.
\newblock {Next-to-next-to-leading order gravitational spin-orbit coupling via
  the effective field theory for spinning objects in the post-Newtonian
  scheme}.
\newblock {\em JCAP}, 1601:011, 2016, 1506.05056.

\bibitem{Levi:2020kvb}
Mich\`ele Levi, Andrew~J. Mcleod, and Matthew Von~Hippel.
\newblock {N$^3$LO gravitational spin-orbit coupling at order $G^4$}.
\newblock 3 2020, 2003.02827.

\bibitem{Hartung:2011ea}
Johannes Hartung and Jan Steinhoff.
\newblock {Next-to-next-to-leading order post-Newtonian spin(1)-spin(2)
  Hamiltonian for self-gravitating binaries}.
\newblock {\em Annalen Phys.}, 523:919--924, 2011, 1107.4294.

\bibitem{Levi:2011eq}
Michele Levi.
\newblock {Binary dynamics from spin1-spin2 coupling at fourth post-Newtonian
  order}.
\newblock {\em Phys. Rev. D}, 85:064043, 2012, 1107.4322.

\bibitem{Levi:2014sba}
Michele Levi and Jan Steinhoff.
\newblock {Equivalence of ADM Hamiltonian and Effective Field Theory approaches
  at next-to-next-to-leading order spin1-spin2 coupling of binary inspirals}.
\newblock {\em JCAP}, 12:003, 2014, 1408.5762.

\bibitem{Levi:2020uwu}
Mich\`ele Levi, Andrew~J. Mcleod, and Matthew Von~Hippel.
\newblock {NNNLO gravitational quadratic-in-spin interactions at the quartic
  order in G}.
\newblock 3 2020, 2003.07890.

\bibitem{Levi:2015ixa}
Michele Levi and Jan Steinhoff.
\newblock {Next-to-next-to-leading order gravitational spin-squared potential
  via the effective field theory for spinning objects in the post-Newtonian
  scheme}.
\newblock {\em JCAP}, 1601:008, 2016, 1506.05794.

\bibitem{Levi:2015msa}
Michele Levi and Jan Steinhoff.
\newblock {Spinning gravitating objects in the effective field theory in the
  post-Newtonian scheme}.
\newblock {\em JHEP}, 09:219, 2015, 1501.04956.

\bibitem{Levi:2016ofk}
Michele Levi and Jan Steinhoff.
\newblock {Complete conservative dynamics for inspiralling compact binaries
  with spins at fourth post-Newtonian order}.
\newblock 7 2016, 1607.04252.

\bibitem{Blanchet:2006gy}
Luc Blanchet, Alessandra Buonanno, and Guillaume Faye.
\newblock {Higher-order spin effects in the dynamics of compact binaries. II.
  Radiation field}.
\newblock {\em Phys. Rev. D}, 74:104034, 2006, gr-qc/0605140.
\newblock [Erratum: Phys.Rev.D 75, 049903 (2007), Erratum: Phys.Rev.D 81,
  089901 (2010)].

\bibitem{Buonanno:2012rv}
Alessandra Buonanno, Guillaume Faye, and Tanja Hinderer.
\newblock {Spin effects on gravitational waves from inspiraling compact
  binaries at second post-Newtonian order}.
\newblock {\em Phys. Rev.}, D87(4):044009, 2013, 1209.6349.

\bibitem{Marsatetal2017}
Sylvain Marsat.
\newblock Private communication.
\newblock {\em Private Communication}, 2018.

\bibitem{1993PhRvD..47.1511C}
Curt {Cutler}, Lee~Samuel {Finn}, Eric {Poisson}, and Gerald~Jay {Sussman}.
\newblock {Gravitational radiation from a particle in circular orbit around a
  black hole. II. Numerical results for the nonrotating case}.
\newblock {\em Phys. Rev. D}, 47(4):1511--1518, February 1993.

\bibitem{Poisson:1995vs}
Eric Poisson.
\newblock {Gravitational radiation from a particle in circular orbit around a
  black hole. 6. Accuracy of the postNewtonian expansion}.
\newblock {\em Phys. Rev. D}, 52:5719--5723, 1995, gr-qc/9505030.
\newblock [Addendum: Phys.Rev.D 55, 7980--7981 (1997)].

\bibitem{Buonanno:1998gg}
A.~Buonanno and T.~Damour.
\newblock {Effective one-body approach to general relativistic two-body
  dynamics}.
\newblock {\em Phys.Rev.}, D59:084006, 1999, gr-qc/9811091.

\bibitem{Buonanno:2000ef}
Alessandra Buonanno and Thibault Damour.
\newblock {Transition from inspiral to plunge in binary black hole
  coalescences}.
\newblock {\em Phys. Rev.}, D62:064015, 2000, gr-qc/0001013.

\bibitem{Thorne:1980ru}
K.~S. Thorne.
\newblock {Multipole Expansions of Gravitational Radiation}.
\newblock {\em Rev. Mod. Phys.}, 52:299--339, 1980.

\bibitem{Baumgarte:2002jm}
Thomas~W. Baumgarte and Stuart~L. Shapiro.
\newblock {Numerical relativity and compact binaries}.
\newblock {\em Phys. Rept.}, 376:41--131, 2003, gr-qc/0211028.

\bibitem{PhysRev.116.1322}
R.~Arnowitt, S.~Deser, and C.~W. Misner.
\newblock Dynamical structure and definition of energy in general relativity.
\newblock {\em Phys. Rev.}, 116:1322--1330, Dec 1959.

\bibitem{Arnowitt:1962hi}
Richard~L. Arnowitt, Stanley Deser, and Charles~W. Misner.
\newblock {The Dynamics of general relativity}.
\newblock {\em Gen. Rel. Grav.}, 40:1997--2027, 2008, gr-qc/0405109.

\bibitem{Pretorius:2005gq}
Frans Pretorius.
\newblock {Evolution of binary black hole spacetimes}.
\newblock {\em Phys. Rev. Lett.}, 95:121101, 2005, gr-qc/0507014.

\bibitem{Campanelli:2005dd}
Manuela Campanelli, C.~O. Lousto, P.~Marronetti, and Y.~Zlochower.
\newblock {Accurate evolutions of orbiting black-hole binaries without
  excision}.
\newblock {\em Phys. Rev. Lett.}, 96:111101, 2006, gr-qc/0511048.

\bibitem{Baker:2005vv}
John~G. Baker, Joan Centrella, Dae-Il Choi, Michael Koppitz, and James van
  Meter.
\newblock {Gravitational wave extraction from an inspiraling configuration of
  merging black holes}.
\newblock {\em Phys. Rev. Lett.}, 96:111102, 2006, gr-qc/0511103.

\bibitem{Jani:2016wkt}
Karan Jani, James Healy, James~A. Clark, Lionel London, Pablo Laguna, and
  Deirdre Shoemaker.
\newblock {Georgia Tech Catalog of Gravitational Waveforms}.
\newblock {\em Class. Quant. Grav.}, 33(20):204001, 2016, 1605.03204.

\bibitem{Healy:2017psd}
James Healy, Carlos~O. Lousto, Yosef Zlochower, and Manuela Campanelli.
\newblock {The RIT binary black hole simulations catalog}.
\newblock {\em Class. Quant. Grav.}, 34(22):224001, 2017, 1703.03423.

\bibitem{Boyle:2019kee}
Michael Boyle et~al.
\newblock {The SXS Collaboration catalog of binary black hole simulations}.
\newblock {\em Class. Quant. Grav.}, 36(19):195006, 2019, 1904.04831.

\bibitem{Healy:2019jyf}
James Healy, Carlos~O. Lousto, Jacob Lange, Richard O'Shaughnessy, Yosef
  Zlochower, and Manuela Campanelli.
\newblock {Second RIT binary black hole simulations catalog and its application
  to gravitational waves parameter estimation}.
\newblock {\em Phys. Rev.}, D100(2):024021, 2019, 1901.02553.

\bibitem{Healy:2020vre}
James Healy and Carlos~O. Lousto.
\newblock {The Third RIT binary black hole simulations catalog}.
\newblock 7 2020, 2007.07910.

\bibitem{Vishveshwara:1970zz}
C.V. Vishveshwara.
\newblock {Scattering of Gravitational Radiation by a Schwarzschild
  Black-hole}.
\newblock {\em Nature}, 227:936--938, 1970.

\bibitem{Press:1971wr}
William~H. Press.
\newblock {Long Wave Trains of Gravitational Waves from a Vibrating Black
  Hole}.
\newblock {\em Astrophys. J. Lett.}, 170:L105--L108, 1971.

\bibitem{Chandrasekhar:1975zza}
S.~Chandrasekhar and Steven~L. Detweiler.
\newblock {The quasi-normal modes of the Schwarzschild black hole}.
\newblock {\em Proc. Roy. Soc. Lond. A}, 344:441--452, 1975.

\bibitem{Teukolsky:1973ha}
Saul~A. Teukolsky.
\newblock {Perturbations of a rotating black hole. 1. Fundamental equations for
  gravitational electromagnetic and neutrino field perturbations}.
\newblock {\em Astrophys.J.}, 185:635--647, 1973.

\bibitem{Buonanno:2006ui}
Alessandra Buonanno, Gregory~B. Cook, and Frans Pretorius.
\newblock {Inspiral, merger and ring-down of equal-mass black-hole binaries}.
\newblock {\em Phys. Rev.}, D75:124018, 2007, gr-qc/0610122.

\bibitem{Berti:2006hb}
Emanuele Berti, Vitor Cardoso, and Clifford~M. Will.
\newblock {Considerations on the excitation of black hole quasinormal modes}.
\newblock {\em AIP Conf. Proc.}, 848(1):687--697, 2006, gr-qc/0601077.

\bibitem{Berti:2006wq}
Emanuele Berti and Vitor Cardoso.
\newblock {Quasinormal ringing of Kerr black holes. I. The Excitation factors}.
\newblock {\em Phys. Rev. D}, 74:104020, 2006, gr-qc/0605118.

\bibitem{Dorband:2006gg}
Ernst~Nils Dorband, Emanuele Berti, Peter Diener, Erik Schnetter, and Manuel
  Tiglio.
\newblock {A Numerical study of the quasinormal mode excitation of Kerr black
  holes}.
\newblock {\em Phys. Rev. D}, 74:084028, 2006, gr-qc/0608091.

\bibitem{Zhang:2013ksa}
Zhongyang Zhang, Emanuele Berti, and Vitor Cardoso.
\newblock {Quasinormal ringing of Kerr black holes. II. Excitation by particles
  falling radially with arbitrary energy}.
\newblock {\em Phys. Rev. D}, 88:044018, 2013, 1305.4306.

\bibitem{Taracchini:2014zpa}
Andrea Taracchini, Alessandra Buonanno, Gaurav Khanna, and Scott~A. Hughes.
\newblock {Small mass plunging into a Kerr black hole: Anatomy of the
  inspiral-merger-ringdown waveforms}.
\newblock {\em Phys. Rev.}, D90(8):084025, 2014, 1404.1819.

\bibitem{Hughes:2019zmt}
Scott~A. Hughes, Anuj Apte, Gaurav Khanna, and Halston Lim.
\newblock {Learning about black hole binaries from their ringdown spectra}.
\newblock {\em Phys. Rev. Lett.}, 123(16):161101, 2019, 1901.05900.

\bibitem{Lim:2019xrb}
Halston Lim, Gaurav Khanna, Anuj Apte, and Scott~A. Hughes.
\newblock {Exciting black hole modes via misaligned coalescences: II. The mode
  content of late-time coalescence waveforms}.
\newblock {\em Phys. Rev. D}, 100(8):084032, 2019, 1901.05902.

\bibitem{Ajith:2007qp}
Parameswaran Ajith et~al.
\newblock {Phenomenological template family for black-hole coalescence
  waveforms}.
\newblock {\em Class. Quant. Grav.}, 24:S689--S700, 2007, 0704.3764.

\bibitem{Ajith:2007kx}
P.~Ajith et~al.
\newblock {A Template bank for gravitational waveforms from coalescing binary
  black holes. I. Non-spinning binaries}.
\newblock {\em Phys. Rev.}, D77:104017, 2008, 0710.2335.
\newblock [Erratum: Phys. Rev.D79,129901(2009)].

\bibitem{Ajith:2009bn}
P.~Ajith et~al.
\newblock {Inspiral-merger-ringdown waveforms for black-hole binaries with
  non-precessing spins}.
\newblock {\em Phys. Rev. Lett.}, 106:241101, 2011, 0909.2867.

\bibitem{Hannam:2013oca}
Mark Hannam, Patricia Schmidt, Alejandro Bohé, Leïla Haegel, Sascha Husa,
  Frank Ohme, Geraint Pratten, and Michael Pürrer.
\newblock {Simple Model of Complete Precessing Black-Hole-Binary Gravitational
  Waveforms}.
\newblock {\em Phys. Rev. Lett.}, 113(15):151101, 2014, 1308.3271.

\bibitem{Schmidt:2014iyl}
Patricia Schmidt, Frank Ohme, and Mark Hannam.
\newblock {Towards models of gravitational waveforms from generic binaries II:
  Modelling precession effects with a single effective precession parameter}.
\newblock {\em Phys. Rev.}, D91(2):024043, 2015, 1408.1810.

\bibitem{Khan:2015jqa}
Sebastian Khan, Sascha Husa, Mark Hannam, Frank Ohme, Michael Pürrer, Xisco
  Jiménez~Forteza, and Alejandro Bohé.
\newblock {Frequency-domain gravitational waves from nonprecessing black-hole
  binaries. II. A phenomenological model for the advanced detector era}.
\newblock {\em Phys. Rev.}, D93(4):044007, 2016, 1508.07253.

\bibitem{Husa:2015iqa}
Sascha Husa, Sebastian Khan, Mark Hannam, Michael P{\"u}rrer, Frank Ohme, Xisco
  Jim{\'e}nez~Forteza, and Alejandro Boh{\'e}.
\newblock {Frequency-domain gravitational waves from nonprecessing black-hole
  binaries. I. New numerical waveforms and anatomy of the signal}.
\newblock {\em Phys. Rev.}, D93(4):044006, 2016, 1508.07250.

\bibitem{London:2017bcn}
Lionel London, Sebastian Khan, Edward Fauchon-Jones, Xisco~Jim{\'e}nez Forteza,
  Mark Hannam, Sascha Husa, Chinmay Kalaghatgi, Frank Ohme, and Francesco
  Pannarale.
\newblock {First higher-multipole model of spinning binary-black-hole
  gravitational waveforms}.
\newblock 2017, 1708.00404.

\bibitem{Khan:2018fmp}
Sebastian Khan, Katerina Chatziioannou, Mark Hannam, and Frank Ohme.
\newblock {Phenomenological model for the gravitational-wave signal from
  precessing binary black holes with two-spin effects}.
\newblock {\em Phys. Rev.}, D100(2):024059, 2019, 1809.10113.

\bibitem{Garcia-Quiros:2020qpx}
Cecilio García-Quirós, Marta Colleoni, Sascha Husa, Héctor Estellés,
  Geraint Pratten, Antoni Ramos-Buades, Maite Mateu-Lucena, and Rafel Jaume.
\newblock {IMRPhenomXHM: A multi-mode frequency-domain model for the
  gravitational wave signal from non-precessing black-hole binaries}.
\newblock 2020, 2001.10914.

\bibitem{Pratten:2020ceb}
Geraint Pratten et~al.
\newblock {Let's twist again: computationally efficient models for the dominant
  and sub-dominant harmonic modes of precessing binary black holes}.
\newblock 2020, 2004.06503.

\bibitem{Field:2013cfa}
Scott~E. Field, Chad~R. Galley, Jan~S. Hesthaven, Jason Kaye, and Manuel
  Tiglio.
\newblock {Fast prediction and evaluation of gravitational waveforms using
  surrogate models}.
\newblock {\em Phys. Rev.}, X4(3):031006, 2014, 1308.3565.

\bibitem{Blackman:2015pia}
Jonathan Blackman, Scott~E. Field, Chad~R. Galley, B\'ela Szil\'agyi, Mark~A.
  Scheel, Manuel Tiglio, and Daniel~A. Hemberger.
\newblock {Fast and Accurate Prediction of Numerical Relativity Waveforms from
  Binary Black Hole Coalescences Using Surrogate Models}.
\newblock {\em Phys. Rev. Lett.}, 115(12):121102, 2015, 1502.07758.

\bibitem{Blackman:2017dfb}
Jonathan Blackman, Scott~E. Field, Mark~A. Scheel, Chad~R. Galley, Daniel~A.
  Hemberger, Patricia Schmidt, and Rory Smith.
\newblock {A Surrogate Model of Gravitational Waveforms from Numerical
  Relativity Simulations of Precessing Binary Black Hole Mergers}.
\newblock {\em Phys. Rev. D}, 95(10):104023, 2017, 1701.00550.

\bibitem{Blackman:2017pcm}
Jonathan Blackman, Scott~E. Field, Mark~A. Scheel, Chad~R. Galley, Christian~D.
  Ott, Michael Boyle, Lawrence~E. Kidder, Harald~P. Pfeiffer, and Béla
  Szilágyi.
\newblock {Numerical relativity waveform surrogate model for generically
  precessing binary black hole mergers}.
\newblock {\em Phys. Rev.}, D96(2):024058, 2017, 1705.07089.

\bibitem{Varma:2018mmi}
Vijay Varma, Scott~E. Field, Mark~A. Scheel, Jonathan Blackman, Lawrence~E.
  Kidder, and Harald~P. Pfeiffer.
\newblock {Surrogate model of hybridized numerical relativity binary black hole
  waveforms}.
\newblock {\em Phys. Rev.}, D99(6):064045, 2019, 1812.07865.

\bibitem{Varma:2019csw}
Vijay Varma, Scott~E. Field, Mark~A. Scheel, Jonathan Blackman, Davide Gerosa,
  Leo~C. Stein, Lawrence~E. Kidder, and Harald~P. Pfeiffer.
\newblock {Surrogate models for precessing binary black hole simulations with
  unequal masses}.
\newblock {\em Phys. Rev. Research.}, 1:033015, 2019, 1905.09300.

\bibitem{Rifat:2019ltp}
Nur~E.M. Rifat, Scott~E. Field, Gaurav Khanna, and Vijay Varma.
\newblock {Surrogate model for gravitational wave signals from comparable and
  large-mass-ratio black hole binaries}.
\newblock {\em Phys. Rev. D}, 101(8):081502, 2020, 1910.10473.

\bibitem{PhysRev.117.306}
J.~Weber.
\newblock Detection and generation of gravitational waves.
\newblock {\em Phys. Rev.}, 117:306--313, Jan 1960.

\bibitem{Weber1961-WEBGRA}
J.~Weber.
\newblock {\em General Relativity and Gravitational Waves}.
\newblock New York: Interscience Publishers, 1961.

\bibitem{Astone:2003mp}
P.~Astone et~al.
\newblock {Methods and results of the IGEC search for burst gravitational waves
  in the years 1997 - 2000}.
\newblock {\em Phys. Rev. D}, 68:022001, 2003, astro-ph/0302482.

\bibitem{1978SvA....22...36S}
M.~V. {Sazhin}.
\newblock {Opportunities for detecting ultralong gravitational waves}.
\newblock {\em Sov. Astronom.}, 22:36--38, February 1978.

\bibitem{1979ApJ...234.1100D}
S.~{Detweiler}.
\newblock {Pulsar timing measurements and the search for gravitational waves}.
\newblock {\em Astrophysical Journal}, 234:1100--1104, December 1979.

\bibitem{1990ApJ...361..300F}
R.~S. {Foster} and D.~C. {Backer}.
\newblock {Constructing a Pulsar Timing Array}.
\newblock {\em Astrophysical Journal}, 361:300, September 1990.

\bibitem{Stodolsky:1978ks}
Leo Stodolsky.
\newblock {Matter and Light Wave Interferometry in Gravitational Fields}.
\newblock {\em Gen. Rel. Grav.}, 11:391--405, 1979.

\bibitem{Dimopoulos:2008sv}
Savas Dimopoulos, Peter~W. Graham, Jason~M. Hogan, Mark~A. Kasevich, and
  Surjeet Rajendran.
\newblock {An Atomic Gravitational Wave Interferometric Sensor (AGIS)}.
\newblock {\em Phys. Rev. D}, 78:122002, 2008, 0806.2125.

\bibitem{Graham:2012sy}
Peter~W. Graham, Jason~M. Hogan, Mark~A. Kasevich, and Surjeet Rajendran.
\newblock {A New Method for Gravitational Wave Detection with Atomic Sensors}.
\newblock {\em Phys. Rev. Lett.}, 110:171102, 2013, 1206.0818.

\bibitem{Martynov:2016fzi}
D.~V. Martynov et~al.
\newblock {Sensitivity of the Advanced LIGO detectors at the beginning of
  gravitational wave astronomy}.
\newblock {\em Phys. Rev.}, D93(11):112004, 2016, 1604.00439.

\bibitem{Meers:1988wp}
B.J. Meers.
\newblock {Recycling in Laser Interferometric Gravitational Wave Detectors}.
\newblock {\em Phys. Rev. D}, 38:2317--2326, 1988.

\bibitem{Mizuno:1993cj}
J.~Mizuno, K.A. Strain, P.G. Nelson, J.M. Chen, R.~Schilling, A.~Ruediger,
  W.~Winkler, and K.~Danzmann.
\newblock {Resonant sideband extraction: A New configuration for
  interferometric gravitational wave detectors}.
\newblock {\em Phys. Lett. A}, 175:273--276, 1993.

\bibitem{LIGOScientific:2019hgc}
Benjamin~P Abbott et~al.
\newblock {A guide to LIGO\textendash{}Virgo detector noise and extraction of
  transient gravitational-wave signals}.
\newblock {\em Class. Quant. Grav.}, 37(5):055002, 2020, 1908.11170.

\bibitem{Hawking:1987en}
S.W. Hawking and W.~Israel, editors.
\newblock {\em {THREE HUNDRED YEARS OF GRAVITATION}}.
\newblock 1987.

\bibitem{1987MNRAS.224..131S}
Bernard~F. {Schutz} and Massimo {Tinto}.
\newblock {Antenna patterns of interferometric detectors of gravitational waves
  - I. Linearly polarized waves.}
\newblock {\em Monthly Notices of the RAS}, 224:131--154, January 1987.

\bibitem{1988MNRAS.234..663D}
Sanjeev~V. {Dhurandhar} and Massimo {Tinto}.
\newblock {Astronomical observations with a network of detectors of
  gravitational waves. I - Mathematical framework and solution of the five
  detector problem}.
\newblock {\em Monthly Notices of the RAS}, 234:663--676, October 1988.

\bibitem{Sathyaprakash:1991mt}
B.~S. Sathyaprakash and S.~V. Dhurandhar.
\newblock {Choice of filters for the detection of gravitational waves from
  coalescing binaries}.
\newblock {\em Phys. Rev.}, D44:3819--3834, 1991.

\bibitem{Owen:1995tm}
Benjamin~J. Owen.
\newblock {Search templates for gravitational waves from inspiraling binaries:
  Choice of template spacing}.
\newblock {\em Phys. Rev.}, D53:6749--6761, 1996, gr-qc/9511032.

\bibitem{TheLIGOScientific:2016pea}
B.~P. Abbott et~al.
\newblock {Binary Black Hole Mergers in the first Advanced LIGO Observing Run}.
\newblock {\em Phys. Rev.}, X6(4):041015, 2016, 1606.04856.
\newblock [erratum: Phys. Rev.X8,no.3,039903(2018)].

\bibitem{DalCanton:2017ala}
Tito Dal~Canton and Ian~W. Harry.
\newblock {Designing a template bank to observe compact binary coalescences in
  Advanced LIGO's second observing run}.
\newblock 5 2017, 1705.01845.

\bibitem{Allen:2004gu}
Bruce Allen.
\newblock {A chi**2 time-frequency discriminator for gravitational wave
  detection}.
\newblock {\em Phys. Rev.}, D71:062001, 2005, gr-qc/0405045.

\bibitem{Babak:2012zx}
S.~Babak et~al.
\newblock {Searching for gravitational waves from binary coalescence}.
\newblock {\em Phys. Rev.}, D87(2):024033, 2013, 1208.3491.

\bibitem{Bose:1999pj}
Sukanta Bose, Archana Pai, and Sanjeev~V. Dhurandhar.
\newblock {Detection of gravitational waves from inspiraling compact binaries
  using a network of interferometric detectors}.
\newblock {\em Int. J. Mod. Phys. D}, 9:325--329, 2000, gr-qc/0002010.

\bibitem{Finn:2000hj}
Lee~Samuel Finn.
\newblock {Aperture synthesis for gravitational wave data analysis:
  Deterministic sources}.
\newblock {\em Phys. Rev. D}, 63:102001, 2001, gr-qc/0010033.

\bibitem{PhysRevD.72.063006}
Curt Cutler and Bernard~F. Schutz.
\newblock Generalized $\mathcal{F}$-statistic: Multiple detectors and multiple
  gravitational wave pulsars.
\newblock {\em Phys. Rev. D}, 72:063006, Sep 2005.

\bibitem{Harry:2010fr}
Ian~W. Harry and Stephen Fairhurst.
\newblock {A targeted coherent search for gravitational waves from compact
  binary coalescences}.
\newblock {\em Phys. Rev. D}, 83:084002, 2011, 1012.4939.

\bibitem{TheLIGOScientific:2017qsa}
B.~P. Abbott et~al.
\newblock {GW170817: Observation of Gravitational Waves from a Binary Neutron
  Star Inspiral}.
\newblock {\em Phys. Rev. Lett.}, 119(16):161101, 2017, 1710.05832.

\bibitem{Abbott:2020uma}
B.~P. Abbott et~al.
\newblock {GW190425: Observation of a Compact Binary Coalescence with Total
  Mass $\sim 3.4 M_{\odot}$}.
\newblock 2020, 2001.01761.

\bibitem{fisher1922mathematical}
Ronald~A Fisher.
\newblock On the mathematical foundations of theoretical statistics.
\newblock {\em Philosophical Transactions of the Royal Society of London.
  Series A, Containing Papers of a Mathematical or Physical Character},
  222(594-604):309--368, 1922.

\bibitem{Cutler:2007mi}
Curt Cutler and Michele Vallisneri.
\newblock {LISA detections of massive black hole inspirals: Parameter
  extraction errors due to inaccurate template waveforms}.
\newblock {\em Phys. Rev. D}, 76:104018, 2007, 0707.2982.

\bibitem{Vallisneri:2007ev}
Michele Vallisneri.
\newblock {Use and abuse of the Fisher information matrix in the assessment of
  gravitational-wave parameter-estimation prospects}.
\newblock {\em Phys. Rev. D}, 77:042001, 2008, gr-qc/0703086.

\bibitem{Flanagan:1997kp}
Eanna~E. Flanagan and Scott~A. Hughes.
\newblock {Measuring gravitational waves from binary black hole coalescences:
  2. The Waves' information and its extraction, with and without templates}.
\newblock {\em Phys. Rev.}, D57:4566--4587, 1998, gr-qc/9710129.

\bibitem{Lindblom:2008cm}
Lee Lindblom, Benjamin~J. Owen, and Duncan~A. Brown.
\newblock {Model Waveform Accuracy Standards for Gravitational Wave Data
  Analysis}.
\newblock {\em Phys. Rev.}, D78:124020, 2008, 0809.3844.

\bibitem{McWilliams:2010eq}
Sean~T. McWilliams, Bernard~J. Kelly, and John~G. Baker.
\newblock {Observing mergers of non-spinning black-hole binaries}.
\newblock {\em Phys. Rev.}, D82:024014, 2010, 1004.0961.

\bibitem{Chatziioannou:2017tdw}
Katerina Chatziioannou, Antoine Klein, Nicolás Yunes, and Neil Cornish.
\newblock {Constructing Gravitational Waves from Generic Spin-Precessing
  Compact Binary Inspirals}.
\newblock {\em Phys. Rev.}, D95(10):104004, 2017, 1703.03967.

\bibitem{Purrer:2019jcp}
Michael Pürrer and Carl-Johan Haster.
\newblock {Ready for what lies ahead? -- Gravitational waveform accuracy
  requirements for future ground based detectors}.
\newblock 2019, 1912.10055.

\bibitem{Veitch:2014wba}
J.~Veitch et~al.
\newblock {Parameter estimation for compact binaries with ground-based
  gravitational-wave observations using the LALInference software library}.
\newblock {\em Phys. Rev.}, D91(4):042003, 2015, 1409.7215.

\bibitem{lalsuite}
{LIGO Scientific Collaboration}.
\newblock {LIGO} {A}lgorithm {L}ibrary - {LALS}uite.
\newblock free software (GPL), 2018.

\bibitem{2020MNRAS.493.3132S}
Joshua~S. {Speagle}.
\newblock {DYNESTY: a dynamic nested sampling package for estimating Bayesian
  posteriors and evidences}.
\newblock {\em Mon. Not. Roy. Astron. Soc.}, 493(3):3132--3158, April 2020,
  1904.02180.

\bibitem{Ashton:2018jfp}
Gregory Ashton et~al.
\newblock {BILBY: A user-friendly Bayesian inference library for
  gravitational-wave astronomy}.
\newblock {\em Astrophys.\ J.\ Suppl.}, 241(2):27, 2019, 1811.02042.

\bibitem{Abbott:2020gyp}
R.~Abbott et~al.
\newblock {Population Properties of Compact Objects from the Second LIGO-Virgo
  Gravitational-Wave Transient Catalog}.
\newblock 10 2020, 2010.14533.

\bibitem{Ghosh:2021aa}
Ghosh et~al.
\newblock Constraints on the quasi-normal mode frequencies of the ligo-virgo
  signals by making full use of gravitational-wave modeling.
\newblock {\em in preparation}.

\bibitem{Abbott:2018lct}
B.~P. Abbott et~al.
\newblock {Tests of General Relativity with GW170817}.
\newblock {\em Phys. Rev. Lett.}, 123(1):011102, 2019, 1811.00364.

\bibitem{Berti:2007fi}
Emanuele Berti, Vitor Cardoso, Jose~A. Gonzalez, Ulrich Sperhake, Mark Hannam,
  Sascha Husa, and Bernd Bruegmann.
\newblock {Inspiral, merger and ringdown of unequal mass black hole binaries: A
  Multipolar analysis}.
\newblock {\em Phys. Rev. D}, 76:064034, 2007, gr-qc/0703053.

\bibitem{Kidder:2007rt}
Lawrence~E. Kidder.
\newblock {Using full information when computing modes of post-Newtonian
  waveforms from inspiralling compact binaries in circular orbit}.
\newblock {\em Phys. Rev. D}, 77:044016, 2008, 0710.0614.

\bibitem{Blanchet:2008je}
Luc Blanchet, Guillaume Faye, Bala~R. Iyer, and Siddhartha Sinha.
\newblock {The Third post-Newtonian gravitational wave polarisations and
  associated spherical harmonic modes for inspiralling compact binaries in
  quasi-circular orbits}.
\newblock {\em Class. Quant. Grav.}, 25:165003, 2008, 0802.1249.
\newblock [Erratum: Class.Quant.Grav. 29, 239501 (2012)].

\bibitem{Healy:2013jza}
James Healy, Pablo Laguna, Larne Pekowsky, and Deirdre Shoemaker.
\newblock {Template Mode Hierarchies for Binary Black Hole Mergers}.
\newblock {\em Phys. Rev.}, D88(2):024034, 2013, 1302.6953.

\bibitem{Cotesta:2018fcv}
Roberto Cotesta, Alessandra Buonanno, Alejandro Bohé, Andrea Taracchini, Ian
  Hinder, and Serguei Ossokine.
\newblock {Enriching the Symphony of Gravitational Waves from Binary Black
  Holes by Tuning Higher Harmonics}.
\newblock {\em Phys. Rev.}, D98(8):084028, 2018, 1803.10701.

\bibitem{Bustillo:2015qty}
Juan Calder{\'o}n~Bustillo, Sascha Husa, Alicia~M. Sintes, and Michael
  P{\"u}rrer.
\newblock {Impact of gravitational radiation higher order modes on single
  aligned-spin gravitational wave searches for binary black holes}.
\newblock {\em Phys. Rev.}, D93(8):084019, 2016, 1511.02060.

\bibitem{Brown:2012nn}
Duncan~A. Brown, Prayush Kumar, and Alexander~H. Nitz.
\newblock {Template banks to search for low-mass binary black holes in advanced
  gravitational-wave detectors}.
\newblock {\em Phys. Rev.}, D87:082004, 2013, 1211.6184.

\bibitem{Pekowsky:2012sr}
Larne Pekowsky, James Healy, Deirdre Shoemaker, and Pablo Laguna.
\newblock {Impact of higher-order modes on the detection of binary black hole
  coalescences}.
\newblock {\em Phys. Rev.}, D87(8):084008, 2013, 1210.1891.

\bibitem{Varma:2014jxa}
Vijay Varma, Parameswaran Ajith, Sascha Husa, Juan~Calderon Bustillo, Mark
  Hannam, and Michael P{\"u}rrer.
\newblock {Gravitational-wave observations of binary black holes: Effect of
  nonquadrupole modes}.
\newblock {\em Phys. Rev.}, D90(12):124004, 2014, 1409.2349.

\bibitem{Capano:2013raa}
Collin Capano, Yi~Pan, and Alessandra Buonanno.
\newblock {Impact of higher harmonics in searching for gravitational waves from
  nonspinning binary black holes}.
\newblock {\em Phys. Rev.}, D89(10):102003, 2014, 1311.1286.

\bibitem{Harry:2017weg}
Ian Harry, Juan Calder{\'o}n~Bustillo, and Alex Nitz.
\newblock {Searching for the full symphony of black hole binary mergers}.
\newblock 2017, 1709.09181.

\bibitem{Shaik:2019dym}
Feroz~H. Shaik, Jacob Lange, Scott~E. Field, Richard O'Shaughnessy, Vijay
  Varma, Lawrence~E. Kidder, Harald~P. Pfeiffer, and Daniel Wysocki.
\newblock {Impact of subdominant modes on the interpretation of
  gravitational-wave signals from heavy binary black hole systems}.
\newblock {\em Phys. Rev. D}, 101(12):124054, 2020, 1911.02693.

\bibitem{Kalaghatgi:2019log}
Chinmay Kalaghatgi, Mark Hannam, and Vivien Raymond.
\newblock {Parameter estimation with a spinning multimode waveform model}.
\newblock {\em Phys. Rev. D}, 101(10):103004, 2020, 1909.10010.

\bibitem{Littenberg:2012uj}
Tyson~B. Littenberg, John~G. Baker, Alessandra Buonanno, and Bernard~J. Kelly.
\newblock {Systematic biases in parameter estimation of binary black-hole
  mergers}.
\newblock {\em Phys. Rev.}, D87(10):104003, 2013, 1210.0893.

\bibitem{Dreyer:2003bv}
Olaf Dreyer, Bernard~J. Kelly, Badri Krishnan, Lee~Samuel Finn, David Garrison,
  and Ramon Lopez-Aleman.
\newblock {Black hole spectroscopy: Testing general relativity through
  gravitational wave observations}.
\newblock {\em Class. Quant. Grav.}, 21:787--804, 2004, gr-qc/0309007.

\bibitem{Berti:2005ys}
Emanuele Berti, Vitor Cardoso, and Clifford~M. Will.
\newblock {On gravitational-wave spectroscopy of massive black holes with the
  space interferometer LISA}.
\newblock {\em Phys. Rev.}, D73:064030, 2006, gr-qc/0512160.

\bibitem{Meidam:2014jpa}
J.~Meidam, M.~Agathos, C.~Van Den~Broeck, J.~Veitch, and B.~S. Sathyaprakash.
\newblock {Testing the no-hair theorem with black hole ringdowns using TIGER}.
\newblock {\em Phys. Rev.}, D90(6):064009, 2014, 1406.3201.

\bibitem{Bhagwat:2016ntk}
Swetha Bhagwat, Duncan~A. Brown, and Stefan~W. Ballmer.
\newblock {Spectroscopic analysis of stellar mass black-hole mergers in our
  local universe with ground-based gravitational wave detectors}.
\newblock {\em Phys. Rev. D}, 94(8):084024, 2016, 1607.07845.
\newblock [Erratum: Phys.Rev.D 95, 069906 (2017)].

\bibitem{Berti:2016lat}
Emanuele Berti, Alberto Sesana, Enrico Barausse, Vitor Cardoso, and Krzysztof
  Belczynski.
\newblock {Spectroscopy of Kerr black holes with Earth- and space-based
  interferometers}.
\newblock {\em Phys. Rev. Lett.}, 117(10):101102, 2016, 1605.09286.

\bibitem{Maselli:2017kvl}
Andrea Maselli, Kostas Kokkotas, and Pablo Laguna.
\newblock {Observing binary black hole ringdowns by advanced gravitational wave
  detectors}.
\newblock {\em Phys. Rev. D}, 95(10):104026, 2017, 1702.01110.

\bibitem{Gossan:2011ha}
S.~Gossan, J.~Veitch, and B.S. Sathyaprakash.
\newblock {Bayesian model selection for testing the no-hair theorem with black
  hole ringdowns}.
\newblock {\em Phys. Rev. D}, 85:124056, 2012, 1111.5819.

\bibitem{Brito:2018rfr}
Richard Brito, Alessandra Buonanno, and Vivien Raymond.
\newblock {Black-hole Spectroscopy by Making Full Use of Gravitational-Wave
  Modeling}.
\newblock {\em Phys. Rev. D}, 98(8):084038, 2018, 1805.00293.

\bibitem{Pang:2018hjb}
Peter~T.H. Pang, Juan Calder\'on~Bustillo, Yifan Wang, and Tjonnie~G.F. Li.
\newblock {Potential observations of false deviations from general relativity
  in gravitational wave signals from binary black holes}.
\newblock {\em Phys. Rev. D}, 98(2):024019, 2018, 1802.03306.

\bibitem{Pan:2011gk}
Yi~Pan, Alessandra Buonanno, Michael Boyle, Luisa~T. Buchman, Lawrence~E.
  Kidder, Harald~P. Pfeiffer, and Mark~A. Scheel.
\newblock {Inspiral-merger-ringdown multipolar waveforms of nonspinning
  black-hole binaries using the effective-one-body formalism}.
\newblock {\em Phys. Rev.}, D84:124052, 2011, 1106.1021.

\bibitem{Ossokine:2020kjp}
Serguei Ossokine et~al.
\newblock {Multipolar Effective-One-Body Waveforms for Precessing Binary Black
  Holes: Construction and Validation}.
\newblock {\em Phys. Rev. D}, 102(4):044055, 2020, 2004.09442.

\bibitem{Mehta:2017jpq}
Ajit~Kumar Mehta, Chandra~Kant Mishra, Vijay Varma, and Parameswaran Ajith.
\newblock {Accurate inspiral-merger-ringdown gravitational waveforms for
  non-spinning black-hole binaries including the effect of subdominant modes}.
\newblock 2017, 1708.03501.

\bibitem{Khan:2019kot}
Sebastian Khan, Frank Ohme, Katerina Chatziioannou, and Mark Hannam.
\newblock {Including higher order multipoles in gravitational-wave models for
  precessing binary black holes}.
\newblock {\em Phys. Rev.}, D101(2):024056, 2020, 1911.06050.

\bibitem{Barausse:2009aa}
Enrico Barausse, Etienne Racine, and Alessandra Buonanno.
\newblock {Hamiltonian of a spinning test-particle in curved spacetime}.
\newblock {\em Phys. Rev.}, D80:104025, 2009, 0907.4745.
\newblock [Erratum: Phys. Rev.D85,069904(2012)].

\bibitem{Vines:2016unv}
Justin Vines, Daniela Kunst, Jan Steinhoff, and Tanja Hinderer.
\newblock {Canonical Hamiltonian for an extended test body in curved spacetime:
  To quadratic order in spin}.
\newblock {\em Phys. Rev.}, D93(10):103008, 2016, 1601.07529.

\bibitem{Hinderer:2013uwa}
Tanja Hinderer et~al.
\newblock {Periastron advance in spinning black hole binaries: comparing
  effective-one-body and Numerical Relativity}.
\newblock {\em Phys. Rev. D}, 88(8):084005, 2013, 1309.0544.

\bibitem{Barausse:2011ys}
Enrico Barausse and Alessandra Buonanno.
\newblock {Extending the effective-one-body Hamiltonian of black-hole binaries
  to include next-to-next-to-leading spin-orbit couplings}.
\newblock {\em Phys. Rev.}, D84:104027, 2011, 1107.2904.

\bibitem{Damour:2001tu}
Thibault Damour.
\newblock {Coalescence of two spinning black holes: an effective one-body
  approach}.
\newblock {\em Phys. Rev.}, D64:124013, 2001, gr-qc/0103018.

\bibitem{Buonanno:2005xu}
Alessandra Buonanno, Yanbei Chen, and Thibault Damour.
\newblock {Transition from inspiral to plunge in precessing binaries of
  spinning black holes}.
\newblock {\em Phys. Rev.}, D74:104005, 2006, gr-qc/0508067.

\bibitem{Peters:1963ux}
P.C. Peters and J.~Mathews.
\newblock {Gravitational radiation from point masses in a Keplerian orbit}.
\newblock {\em Phys.Rev.}, 131:435--439, 1963.

\bibitem{Arun:2007sg}
K.G. Arun, Luc Blanchet, Bala~R. Iyer, and Moh'd~S.S. Qusailah.
\newblock {Inspiralling compact binaries in quasi-elliptical orbits: The
  Complete 3PN energy flux}.
\newblock {\em Phys. Rev. D}, 77:064035, 2008, 0711.0302.

\bibitem{Arun:2007rg}
K.G. Arun, Luc Blanchet, Bala~R. Iyer, and Moh'd~S.S. Qusailah.
\newblock {Tail effects in the 3PN gravitational wave energy flux of compact
  binaries in quasi-elliptical orbits}.
\newblock {\em Phys. Rev. D}, 77:064034, 2008, 0711.0250.

\bibitem{Arun:2009mc}
K.G. Arun, Luc Blanchet, Bala~R. Iyer, and Siddhartha Sinha.
\newblock {Third post-Newtonian angular momentum flux and the secular evolution
  of orbital elements for inspiralling compact binaries in quasi-elliptical
  orbits}.
\newblock {\em Phys. Rev. D}, 80:124018, 2009, 0908.3854.

\bibitem{Maia:2017yok}
Natalia~T. Maia, Chad~R. Galley, Adam~K. Leibovich, and Rafael~A. Porto.
\newblock {Radiation reaction for spinning bodies in effective field theory II:
  Spin-spin effects}.
\newblock {\em Phys. Rev. D}, 96(8):084065, 2017, 1705.07938.

\bibitem{Bohe:2013cla}
Alejandro Bohé, Sylvain Marsat, and Luc Blanchet.
\newblock {Next-to-next-to-leading order spin--orbit effects in the
  gravitational wave flux and orbital phasing of compact binaries}.
\newblock {\em Class. Quant. Grav.}, 30:135009, 2013, 1303.7412.

\bibitem{Bohe:2015ana}
Alejandro Bohé, Guillaume Faye, Sylvain Marsat, and Edward~K. Porter.
\newblock {Quadratic-in-spin effects in the orbital dynamics and
  gravitational-wave energy flux of compact binaries at the 3PN order}.
\newblock {\em Class. Quant. Grav.}, 32(19):195010, 2015, 1501.01529.

\bibitem{Damour:2007xr}
Thibault Damour and Alessandro Nagar.
\newblock {Faithful effective-one-body waveforms of small-mass-ratio coalescing
  black-hole binaries}.
\newblock {\em Phys. Rev.}, D76:064028, 2007, 0705.2519.

\bibitem{Damour:2008gu}
Thibault Damour, Bala~R. Iyer, and Alessandro Nagar.
\newblock {Improved resummation of post-Newtonian multipolar waveforms from
  circularized compact binaries}.
\newblock {\em Phys. Rev.}, D79:064004, 2009, 0811.2069.

\bibitem{Pan:2010hz}
Yi~Pan, Alessandra Buonanno, Ryuichi Fujita, Etienne Racine, and Hideyuki
  Tagoshi.
\newblock {Post-Newtonian factorized multipolar waveforms for spinning,
  non-precessing black-hole binaries}.
\newblock {\em Phys. Rev.}, D83:064003, 2011, 1006.0431.
\newblock [Erratum: Phys. Rev.D87,no.10,109901(2013)].

\bibitem{Nagar:2016ayt}
Alessandro Nagar and Abhay Shah.
\newblock {Factorization and resummation: A new paradigm to improve
  gravitational wave amplitudes}.
\newblock {\em Phys. Rev.}, D94(10):104017, 2016, 1606.00207.

\bibitem{Messina:2018ghh}
Francesco Messina, Alberto Maldarella, and Alessandro Nagar.
\newblock {Factorization and resummation: A new paradigm to improve
  gravitational wave amplitudes. II: the higher multipolar modes}.
\newblock 2018, 1801.02366.

\bibitem{Nagar:2019wrt}
Alessandro Nagar, Francesco Messina, Chris Kavanagh, Georgios
  Lukes-Gerakopoulos, Niels Warburton, Sebastiano Bernuzzi, and Enno Harms.
\newblock {Factorization and resummation: A new paradigm to improve
  gravitational wave amplitudes. III: the spinning test-body terms}.
\newblock {\em Phys. Rev. D}, 100(10):104056, 2019, 1907.12233.

\bibitem{Poisson:1993zr}
Eric Poisson.
\newblock {Gravitational radiation from a particle in circular orbit around a
  black hole. 4: Analytical results for the slowly rotating case}.
\newblock {\em Phys. Rev. D}, 48:1860--1863, 1993.

\bibitem{1957PhRv..108.1063R}
Tullio {Regge} and John~A. {Wheeler}.
\newblock {Stability of a Schwarzschild Singularity}.
\newblock {\em Physical Review}, 108(4):1063--1069, November 1957.

\bibitem{Zerilli:1970se}
Frank~J. Zerilli.
\newblock {Effective potential for even parity Regge-Wheeler gravitational
  perturbation equations}.
\newblock {\em Phys. Rev. Lett.}, 24:737--738, 1970.

\bibitem{Zerilli:1971wd}
F.J. Zerilli.
\newblock {Gravitational field of a particle falling in a schwarzschild
  geometry analyzed in tensor harmonics}.
\newblock {\em Phys. Rev. D}, 2:2141--2160, 1970.

\bibitem{Ossokine:2017dge}
Serguei Ossokine, Tim Dietrich, Evan Foley, Reza Katebi, and Geoffrey Lovelace.
\newblock {Assessing the Energetics of Spinning Binary Black Hole Systems}.
\newblock {\em Phys. Rev. D}, 98(10):104057, 2018, 1712.06533.

\bibitem{Nagar:2015xqa}
Alessandro Nagar, Thibault Damour, Christian Reisswig, and Denis Pollney.
\newblock {Energetics and phasing of nonprecessing spinning coalescing black
  hole binaries}.
\newblock {\em Phys. Rev.}, D93(4):044046, 2016, 1506.08457.

\bibitem{Antonelli:2020aeb}
Andrea Antonelli, Chris Kavanagh, Mohammed Khalil, Jan Steinhoff, and Justin
  Vines.
\newblock {Gravitational spin-orbit coupling through third-subleading
  post-Newtonian order: from first-order self-force to arbitrary mass ratios}.
\newblock {\em Phys. Rev. Lett.}, 125(1):011103, 2020, 2003.11391.

\bibitem{Antonelli:2019ytb}
Andrea Antonelli, Alessandra Buonanno, Jan Steinhoff, Maarten van~de Meent, and
  Justin Vines.
\newblock {Energetics of two-body Hamiltonians in post-Minkowskian gravity}.
\newblock {\em Phys. Rev. D}, 99(10):104004, 2019, 1901.07102.

\bibitem{Dietrich:2016lyp}
Tim Dietrich, Sebastiano Bernuzzi, Maximiliano Ujevic, and Wolfgang Tichy.
\newblock {Gravitational waves and mass ejecta from binary neutron star
  mergers: Effect of the stars' rotation}.
\newblock {\em Phys. Rev. D}, 95(4):044045, 2017, 1611.07367.

\bibitem{Taracchini:2012ig}
Andrea Taracchini, Yi~Pan, Alessandra Buonanno, Enrico Barausse, Michael Boyle,
  Tony Chu, Geoffrey Lovelace, Harald~P. Pfeiffer, and Mark~A. Scheel.
\newblock {Prototype effective-one-body model for nonprecessing spinning
  inspiral-merger-ringdown waveforms}.
\newblock {\em Phys. Rev.}, D86:024011, 2012, 1202.0790.

\bibitem{Taracchini:2013rva}
Andrea Taracchini et~al.
\newblock {Effective-one-body model for black-hole binaries with generic mass
  ratios and spins}.
\newblock {\em Phys. Rev.}, D89(6):061502, 2014, 1311.2544.

\bibitem{Antonelli:2019fmq}
Andrea Antonelli, Maarten van~de Meent, Alessandra Buonanno, Jan Steinhoff, and
  Justin Vines.
\newblock {Quasicircular inspirals and plunges from nonspinning
  effective-one-body Hamiltonians with gravitational self-force information}.
\newblock {\em Phys.\ Rev.\ D}, 101(2):024024, 2020, 1907.11597.

\bibitem{Damour:1997ub}
Thibault Damour, Bala~R. Iyer, and B.~S. Sathyaprakash.
\newblock {Improved filters for gravitational waves from inspiralling compact
  binaries}.
\newblock {\em Phys. Rev.}, D57:885--907, 1998, gr-qc/9708034.

\bibitem{Kokkotas:1999bd}
Kostas~D. Kokkotas and Bernd~G. Schmidt.
\newblock {Quasinormal modes of stars and black holes}.
\newblock {\em Living Rev. Rel.}, 2:2, 1999, gr-qc/9909058.

\bibitem{Barausse:2012qz}
Enrico Barausse, Viktoriya Morozova, and Luciano Rezzolla.
\newblock {On the mass radiated by coalescing black-hole binaries}.
\newblock {\em Astrophys. J.}, 758:63, 2012, 1206.3803.
\newblock [Erratum: Astrophys. J.786,76(2014)].

\bibitem{Hemberger:2013hsa}
Daniel~A. Hemberger, Geoffrey Lovelace, Thomas~J. Loredo, Lawrence~E. Kidder,
  Mark~A. Scheel, B\'{e}la Szil\'{a}gyi, Nicholas~W. Taylor, and Saul~A.
  Teukolsky.
\newblock {Final spin and radiated energy in numerical simulations of binary
  black holes with equal masses and equal, aligned or anti-aligned spins}.
\newblock {\em Phys. Rev.}, D88:064014, 2013, 1305.5991.

\bibitem{Nagar:2016iwa}
Walter Del~Pozzo and Alessandro Nagar.
\newblock {Analytic family of post-merger template waveforms}.
\newblock {\em Phys. Rev.}, D95(12):124034, 2017, 1606.03952.

\bibitem{Giesler:2019uxc}
Matthew Giesler, Maximiliano Isi, Mark~A. Scheel, and Saul Teukolsky.
\newblock {Black Hole Ringdown: The Importance of Overtones}.
\newblock {\em Phys. Rev. X}, 9(4):041060, 2019, 1903.08284.

\bibitem{PhysRevD.31.1815}
Kip~S. Thorne and James~B. Hartle.
\newblock Laws of motion and precession for black holes and other bodies.
\newblock {\em Phys. Rev. D}, 31:1815--1837, Apr 1985.

\bibitem{Buonanno:2002fy}
Alessandra Buonanno, Yan-bei Chen, and Michele Vallisneri.
\newblock {Detecting gravitational waves from precessing binaries of spinning
  compact objects: Adiabatic limit}.
\newblock {\em Phys. Rev.}, D67:104025, 2003, gr-qc/0211087.
\newblock [Erratum: Phys. Rev.D74,029904(2006)].

\bibitem{Kesden:2014sla}
Michael Kesden, Davide Gerosa, Richard O'Shaughnessy, Emanuele Berti, and
  Ulrich Sperhake.
\newblock {Effective potentials and morphological transitions for binary
  black-hole spin precession}.
\newblock {\em Phys. Rev. Lett.}, 114(8):081103, 2015, 1411.0674.

\bibitem{Ossokine:2015phd}
Serguei Ossokine.
\newblock {\em Modelling Precessing Binary Black Hole Systems}.
\newblock PhD thesis, University of Toronto, 11 2015.

\bibitem{Zhao:2017tro}
Xinyu Zhao, Michael Kesden, and Davide Gerosa.
\newblock {Nutational resonances, transitional precession, and
  precession-averaged evolution in binary black-hole systems}.
\newblock {\em Phys. Rev. D}, 96(2):024007, 2017, 1705.02369.

\bibitem{Gerosa:2015hba}
Davide Gerosa, Michael Kesden, Richard O'Shaughnessy, Antoine Klein, Emanuele
  Berti, Ulrich Sperhake, and Daniele Trifir\`o.
\newblock {Precessional instability in binary black holes with aligned spins}.
\newblock {\em Phys. Rev. Lett.}, 115:141102, 2015, 1506.09116.

\bibitem{Gerosa:2018mwg}
Davide Gerosa, Alicia Lima, Emanuele Berti, Ulrich Sperhake, Michael Kesden,
  and Richard O'Shaughnessy.
\newblock {Wide nutation: binary black-hole spins repeatedly oscillating from
  full alignment to full anti-alignment}.
\newblock {\em Class. Quant. Grav.}, 36(10):105003, 2019, 1811.05979.

\bibitem{Pan:2013rra}
Yi~Pan, Alessandra Buonanno, Andrea Taracchini, Lawrence~E. Kidder, Abdul~H.
  Mroué, Harald~P. Pfeiffer, Mark~A. Scheel, and Béla Szilágyi.
\newblock {Inspiral-merger-ringdown waveforms of spinning, precessing
  black-hole binaries in the effective-one-body formalism}.
\newblock {\em Phys. Rev.}, D89(8):084006, 2014, 1307.6232.

\bibitem{Babak:2016tgq}
Stanislav Babak, Andrea Taracchini, and Alessandra Buonanno.
\newblock {Validating the effective-one-body model of spinning, precessing
  binary black holes against numerical relativity}.
\newblock {\em Phys. Rev.}, D95(2):024010, 2017, 1607.05661.

\bibitem{Harry:2009ea}
Ian~W. Harry, Bruce Allen, and B.S. Sathyaprakash.
\newblock {A Stochastic template placement algorithm for gravitational wave
  data analysis}.
\newblock {\em Phys.Rev.}, D80:104014, 2009, 0908.2090.

\bibitem{Manca:2009xw}
Gian~Mario Manca and Michele Vallisneri.
\newblock {Cover art: Issues in the metric-guided and metric-less placement of
  random and stochastic template banks}.
\newblock {\em Phys.Rev.}, D81:024004, 2010, 0909.0563.

\bibitem{Devine:2016ovp}
Caleb Devine, Zachariah~B. Etienne, and Sean~T. McWilliams.
\newblock {Optimizing spinning time-domain gravitational waveforms for Advanced
  LIGO data analysis}.
\newblock {\em Class. Quant. Grav.}, 33(12):125025, 2016, 1601.03393.

\bibitem{Nagar:2018gnk}
Alessandro Nagar and Piero Rettegno.
\newblock {Efficient effective one body time-domain gravitational waveforms}.
\newblock {\em Phys.\ Rev.\ D}, 99(2):021501, 2019, 1805.03891.

\bibitem{Purrer:2014fza}
Michael Pürrer.
\newblock {Frequency domain reduced order models for gravitational waves from
  aligned-spin compact binaries}.
\newblock {\em Class. Quant. Grav.}, 31(19):195010, 2014, 1402.4146.

\bibitem{Purrer:2015tud}
Michael Pürrer.
\newblock {Frequency domain reduced order model of aligned-spin
  effective-one-body waveforms with generic mass-ratios and spins}.
\newblock {\em Phys. Rev.}, D93(6):064041, 2016, 1512.02248.

\bibitem{Lackey:2018zvw}
Benjamin~D. Lackey, Michael Pürrer, Andrea Taracchini, and Sylvain Marsat.
\newblock {Surrogate model for an aligned-spin effective one body waveform
  model of binary neutron star inspirals using Gaussian process regression}.
\newblock {\em Phys. Rev.}, D100(2):024002, 2019, 1812.08643.

\bibitem{Doctor:2017csx}
Zoheyr Doctor, Ben Farr, Daniel~E. Holz, and Michael Pürrer.
\newblock {Statistical Gravitational Waveform Models: What to Simulate Next?}
\newblock {\em Phys. Rev.}, D96(12):123011, 2017, 1706.05408.

\bibitem{Setyawati:2019xzw}
Yoshinta~Eka Setyawati, Michael Pürrer, and Frank Ohme.
\newblock Regression methods in waveform modeling: a comparative study.
\newblock {\em Classical and Quantum Gravity}, 2020.

\bibitem{Buonanno:2009zt}
Alessandra Buonanno, Bala Iyer, Evan Ochsner, Yi~Pan, and B.~S. Sathyaprakash.
\newblock {Comparison of post-Newtonian templates for compact binary inspiral
  signals in gravitational-wave detectors}.
\newblock {\em Phys. Rev.}, D80:084043, 2009, 0907.0700.

\bibitem{Gadre:2021aa}
Gadre et~al.
\newblock A fully precessing higher-mode surrogate model of effective-one-body
  waveforms.
\newblock {\em in preparation}.

\bibitem{Barsotti:2018}
Lisa Barsotti, Peter Fritschel, Matthew Evans, and Slawomir Gras.
\newblock Updated advanced ligo sensitivity design curve, 2018.
\newblock {LIGO} Document T1800044-v5.

\bibitem{Santamaria:2010yb}
L.~Santamaria et~al.
\newblock {Matching post-Newtonian and numerical relativity waveforms:
  systematic errors and a new phenomenological model for non-precessing black
  hole binaries}.
\newblock {\em Phys. Rev.}, D82:064016, 2010, 1005.3306.

\bibitem{Cutler:1994ys}
Curt Cutler and Eanna~E. Flanagan.
\newblock {Gravitational waves from merging compact binaries: How accurately
  can one extract the binary's parameters from the inspiral wave form?}
\newblock {\em Phys. Rev. D}, 49:2658--2697, 1994, gr-qc/9402014.

\bibitem{Poisson:1995ef}
Eric Poisson and Clifford~M. Will.
\newblock {Gravitational waves from inspiraling compact binaries: Parameter
  estimation using second postNewtonian wave forms}.
\newblock {\em Phys. Rev. D}, 52:848--855, 1995, gr-qc/9502040.

\bibitem{Baird:2012cu}
Emily Baird, Stephen Fairhurst, Mark Hannam, and Patricia Murphy.
\newblock {Degeneracy between mass and spin in black-hole-binary waveforms}.
\newblock {\em Phys. Rev. D}, 87(2):024035, 2013, 1211.0546.

\bibitem{Hannam:2013uu}
Mark Hannam, Duncan~A. Brown, Stephen Fairhurst, Chris~L. Fryer, and Ian~W.
  Harry.
\newblock {When can gravitational-wave observations distinguish between black
  holes and neutron stars?}
\newblock {\em Astrophys. J. Lett.}, 766:L14, 2013, 1301.5616.

\bibitem{Ohme:2013nsa}
Frank Ohme, Alex~B. Nielsen, Drew Keppel, and Andrew Lundgren.
\newblock {Statistical and systematic errors for gravitational-wave inspiral
  signals: A principal component analysis}.
\newblock {\em Phys. Rev. D}, 88(4):042002, 2013, 1304.7017.

\bibitem{Cotesta:2020qhw}
Roberto Cotesta, Sylvain Marsat, and Michael P\"urrer.
\newblock {Frequency domain reduced order model of aligned-spin
  effective-one-body waveforms with higher-order modes}.
\newblock {\em Phys. Rev. D}, 101(12):124040, 2020, 2003.12079.

\bibitem{Kumar:2018hml}
Prayush Kumar, Jonathan Blackman, Scott~E. Field, Mark Scheel, Chad~R. Galley,
  Michael Boyle, Lawrence~E. Kidder, Harald~P. Pfeiffer, Bela Szilagyi, and
  Saul~A. Teukolsky.
\newblock {Constraining the parameters of GW150914 and GW170104 with numerical
  relativity surrogates}.
\newblock {\em Phys. Rev. D}, 99(12):124005, 2019, 1808.08004.

\bibitem{Chatziioannou:2019dsz}
Katerina Chatziioannou et~al.
\newblock {On the properties of the massive binary black hole merger GW170729}.
\newblock {\em Phys. Rev. D}, 100(10):104015, 2019, 1903.06742.

\bibitem{Colleoni:2020tgc}
Marta Colleoni, Maite Mateu-Lucena, H\'ector Estell\'es, Cecilio
  Garc\'\i{}a-Quir\'os, David Keitel, Geraint Pratten, Antoni Ramos-Buades, and
  Sascha Husa.
\newblock {Towards the routine use of subdominant harmonics in
  gravitational-wave inference: re-analysis of GW190412 with generation X
  waveform models}.
\newblock 10 2020, 2010.05830.

\bibitem{Islam:2020reh}
Tousif Islam, Scott~E. Field, Carl-Johan Haster, and Rory Smith.
\newblock {Improved analysis of GW190412 with a precessing numerical relativity
  surrogate waveform model}.
\newblock 10 2020, 2010.04848.

\bibitem{Kovetz:2016kpi}
Ely~D. Kovetz, Ilias Cholis, Patrick~C. Breysse, and Marc Kamionkowski.
\newblock {Black hole mass function from gravitational wave measurements}.
\newblock {\em Phys. Rev. D}, 95(10):103010, 2017, 1611.01157.

\bibitem{Fishbach:2017dwv}
Maya Fishbach, Daniel~E. Holz, and Ben Farr.
\newblock {Are LIGO's Black Holes Made From Smaller Black Holes?}
\newblock {\em Astrophys. J.}, 840(2):L24, 2017, 1703.06869.

\bibitem{Talbot:2018cva}
Colm Talbot and Eric Thrane.
\newblock {Measuring the binary black hole mass spectrum with an
  astrophysically motivated parameterization}.
\newblock {\em Astrophys. J.}, 856(2):173, 2018, 1801.02699.

\bibitem{LIGOScientific:2018jsj}
B.~P. Abbott et~al.
\newblock {Binary Black Hole Population Properties Inferred from the First and
  Second Observing Runs of Advanced LIGO and Advanced Virgo}.
\newblock {\em Astrophys. J.}, 882(2):L24, 2019, 1811.12940.

\bibitem{Nitz:2017svb}
Alexander~H. Nitz, Thomas Dent, Tito Dal~Canton, Stephen Fairhurst, and
  Duncan~A. Brown.
\newblock {Detecting binary compact-object mergers with gravitational waves:
  Understanding and Improving the sensitivity of the PyCBC search}.
\newblock {\em Astrophys. J.}, 849(2):118, 2017, 1705.01513.

\bibitem{Usman:2015kfa}
Samantha~A. Usman et~al.
\newblock {The PyCBC search for gravitational waves from compact binary
  coalescence}.
\newblock {\em Class. Quant. Grav.}, 33(21):215004, 2016, 1508.02357.

\bibitem{Canton:2014ena}
Tito Dal~Canton et~al.
\newblock {Implementing a search for aligned-spin neutron star-black hole
  systems with advanced ground based gravitational wave detectors}.
\newblock {\em Phys. Rev.}, D90(8):082004, 2014, 1405.6731.

\bibitem{Cannon:2011vi}
Kipp Cannon et~al.
\newblock {Toward Early-Warning Detection of Gravitational Waves from Compact
  Binary Coalescence}.
\newblock {\em Astrophys. J.}, 748:136, 2012, 1107.2665.

\bibitem{Cannon:2012zt}
Kipp Cannon, Chad Hanna, and Drew Keppel.
\newblock {Method to estimate the significance of coincident gravitational-wave
  observations from compact binary coalescence}.
\newblock {\em Phys. Rev.}, D88(2):024025, 2013, 1209.0718.

\bibitem{Arun:2008kb}
K.~G. Arun, Alessandra Buonanno, Guillaume Faye, and Evan Ochsner.
\newblock {Higher-order spin effects in the amplitude and phase of
  gravitational waveforms emitted by inspiraling compact binaries: Ready-to-use
  gravitational waveforms}.
\newblock {\em Phys. Rev.}, D79:104023, 2009, 0810.5336.
\newblock [Erratum: Phys. Rev.D84,049901(2011)].

\bibitem{Mishra:2016whh}
Chandra~Kant Mishra, Aditya Kela, K.~G. Arun, and Guillaume Faye.
\newblock {Ready-to-use post-Newtonian gravitational waveforms for binary black
  holes with nonprecessing spins: An update}.
\newblock {\em Phys. Rev.}, D93(8):084054, 2016, 1601.05588.

\bibitem{Damour:2008qf}
Thibault Damour, Piotr Jaranowski, and Gerhard Schaefer.
\newblock {Effective one body approach to the dynamics of two spinning black
  holes with next-to-leading order spin-orbit coupling}.
\newblock {\em Phys. Rev.}, D78:024009, 2008, 0803.0915.

\bibitem{Damour:2009kr}
Thibault Damour and Alessandro Nagar.
\newblock {An Improved analytical description of inspiralling and coalescing
  black-hole binaries}.
\newblock {\em Phys. Rev.}, D79:081503, 2009, 0902.0136.

\bibitem{Barausse:2009xi}
Enrico Barausse and Alessandra Buonanno.
\newblock {An Improved effective-one-body Hamiltonian for spinning black-hole
  binaries}.
\newblock {\em Phys. Rev.}, D81:084024, 2010, 0912.3517.

\bibitem{TheLIGOScientific:2016wfe}
B.~P. Abbott et~al.
\newblock {Properties of the Binary Black Hole Merger GW150914}.
\newblock {\em Phys. Rev. Lett.}, 116(24):241102, 2016, 1602.03840.

\bibitem{TheLIGOScientific:2016src}
B.~P. Abbott et~al.
\newblock {Tests of general relativity with GW150914}.
\newblock {\em Phys. Rev. Lett.}, 116(22):221101, 2016, 1602.03841.

\bibitem{Graff:2015bba}
Philip~B. Graff, Alessandra Buonanno, and B.~S. Sathyaprakash.
\newblock {Missing Link: Bayesian detection and measurement of
  intermediate-mass black-hole binaries}.
\newblock {\em Phys. Rev.}, D92(2):022002, 2015, 1504.04766.

\bibitem{Varma:2016dnf}
Vijay Varma and Parameswaran Ajith.
\newblock {Effects of non-quadrupole modes in the detection and parameter
  estimation of black hole binaries with nonprecessing spins}.
\newblock 2016, 1612.05608.

\bibitem{Bustillo:2016gid}
Juan Calder{\'o}n~Bustillo, Pablo Laguna, and Deirdre Shoemaker.
\newblock {Detectability of gravitational waves from binary black holes: Impact
  of precession and higher modes}.
\newblock {\em Phys. Rev.}, D95(10):104038, 2017, 1612.02340.

\bibitem{Abbott:2016wiq}
Benjamin~P. Abbott et~al.
\newblock {Effects of waveform model systematics on the interpretation of
  GW150914}.
\newblock {\em Class. Quant. Grav.}, 34(10):104002, 2017, 1611.07531.

\bibitem{Fujita:2012cm}
Ryuichi Fujita.
\newblock {Gravitational Waves from a Particle in Circular Orbits around a
  Schwarzschild Black Hole to the 22nd Post-Newtonian Order}.
\newblock {\em Prog. Theor. Phys.}, 128:971--992, 2012, 1211.5535.

\bibitem{Zilhao:2013hia}
Miguel Zilh{\~a}o and Frank L{\"o}ffler.
\newblock {An Introduction to the Einstein Toolkit}.
\newblock {\em Int. J. Mod. Phys.}, A28:1340014, 2013, 1305.5299.

\bibitem{Loffler:2011ay}
Frank Loffler et~al.
\newblock {The Einstein Toolkit: A Community Computational Infrastructure for
  Relativistic Astrophysics}.
\newblock {\em Class. Quant. Grav.}, 29:115001, 2012, 1111.3344.

\bibitem{Barausse:2011kb}
Enrico Barausse, Alessandra Buonanno, Scott~A. Hughes, Gaurav Khanna, Stephen
  O'Sullivan, and Yi~Pan.
\newblock {Modeling multipolar gravitational-wave emission from small
  mass-ratio mergers}.
\newblock {\em Phys. Rev.}, D85:024046, 2012, 1110.3081.

\bibitem{Mroue2013}
Abdul~H. Mroue et~al.
\newblock {Catalog of 174 Binary Black Hole Simulations for Gravitational Wave
  Astronomy}.
\newblock {\em Phys. Rev. Lett.}, 111(24):241104, 2013, 1304.6077.

\bibitem{Kumar:2016dhh}
Prayush Kumar, Tony Chu, Heather Fong, Harald~P. Pfeiffer, Michael Boyle,
  Daniel~A. Hemberger, Lawrence~E. Kidder, Mark~A. Scheel, and Bela Szilagyi.
\newblock {Accuracy of binary black hole waveform models for aligned-spin
  binaries}.
\newblock {\em Phys. Rev.}, D93(10):104050, 2016, 1601.05396.

\bibitem{Bruegmann:2006at}
Bernd Bruegmann, Jose~A. Gonzalez, Mark Hannam, Sascha Husa, Ulrich Sperhake,
  and Wolfgang Tichy.
\newblock {Calibration of Moving Puncture Simulations}.
\newblock {\em Phys. Rev.}, D77:024027, 2008, gr-qc/0610128.

\bibitem{Husa:2007hp}
Sascha Husa, Jose~A. Gonzalez, Mark Hannam, Bernd Bruegmann, and Ulrich
  Sperhake.
\newblock {Reducing phase error in long numerical binary black hole evolutions
  with sixth order finite differencing}.
\newblock {\em Class. Quant. Grav.}, 25:105006, 2008, 0706.0740.

\bibitem{Baker:2008mj}
John~G. Baker, William~D. Boggs, Joan Centrella, Bernard~J. Kelly, Sean~T.
  McWilliams, and James~R. van Meter.
\newblock {Mergers of non-spinning black-hole binaries: Gravitational radiation
  characteristics}.
\newblock {\em Phys. Rev.}, D78:044046, 2008, 0805.1428.

\bibitem{Damour:2014yha}
Thibault Damour and Alessandro Nagar.
\newblock {A new analytic representation of the ringdown waveform of coalescing
  spinning black hole binaries}.
\newblock {\em Phys. Rev.}, D90(2):024054, 2014, 1406.0401.

\bibitem{London:2014cma}
Lionel London, Deirdre Shoemaker, and James Healy.
\newblock {Modeling ringdown: Beyond the fundamental quasinormal modes}.
\newblock {\em Phys. Rev.}, D90(12):124032, 2014, 1404.3197.
\newblock [Erratum: Phys. Rev.D94,no.6,069902(2016)].

\bibitem{London:2018gaq}
L.~T. London.
\newblock {Modeling ringdown II: non-precessing binary black holes}.
\newblock 2018, 1801.08208.

\bibitem{Yang:2017zxs}
Huan Yang, Kent Yagi, Jonathan Blackman, Luis Lehner, Vasileios Paschalidis,
  Frans Pretorius, and Nicol{\'a}s Yunes.
\newblock {Black hole spectroscopy with coherent mode stacking}.
\newblock {\em Phys. Rev. Lett.}, 118(16):161101, 2017, 1701.05808.

\bibitem{Shoemaker:2010}
David Shoemaker.
\newblock Advanced {LIGO} anticipated sensitivity curves, 2010.
\newblock {LIGO} Document T0900288-v3.

\bibitem{Kamaretsos:2012bs}
Ioannis Kamaretsos, Mark Hannam, and B.~Sathyaprakash.
\newblock {Is black-hole ringdown a memory of its progenitor?}
\newblock {\em Phys. Rev. Lett.}, 109:141102, 2012, 1207.0399.

\bibitem{Finn:1992xs}
Lee~Samuel Finn and David~F. Chernoff.
\newblock {Observing binary inspiral in gravitational radiation: One
  interferometer}.
\newblock {\em Phys. Rev.}, D47:2198--2219, 1993, gr-qc/9301003.

\bibitem{Harry:2016ijz}
Ian Harry, Stephen Privitera, Alejandro Bohé, and Alessandra Buonanno.
\newblock {Searching for Gravitational Waves from Compact Binaries with
  Precessing Spins}.
\newblock {\em Phys. Rev.}, D94(2):024012, 2016, 1603.02444.

\bibitem{Hinder:2013oqa}
Ian Hinder et~al.
\newblock {Error-analysis and comparison to analytical models of numerical
  waveforms produced by the NRAR Collaboration}.
\newblock {\em Class. Quant. Grav.}, 31:025012, 2014, 1307.5307.

\bibitem{Taylor:2013zia}
Nicholas~W. Taylor, Michael Boyle, Christian Reisswig, Mark~A. Scheel, Tony
  Chu, Lawrence~E. Kidder, and B{\'e}la Szil{\'a}gyi.
\newblock {Comparing Gravitational Waveform Extrapolation to
  Cauchy-Characteristic Extraction in Binary Black Hole Simulations}.
\newblock {\em Phys. Rev.}, D88(12):124010, 2013, 1309.3605.

\bibitem{Kelly:2012nd}
Bernard~J. Kelly and John~G. Baker.
\newblock {Decoding mode mixing in black-hole merger ringdown}.
\newblock {\em Phys. Rev.}, D87(8):084004, 2013, 1212.5553.

\bibitem{Berti:2014fga}
Emanuele Berti and Antoine Klein.
\newblock {Mixing of spherical and spheroidal modes in perturbed Kerr black
  holes}.
\newblock {\em Phys.\ Rev.\ D}, 90(6):064012, 2014, 1408.1860.

\bibitem{Berti:2009kk}
Emanuele Berti, Vitor Cardoso, and Andrei~O. Starinets.
\newblock {Quasinormal modes of black holes and black branes}.
\newblock {\em Class. Quant. Grav.}, 26:163001, 2009, 0905.2975.

\bibitem{Hofmann:2016yih}
Fabian Hofmann, Enrico Barausse, and Luciano Rezzolla.
\newblock {The final spin from binary black holes in quasi-circular orbits}.
\newblock {\em Astrophys. J.}, 825(2):L19, 2016, 1605.01938.

\bibitem{Faye:2014fra}
Guillaume Faye, Luc Blanchet, and Bala~R. Iyer.
\newblock {Non-linear multipole interactions and gravitational-wave octupole
  modes for inspiralling compact binaries to third-and-a-half post-Newtonian
  order}.
\newblock {\em Class. Quant. Grav.}, 32(4):045016, 2015, 1409.3546.

\bibitem{OShaughnessy:2014shr}
R.~O'Shaughnessy, Benjamin Farr, E.~Ochsner, Hee-Suk Cho, V.~Raymond, Chunglee
  Kim, and Chang-Hwan Lee.
\newblock {Parameter estimation of gravitational waves from precessing black
  hole-neutron star inspirals with higher harmonics}.
\newblock {\em Phys. Rev.}, D89(10):102005, 2014, 1403.0544.

\bibitem{Taracchini:2013wfa}
Andrea Taracchini, Alessandra Buonanno, Scott~A. Hughes, and Gaurav Khanna.
\newblock {Modeling the horizon-absorbed gravitational flux for
  equatorial-circular orbits in Kerr spacetime}.
\newblock {\em Phys. Rev.}, D88:044001, 2013, 1305.2184.
\newblock [Erratum: Phys. Rev.D88,no.10,109903(2013)].

\bibitem{Harms:2015ixa}
Enno Harms, Georgios Lukes-Gerakopoulos, Sebastiano Bernuzzi, and Alessandro
  Nagar.
\newblock {Asymptotic gravitational wave fluxes from a spinning particle in
  circular equatorial orbits around a rotating black hole}.
\newblock {\em Phys. Rev.}, D93(4):044015, 2016, 1510.05548.

\bibitem{Harms:2016ctx}
Enno Harms, Georgios Lukes-Gerakopoulos, Sebastiano Bernuzzi, and Alessandro
  Nagar.
\newblock {Spinning test body orbiting around a Schwarzschild black hole:
  Circular dynamics and gravitational-wave fluxes}.
\newblock {\em Phys. Rev.}, D94(10):104010, 2016, 1609.00356.

\bibitem{Abbott:2016izl}
B.~P. Abbott et~al.
\newblock {Improved analysis of GW150914 using a fully spin-precessing waveform
  Model}.
\newblock {\em Phys. Rev.}, X6(4):041014, 2016, 1606.01210.

\bibitem{Zackay:2019tzo}
Barak Zackay, Tejaswi Venumadhav, Liang Dai, Javier Roulet, and Matias
  Zaldarriaga.
\newblock {Highly spinning and aligned binary black hole merger in the Advanced
  LIGO first observing run}.
\newblock {\em Phys. Rev.}, D100(2):023007, 2019, 1902.10331.

\bibitem{Venumadhav:2019lyq}
Tejaswi Venumadhav, Barak Zackay, Javier Roulet, Liang Dai, and Matias
  Zaldarriaga.
\newblock {New Binary Black Hole Mergers in the Second Observing Run of
  Advanced LIGO and Advanced Virgo}.
\newblock 2019, 1904.07214.

\bibitem{Nitz:2019hdf}
Alexander~H. Nitz, Thomas Dent, Gareth~S. Davies, Sumit Kumar, Collin~D.
  Capano, Ian Harry, Simone Mozzon, Laura Nuttall, Andrew Lundgren, and Márton
  Tápai.
\newblock {2-OGC: Open Gravitational-wave Catalog of binary mergers from
  analysis of public Advanced LIGO and Virgo data}.
\newblock 2019, 1910.05331.

\bibitem{GBM:2017lvd}
B.~P. Abbott et~al.
\newblock {Multi-messenger Observations of a Binary Neutron Star Merger}.
\newblock {\em Astrophys. J.}, 848(2):L12, 2017, 1710.05833.

\bibitem{Monitor:2017mdv}
B.~P. Abbott et~al.
\newblock {Gravitational Waves and Gamma-rays from a Binary Neutron Star
  Merger: GW170817 and GRB 170817A}.
\newblock {\em Astrophys. J.}, 848(2):L13, 2017, 1710.05834.

\bibitem{Aasi:2013wya}
B.~P. Abbott et~al.
\newblock {Prospects for Observing and Localizing Gravitational-Wave Transients
  with Advanced LIGO, Advanced Virgo and KAGRA}.
\newblock {\em Living Rev. Rel.}, 21(1):3, 2018, 1304.0670.

\bibitem{Schnittman:2004vq}
Jeremy~D. Schnittman.
\newblock {Spin-orbit resonance and the evolution of compact binary systems}.
\newblock {\em Phys. Rev.}, D70:124020, 2004, astro-ph/0409174.

\bibitem{Gerosa:2015tea}
Davide Gerosa, Michael Kesden, Ulrich Sperhake, Emanuele Berti, and Richard
  O'Shaughnessy.
\newblock {Multi-timescale analysis of phase transitions in precessing
  black-hole binaries}.
\newblock {\em Phys. Rev.}, D92:064016, 2015, 1506.03492.

\bibitem{Schmidt:2010it}
Patricia Schmidt, Mark Hannam, Sascha Husa, and P.~Ajith.
\newblock {Tracking the precession of compact binaries from their
  gravitational-wave signal}.
\newblock {\em Phys. Rev.}, D84:024046, 2011, 1012.2879.

\bibitem{OShaughnessy:2012iol}
R.~O'Shaughnessy, L.~London, J.~Healy, and D.~Shoemaker.
\newblock {Precession during merger: Strong polarization changes are
  observationally accessible features of strong-field gravity during binary
  black hole merger}.
\newblock {\em Phys. Rev.}, D87(4):044038, 2013, 1209.3712.

\bibitem{Pekowsky:2013ska}
L.~Pekowsky, R.~O'Shaughnessy, J.~Healy, and D.~Shoemaker.
\newblock {Comparing gravitational waves from nonprecessing and precessing
  black hole binaries in the corotating frame}.
\newblock {\em Phys. Rev.}, D88(2):024040, 2013, 1304.3176.

\bibitem{Campanelli:2006fy}
Manuela Campanelli, Carlos~O. Lousto, Yosef Zlochower, Badri Krishnan, and
  David Merritt.
\newblock {Spin Flips and Precession in Black-Hole-Binary Mergers}.
\newblock {\em Phys. Rev.}, D75:064030, 2007, gr-qc/0612076.

\bibitem{Lousto:2015uwa}
Carlos~O. Lousto, James Healy, and Hiroyuki Nakano.
\newblock {Spin flips in generic black hole binaries}.
\newblock {\em Phys. Rev.}, D93(4):044031, 2016, 1506.04768.

\bibitem{Ossokine:2015vda}
Serguei Ossokine, Michael Boyle, Lawrence~E. Kidder, Harald~P. Pfeiffer,
  Mark~A. Scheel, and Béla Szilágyi.
\newblock {Comparing Post-Newtonian and Numerical-Relativity Precession
  Dynamics}.
\newblock {\em Phys. Rev.}, D92(10):104028, 2015, 1502.01747.

\bibitem{Lewis:2016lgx}
Adam G.~M. Lewis, Aaron Zimmerman, and Harald~P. Pfeiffer.
\newblock {Fundamental frequencies and resonances from eccentric and precessing
  binary black hole inspirals}.
\newblock {\em Class. Quant. Grav.}, 34(12):124001, 2017, 1611.03418.

\bibitem{Afle:2018slw}
Chaitanya Afle et~al.
\newblock {Detection and characterization of spin-orbit resonances in the
  advanced gravitational wave detectors era}.
\newblock {\em Phys. Rev.}, D98(8):083014, 2018, 1803.07695.

\bibitem{Pratten:2020fqn}
Geraint Pratten, Sascha Husa, Cecilio Garcia-Quiros, Marta Colleoni, Antoni
  Ramos-Buades, Hector Estelles, and Rafel Jaume.
\newblock {Setting the cornerstone for the IMRPhenomX family of models for
  gravitational waves from compact binaries: The dominant harmonic for
  non-precessing quasi-circular black holes}.
\newblock 2020, 2001.11412.

\bibitem{Estelles:2020c}
H\'ector Estell\'es et~al.
\newblock {\em in preparation}, 2020.

\bibitem{Boyle:2011gg}
Michael Boyle, Robert Owen, and Harald~P. Pfeiffer.
\newblock {A geometric approach to the precession of compact binaries}.
\newblock {\em Phys. Rev.}, D84:124011, 2011, 1110.2965.

\bibitem{O'Shaughnessy:2011fx}
R.~O'Shaughnessy, B.~Vaishnav, J.~Healy, Z.~Meeks, and D.~Shoemaker.
\newblock {Efficient asymptotic frame selection for binary black hole
  spacetimes using asymptotic radiation}.
\newblock {\em Phys. Rev.}, D84:124002, 2011, 1109.5224.

\bibitem{Schmidt:2012rh}
Patricia Schmidt, Mark Hannam, and Sascha Husa.
\newblock {Towards models of gravitational waveforms from generic binaries: A
  simple approximate mapping between precessing and non-precessing inspiral
  signals}.
\newblock {\em Phys. Rev.}, D86:104063, 2012, 1207.3088.

\bibitem{Huerta:2019oxn}
E.~A. Huerta et~al.
\newblock {Physics of eccentric binary black hole mergers: A numerical
  relativity perspective}.
\newblock {\em Phys. Rev.}, D100(6):064003, 2019, 1901.07038.

\bibitem{Hinder:2018fsy}
Ian Hinder, Serguei Ossokine, Harald~P. Pfeiffer, and Alessandra Buonanno.
\newblock {Gravitational waveforms for high spin and high mass-ratio binary
  black holes: A synergistic use of numerical-relativity codes}.
\newblock {\em Phys. Rev.}, D99(6):061501, 2019, 1810.10585.

\bibitem{Lovelace:2014twa}
Geoffrey Lovelace et~al.
\newblock {Nearly extremal apparent horizons in simulations of merging black
  holes}.
\newblock {\em Class. Quant. Grav.}, 32(6):065007, 2015, 1411.7297.

\bibitem{Szilagyi:2015rwa}
Béla Szilágyi, Jonathan Blackman, Alessandra Buonanno, Andrea Taracchini,
  Harald~P. Pfeiffer, Mark~A. Scheel, Tony Chu, Lawrence~E. Kidder, and Yi~Pan.
\newblock {Approaching the Post-Newtonian Regime with Numerical Relativity: A
  Compact-Object Binary Simulation Spanning 350 Gravitational-Wave Cycles}.
\newblock {\em Phys. Rev. Lett.}, 115(3):031102, 2015, 1502.04953.

\bibitem{Lovelace:2016uwp}
Geoffrey Lovelace et~al.
\newblock {Modeling the source of GW150914 with targeted numerical-relativity
  simulations}.
\newblock {\em Class. Quant. Grav.}, 33(24):244002, 2016, 1607.05377.

\bibitem{Foucart:2015gaa}
Francois Foucart, Roland Haas, Matthew~D. Duez, Evan O'Connor, Christian~D.
  Ott, Luke Roberts, Lawrence~E. Kidder, Jonas Lippuner, Harald~P. Pfeiffer,
  and Mark~A. Scheel.
\newblock {Low mass binary neutron star mergers : gravitational waves and
  neutrino emission}.
\newblock {\em Phys. Rev.}, D93(4):044019, 2016, 1510.06398.

\bibitem{Haas:2016cop}
Roland Haas et~al.
\newblock {Simulations of inspiraling and merging double neutron stars using
  the Spectral Einstein Code}.
\newblock {\em Phys. Rev.}, D93(12):124062, 2016, 1604.00782.

\bibitem{Foucart:2018lhe}
Francois Foucart et~al.
\newblock {Gravitational waveforms from spectral Einstein code simulations:
  Neutron star-neutron star and low-mass black hole-neutron star binaries}.
\newblock {\em Phys. Rev.}, D99(4):044008, 2019, 1812.06988.

\bibitem{Vincent:2019kor}
Trevor Vincent, Francois Foucart, Matthew~D. Duez, Roland Haas, Lawrence~E.
  Kidder, Harald~P. Pfeiffer, and Mark~A. Scheel.
\newblock {Unequal Mass Binary Neutron Star Simulations with Neutrino
  Transport: Ejecta and Neutrino Emission}.
\newblock {\em Phys. Rev.}, D101(4):044053, 2020, 1908.00655.

\bibitem{Okounkova:2018abo}
Maria Okounkova, Mark~A. Scheel, and Saul~A. Teukolsky.
\newblock {Numerical black hole initial data and shadows in dynamical
  Chern–Simons gravity}.
\newblock {\em Class. Quant. Grav.}, 36(5):054001, 2019, 1810.05306.

\bibitem{Okounkova:2019dfo}
Maria Okounkova, Leo~C. Stein, Mark~A. Scheel, and Saul~A. Teukolsky.
\newblock {Numerical binary black hole collisions in dynamical Chern-Simons
  gravity}.
\newblock {\em Phys. Rev.}, D100(10):104026, 2019, 1906.08789.

\bibitem{Okounkova:2019zjf}
Maria Okounkova, Leo~C. Stein, Jordan Moxon, Mark~A. Scheel, and Saul~A.
  Teukolsky.
\newblock {Numerical relativity simulation of GW150914 beyond general
  relativity}.
\newblock 2019, 1911.02588.

\bibitem{Okounkova:2020rqw}
Maria Okounkova.
\newblock {Numerical relativity simulation of GW150914 in Einstein dilaton
  Gauss-Bonnet gravity}.
\newblock 2020, 2001.03571.

\bibitem{Lindblom:2005qh}
Lee Lindblom, Mark~A. Scheel, Lawrence~E. Kidder, Robert Owen, and Oliver
  Rinne.
\newblock {A New generalized harmonic evolution system}.
\newblock {\em Class. Quant. Grav.}, 23:S447--S462, 2006, gr-qc/0512093.

\bibitem{Lindblom:2009tu}
Lee Lindblom and Bela Szilagyi.
\newblock {An Improved Gauge Driver for the GH Einstein System}.
\newblock {\em Phys. Rev.}, D80:084019, 2009, 0904.4873.

\bibitem{Szilagyi:2009qz}
Bela Szilagyi, Lee Lindblom, and Mark~A. Scheel.
\newblock {Simulations of Binary Black Hole Mergers Using Spectral Methods}.
\newblock {\em Phys. Rev.}, D80:124010, 2009, 0909.3557.

\bibitem{Hemberger:2012jz}
Daniel~A. Hemberger, Mark~A. Scheel, Lawrence~E. Kidder, Béla Szilágyi,
  Geoffrey Lovelace, Nicholas~W. Taylor, and Saul~A. Teukolsky.
\newblock {Dynamical Excision Boundaries in Spectral Evolutions of Binary Black
  Hole Spacetimes}.
\newblock {\em Class. Quant. Grav.}, 30:115001, 2013, 1211.6079.

\bibitem{Buonanno:2010yk}
Alessandra Buonanno, Lawrence~E. Kidder, Abdul~H. Mroue, Harald~P. Pfeiffer,
  and Andrea Taracchini.
\newblock {Reducing orbital eccentricity of precessing black-hole binaries}.
\newblock {\em Phys. Rev.}, D83:104034, 2011, 1012.1549.

\bibitem{Boyle:2014ioa}
Michael Boyle, Lawrence~E. Kidder, Serguei Ossokine, and Harald~P. Pfeiffer.
\newblock {Gravitational-wave modes from precessing black-hole binaries}.
\newblock 2014, 1409.4431.

\bibitem{Boyle:2013nka}
Michael Boyle.
\newblock {Angular velocity of gravitational radiation from precessing binaries
  and the corotating frame}.
\newblock {\em Phys. Rev.}, D87(10):104006, 2013, 1302.2919.

\bibitem{LAL}
\url{https://www.lsc-group.phys.uwm.edu/daswg/projects/lalsuite.html}.

\bibitem{Damour:2000we}
Thibault Damour, Piotr Jaranowski, and Gerhard Schaefer.
\newblock {On the determination of the last stable orbit for circular general
  relativistic binaries at the third postNewtonian approximation}.
\newblock {\em Phys. Rev.}, D62:084011, 2000, gr-qc/0005034.

\bibitem{Hinderer:2017jcs}
Tanja Hinderer and Stanislav Babak.
\newblock {Foundations of an effective-one-body model for coalescing binaries
  on eccentric orbits}.
\newblock {\em Phys. Rev.}, D96(10):104048, 2017, 1707.08426.

\bibitem{Liu:2019jpg}
Xiaolin Liu, Zhoujian Cao, and Lijing Shao.
\newblock {Validating the Effective-One-Body Numerical-Relativity Waveform
  Models for Spin-aligned Binary Black Holes along Eccentric Orbits}.
\newblock {\em Phys. Rev.}, D101(4):044049, 2020, 1910.00784.

\bibitem{Chiaramello:2020ehz}
Danilo Chiaramello and Alessandro Nagar.
\newblock {A faithful analytical effective one body waveform model for
  spin-aligned, moderately eccentric, coalescing black hole binaries}.
\newblock 2020, 2001.11736.

\bibitem{Bernuzzi:2011aj}
Sebastiano Bernuzzi, Alessandro Nagar, and Anil Zenginoglu.
\newblock {Binary black hole coalescence in the large-mass-ratio limit: the
  hyperboloidal layer method and waveforms at null infinity}.
\newblock {\em Phys. Rev.}, D84:084026, 2011, 1107.5402.

\bibitem{Ossokine:2020}
Serguei Ossokine.
\newblock Private communication.
\newblock {\em Private Communication}, 2020.

\bibitem{Varma:2018aht}
Vijay Varma, Davide Gerosa, Leo~C. Stein, François Hébert, and Hao Zhang.
\newblock {High-accuracy mass, spin, and recoil predictions of generic
  black-hole merger remnants}.
\newblock {\em Phys. Rev. Lett.}, 122(1):011101, 2019, 1809.09125.

\bibitem{Bardeen:1972fi}
James~M. Bardeen, William~H. Press, and Saul~A Teukolsky.
\newblock {Rotating black holes: Locally nonrotating frames, energy extraction,
  and scalar synchrotron radiation}.
\newblock {\em Astrophys. J.}, 178:347, 1972.

\bibitem{Damour:2009sm}
Thibault Damour.
\newblock {Gravitational Self Force in a Schwarzschild Background and the
  Effective One Body Formalism}.
\newblock {\em Phys.\ Rev.\ D}, 81:024017, 2010, 0910.5533.

\bibitem{Bini:2018ylh}
Donato Bini, Thibault Damour, Andrea Geralico, Chris Kavanagh, and Maarten
  van~de Meent.
\newblock {Gravitational self-force corrections to gyroscope precession along
  circular orbits in the Kerr spacetime}.
\newblock {\em Phys.\ Rev.\ D}, 98(10):104062, 2018, 1809.02516.

\bibitem{Rettegno:2019tzh}
Piero Rettegno, Fabio Martinetti, Alessandro Nagar, Donato Bini, Gunnar
  Riemenschneider, and Thibault Damour.
\newblock {Comparing Effective One Body Hamiltonians for spin-aligned
  coalescing binaries}.
\newblock 11 2019, 1911.10818.

\bibitem{Khalil:2020mmr}
Mohammed Khalil, Jan Steinhoff, Justin Vines, and Alessandra Buonanno.
\newblock {Fourth post-Newtonian effective-one-body Hamiltonians with generic
  spins}.
\newblock 3 2020, 2003.04469.

\bibitem{alex_nitz_2020_3630601}
Alex Nitz, Ian Harry, Duncan Brown, Christopher~M. Biwer, Josh Willis, Tito~Dal
  Canton, Collin Capano, Larne Pekowsky, Thomas Dent, Andrew~R. Williamson,
  Soumi De, Gareth Davies, Miriam Cabero, Duncan Macleod, Bernd Machenschalk,
  Steven Reyes, Prayush Kumar, Thomas Massinger, Francesco Pannarale, dfinstad,
  Márton Tápai, Stephen Fairhurst, Sebastian Khan, Leo Singer, Sumit Kumar,
  Alex Nielsen, shasvath, idorrington92, Amber Lenon, and Hunter Gabbard.
\newblock gwastro/pycbc: Pycbc release v1.15.4, January 2020.

\bibitem{Smith:2019ucc}
Rory Smith and Gregory Ashton.
\newblock {Expediting Astrophysical Discovery with Gravitational-Wave
  Transients Through Massively Parallel Nested Sampling}.
\newblock 9 2019, 1909.11873.

\bibitem{Mehta:2019wxm}
Ajit Kumar~Mehta, Praveer Tiwari, Nathan~K. Johnson-McDaniel, Chandra~Kant
  Mishra, Vijay Varma, and Parameswaran Ajith.
\newblock {Including mode mixing in a higher-multipole model for gravitational
  waveforms from nonspinning black-hole binaries}.
\newblock {\em Phys.\ Rev.\ D}, 100(2):024032, 2019, 1902.02731.

\bibitem{Gadre:2020a}
Bhooshan Gadre et~al.
\newblock {\em in preparation}, 2020.

\bibitem{GWFrames}
\url{https://github.com/moble/GWFrames}.

\bibitem{gracedb-O3-superevents}
LIGO~Scientific Collaboration and Virgo Collaboration.
\newblock Gracedb — gravitational-wave candidate event database, ligo/virgo
  o3 public alerts.

\bibitem{Damour:2014sva}
Thibault Damour and Alessandro Nagar.
\newblock {New effective-one-body description of coalescing nonprecessing
  spinning black-hole binaries}.
\newblock {\em Phys. Rev.}, D90(4):044018, 2014, 1406.6913.

\bibitem{Nagar:2018zoe}
Alessandro Nagar et~al.
\newblock {Time-domain effective-one-body gravitational waveforms for
  coalescing compact binaries with nonprecessing spins, tides and self-spin
  effects}.
\newblock {\em Phys. Rev.}, D98(10):104052, 2018, 1806.01772.

\bibitem{Ajith:2011}
P.~Ajith et~al.
\newblock Inspiral-merger-ringdown waveforms for black-hole binaries with
  nonprecessing spins.
\newblock {\em Phys. Rev. Lett.}, 106:241101, Jun 2011, 0909.2867.

\bibitem{Centrella:2010mx}
Joan Centrella, John~G. Baker, Bernard~J. Kelly, and James~R. van Meter.
\newblock {Black-hole binaries, gravitational waves, and numerical relativity}.
\newblock {\em Rev.Mod.Phys.}, 82:3069, 2010, 1010.5260.

\bibitem{Gadre:2020}
Bhooshan Gadre, Michael Pürrer, Scott Field, and Sergei Ossokine.
\newblock {A fully precessing surrogate model of effective-one-body waveforms}.
\newblock {\em in prep}, 2020.

\bibitem{Huerta:2017kez}
E.~A. Huerta et~al.
\newblock {Eccentric, nonspinning, inspiral, Gaussian-process merger
  approximant for the detection and characterization of eccentric binary black
  hole mergers}.
\newblock {\em Phys. Rev.}, D97(2):024031, 2018, 1711.06276.

\bibitem{Hinder:2017sxy}
Ian Hinder, Lawrence~E. Kidder, and Harald~P. Pfeiffer.
\newblock {Eccentric binary black hole inspiral-merger-ringdown gravitational
  waveform model from numerical relativity and post-Newtonian theory}.
\newblock {\em Phys. Rev.}, D98(4):044015, 2018, 1709.02007.

\bibitem{Cao:2017ndf}
Zhoujian Cao and Wen-Biao Han.
\newblock {Waveform model for an eccentric binary black hole based on the
  effective-one-body-numerical-relativity formalism}.
\newblock {\em Phys. Rev.}, D96(4):044028, 2017, 1708.00166.

\bibitem{PhysRevD.79.064004}
Thibault Damour, Bala~R. Iyer, and Alessandro Nagar.
\newblock Improved resummation of post-newtonian multipolar waveforms from
  circularized compact binaries.
\newblock {\em Phys. Rev. D}, 79:064004, Mar 2009.

\bibitem{PhysRevD.76.064028}
Thibault Damour and Alessandro Nagar.
\newblock Faithful effective-one-body waveforms of small-mass-ratio coalescing
  black hole binaries.
\newblock {\em Phys. Rev. D}, 76:064028, Sep 2007.

\bibitem{Nagar:2020pcj}
Alessandro Nagar, Gunnar Riemenschneider, Geraint Pratten, Piero Rettegno, and
  Francesco Messina.
\newblock {A multipolar effective one body waveform model for spin-aligned
  black hole binaries}.
\newblock 2020, 2001.09082.

\bibitem{lrr-2006-4}
Luc Blanchet.
\newblock Gravitational radiation from post-newtonian sources and inspiralling
  compact binaries.
\newblock {\em Living Reviews in Relativity}, 9(4), 2006.

\bibitem{LALSuiteGit}
\mbox{LIGO Scientific Collaboration, Virgo Collaboration}.
\newblock \mbox{LALSuite}.
\newblock
  \href{https://git.ligo.org/lscsoft/lalsuite}{https://git.ligo.org/lscsoft/lalsuite},
  2018.

\bibitem{GolubVanLoan}
Gene~H. Golub and Charles F.~Van Loan.
\newblock {\em Matrix Computations (3rd ed.)}.
\newblock Johns Hopkins, 1996.

\bibitem{Demmel}
James~W. Demmel.
\newblock {\em Applied Numerical Linear Algebra}.
\newblock SIAM, 1997.

\bibitem{githubTPI}
Michael Pürrer and Jonathan Blackman.
\newblock \mbox{TPI - Tensor Product Interpolation Package for Python}, 2018.

\bibitem{deBoor}
Carl de~Boor.
\newblock {\em A Practical Guide to Splines}.
\newblock Springer, 2001.

\bibitem{LALInference}
LIGO~Scientific Collaboration.
\newblock The lalinference package of the lal software suite is available from.
\newblock \url{https://wiki.ligo.org/DASWG/LALSuite}.

\bibitem{GWOSC:GWTC}
{LIGO Scientific Collaboration, Virgo Collaboration}.
\newblock {GWTC-1}.
\newblock
  \href{https://doi.org/10.7935/82H3-HH23}{https://doi.org/10.7935/82H3-HH23},
  2018.

\bibitem{Messick:2016aqy}
Cody Messick et~al.
\newblock {Analysis Framework for the Prompt Discovery of Compact Binary
  Mergers in Gravitational-wave Data}.
\newblock {\em Phys. Rev. D}, 95(4):042001, 2017, 1604.04324.

\bibitem{Nitz_2017}
Alexander~H. Nitz, Thomas Dent, Tito~Dal Canton, Stephen Fairhurst, and
  Duncan~A. Brown.
\newblock Detecting binary compact-object mergers with gravitational waves:
  Understanding and improving the sensitivity of the {PyCBC} search.
\newblock {\em The Astrophysical Journal}, 849(2):118, nov 2017.

\bibitem{Klimenko:2015ypf}
S.~Klimenko et~al.
\newblock {Method for detection and reconstruction of gravitational wave
  transients with networks of advanced detectors}.
\newblock {\em Phys. Rev. D}, 93(4):042004, 2016, 1511.05999.

\bibitem{sharm}
Goldberg et~al.
\newblock {\em J. Math. Phys. 8:2155--2161}, 1967.

\bibitem{VanDenBroeck:2006ar}
Chris Van Den~Broeck and Anand~S. Sengupta.
\newblock {Binary black hole spectroscopy}.
\newblock {\em Class. Quant. Grav.}, 24:1089--1114, 2007, gr-qc/0610126.

\bibitem{2018arXiv181205121M}
Debnandini {Mukherjee}, Sarah {Caudill}, et~al.
\newblock {The GstLAL template bank for spinning compact binary mergers in the
  second observation run of Advanced LIGO and Virgo}.
\newblock {\em arXiv e-prints}, page arXiv:1812.05121, December 2018,
  1812.05121.

\bibitem{CalderonBustillo:2017skv}
Juan Calder{\'o}n~Bustillo, Francesco Salemi, Tito Dal~Canton, and Karan~P.
  Jani.
\newblock {Sensitivity of gravitational wave searches to the full signal of
  intermediate-mass black hole binaries during the first observing run of
  Advanced LIGO}.
\newblock {\em Phys. Rev.}, D97(2):024016, 2018, 1711.02009.

\bibitem{Abbott:2016apu}
B.~P. Abbott et~al.
\newblock {Directly comparing GW150914 with numerical solutions of Einstein's
  equations for binary black hole coalescence}.
\newblock {\em Phys. Rev.}, D94:064035, 2016, 1606.01262.

\bibitem{CalderonBustillo:2018zuq}
Juan Calder{\'o}n~Bustillo, James~A. Clark, Pablo Laguna, and Deirdre
  Shoemaker.
\newblock {Tracking black hole kicks from gravitational wave observations}.
\newblock {\em Phys. Rev. Lett.}, 121(19):191102, 2018, 1806.11160.

\bibitem{2013PhRvL.111x1104M}
A.~H. {Mrou{\'e}}, M.~A. {Scheel}, B.~{Szil{\'a}gyi}, H.~P. {Pfeiffer},
  M.~{Boyle}, D.~A. {Hemberger}, L.~E. {Kidder}, G.~{Lovelace}, S.~{Ossokine},
  N.~W. {Taylor}, A.~{Zengino{\u g}lu}, L.~T. {Buchman}, T.~{Chu}, E.~{Foley},
  M.~{Giesler}, R.~{Owen}, and S.~A. {Teukolsky}.
\newblock {Catalog of 174 Binary Black Hole Simulations for Gravitational Wave
  Astronomy}.
\newblock {\em Physical Review Letters}, 111(24):241104, December 2013,
  1304.6077.

\bibitem{2016CQGra..33t4001J}
K.~{Jani}, J.~{Healy}, J.~A. {Clark}, L.~{London}, P.~{Laguna}, and
  D.~{Shoemaker}.
\newblock {Georgia tech catalog of gravitational waveforms}.
\newblock {\em Classical and Quantum Gravity}, 33(20):204001, October 2016,
  1605.03204.

\bibitem{2017CQGra..34v4001H}
J.~{Healy}, C.~O. {Lousto}, Y.~{Zlochower}, and M.~{Campanelli}.
\newblock {The RIT binary black hole simulations catalog}.
\newblock {\em Classical and Quantum Gravity}, 34(22):224001, November 2017,
  1703.03423.

\bibitem{2015PhRvL.115l1102B}
J.~{Blackman}, S.~E. {Field}, C.~R. {Galley}, B.~{Szil{\'a}gyi}, M.~A.
  {Scheel}, M.~{Tiglio}, and D.~A. {Hemberger}.
\newblock {Fast and Accurate Prediction of Numerical Relativity Waveforms from
  Binary Black Hole Coalescences Using Surrogate Models}.
\newblock {\em Physical Review Letters}, 115(12):121102, September 2015,
  1502.07758.

\bibitem{2017PhRvD..95j4023B}
J.~{Blackman}, S.~E. {Field}, M.~A. {Scheel}, C.~R. {Galley}, D.~A.
  {Hemberger}, P.~{Schmidt}, and R.~{Smith}.
\newblock {A Surrogate model of gravitational waveforms from numerical
  relativity simulations of precessing binary black hole mergers}.
\newblock {\em Physical Review D}, 95(10):104023, May 2017, 1701.00550.

\bibitem{2017PhRvD..96b4058B}
J.~{Blackman}, S.~E. {Field}, M.~A. {Scheel}, C.~R. {Galley}, C.~D. {Ott},
  M.~{Boyle}, L.~E. {Kidder}, H.~P. {Pfeiffer}, and B.~{Szil{\'a}gyi}.
\newblock {Numerical relativity waveform surrogate model for generically
  precessing binary black hole mergers}.
\newblock {\em Physical Review D}, 96(2):024058, July 2017, 1705.07089.

\bibitem{Woosley_2017}
S.~E. Woosley.
\newblock Pulsational pair-instability supernovae.
\newblock {\em The Astrophysical Journal}, 836(2):244, feb 2017.

\bibitem{Marchant:2018kun}
Pablo Marchant, Mathieu Renzo, Robert Farmer, Kaliroe M.~W. Pappas, Ronald~E.
  Taam, Selma de~Mink, and Vassiliki Kalogera.
\newblock {Pulsational pair-instability supernovae in very close binaries}.
\newblock 2018, 1810.13412.

\bibitem{2002ApJ...567..532H}
A.~{Heger} and S.~E. {Woosley}.
\newblock {The Nucleosynthetic Signature of Population III}.
\newblock {\em Astrophysical Journal}, 567:532--543, March 2002,
  astro-ph/0107037.

\bibitem{davide2g}
Davide {Gerosa} and Emanuele {Berti}.
\newblock {Are merging black holes born from stellar collapse or previous
  mergers?}
\newblock {\em Physical Review D}, 95:124046, June 2017, 1703.06223.

\bibitem{Kimball:2019mfs}
Chase Kimball, Christopher P.~L. Berry, and Vicky Kalogera.
\newblock {What GW170729's exceptional mass and spin tells us about its family
  tree}.
\newblock 2019, 1903.07813.

\bibitem{Cornish:2014kda}
Neil~J. Cornish and Tyson~B. Littenberg.
\newblock {BayesWave: Bayesian Inference for Gravitational Wave Bursts and
  Instrument Glitches}.
\newblock {\em Class. Quant. Grav.}, 32(13):135012, 2015, 1410.3835.

\bibitem{TheLIGOScientific:2016uux}
B.~P. Abbott et~al.
\newblock {Observing gravitational-wave transient GW150914 with minimal
  assumptions}.
\newblock {\em Phys. Rev.}, D93(12):122004, 2016, 1602.03843.
\newblock [Addendum: Phys. Rev.D94,no.6,069903(2016)].

\bibitem{170729SamplesRelease}
K.~Chatziioannou, R.~Cotesta, S.~Ghonge, J.~Lange, K.~Ng, J.~C. Bustillo,
  J.~Clark, C.-J. Haster, S.~Khan, P{\"u}rrer M., V~Raymond, S.~Vitale, et~al.
\newblock {GW170729 Sample Release}, 2019.

\bibitem{GWOSC}
{Gravitational Wave Open Science Center (GWOSC)}.
\newblock \url{https://www.gw-openscience.org/}.

\bibitem{Vallisneri:2014vxa}
Michele Vallisneri, Jonah Kanner, Roy Williams, Alan Weinstein, and Branson
  Stephens.
\newblock {The LIGO Open Science Center}.
\newblock {\em J. Phys. Conf. Ser.}, 610(1):012021, 2015, 1410.4839.

\bibitem{Driggers:2018gii}
J.~C. Driggers et~al.
\newblock {Improving astrophysical parameter estimation via offline noise
  subtraction for Advanced LIGO}.
\newblock 2018, 1806.00532.

\bibitem{Davis:2018yrz}
D.~Davis, T.~J. Massinger, A.~P. Lundgren, J.~C. Driggers, A.~L. Urban, and
  L.~K. Nuttall.
\newblock {Improving the Sensitivity of Advanced LIGO Using Noise Subtraction}.
\newblock 2018, 1809.05348.

\bibitem{Littenberg:2014oda}
Tyson~B. Littenberg and Neil~J. Cornish.
\newblock {Bayesian inference for spectral estimation of gravitational wave
  detector noise}.
\newblock {\em Phys. Rev.}, D91(8):084034, 2015, 1410.3852.

\bibitem{GW170729PSD}
{Power Spectral Densities (PSD) release for GWTC-1}.
\newblock \url{https://dcc.ligo.org/LIGO-P1900011/public}.

\bibitem{2014PhRvD..89f1502T}
A.~{Taracchini}, A.~{Buonanno}, Y.~{Pan}, T.~{Hinderer}, M.~{Boyle}, D.~A.
  {Hemberger}, L.~E. {Kidder}, G.~{Lovelace}, A.~H. {Mrou{\'e}}, H.~P.
  {Pfeiffer}, M.~A. {Scheel}, B.~{Szil{\'a}gyi}, N.~W. {Taylor}, and
  A.~{Zenginoglu}.
\newblock {Effective-one-body model for black-hole binaries with generic mass
  ratios and spins}.
\newblock {\em Physical Review D}, 89(6):061502, March 2014, 1311.2544.

\bibitem{2014PhRvD..89h4006P}
Y.~{Pan}, A.~{Buonanno}, A.~{Taracchini}, L.~E. {Kidder}, A.~H. {Mrou{\'e}},
  H.~P. {Pfeiffer}, M.~A. {Scheel}, and B.~{Szil{\'a}gyi}.
\newblock {Inspiral-merger-ringdown waveforms of spinning, precessing
  black-hole binaries in the effective-one-body formalism}.
\newblock {\em Physical Review D}, 89(8):084006, April 2014, 1307.6232.

\bibitem{Millhouse:2018dgi}
Margaret Millhouse, Neil~J. Cornish, and Tyson Littenberg.
\newblock {Bayesian reconstruction of gravitational wave bursts using
  chirplets}.
\newblock {\em Phys. Rev.}, D97(10):104057, 2018, 1804.03239.

\bibitem{Blanchet:2014zz}
Luc Blanchet.
\newblock {Gravitational Radiation from Post-Newtonian Sources and Inspiralling
  Compact Binaries}.
\newblock {\em Living Rev. Rel.}, 17:2, 2014, 1310.1528.

\bibitem{2010RvMP...82.3069C}
J.~{Centrella}, J.~G. {Baker}, B.~J. {Kelly}, and J.~R. {van Meter}.
\newblock {Black-hole binaries, gravitational waves, and numerical relativity}.
\newblock {\em Reviews of Modern Physics}, 82:3069--3119, October 2010,
  1010.5260.

\bibitem{GW170729Samples}
{Parameter estimation sample release for GWTC-1}.
\newblock \url{https://dcc.ligo.org/LIGO-P1800370/public}.

\bibitem{salvoprior}
Salvatore {Vitale}, Davide {Gerosa}, Carl-Johan {Haster}, Katerina
  {Chatziioannou}, and Aaron {Zimmerman}.
\newblock {Impact of Bayesian Priors on the Characterization of Binary Black
  Hole Coalescences}.
\newblock {\em Physical Review Letters}, 119:251103, December 2017, 1707.04637.

\bibitem{lalinference_o2}
{LIGO Scientific Collaboration and Virgo Collaboration}.
\newblock {LALSuite},
  https://git.ligo.org/lscsoft/lalsuite/tree/lalinference\_o2, 2017.

\bibitem{2015PhRvD..92b3002P}
C.~{Pankow}, P.~{Brady}, E.~{Ochsner}, and R.~{O'Shaughnessy}.
\newblock {Novel scheme for rapid parallel parameter estimation of
  gravitational waves from compact binary coalescences}.
\newblock {\em Physical Review D}, 92(2):023002, July 2015, 1502.04370.

\bibitem{2018arXiv180510457L}
J.~{Lange}, R.~{O'Shaughnessy}, and M.~{Rizzo}.
\newblock {Rapid and accurate parameter inference for coalescing, precessing
  compact binaries}.
\newblock {\em arXiv e-prints}, May 2018, 1805.10457.

\bibitem{2019arXiv190204934W}
D.~{Wysocki}, R.~{O'Shaughnessy}, Y.~L. {Fang}, and J.~{Lange}.
\newblock {Accelerating parameter inference with graphics processing units}.
\newblock {\em arXiv e-prints}, February 2019, 1902.04934.

\bibitem{GW170729CalEnv}
{Calibration uncertainty envelope release for GWTC-1}.
\newblock \url{https://dcc.ligo.org/LIGO-P1900040/public}.

\bibitem{SplineCalMarg-T1400682}
Will~M Farr, Ben Farr, and Tyson Littenberg.
\newblock {Modelling Calibration Errors In CBC Waveforms}.
\newblock
  \href{https://dcc.ligo.org/T1400682/public}{https://dcc.ligo.orgT1400682/public},
  2015.
\newblock {Tech. Rep. {LIGO}-T1400682 (LIGO Scientific Collaboration and Virgo
  Collaboration)}.

\bibitem{Vitale:2011wu}
Salvatore Vitale, Walter Del~Pozzo, Tjonnie G.~F. Li, Chris Van Den~Broeck,
  Ilya Mandel, Ben Aylott, and John Veitch.
\newblock {Effect of calibration errors on Bayesian parameter estimation for
  gravitational wave signals from inspiral binary systems in the Advanced
  Detectors era}.
\newblock {\em Phys. Rev.}, D85:064034, 2012, 1111.3044.

\bibitem{bayeswave}
{LIGO Scientific Collaboration and Virgo Collaboration}.
\newblock {BayesWave}, https://git.ligo.org/lscsoft/bayeswave, 2018.

\bibitem{10.1093/biomet/82.4.711}
PETER~J. GREEN.
\newblock {Reversible jump Markov chain Monte Carlo computation and Bayesian
  model determination}.
\newblock {\em Biometrika}, 82(4):711--732, 12 1995,
  http://oup.prod.sis.lan/biomet/article-pdf/82/4/711/699533/82-4-711.pdf.

\bibitem{Apostolatos:1995pj}
T.A. Apostolatos.
\newblock {Search templates for gravitational waves from precessing,
  inspiraling binaries}.
\newblock {\em Phys.Rev.}, D52:605--620, 1995.

\bibitem{Ade:2015xua}
P.~A.~R. Ade et~al.
\newblock {Planck 2015 results. XIII. Cosmological parameters}.
\newblock {\em Astron. Astrophys.}, 594:A13, 2016, 1502.01589.

\bibitem{Ng:2018neg}
Ken K.~Y. Ng, Salvatore Vitale, Aaron Zimmerman, Katerina Chatziioannou, Davide
  Gerosa, and Carl-Johan Haster.
\newblock {Gravitational-wave astrophysics with effective-spin measurements:
  asymmetries and selection biases}.
\newblock {\em Phys. Rev.}, D98(8):083007, 2018, 1805.03046.

\bibitem{DiCarlo:2019pmf}
Ugo~N. Di~Carlo, Nicola Giacobbo, Michela Mapelli, Mario Pasquato, Mario Spera,
  Long Wang, and Francesco Haardt.
\newblock {Merging black holes in young star clusters}.
\newblock 2019, 1901.00863.

\bibitem{Gonzalez:2006md}
Jose~A. Gonzalez, Ulrich Sperhake, Bernd Bruegmann, Mark Hannam, and Sascha
  Husa.
\newblock {Total recoil: The Maximum kick from nonspinning black-hole binary
  inspiral}.
\newblock {\em Phys. Rev. Lett.}, 98:091101, 2007, gr-qc/0610154.

\bibitem{Buonanno:2007sv}
Alessandra Buonanno, Lawrence~E. Kidder, and Luis Lehner.
\newblock {Estimating the final spin of a binary black hole coalescence}.
\newblock {\em Phys. Rev.}, D77:026004, 2008, 0709.3839.

\bibitem{Campanelli:2006gf}
Manuela Campanelli, C.~O. Lousto, and Y.~Zlochower.
\newblock {The Last orbit of binary black holes}.
\newblock {\em Phys. Rev.}, D73:061501, 2006, gr-qc/0601091.

\bibitem{Baker:2003ds}
John~G. Baker, Manuela Campanelli, Carlos~O. Lousto, and R.~Takahashi.
\newblock {Coalescence remnant of spinning binary black holes}.
\newblock {\em Phys. Rev.}, D69:027505, 2004, astro-ph/0305287.

\bibitem{Bouffanais:2019nrw}
Yann Bouffanais, Michela Mapelli, Davide Gerosa, Ugo~N. Di~Carlo, Nicola
  Giacobbo, Emanuele Berti, and Vishal Baibhav.
\newblock {Constraining the fraction of binary black holes formed in isolation
  and young star clusters with gravitational-wave data}.
\newblock 2019, 1905.11054.

\bibitem{kullback1951}
S.~Kullback and R.~A. Leibler.
\newblock On information and sufficiency.
\newblock {\em Ann. Math. Statist.}, 22(1):79--86, 03 1951.

\bibitem{Hunter:2007}
J.~D. Hunter.
\newblock Matplotlib: A 2d graphics environment.
\newblock {\em Computing In Science \& Engineering}, 9(3):90--95, 2007.

\bibitem{corner}
Daniel Foreman-Mackey.
\newblock corner.py: Scatterplot matrices in python.
\newblock {\em The Journal of Open Source Software}, 24, 2016.

\bibitem{Schutz:1986gp}
Bernard~F. Schutz.
\newblock {Determining the Hubble Constant from Gravitational Wave
  Observations}.
\newblock {\em Nature}, 323:310--311, 1986.

\bibitem{Abbott:2019yzh}
B.P. Abbott et~al.
\newblock {A gravitational-wave measurement of the Hubble constant following
  the second observing run of Advanced LIGO and Virgo}.
\newblock 8 2019, 1908.06060.

\bibitem{Soares-Santos:2019irc}
M.~Soares-Santos et~al.
\newblock {First Measurement of the Hubble Constant from a Dark Standard Siren
  using the Dark Energy Survey Galaxies and the LIGO/Virgo
  Binary\textendash{}Black-hole Merger GW170814}.
\newblock {\em Astrophys. J. Lett.}, 876(1):L7, 2019, 1901.01540.

\bibitem{Fishbach:2018gjp}
M.~Fishbach et~al.
\newblock {A Standard Siren Measurement of the Hubble Constant from GW170817
  without the Electromagnetic Counterpart}.
\newblock {\em Astrophys. J. Lett.}, 871(1):L13, 2019, 1807.05667.

\bibitem{Abbott:2017xzu}
B.P. Abbott et~al.
\newblock {A gravitational-wave standard siren measurement of the Hubble
  constant}.
\newblock {\em Nature}, 551(7678):85--88, 2017, 1710.05835.

\bibitem{Borhanian:2020vyr}
Ssohrab Borhanian, Arnab Dhani, Anuradha Gupta, K.G. Arun, and B.S.
  Sathyaprakash.
\newblock {Dark Sirens to Resolve the
  Hubble-Lema\textbackslash{}\textasciicircum{}itre Tension}.
\newblock 7 2020, 2007.02883.

\bibitem{Abbott:2020jks}
R.~Abbott et~al.
\newblock {Tests of General Relativity with Binary Black Holes from the second
  LIGO-Virgo Gravitational-Wave Transient Catalog}.
\newblock 10 2020, 2010.14529.

\bibitem{Abbott:2020mjq}
R.~Abbott et~al.
\newblock {Properties and Astrophysical Implications of the 150 M$_\odot$
  Binary Black Hole Merger GW190521}.
\newblock {\em Astrophys. J. Lett.}, 900(1):L13, 2020, 2009.01190.

\bibitem{Aasi:2014iwa}
J.~Aasi et~al.
\newblock {Search for gravitational radiation from intermediate mass black hole
  binaries in data from the second LIGO-Virgo joint science run}.
\newblock {\em Phys. Rev.}, D89(12):122003, 2014, 1404.2199.

\bibitem{Aasi:2014bqj}
J.~Aasi et~al.
\newblock {Search for gravitational wave ringdowns from perturbed intermediate
  mass black holes in LIGO-Virgo data from 2005--2010}.
\newblock {\em Phys. Rev.}, D89(10):102006, 2014, 1403.5306.

\bibitem{Aasi:2012rja}
J.~Aasi et~al.
\newblock {Search for gravitational waves from binary black hole inspiral,
  merger, and ringdown in LIGO-Virgo data from 2009--2010}.
\newblock {\em Phys. Rev.}, D87(2):022002, 2013, 1209.6533.

\bibitem{Virgo:2012aa}
J.~Abadie et~al.
\newblock {Search for Gravitational Waves from Intermediate Mass Binary Black
  Holes}.
\newblock {\em Phys. Rev.}, D85:102004, 2012, 1201.5999.

\bibitem{Schmidt:2017btt}
Patricia Schmidt, Ian~W. Harry, and Harald~P. Pfeiffer.
\newblock {Numerical Relativity Injection Infrastructure}.
\newblock 2017, 1703.01076.

\bibitem{Lange:2018}
R.~O'Shaughnessy and J.~Lange.
\newblock {Technical document on 170729 RIFT analysis stability}.
\newblock
  \href{https://dcc.ligo.org/T1900096/public}{https://dcc.ligo.org/T1900096/public},
  2018.
\newblock {Tech. Rep. {LIGO}-T1900096}.

\bibitem{Abbott:2018wiz}
B.~P. Abbott et~al.
\newblock {Properties of the binary neutron star merger GW170817}.
\newblock {\em Phys. Rev.}, X9(1):011001, 2019, 1805.11579.

\end{thebibliography}
\end{document}